\documentclass[letter,12pt]{article}
\usepackage[english]{babel}
\usepackage{amsmath, amsthm}
\usepackage{graphicx}
\usepackage{placeins}
\usepackage{subfig}

\usepackage{epsfig}
\usepackage{rotating}
\usepackage{amssymb}
\usepackage{geometry}
\usepackage[usenames, dvipsnames, table]{xcolor}
\usepackage{inputenc}
\usepackage[T1]{fontenc}
\usepackage{mathpazo}
\usepackage{color, soul}
\usepackage{booktabs}
\usepackage{xpatch}
\usepackage{verbatim}

\usepackage{accents}

\newcommand{\Exp}{\mathbb{E}}

\newcommand{\non}{\nonumber}

\newtheorem{proposition}{Proposition}

\newtheorem{lemma}{Lemma}

\renewcommand{\vec}[1]{\mathbf{#1}}

\usepackage{setspace}
\makeatletter
\g@addto@macro\normalsize{%
  \setlength\abovedisplayskip{6pt}
  \setlength\belowdisplayskip{6pt}
  \setlength\abovedisplayshortskip{6pt}
  \setlength\belowdisplayshortskip{6pt}
}
\makeatother

\let\oldabstract\abstract
\let\oldendabstract\endabstract
\makeatletter
\renewenvironment{abstract}
{%
               {\list{}{\addtolength{\leftmargin}{-2em} 
                        \listparindent 1.5em%
                        \itemindent    \listparindent%
                        \rightmargin   \leftmargin%
                        \parsep        \z@ \@plus\p@}%
                \item\relax}%
               {\endlist}%
\oldabstract}
{\oldendabstract}
\makeatother

\begin{document}

\begin{titlepage}
\setcounter{footnote}{3}
\title{ \vspace{-2cm} Unemployment and Endogenous Reallocation over the Business Cycle\thanks{This working paper collects the main text and online appendix of ``Unemployment and Endogenous Reallocation over the Business Cycle'' (accepted in Econometrica in February 2023), and the supplementary appendices that provide further background and investigation. The online appendix starts on page \pageref{onlineappx_firstpage}, supplementary appendix A (code error correction) on page \pageref{suppappx_A_firstpage}, supplementary appendix B (data) on page \pageref{suppappx_B_firstpage}, and supplementary appendix C (theory) on page \pageref{suppappx_C_firstpage}.} $\ ^{,}$\thanks{
We would like to thank Melvyn G. Coles, Leo Kaas, David Wiczer and many other colleagues in workshops and seminars whose comments and suggestions helped improved the paper. We also thank Rahel Felder and Junko Iwagami who provided excellent research assistance with the data analysis. The authors acknowledge financial support from the UK Economic and Social Research Council (ESRC), award reference ES/V016970/1. Ludo Visschers further acknowledges financial support from the ESRC, award reference ES/L009633/1 (MacCaLM project at the University of Edinburgh) and Fundaci\'on BBVA (Ayudas a Equipos de Investigación Científica Sars-Cov-2 y Covid-19). The usual disclaimer applies. \textcopyright 2023 The authors.} }
\author{Carlos Carrillo-Tudela \thanks{Correspondence: Department of Economics, University of Essex, Wivenhoe Park, Colchester, CO4 3SQ, UK. Email: cocarr@essex.ac.uk.} \\
University of Essex, \\
CEPR, CESifo \& IZA
\and Ludo Visschers \thanks{%
Correspondence: School of Economics, The University of Edinburgh, 30 Buccleuch Place, Edinburgh, UK, EH8 9JT, UK. Email: ludo.visschers@ed.ac.uk.} \\
University of Edinburgh, \\
UC3M, CESifo \& IZA}
\date{March 2023}

\maketitle

\thispagestyle{empty}

\begin{abstract}
\begin{singlespace}
\small{This paper studies the extent to which the cyclicality of occupational mobility shapes that of aggregate unemployment and its duration distribution. We document the relation between workers' occupational mobility and unemployment duration over the long run and business cycle. To interpret this evidence, we develop a multi-sector business cycle model with heterogenous agents. The model is quantitatively consistent with several important features of the US labor market: procyclical gross and countercyclical net occupational mobility, the large volatility of unemployment and the cyclical properties of the unemployment duration distribution, among many others. Our analysis shows that occupational mobility due to workers' changing career prospects, and not occupation-wide differences, interacts with aggregate conditions to drive the fluctuations of the unemployment duration distribution and the aggregate unemployment rate.}
\end{singlespace}
\end{abstract}



\end{titlepage}

\section{Introduction}

\vspace{-0.15cm}

Occupational mobility is an important part of unemployed workers' job finding process. On average 44\% of workers who went through a spell of unemployment in the US changed ``major occupational groups'' (MOGs) at re-employment.\footnote{Major occupational groups are broad categories that can be thought of as representing one-digit occupations. For example, managers, sales, mechanic and repairers, construction/extraction, office/admin support, elementary trades, etc. The above proportion is obtained after correcting for measurement error.} These occupation movers also take longer to find a job and contribute to the cyclical changes in long-term unemployment. When in downturns the average unemployment duration for occupation stayers increases, for occupation movers the increase is around 35\% larger. This suggests that the willingness and ability of individuals to move across different sectors of the economy can have important consequences for aggregate labor market fluctuations. This paper builds on this evidence and studies the implications of unemployed workers' occupational mobility for the cyclical behavior of the unemployment duration distribution and the aggregate unemployment rate.

We propose and quantitatively assess a multi-sector business cycle model in which the unemployed face search frictions in, and reallocation frictions across, heterogeneous occupations. The economy we consider further exhibits idiosyncratic worker-occupation productivity shocks, orthogonal to occupation-wide productivities, to capture the evolving career prospects of a worker within an occupation. Workers accumulate occupation-specific human capital through learning-by-doing, but face skill loss during unemployment. Even with this rich level of heterogeneity, workers' job separations and reallocation decisions can be characterised by simple reservation (idiosyncratic) productivity cutoffs that respond to aggregate and occupational-wide productivities.

A key success of the framework is that it can generate a wide range of cross-sectional occupational mobility and unemployment duration patterns, as well as the observed cyclical fluctuations of aggregate unemployment, its duration distribution and a strongly downward-sloping Beveridge curve. The cyclical responses of the model's aggregate job separation and job finding rates are also in line with the data (see Shimer, 2005). In addition, the model generates the observed procyclicality of gross occupational mobility among the unemployed and the stronger countercyclicality of unemployment duration among occupational movers. It also generates the observed increase in net reallocation of workers across occupations during recessions (see Dvorkin, 2014, Pilossoph, 2014 and Chodorow-Reich and Wieland, 2020).

Our approach provides a novel insight. It is the interaction between workers' idiosyncratic career productivities and aggregate conditions, and not occupation-wide differences, that drive cyclical unemployment. The main mechanism is as follows. The estimation yields within each occupation a job separation cutoff that is above the reallocation cutoff. This captures that with uncertain career prospects and costly reallocation, those unemployed with idiosyncratic productivities between the cutoffs prefer the option of waiting and remaining attached to their pre-separation occupations instead of reallocating. During recessions the area between these cutoffs widens endogenously and workers spend a longer period of their jobless spells waiting even though there are currently no jobs they could fill. The higher option value of waiting drives up (long-term) unemployment more for occupation movers than stayers and helps create the observed cyclical amplification and persistence in the aforementioned aggregate labor market variables.

The importance of idiosyncratic career productivities in the model's mechanism reflects the prominence of \emph{excess} mobility, i.e. moves that cancel each other out at the occupation level, in driving key occupation mobility patterns in the data. We use the observed high propensity to change occupations and its increase with unemployment duration to uncover the stochastic process of idiosyncratic career shocks. The estimated process then shapes workers' incentive to wait. This waiting motive is evidenced by the observation that even after a year in unemployment about 45\% of workers still re-gain employment in their previous occupations. As the incentive to wait changes over the cycle, the model generates procyclical excess and gross mobility, inline with the data.

A prominent literature of multi-sector models in the spirit of Lucas and Prescott (1974) ``islands'' framework typically emphasises countercyclical net reallocation of unemployed workers across sectors as the main underlying force behind unemployment fluctuations (see Lilien, 1982, Rogerson, 1987). Countercyclical unemployment can arise when in recessions more workers engage in time consuming switches from hard hit sectors to those which offer relatively higher job finding prospects. To capture the role of occupation heterogeneity we use an imperfect directed search approach to model search across occupations over the business cycle (see also Cheremukhin et al. 2020 and Wu, 2020). Nevertheless, as gross flows are an order of magnitude greater than net flows, adding this dimension does not change the importance of workers' career shocks over occupation-wide productivities in explaining labor market fluctuations or the procyclical nature of gross occupational mobility. This occurs because the option value of waiting remains important within (cyclically) declining and expanding occupations. We show that there is no contradiction between changing career prospects playing a very important role in shaping cyclical unemployment, and worker flows through unemployment contributing meaningfully to the changing sizes of occupations particularly during recessions.

The empirical study of occupation (or industry) mobility focused exclusively on workers who went through unemployment has received relatively little attention. This is in contrast to the larger amount of research investigating occupational mobility among pooled samples of employer movers and stayers (see Kambourov and Manovskii, 2008, and Moscarini and Thomsson, 2007, among others). There is no reason, a priori, to conclude that the mobility patterns uncovered by these studies apply to the unemployed. We use data from the Survey of Income and Programme Participation (SIPP) between 1983-2014 to document relevant patterns linking individuals' occupational mobility with their unemployment duration outcomes. We use the Panel Survey for Income Dynamics (PSID) and the Current Population Survey (CPS) to corroborate our results.

We calibrate our model using simulated method of moments. The calibration finds that the nature of unemployment changes over the cycle. Rest/wait unemployment becomes relative more prominent in recession and search unemployment in expansions. Alvarez and Shimer (2011) also study the relative importance of rest and search unemployment using a multi-sector model, but in an aggregate steady state. Their analysis implies that transitions between work, rest and search are not determined. In contrast, the dynamics of workers' idiosyncratic career shocks in our framework determines the transitions between employment and the different types of unemployment. This enables the joint analysis of unemployment duration and occupational mobility, both in the long-run and over the cycle.

The large and persistent rise in unemployment observed during and in the aftermath of the Great Recession generated a renewed interest in multi-sector business cycle models as useful frameworks to investigate cyclical unemployment. Like our paper, Pilossoph (2014) finds a muted effect of net reallocation on aggregate unemployment. Chodorow-Reich and Wieland (2020) build on this work and find that net reallocation co-moves with unemployment most strongly during the recession-to-recovery phase of the cycle. In these papers, gross mobility is constant or countercyclical, which is at odds with the data.\footnote{To the best of our knowledge Dvorkin (2014) is the only one who attempts to reproduce the procyclicality of gross mobility together with the countercyclicality of net mobility. However, his calibrated model remains far from the data (see his Table 9).} These papers also do not focus on the relation between individuals' unemployment duration and their occupational mobility, how this relation changes over the cycle or results in cyclical shifts of the unemployment duration distribution, where the rise of long-term unemployment is shared across occupations (see Kroft et al., 2016).\footnote{Closer to our analysis is Wiczer (2015). An important difference is that in our framework workers take into account the potential recovery of their idiosyncratic productivities when making job separations and reallocation decisions. This feature is crucial for the cyclical properties of our model.}

The rest of the paper proceeds as follows. Section 2 presents the empirical evidence motivating our paper. Section 3 presents the model and its main implications. Sections 4 and 5 provide its quantitative analysis. Section 6 concludes. All proofs, detailed data, quantitative analysis and extensive robustness exercises are relegated to the online appendix and several additional supplementary appendices.\footnote{The online appendix and the additional supplementary appendices are included below, after the main text. The online appendix starts on page \pageref{onlineappx_firstpage}, supplementary appendix A (code error correction) on page \pageref{suppappx_A_firstpage}, supplementary appendix B (data) on page \pageref{suppappx_B_firstpage}, and supplementary appendix C (theory) on page \pageref{suppappx_C_firstpage}.}

\vspace{-0.55cm}

\section{Occupational Mobility of the Unemployed}
\label{s:data}

\vspace{-0.15cm}

Our main statistical analysis is based on the sequence of 1984-2008 SIPP panels, covering the 1983-2014 period. The sample restricts attention to those workers who were observed transiting from employment to unemployment and back within a given panel ($EUE$ flows), and excludes those in self-employment, the armed forces, or agricultural occupations.\footnote{The self-employed are not included as they might face a very different frictional environment, one were vacancies are not needed to gain employment. These differences also seem to persist over time. We find that 50\% of those who transited from self-employment to unemployment in the SIPP went back to self-employment. This suggests that self-employment begets self-employment, a feature not captured in our model. On the other hand, 96\% of those who transited from paid employment into unemployment returned to paid employment and are captured in our model. Not including individuals that at some point during a SIPP panel were self-employed implies dropping 11\% of person-month observations from our data.} In our main analysis we consider workers who have been unemployed throughout their jobless spells, but show that our results hold when using mixed unemployment / out-of-labor-force spells. To minimize the effects of censoring due to the SIPP structure, we consider $EUE$ spells for which re-employment occurs as from month 16 since the start of the corresponding panel and impose that workers at the moment of re-employment have at least 14 months of continuous labor market history within their panel. In Supplementary Appendix B.7 we provide further details of the data construction and analyse the implications of these restrictions.

An individual is considered unemployed if he/she has not been working for at least a month after leaving employment and reported ``no job - looking for work or on layoff''. Since we want to focus on workers who have become unattached from their previous employers, we consider those who report to be ``with a job - on layoff'', as employed.\footnote{\label{FM}Fujita and Moscarini (2017) find that the unemployed consist of ``temporary laid-off workers'' and ``permanent separators''. In the latter group are those who lost their job with no indication of recall. Similarly, Hornstein (2013) and Ahn and Hamilton (2020) consider two groups among the unemployed: those with high and those with low job finding rates. Excluding those ``with a job - on layoff'' and those who find employment within a month means that our unemployment sample is close to Fujita and Moscarini's ``permanent separators'' and to Hornstein's and Ahn and Hamilton's ``low job finding rate'' workers. In Supplementary Appendix B.4.4, we discuss this issue further.} After dropping all observations with imputed occupations, we compare reported occupations before and after the jobless spell. To capture meaningful career changes we use the 21 ``major'' occupational groups of the 2000 Census Occupational Classification (2000 SOC) as well as their aggregation into task-based occupational categories (see Autor et al., 2003). In the SIPP, however, the occupation information of employer movers is collected under independent interviewing, which is known to inflate the importance of occupational mobility. We address this issue by developing a novel classification error model that corrects for coding errors in the flows between particular occupations, and thereby capture more accurately coding errors for those occupations that weigh more among the unemployed.

\subsection{Correcting for Coding Errors in Occupation Mobility}

Suppose that coding errors are made according to a garbling matrix $\mathbf{\Gamma}$ of size $O$x$O$, where $O$ denotes the number of occupational categories. The element $\gamma_{ij}$ is the probability that the true occupation $i=1,2,...,O$ is coded as occupation $j=1,2,...,O$, such that $\sum_{j=1}^{O} \gamma_{ij}=1$. Let $\mathbf{M}$ denote the matrix that contains workers' \emph{true} occupational flows, where element $m_{ij}$ is the flow of workers from occupation $i$ to occupation $j$. Under independent interviewing such a matrix appears as $\mathbf{M^{I}}=\mathbf{\Gamma' M \Gamma}$, where the pre- and post-multiplication by $\mathbf{\Gamma}$ takes into account that the observed occupations of origin and destination would be subject to coding error. Knowledge of $\mathbf{\Gamma}$ (and of its invertibility) allows us to de-garble $\mathbf{M}$ as $\mathbf{\Gamma^{-1}{'} }\mathbf{M^{I} }\mathbf{\Gamma^{-1}}$.

Online Appendix A describes this correction methodology formally. Supplementary Appendix A provides all the proofs and detailed discussion. There we prove that $\mathbf{\Gamma}$ can be identified and estimated by making three assumptions. (\emph{A1}) \emph{Independent classification errors}: conditional on the true occupation, the realization of an occupational code does not depend on workers' labor market histories, demographic characteristics or the time it occurred in our sample. (\emph{A2}) \emph{``Detailed balance'' in miscoding}: coding mistakes are symmetric in that the number of workers whose true occupation $i$ gets mistakenly coded as $j$ is the same as the number of workers whose true occupation $j$ gets mistakenly coded as $i$. (\emph{A3}) \emph{Strict diagonal dominance}: It is more likely to correctly code occupation $i$ than to miscode it. In Supplementary Appendix A we use SIPP, PSID and CPS data to evaluate the plausibility of these assumptions, investigate the implications of the error correction model and verify that the resulting patterns hold under alternative correction methods (see also Supplementary Appendix B).

We implement our method using the change from independent to dependent interviewing that occurred between the 1985 and 1986 SIPP panels. This shows that at re-employment true occupational stayers have on average about a 20\% chance of appearing as occupational movers, based on the 2000 SOC. Further, different occupations have very different propensities to be miscoded and, given a true occupation, some mistakes are much more likely than others. This matters for measuring net mobility (defined below), where we find a sizeable \textit{relative increase} in net mobility after correction.

\vspace{-0.55cm}

\subsection{Gross Occupational Mobility and Unemployment Duration}

\begin{figure}[t]
\begin{centering}
\caption{Extent of Occupational Mobility by Unemployment Duration}
 \label{f:main_duration}
  \resizebox{1.0\textwidth}{!}{
\subfloat[Overall]{\label{f:MO} \includegraphics [width=0.50 \textwidth]{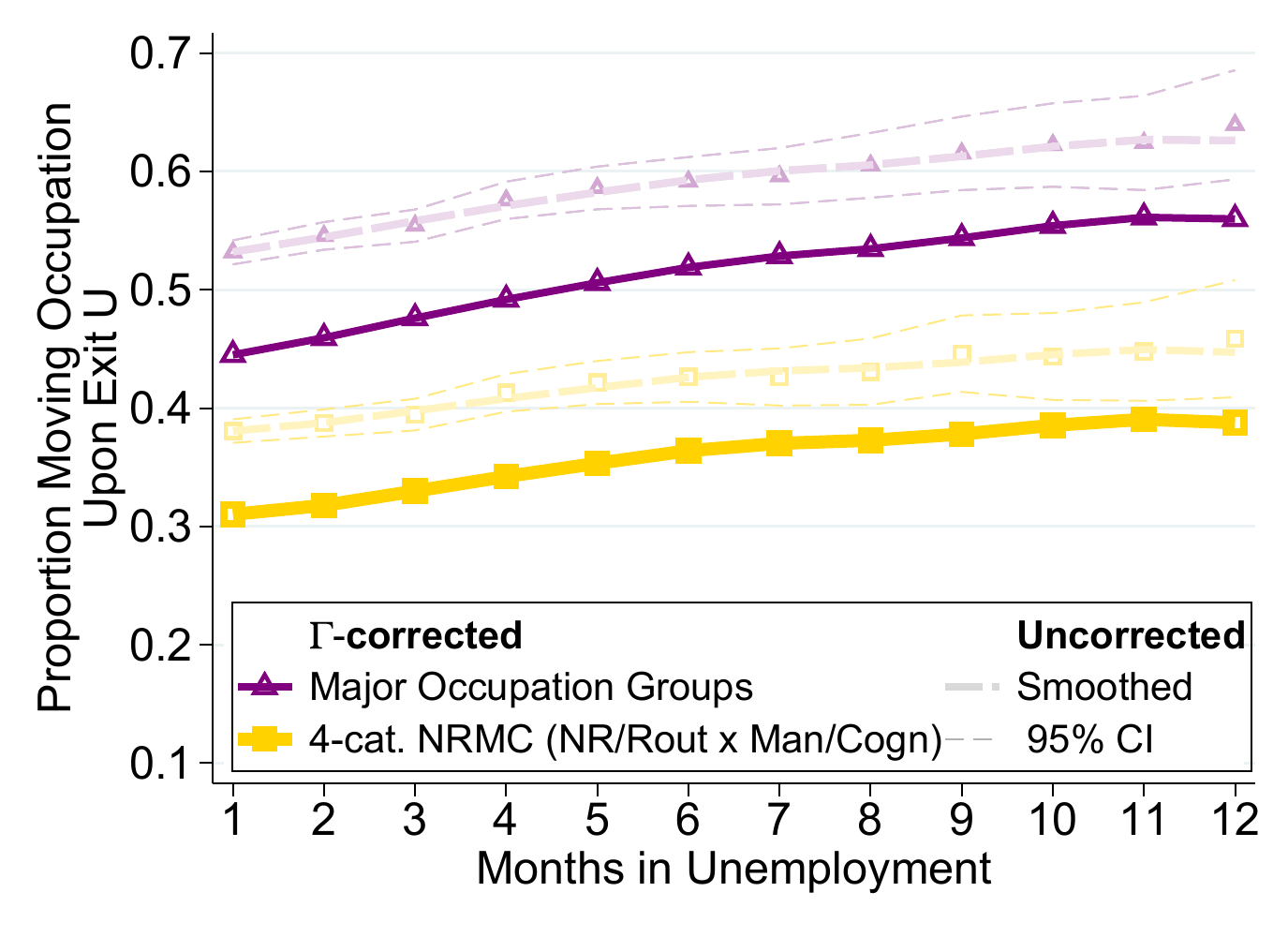}}
\subfloat[By Age Groups (Major Occupational Groups)]{\label{f:Age}\includegraphics [width=0.50 \textwidth]{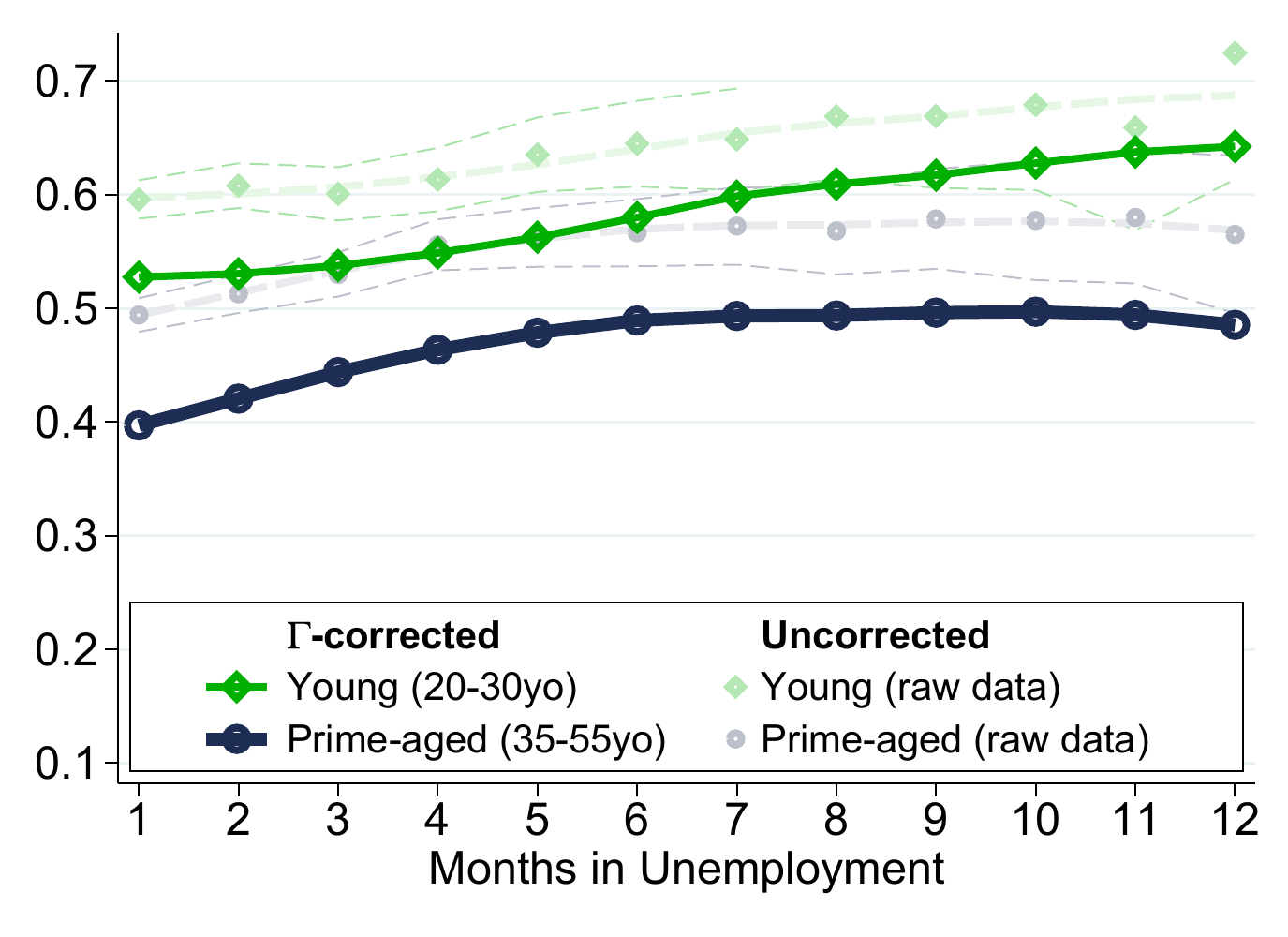}}}
  \vspace{-0.15cm}
   \resizebox{1\textwidth}{!}{
    \begin{tabular}{lccccccc}
  \multicolumn{8}{p{1\textwidth}}{\scriptsize{{\bf{Notes:}} Each mobility-duration profile shows for a given unemployment duration $x$, the proportion of workers who changed occupations at re-employment among all workers who had unemployment spells which lasted at least $x$ months.}}
    \end{tabular}}
     \par\end{centering}
\end{figure}

Figure \ref{f:main_duration} presents a key empirical pattern for our analysis: the mobility-duration profile. It shows the degree of attachment workers have to their pre-separation occupation in relation to their unemployment duration. Each profile shows, for duration $x$, the proportion of workers who changed occupations at re-employment among all workers who had unemployment spells that lasted at least $x$ months.

Figure \ref{f:MO} shows that 44.5\% of workers who had at least one month in unemployment changed occupation at re-employment, while 54.6\% of workers who had at least 9 months in unemployment changed occupation at re-employment. This evidence thus shows that gross occupational mobility at re-employment is \emph{high} and \emph{increases moderately} with unemployment duration. The moderate increase implies that a large proportion of long-term unemployed, over 40\%, still return to their previous occupation at re-employment.\footnote{Kambourov and Manovskii (2008) compare two measures of year-to-year occupational mobility of pooled employer movers and stayers using the PSID, one that includes and one that excludes the unemployed. They find that the inclusion of unemployed workers raises the year-to-year occupational mobility rate by 2.5 percentage points, using a two-digit aggregation. Supplementary Appendices A and B.5 relate in more detail our analysis to theirs.} The figure shows that a similar pattern arises when using the task-based occupational categories: non-routine cognitive $(NRC)$, routine cognitive $(RC)$, non-routine manual $(NRM)$ and routine manual $(RM)$ occupations. Supplementary Appendix B.1 shows this pattern also holds when using non-employment spells, simultaneous industry/occupation mobility or self-reported duration of occupational tenure.

\vspace{-0.55cm}

\paragraph{Demographics} Supplementary Appendix B.1 shows the same patterns across gender, education and race groups. The level of gross mobility, however, decreases substantially with age, from 52.6\% when young (20-30yo) to 39.7\% when prime-aged (35-55yo). Figure \ref{f:Age} shows that the profile of prime-aged workers is below that of the young by about 9-13 percentage points but has a very similar slope. Thus, prime-aged workers display more attachment to their occupation but lose it in a similar way with duration as young workers.

\vspace{-0.55cm}

\begin{figure}[t]
\begin{centering}
  \caption{Gross and Net Occupational Mobility per Occupation}
  \resizebox{0.9\textwidth}{!}{
  \includegraphics[width=0.85 \textwidth]{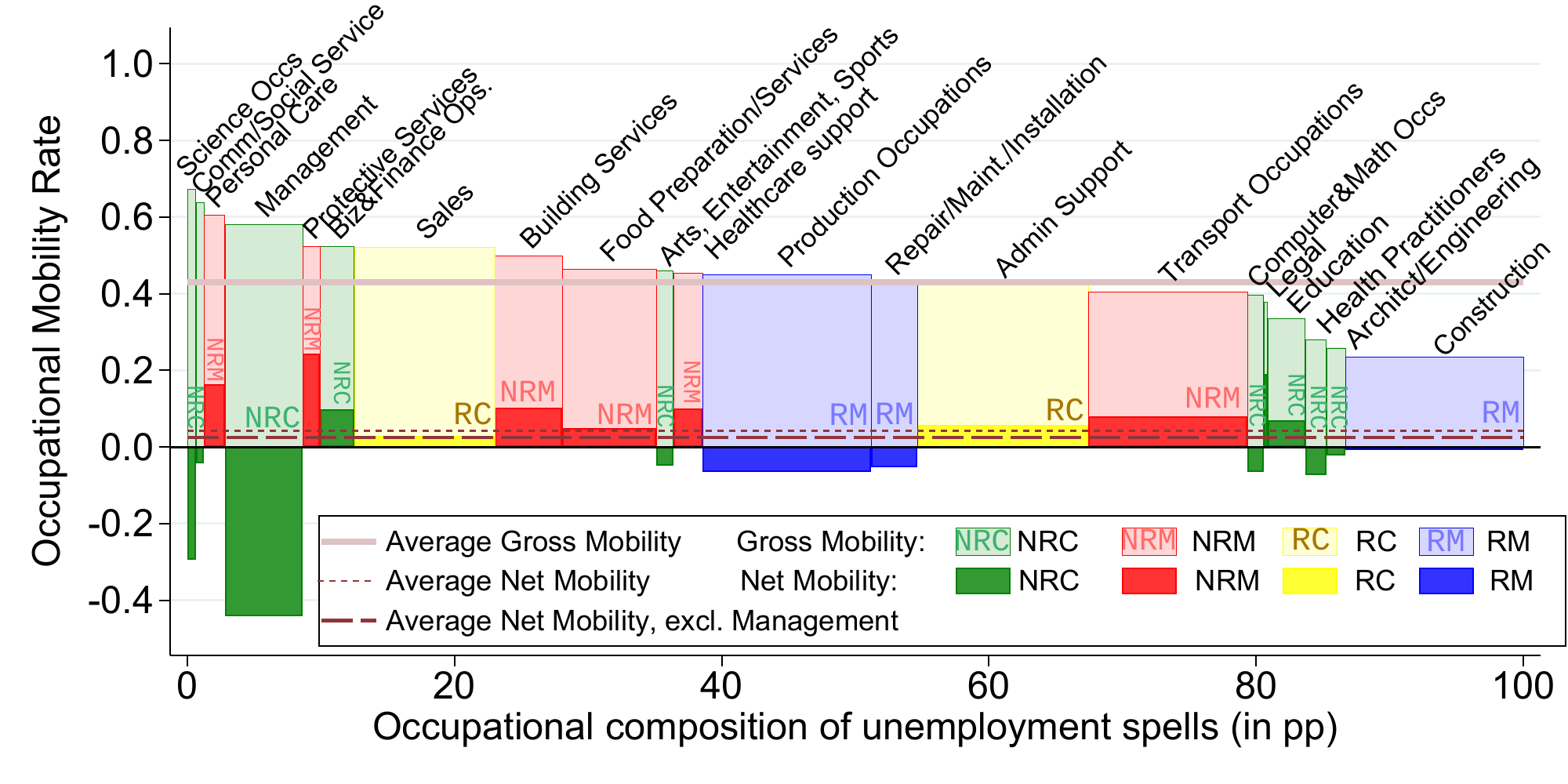}}
   \label{f:grossnet_occmob_per_occ}
   \vspace{0.1cm}
   \resizebox{0.9\textwidth}{!}{
    \begin{tabular}{lccccccc}
   \multicolumn{8}{p{1\textwidth}}{\scriptsize{{\bf{Notes:}} {\emph{Gross mobility:}} The height of each light-shaded bar measures the gross mobility. Occupations are sorted in decreasing order by their gross mobility. {\emph{Net mobility:}} The height of each dark-shaded bar measures net mobility of occupation $i$. A positive value refers to net inflows, while a negative value to net outflows. The area of each of these bars gives the occupation-specific net flows. The solid line correspond to the average gross occupational mobility rate. The dashed lines correspond to the average net mobility rate with and without managerial occupations. All data are corrected for miscoding.}}
      \end{tabular}}
     \par\end{centering}
   \end{figure}

\paragraph{Mobility by occupation} Figure \ref{f:grossnet_occmob_per_occ} shows that most occupations share high mobility rates. Occupation $i$ gross mobility rate (height of each light-shaded bar) is defined as $E_{i}UE_{-i}/E_{i}UE$, where the numerator denotes the $EUE$ spells of workers previously employed in $i$ finding employment in a different occupation and the denominator captures all $EUE$ spells that originate from occupation $i$.\footnote{We define our measures of gross, excess and net occupational mobility based on $EUE$ spells as the longitudinal dimension of the SIPP implies that a worker may have more than one $EUE$ spell. We consider each spell separately when constructing these mobility measures.} Occupations with mobility rates above 40\% cover more than 80\% of all $EUE$ spells in our data. Apart from small and specialized occupations (as engineers, architects, and doctors), construction is the only large occupation with a rate of about 25\%. Further, the slope of the mobility-duration profile does not arise because some occupations with relatively high unemployment durations have particularly high occupational outflows -- rather, it appears that the unemployed across all occupations lose their attachment gradually (see Supplementary Appendix B.1 for details).

\vspace{-0.55cm}

\subsection{Excess and Net Mobility}

To assess the importance of moves that result in certain occupations experiencing net inflows (outflows) through unemployment, we divide gross occupational mobility into net and excess mobility. The dark bars in Figure \ref{f:grossnet_occmob_per_occ} depict the net mobility rate per occupation, defined as $(E_{-i}UE_{i}-E_{i}UE_{-i})/E_{i}UE$, where the numerator denotes the difference between gross inflows and outflows for occupation $i$. It is evident that net flows are an order of magnitude smaller than gross flows across almost all occupations, the main exception being managerial occupations. The average net mobility rate, $0.5 \sum_i |E_{-i}UE_{i}-E_{i}UE_{-i}| / EUE$ (where $EUE=\sum_i E_iUE$) equals 4.3\% (uncorrected for miscoding, 3\%).\footnote{The pre-multiplication by 0.5 reflects that each net outflow in some occupation is simultaneously also counted as a net inflow in other occupations. Kambourov and Manovskii (2008) have also highlighted the small relative importance of net mobility across occupations in pooled samples of employer mover and stayers.} Figure \ref{f:grossnet_occmob_per_occ} also shows a clear directional pattern: net outflows from the $RM$ occupations and net inflows into the $NRM$ occupations.

Excess mobility is the most important component of occupational mobility, except for management. The average excess mobility rate $\sum \min \{E_{-i}UE_{i}, E_{i}UE_{-i}\}/EUE$ implies that 40.1\% of all $EUE$ spells represent excess mobility, about 90\% of all gross mobility. In Supplementary Appendix B.2 we show that these results are robust to alternative occupational classifications and using non-employment spells.

The increase of occupational mobility with duration documented in Figure \ref{f:main_duration} is also driven predominantly by excess mobility. We re-compute the average net and excess mobility rates defined above on the subset of $EUE$ spells of at least duration $x=1,2,3,...,12$. This shows that the rise of excess mobility with duration does not support the notion that long-term unemployment is primarily driven by a subset of occupations in which workers are particularly eager to leave for another set of occupations with better conditions (see Supplementary Appendix B.2).

\vspace{-0.55cm}

\subsection{Repeat Mobility}

The SIPP allows us to investigate the evolution of a worker's attachment to occupations across multiple unemployment spells. These ``repeat mobility'' statistics tell us whether workers who changed (did not change) occupations after an unemployment spell, will change occupation subsequently after a following unemployment spell. Here we can also use the $\mathbf{\Gamma}$-correction to counteract coding errors in three-occupation histories (surrounding two unemployment spells).\footnote{Let the matrix $\mathbf{M^r}$ (with elements $m^r_{ijk}$) be the $O \times O \times O$ matrix of true repeat flows. Then, this matrix relates to the \emph{observed} repeat flow matrix $\mathbf{M^{r, obs}}$ through $\textbf{vec}(\mathbf{M^r})'=\textbf{vec}(\mathbf{M^{r, obs}})'(\mathbf{\Gamma}\otimes\mathbf{\Gamma}\otimes\mathbf{\Gamma})^{-1},$ where $\textbf{vec}(\mathbf{M})$ is the vectorization of matrix $\mathbf{M}$, and $\otimes$ denotes the Kronecker product. Since $\mathbf{\Gamma}$ is invertible, $\mathbf{\Gamma}\otimes\mathbf{\Gamma}\otimes\mathbf{\Gamma}$ is also invertible. The repeat mobility statistics are then measured \textit{within SIPP 3.5 to 5 years windows} and are based on 610 of observations of individuals with multiple spells across all panels when considering only pure unemployment spells and 1,306 when considering non-employment spells that include months of unemployment. For further details see Supplementary Appendix B.7. Note that workers with two consecutive unemployment spells within this window are not necessarily representative of all unemployed workers, nor of behavior in unemployment spells that are further apart. Nevertheless, these statistics are valuable and will inform our modelling choices and quantitative analysis (by indirect inference).}

We find that of all those stayers who became unemployed once again, 63.4\% remain in the same occupation after concluding their second unemployment spell. This percentage is higher for prime-aged workers, 65.6\%, and lower for young workers, 61.7\%. However, the loss of occupational attachment itself also persists. Among workers who re-enter unemployment after changing occupations in the preceding unemployment spell, 54.4\% move again. This is lower for prime-aged workers, 49.1\%, and higher for the young, 64.5\%. Supplementary Appendix B.5 shows a similar pattern in the PSID.
				
\vspace{-0.55cm}

\subsection{Occupational Mobility of the Unemployed over the Cycle}

Unemployed workers' attachment to their previous occupations changes over the business cycle. In expansions unemployed workers change occupations more frequently than in recessions. Panel A of Table \ref{tab:2} investigates the cyclicality of occupational mobility by regressing the (log) gross mobility rate on the (log) unemployment rate. Columns (i) and (ii) relate the HP-filtered quarterly series of the $\mathbf{\Gamma}$-corrected and uncorrected occupational mobility rates obtained from the SIPP to HP-filtered series of the unemployment rate, with a filtering parameter of 1600. Because there are proportionally more stayers and hence more spurious mobility in recessions, the $\mathbf{\Gamma}$-corrected series yields a somewhat stronger cyclicality than the uncorrected one. Column (iii) presents the regression results based on (uncorrected) occupational mobility data from the CPS for the period 1979-2019 (see Supplementary Appendix B.5). We use the CPS as its quarterly mobility series does not suffer from gaps. We observe that the uncorrected SIPP and CPS series have nearly the same degree of procyclicality, suggesting that data gaps do not meaningfully affect our conclusion.\footnote{Restricting the CPS series to start after the 1994-redesign does not change our results. See Supplementary Appendix B.5. The SIPP series have data missing due to non-overlapping panels combined with our sampling restrictions (to avoid censoring issues), as described in Supplementary Appendix B.7. To deal with these gaps, we use TRAMO-SEATS for interpolation, HP-filter the series and then discard all quarters that were interpolated.}

\begin{table}[t]
  \centering
  \small
  \caption{Occupational Mobility and Unemployment Duration over the Business Cycle}
  \resizebox{1\textwidth}{!}{
    \begin{tabular}{lccccccc}
    \toprule
    \toprule
          & \multicolumn{3}{c}{HP-filtered Qtrly  Occ. Mobility}   &   \multicolumn{4}{c}{Unfiltered Occ Mobility} \\
\cmidrule(lr){2-4} \cmidrule(lr){5-8}      & (i)   & (ii)   & (iii)   & (iv)   & (v)   & (vi) & (vii) \\
          &  SIPP & SIPP & CPS &  SIPP & SIPP & CPS &  SIPP  \\
          &  ${\bf{\Gamma}}$-corrected & uncorrected  & uncorrected &  ${\bf{\Gamma}}$-corrected & uncorrected & uncorrected &  uncorrected \\
    \midrule
    \midrule
    & \multicolumn{7}{c}{Panel A: Mobility regression, not controlling for non-employment duration}  \\ \hline
 \multicolumn{1}{l}{HP U} & -0.145 & -0.088 & -0.087    & \ \ \ -0.122 & \ \ \ -0.114 & -0.077   & \ \ \ -0.132 \\
          & (0.041) & (0.026) & (0.024) & \ \ \ (0.059) & \ \ \ (0.046) & \ \ \ (0.022) & \ \ \ (0.041) \\
          \hline
          Controls & -     & -  & -&  T  & T & T & D,T, C, S.O. \\

\hline \hline
      & \multicolumn{7}{c}{Panel B: Mobility regression, controlling for non-employment duration}  \\
\hline
        \multicolumn{1}{l}{HP U} & - & - & - & \ \ \ -0.171 & \ \ \ -0.149 & -0.112 & \ \ \ -0.176 \\
          & - &  - & -  & \ \ \ (0.061) & \ \ \ (0.048) & (0.024) & \ \ \ (0.042) \\
    \multicolumn{1}{l}{Dur. coef} &- & - & - &  \ \ \ 0.0126 & \ \ \ 0.0138 & 0.0116 & \ \ \ 0.0145 \\
          & - & - & - & \ \ \ (0.002) & \ \ \ (0.002) & (0.001) & \ \ \ (0.002) \\
\hline
          Controls & -     & -  & -&  T  & T & T & D,T, C, S.O. \\
    \bottomrule
    \bottomrule
    \multicolumn{8}{p{1.2\textwidth}}{\scriptsize{{\bf{Notes:}} SIPP sample is restricted to quarters where the data allows the full spectrum of durations between 1-12 months to be measured. Standard errors clustered on quarters and shown in parenthesis. See Supplementary Appendix B.7 for details. CPS data described in Supplementary Appendix B.5. \textbf{\emph{Controls}}: D=demographic controls (gender, race, education, and a quartic in age); T=linear trend, C=dummies for the classification in which data was originally reported; S.O.= source occupation.
    }}
   \end{tabular}}
  \label{tab:2}
\end{table}

Columns (iv)-(vii) present the results of regressing unfiltered occupational mobility series on the HP-filtered unemployment rate for further robustness. Again, both SIPP and CPS give a broadly similar procyclicality. The last column adds further individual-level controls and shows that these do not meaningfully change our results. The procyclicality of occupational mobility is thus not the result of a compositional shift towards occupations or demographics characteristics that are associated with higher mobility during an expansion. In Supplementary Appendix B.3 we provide an extensive set of robustness exercises based on the SIPP, all showing the procyclicality of gross occupational mobility. Supplementary Appendix B.5 further shows procyclical occupational mobility using the PSID for the period 1968-1997.

\vspace{-0.55cm}

\begin{figure}[t]
\begin{centering}
\caption{Cyclicality of Occupational Mobility, 1985-2014}
\label{f:aggregate_main_cycle}
\resizebox{1\textwidth}{!}{
\subfloat[Occ. Mobility - Duration Profile Shift]{\label{sf:durprofile_shift} \includegraphics [height=0.2 \textheight, width=0.45 \textwidth] {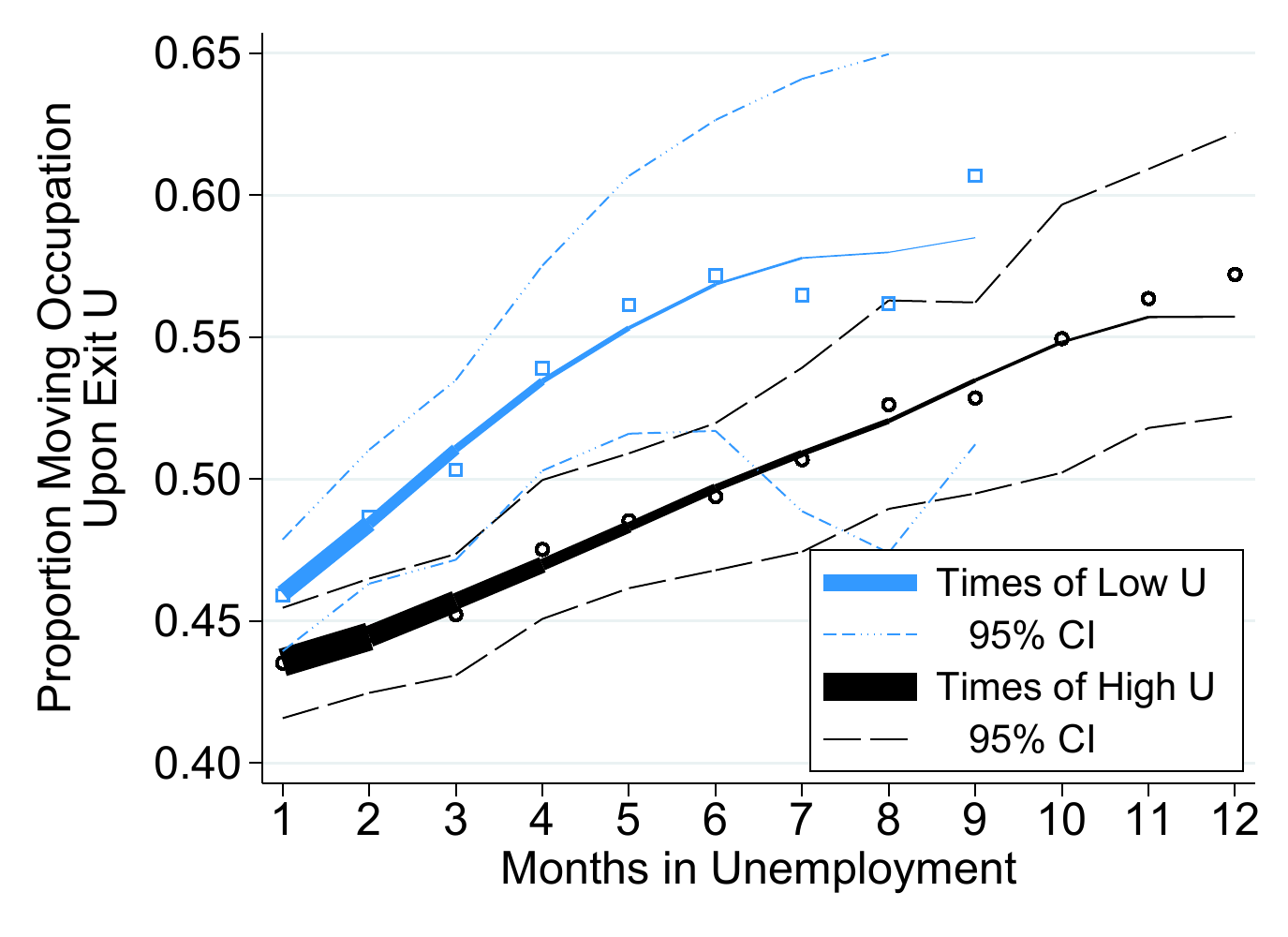}}
\subfloat[Net Occ. Mobility - Task-based categories]{\label{f:app_netmob_baseline} \includegraphics [height=0.2 \textheight, width=0.45 \textwidth] {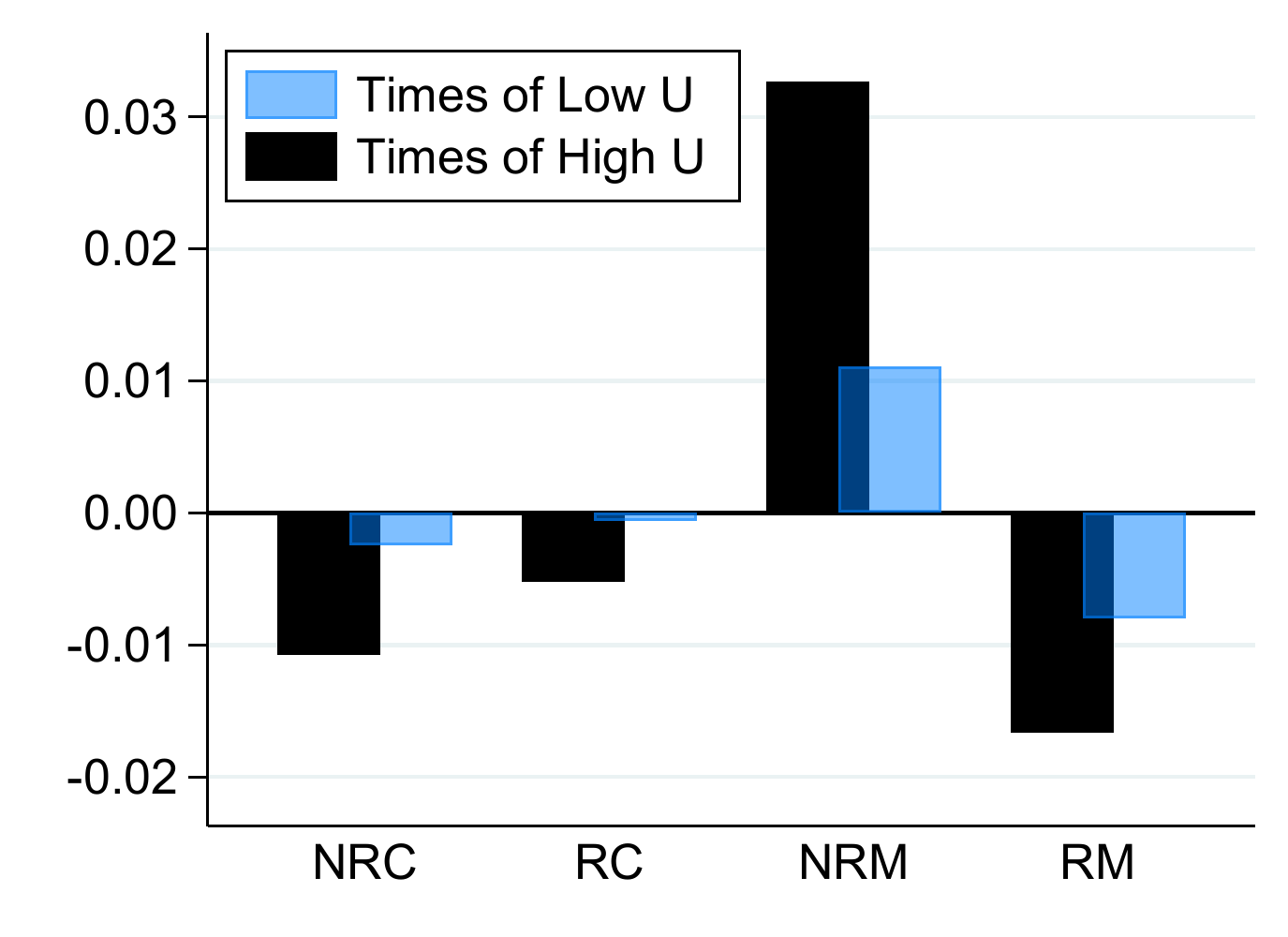}}}
  \vspace{-0.3cm}
   \resizebox{1\textwidth}{!}{
    \begin{tabular}{lccccccc}
   \multicolumn{8}{p{1\textwidth}}{\scriptsize{{\bf{Notes:}} {\emph{Left panel:}} The circular markets depict non-smoothed data and the solid curves represent the smoothed mobility-duration profile. The thickness of the profiles indicates the amount of spells surviving at a given duration. {\emph{Right panel:}} The net mobility rate for each task-based category is computed excluding Managers, separately for periods of high and low unemployment. The version including Managers can be found in the Supplementary Appendix B.3.}}
    \end{tabular}}
     \par\end{centering}
\end{figure}

\paragraph{Cyclicality of the mobility-duration profile} Figure \ref{sf:durprofile_shift} depicts the cyclical shift of the mobility-duration profile. It plots the profile separately for those $EUE$ spells that ended in times of high unemployment and those that ended in times of low unemployment. Times of high (low) unemployment are defined as periods in which the de-trended (log) unemployment rate was within the bottom (top) third of the de-trended (log) unemployment distribution. Occupational mobility at any duration is lower in recessions, corroborating the procyclicality of gross occupational mobility. Both in times of high and low unemployment, an increase in unemployment duration is associated with a moderate loss of attachment to workers' previous occupation. Panel B in Table \ref{tab:2} similarly shows the vertical shift of the mobility-duration profile over the cycle and that this is robust to demographics and (origin) occupation controls.

\vspace{-0.55cm}

\paragraph{The cyclicality of net occupational mobility} Figure \ref{f:app_netmob_baseline} shows the cyclical behavior of the net mobility rate for each of the task-based categories. The net mobility rate is computed as $(E_{-i}UE_{i}-E_{i}UE_{-i})/EUE$, separately for periods of high and low unemployment.\footnote{Differently from Section 2.3 we normalise net flows in each task-based category by the total number of $EUE$ spells observed in periods of either high or low unemployment. Here we also exclude managers. Supplementary Appendix B.3 shows that this exclusion implies that $RC$ occupations are now experiencing net outflows instead of net inflows as suggested by Figure \ref{f:grossnet_occmob_per_occ}.} Across all task-based categories the net mobility rate increases when unemployment is high, even though $EUE$ also increases. In particular, $RM$ occupations increase their net outflows in downturns relative to expansions, while $NRM$ occupations increase their net inflows in downturns relative to expansions. The countercyclicality of net mobility therefore implies that the stronger procyclicality of excess mobility is the main driver of the procyclical behavior of gross mobility among unemployed workers.\footnote{Kambourov and Manovskii (2008) using PSID data also find countercyclical net mobility and procyclical gross mobility among a pooled sample of employer stayers and movers.}

\vspace{-0.55cm}

\paragraph{Comparing unemployment spells between movers and stayers} The mobility-duration profile implies that occupational movers have on average longer spells than stayers. From an expansion to a recession, this difference grows from 0.4 to 0.9 months.\footnote{Averaging all spells in our sample, the difference is 0.7 month. This amount is economically significant, 40\% of the difference between the average duration of unemployment spells in periods of high versus low unemployment.} This increase does not result from cyclically different demographics of unemployed movers or because they are more likely to come from (move to) occupations with long unemployment durations in recessions (see Supplementary Appendix B.4). Although occupational mobility decreases in recessions, the lengthening of unemployment spells among movers is proportionally stronger. Occupational movers thus contribute meaningfully to the increase in aggregate unemployment, and especially to the increase in long-term unemployment.

\vspace{-0.55cm}

\section{Theoretical Framework}
\vspace{-0.15cm}

We now develop a theory of occupational mobility of the unemployed to explain the above empirical patterns and link them to the cyclical behavior of long and short term unemployment as well as the aggregate unemployment rate.

\vspace{-0.55cm}

\subsection{Environment}

Time is discrete $t=0, 1, 2, \ldots$ A mass of infinitely-lived, risk-neutral workers is distributed over a finite number of occupations $o=1,\ldots,O$. At any time $t$, workers within a given occupation can be either employed or unemployed and differ in two components: an idiosyncratic productivity, $z_{t}$, and human capital, $x_{t}$. We interpret the $z$-productivity as a ``career match'' which captures in a reduced form the changing career prospects workers have in their occupations (see Neal, 1999). These $z$-productivities follow a common and bounded first-order stationary Markov process, with transition law $F(z_{t+1}|z_{t})$.\footnote{The assumption that the $z$ process is common across workers and occupations is motivated by our evidence showing that the change in occupational mobility with unemployment duration does not seem to differ across occupations or demographic groups.} Their realizations affect a worker both in employment and in unemployment and will drive excess occupational mobility. To capture the different levels of attachment to occupations found across age groups, workers' accumulate occupational human capital through a learning-by-doing process while employed, and are subject to human capital depreciation while unemployed. Conditional on the worker's employment status, his human capital $x_{t}$ is assumed to evolve stochastically following a Markov chain with values $x_{t}\in \{x^{1},..., x^{H}\}$, $x^{1}>0$ and $x^{H}<\infty$.

Each occupation is subject to occupation-wide productivity shocks. Let $p_{o,t}$ denote the productivity of occupation $o$ at time $t$ and $p_{t} =\{p_{o,t}\}_{o=1}^{O}$ the vector that contains all occupation productivities at time $t$. Differences across $p_{o,t}$ will drive net mobility. Business cycle fluctuations occur due to changes in aggregate productivity, $A_{t}$. We allow the occupation-wide productivity process to depend on $A_{t}$. Both $p_{o,t}$ and $A_{t}$ follow bounded first-order stationary Markov processes.

There is a mass of infinitely-lived risk-neutral firms distributed across occupations. All firms are identical and operate under a constant return to scale technology, using labor as the only input. Each firm consists of only one job that can be either vacant or filled. The output of an employed worker characterised by $(z,x,o)$ in period $t$ is given by the production function $y(A_{t},p_{o,t}, z_{t},x_{t})$, which is strictly increasing and continuous in all of its arguments and differentiable in the first three.

All agents discount the future at rate $\beta$. Workers retire stochastically, receiving a fixed utility flow normalized to zero. They are replaced by new entrants, unemployed and inexperienced workers with $x^{1}$ that are allocated across occupations following an exogenous distribution $\psi$. We rescale $\beta$ to incorporate this retirement risk. Match break-up can occur with an exogenous (and constant) probability $\delta$, but also if the worker and firm decide to do so, and after a retirement shock. Once the match is broken, the firm decides to reopen the vacancy and, unless retired, the worker stays unemployed until the end of the period. An unemployed worker receives $b$ each period. Wages will be determined below.

To study business cycle behavior in a tractable way, we focus on Block Recursive Equilibria (BRE). In this type of equilibria the value functions and decisions of workers and firms only depend on $\omega_{t}=\{z_{t}, x_{t}, o, A_{t}, p_{t}\}$ and not on the joint productivity distribution of unemployed and employed workers over all occupations. An occupation can be segmented into many labor markets, one for each pair $(z,x)$ such that workers in different markets do not congest each other in the matching process. Each of these $(z,x)$ labor market has the Diamond-Mortensen-Pissarides (DMP) structure. Each has a constant returns to scale matching function which governs the meetings of unemployed workers and vacancies within a market. We assume that all these markets have the same random matching technology. Each market exhibits free entry of firms, where posting a vacancy costs $k$ per period. Once an unemployed worker's $z$ or $x$ changes, his relevant labor market changes accordingly.\footnote{In Supplementary Appendix C we show that a competitive search model in the spirit of Menzio and Shi (2010) endogenously generates this sub-market structure, such that in equilibrium unemployed workers with current productivities $(z,x)$ decide to participate only in the $(z,x)$ market. Here we proceed by assuming the sub-market structure from the start in order to reduce unnecessary complexity in the analysis. The allocations and equilibrium outcomes are the same under both approaches.}

\vspace{-0.55cm}

\paragraph{Searching across occupations} Instead of searching for jobs in their own occupation, unemployed workers can decide to search for jobs in different occupations. This comes at a per-period cost $c$ and entails re-drawing their $z$-productivity. Workers rationally expect their initial career match in any occupation to be a draw from $F(.)$, the ergodic distribution associated with the Markov process $F(z_{t+1}|z_{t})$. The i.i.d. nature of the re-draws allows us to capture that some occupational movers end up changing occupations again after a subsequent jobless spell, as suggested by the repeat mobility patterns documented earlier.

Differences in $p_{o}$ imply that workers are not indifferent from which occupation the draw of $z$ comes from. To capture that in the data excess mobility is much larger than net mobility and hence that workers not always specialise their search in the occupation with the highest $p_{o}$, we model the choice of occupation following an imperfectly directed search approach in the spirit of Fallick (1993). During a period, workers have a unit of search effort to investigate their employment prospects in the remaining occupations. They can only receive at most one new draw of $z$ per period without recall. A worker must then chose how much effort to allocate to each one of these occupations to maximise the probability of receiving a $z$. Let $s_{\tilde{o}}$ denote the search effort devoted to occupation $\tilde{o}$ such that $\sum_{\tilde{o} \in O^{-}} s_{\tilde{o}}=1$, where $O^{-}$ denotes the set of remaining occupations. Each $s_{\tilde{o}}$ maps into a probability of receiving the new $z$ from occupation $\tilde{o}$. Conditional on switching from $o$, this probability is denoted by $\alpha(s_{\tilde{o}};o)$, where $\alpha(.;o)$ is a continuous, weakly increasing and weakly concave function of $s$ with $\alpha(0;o)=0$. The concavity creates a trade-off between concentrating search effort on desirable occupations and the total probability that the worker draws some $z$, given by $\sum_{\tilde{o} \in O^{-}} \alpha(s_{\tilde{o}};o) \leq 1$. With probability $1-\sum_{\tilde{o} \in O^{-}} \alpha(s_{\tilde{o}};o)$ no $z$ is received and the above process is repeated the following period.

If a $z$ is received, the worker must sit out one period unemployed in the new occupation $\tilde{o}$ before deciding whether to sample another $z$ from a different occupation.\footnote{This implies that the worker is forced to move to the new occupation even if the $z$ turns out to be low enough. To further simplify we also assume that after the worker is in the new occupation, he can sample $z$-productivities from previous occupations. This way we avoid carrying around in the state space the histories of occupations ever visited by a worker.} If the worker decides to sample once again, the above process is repeated. However, if the worker decides to accept the $z$, he starts with human capital $x^{1}$ in the new occupation. The worker's $z$ and $x$ then evolve as described above.

\vspace{-0.55cm}
\subsection{Agents' Decisions}

The timing of the events is summarised as follows. At the beginning of the period the new values of $A$, $p$, $z$ and $x$ are realised. The period is then subdivided into four stages: separation, reallocation, search and matching, and production. To reduce notation complexity, we leave implicit the time subscripts, denoting the following period with a prime.

\vspace{-0.55cm}

\paragraph{Worker's Problem}

Consider an unemployed worker currently characterised by $(z,x,o)$. The value function of this worker at the beginning of the production stage is given by
\begin{align} \label{wushort}
W^U\!(\omega)\!=\! b\! +\! \beta \mathbb{E}_{\omega'}\!\bigg[\!\! \max_{\rho(\omega')}\! \Big\{ & \rho(\omega')R(\omega')\! +\!(1\!-\!\rho(\omega')) \Big[\lambda (\theta (\omega'))W^{E}(\omega')\!+\!(1\!-\!\lambda (\theta (\omega')))W^{U}(\omega')\Big]\! \Big\} \! \bigg],
\end{align}
where $\theta(\omega)$ denotes the ratio between vacancies and unemployed workers currently in labor market $(z,x)$ of occupation $o$, with $\lambda(.)$ the associated job finding probability. The value of unemployment consists of the flow benefit of unemployment $b$, plus the discounted expected value of being unemployed at the beginning of next period's reallocation stage, where $\rho(\omega)$ takes the value of one when the worker decides to search across occupations and zero otherwise. The worker's decision to reallocate is captured by the choice between the expected net gains from drawing a new $\tilde{z}$ in another occupation and the expected payoff of remaining in the current occupation. The latter is given by the expression within the inner squared brackets in \eqref{wushort}. The term $R(\omega)$ denotes the expected net value of searching across occupations and is given by
\begin{equation}\label{eq:reall}
R(\omega)=\max_{\mathcal{S}(\omega)}\Big( \sum_{\tilde{o} \in O^{-}} \alpha(s_{\tilde{o}}(\omega)) \int_{\underline{z}}^{\overline{z}}W^{U}(\tilde{z},x^{1},\tilde{o},A,p)dF(\tilde{z}) + (1-\sum_{\tilde{o} \in O^{-}} \alpha(s_{\tilde{o}}(\omega)))\hat{W}^U(\omega) - c \Big),
\end{equation}
where $\hat{W}^U(\omega)=b+ \beta \mathbb{E}_{\omega'} R(\omega')$, $\mathcal{S}$ denote a vector of $s_{\tilde{o}}$ for all $\tilde{o}\in O^{-}$ and the maximization is subject to $s_{\tilde{o}}\in[0,1]$ and $\sum_{\tilde{o} \in O^{-}} s_{\tilde{o}}=1$. The first term denotes the expected value of drawing a new $z$ and losing any accumulated human capital, while the second term denotes the value of not obtaining a $z$ and waiting until the following period to search across occupations once again. The formulation of $\hat{W}^U(\omega)$ is helpful as it implies that $R(\omega)$ and $\{s_{\tilde{o}}\}$ become independent of $z$. It is through $R(\omega)$ that expected labor market conditions in other occupations affect the value of unemployment, and indirectly the value of employment.

Now consider an employed worker currently characterised by $(z,x,o)$. The expected value of employment at the beginning of the production stage, given wage $w(\omega)$, is
\begin{eqnarray}  \label{WEshort}
W^{E}(\omega)=w(\omega) +\beta \mathbb{E}_{\omega'}\Big[ \max_{d(\omega')} \{ (1-d(\omega')) W^{E}(\omega')+ d(\omega')W^U(\omega')\}\Big].
\end{eqnarray}
The second term describes the worker's option to quit into unemployment in next period's separation stage. The job separation decision is summarised in $d(\omega')$, such that it take the value of $\delta$ when $W^{E}(\omega')\geq W^U(\omega')$ and the value of one otherwise.

\vspace{-0.55cm}

\paragraph{Firm's Problem}

Consider a firm posting a vacancy in sub-market $(z,x)$ in occupation $o$ at the start of the search and matching stage. The expected value of a vacancy solves the entry equation
\begin{equation}  \label{vshort}
V(\omega)=-k+ q(\theta (\omega))J(\omega),
\end{equation}
where $q(.)$ denotes firms' probability of finding an unemployed worker and $J(\omega)$ denotes the expected value of a filled job. Free entry implies that $V(\omega)=0$ for all those sub-markets that yield a $\theta(\omega)>0$, and $V(\omega)\leq0$ for all those sub-markets that yield a $\theta(\omega)\leq0$. In the former case, the entry condition simplifies \eqref{vshort} to $k= q(\theta (\omega))J(\omega)$.

Now consider a firm employing a worker currently characterized by $(z,x,o)$ at wage $w(\omega)$. The expected lifetime discounted profit of this firm at the beginning of the production stage can be described recursively as
\begin{eqnarray}  \label{jshort}
J(\omega) = y(A,p_{o},z,x)-w(\omega) + \beta \mathbb{E}_{\omega'}\Big[\max_{\sigma (\omega')}\Big\{(1-\sigma (\omega'))J(\omega') +\sigma (\omega')V(\omega')\Big\}\Big],
\end{eqnarray}
where $\sigma (\omega')$ takes the value of $\delta$ when $J(\omega')\geq V(\omega')$ and the value of one otherwise.

\vspace{-0.55cm}

\paragraph{Wages}
We assume that wages are determined by Nash Bargaining. Consider a firm-worker match currently characterised by $(z,x,o)$ such that it generates a positive surplus. Nash Bargaining implies that the wage, $w(\omega)$, solves
\begin{equation}\label{wages}
(1-\zeta) \Big(W^{E}(\omega)-W^{U}(\omega) \Big)=\zeta \Big(J(\omega)-V(\omega)\Big),
\end{equation}
where $\zeta \in[0,1]$ denotes the worker's exogenous bargaining power. This guarantees that separation decisions are jointly efficient, $d(\omega)=\sigma(\omega)$.

In what follows we impose a Cobb-Douglas matching function and the Hosios condition, such that $1- \zeta=\eta$, where $\eta$ denotes the elasticity of the job finding probability with respect to labor market tightness within sub-market $(z,x)$. In our framework this will guarantee firms post the efficient number of vacancies within sub-markets and the constraint efficiency of our decentralized economy (see Supplementary Appendix C).

\vspace{-0.55cm}
\subsection{Equilibrium and Characterization}
\label{s:equilibrium}

In a BRE outcomes can be derived in two steps. Decision rules are first solved using \eqref{wushort}-\eqref{jshort}. We then fully describe the dynamics of the workers' distribution, using the workers' flow equations. To prove existence and uniqueness we build on the proofs of Menzio and Shi (2010) but incorporate the value of reallocation across occupations and show it preserves the block recursive structure. The formal definition of the BRE is relegated to Supplementary Appendix C, where we also present the derivation of the flow equations and the proofs of all the results of this section.

\vspace{-0.55cm}

\paragraph{Existence} Let $M(\omega) \equiv W^E(\omega)+J(\omega)$ denote the joint value of the match. To prove existence and uniqueness of the BRE we define an operator $T$ that is shown to map $\{M(\omega)$, $W^{U}(\omega)$, $R(\omega)\}$ from the appropriate functional space into itself, with a fixed point that implies a BRE. The key step to proof efficiency is to ensure that a worker's value of searching across occupations coincides with the planner's value of making the worker search across occupations.

\begin{proposition} \label{prop existence}
Given $F(z'|z)<F(z'|\tilde{z})\ $for\ all $z,z'$ when $z>\tilde{z}$: (i) a BRE exists and it is the unique equilibrium; and (ii) the BRE is constrained efficient.
\end{proposition}

\vspace{-0.55cm}

\paragraph{Characterization} The decision to separate from a job and the decision to search across occupations can be characterised by $z$-productivity cutoffs, which are themselves functions of $A$, $p$, $o$ and $x$. The job separation cutoff function, $z^s(.)$ is obtained when the match surplus becomes zero, $M(\omega)-W^{U}(\omega)=0$. In contrast to Mortensen and Pissarides (1994), $z$ refers to the worker's idiosyncratic productivity in an occupation and not to a match-specific productivity with a firm. This difference implies that when the worker becomes unemployed, his $z$ is not lost or is reset when re-entering employment in the same occupation. Instead, the worker's $z$ continuous evolving during the unemployment spell. It is only when the worker searches across occupations that he can reset his $z$. This occurs if and only if $z<z^r(.)$, where the reallocation cutoff function $z^r(.)$ solves $R(\omega)=W^{U}(\omega)$.

\begin{figure}[t]
\centering
\caption{Relative Positions of the Reservation Productivities}
 \resizebox{0.85\textwidth}{!}{
\subfloat[$z^r > z^s$]{\label{no rest} \includegraphics [height=0.2 \textheight, width=0.4 \textwidth] {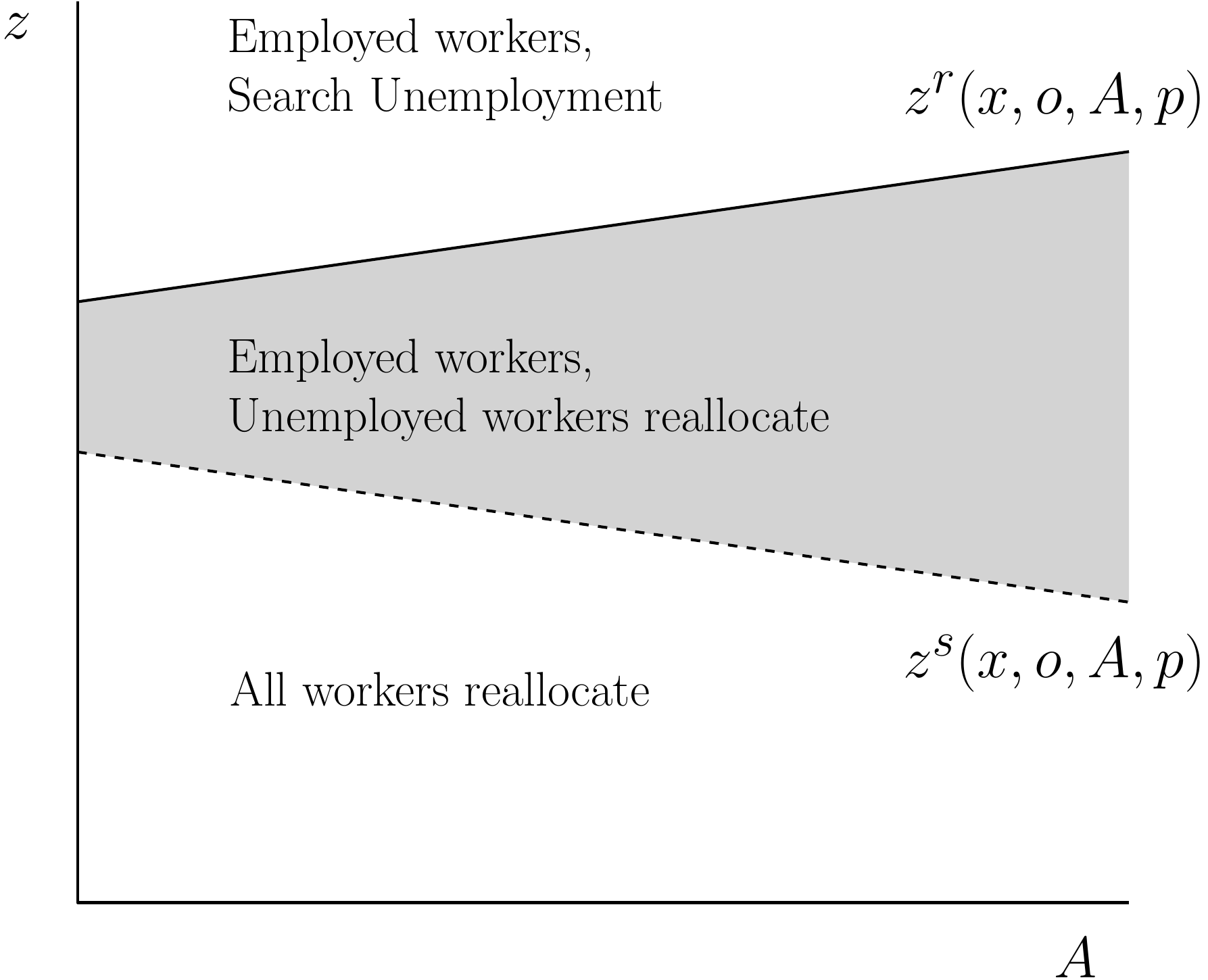}}
\subfloat[$z^s > z^r$]{\label{rest} \includegraphics [height=0.2 \textheight, width=0.4 \textwidth] {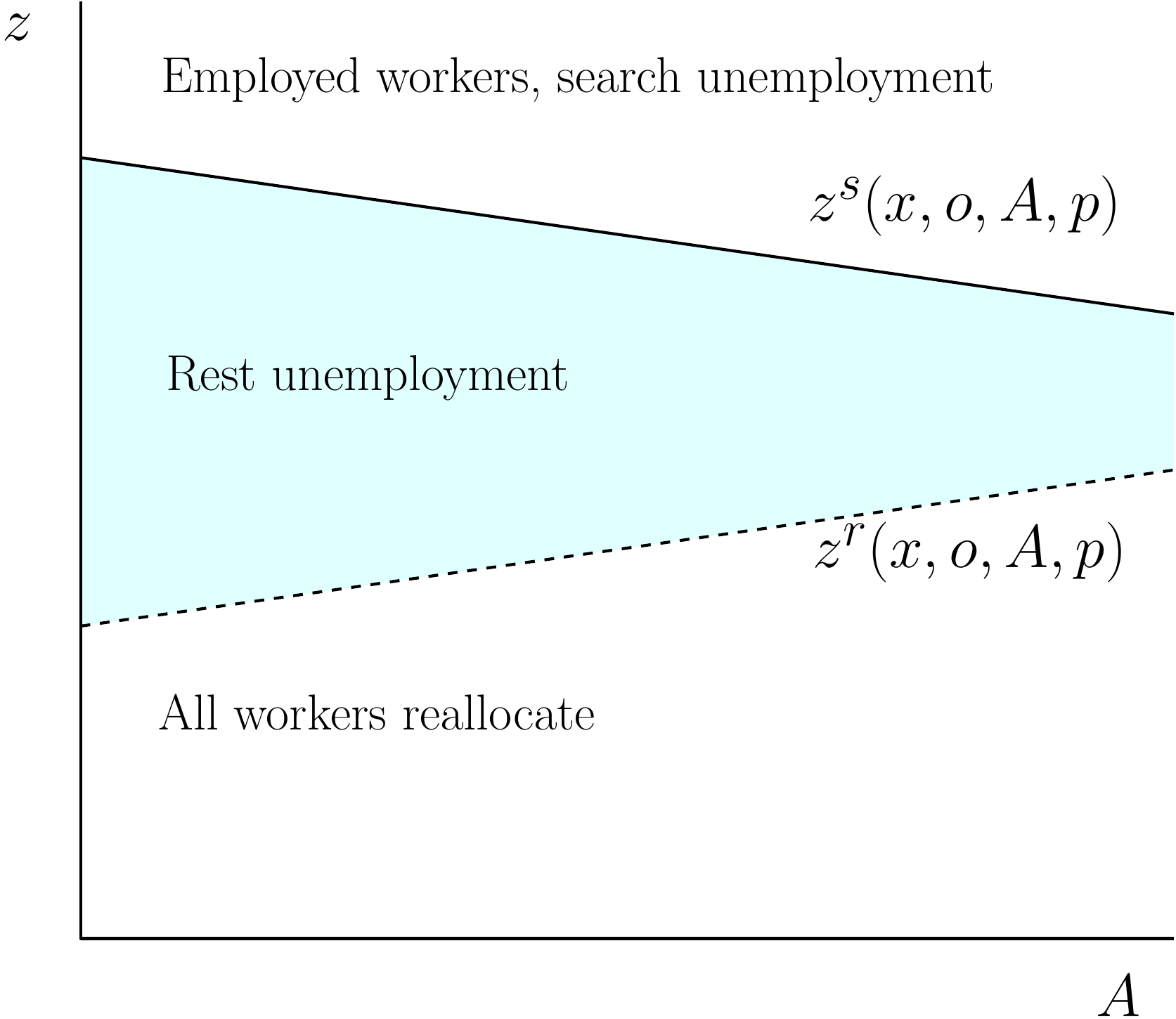}}
}
\label{f:cutoffprod}
\end{figure}

The relative position and the slopes of $z^r(.)$ and $z^s(.)$ are crucial determinants of the long-run and cyclical outcomes in our model. To show this, we first discuss the implications of their relative position and then those of their slopes. Figure \ref{no rest} illustrates the case in which $z^{r}>z^{s}$ for all $A$, holding constant $p$, $o$ and $x$. Here having a job makes a crucial difference on whether a worker stays or leaves his occupation. When an employed worker has a $z\in[z^s,z^r)$, the match surplus is enough to keep him attached to his occupation. For an unemployed worker with a $z$ in the same interval, however, the probability of finding a job is sufficiently small to make searching across occupations the more attractive option, even though this worker could generate a positive match surplus if he were to become employed in his pre-separation occupation. For values of $z<z^s$, all workers search across occupations. For values of $z\geq z^r$, firms post vacancies and workers remain in their occupations, flowing between unemployment and employment as in the canonical DMP model.

Figure \ref{rest} instead shows the case in which $z^{s}>z^{r}$ for all $A$. Workers who endogenously separate into unemployment, at least initially, do not search across occupations, while firms do not create vacancies in sub-markets associated with values of $z<z^s$. These two cutoffs create an area of inaction, in which workers become \emph{rest unemployed} during the time their $z$ lies in $[z^r,z^s)$: they face a very low -- in the model (starkly) zero --  contemporaneous job finding probability, but still choose to remain attached to their occupations. The stochastic nature of the $z$ process, however, implies that these workers face a positive expected job finding probability for the following period. Only after the worker's $z$ has declined further, such that $z<z^r$, the worker searches across occupations. For values of $z \geq z^s$, the associated sub-markets function as in the DMP model.

An unemployed worker is considered \emph{search unemployed} during the time in which his $z\geq z^s$, as in the associated labor markets firms are currently posting vacancies. A worker whose current $z<z^r$ is considered \emph{reallocation unemployed} only during the time in which he is trying to find another occupation that offers him a $z>z^r$. Once he finds such an occupation, he continues his unemployment spell potentially with periods in search and rest unemployment, depending on the relative position of $z^s$ and $z^r$ and the initial draw and evolution of his $z$ in such an occupation. The stochastic nature of the $z$ process implies that search, rest and reallocation unemployment are not fixed characteristics, but transient states during an unemployment spell. Therefore, to be consistent with the analysis of Section 2, an \emph{occupational mover} is a worker who left his old occupation, went through a spell of unemployment (which could encompass all three types of unemployment) and found a job in a different occupation.

A key decision for an unemployed worker is whether to remain in his occupation, waiting for his $z$ to improve, or to search across occupations, drawing a new $z$. Periods of rest unemployment arise when the option value of waiting in unemployment is sufficiently large. However, search frictions imply that there is also an option value associated with waiting in employment in an existing job match. In the face of irreversible match destruction, workers remain employed at lower output levels relative to the frictionless case because of potential future improvements in their $z$-productivities. This drives the separation cutoff function down. The tension lies in that these two waiting motives work against each other. Which one dominates depends on parameter values.

Using a simplified version of the model without aggregate or occupation-specific shocks, we show that the difference $z^s - z^r$ increases when $c$, $b$ or $x$ increase (see Supplementary Appendix C.1). Although it is intuitive that a higher $c$ or $x$ reduces $z^r$ by making occupational mobility more costly, they also reduce $z^s$ by increasing the match surplus and making employed workers less likely to separate. We show that, overall, the first effect dominates. A rise in $b$ decreases $z^r$ by lowering the effective cost of waiting, while decreasing the match surplus by increasing $W^U(.)$ and hence increasing $z^s$, pushing towards rest unemployment. We also show that a higher degree of persistence in the $z$ process decreases $z^s - z^r$ as it decreases the option value of waiting.

Figure \ref{f:cutoffprod} shows the case of countercyclical job separation decisions ($\partial z^s(.) / \partial A<0$) and procyclical occupational mobility decisions ($\partial z^r(.) / \partial A>0$), as suggested by the data. The relative position of $z^s$ and $z^r$ is an important determinant of the cyclicality of occupational mobility decisions. Using a simplified version of the model without occupation-specific shocks, we show that when $z^s>z^r$ one obtains procyclical occupational mobility decisions without the need of complementarities in the production function (see Supplementary Appendix C.1). This arises as with search frictions wages and job finding probabilities increase with $A$, and complement each other to increase the expected value of occupational mobility (relative more than in the frictionless case). In addition, the presence of rest unemployment reduces the opportunity cost of mobility, making the latter less responsive to $A$. This occurs as any change in $A$ does not immediately affect the utility flow of the rest unemployed.

The relative position of $z^s$ and $z^r$ also affects the cyclicality of job separation decisions. When $z^s-z^r>0$ is sufficiently large, job separations decisions mainly reflect whether or not an employed worker should wait unemployed in his current occupation for potential improvements in his $z$. As occupational mobility is uncertain and only a potential future outcome, it is discounted. Thus rest unemployment moderates the feedback of procyclical occupational mobility decisions on the cyclicality of job separation decisions.

\vspace{-0.55cm}

\section{Quantitative Analysis}

\vspace{-0.15cm}

As the relative position and the slope of the $z^s$ and $z^r$ cutoffs can only be fully determined through quantitative analysis, we now turn to estimate the model and investigate its resulting cyclical properties.

\vspace{-0.55cm}

\subsection{Calibration Strategy} We set the model's period to a week and the discount factor $\beta=(1-d)/(1+r)$ is such that the exit probability, $d$, is chosen to match an average working life of 40 years and $r$ is chosen such that $\beta$ matches an annual real interest rate of 4\%. We target data based on major occupational groups and task-based categories as done in Section 2. Our classification error model allows us to easily correct for aggregate and occupation-specific levels of miscoding by imposing the $\mathbf{\Gamma}$-correction matrix on simulated worker occupational flows at the required level of aggregation.

\vspace{-0.55cm}

\paragraph{Aggregate and occupation productivities} The production function is assumed multiplicative and given by $y_{o}=Ap_{o}xz$ for all $o \in O$, chosen to keep close to a ``Mincerian'' formulation. The logarithm of aggregate productivity, $ \ln A_{t}$, follows an AR(1) process with persistence and dispersion parameters $\rho_A$ and $\sigma_A$. For a given occupation $o$, the logarithm of the occupation-wide productivity is given by $\ln p_{o,t}= \ln \overline{p}_{o}+ \epsilon_{o} \ln A_{t}$, where $\overline{p}_{o}$ denotes this occupation's constant productivity level and $\epsilon_{o}$ its cyclical loading. This formulation implies that different occupations can have different sensitivities to the aggregate shock and hence different relative attractiveness to workers over the business cycle.\footnote{The evidence presented in Supplementary Appendix B.3 suggests that our approach is consistent with the observed cyclical behavior of net occupational flows, where the majority of occupations exhibit a very similar cyclical pattern across several recession/expansion periods.} We consider occupation-wide productivity differences at the level of task-based categories, $O=\{NRC, RC, NRM, RM \} $. All major occupations within a task-based category $o \in O $ then share the same $p_{o,t}$. This approach not only simplifies the computational burden by reducing the state space, but is also consistent with the evidence presented in Figure \ref{f:grossnet_occmob_per_occ} showing that within the majority of task-based categories all major occupations' net flows exhibit the same sign. To further simplify we normalize the employment weighted average of $\overline{p}_{o}$ and of $\epsilon_{o}$ across $o \in O$ to one.

\vspace{-0.55cm}

\paragraph{Worker heterogeneity within occupations} The logarithm of the worker's idiosyncratic productivity, $\ln z_{t}$, is also modelled as an AR(1) process with persistence and dispersion parameters $\rho_z$ and $\sigma_z$. The normalization parameter $\underline{z}_{norm}$ moves the entire distribution of $z$-productivities such that measured economy-wide productivity averages one. Occupational human capital is parametrized by a three-level process $h=1,2,3$, where $x^{1}=1$. Employed workers stochastically increase their human capital one level after five years on average. With probability $\gamma _{d}$ the human capital of an unemployed worker depreciates one level until it reaches $x^{1}$.

To allow for differences in the separation rates across young and prime-age workers that are not due to the interaction between $z$ and $x$, we differentiate the exogenous job separation probability between low $(x^{1})$ and high human capital $(x^{2},x^{3})$ workers: $\delta_{L}$ and $\delta_{H}$. The matching function within each sub-market $(z,x)$ is given by $m(\theta)=\theta^\eta$.

\vspace{-0.55cm}

\paragraph{Search across occupations} The probability that a worker in a major occupation within task-based category $o$ receives the new $z$ from a different major occupation in task-based $\tilde{o}$ is parametrized as $\alpha(s_{\tilde{o}};o)=\overline{\alpha}_{o,\tilde{o}}^{(1-\nu)} s_{\tilde{o}}^{\nu}$ for all $o,\tilde{o}$ pairs in $O$ and $s_{\tilde{o}} \in [0,1]$. The parameter $\nu \in[0,1]$ governs the responsiveness of the direction of search across occupations due to differences in $p_{o}$. The parameter $\overline{\alpha}_{o,\tilde{o}}$ is a scaling factor such that $\sum_{\tilde{o} \in O} \overline{\alpha}_{o,\tilde{o}}=1$. It captures the extent to which an unemployed worker in task-based category $o$ has access to job opportunities in another task-based category $\tilde{o}$. Since $\sum_{\tilde{o} \in O} \alpha(s_{\tilde{o}};o) \leq 1$, this formulation implies that if a worker in $o$ wants to obtain a new $z$ with probability one, he will choose $s_{\tilde{o}}=\overline{\alpha}_{o,\tilde{o}}$ for all $\tilde{o} \in O$. If a worker wants to take into account current occupation-wide productivity differences, he will choose $s_{\tilde{o}} \neq \overline{\alpha}_{o,\tilde{o}}$ for at least some $\tilde{o}$. The cost of doing so is the possibility of not receiving a new $z$ at all (i.e. $\sum_{\tilde{o} \in O} \alpha(s_{\tilde{o}};o)< 1$) and paying $c$ again the following period. The concavity parameter $\nu$ determines the extent of this cost, with higher values of $\nu$ leading to lower probabilities of not receiving a new $z$.

The formulation of $\alpha(s_{\tilde{o}};o)$ is convenient for it implies that the optimal value of $s_{\tilde{o}}$ can be solved explicitly,
\begin{equation*}
s^{*}_{\tilde{o}}(\omega)=\frac{e^{\frac{1}{1-\nu} \log [\overline{\alpha}_{o,\tilde{o}}^{(1-\nu)} (\int_{\underline{z}}^{\overline{z}}W^{U}(\tilde{z},x_{1},\tilde{o},A,p)dF(\tilde{z})-\hat{W}^U(\omega))]}}{\sum_{\tilde{o}\in O^{-}}e^{\frac{1}{1-\nu} \log [\overline{\alpha}_{o,\tilde{o}}^{(1-\nu)} (\int_{\underline{z}}^{\overline{z}}W^{U}(\tilde{z},x_{1},\tilde{o},A,p)dF(\tilde{z})-\hat{W}^U(\omega))]} }
\end{equation*}
with $\sum_{\tilde{o} \in O^{-}} s^{*}_{\tilde{o}}(\omega)=1$ and takes a similar form as the choice probabilities obtained from a multinomial logit model.\footnote{To derive this result note that for each $s_{\tilde{o};o}$ equation \eqref{eq:reall} yields the first order condition $s^{*}_{\tilde{o}}(\omega)=\Big[\frac{\nu\overline{\alpha}_{o,\tilde{o}}^{(1-\nu)}}{\mu} \int_{\underline{z}}^{\overline{z}}W^{U}(\tilde{z},x_{1},\tilde{o},A,p)dF(\tilde{z})-\hat{W}^U(\omega)\Big]^{1/(1-\nu)}$, where $\mu$ is the multiplier of the constraint $\sum_{\tilde{o} \in O^{-}} s^{*}_{\tilde{o}}(\omega)=1$. Substituting out $s^{*}_{\tilde{o}}(\omega)$ in the constraint and using the change of variable $X^\frac{1}{1/(1-\nu)}=e^{\frac{1}{1/(1-\nu)}\log(X)}$ leads to the above expression. See Carrillo-Tudela, et al. (2022) for a detailed discussion.} Note that $\overline{\alpha}_{o,\tilde{o}}$ appears inside the closed form and can freely shape \emph{bilateral} flows between occupations. This leaves $\nu$ free to capture the responsiveness to cyclically changing occupation-wide productivities, which in turn allows us to capture net mobility flows over the cycle. It also leaves free the \textit{persistent} career match $z$ process to drive excess mobility in a way that is consistent with the patterns documented in Section 2.\footnote{Many multi-sector models use the random utility model to drive excess mobility, where additive taste shocks are distributed i.i.d Type 1 Extreme Value (see Chodorow-Reich and Wieland, 2020, Wiczer, 2015, Dvorkin, 2014 and Pilossoph, 2014, among others). In the most tractable of such settings, underlying gross flows are constant at all times (e.g. Chodorow-Reich and Wieland, 2020). More generally, when the reallocation decision involves $\max_{o \in O} \{U_o(.) + \epsilon_o\}$, where $U_o(.)$ is the value of being in occupation $o$ and $\epsilon_o$ is the taste shock, this imposes a symmetry. All mobile workers who are considering occupations in set $O$ have the same distribution over the destinations in $O$, independently of where they originated. Here we want to explicitly break this symmetry to be consistent with the bilateral flows of the transition matrix, a feature we can do through $\overline{\alpha}_{o,\tilde{o}}$ without giving up on a convenient closed form. Our formulation also decouples the cyclical responsiveness from the cross-sectional flows, again without giving up on the closed form. In contrast, in the additive taste shock setting fitting cross-sectional patterns constrains the mobility response to cyclical shifts in $U_o(.)$: both dimensions rely on how differences in $U_o(.)$ translate into differences in the cdf of $\epsilon_o$ (or a transformation of the latter).}

Since our data analysis covers three decades, we need to distinguish the observed long-run changes in the employment-size distribution from their cyclical changes. For this we first externally calibrate the initial size distribution to match the one observed in the SIPP in 1984. This results in setting the employment proportions for $NRC$, $RC$, $NRM$, $RM$ to 0.224, 0.292, 0.226 and 0.258, respectively, at the start of the simulation. This size distribution then changes over time due to unemployed workers' mobility decisions. Let $\psi_{o}$ denote the exogenous probability that a new entrant is allocated to task-based category $o$ such that $\sum_{o \in O} \psi_{o}=1$. This worker is then randomly allocated to a major occupation within the drawn task-based category at the point of entry, and is allowed to search across occupations to obtain first employment somewhere else.

\vspace{-0.55cm}

\paragraph{Simulated method of moments} In the above parametrization $[c,$ $\rho_z$, $\sigma_z$, $\underline{z}_{norm}]$ govern occupational mobility due to idiosyncratic reasons (excess mobility); $[x^{2},$ $x^{3}, \gamma _{d},\delta_{L}, \delta_{H}]$ govern differences in occupational human capital; $\left[\overline{p}_{o}, \epsilon_{o}, \overline{\alpha}_{o,\tilde{o}}, \nu, \psi_{o} \right]$ for all $o,\tilde{o} \in$ $\{NRC$, $RC$, $NRM$, $RM\}$ govern occupational mobility due to occupation-wide productivity differences (net mobility); and the remainder parameters $[ k, b, \eta,$ $\rho_A, \sigma_A]$ are shared with standard DMP calibrations. All these parameters are estimated by minimising the sum of squared distances between a set of model simulated moments and their data counterparts. For consistent measurement we generate `pseudo-SIPP panels' within one hundred time-windows each of 30 year length and follow the same procedures and definitions to construct the moments in data and in model simulations.

Figures \ref{f:mdur_all}-\ref{f:survival_youngprime} and Table \ref{t:minimumdistance} show the set of moments used to recover these parameters as well as the fit of the model. The calibrated model provides a very good fit to the data across all the targeted dimensions. The mobility-duration profiles and survival functions primarily inform the excess mobility and the human capital parameters. Employer separations patterns inform the parameters shared with DMP calibrations, except for the persistence and standard deviation of the aggregate productivity process, $\rho_A$ and $\sigma_A$, which are informed by the corresponding parameters of the series of output per worker ($outpw$) obtained from the BLS, $\rho_{outpw}$ and $\sigma_{outpw}$, and measured quarterly for the period 1983-2014.\footnote{We cannot set $\rho_A$ and $\sigma_A$ directly because the composition of the economy changes with the cycle due to workers' endogenous separation and reallocation decisions. We measure output in the model and data on a quarterly basis (aggregating the underlying weekly process in the model). For the data, we HP-filtered the series of (log) output per worker for the period 1970 to 2016, with a filtering parameter of 1600. Then, we use the persistence and the variance parameters of this series calculated over the period 1983-2014, which is the period that the SIPP and the BLS series overlap.} The net mobility patterns inform the occupation-specific productivities, occupation distribution for new entrants and the imperfect direct search technology. The latter adds a number of extra parameters to the estimation, particularly the scale parameters $\overline{\alpha}_{o,\tilde{o}}$. As mentioned above these allow us to capture very well the relevant differences observed across occupations. We now present the arguments that justify the choice of moments, keeping in mind that all parameters need to be estimated jointly.

\vspace{-0.55cm}

\subsection{Gross occupational mobility and unemployment duration}

\begin{figure}[t]
\caption{Data and Model Comparison (including Targeted Moments)}\label{f:minimumdistance}
\centering
\resizebox{1\textwidth}{!}{
\subfloat[Mob./Unemp. Duration (All)]{\centering\includegraphics[width=0.35 \textwidth]{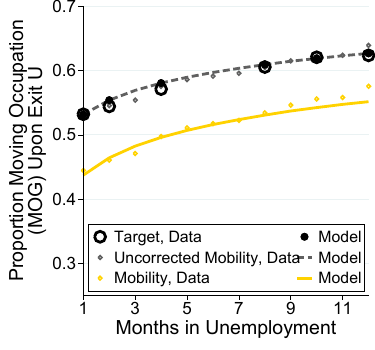}
\label{f:mdur_all} }
\subfloat[Mob./Duration Young, Prime]{\centering\includegraphics[width=0.35 \textwidth]{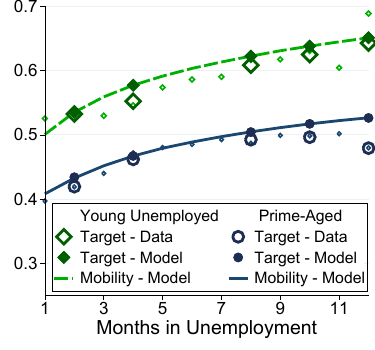}
\label{f:mdur_young} }
\subfloat[Cycl. Mobility/Duration]{\centering\includegraphics[width=0.35 \textwidth]{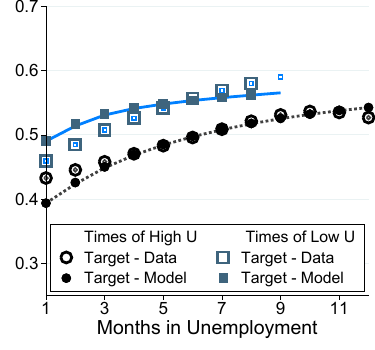}
\label{f:mdur_cyclical} }
}
\resizebox{1\textwidth}{!}{\subfloat[{Survival Profile, All}]{\centering\includegraphics[height=0.2 \textheight]{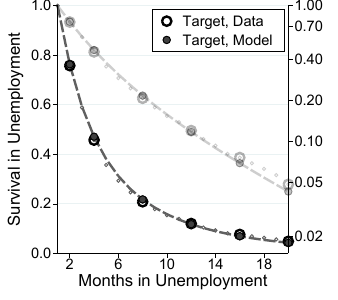}

\label{f:survival_all} }
\subfloat[{Survival Young, Prime}]{\centering\includegraphics[height=0.2 \textheight]{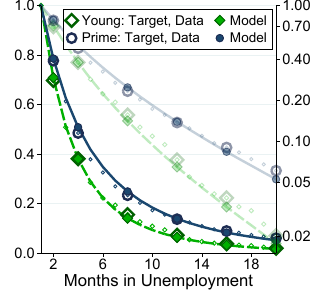}

\label{f:survival_youngprime} }
\subfloat[{Survival Movers, Stayers}]{\centering\includegraphics[height=0.2 \textheight]{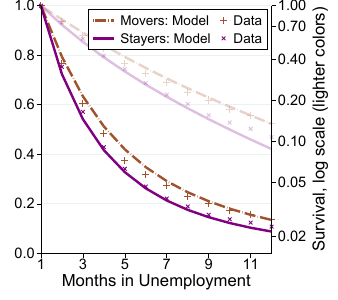}

\label{f:survival_moverstayer} }}
\end{figure}

A worker's attachment to his pre-separation occupation during an unemployment spell depends on the properties of the $z$ process, the human capital process and the reallocation cost $c$. The aggregate and age-group mobility-duration profiles depicted in Figures \ref{f:mdur_all} and \ref{f:mdur_young} (see also Section 2) play an important role in informing these parameters.

The aggregate mobility-duration profile contains information about $c$ and $\rho_z$. As shown in Lemma 1 (see Supplementary Appendix C.1) changes in the overall level of mobility lead to opposite changes in $c$. The slope of the profile informs $\rho_z$ primarily through the time it takes unemployed workers to start searching across occupations. A lower $\rho_{z}$ increases the relative number of unemployed workers deciding to reallocate at shorter durations, decreasing the slope of the model's mobility-duration profile. Lemma 1, however, also implies that a lower $\rho_{z}$ reduces overall mobility (ceteris paribus), creating a tension between $c$ and $\rho_{z}$ such that an increase in $\rho_{z}$ must go together with an increase in $c$ to fit the observed mobility-duration profile as depicted in Figure \ref{f:mdur_all}.

To help identify $\sigma_{z}$ we match instead the mobility-duration profiles of young and prime-aged workers. For given values of $x$, a larger value of $\sigma_{z}$ leads to a smaller importance of human capital differences relative to $z$ differences in workers' output. This brings the simulated occupational mobility patterns across age groups closer together, creating a negative relationship between $\sigma_{z}$ and the difference between the mobility-duration profiles of young and prime-aged workers. Figure \ref{f:mdur_young} shows that the model is able to resolve this tension very well. The model also remains fully consistent with the much larger contribution of excess mobility relative to net mobility in accounting for the mobility-duration profile at all durations (see Figure 3a, Online Appendix B.1) .

The parameters $x^2$ and $x^3$ are informed by the observed five and ten-year returns to occupational experience. As it is difficult to accurately estimate the later with the SIPP, we use the OLS estimates for 1-digit occupations reported in Kambourov and Manovskii (2009) from the PSID and estimate the same OLS regression in simulated data.\footnote{We use the OLS estimates because occupation selection occurs both in the model and in the data, where selection arises as measured returns are a result of two opposing forces: human capital acquisition and $z$-productivity mean reversion.}

Calibrations with or without occupational human capital depreciation yield very similar long-run moments (see Online Appendix B.2). This occurs as the gradual loss of occupational attachment with unemployment duration underlying the mobility-duration profile can be generated by human capital depreciation or the $z$ process. To differentiate these two forces we use the cyclical shift of the mobility-duration profile. During recessions longer unemployment spells imply that expected depreciation is higher, making employed workers more attached to their jobs and unemployed workers less attached to their occupations. At the same time low aggregate productivity interacted with $z$ typically makes employed workers less attached to their jobs and unemployed workers more attached to their occupations. To inform this tension and recover $\gamma_{d}$ we fit the mobility-duration profile in recessions and expansions as depicted in Figure \ref{f:mdur_cyclical}.

\begin{table}[t]
\centering
\caption{Targeted Moments. Data and Model Comparison}\label{t:minimumdistance}
  \resizebox{1.0
  \textwidth}{!}{
\small{
\begin{tabular}{lcclcc}
\toprule
\toprule \multicolumn{6}{c}{Panel A: Economy-wide moments} \\ \midrule
Moment      & Model & Data & Moment      & Model     & Data \\ \cmidrule(lr){1-3} \cmidrule(lr){4-6}
Agg. output per worker mean & 0.999 & 1.000 & Rel. separation rate young/prime-aged & 2.002 & 1.994 \\
Agg. output per worker persistence, $\rho_{outpw}$ & 0.764 & 0.753 &  Rel. separation rate recent hire/all  & 5.169 & 4.944\\
Agg. output per worker st. dev., $\sigma_{outpw}$ & 0.009 & 0.009  &  Prob (unemp. within 3 yr for empl.) & 0.151  & 0.122 \\
Mean unemployment & 0.036 & 0.035 & Empirical elasticity matching function & 0.532 & 0.500 \\
Task-based gross occ. mobility rate &  0.280  & 0.289 & 5-year OLS return to occ. tenure & 0.143 & 0.154 \\
Repeat mobility: occ. stay after stay & 0.598 & 0.634   & 10-year OLS return to occ. tenure & 0.219  & 0.232 \\
Occ. mobility young/prime-aged & 1.173 & 1.163 & Average u. duration movers/stayers & 1.184 & 1.139 \\
Occ. mobility-duration profiles & \multicolumn{2}{c}{Fig 5a,b,c} & U. survival profiles & \multicolumn{2}{c}{Fig 5d,e} \\
\end{tabular}
}
}

\resizebox{1.0
\textwidth}{!}{
\small{
\begin{tabular}{lccrccrccccrcccc}
\toprule
\multicolumn{16}{c}{\small{Panel B: Occupation-Specific Moments, Long-run}} \\
\midrule
& \multicolumn{2}{c}{Proportion} & & \multicolumn{2}{c}{Net mobility} & & & \multicolumn{8}{c}{Transition Matrix \ \ \ \ \ \ \ \ \ \ \ \ \ \ } \\
& \multicolumn{2}{c}{empl. size $o_{2014}$} &       & \multicolumn{2}{c}{\emph{Mean}} &       & \multicolumn{4}{c}{Model} &       & \multicolumn{4}{c}{Data} \\
\cmidrule{2-3}\cmidrule{5-6}\cmidrule{8-11}\cmidrule{13-16}          & Model & Data &       & Model & Data     &       & NRC   & RC    & NRM   & RM    &       & NRC   & RC    & NRM   & RM \\
\cmidrule{2-3}\cmidrule{5-6}\cmidrule{8-11}\cmidrule{13-16}
NRC   & 0.337 & 0.328 &     & 0.008 & 0.006 &      & 0.763 & 0.163 &0.055 & 0.018&       & 0.721 & 0.167 & 0.084 & 0.028 \\
RC    & 0.246 & 0.258 &      & 0.007 & 0.000 &       & 0.175 & 0.681 & 0.108& 0.036 &       & 0.078 & 0.680 & 0.168 & 0.074 \\
NRM  & 0.260 & 0.260 &     & -0.027 & -0.021&     & 0.034 & 0.064 & 0.760  & 0.141 &       & 0.020 & 0.115 & 0.710 & 0.155 \\
RM    & 0.157 & 0.154 &      & 0.011 & 0.015 &       & 0.037 & 0.069 & 0.246 & 0.647 &       & 0.013 & 0.066 & 0.188 & 0.733 \\
\end{tabular}%
}
}

\resizebox{1.0
\textwidth}{!}{
\small{
\begin{tabular}{lccrccrccrccrcc}
\toprule
\multicolumn{15}{c}{Panel C: Occupation-Specific Moments, Cyclical } \\
\midrule
& \multicolumn{8}{c}{Net mobility} & & \multicolumn{2}{c}{$\Delta_{exp - rec}$} & & \multicolumn{2}{c}{}\\
& \multicolumn{2}{c}{\emph{Recessions}} &       & \multicolumn{2}{c}{\emph{Expansions}} &       & \multicolumn{2}{c}{\emph{Rec-Exp}} &       & \multicolumn{2}{c}{ (\text{inflow} $o / \text{all flows}$)} &       & \multicolumn{2}{c}{$\varepsilon_{UD_{o},u}/\varepsilon_{UD_{avg},u}$} \\
\cmidrule{2-3}\cmidrule{5-6}\cmidrule{8-9}\cmidrule{11-12}\cmidrule{14-15}
& Model & Data    &       &Model & Data   &       & Model & Data    &       & Model & Data   &       &Model & Data\\
\cmidrule{2-3}\cmidrule{5-6}\cmidrule{8-9}\cmidrule{11-12}\cmidrule{14-15}
NRC   & -0.008 & -0.011 &       & -0.009 & -0.002 &       & 0.002 & -0.008 &       & -0.003 & -0.010  &       & 0.988 & 1.096  \\
RC     & -0.009 & -0.005 &       & -0.005 & -0.001 &       & -0.003 & -0.005 &       &  0.004 & 0.003    &  & 1.055 & 1.026 \\
NRM  & 0.033 & 0.033  &        & 0.020 & 0.011 &          & 0.013 & 0.022 &            &  -0.028 & -0.054  &              &  0.890 &  0.759\\
RM    & -0.017 & -0.017 &       & -0.006 &  -0.008 &       & -0.011 & -0.009 &       & 0.026 & 0.060 &         & 1.072 & 1.119 \\
    \bottomrule
    \end{tabular}%
}
}
\end{table}

The unemployment survival function depicted in Figure \ref{f:survival_all} additionally inform the $z$ and $x$ processes. The extent of duration dependence is linked to the properties of the $z$ process (and the importance of search frictions) through its effect on the extent of true duration dependence and dynamic selection in our model, where the latter is driven by worker heterogeneity in $x$ and $z$ at the moment of separation. We use the cumulative survival rates at intervals of 4 months to reduce the seam bias found in the SIPP. The model also reproduces well the associated hazard functions (see Figures 1 and 2, Online Appendix B.1). The model captures that duration dependence is different across occupational stayers and movers and across age groups, where duration dependence is stronger among occupational stayers relative to movers and among young relative to prime-aged workers. Young occupational stayers have especially high job finding at low durations, which decrease faster with duration. In addition, the model replicates the (untargeted) unemployment duration distribution among all workers and separately by age groups. In particular the empirical amount of long-term unemployment that occurs in the face of high occupational mobility (see Table 1, Online Appendix B.1). Finally, we target the ratio between the average unemployment durations of occupational movers and stayers.

The elasticity of the matching function, $\eta$, at the sub-market $(z,x)$ level is obtained by estimating through OLS a log-linear relation between the aggregate job finding rate (the proportion of all unemployed workers in the economy who have a job next month) and aggregate labor market tightness (aggregate vacancies over aggregate unemployed) across quarters, in simulated data. The estimated elasticity $\hat{\eta}$ is targeted to the standard value of $0.5$ and allows us to indirectly infer $\eta$.

\vspace{-0.55cm}

\subsection{Employer separations}

A worker's attachment to employment depends on the size of search frictions. A higher value of $k$ leads to stronger search frictions through its effect on firm entry and labor market tightness. Larger search frictions push down the $z^s$ cutoff relative to $z^r$, reducing the extent of endogenous separations.\footnote{Intuitively, note that with $z^s<z^r$ and a persistent $z$-process, workers who endogenously separate will immediately change occupation (see Figure \ref{f:cutoffprod}). Since these workers will be above their $z^r$ cutoffs in the new occupation, they face a lower risk of further endogenous separations damping down this margin. However, with $z^s>z^r$ workers who endogenously separate and managed to become re-employed in the same occupation remain close to $z^s$, facing once again a high job separation probability. Among those who changed occupations, there will still be a mass of workers close to their $z^s$ cutoffs who face a high risk of future job separation. This leads to a larger amount of endogenous separations for both stayers and movers. As shown below, in the calibrated model $z^s>z^r$ and the hazard rate of job separations among new hires out of unemployment is greater for occupational stayers, 0.037, than for occupational movers, 0.027, as suggested by the previous arguments. This is qualitatively consistent with SIPP data, where we find a hazard rate among new hires of 0.026 for stayers and 0.024 for movers.} Therefore to inform $k$ (and the relative position of $z^s$ and $z^r$) we use the proportion of separations observed within a year of workers leaving unemployment relative to the overall yearly separation rate (``Rel. separation rate recent hire/all'') and the concentration of unemployment spells over a SIPP panel among the subset of workers who start employed at the beginning of the panel (``Prob (unemp. within 3yr for empl.)''). The probability that an occupational stayer becomes an occupational mover in the next unemployment spell (``Repeat mobility'') also informs endogenous separations and how these relate to occupational mobility. Although not shown here, the model is also consistent with the probability that an occupational mover remains a mover in the next unemployment spell, as documented in Section 2.4.

Given the job-finding moments, the overall job separation rate follows from targeting the average unemployment rate. As we focus on those who held a job previously, we use the most direct counterpart and construct the unemployment rate only for those who were employed before and satisfied our definition of unemployment (see Section 2). Note that this unemployment rate (3.6\%) is lower than the BLS unemployment rate, but we find it responsible for more than 0.75 for every one percentage point change in the BLS unemployment rate (see Online Appendix B.1 and Supplementary Appendix B.7 for details), consistent with the results of Hornstein (2013), Fujita and Moscarini (2017) and Ahn and Hamilton (2020).

The ratio of separation rates between young and prime-aged workers (``Rel. separation rate young/prime-aged'') as well as their survival functions in Figure \ref{f:survival_youngprime} inform $\delta_L, \delta_H$ and $b$. The extent of separations for young and prime-aged workers also informs us about $b$ through the positions of the $z^s$ cutoffs of low and high human capital workers relative to the average of these workers' productivities.

\vspace{-0.55cm}

\subsection{Net occupational mobility}

Variation over the business cycle can naturally inform the loading parameters $\epsilon_{o}$. We target the level of net mobility each task-based category exhibits in recessions and expansions (``Net mobility $o$, \emph{Recessions} and Net mobility $o$, \emph{Expansions}'') as well as their implied difference (``Net mobility $o$, \emph{Rec-Exp}''). We also regress (for each $o$) the completed (log) unemployment durations of those workers whose pre-separation task-based category was $o$ on the (log) unemployment rate and a time trend, and target the ratio between the estimated unemployment duration elasticity and the average elasticity across task-based categories, $\varepsilon_{UD_{o},u}/\varepsilon_{UD_{avg},u}$  (see Online Appendix B.1 for details). The advantage of this approach is that it allows us to leave untargeted the cyclicality of aggregate unemployment, which we separately evaluate in Section 5. To inform the values of $\overline{p}_{o}$ we target the average net mobility level of each $o$ (``Net mobility $o$,  \emph{Mean}'').

To recover $\nu$ we exploit the observed differences in the cyclicality of inflows across task-based categories. As $\nu$ increases, workers should be more sensitive (ceteris paribus) to cyclical differences in $p_{o}$ when choosing occupations, making the inflows to occupations with the higher $p_{o}$ respond stronger. To capture how cyclically sensitive are the inflows we compute, separately for expansions and recessions, the ratio of inflows into task-based category $o$ over the sum of all flows. For each $o$ we target the difference between the expansion and recession ratios, $\Delta_{exp - rec}$ (\text{inflow} $o / \text{all flows}$). To recover $\overline{\alpha}_{o,\tilde{o}}$ we target the observed task-based occupation transition matrix. To recover $\psi_{o}$ we use the employment-size distribution of task-based categories observed in 2014, the end of our sample period. We target the average gross mobility rate across task-based categories so that the model remains consistent with gross mobility at this level of aggregation.

\vspace{-0.55cm}

\subsection{Estimated parameters}

Table \ref{t:calibration} reports the resulting parameter values implied by the calibration. The estimated value of $b$ represents about $80\%$ of total average output, $y$. Vacancy cost $k$ translates to a cost of about 30\% of weekly output to fill a job. The elasticity of the matching function in each sub-market $(z,x)$ within an occupation is estimated to be $\eta=0.24$, about half of $\widehat{\eta}=0.5$ when aggregating all sub-markets across occupations.\footnote{The difference between $\eta$ and $\widehat{\eta}$ is mainly due to the effect of aggregation across sub-markets that exhibit rest unemployment. Workers in episodes of rest unemployed entail no vacancies, have zero job finding rates, do not congest matching in other sub-market, but are included in the aggregate number of unemployed. Hence they are included in the denominator of the aggregate labor market tightness and the aggregate job finding rate. It can be shown that this creates a wedge between $\eta$ and $\hat{\eta}=0.5$ that is governed by $\frac{0.5-\eta}{1-\eta} \varepsilon_{\hat{\theta},A}  =\varepsilon_{u^{s}/u,A}$, where $\varepsilon_{\hat{\theta},A}$ and $\varepsilon_{u^{s}/u,A}$ denote the cyclical elasticity of aggregate labor market tightness, $\hat{\theta}$, and the proportion of search unemployment over total unemployment, $u^{s}/u$, respectively. Since in the calibrated model both elasticities are positive, $\frac{0.5-\eta}{1-\eta}$ must also be strictly positive and hence $\eta< \widehat{\eta}=0.5$. In addition, each sub-market within an occupation has its own concave matching function and hence aggregating these concave functions across sub-markets also imply that the calibrated value of $\eta$ will further diverge from $0.5$.}

\begin{table}[t]
\caption{Calibrated Parameters}\label{t:calibration}
\vspace{-0.5cm}
\begin{center}
\resizebox{0.9\textwidth}{!}{
{\small
\begin{tabular}{|l|ccccccc|}
\hline
{\bf{Agg. prod. and search frictions}} & & $\rho_A$ &  $\sigma_A$  & $b$ & $k$  & $\eta$    & \\
\vspace{-0.35cm}& & & & & & &\\
& & $0.9985$ & $0.0020$ & $0.830$  & $124.83$ & $0.239$ &  \\
\hline
{\bf{Occ. human capital process}}&  & $x^{2}$ &  $x^{3}$  & $\gamma _{d}$ & $\delta_{L}$ & $\delta_{H}$ &  \\
\vspace{-0.35cm}& & & & & & &\\
                        &  &	$1.171$ & $1.458$ & $0.0032$ & $0.0035$ & $0.0002$ &  \\

\hline
{\bf{Occupational mobility}} & & $c$         &$\rho_{z}$ &	$\sigma_{z}$ & $\underline{z}_{norm}$ & $\nu$  &   \\
\vspace{-0.35cm}& & & & & & &\\

     			 & & $7.604$ & $0.9983$    & $0.0072$     & $0.354$                        & 0.04&  \\
\hline
 {\bf{Occupation-specific}}                &  $\overline{p}_{o}$ 	&	$\epsilon_{o}$    &	$\psi_{o}$ & $\overline{\alpha}_{o,NRC}$   & $\overline{\alpha}_{o,RC}$  &  $\overline{\alpha}_{o,NRM}$  & $\overline{\alpha}_{o,RM}$\\
\vspace{-0.35cm}& & & & & & &\\
\emph{Non-routine Cognitive}  &  1.019	&	1.081  &	0.620  &   0.436   &  0.560  & 0.004 & 0.000  \\
\emph{Routine Cognitive} &   0.988	&	1.120  &	  0.145 &    0.407  & 0.383 &  0.210 & 0.000 \\
\emph{Non-routine Manual} &   1.004	& 0.532  &	0.087 &  0.000  &   0.093  &  0.384 & 0.524 \\
\emph{Routine Manual} &    0.988	&	1.283  &	 0.147 &  0.000    &  0.140 &  0.767&  0.094 \\
\hline
\end{tabular}}}
\par\end{center}
\end{table}

The actual returns to occupational experience $x^{2}$ and $x^{3}$ are higher than the OLS returns, because occupational entrants select better $z$-productivities that typically mean-revert over time, dampening the average evolution of composite $xz$-productivity. The parameter $\gamma_{d}$ implies that a year in unemployment costs an experienced worker in expectation about 5\% of his productivity. The estimated value of $c$ and the sampling process imply that upon starting a job in a new occupation, a worker has paid on average a reallocation cost of 15.18 weeks (or about 3.5 months) of output. This suggests that reallocation frictions are important and add to the significant lose in occupational human capital when changing occupation.\footnote{The average reallocation cost is computed as the product of $c$ and the number of times workers sample a new occupation, which is 1.996 times.}

The $z$ process has a broadly similar persistence (at a weekly basis) as the aggregate shock process. Its larger variance implies there is much more dispersion across workers' $z$-productivities than there is across values of $A$. They are also much more dispersed than occupation-wide productivities. For example, the max-min ratio of $p_{o}$ is 1.13 (1.09) at the highest (lowest) value of $A$, where the $RM$ task-based category is the most responsive to aggregate shocks and $NRM$ the least. In contrast, the max-min ratio among $z$-productivities is 2.20. To gauge whether the dispersion across $z$-productivities is reasonable we calculate the implied amount of frictional wage dispersion using Hornstein et al. (2012) $Mm$ ratio. These authors find an $Mm$ between 1.46 and 1.90 using the PSID, while the estimated $z$-dispersion yields 1.41.

The estimated value of $\nu$ implies that the ability of workers to access job opportunities in other task-based categories plays an important role in shaping the direction of their search. The estimated values of $\overline{\alpha}_{o,\tilde{o}}$ imply that on average workers in $NRC$ have a low probability of drawing a new $z$ from manual occupations and vice versa; while workers in $NRM$ and $RM$ occupations mostly draw a new $z$ from these same two categories, although drawing from $RC$ is not uncommon. The value of $\nu$ also implies workers significantly adjust their direction of search as a response to cyclical differences in $p_{o}$. This is evidenced by the ability of the model to reproduce the observed cyclical changes in the net mobility patterns presented in Section 2 and Table \ref{t:minimumdistance}. Taken together, these estimates show a high degree of directedness when workers search across task-based categories.

\vspace{-0.55cm}

\section{Cyclical Unemployment Outcomes}

\vspace{-0.15cm}

We now evaluate the cyclicality of aggregate unemployment and its duration distribution in the model, noting that these were not targeted in the estimation procedure. Our aim is to evaluate the importance of excess and net mobility in generating these patterns. We first present the implications of the full model as estimated above. Then we discuss the implications of a re-estimated version where we shut down the heterogeneity in occupation-wide productivities.\footnote{In this version the observed net mobility patterns can be imposed exogenously to keep the model's gross occupational mobility patterns consistent with the evidence presented in Section 2 and Supplementary Appendix B.} With a slight abuse of terminology, we label this version ``excess mobility model'' as unemployed workers' occupational mobility decisions are based solely on the changing nature of their $z$-productivities and their interaction with $A$ and $x$. Online Appendix B.2 presents the estimation results of this model.

\begin{table}[t]
\centering
\caption{Business Cycle Statistics. Data (1983-2014) and Model }\label{t:volatilities1}
\resizebox{1\textwidth}{!}{
\begin{tabular}{ccccccccccccccccc}
\cline{1-8} \cline{10-17}
& \multicolumn{7}{c}{\textsc{Volatility and Persistence}} && & \multicolumn{7}{c}{\textsc{Correlations with u and output per worker}} \\ \cline{1-8} \cline{10-17}
& $u$ & $v$ & $\theta$ & $s$ & $f$ & $outpw$ & $occ m$ && & $u$ & $v$ & $\theta$ & $s$ & $f$  & $outpw$ & $occ m$\\ \cline{1-8} \cline{10-17} \cline{1-8} \cline{10-17}
&\multicolumn{7}{c}{\textbf{Data}} & & & \multicolumn{7}{c}{\textbf{Data}}\\\cline{1-8} \cline{10-17} \cline{1-8} \cline{10-17}
$\sigma$ &  0.15 &	0.11 &	0.25 &	0.10 &	0.09 &	0.01 & 0.03 &  &$u$ & 1.00 & -0.95	& -0.99 &0.83 &-0.86 &-0.50 & -0.56  \\
$\rho_{t-1}$ & 0.98 &	0.99 &	0.99 &	0.94 &	0.93 &	0.92 & 0.94 & & $outpw$ &  &  0.61 & 0.55 & -0.43 & 0.40 & 1.00 &0.46  \\ \cline{1-8} \cline{10-17} \cline{1-8} \cline{10-17}
&\multicolumn{7}{c}{\textbf{Full Model}} & & &\multicolumn{7}{c}{\textbf{Full Model}}\\ \cline{1-8} \cline{10-17} \cline{1-8} \cline{10-17}
$\sigma$ & 0.14  &  0.04 &  0.17 &  0.07 &  0.09 &   0.01 &  0.04 && $u$ &1.00 &  -0.66  &  -0.97  &   0.79  &  -0.89 &   -0.94  &  -0.85 \\
$\rho_{t-1}$ &0.93 &  0.91 &   0.92 &  0.88 &  0.93 &  0.88 & 0.94 && $outpw$ &  & 0.81 & 0.97 &   -0.91 &   0.94 & 1.00&  0.85 \\ \cline{1-8} \cline{10-17} \cline{1-8} \cline{10-17}
&\multicolumn{7}{c}{{\textbf{Excess Mobility Model}}} & & & \multicolumn{7}{c}{\textbf{Excess Mobility Model}}\\\cline{1-8} \cline{10-17} \cline{1-8} \cline{10-17}
$\sigma$ & 0.14  &  0.05 &  0.17 &  0.06 &  0.10 &   0.01 &  0.04 && $u$ &1.00 & -0.72 &  -0.98 &  0.80 &  -0.89 &  -0.95 & -0.87 \\
$\rho_{t-1}$ &0.95 &  0.91 &  0.94 &  0.89  & 0.94 &  0.94   & 0.93 && $outpw$  & & 0.85 & 0.97 &  -0.90 &   0.95 & 1.00&  0.89 \\
\cline{1-8} \cline{10-17}
\multicolumn{17}{p{1.37\textwidth}}{\footnotesize{Note: All variables are obtained from the SIPP, except output per worker, obtained from the BLS, and the vacancy rate where we use the composite help-wanted index developed by Barnichon and Nekarda (2012). All times series are centered 5Q-MA series of quarterly data, smoothing out the discreteness in the relatively flat cutoffs (relative to the grid) and noisy observation of especially occupational mobility. The cyclical components of the (log) of these time series are obtained by using an HP filter with parameter 1600. See Online Appendix B.1 for more details and results without the 5Q-MA smoothing.}}
\end{tabular}}
\end{table}

\vspace{-0.55cm}

\paragraph{Aggregate unemployment} Table \ref{t:volatilities1} shows the cyclical properties of the aggregate unemployment, vacancy, job finding, job separation and gross occupational mobility rates, computed from the data and the simulations. It shows that the full model is able to generate a countercyclical unemployment rate, together with a countercyclical job separation rate, procyclical job finding and gross occupational mobility rates. Table \ref{t:volatilities1} also shows that the cyclical volatilities and persistence of all these rates are very close to the data.

This aggregate behavior is not driven by a higher cyclicality of young workers' unemployment rate. In Online Appendix B.1 we show that the responsiveness of the unemployment rate to aggregate output per worker is slightly stronger for prime-aged workers than for young workers, leading to a countercyclical ratio of unemployment rates between young and prime-aged workers. Therefore, in the model the pool of unemployment shifts towards high human-capital, prime-aged workers during recessions, a feature noted by Mueller (2017).

The model also generates a strongly negatively-sloped Beveridge curve. The latter stands in contrast with the canonical DMP model, where it is known that endogenous separations hamper it from achieving a Beveridge curve consistent with the data. Because all $p_{o}$ co-move with $A$ and the loadings $\epsilon_{o}$ only create relative productivity differences, it also stands in contrasts with many multi-sector models that predict an upward sloping Beveridge curve. In these models unemployment fluctuations arise from the time-consuming reallocation of workers from sectors that experienced a negative shock to the ones that experienced a positive shock and lead to more vacancies created in the latter sector (see Chodorow-Reich and Wieland, 2020, for a recent exception).

\begin{figure}[!t]
    \centering
\begin{minipage}{.55\textwidth}
 \captionof{table}{Cyclicality of Duration Distribution} \label{t:semi-elast1}
\resizebox{1\textwidth}{!}{
\begin{tabular}{lcccccc}
\\
\toprule\toprule
\multicolumn{7}{c}{Panel A: Cyclicality of Duration Distribution} \\
\midrule
& \multicolumn{3}{c}{Elasticity wrt $u$} & \multicolumn{3}{c}{HP-filt. semi-el. wrt $u$} \\
\cmidrule(lr){2-4} \cmidrule(lr){5-7}
Unemp.  & Full  & Excess    &       &   Full & Excess &  \\
Duration & Model & Model  & Data   &  Model & Model  & Data  \\ \hline
$1-2$m  & -0.451 & -0.449 & -0.464 & -0.155 & -0.169 & -0.167 \\
$1-4$m  &-0.321  & -0.330 &  -0.363 & -0.168 &  -0.184 &  -0.184  \\
$5-8$m  & 0.415  & 0.346  &  0.320 &  0.067  & 0.071 & 0.076  \\
$9-12$m & 1.10 & 1.000  & 0.864  & 0.058  & 0.063 & 0.072  \\
$>13$m  & 1.817 & 1.742  &  1.375     & 0.044 &0.050 & 0.043\\
 \hline
 \end{tabular}}\\

\resizebox{1\textwidth}{!}{
 \begin{tabular}{lcccccc}
\midrule
\multicolumn{7}{c}{Panel B: Semi-Elasticity Duration wrt $u$, by Occupational Mobility} \\ \midrule
& \multicolumn{3}{c}{HP-filtered} &  \multicolumn{3}{c}{Log $u$ linearly detrended} \\
\cmidrule(lr){2-4}  \cmidrule(lr){5-7}
& Full & Excess &  &  Full & Excess & \\
& Model & Model & Data  &  Model & Model & Data  \\ \hline
 Movers &$2.8$ & $3.0$ & $3.2 $ &$2.2$ & $2.3$ & $2.0 $\\
 Stayers &$1.3$ & $1.5$ & $2.5 $ &$1.2$ & $1.2$ & $1.7 $\\
 \bottomrule
 \bottomrule
 \multicolumn{7}{p{1.3\textwidth}}{\footnotesize{Note: The elasticities are constructed using the cyclical components (after HP filtering or linear detrending) of the shares of unemployed workers by duration category and the aggregate unemployment rate.}} \\
 \end{tabular}}
\end{minipage}\hfill
\begin{minipage}{.413\textwidth}
\captionof{figure}{Cyclical Shift of the U. Duration Distribution}\label{f:cycl_distr_shift}
\centering
\includegraphics [width=1.1 \textwidth]{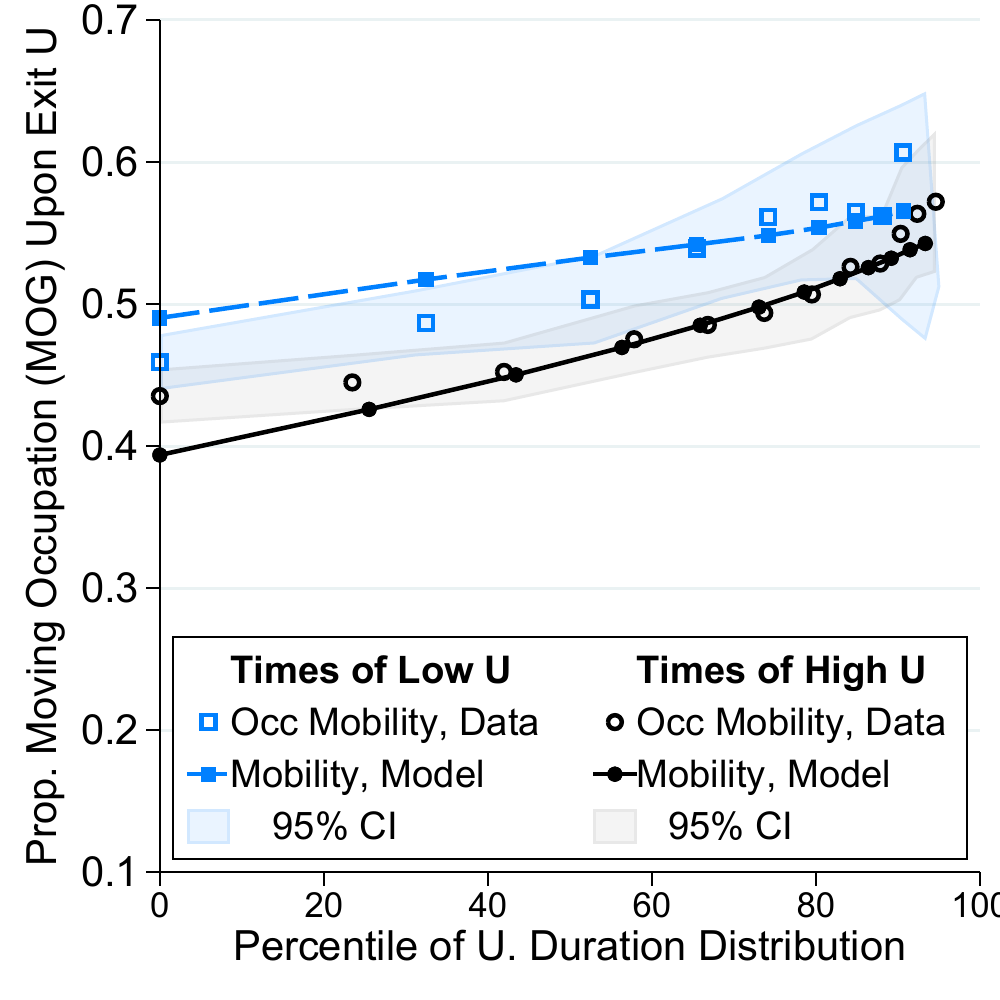}
\end{minipage}
\end{figure}

\vspace{-0.55cm}

\paragraph{Unemployment duration distribution}

Panel A in Table \ref{t:semi-elast1} evaluates the ability of the model to reproduce the shifts in the incomplete unemployment duration distribution with respect to changes in the unemployment rate. It shows that the shares of unemployed workers by duration exhibit a very similar degree of responsiveness with cyclical unemployment as in the data. Crucially the elasticity measure shows that the model creates a strong response in the shares of unemployment at long durations. When using the semi-elasticity the model generates a nearly perfect fit. Thus, in our model as in the data cyclical changes in the aggregate unemployment rate are driven by particularly strong cyclical changes in long-term unemployment.

An important force behind the increase in long-term unemployment during recessions is the larger increase in the unemployment duration of occupational movers relative to stayers. Panel B in Table \ref{t:semi-elast1} shows the cyclical responses of the average unemployment duration of movers and stayers using different measures. Along all of these measures the model's average unemployment duration of occupational movers increases more than that of stayers, an increase that is consistent with the data. Stayers' durations respond somewhat less relative to the data, between 56\% (relative to log HP-filtered unemployment) and 71\% (relative to linearly detrended unemployment). Relative to the lack of amplification in conventional DMP models, this still constitutes a large response. As in the data, the lengthening of movers' unemployment duration contributes meaningfully to the increase in long-term unemployment during recessions.

Figure \ref{f:cycl_distr_shift} shows how the untargeted shift in unemployment durations combines with the targeted shift of the mobility-duration profile. At any percentile of the unemployment duration distribution, the model generates a drop in occupational mobility in recessions. By comparing the observations' x-coordinates, this figure also illustrates that the cyclical shift of the model's duration distribution follows the data.

\vspace{-0.55cm}

\paragraph{Excess vs. net mobility} A key insight from Tables \ref{t:volatilities1} and \ref{t:semi-elast1} is that the aforementioned cyclical patterns are nearly identical to the ones generated by the excess mobility model. Online Appendix B.2 shows that this model also fits very well the economy-wide targets described in Table \ref{t:minimumdistance} and the estimated values of the corresponding parameters are nearly identical to those in the full model. We further show that this conclusion holds when considering non-employment spells. This comparison demonstrates that allowing workers to chose in which occupations to search due to difference in $p_{o}$ is not the reason why the model is able to replicate the cyclicality of unemployment or its duration distribution. Instead, it highlights the importance of workers' idiosyncratic career shocks and its interaction with $A$.

The two versions are successful in these dimensions because they yield similar implications for search, rest and reallocation unemployment. Section \ref{ss:rest unemployment} first demonstrates this claim using the excess mobility model. This shows in more detail the importance of having a persistent $z$ process for the cyclical performance of the model. Section 5.2 shows that the same forces occur within each $o$, although modulated by differences in the level and cyclical responsiveness of $p_{o}$ across occupations.

\vspace{-0.55cm}
\subsection{Main mechanism}\label{ss:rest unemployment}

As argued in Section 3.3, the relative position and slopes of $z^{s}$ and $z^{r}$ help determine the long-run and cyclical implications of our model.

\vspace{-0.55cm}

\paragraph{Relative position of $z^{s}$ and $z^{r}$}

\begin{figure}[t]
\centering
\caption{Cutoffs, Unemployment Distribution and Decomposition}
\label{f:calibration_reservation}
\resizebox{1.03\textwidth}{!}{\subfloat[Unemployment Distribution]{\includegraphics[width=0.35 \textwidth]{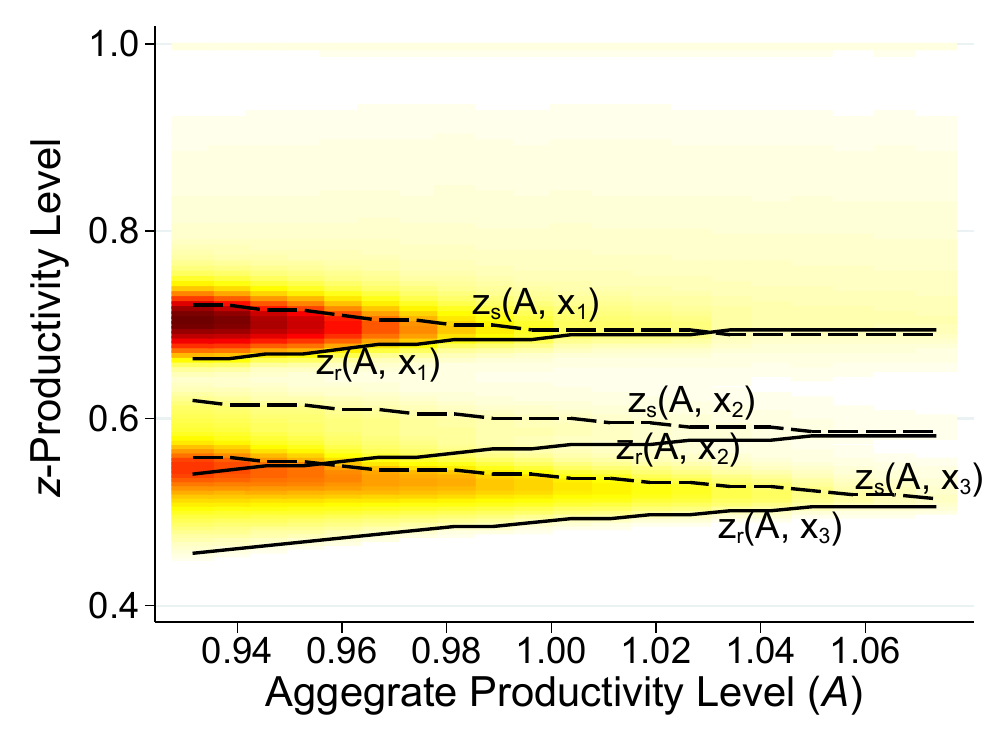}
}\subfloat[Employment Distribution]{\includegraphics[width=0.33 \textwidth]{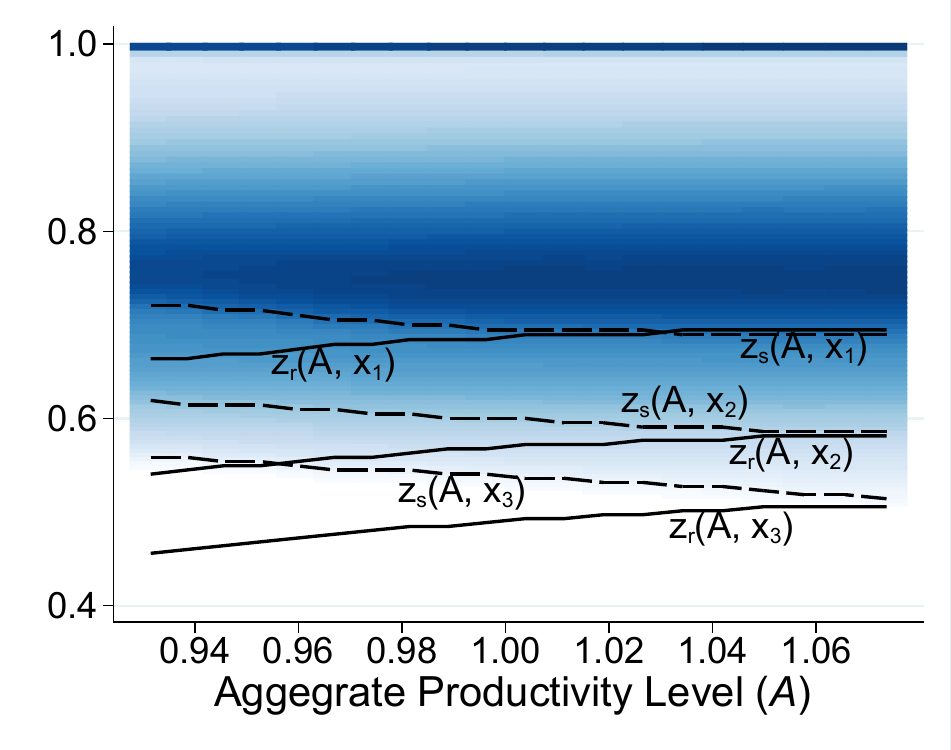}} \hspace{0.01cm}
\subfloat[Unemployment Decomposition]{\includegraphics[width=0.35 \textwidth]{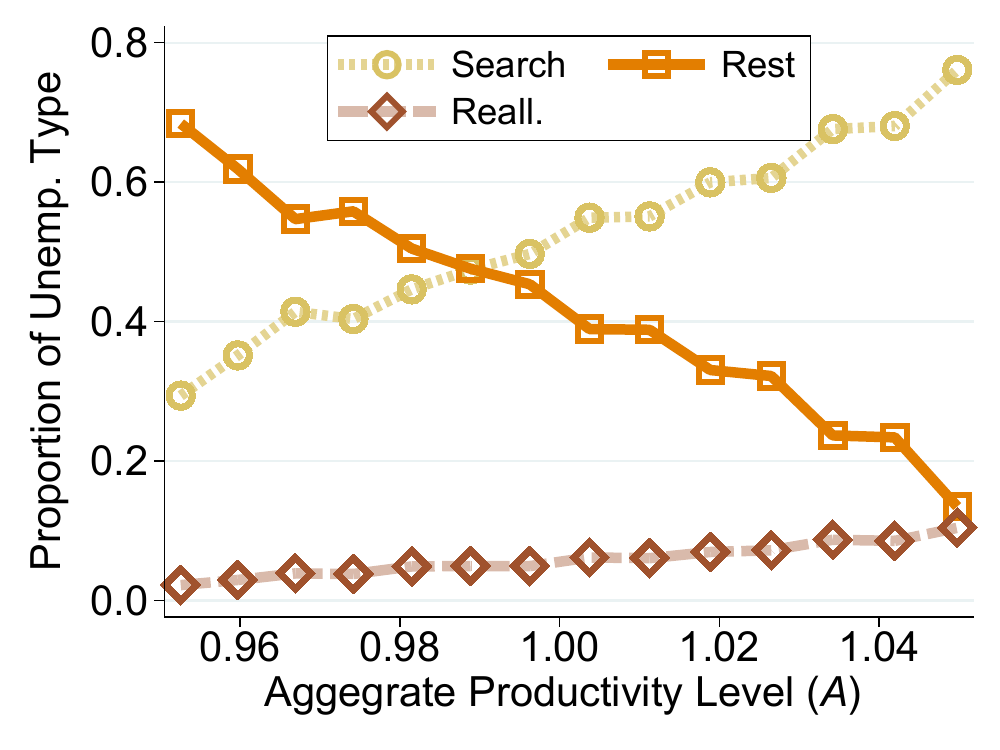}}}
\end{figure}

Figure \ref{f:calibration_reservation}a depicts the cutoff functions generated by the excess mobility model calibration as a function $A$ given $x$, where all occupations share the same cutoff functions. It shows that $z^s\geq z^r$ for nearly all $A$ and $h=1,2,3$.\footnote{As predicted by our theory, workers with higher human capital are less likely to change occupations relative to those with lower human capital. As $z^s(.,x^3)<z^s(.,x^1)$ the average level of separations is also lower for high human capital workers (noting that $\delta_{L}$ and $\delta_{H}$ also contribute to this difference). Once separated, high human capital workers spend on average a longer time in unemployment due to the larger distance between their $z^s$ and $z^r$ cutoffs.} This implies that periods of rest unemployment can occur together with periods of search and reallocation unemployment within the same unemployment spell as $A$ and $z$ evolve. Thus our calibration shows that the option value of waiting (as opposed to immediate reallocation) is important to explain the data. The importance of rest unemployment is grounded empirically in the mobility-duration profile and the unemployment survival functions.

The moderate increase of occupational mobility with unemployment duration implies that even though overall mobility is high, there is still a sizeable proportion of unemployed workers that regain employment in their pre-separation occupations after 12 months. The model rationalizes this feature with a $z$ process that, while persistent, has still meaningful uncertainty. An occupational stayer, even after a long time unemployed, is interpreted as the realization of the worker's earlier ``hope'' for the recovery of his $z$-productivity.

The presence of rest unemployment also rationalizes the moderate duration dependence observed in our unemployment sample and the relatively stronger duration dependence among occupational stayers. To illustrate this consider a set of workers with the same $x$ who just endogenously separated. Given $z^s \geq z^r$ and a persistent $z$ process, these workers will be initially close to $z^s$. A small positive shock would then suffice to move them above $z^s$, while only large negative shocks would take them below $z^r$. Hence at short durations these workers face relatively high job finding rates and, if re-employed, they will be most likely occupational stayers. Those who stayed unemployed for longer would have on average experienced further negative $z$ shocks and would face a higher probability of crossing $z^r$.

\vspace{-0.55cm}

\paragraph{Slope of $z^{s}$ and $z^{r}$}

As discussed in Section 3.3, the presence of rest unemployment makes it more likely for the model to generate countercyclical job separation decisions and procyclical occupational mobility decisions. Figures \ref{f:calibration_reservation}a,b shows that in the calibration this is indeed the case; i.e. $\partial z^s / \partial A <0$ and $\partial z^r / \partial A >0$ for each $x$. This property implies that during recessions there is an increased scope for episodes of rest unemployment; while in expansions there is an increased scope for episodes of search unemployment. Figure \ref{f:calibration_reservation}c illustrate this last feature by showing the proportion of workers facing search, rest or reallocation unemployment for a given value of $A$. Although both rest and search unemployment are countercyclical, search unemployment episodes are relatively more common when the economy moves from mild recessions up to strong expansions. It is only as recessions get stronger that rest unemployment episodes become more common.

The position and slopes of the cutoffs reveal a cyclical area of inaction, $[z^r(A;x),$ $z^s(A;x)]$ for each $x$. The cyclical change of the areas of inaction is important for it captures that workers' option value of waiting unemployed in their pre-separation occupations is higher in recessions than expansions, and is a key determinant of the cyclical performance of unemployment and vacancies in our model. The negative slope of the $z^s$ cutoffs together with the large mass of workers right above them (see Figure \ref{f:calibration_reservation}b) imply that a decrease in $A$ leads to a large increase in the inflow of workers into rest unemployment. The positive slope of the $z^r$ cutoffs implies that the same decrease in $A$ also leads to a large decrease in the outflow from rest unemployment via reallocation. These forces significantly add to the density of unemployed workers already ``trapped'' within these areas (see Figure \ref{f:calibration_reservation}a). Given that no firm in an occupation expects to be able to make a profit by hiring these workers, vacancy creation falls as well. As conditions improve the areas of inaction narrow considerably such that rest unemployed workers are now much more likely to get a $z$ shock that takes them below $z^r$ or above $z^s$.\footnote{In recessions that involve a 5\% reduction in $A$ relative to the mean, workers still face an average probability of about 25\% of transitioning out of rest unemployment within a month; and this probability sharply increases with $A$.} As the surplus from hiring these workers becomes positive and higher occupational mobility flows help workers increase their $z$-productivities, vacancy creation goes up across all occupations. The strong cyclical responses of rest and search unemployment, reflecting the changes in the areas of inaction, imply that aggregate unemployment also becomes highly responsive to $A$.\footnote{Episodes of reallocation unemployment make a small contribution to the cyclicality of $u$ because they only capture the time spent transiting between occupations, which is about 2 weeks on average, after which workers continue their jobless spell in episodes of rest or search before finding a job in a new occupation.} Online Appendix B.2 shows that these patterns occur across all human capital levels, explaining why we obtain unemployment, job finding and separations rates across age groups with similar cyclical responses.

The widening of the area of inaction during recessions also helps capture the cyclical behavior of the duration distribution. In recessions, long-term rest unemployed workers typically require a sequence of more and larger good $z$ shocks before becoming search unemployed in their pre-separation occupation. They would typically also require a sequence of more and larger bad shocks before deciding to reallocate. In contrast, for those workers who have just endogenously separated, $z^s$ is the cutoff that weighs most on their future outcomes. For these workers the distance to the nearest cutoff is therefore not as responsive to $A$ as it is for the long-term unemployed. Hence over the cycle we observe that the outflow rate of long-term unemployed workers responds more to changes in $A$ relative to the outflow rate of shorter-term unemployed workers. This mechanism then translates into a stronger increase in the share of long-term unemployed in recessions as shown in Table \ref{t:semi-elast1}, stronger than the one predicted based on the decline of $f$ alone. The widening of the area of inaction in recessions implies that the expected time spent in rest unemployment increases for (ex-post) occupations stayers as well as for (ex-post) movers, but more so for the latter. This rationalizes the stronger increase in average unemployment duration among occupational movers relative to stayers during recessions documented in Section 2.4.

\vspace{-0.55cm}

\paragraph{The role of human capital depreciation} Online Appendix B.2 shows that human capital depreciation is important in determining these dynamics as it affects the cyclical changes in the areas of inaction. As discussed in Section 4.2, without it the model generates aggregate unemployment, job finding and occupational mobility rates that are too volatile. This occurs as a potential loss of $x$ during unemployment decreases the option value of waiting in the occupation and flattens the $z^r$ cutoff. It also flattens the $z^s$ cutoff as it increases the option value of staying employed.

\vspace{-0.55cm}

\paragraph{The role of occupational mobility}

The cyclical sensitivity of the areas of inaction is also tightly linked with the existence of the $z^r$ cutoff and the properties of the $z$ process. To show this we re-estimate the model not allowing workers to change occupations. Online Appendix B.3 shows that this version of our model appears unable to reconcile the observe cyclical fluctuations of the unemployment duration distribution with those of the aggregate unemployment rate. This occurs as it cannot resolve a key trade-off. In the absence of the $z^r$ cutoff the estimated $z$ process is less persistent and exhibits a larger standard deviation, which creates enough heterogeneity in unemployment durations to allow it to match the empirical unemployment survival functions. However, this $z$ process also increases the heterogeneity of $z$ relative to the cyclical range of $A$. This makes the $z^s$ cutoffs less responsive and weakens the cyclical responses of job separations and the rate at which workers leave the new area of inaction, $[\underline{z}, z^s(A;x)$], where $\underline{z}$ denotes the lowest value of $z$.

\vspace{-0.55cm}

\subsection{Occupation Heterogeneity and Cyclical Unemployment}


The same mechanism described above also holds within each task-based category but its strength varies across these occupational groups. Consequently, unemployed workers face different unemployment outcomes that depend also on the identity of the occupation. Both the long-run and cyclical dimensions of occupation-wide productivity differences are relevant. To understand the former, column 5 in Table \ref{tab:TBOCshift} shows the contribution of unemployed occupational switchers in changing the observed sizes of the task-based categories in our calibration. This is compared to the contribution of the exogenous entry and exit process as captured by $d$ and $\psi_{o}$ (column 4 ``Entrants''), such that for each task-based category the two values add up to the change in the employment stock (column 3 - column 1). The calibration shows that $NRM$ occupations increased in size due to more unemployed workers switching to these occupations than away from them. In contrast, $RM$ and $RC$ decrease in size as more unemployed workers move away from these occupations than to them.

\begin{table}[t!]
  \centering
  \caption{Role Unemployment in the Changing Size of Occupations}
    \resizebox{1\textwidth}{!}{
\small{

    \begin{tabular}{lcccccccc}

    \toprule
    \toprule
    & \multicolumn{3}{c}{Distributions} &       & \multicolumn{4}{c}{Model Decomposition of Distribution Change } \\ \cmidrule(lr){2-4} \cmidrule(lr){6-9}
    Task-Based  & \multicolumn{1}{c}{Initial} & \multicolumn{2}{c}{End Distribution} &       & \multicolumn{1}{c}{Entrants} & \multicolumn{3}{c}{Endogenous Occ. Mob in Unemployment} \\ \cmidrule(lr){3-4} \cmidrule(lr){6-6} \cmidrule(lr){7-9}
    Occupational Categories & \multicolumn{1}{c}{Distribution} & \multicolumn{1}{c}{Data} & \multicolumn{1}{c}{Model} &       & \multicolumn{1}{c}{All Qtrs} & \multicolumn{1}{c}{All Qtrs} & \multicolumn{1}{c}{Qtrs $u < u^{median}$} & \multicolumn{1}{c}{Qtrs $u \geq u^{median}$} \\
    \midrule
    Non-routine Cognitive & 0.224 & 0.329 & 0.337 &       & \phantom{-}0.125 & -0.011 & -0.006 & -      0.005  \\
    Routine Cognitive & 0.292 & 0.258 & 0.246 &       & -0.012 & -0.034 & -0.012 & -      0.023  \\
    Non-routine Manual & 0.226 & 0.260 & 0.260 &       & -0.054 & \phantom{-}0.088 & \phantom{-}0.031 &         \phantom{-}0.057  \\
    Routine Manual & 0.258 & 0.154 & 0.157 &       & -0.062 & -0.039 & -0.009 & -      0.030  \\
    \bottomrule
    \bottomrule
     \multicolumn{9}{p{1.33\textwidth}}{\footnotesize{Note: Columns 1 to 3 show the initial and end distributions of workers across tasked-based occupations. Column 4 shows the contribution of the exogenous entry process in changing the initial distribution, while column 5 shows the contribution of unemployed workers switching occupations. The sum of columns 4 and 5 is equal to the difference of columns 3 and 1. The last two columns show the contributions of occupation switchers in times when cyclical unemployment is above or below its median.}} \\
    \end{tabular}%
}}
  \label{tab:TBOCshift}%
\end{table}

The last two columns of Table \ref{tab:TBOCshift} show the contribution of mobility through unemployment separately by periods of high and low unemployment, where we categorise these periods by comparing the HP-filtered unemployment rate to its median. We observe that it is during recessions that mobility through unemployment particularly accelerates the changing size of $NRM$ and $RM$ occupations, representing about two-thirds and three-quarters of the total contribution of this channel, respectively. Jaimovich and Siu (2020) already documented the importance of recessions in changing the size of routine occupations. Here we show that the net mobility patterns described in Section 2 together with the endogenous response in unemployment yield precisely such a pattern within our model.

\begin{figure}[t]
\centering
\caption{Heterogeneity across Occupations over the Business Cycle}
\label{f:OccHet_Decomp}
\resizebox{1.0\textwidth}{!}{
\subfloat[Cycl. Occ. Productivty]{\includegraphics[width=0.35 \textwidth]{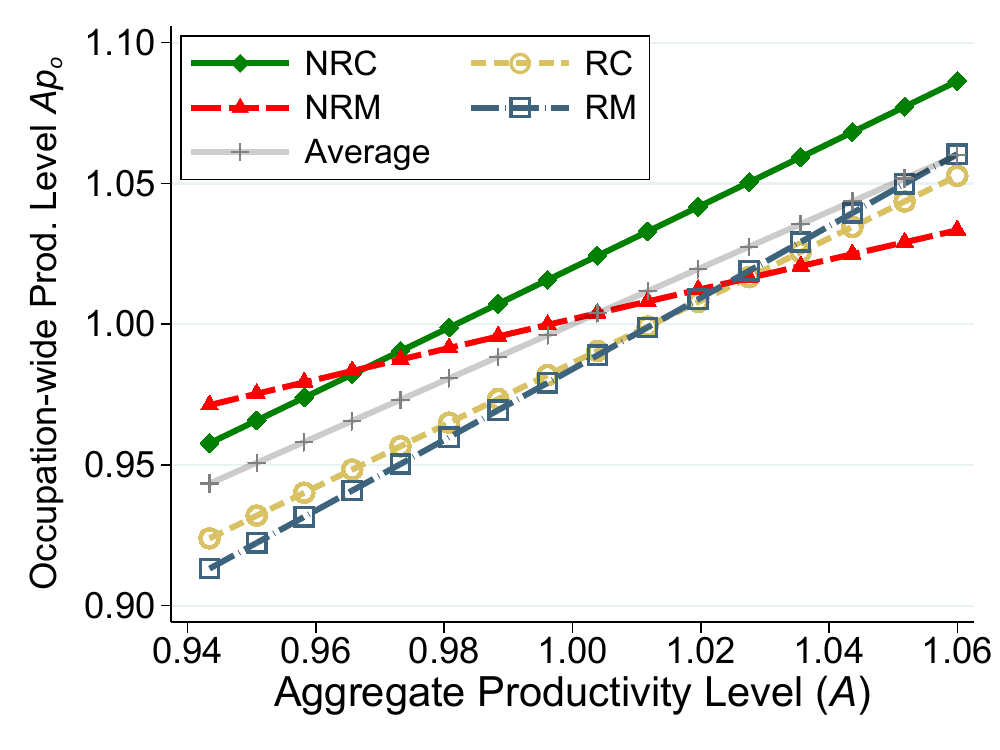}\label{f:occprod_wcycle}}
\subfloat[NRC Cutoffs \& U. Distr.]{\includegraphics[width=0.35 \textwidth]{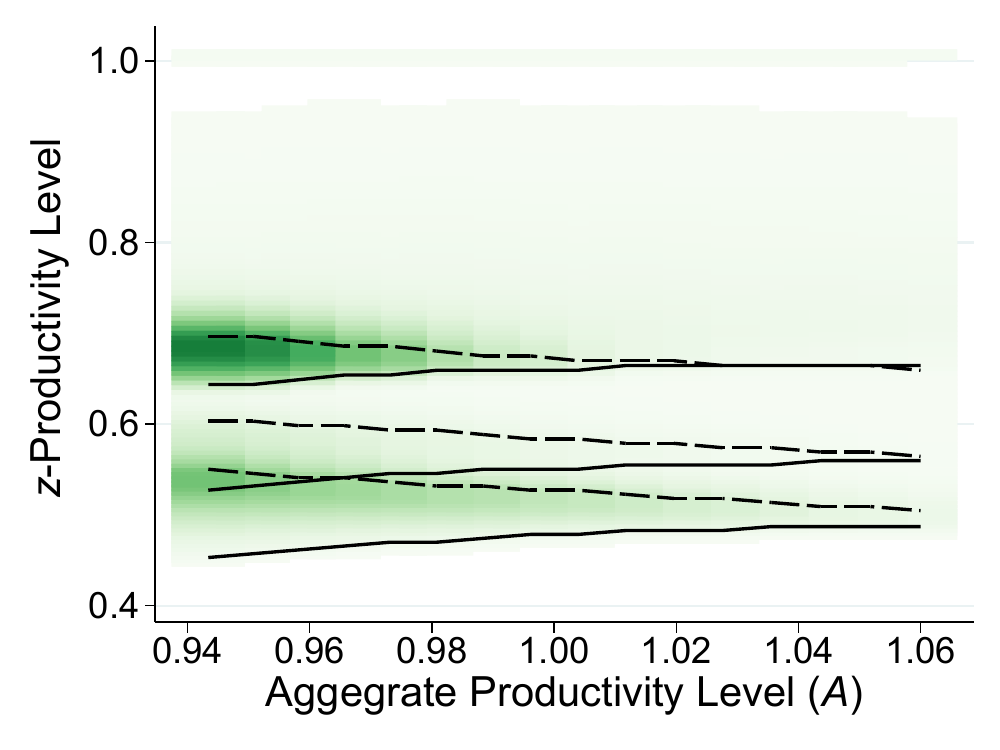}\label{f:NRC_Uheat}}
\subfloat[RC Cutoffs \& U. Distr.]{\includegraphics[width=0.35 \textwidth]{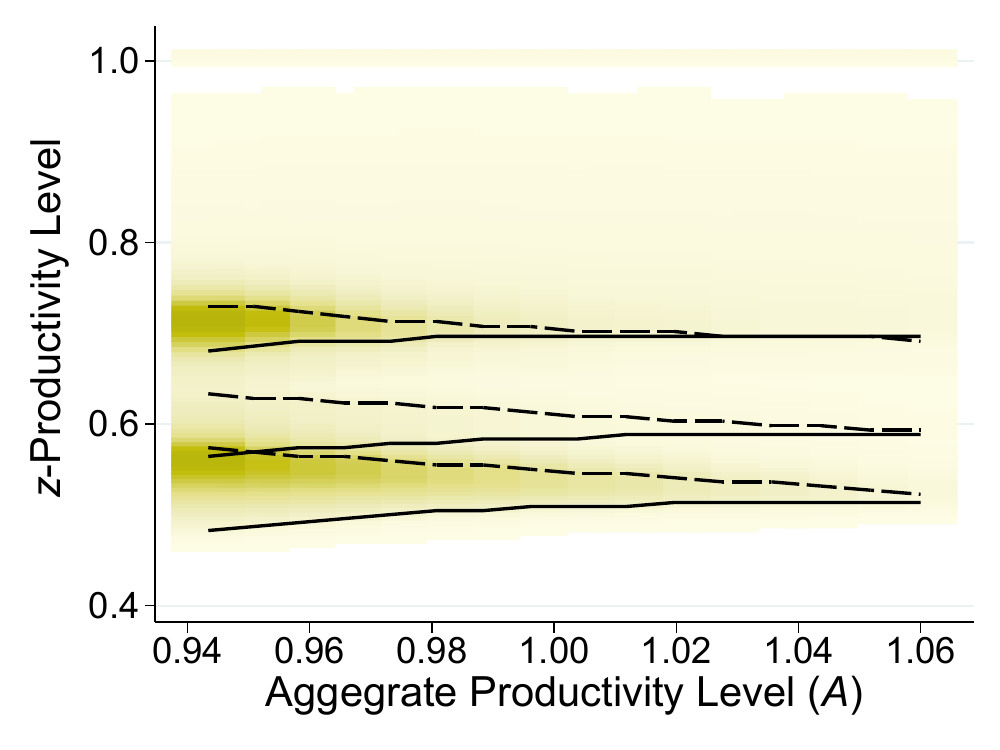}\label{f:RC_Uheat}}
}

\resizebox{1.0\textwidth}{!}{
\subfloat[Cyclical U. Decomposition]{\includegraphics[width=0.35 \textwidth]{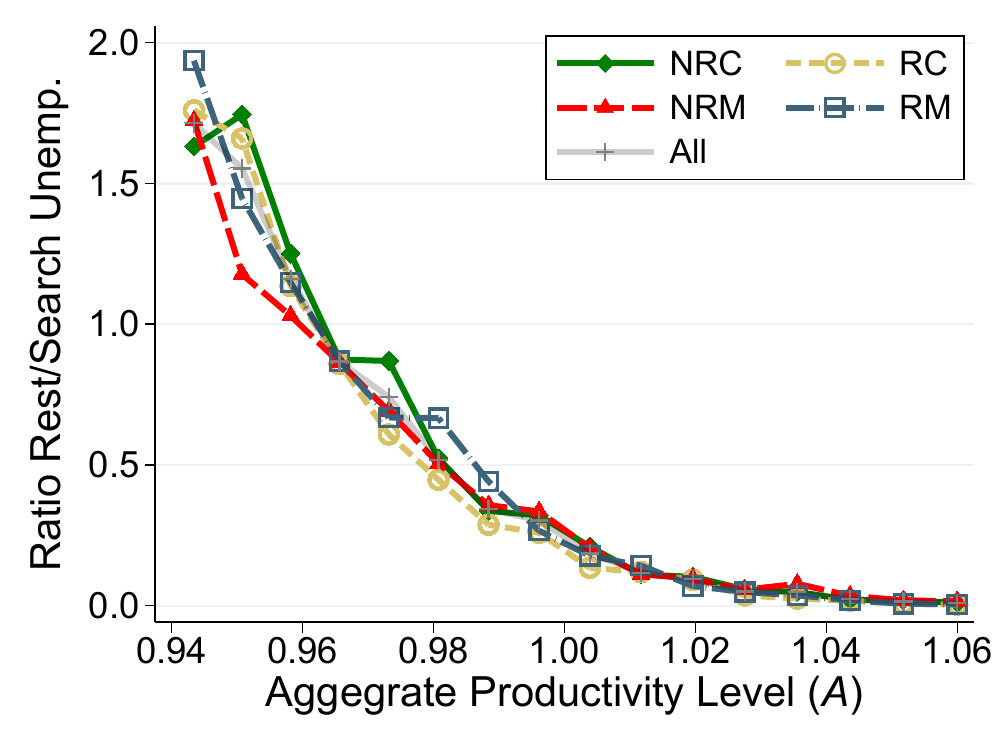}\label{f:udecomp_wcycle}}
\subfloat[NRM Cutoffs \& U. Distr.]{\includegraphics[width=0.35 \textwidth]{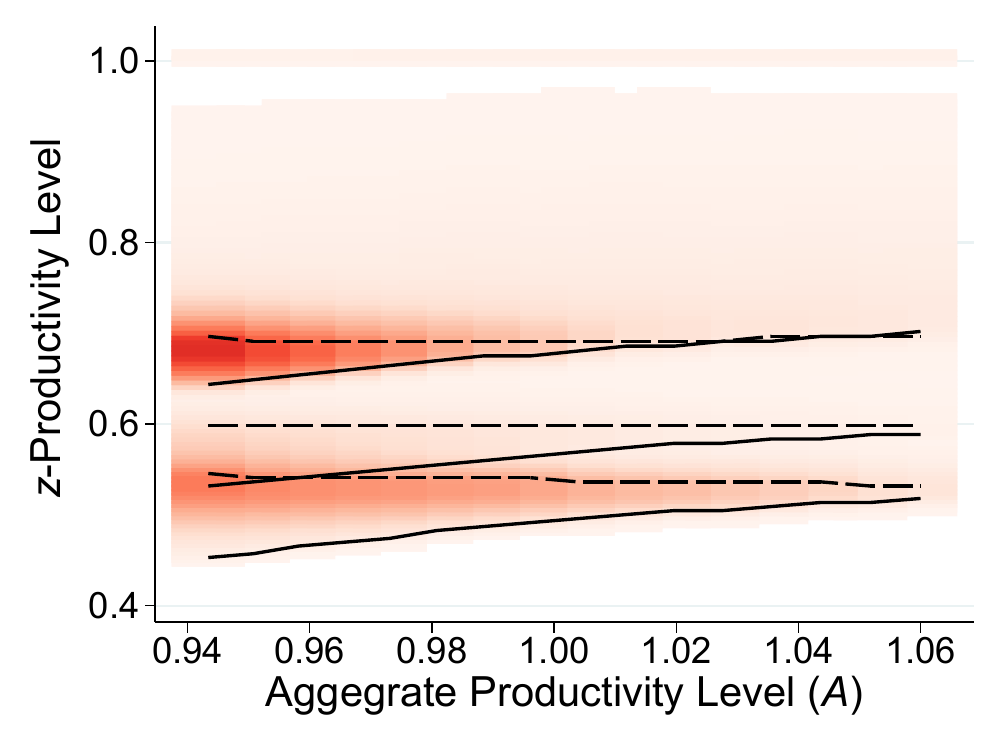}\label{f:NRM_Uheat}}
\subfloat[RM Cutoffs \& U. Distr.]{\includegraphics[width=0.35 \textwidth]{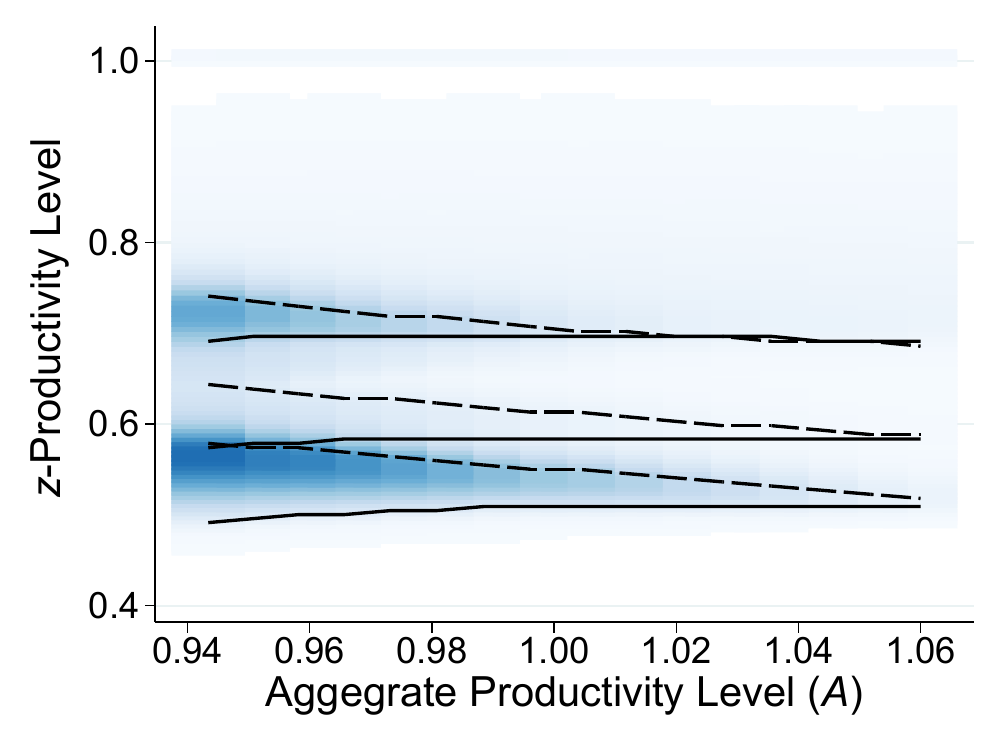}\label{f:RM_Uheat}}
}
\end{figure}

Figure \ref{f:OccHet_Decomp} illustrates the mechanism. Figure \ref{f:occprod_wcycle} shows the levels and cyclicalities of the estimated occupation-wide productivities for the range of $A$. Reflecting the estimated values of $\epsilon_{o}$, it shows that $RM$ and $RC$ occupations are strongly negatively affected in recessions, but catch up with the average in expansions. In contrast, $NRM$ occupations are the least attractive in expansions but become more attractive in recessions. $NRC$ occupations are consistently above average over the cycle (more so in expansions).

Figures \ref{f:NRC_Uheat}, \ref{f:RC_Uheat}, \ref{f:NRM_Uheat}, \ref{f:RM_Uheat} show that these different cyclical productivities result in different separation and reallocation cutoffs. Although their levels are not that different across task-based categories, in $RM$ occupations the separation cutoffs decreases more steeply, while the reallocation cutoffs are nearly horizontal. In $NRM$ occupations the separation cutoffs are nearly horizontal and the reallocation cutoffs are strongly upward-sloping. This implies that in recessions job separations are more prominent in $RM$ than in $NRM$ occupations.

Despite the differences in slopes, all task-based categories exhibit cutoffs with the $z^s>z^r$ property. Further, the distance between these cutoffs creates areas of inaction that increase in recessions and narrow in expansions as described earlier. Figure \ref{f:udecomp_wcycle} shows that as a result rest unemployment episodes are more common than search unemployment episodes in recessions within each task-based category. As the economy recovers search unemployment episodes are the most common ones.

The observed countercyclical net mobility patterns then occur for mainly two reasons: (i) a differential cyclical response in the outflows across task-based categories, such that some task-based categories shed more workers during recessions relative to the average; and (ii) a differential cyclical response in the inflows, such that those workers who have decided to change occupations choose their destination task-based category differently in recessions than in expansions. The widening of the area of inactions as $A$ decreases implies that overall occupational mobility falls during recessions in all task-based categories. However, the differential responses in $p_{o}$ across the cycle imply that the decrease in outflows is stronger in $NRM$ occupations and weaker in $RM$ occupations relative to the average, as observed in the data. At the same time, Table \ref{t:minimumdistance} shows that the model is also able to reproduce the shift in the inflow distribution towards $RM$ and away from $NRM$ occupations that occurs in recessions.

\vspace{-0.55cm}

\section{Conclusions}

\vspace{-0.35cm}

We have argued that workers' option value of remaining attached to their careers (occupations) while unemployed is relatively larger in recessions than in expansions. The cyclical variation in this option value creates more wait/rest unemployment episodes than search unemployment episodes in recessions, and it can jointly explain many features of cyclical unemployment, its duration distribution and occupational mobility. While idiosyncratic uncertainty regarding a worker's career is the main force shaping this option value, the latter is also affected by occupation-wide differences that create net mobility across occupations. We find no tension between the cyclical behavior of individual unemployment outcomes, procyclical gross occupational mobility and countercyclical net mobility through unemployment, where $EUE$ transitions play a meaningful role in shaping the changing size of $RM$, $RC$ and $NRM$ occupations. The potential responsiveness of this option value (or its relative importance in job search, separation and reallocation decisions) to policy opens the door for normative investigations.

\bigskip
\bigskip

\clearpage

\appendix
\section*{\textsc{Online Appendix}}
\section{Correcting for Occupational Coding Errors}
\vspace{-0.25cm}
\label{onlineappx_firstpage}

This Appendix complements Section 2.1 of the paper. Supplementary Appendix A provides the full version of this appendix. There we provide all results and proofs, present the estimate of ${\bf{\Gamma}}$-correction matrix, show that our method is successful out of sample, and that it affects differently employer/activity movers and pooled samples of all workers, also when the same (in)dependent interviewing procedure is applied to both groups in the survey. We further compare the implied extent of coding error across different occupation (and industry) categories. We also show that our correction method implies an average occupational mobility rate at re-employment that is in line with the one derived from the PSID retrospective occupation-industry supplementary data files. Finally, we discuss the plausibility of the assumption used to recover ${\bf{\Gamma}}$. To save space, here we only summarise the main mathematical results. We use this error correction model to produce the results in the main text and Supplementary Appendix B.

The elements of garbling matrix $\mathbf{\Gamma}$ are defined to be the probabilities that an occupation $i$ is miscoded as an occupation $j$, for all $i,j=1,2,...,O$. We make three assumptions that allows us to identify and estimate $\mathbf{\Gamma}$. (\emph{A1}) \emph{Independent classification errors}: conditional on the true occupation, the realization of the occupational code is independent of worker history, worker characteristics or time. (\emph{A2}) \emph{``Detailed balance'' in miscoding}: $\mathbf{diag(c)\Gamma}$ is symmetric, where $\mathbf{c}$ is a $O$x$1$ vector that describes the distribution of workers across occupations and $\mathbf{diag(c)}$ is the diagonal matrix of $\mathbf{c}$. (\emph{A3}) \emph{Strict diagonal dominance}: $\mathbf{\Gamma}$ is strictly diagonally dominant in that $\gamma_{ii}>0.5$ for all $i=1,2,...,O$.

To estimate $\mathbf{\Gamma}$ we exploit the change of survey design between the 1985 and 1986 SIPP panels. Until the 1985 panel the SIPP used independent interviewing for all workers: in each wave all workers were asked to describe their job anew, without reference to answers given at an earlier date. Subsequently, a coder would consider that wave's verbatim descriptions and allocate occupational codes. This practise is known to generate occupational coding errors. In the 1986 panel, instead, the practise changed to one of dependent interviewing. Respondents were only asked ``independently'' to describe their occupation if they reported a change in employer or if they reported a change in their main activities without an employer change. If respondents declared no change in employer \emph{and} in their main activities, the occupational code assigned to the respondent in the previous wave is carried forward.

To identify $\mathbf{\Gamma}$ it is important to note that during February 1986 to April 1987, the 1985 and 1986 panels overlap, representing the \emph{same} population under different survey designs. The identification theory we develop in the next section refers to this population. We then show how to consistently estimate $\mathbf{\Gamma}$ using the population samples.

\vspace{-0.5cm}
\subsubsection*{A.1 Identification of the $\mathbf{\Gamma}$ matrix}

\vspace{-0.25cm}

Consider the population represented by 1985/86 SIPP panels during the overlapping period and divide it into two groups of individuals across consecutive interviews by whether or not they changed employer or activity. Label those workers who stayed with their employers in both interviews and did not change activity as ``employer/activity stayers''. By design this group \emph{only} contains true occupational stayers. Similarly, label those workers who changed employers or changed activity within their employers as ``employer/activity changers''. By design this group contains all true occupational movers and the set of true occupational stayers who changed employers.

Suppose that we were to subject the employer/activity stayers in this population to dependent interviewing as applied in the 1986 panel. Let $\mathbf{c_{s}}$ denote the $O$x$1$ vector that describes their \emph{true} distribution across occupations and let $\mathbf{M_{s}}=\mathbf{diag(c_{s})}$. Let $\mathbf{c^{D}_{s}}$ denote the $O$x$1$ vector that describes their \emph{observed} distribution across occupations under dependent interviewing and let $\mathbf{M^{D}_{s}}=\mathbf{diag(c^{D}_{s})}$. Note that $\mathbf{c^{D}_{s}}=\mathbf{\Gamma^{'} M_{s} \overrightarrow{\bf{1}}}$, where $\overrightarrow{\bf{1}}$ describes a vector of ones. $\mathbf{M_{s}}$ is pre-multiplied by $\mathbf{\Gamma^{'}}$ as true occupations would have been miscoded in the first of the two consecutive interviews. Assumption \emph{A2} implies that $\mathbf{c^{D}_{s}}=\mathbf{diag(c_{s})\Gamma \overrightarrow{\bf{1}}}=\mathbf{c_{s}}$ and hence $\mathbf{M^{D}_{s}=M_{s}}$.

Next suppose that instead we were to subject the employer/activity stayers in this population to independent interviewing as applied in the 1985 panel. Let $\mathbf{M^{I}_{s}}$ denote the matrix that contains these workers' \emph{observed} occupational transition \emph{flows} under independent interviewing. In this case $\mathbf{M^{I}_{s}}=\mathbf{\Gamma^{\prime} M_{s} \Gamma}$. Here $\mathbf{M_{s}}$ is pre-multiplied by $\mathbf{\Gamma^{'}}$ and post-multiplied by $\mathbf{\Gamma}$ to take into account that the observed occupations of origin and destination would be subject to coding error.

Let $\mathbf{M_{m}}$ denote the matrix that contains the \emph{true} occupational transition \emph{flows} of employer/activity changers in this population. The diagonal of $\mathbf{M_{m}}$ describes the distribution of true occupational stayers across occupations among employer/activity changers. The off-diagonal elements contain the flows of all true occupational movers. Under independent interviewing $\mathbf{M^{I}_{m}}=\mathbf{\Gamma^{\prime} M_{m} \Gamma}$. Once again $\mathbf{M_{m}}$ is pre-multiplied by $\mathbf{\Gamma^{'}}$ and post-multiplied by $\mathbf{\Gamma}$ as the observed occupations of origin and destination would be subject to coding error.

Letting $\mathbf{M^{I}}=\mathbf{M^{I}_{m}+M^{I}_{s}}$ denote the matrix that contains the aggregate occupational transition flows across two interview dates under independent interviewing, it follows that $\mathbf{M^{I}_{s}}=\mathbf{M^{I}-M^{I}_{m}}=\mathbf{\Gamma^{\prime} M_{s} \Gamma}$. By virtue of the symmetry of $\mathbf{M_{s}}$ and assumption \emph{A2}, $\mathbf{M_{s} \Gamma=\Gamma^{\prime} {M_{s}}^{'}=\Gamma^{'} M_{s}}$. Substituting back yields $\mathbf{M^{I}_{s}}=\mathbf{M_{s}\Gamma \Gamma}$. Next note that $\mathbf{M^{I}_{s}=M_{s}T^{I}_{s}}$, where $\mathbf{T^{I}_{s}}$ is the occupational transition probability matrix of the employer/activity stayers in this population \emph{observed} under independent interviewing. Substitution yields $\mathbf{M_{s} T^{I}_{s}}=\mathbf{M_{s} \Gamma \Gamma}$. Multiply both sides by $\mathbf{M_{s}^{-1}}$, which exists as long as all the diagonal elements of $\mathbf{M_{s}}$ are non-zero, yields the key relationship we exploit to estimate $\mathbf{\Gamma}$,
\begin{equation}\label{eq:est_gamma_2}
\mathbf{T^{I}_{s}}=\mathbf{\Gamma \Gamma}.
\end{equation}

To use this equation we first need to show that it implies a unique solution for $\mathbf{\Gamma}$. Towards this result, we now establish that $\mathbf{\Gamma}$ and $\mathbf{T^{I}_{s}}$ are diagonalizable. For the latter it is useful to interpret the coding error process described above as a Markov chain such that $\mathbf{\Gamma}$ is the one-step probability matrix associated with this process.

\vspace{0.10cm}

\noindent{\bf{Lemma A.1:}} \emph{Assumptions A2 and A3 imply that $\mathbf{\Gamma}$ and $\mathbf{T^{I}_{s}}$ are diagonalizable.}

\vspace{0.10cm}

In general one cannot guarantee the uniqueness, or even existence, of a transition matrix that is the $(nth)$ root of another transition matrix. Here, however, existence is obtained by construction: $\mathbf{T_s}$ is constructed from $\mathbf{\Gamma}$, and in reverse, we can find its roots. The next result shows that $\mathbf{T_s}$ has a unique root satisfying assumptions \emph{A2} and \emph{A3}.

\vspace{0.10cm}

\noindent{\bf{Proposition A.1:}} $\mathbf{\Gamma}$ \emph{is the unique solution to $\mathbf{T^{I}_{s}=\Gamma \ \Gamma}$ that satisfies assumptions A2 and A3. It is given by $\mathbf{P \Lambda^{0.5} P^{-1}}$, where $\mathbf{\Lambda}$ is the diagonal matrix with eigenvalues of $\mathbf{T^{I}_{s}}$, $0<\lambda_i\leq 1$, and $\mathbf{P}$ is the orthogonal matrix with the associated (normalized) eigenvectors.}

 \vspace{0.10cm}

The above results imply that under \emph{A2} and \emph{A3}, $\mathbf{\Gamma}$ is uniquely identified from the transition matrix of true occupational stayers under independent interviewing, $\mathbf{T^{I}_{s}}$.

\vspace{-0.5cm}

\subsubsection*{A.2 Estimation of $\mathbf{\Gamma}$}

\vspace{-0.25cm}

The next lemma provides an intermediate step towards estimating $\mathbf{\Gamma}$. For this purpose let $PDT(.)$ denote the space of transition matrices that are similar, in the matrix sense, to positive definite matrices.

\vspace{0.10cm}

\noindent{\bf{Lemma A.2:}} \emph{The function} $f:PDT(\mathbb{R}^{O\times O})\to PDT(\mathbb{R}^{O\times O})$ \emph{given by} $f(\mathbf{T})=\mathbf{T^{0.5}}$ \emph{exists and is continuous with} $f(\mathbf{T^{I}_{s}})=\mathbf{\Gamma}$ in the spectral matrix norm.

\vspace{0.10cm}

Let $\mathbf{\hat{T}^{I}_{s}}$ denote the sample estimate of $\mathbf{T^{I}_{s}}$ and let $\mathbf{\hat{\Gamma}}$ be estimated by the root $\mathbf{(\hat{T}^{I}_{s})}^{0.5} \in PDT(\mathbb{R}^{O \times O})$ such that $\mathbf{\hat{\Gamma}}=\mathbf{(\hat{T}^{I}_{s})^{0.5}}=\mathbf{\hat{P} \hat{\Lambda}^{0.5} \hat{P}^{-1}}$, where $\mathbf{\hat{\Lambda}}$ is the diagonal matrix with eigenvalues of $\mathbf{\hat{T}^{I}_{s}}$, $0<\hat{\lambda_i}^{0.5}\leq 1$ and $\mathbf{\hat{P}}$ the orthogonal matrix with the associated (normalized) eigenvectors. We then have the following result.

\vspace{0.10cm}

\noindent{\bf{Proposition A.2:}} \emph{$\mathbf{\Gamma}$ is consistently estimated from $\mathbf{(\hat{T}^{I}_{s})}^{0.5} \in PDT(\mathbb{R}^{O \times O})$ such that $\mathbf{\hat{\Gamma}}=\mathbf{(\hat{T}^{I}_{s})^{0.5}}=\mathbf{\hat{P} \hat{\Lambda}^{0.5} \hat{P}^{-1}}$. That is, $plim_{n \to \infty} \mathbf{\hat{\Gamma}}=\mathbf{\Gamma}$.}

\vspace{0.10cm}

To identify and estimate $\mathbf{\Gamma}$ in the SIPP it is not sufficient to directly compare the aggregate occupational transition flows under independent interviewing with the aggregate occupational transition flows under dependent interviewing. Let $\mathbf{M^{D}}=\mathbf{M^{I}_{m}+M^{D}_{s}}$ denote the matrix that contains the aggregate occupational transition flows across two interview dates under dependent interviewing for employer/activity stayers and under independent interviewing for employer/activity movers. Subtracting $\mathbf{M^{I}}=\mathbf{M^{I}_{m}+M^{I}_{s}}$ from $\mathbf{M^{D}}$ yields $\mathbf{M^{D}_{s}-M^{I}_{s}}=\mathbf{M_{s} -\Gamma' M_{s} \Gamma}$. Given the symmetry assumed in \emph{A2}, the latter expression has $0.5n(n-1)$ exogenous variables on the LHS and $0.5n(n+1)$ unknowns (endogenous variables) on the RHS, leaving $\mathbf{\Gamma}$ (and $\mathbf{M_s}$) unidentified.

In addition to $\mathbf{M^{D}-M^{I}}=\mathbf{M_{s} -\Gamma' M_{s} \Gamma}$ one can use $\mathbf{M^D}=\mathbf{\Gamma^{'} M_{m} \Gamma + M_{s}}$, which contains the remainder information. When $\mathbf{M_{m}}$ has mass on its diagonal, however, this additional system of equations has $n^{2}$ exogenous variables on the LHS and $n^{2}$ unknowns (arising from $\mathbf{M_m}$) on the RHS. This implies that with the $n$ unknowns remaining from $\mathbf{M^{D}-M^{I}}=\mathbf{M_{s} -\Gamma' M_{s} \Gamma}$, one is still unable to identify $\mathbf{\Gamma}$ and $\mathbf{M_s}$.

 \vspace{0.10cm}

\noindent{\bf{Corollary A.1:}}
\emph{If $\mathbf{M_{m}}$ has mass on its diagonal, $\mathbf{\Gamma}$ cannot be identified from $\mathbf{M^I}$ and $\mathbf{M^D}$ alone.}

 \vspace{0.10cm}

The intuition behind this result is that by comparing aggregate occupational transition flows under dependent and independent interviewing, it is unclear how many workers are `responsible' for the change in occupational mobility between $\mathbf{M^D}$ and $\mathbf{M^I}$. Only when the diagonal of $\mathbf{M_{m}}$ contains exclusively zeros, identification could be resolved and one can recover $\mathbf{M_s}$, $\mathbf{\Gamma}$ and $\mathbf{M_m}$ as the number of equations equals the number of unknowns.\footnote{However, in the SIPP this case is empirically unreasonable as it requires that all employer/activity changers be true occupational movers.} An implication of the above corollary is that interrupted time-series analysis that is based on the difference in occupational mobility at the time of a switch from independent to dependent interviewing, does not identify the precise extent of the average coding error, but provides a downwards biased estimate.

To identify $\mathbf{\Gamma}$, however, Proposition A.2 implies that one can use the observed occupational transition flows of a sample of \emph{true} occupational stayers that are subject to two rounds of independent interviewing. Some of these workers will appear as occupational stayers and some of them as occupational movers. Ideally, such a sample of workers should be isolated directly from the 1985 panel. Unfortunately, the questions on whether the individual changed activity or employer were only introduced in the 1986 panel, as a part of the switch to dependent interviewing. As a result, the 1985 panel by itself does not provide sufficient information to separate employer/activity stayers from employer/activity movers. Instead we use 1986 panel to estimate $\mathbf{\hat{M}^I_{m}}$. We can infer $\mathbf{M^I_{s}}$ indirectly by subtracting the observed occupational transition flow matrix $\mathbf{\hat{M}^I_m}$ in the 1986 panel from the observed occupational transition flow matrix $\mathbf{\hat{M}^I}$ in the 1985 panel. This is possible as the 1986 panel refers to the same underlying population as the 1985 panel and separates the employer/activity changers, who are independently interviewed.

\vspace{0.10cm}

\noindent{\bf{Corollary A.2:}}
\emph{$\mathbf{\hat{\Gamma}}$ is consistently estimated from $\mathbf{\hat{T}_s^I}$ when the latter is estimated from $\mathbf{\hat{M}^{I}}-\mathbf{\hat{M}^{I}_{m}}$}

\vspace{0.10cm}

This result is important to implement our approach. It follows as the population proportions underlying each cell of $\mathbf{\hat{M}_{s}}$, the sample estimate of $\mathbf{M_{s}}$, are consistently estimated. In turn, the latter follows from the standard central limit theorem for estimating proportions, which applies to $\mathbf{\hat{M}^{I}}$, $\mathbf{\hat{M}^{I}_{m}}$ and its difference. Proposition A.2 then implies that $\mathbf{\hat{\Gamma}}$ is consistently estimated.

\vspace{-0.5cm}
\setcounter{section}{1}
\section{Quantitative Analysis}
\vspace{-0.2cm}

This Appendix is divided into three parts that complement Sections 4 and 5 of the paper. The first part provides further details of the full model calibration done in Section 4. The second part presents the calibration results from the ``excess mobility model'', where we analyse its ability to reproduce the long-run and cyclical patterns of several labor market variables.

\vspace{-0.4cm}
\subsection{Full Model: Gross and Net Mobility}\label{app:c1}
\vspace{-0.2cm}

In the main text we show that the calibrated version of the full model is able to replicate well all the targeted long-run occupational mobility, job separation, job finding and unemployment patterns of the US labor market. It does so by generating within each task-based category periods of search, rest and reallocation unemployment as $A$, $p_{o}$ and $z$ evolve.

Here we expand on the analysis presented in Sections 4 and 5 along three dimensions. First, we further show the model's implied unemployment durations by presenting (i) the job finding rates as a function of duration, (ii) the (incomplete) unemployment duration distribution and (iii) the mobility-duration profile decomposed by excess and net mobility. Second, we provide further details of the differences between occupational categories with respect to their relative cyclical unemployment responses, and the cyclical inflow and net flow responses that are used to estimate occupation-specific cyclical differences in the model. Third, we present the full correlation tables describing the cyclical performance of the model using the 5Q-MA smoothed and quarterly HP-filtered measures. We also discuss the cyclicality of an alternative unemployment measure that includes entrants; show the cyclicality of the unemployment, job finding and job separation rates by age groups; and present the decomposition of search, rest and reallocation unemployment episodes for a given value of $A$ in a comparable way to the one derived for the excess mobility model discussed in Section 5.1 of the main text.

\vspace{-0.55cm}

\begin{figure}[t]
\caption{Hazard Functions. Data and Model Comparison}
\label{f:hazard}
\begin{centering}
\resizebox{1.0\textwidth}{!}{\subfloat[All Workers]{\centering\includegraphics[height=0.25 \textheight]{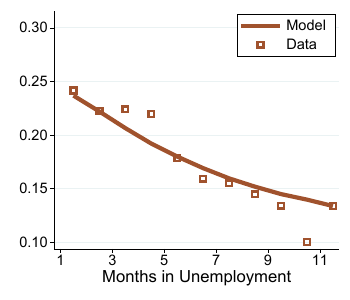}
\label{f:hazard_all} } \hspace{-0.08cm}

\subfloat[Young Workers]{\centering\includegraphics[height=0.25 \textheight ]{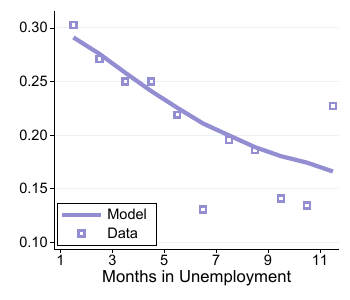}
\label{f:hazard_young} }\hspace{0.01cm}

\subfloat[Prime-aged Workers]{\centering\includegraphics[height=0.25 \textheight]{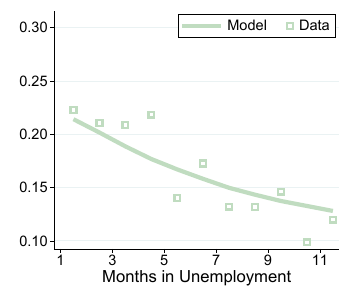}

\label{f:hazard_prime} }}
\par\end{centering}

\end{figure}

\paragraph{Unemployment duration moments} Figure \ref{f:hazard} shows the aggregate and age-specific unemployment hazard functions, comparing the model to the data.
The observed duration dependence patterns are captured very well, where the young exhibit stronger negative duration duration dependence than the prime-aged. In our sample (and in the calibration) the degree of duration dependence in the unemployment hazard is relatively weak as we have tried to minimise the presence of unemployed workers who were in temporary layoff and/or returned to their previous employers (see Supplementary Appendix B.4 for a further discussion of this issue).

\begin{figure}[ht!]
\caption{Occupational Movers/Stayers Hazard Functions. Data and Model Comparison}
\label{f:survival_occ}
\resizebox{1\textwidth}{!}{\subfloat[All Workers] {\includegraphics[height=0.25 \textheight]{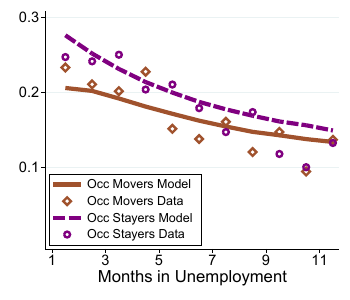}} \hspace{0.01cm}
\subfloat[Young Workers] {\includegraphics[height=0.25 \textheight]{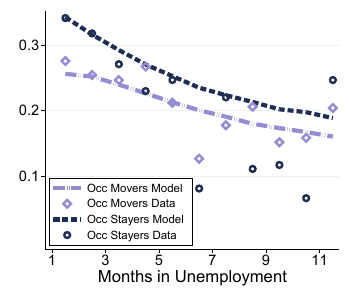}} \hspace{0.01cm}
\subfloat[Prime-aged Workers] {\includegraphics[height=0.25 \textheight]{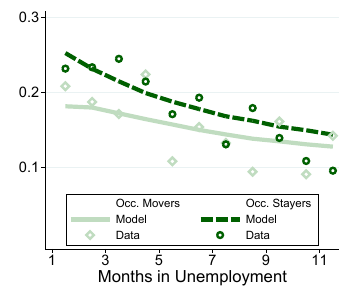}}}
\end{figure}

Figure \ref{f:survival_occ} shows the aggregate and age-specific hazard functions separately for occupational movers and stayers. The model captures these well, where we find a stronger degree of negative duration dependence among occupational stayers than occupational movers, particularly among young workers, as in the data.

The unemployment duration distribution is also well matched. Table \ref{t:dur_unemp1} shows it reproduces very well both the proportion of short and long durations spells across the distribution. This fit is achieved when pooling together all workers and when separating them between young and prime-aged workers. Crucially, the fit of the duration distribution is not implied by targeting the empirical unemployment survival functions. The reported duration distribution is constructed by averaging duration distributions across quarters, while the survival functions are derived from pooling all observations. For example, the observed long-term unemployment in the pooled survival functions occur mainly in recessions and these observations get down-weighted when averaging across quarters (instead of counting each observation equally). Indeed, we show below that matching the survival functions does not imply also matching the duration distribution.

\begin{table}[t]
\centering
\caption{Incomplete Unemployment Duration Distribution Behavior (1-18 months)}\label{t:dur_unemp1}
\resizebox{0.8\textwidth}{!}{
\begin{tabular}{lccccccccc}
\toprule
\toprule
 & \multicolumn{3}{c}{All workers} & \multicolumn{3}{c}{Young workers} & \multicolumn{3}{c}{Prime-aged workers} \\
 \cmidrule(lr){2-4} \cmidrule(lr){5-7} \cmidrule(lr){8-10}
Unemp.  & Full & Excess&  Data  & Full & Excess &   Data      &  Full & Excess  & Data \\
 Duration   & Model  & Model  &   & Model & Model & & Model & Model   &   \\
 \cmidrule(lr){2-4} \cmidrule(lr){5-7} \cmidrule(lr){8-10}
1-2 m  & 0.43 & 0.42    &   0.43  & 0.53 & 0.52 & 0.47 & 0.41 & 0.39   &  0.41 \\
1-4 m  & 0.65 & 0.64 &  0.67  & 0.76 & 0.75 &   0.71 & 0.63 & 0.61  &  0.65 \\
5-8 m  & 0.20 & 0.21 &  0.20  & 0.16 & 0.17&   0.19  & 0.21 & 0.22 &  0.21  \\
9-12 m & 0.09 & 0.09 &  0.08  & 0.05 & 0.05 & 0.07  &  0.10& 0.10&  0.09  \\
13-18m  & 0.06 & 0.06 &  0.05 & 0.03 & 0.03 &   0.03 & 0.07 & 0.07 &  0.06 \\
\bottomrule
 \bottomrule
  \end{tabular}}
\end{table}

\begin{figure}[!t]
\caption{Task-Based Occupational Mobility}
\label{f:net_mobility}
\begin{centering}
\resizebox{1\textwidth}{!}{\subfloat[Decomposition of the Mobility-Duration Profile]{\centering\includegraphics[height=2.5in]{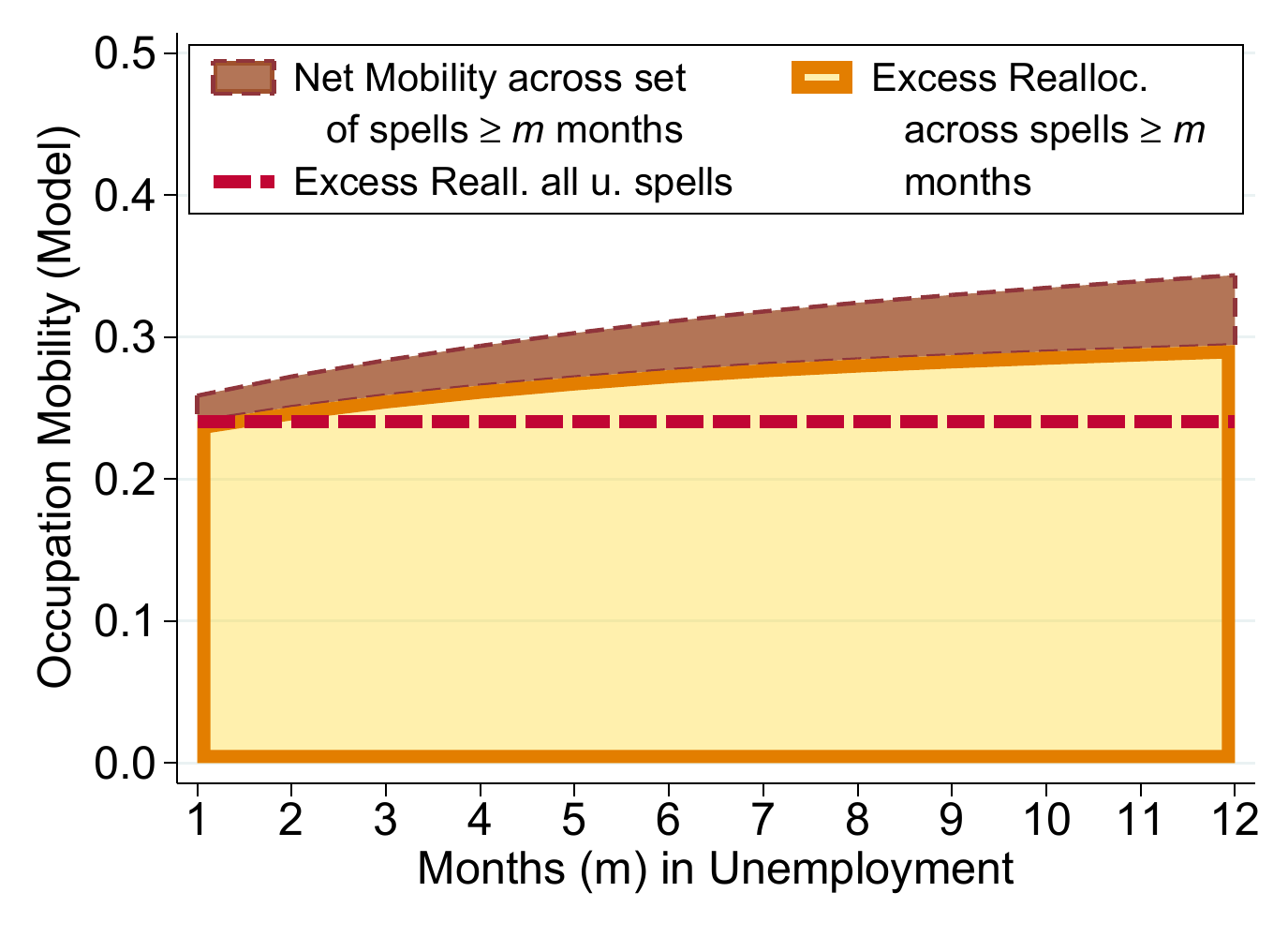}
\label{f:mobility_duration} }
\quad
\subfloat[Cyclical Shifts across Occupation Categories in Netflows vs Inflows]{\centering\includegraphics[height=2.5in]{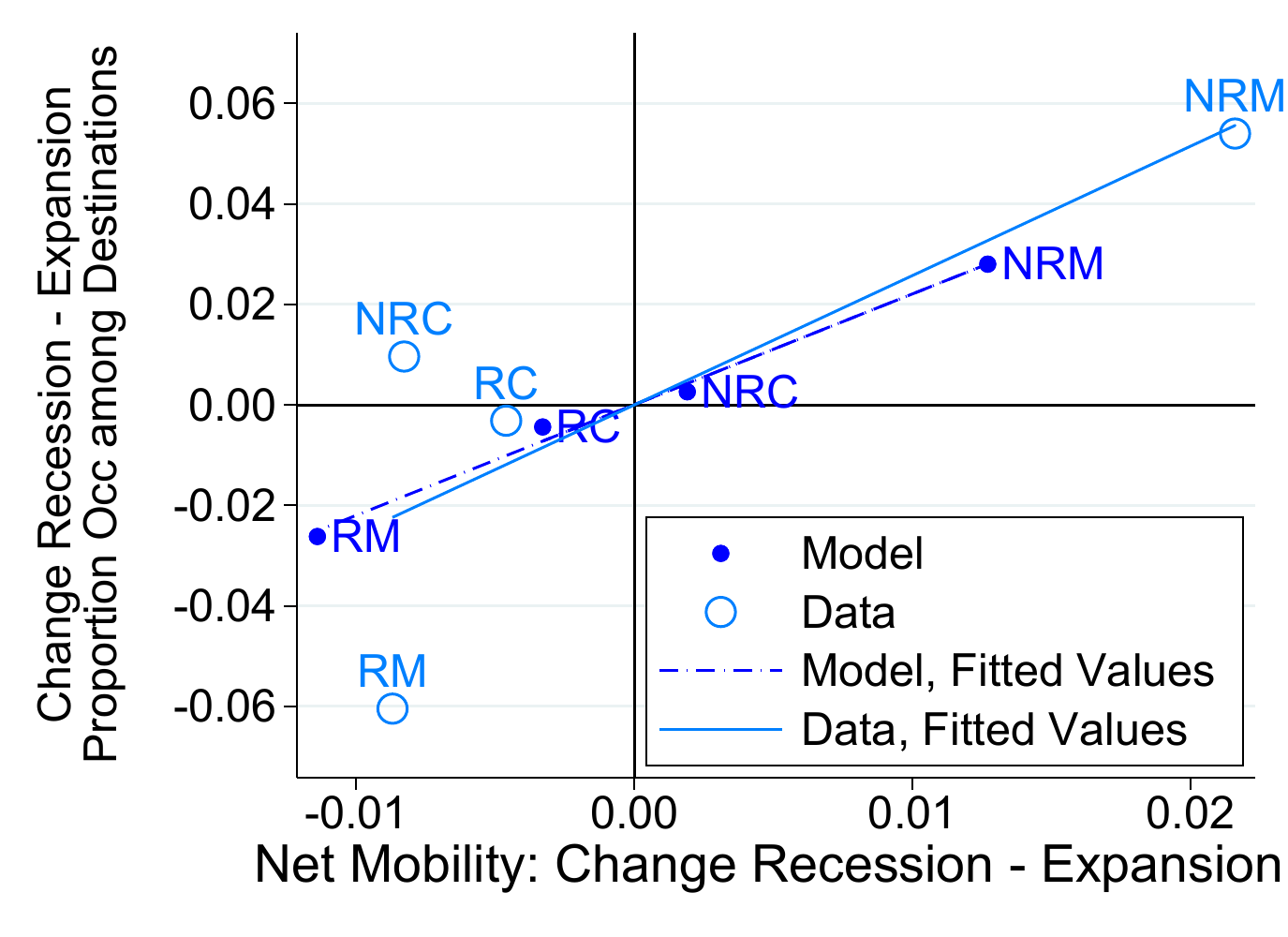}
\label{f:netflow_inflow_shift} } }
\par\end{centering}

\end{figure}

Supplementary Appendix B (Figures 9 and 10) shows that both excess and net mobility increase with unemployment duration and that the former is the main driver of the overall increase in the mobility-duration profile. Figure \ref{f:mobility_duration} (below) depicts the model's equivalent decomposition using task-based categories and without considering the ``management'' occupation. In the model both excess and net mobility increase with unemployment duration. Given the countercyclicality of net mobility, the latter occurs as net mobility is more prominent in recessions where workers' unemployment durations are typically longer. As in the data, excess mobility is the main driver of the mobility-duration profile.

\vspace{-0.55cm}

\paragraph{Task-based occupational categories over the cycle}

The cyclical productivity loadings $\epsilon_o$ are the only four cyclical parameters that explicitly differ across task-based categories $o \in \{NRC, RC, NRM, RM \}$. Together with the elasticity of the cross-occupation search, $\nu$, these parameters shape the differential cyclical response of each category $o$ along three dimensions, summarised by 12 moments in the main text. (i) The cyclical response of net mobility for each task-based category, (ii) the cyclical change in the proportion of occupational movers that choose an occupation category $o$, and (iii) the strength of each category's unemployment durations response relative to the economy-wide average response to the aggregate unemployment rate.

\begin{table}[t]
  \centering
  \caption{Task-based Unemployment Duration Elasticities}
  \resizebox{0.6\textwidth}{!}{
    \begin{tabular}{lcccc}
    \toprule
    \toprule
          & \multicolumn{1}{c}{NRC} & \multicolumn{1}{c}{RC} & \multicolumn{1}{c}{NRM} & \multicolumn{1}{c}{RM} \\
    \midrule
    $\varepsilon^{Data }_{UD_{o},u}$ & 0.410 & 0.384 & 0.284 & 0.418 \\
    (s.e.) & (0.068) & (0.050) & (0.045) & (0.053) \\
    $\varepsilon^{Model}_{UD_{o},u}$ & \emph{0.390} & \emph{0.413} & \emph{0.342} & \emph{0.423} \\ \midrule
    $\varepsilon^{Data}_{UD_{o},u}/\varepsilon^{Data}_{UD_{avg},u}$ (targeted) & 1.096 & 1.026 & 0.759 & 1.119 \\
    (s.e.) & (0.183) & (0.132) & (0.119) & (0.141) \\
    $\varepsilon^{Model}_{UD_{o},u}/\varepsilon^{Model}_{UD_{avg},u}$& \emph{0.988} & \emph{1.055} & \emph{0.890} & \emph{1.072} \\
    \bottomrule
    \bottomrule
    \end{tabular}}
  \label{tab:cycl_sens_udur}%
\end{table}%

We highlight that in (iii) we target the unemployment duration elasticities for each task-based category \emph{relative} to the economy-wide elasticity. We do this as we want to leave untargeted the amplification of aggregate unemployment. In particular, as a first step to derive these elasticities in the SIPP we regress for each task-based category $o$ the log unemployment durations of workers who lost their job in $o$ on the log (aggregate) unemployment rate and a linear trend. Let $\varepsilon_{UD_{o},u}$ for $o \in \{NRC, RC, NRM, RM \}$ denote the resulting unemployment duration elasticities with respect to aggregate unemployment. The first row of Table \ref{tab:cycl_sens_udur} presents these elasticities and compares them to the simulated ones in the calibration. These elasticities show that NRM occupations have a more muted cyclical response than RM occupations. This differential response is also statistically significant: a Wald test on equality of the two corresponding coefficients has an associated p-value of 0.02. In the second step, we normalize each elasticity by the (occupation size-weighted) average of all four elasticities. The resulting normalized elasticities are the ones we target in the model. The last two rows of Table \ref{tab:cycl_sens_udur} shows these ratios, showing that the model fits the data well. In particular, it shows that RM occupations are the most cyclically sensitive in terms of unemployment durations; while NRM occupations are the least cyclically sensitive. These elasticities are in line with the data. Below we show that the model is also successful in generating the untargeted aggregate unemployment amplification.

As shown in the main text, the model is consistent with the cyclical changes in net mobility as well as the cyclical changes in the inflows for each task-based category. This occurs as differences in $\epsilon_o$ translate into cyclically changing incentives for workers to leave an occupation in category $o$ and, depending on $\nu$, to sample $z$-productivities from another occupation in category $o'$. Figure \ref{f:netflow_inflow_shift} displays the relationship between these two set of moments. For each task-based category, it shows the relationship between the cyclical changes in net mobility on the x-axis (``Net mobility $o$, \emph{Expansions} - Net mobility $o$, \emph{Recessions}'') and the cyclical changes in inflows as a proportion of all occupational movers on the y-axis ($\Delta_{exp - rec}$ (\text{inflow} $o / \text{all flows}$)). RM occupations have the strongest cyclical response of net outflows, increasing in recessions, as well as the strongest response in the inflow proportion, also larger in recessions. In contrast, NRM occupations are the ones which experience the largest increase in net inflows in recessions and the largest increase in inflows as destination category.

As Figure \ref{f:netflow_inflow_shift} and Table \ref{tab:cycl_sens_udur} show, the model captures well the co-movement along dimensions (i), (ii) and (iii) discussed above. In particular, with only one set of parameters indexed by task-based occupation category, $\epsilon_{o}$, the model reproduces well the average co-movement of the cyclical inflow shift with the cyclical changes in net flows. Figure \ref{f:netflow_inflow_shift} shows that when comparing the fitted regression lines in the data and the model both display very similar slopes, where changing the net flow by 1\% goes together with an inflow response of more than 2\%.

\vspace{-0.55cm}

\paragraph{Economy-wide cyclical outcomes}

In terms of the cyclical properties of the unemployment, vacancies, job finding and separation rates, Table \ref{t:volatilities} show the full set of correlations for the model and data. The model's aggregate time series arise from the distributions of employed and unemployed workers across all labor markets, combined with agents' decisions. The top panel compares the data and model using centered 5Q-MA time series of quarterly data. The cyclical components of the (log) of these time series are obtained by using an HP filter with parameter 1600. It shows that the model matches the data very well. The bottom panel compares the data and model without using this smoothing procedure. The version now yields vacancy and job separation rates that are much less persistent than their data counterparts. This happens because we have used a relative coarse grid for the simulated productivity process, as making the productivity grid finer will make the computational time of the calibration unmanageable. This implies that the discreteness of the $z^s$ and $z^r$ cutoffs functions (relative to the productivity grid) makes the vacancy and job separation rates change value too often. Using a centered 5Q-MA on quarterly data alleviates this feature without further compromising on computation time. Note, however, that this comes at the cost of reducing the volatility of the vacancy rate (and labor market tightness) in the model from 0.07 to 0.05 (0.21 to 0.17), while in the data they remain stable. Similarly on the data side, the job finding rate, measured in a consistent way with the model while taking into account censoring in the SIPP, is somewhat noisy at quarterly frequency. Smoothing this time-series using the 5Q-MA helps diminish this noise.

\begin{table}[t]
\centering
\caption{Full Model: Logged and HP-filtered Business Cycle Statistics}\label{t:volatilities}
\resizebox{0.9\textwidth}{!}{
{\footnotesize
\begin{tabular}{|c|cccccc||cccccc|}
\hline
\multicolumn{13}{|l|}{{\bf{Smoothed data: centred 5Q MA time series of quarterly data}}} \\ \hline
& \multicolumn{6}{|c||}{\textbf{Data (1983-2014)}} & \multicolumn{6}{c|}{\textbf{Full Model}} \\ \hline
& $u$ & $v$ & $\theta$ & $s$ & $f$ & $outpw$ & $u$ & $v$ & $\theta$ & $s$ & $f$  & $outpw$ \\ \hline\hline
$\sigma$ &  0.15 & 0.11	& 0.25	& 0.10 & 	0.09 &	0.01  &0.14	& 0.04 &	0.17	& 0.07	& 0.09	& 0.01 \\
$\rho_{t-1}$ & 0.98 &	0.99 &0.99 &	0.94 &	0.93 &	0.92  &0.93 &	0.91 &	0.92 &	0.88 &	0.93 &	0.88 \\ \hline\hline
& \multicolumn{12}{|c|}{\textbf{Correlation Matrix}} \\ \hline
$u$ & 1.00 &	-0.95 &	-0.99 &	0.83 &	-0.86 &	-0.50 &  1.00 &	-0.66 &	-0.97 &	0.79 &	-0.89 &	-0.94  \\
$v$ &  & 1.00 & 0.98 & 	-0.79 &	0.81 & 	0.61 & 	& 1.00 & 	0.80 & 	-0.83 & 	0.90 & 	0.81 \\
$\theta$ &  &  & 1.00 &	-0.82 & 	0.85 & 	0.55 &  &  & 1.00 &	-0.84 & 	0.96 &	0.97 \\
$s$ &   &   &   & 1.00 & -0.71 & 	-0.43  & &  & & 1.00  & -0.87 & 	-0.91 \\
$f$ &  &  &    &      & 1.00 &  0.40 & &   & & & 1.00  &  0.94 \\
$outpw$ &  &  &  &  & & 1.00 & &  & & & & 1.00 \\ \hline
\multicolumn{13}{|l|}{{\bf{Un-smoothed data}}} \\ \hline
& $u$ & $v$ & $\theta$ & $s$ & $f$ & $outpw$& $u$ & $v$ & $\theta$ & $s$ & $f$  & $outpw$ \\ \hline\hline
$\sigma$ &  0.16 & 	0.11 &	0.26 &	0.12 & 	0.13 & 	0.01 & 0.16 & 	0.07 &	0.20 & 	0.11 & 	0.12 &	0.01 \\
$\rho_{t-1}$ & 0.85 &	0.96 &	0.94 &	0.58 &	0.33 &	0.75 & 0.88 &	0.54 & 	0.83 & 	0.39 &	0.78 &	0.76 \\ \hline\hline
& \multicolumn{12}{|c|}{\textbf{Correlation Matrix}} \\ \hline
$u$ & 1.00 &	-0.86 & 	-0.98 &	0.69 &	-0.65 &	-0.38 &1.00 &	-0.54 & 	-0.96 & 	0.52 & 	-0.80 &	-0.89 \\
$v$ &  & 1.00 & 0.95 & 	-0.75 & 	0.59 &	0.50 & & 1.00 & 0.75 &	-0.74 &	0.85 &	0.78 \\
$\theta$ &  &  & 1.00 & -0.74 & 	0.65 &	0.44 &  &  & 1.00 & -0.65 & 0.90 &	0.95\\
$s$ &   &   &   & 1.00 & -0.57 &	-0.32 & &  & & 1.00  & -0.79 &	-0.79 \\
$f$ &  &  &    &      & 1.00 &  0.26 & &   & & & 1.00  &  0.91 \\
$outpw$ &  &  &  &  & & 1.00 & &  & & & & 1.00 \\ \hline

\end{tabular}}
}
\end{table}

As argued in main text (and in Supplementary Appendix B.7) we consider the unemployment rate of those workers who are unemployed between jobs $(EUE)$, so that the occupational mobility of these workers can be straightforwardly measured. The resulting $EUE$ unemployment rate $(EUE/(EUE+E))$, under the definitions and restrictions we explained in the main text, is significantly lower than the BLS at 3.5\% (vs 6.3\%), but drives much of its changes. In particular, for every one percentage point change in the BLS unemployment rate, we find that about 0.75 percentage points originate from the response of the $EUE$ unemployment rate. This means that the relative cyclical response of the $EUE$ unemployment rate is much stronger than the relative response of the BLS unemployment rate. Indeed, the volatility of the HP-filtered logged quarterly $EUE$ unemployment rate is 0.16 while the corresponding BLS unemployment measure (which includes inflows from non-participation) over the same period is 0.11. For the 5Q-MA smoothed time series, the difference is from 0.15 $(EUE)$ to 0.10 (BLS). The above also means that the focus on $EUE$ unemployment raises the bar further to achieve sufficient amplification. Nevertheless, Table \ref{t:volatilities} shows that our model performs well.

In the model we also can calculate a measure of unemployment that includes unemployment following first entry into the labor market. Relative to the BLS measure, this measure still excludes unemployment associated with workers who re-enter the labor market during their working life or who subsequently leave the labor force but not before spending time in unemployment. Including entrants in unemployment raises the average total unemployment rate to 5.2\% in the model, exhibiting a lower volatility of 0.12 (5Q-MA smoothed). The latter arises as with this unemployment measure, roughly 60\% of the way from the $EUE$ to BLS unemployment measures, the volatility gets closer to that of the BLS measure. Cross-correlation and autocorrelation statistics of this alternative unemployment measure are very similar to the $EUE$ unemployment measure.

\begin{figure}[t]
\caption{Unemployment Decomposition - Full model}
\label{f:Decomp_full}
\resizebox{1\textwidth}{!}{\subfloat[All Workers]{\includegraphics[width=0.35 \textwidth]{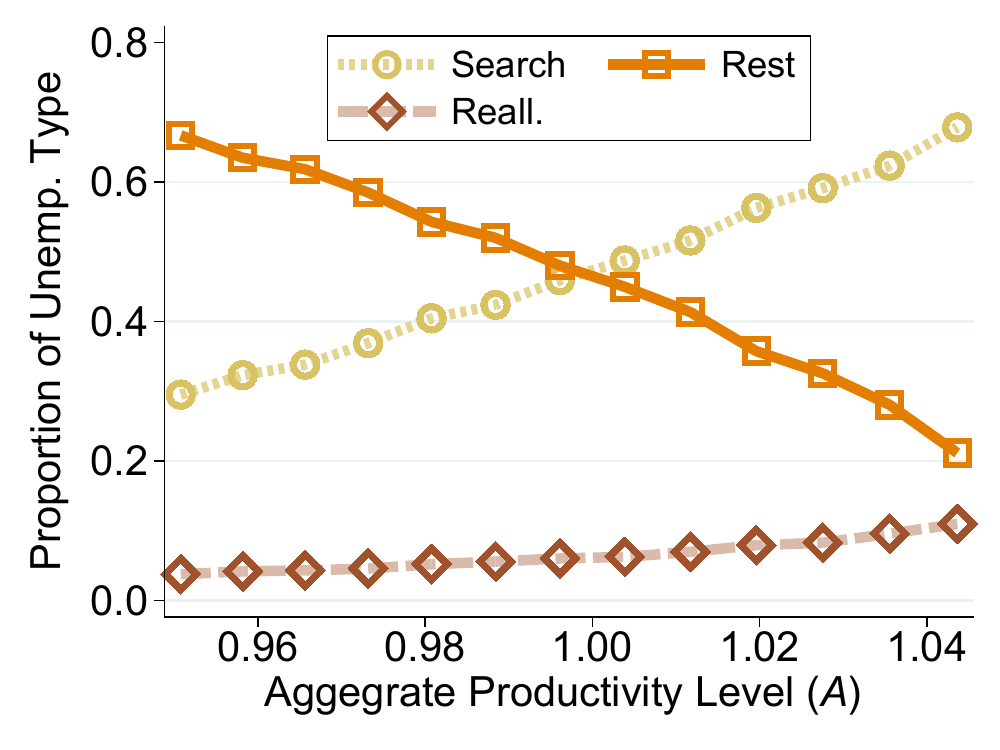}
}\subfloat[Young Workers]{\includegraphics[width=0.35 \textwidth]{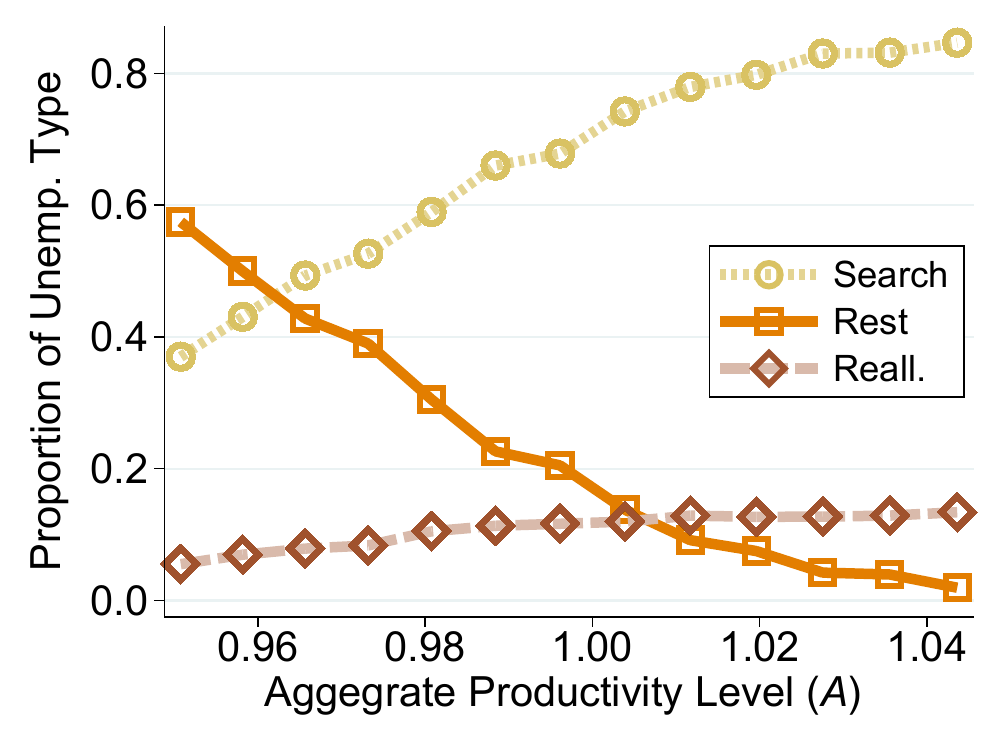}} \hspace{0.01cm}
\subfloat[Prime-aged Workers]{\includegraphics[width=0.35 \textwidth]{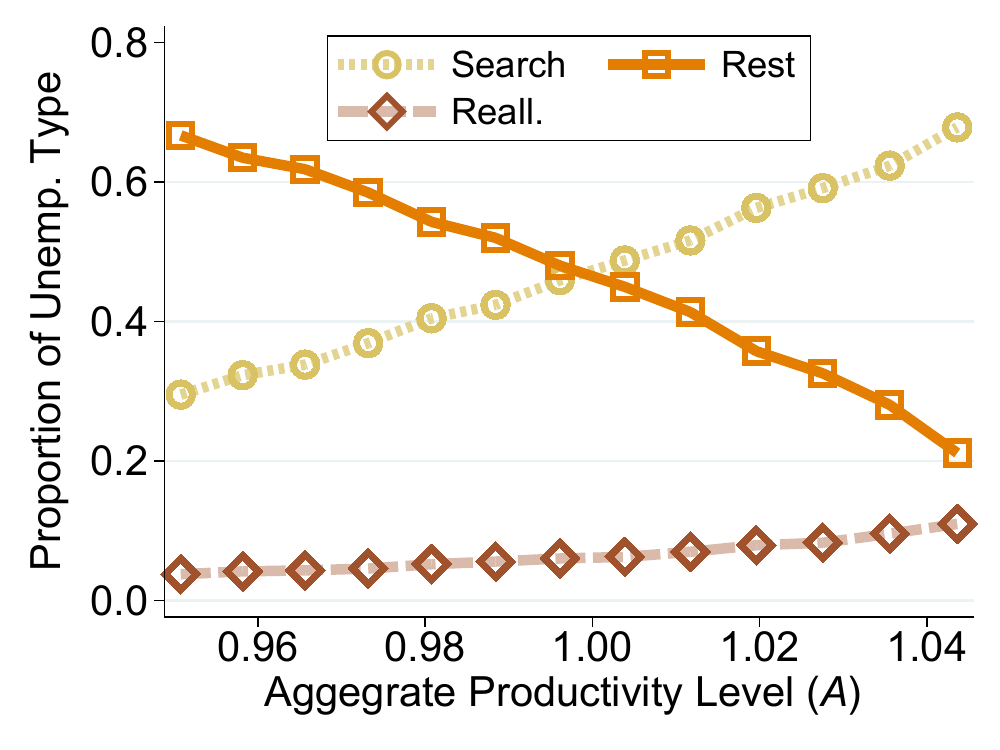}}}
\end{figure}

The ability of the model to replicate the cyclical behavior of many labor market variables is down to the coexistence of episodes of search, rest and reallocation unemployment during workers' jobless spells. Figure \ref{f:Decomp_full} shows that when aggregating across all occupations the distribution of these types of unemployment episodes across values of $A$ is very similar to the one generated by the excess mobility model depicted in the main text. That is, search unemployment episodes are the most common when the economy moves from mild recessions up to strong expansions. It is only as recessions get stronger that rest unemployment episodes become more common.

The middle and right panels of Figure \ref{f:Decomp_full} shows that among young and prime-aged workers the calibration generates similar search and rest unemployment dynamics over the business cycle. This yields high and similar cyclical volatilities for the unemployment, job finding and separation rates across age groups. In particular, the $u$ volatilities for the young and the prime-aged are 0.139 and 0.141, the volatilities of $f$ for young and prime-aged workers are 0.099 and 0.096; and the volatilities of $s$ are 0.059 for young workers and 0.063 for prime-aged workers. We return to this point in the next section when presenting the calibration details of the excess mobility model.

\vspace{-0.55cm}
\subsection{Excess Mobility and Cyclical Unemployment}\label{app:c2}
\vspace{-0.25cm}

To show the importance of idiosyncratic occupation-worker ($z$) productivity shocks in allowing the full model to replicate the cyclical behavior of many labor market variables, we re-estimate the model by shutting down occupation-wide heterogeneity (level and business cycle loadings), effectively setting $p_{o,t}=1$ at all $t$. In this case, a worker's productivity at time $t$ in an occupation $o$ is completely described by aggregate productivity $A$, worker-occupation match productivity $z$ and occupation-specific human capital $x$. Workers do not (and do not want to) prefer a new occupation over another before knowing their $z$. Note that although we label this model as the ``excess mobility model'', it can easily be made consistent with the observed average net flows by imposing an exogenous transition matrix that governs the probabilities with which a worker in occupation $o$ observes a $z$ in a different occupation $o'$.\footnote{With a cyclically varying exogenous transition matrix, we would also be able to match the observed cyclical net flows.} This is in contrast to our full model, where occupational productivities $p_o, p_{o'}$ differ and change relative to one another over the cycle and in response workers change the direction of their cross-occupation search. As such the full model can be considered as the ``endogenous net mobility'' model, while the excess mobility model as the ``exogenous net mobility'' model.

\vspace{-0.55cm}
\subsubsection{Benchmark Excess Mobility Model}\label{app:c2benchmark}
\vspace{-0.2cm}

\begin{figure}[t]
\caption{Excess Mobility Model: Data and Model Comparison} \label{f:minimumdistance_xs1}
\centering
\resizebox{1.0\textwidth}{!}{
\subfloat[Mobility-Duration Profile (All)]{\centering\includegraphics[width=0.4 \textwidth]{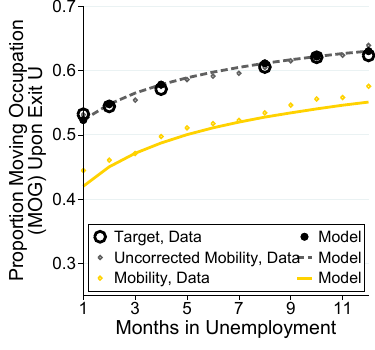}
\label{f:xs_mdur_all} }
\subfloat[Mobility-Duration Young-Prime]{\centering\includegraphics[width=0.4 \textwidth]{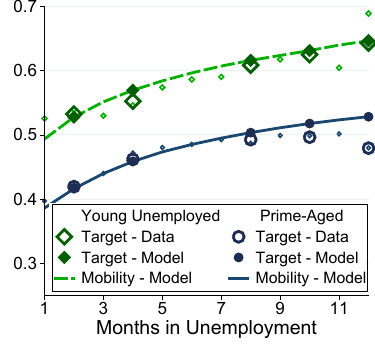}
\label{f:xs_mdur_young} }
\subfloat[Cyclical Mobility-Duration Profile]{\centering\includegraphics[width=0.4 \textwidth]{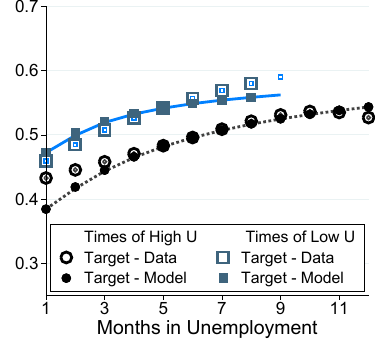}
\label{f:xs_mdur_cyclical} }
}

\resizebox{1.0\textwidth}{!}{\subfloat[\large{Survival Profile (All)}]{\centering\includegraphics[width=0.4 \textwidth]{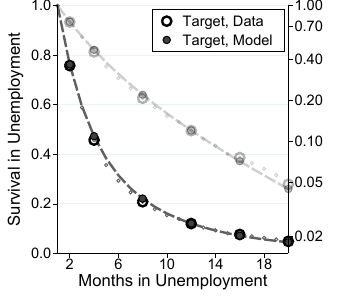}
\label{f:xs_survival_all} }
\subfloat[\large{Survival Profile Young-Prime}]{\centering\includegraphics[width=0.4 \textwidth]{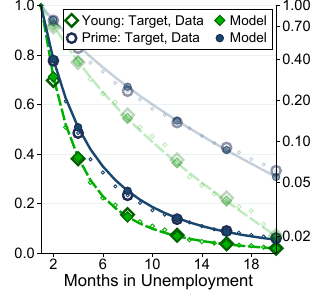}
\label{f:xs_survival_youngprime} }
\subfloat[\large{Survival Profile Movers-Stayers}]{\centering\includegraphics[width=0.4 \textwidth]{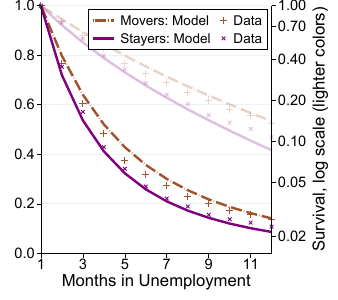}
\label{f:xs_survival_moverstayer} }}
\end{figure}

This version of the model corresponds to the excess mobility model in the main text. Except for occupation-wide productivity differences and a cross-occupation search decisions, everything else remains as described in Section 3 of the main text. We use the same functional forms as done to calibrate the full model in Section 4 of the main text. This implies that to capture economic choices and gross mobility outcomes, we now have a set of 14 parameters to recover, where $\left[c, \rho_z, \sigma_z, \underline{z}_{norm} \right]$ governs occupational mobility due to idiosyncratic reasons (excess mobility); $\left[x^2, x^3, \gamma _{d},\delta_{L}, \delta_{H} \right]$ governs differences in occupational human capital; and the remainder parameters $\left[ k, b, \eta, \rho_A, \sigma_A \right]$ are shared with standard DMP calibrations. We jointly calibrate these parameters by matching the moments reported in Table \ref{t:excess_full} and Figures \ref{f:xs_mdur_all}-\ref{f:xs_survival_youngprime}.

\begin{table}[t]
\centering
\caption{Targeted Moments, Excess Mobility Model}\label{t:excess_full}
  \resizebox{1.0
  \textwidth}{!}{
\small{
\begin{tabular}{lcclcc}
\toprule
\toprule
Moment      & Model & Data & Moment      & Model     & Data \\ \cmidrule(lr){1-3} \cmidrule(lr){4-6}
Agg. output per worker mean & 1.005 & 1.000 & Rel. separation rate young/prime-aged & 2.134 & 1.994 \\
Agg. output per worker persistence, $\rho_{outpw}$ & 0.761 & 0.753 &  Rel. separation rate recent hire/all  & 5.230 & 4.944\\
Agg. output per worker st. dev., $\sigma_{outpw}$ & 0.0094 & 0.0094  &  Prob (unemp. within 3 yr for empl.) & 0.148  & 0.122 \\
Mean unemployment & 0.0355 & 0.0355 & Empirical elasticity matching function & 0.522 & 0.500 \\
Average u. duration movers/stayers & 1.211 & 1.139 & 5-year OLS return to occ. tenure & 0.150 & 0.154 \\
Repeat mobility: occ. stay after stay & 0.600 & 0.634   & 10-year OLS return to occ. tenure & 0.230  & 0.232 \\
Occ. mobility young/prime-aged & 1.162 & 1.163  \\
Occ. mobility-duration profiles & \multicolumn{2}{c}{Fig 5a,b,c} & U. survival profiles & \multicolumn{2}{c}{Fig 5d,e} \\
\bottomrule
\bottomrule
\end{tabular}
}
}
\end{table}

The excess mobility model matches very well the targeted occupational mobility moments as well as the job finding and job separation moments. The fit is comparable with the one of the full model. This model also matches well the untargeted moments pertaining to workers' gross occupational mobility and job finding hazards discussed in the previous section. The fit of other untargeted moments is not shown here to save space, but available upon request. The estimated parameter values in this calibration are also very similar to the ones obtained in the full model. These are $c=7.549$, $k=125.733$, $b=0.843$, $\eta=0.241$, $\delta_{L}=0.0034$, $\delta_{H}=0.0004$, $\underline{z}_{corr}=0.349$, $\rho_{A}=0.998$, $\sigma_{A}=0.00198$, $\rho_{z}=0.998$, $\sigma_{z}=0.00707$, $x^{2}=1.181$, $x^{3}=1.474$ and $\gamma_{h}=0.0039$.

As shown in the main text, the excess mobility calibration is also able to fit a wide range of cyclical features of the labor market. The left panel of Table \ref{t:vol_excess_full} (below) shows the time series properties of the unemployment, vacancy, job finding and job separation rates and labor market tightness as well as the full set of correlations between them, obtained from the excess mobility calibration. Here we find that the cyclical implications of the excess mobility model are very similar to that of the full model, as shown and discussed in the main text.

\begin{figure}[t]
\caption{Unemployment Decomposition - Excess Mobility Model}
\label{f:Decomp_excess}
\resizebox{1\textwidth}{!}{\subfloat[All workers]{\includegraphics[width=0.35 \textwidth]{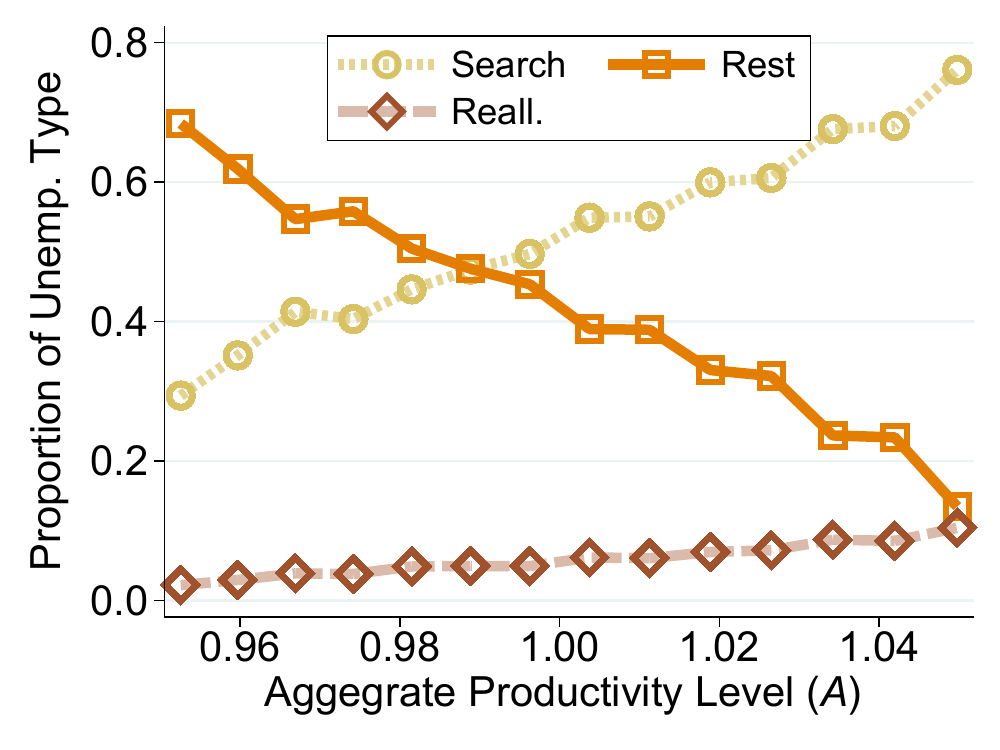}\hspace{0.1cm}}
\subfloat[Young Workers]{\includegraphics[width=0.35 \textwidth]{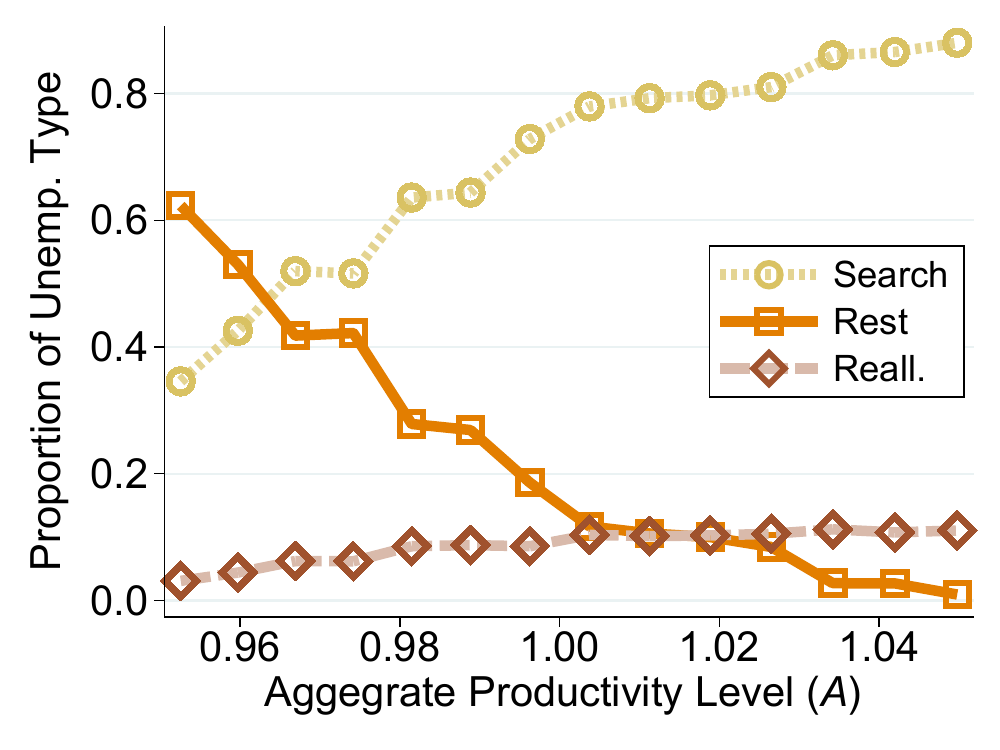} \hspace{0.01cm}}
\subfloat[Prime-aged Workers]{\includegraphics[width=0.35 \textwidth]{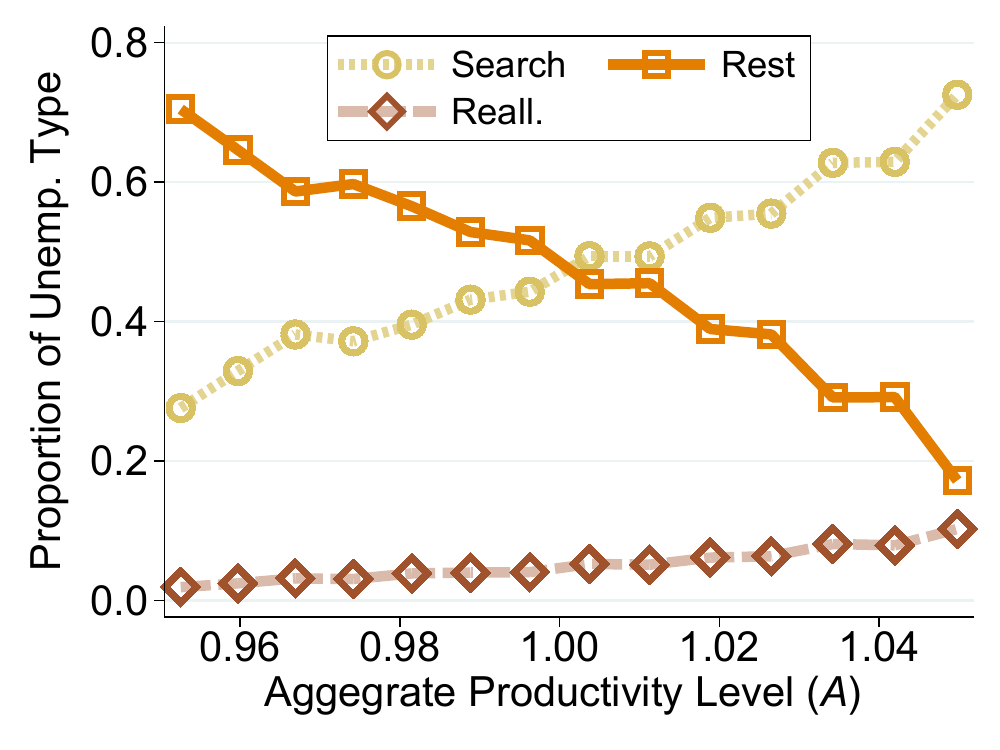}}}
\end{figure}

\begin{figure}[ht!]
\caption{(Un)Employment Distributions and Aggregate Productivity by Age Groups}
\label{f:Decomp_heatmaps}
\centering
\resizebox{0.75\textwidth}{!}{\subfloat[\normalsize{Young Unemployed}]{\includegraphics[width=0.368 \textwidth]{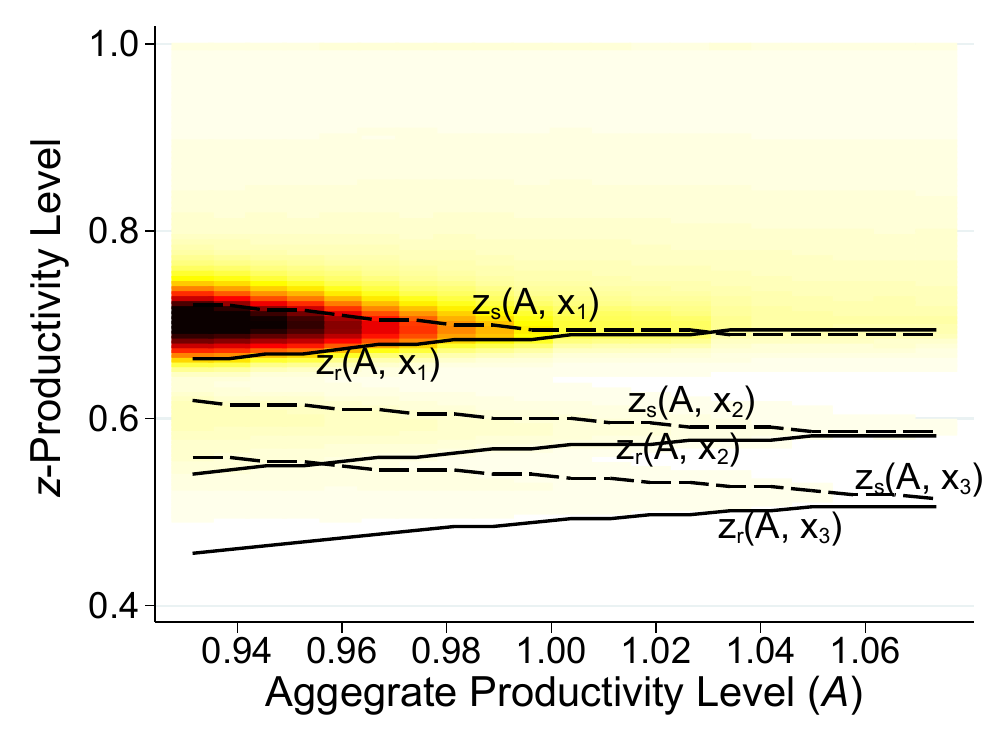}
}\subfloat[\normalsize{Young Employed}]{\includegraphics[width=0.35 \textwidth]{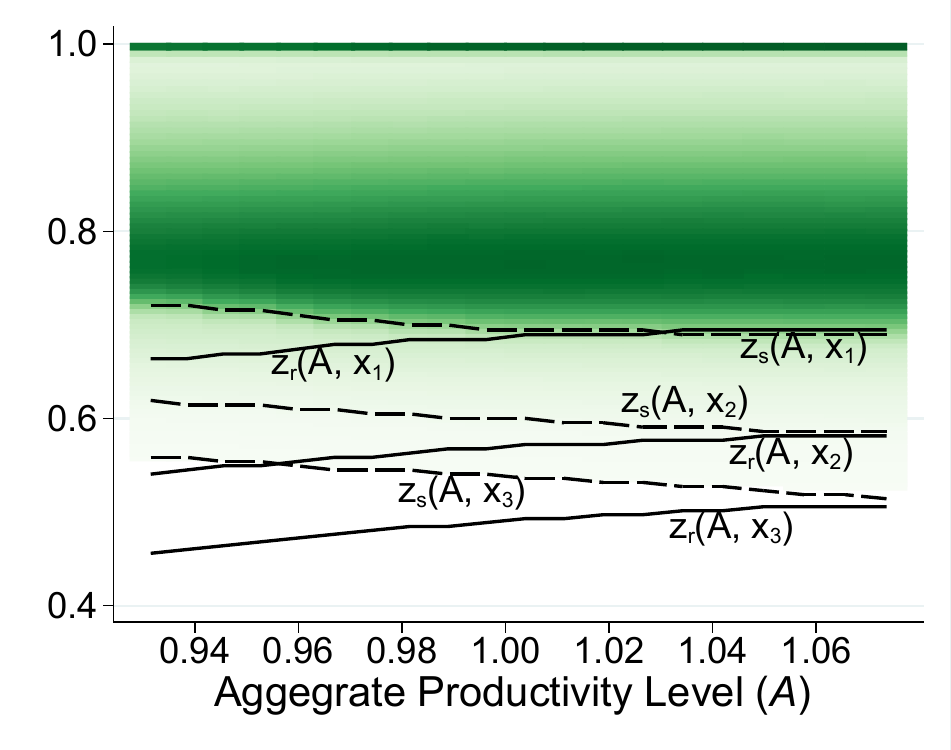}}
}
\vspace{0.2cm}
\resizebox{0.75\textwidth}{!}{\subfloat[\normalsize{Prime-Aged Unemployed}]{\includegraphics[width=0.368 \textwidth]{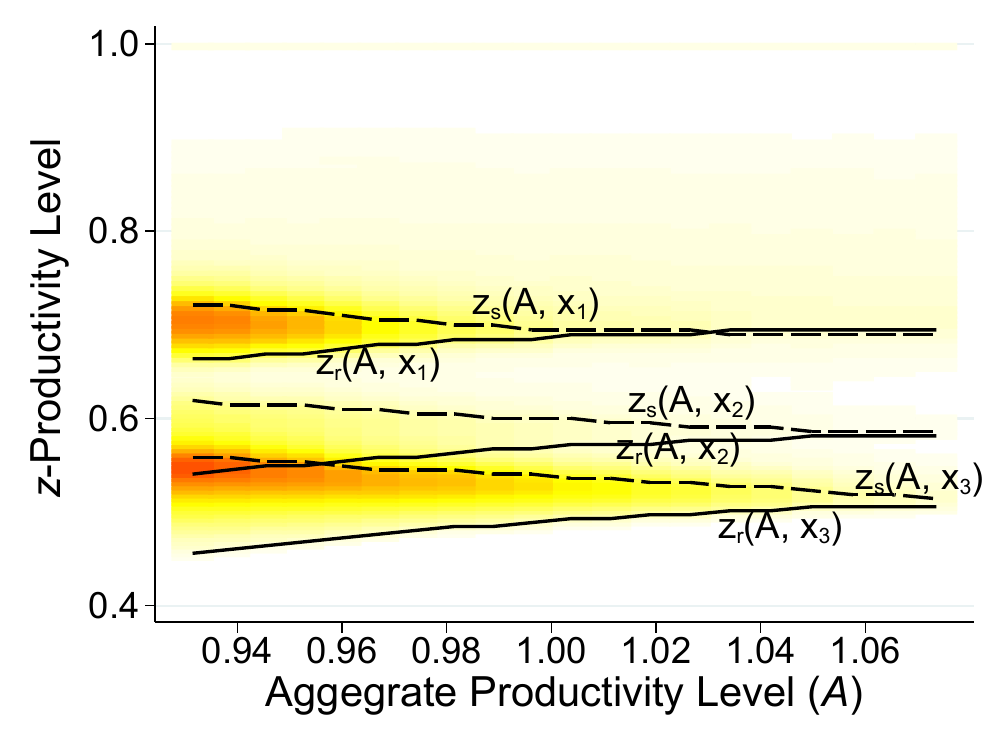}
}\subfloat[\normalsize{Prime-Aged Employed}]{\includegraphics[width=0.35 \textwidth]{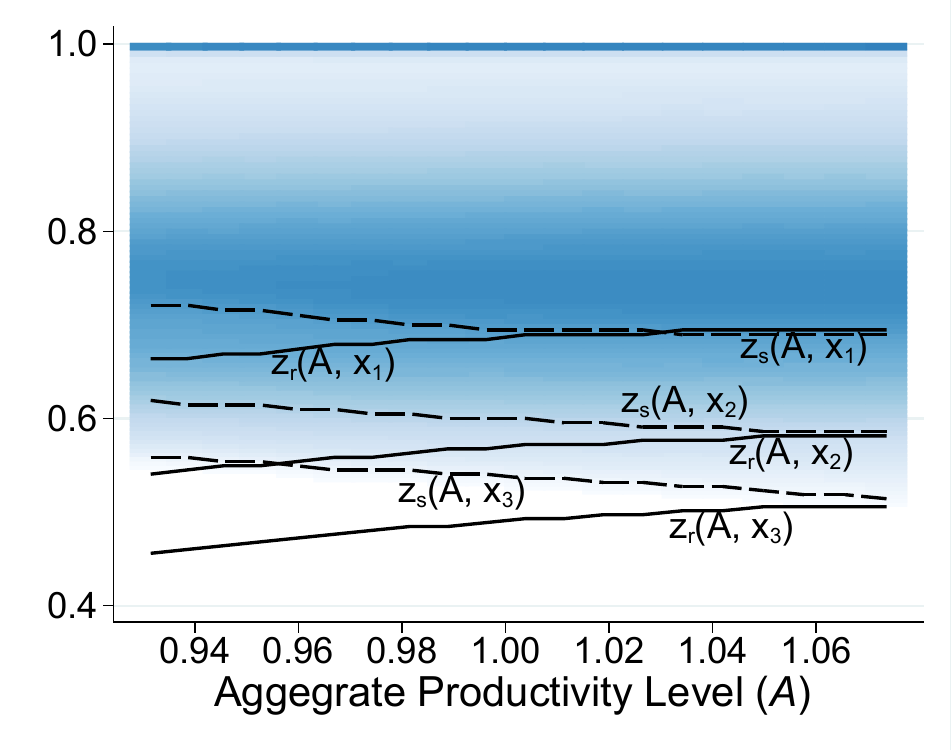}}
}
\end{figure}

\vspace{-0.55cm}

\paragraph{Age patterns} Figure \ref{f:Decomp_excess} shows that the occurrence of search, rest and reallocation unemployment episodes not only happens when pooling all workers together but also for each age group, as in the full model. Figure \ref{f:Decomp_heatmaps} shows these age group dynamics more clearly by depicting the distribution of unemployed and employed workers among young and prime-aged workers. It shows that during recessions unemployment among young workers is concentrated slightly above $z^s(.,x^{1})$ and between $z^s(.,x^{1})$ and $z^r(.,x^{1})$. During expansions, however, unemployment is located above the $z^s(.,x^{1})$ cutoff. In the case of prime-aged workers, the concentration of unemployment during recessions and expansions occurs mostly above $z^s(.,x^{3})$ and between $z^s(.,x^{3})$ and $z^r(.,x^{3})$, but also between the $z^s(.,x^{1})$ and $z^r(.,x^{1})$ cutoffs. This difference implies that during expansion episodes of rest unemployment are still prevalent among prime-aged workers, while for young workers these episodes basically disappear (as shown in Figure \ref{f:Decomp_excess}) and are consistent with a lower occupational mobility rate among prime-aged workers.

As in the full model, the excess mobility calibration obtains a similar cyclicality for the unemployment, job finding and separations rates across age-groups. In both models this occurs because the estimated $z$-productivity process places enough workers on the $z^s$ cutoffs across the respective human capital levels. Figure \ref{f:Decomp_heatmaps} shows that, as is the case for low human capital workers, many high human capital workers enjoy high $z$-productivities, but for a number of them their $z$-productivities have drifted down, positioning themselves close to $z^s(.,x^3)$. Some of these high human capital workers will subsequently leave the occupation, but over time the stock of workers close to $z^s(.,x^3)$ will be replenished by those workers who currently have high $z$-productivities but will suffer bad $z$-realizations in the near future. As $z^s(.,x^3)<z^s(.,x^1)$ the average level of separations is lower for high human capital workers, but this nevertheless does not preclude the similarity in the aforementioned cyclical responsiveness.

Given that it is clear the excess mobility model is able to replicate on its own many critical features of the full model, in what follows we use it to perform two key exercises. The first one highlights the effect of human capital depreciation in attenuating the cyclical properties of the above labor market variables and motivates our use of the cyclical shift of the mobility-duration profile as a target. The second exercise investigates the quantitative implications of our model when considering that a worker's varying job finding prospects (due to the stochastic nature of the $z$-productivity process) during a jobless spell can be linked to observed transitions between the states of unemployment and non-participation (or marginally attached to the labor force) as defined in the SIPP. For this exercise we recompute all of the relevant empirical targets using non-employment spells that contain a mix of periods of unemployment and non-participation. To save space we refer the reader to our working paper Carrillo-Tudela and Visschers (2020) for the list of targeted moments, the fit and the estimated parameter values under both exercises.

\vspace{-0.55cm}
\subsubsection{The Importance of Human Capital Depreciation}\label{app:c2nohc}
\vspace{-0.25cm}

To estimate the full and the excess mobility models we used the mobility-duration profiles at different durations during recessions and expansions. These patterns informed us about the rate of occupational human capital depreciation during spells of unemployment. In the main text, we argued that these profiles were crucial in helping us identify the depreciation parameter, $\gamma_{h}$. The reason for the latter is that a model which did not incorporate human capital depreciation will generate very similar long-run moments as a model which did incorporate depreciation, but generate different cyclical predictions. To show this, we now discuss the result of the excess mobility model without human capital depreciation. To estimate the latter we target the same \textit{long-run} moments as in the calibration described above, but do not target the cyclical behaviour of the mobility-duration profile.

We find that the fit is very good, similar to the models which incorporates human capital depreciation. It also does well in matching the same untargeted long-run moments described above. The estimated parameter values are also similar. Further, this calibration finds that periods of search, rest and reallocation unemployment can arise during a worker's jobless spell across all levels of occupational human capital.

\begin{figure}[t]
\caption{The Cyclical Mobility-Duration Profile in Relation to Human Capital Depreciation}
\label{f:mob_dur}
\begin{centering}
\resizebox{0.80\textwidth}{!}{\subfloat[Model with HC Depreciation]{\centering\includegraphics[height=0.2 \textheight, width=0.35 \textwidth]{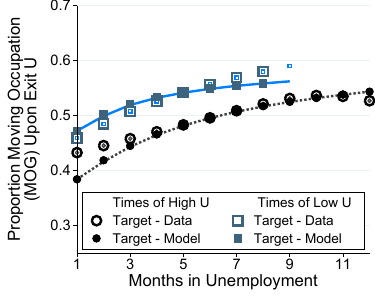}\label{f:mob_dur_cyc_HCD} }
\subfloat[Model without HC Depreciation]{\centering\includegraphics[height=0.2 \textheight, width=0.35 \textwidth]{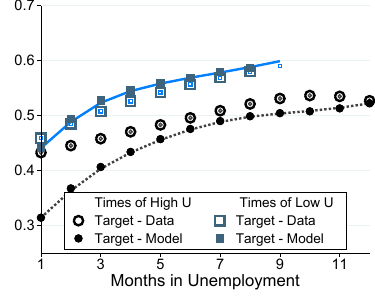}
\label{f:mob_dur_cyc_NHCD} }}
\par\end{centering}
\end{figure}

Figure \ref{f:mob_dur} shows the first key difference between the excess mobility model with and without occupational human capital depreciation. We plot the mobility-duration profile in times of expansions and recessions (low and high unemployment, respectively), where Figure \ref{f:mob_dur_cyc_HCD} shows the mobility-duration profiles from the model with human capital depreciation and Figure \ref{f:mob_dur_cyc_NHCD} shows the ones for the model without human capital depreciation. The latter finds that the lack of human capital depreciation implies the model completely misses the profile at nearly all durations during recessions, particularly at low durations. Is precisely this lack of fit that motivated us to add the cyclical patterns of the mobility-duration profile as targets in order to help identify the rate of human capital depreciation.

\begin{table}[t]
\centering
\caption{Excess Mobility Models: HP-filtered Business Cycle Statistics}\label{t:vol_excess_full}
 \resizebox{0.9\textwidth}{!}{
{\footnotesize
\begin{tabular}{|c|cccccc||cccccc|}
\hline
& \multicolumn{6}{|c||}{\textbf{Excess Mobility Model with HC dep.}}  & \multicolumn{6}{c|}{\textbf{Excess Mobility Model with No HC dep.}}\\ \hline
   & $u$ & $v$ & $\theta$ & $s$ & $f$ & $outpw$ & $u$ & $v$ & $\theta$ & $s$ & $f$ & $outpw$ \\ \hline\hline
$\sigma$ &   0.14 & 0.05 & 0.17 & 0.06 &  0.10 &0.01&  0.20 & 0.06 & 0.25 & 0.10 & 0.14 &0.01 \\
$\rho_{t-1}$ & 0.95 & 0.91 & 0.94 & 0.89 & 0.94 & 0.94 & 0.94 & 0.87 & 0.94 & 0.88 & 0.93 & 0.94 \\ \hline\hline
& \multicolumn{6}{|c||}{\textbf{Correlation Matrix}} & \multicolumn{6}{c|}{\textbf{Correlation Matrix}}\\ \hline
$u$ & 1.00 & -0.72 & -0.98 & 0.80 & -0.89 &  -0.95  & 1.00 & -0.62 & -0.97 & 0.78 & -0.89 & -0.92  \\
$v$ &  & 1.00 &  0.84 & -0.81 & 0.94 & 0.85 &  & 1.00 &  0.76 & -0.61 & 0.77 & 0.72  \\
$\theta$ &  &  & 1.00 & -0.84 & 0.96 & 0.97    &  &  & 1.00 & -0.78 & 0.94 & 0.93 \\
$s$ &   &   &   & 1.00 & -0.87& -0.90 &   &   &   & 1.00 & -0.81 & -0.84  \\
$f$ &  &  &    &      & 1.00 &  0.95  &  &  &    &      & 1.00 &  0.89   \\
$outpw$ &  &  &  &  & & 1.00 &  &  &  &  & & 1.00 \\ \hline
\multicolumn{13}{p{0.85\textwidth}}{\scriptsize{Note: Each model's aggregate time series arise from the distributions of employed and unemployed workers across all labor markets, combined with agents' decisions. Times series are centered 5Q-MA series of quarterly data to smooth out the discreteness in the relatively flat cutoffs (relative to the grid). The cyclical components of the logged time series are obtained by using an HP filter with parameter 1600.}}
\end{tabular}}
}
\end{table}

Table \ref{t:vol_excess_full} shows the second key difference. The model without depreciation generates a larger amount of cyclical volatility in the aggregate unemployment, job finding and job separation rates in relation to the model with human capital depreciation. Relative to the data, Table \ref{t:volatilities} shows an overshooting in the volatilities of the unemployment and job finding rates. This occurs as the area of inaction between $z^r$ and $z^s$ becomes too cyclical. Without human capital depreciation there is a sharp drop in the proportion of rest unemployment and a sharper rise in the proportions of search and reallocation unemployment as the economy improves. Human capital depreciation attenuates these effects, improving the model's fit.

\vspace{-0.55cm}
\subsubsection{The Unemployed and Marginally Attached}\label{app:c2nun}
\vspace{-0.25cm}

Previous calibrations build the analysis based on the interpretation that, although a worker who is currently in rest unemployment cannot find a job, he would want to search for jobs (as opposed to stay idle at home) because he still faces a positive expected job finding probability in the near future. Episodes of rest unemployment, however, could conceptually be extended to incorporate marginally attached workers. To investigate the latter we expand our analysis to capture more broadly the occupational mobility decisions of the unemployed and marginally attached in shaping the cyclicality of aggregate unemployment.

We do this by re-estimating the excess mobility model, recomputing the targets using non-employment spells in which workers transition between unemployment and non-participation as labelled in the SIPP. We consider non-employment spells with at least one period of unemployment, which we label `NUN' spells. To avoid maternity and related issues in non-participation we restrict the focus to men. We show that when considering NUN spells the model still reproduces the observed cyclical amplification in the non-employment rate.\footnote{Here we also focus on spells of at least one month and workers who say that they are ``without a job'', mirroring these sample restrictions for unemployment spells.}

The fit is once again very good in this case even though the survival probability in NUN spells shifts up significantly at longer durations, compared to the corresponding patterns for unemployment spells, both for all workers and across age groups. The estimated parameter values are also broadly similar to the ones in the previous versions of the excess mobility model, changing in expected directions. The $z$ process is now somewhat more volatile, but the higher reallocation cost implies that the area of inaction between $z^s$ and $z^r$ is (in relative terms) also larger. The latter leaves more scope for workers to get ``trapped'' for longer periods in rest unemployment episodes, thus creating an increase in the survival functions across all, young and prime-aged workers as observed in the data.

\begin{table}[t]
\centering
\caption{NUN Spells: HP-filtered Business Cycle Statistics}\label{t:volatilities12}
 \resizebox{0.9\textwidth}{!}{
{\footnotesize
\begin{tabular}{|c|cccccc||cccccc|}
\hline
& \multicolumn{6}{|c||}{\textbf{Data (1983-2014) - NUN spells}}  & \multicolumn{6}{c|}{\textbf{Excess Mobility Model - NUN spells}}\\ \hline
   & $u$ & $v$ & $\theta$ & $s$ & $f$ & $outpw$ & $u$ & $v$ & $\theta$ & $s$ & $f$ & $outpw$ \\ \hline\hline
$\sigma$ &   0.11 &	0.11	& 0.21 &0.10 &0.08 &0.01&  0.09 &	0.03	& 0.11 &	0.04 &	0.07 &	0.01 \\
$\rho_{t-1}$ & 0.97& 0.99	& 0.94 &	0.99	& 0.94 &	0.93	& 0.95& 0.85	& 0.94 &	0.87	& 0.93	& 0.94 \\ \hline\hline
& \multicolumn{6}{|c||}{\textbf{Correlation Matrix}} & \multicolumn{6}{c|}{\textbf{Correlation Matrix}}\\ \hline
$u$ & 1.00 &-0.92	& -0.98	& 0.79 &	-0.96	 &-0.51  & 1.00 & -0.51 &	-0.96	 &0.65 &	-0.81 &	-0.89  \\
$v$ &  & 1.00 &  0.98 &	-0.79 &	0.94	& 0.61 &  & 1.00 &  0.72 &	-0.50 &	0.75	& 0.65  \\
$\theta$ &  &  & 1.00 & -0.80 &	0.96	& 0.57   &  &  & 1.00 & -0.68 &	0.89	& 0.92 \\
$s$ &   &   &   & 1.00 & -0.87 &	-0.43 &   &   &   & 1.00 & -0.64	& -0.80  \\
$f$ &  &  &    &      & 1.00 &  0.46 &  &  &    &      & 1.00 &  0.86  \\
$outpw$ &  &  &  &  & & 1.00 &  &  &  &  & & 1.00 \\ \hline
\multicolumn{13}{p{0.85\textwidth}}{\scriptsize{See description under Table \ref{t:vol_excess_full} for details.}}
\end{tabular}}
}
\end{table}

Table \ref{t:volatilities12} shows the main takeaway of this exercise. The model remains able to generate cyclical movements of the non-employment, job finding and job separation rates as well as a relatively strong Beveridge curve. In particular, the cyclical volatilities of the non-employment and job finding rates are the same as in the data. As in the previous estimations, here we also find that the reason for the amplification of the non-employment rate is that the model generates period of search, rest and reallocation unemployment, whose relative importance changes over the cycle.

Including marginally attached workers in our analysis increases the overall importance of rest unemployment in normal times. This is consistent with the fact that in these times the non-employment rate is higher and the associated job finding rate lower, compared to our benchmark unemployment and job finding rates measures. Further, this version of the model still needs to accommodate short-term outflows as before and does so mostly through search unemployment episodes. As a result, the proportion of rest unemployment decreases at a slower rate with $A$. Even at the highest aggregate productivity levels rest unemployment is very prevalent, representing about 40\% of all episodes during a non-employment spell, with a large role for prime-aged workers. Overall we find that a version of the excess mobility model that considers NUN spells exhibits a higher non-employment rate but a lower cyclicality than in our benchmark model. This is consistent with the data, where we observe a lower cyclicality among the non-employment than among the unemployment.

\vspace{-0.55cm}
\subsection{The Importance of Occupational Mobility}\label{app:c3}
\vspace{-0.25cm}

To demonstrate the importance of including the $z^r$ cutoff to \emph{simultaneously} replicate the cyclical behaviour of the unemployment duration distribution and the aggregate unemployment rate, we re-estimate the model by shutting down occupational mobility. This is done by exogenously setting $c$ to a prohibiting level. We present two calibrations. Model I targets the same moments as in the full model with the exception of those pertaining to occupational mobility.\footnote{To make the estimation of Model I as comparable as possible with the previous ones, we continue targeting the returns to occupational mobility to inform the human capital levels, $x^{2}$ and $x^{3}$. Under no occupational mobility, human capital could also be interpreted as general and not occupation specific, depending on the aim of the exercise. In this sense it would be more appropriate to target the returns to general experience. However, a comparison between the OLS returns to general experience and the OLS returns to occupational human capital estimated by Kambourov and Manovskii (2009) from the PSID (see their Table 3 comparing columns 1 and 3 or 6 and 8), suggests that this bias should be moderate. Using their estimates, the 5 year return to general experience is about 0.19, while the 10 years returns is about 0.38.} Model 2 is chosen to achieve a higher cyclical volatility in the aggregate unemployment rate but is more permissive of deviations from the targets.

\vspace{-0.55cm}

\paragraph{Model I} To save space our working paper, Carrillo-Tudela and Visschers (2020), shows the table of targeted moments and the fit as well as the estimated parameter values. The fit is largely comparable along nearly all corresponding dimensions to the full and excess mobility models. The estimated parameters are largely sensible. Two key features stand out: (i) there is now a higher role for search frictions as the model estimates a higher value $k=195.58$; and (ii) the $z$ process is now less persistent $\rho_{z}=0.9923$ and exhibits a much larger variance $\sigma_{z}=0.0300$ in the stationary distribution, generating a perhaps too large $Mm$ ratio of 2.28.\footnote{In this context the $z$-productivity process can be interpreted as an idiosyncratic productivity shock affecting a worker's overall productivity, rather than a worker's idiosyncratic productivity within an occupation.} Table \ref{t:volatilities2} under ``No Occupational Mobility - Model I'', however, shows that this model cannot generate enough cyclical volatility on all the relevant labor market variables.

\begin{table}[t]
\centering
\caption{Models Without Reallocation: HP-filtered Business Cycle Statistics}\label{t:volatilities2}
 \resizebox{0.9\textwidth}{!}{
{\footnotesize
\begin{tabular}{|c|cccccc||cccccc|}
\hline
& \multicolumn{6}{|c||}{\textbf{No Occupational Mobility - Model I}}  & \multicolumn{6}{c|}{\textbf{No Occupational Mobility - Model II}}\\ \hline
   & $u$ & $v$ & $\theta$ & $s$ & $f$ & $outpw$ & $u$ & $v$ & $\theta$ & $s$ & $f$ & $outpw$ \\ \hline\hline

$\sigma$ & 0.04 &   0.02  &  0.06 &   0.03  &  0.03 &   0.01 &  0.10 & 0.03  & 0.12 & 0.07 & 0.05 & 0.01\\
$\rho_{t-1}$ & 0.94 &   0.85  &  0.93 &   0.83  &  0.87  &  0.94 & 0.94 & 0.85 & 0.94 & 0.89 &  0.89 & 0.94 \\ \hline\hline
& \multicolumn{6}{|c||}{\textbf{Correlation Matrix}} & \multicolumn{6}{c|}{\textbf{Correlation Matrix}}\\ \hline
$u$ & 1.00 &  -0.42 &  -0.93  &  0.71 &  -0.77 &  -0.86 & 1.000 & -0.57 &   -0.98 &    0.83 &   -0.79 &   -0.96  \\
$v$ &  & 1.000 &  0.72 &  -0.26  &  0.58 &   0.60. &  & 1.00 &  0.73 &   -0.60 &   0.72 &    0.66 \\
$\theta$ &  &  & 1.00 & -0.65  &  0.82 &   0.90 &  &  & 1.00 & -0.85 &  0.84 &    0.97  \\
$s$ &   &   &   & 1.00 & -0.52  &  -0.71 &   &   &   & 1.00 & -0.85 &   -0.89 \\
$f$ &  &  &    &      & 1.00 & 0.74 &  &  &    &      & 1.00 &  0.84  \\
$outpw$ &  &  &  &  & & 1.00 &  &  &  &  & & 1.00 \\ \hline
\multicolumn{13}{p{0.89\textwidth}}{\scriptsize{See description under Table \ref{t:vol_excess_full} for details.}}
\end{tabular}}
}
\end{table}

 \begin{table}[t]
\centering
\caption{Incomplete Unemployment Duration Distribution Behavior}\label{t:dur_unemp}
\resizebox{1\textwidth}{!}{
\begin{tabular}{lccclrcclrccc}
\toprule
\toprule
\multicolumn{2}{c}{} & \multicolumn{10}{c}{Panel A: Incomplete Unemployment Distribution (1-18 months)} &\multicolumn{1}{c}{}\\ \midrule
 & \multicolumn{4}{c}{All workers} & \multicolumn{4}{c}{Young workers} & \multicolumn{4}{c}{Prime-aged workers} \\
 \cmidrule(lr){2-5} \cmidrule(lr){6-9} \cmidrule(lr){10-13}
Unemp.  & Occ & No Occ.& No Occ. & Data  & Occ. & No Occ. & No Occ. &   Data      &  Occ. & No Occ. & No Occ. & Data \\
 Duration   & Model  & Model I & Model II     &  & Model & Model I & Model II     & & Model & Model I & Model II  &   \\ \cmidrule(lr){2-5} \cmidrule(lr){6-9} \cmidrule(lr){10-13}
1-2 m  & 0.43 & 0.38    &  0.36     &  0.43  & 0.53 & 0.45 & 0.36 & 0.47 & 0.41 & 0.35    &  0.36     &  0.41 \\
1-4 m  & 0.65 & 0.58 & 0.56 &  0.67  & 0.76 & 0.67 & 0.55 &   0.71 & 0.63 & 0.56  & 0.56  &  0.65 \\
5-8 m  & 0.20 & 0.22 & 0.22 &  0.20  & 0.16 & 0.19 & 0.22  &  0.19  & 0.21 & 0.23  & 0.22 & 0.21  \\
9-12 m & 0.09 & 0.11 & 0.12 & 0.08  & 0.05 & 0.08  &0.12 & 0.07  &  0.10& 0.12 & 0.12& 0.09  \\
13-18m  & 0.06 & 0.09 & 0.10 & 0.05 & 0.03 & 0.06 & 0.11 &  0.03 & 0.07 & 0.09 & 0.09 & 0.06 \\ \toprule
\multicolumn{2}{c}{} & \multicolumn{10}{c}{Panel B: Cyclical Changes of the Incomplete Unemployment Distribution (1-18 months)} &\multicolumn{1}{c}{} \\ \midrule
 & \multicolumn{2}{c}{} & \multicolumn{4}{c}{Elasticity wrt $u$} & \multicolumn{4}{c}{HP-filtered Semi-elasticity wrt $u$} & \multicolumn{2}{c}{}\\  \cmidrule(lr){4-7} \cmidrule(lr){8-11}
 Unemp. & & & Occ & No Occ.& No Occ. & Data & Occ & No Occ.& No Occ. & Data  & \multicolumn{2}{c}{}\\
 Duration & & & Model & Model I & Model II &  & Model & Model I & Model II &   \multicolumn{2}{c}{}\\  \cmidrule(lr){4-7} \cmidrule(lr){8-11}
1-2 m  & &  & -0.450 &\multicolumn{1}{c}{-0.328} & \multicolumn{1}{c}{-0.265} & -0.464 & -0.155 & \multicolumn{1}{c}{-0.122} & \multicolumn{1}{c}{-0.099} & -0.169 & \multicolumn{2}{c}{}\\
1-4 m & & & -0.321 & \multicolumn{1}{c}{-0.241} & \multicolumn{1}{c}{-0.186} &  -0.363 & -0.168 & \multicolumn{1}{c}{-0.134} & \multicolumn{1}{c}{-0.107} &  -0.184 & \multicolumn{2}{c}{} \\
5-8 m &  &  & 0.414 & \multicolumn{1}{c}{0.148} &  \multicolumn{1}{c}{0.119} & 0.320 & 0.067 & \multicolumn{1}{c}{0.046} &  \multicolumn{1}{c}{0.040}  & 0.076  & \multicolumn{2}{c}{}\\
9-12 m & &  & 1.102 & \multicolumn{1}{c}{0.422} &  \multicolumn{1}{c}{0.300}  & 0.864 & 0.058 & \multicolumn{1}{c}{0.049}  &  \multicolumn{1}{c}{0.039}  & 0.072  & \multicolumn{2}{c}{}\\
$>$13 m & &     &  1.817 & \multicolumn{1}{c}{0.752} & \multicolumn{1}{c}{0.504} & 1.375 & 0.044 & \multicolumn{1}{c}{0.039}& \multicolumn{1}{c}{0.029} & 0.043 & \multicolumn{2}{c}{}\\
 \bottomrule
 \bottomrule
 \end{tabular}}
\end{table}

Table \ref{t:dur_unemp} further shows that Model I is not able to reproduce the observed average quarterly unemployment duration distribution at short or long durations (Panel A), nor does it capture the cyclical behavior of this distribution (Panel B). While Model I and the occupational mobility models replicate the same unemployment survival functions, they generate different incomplete duration distributions. This occurs because the survival functions are computed pooling the entire sample, while the distribution of incomplete spells between 1 and 18 months is calculated for each quarter and then averaged across quarters. Model I generates about 50\% more long-term unemployment (9-12 months) relative to the data and the discrepancy is even stronger, about 80\%, at longer durations. That is, this model matches the unemployment survival functions by creating too dispersed unemployment durations within a typical quarter, in particular too many long spells, but its distribution then responds too little to the cycle.

\vspace{-0.55cm}

\paragraph{Model II} If one is willing to compromise on replicating the aggregate and age-group survival functions, however, the model without occupational mobility is able to generate larger cyclical volatilities. To show this we re-estimated the model by de-emphasising these survival functions. Once again we refer to Carrillo-Tudela and Visschers (2020) for the fit of this model and the estimated parameter values. There we show that the unemployment survival functions of young and prime-aged workers are no longer well matched. This model misses the distribution at longer durations for young workers and at shorter durations for prime-aged workers, such that age differences in job finding hazards have nearly disappeared. However, there is now a lower degree of search frictions, $k=102.019$ and the $z$ process is more persistent, $\rho_{z}=0.993$, and its overall dispersion is much lower than in Model I, $\sigma_{z}=0.0132$. It is these last properties that allow Model II to create more cyclical volatility as shown in the right panel of Table \ref{t:volatilities12}.

Panel A of Table \ref{t:dur_unemp} - No Occ. Model II shows that this model still creates too much long-term unemployment (13-18 months) in the average quarter, where the proportion of long-term unemployed (among those with spells between 1-18 months) is missed by a large margin for all, young and prime-aged workers. Further, Panel B of Table \ref{t:dur_unemp} shows that this feature is also reflected in a muted cyclical response of the unemployment duration distribution. Here we also find that Model II misses the semi-elasticity with respect to the unemployment rate by an average of about 40\% across the duration distribution.\footnote{In these versions of the model the behavior of spells with durations beyond 18 months might also impact the overall unemployment rate more than empirically warranted, especially as persistence can create a ``first-in last-out pattern'' when entering a recession in the form of very long unemployment spells for those who lost their job early on in the recession. We focus on the distribution of spells up to 18 months because censoring issues in the SIPP restrict how accurately we can investigate the behavior of very long spells over the cycle.}

\vspace{-0.25cm}

\paragraph{Discussion} These calibrations show that without the $z^r$ cutoff versions of our model with no occupational mobility cannot resolve the tension between individual unemployment outcomes and aggregate unemployment volatility. This arises as without the possibility of occupational mobility the new area of inaction is defined by the set of $z \in[\underline{z},z^s]$, where $\underline{z}$ is the lowest value of $z$, and its cyclical response now solely depends on $\partial z^s / \partial A$.

\begin{figure}[t]
\caption{Unemployment Decomposition - Models Without Occupational Mobility}
\label{f:Decomp2}
\centering
\resizebox{0.85\textwidth}{!}{\subfloat[\normalsize{All Workers - Model I}]{\includegraphics[width=0.5 \textwidth]{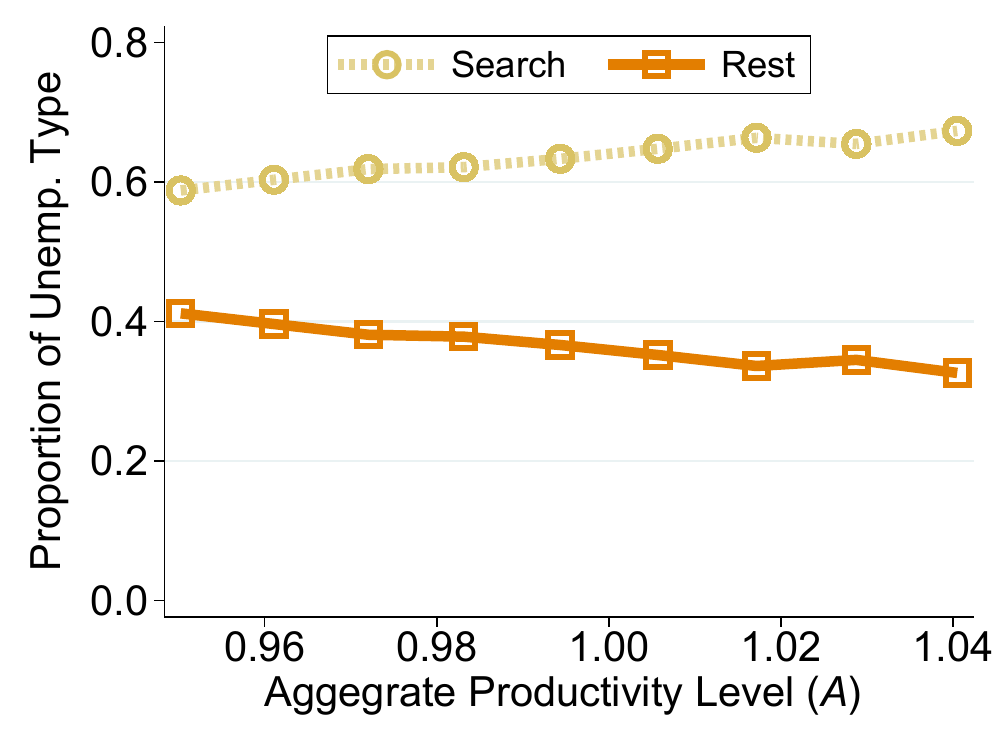}}
\hspace{0.02cm}
\subfloat[\normalsize{All Workers - Model II}]{\includegraphics[width=0.5 \textwidth]{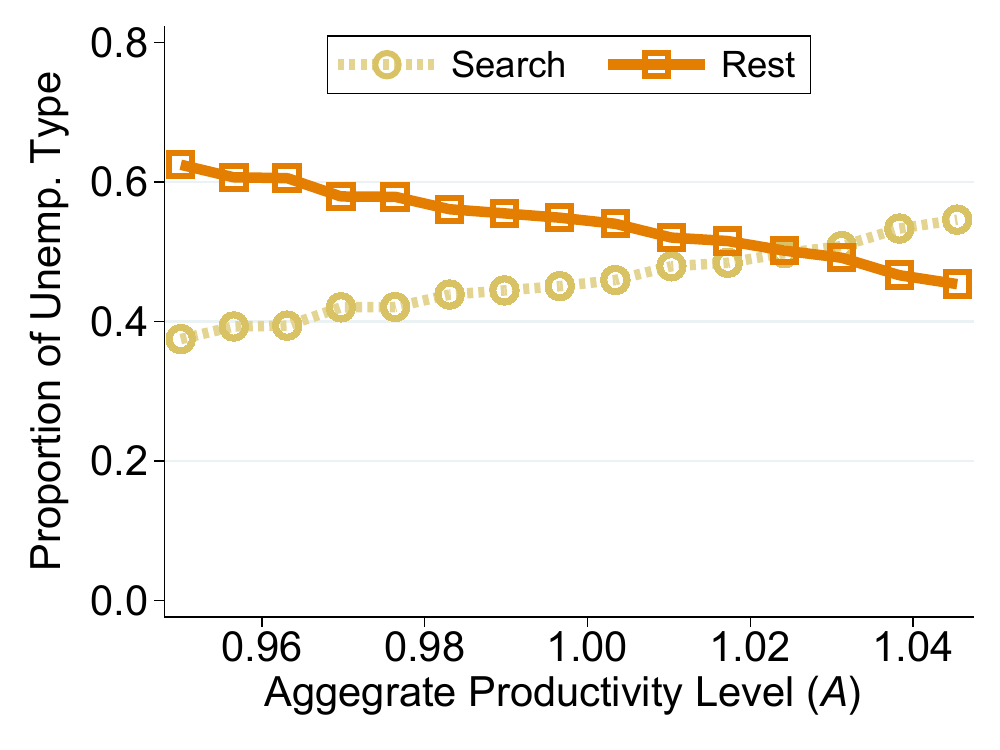}}}
\end{figure}

In the case of Model I, a less persistent and much more volatile $z$ process creates enough heterogeneity in unemployment durations that allows it to match the empirical unemployment survival functions at the aggregate and across age groups. However, it also increases the heterogeneity in $z$-productivities relative to the cyclical range of $A$. This dampens the model's cyclical performance as it implies less responsive $z^s$ cutoffs relative to the workers' $z$ distribution, weakening the cyclical responses of job separations and the rate at which workers leave the area of inaction. Moreover, with a larger vacancy posting cost, Figure \ref{f:Decomp2}.a shows that search unemployment is now more prominent than rest unemployment at any point of the cycle. Larger search frictions imply larger surpluses and therefore further reducing the cyclical responsiveness of the model.

The increased cyclical performance of Model II arises as its estimated $z$ process becomes more persistent and less volatile, creating more responsive $z^s$ cutoffs leading to stronger cyclical responses in job separations, as well as much more episodes of rest unemployment over all $A$. Figure \ref{f:Decomp2}.b shows that rest unemployment episodes are now more prominent than search unemployment episodes even during economic recoveries. It is only for the highest values of $A$ that search unemployment is more prominent, but only by a relatively small margin. With more responsive $z^s$ cutoffs, an aggregate shock can now move somewhat larger masses of workers from rest into search unemployment episodes creating more amplification. This result is in line with Chassamboulli (2013), who extends the DMP model by adding permanent productivity differences among workers and shows that this feature increases it cyclical performance relative to the data. Table \ref{t:dur_unemp} demonstrates, however, that this comes at the cost of not being able to match the distribution of unemployment durations nor the dynamic behavior of this distribution over the cycle.\footnote{An extrapolation from the above discussion appears to suggest a model with permanent heterogeneity, where all moves in and out of rest unemployment would be because of aggregate productivity changes, cannot resolve the tension between cyclical performance and fitting the unemployment duration distribution moments. In principle our estimation allows and has evaluated in its procedure parameter tuples with a near-permanent $z$-productivity process (i.e. a persistence approximating $1$). However, such parameter tuples yield larger deviations from the individual-level unemployment outcomes we targeted compared to Model II.} Thus, in Model II the average unemployment rate responds more to the cycle as aggregate productivity takes on a more prominent role in shaping the amount of rest unemployment, but the individual unemployment outcomes that underlie these dynamics become counterfactual.

The possibility of occupational mobility would have given workers in rest unemployment episodes another margin through which they can escape the area of inaction and get re-employed faster. As a result, the individual-level unemployment duration dependence (given an aggregate state) is not only affected by $z^s$ but also by the distance to $z^r$. Further, since the $z^r$ cutoff is at different distances from the $z^s$ across the cycle, it creates a more cyclically sensitive area of inaction. That is, occupational mobility creates a more responsive area of inaction with respect to both worker heterogeneity and the business cycle that resolves the tensions discussed above.

\clearpage
\topskip0pt
\vspace*{\fill}
{\Huge \textsc{Supplementary Appendices}}
\vspace*{\fill}
\clearpage


\noindent {\Large \bf{Supplementary Appendix A: Occupational Coding Error and its Correction}}
\renewcommand{\thesection}{A.\arabic{section}}
\setcounter{section}{0}
\label{suppappx_A_firstpage}

\bigskip

This appendix complements Section 2 of the paper and expands Online Appendix A. Here we quantify the extent to which coding errors affect the probability of an occupational change after a non-employment spell. This is done in two ways. (1) We exploit the change from independent to dependent interviewing that occurred across the 1985 and 1986 SIPP panels, which overlap between February 1986 to April 1987. The change of interviewing technique allows us to estimate a garbling matrix $\mathbf{\Gamma}$ whose elements denote the probability that a worker's true occupation $i$ gets miscoded as another occupation $j$. (2) We then take advantage of the retrospective coding exercise done to occupational codes in the PSID. Retrospective coding improved the assignment of the occupational codes obtained during the 1970s, but did not affect the codes obtained during later years (see also Kambourov and Manovskii, 2008). We use probabilistic models to estimate the average effect retrospective coding had on the probability of an occupational change.

Across these data sets we obtain a very similar conclusion. Our correction method implies that on average 82\% of the observed wave-to-wave  occupational transitions after re-employment in the SIPP are genuine, were each wave covers a 4 months interval. In stark contrast, when pooling together workers who changed and did not change employers, we find that 40\% of the observed changes among all workers are genuine. Similarly, retrospective coding in the PSID implies that 84\% of the observed year-to-year occupational transitions after re-employment are genuine. When pooling together workers who changed and did not change employers, retrospective coding implies that only 44\% are genuine. A key insight is that the same propensities to miscode occupations affect very differently the measured occupational change of employer movers and employer stayers. This is important as it implies one cannot use the error correction estimates obtained from pooled samples to adjust the occupational mobility statistics of those who changed employers through spells of unemployment.

We use the estimated $\mathbf{\Gamma}$-correction matrix to eliminate the impact of coding errors on the occupational mobility statistics of the unemployed obtained from the SIPP. We show that occupational miscoding increases observed excess mobility and reduces the importance of net mobility (see also Kambourov and Manovskii, 2013). It also makes occupational mobility appear less responsive to unemployment duration and the business cycle.
In Supplementary Appendix B we use the SIPP to evaluate robustness of the occupational mobility patterns derived from our $\Gamma$-correction method. To do so, we follow two alternative approaches of measuring occupational mobility: (i) simultaneously mobility of major occupational and major industrial groups at re-employment and (ii) self-reported duration of occupational tenure obtained from the topical modules. The first measure is considered less sensitive to miscoding as it typically requires errors to be made simultaneously along two dimensions. The second captures the worker's own perception of occupational mobility and is not based on occupational coding. We find that the occupational mobility patterns obtained through our classification error model applied to the SIPP, the retrospective coding in the PSID and these alternative measures provide a very consistent picture.

\vspace{-0.5cm}

\section{A classification error model}

In Section 2.1 of the main text we defined the matrix $\mathbf{\Gamma}$, where its elements are the probabilities that an occupation $i$ is miscoded as an occupation $j$, for all $i,j=1,2,...,O$. There is an important literature that investigates classification error models. Magnac and Visser (1999) identify two main branches. The first one uses assumptions on the measurement error process and auxiliary data on the error rates, where it is assumed that the error rates can be directly observe from the auxiliary data (see Abowd and Zellner, 1985, Poterba and Summers, 1986, and Magnac and Visser, 1999, among others). The second one does not rely on auxiliary data on error rates, but estimates parametric models in the presence of misclassified data, where these models are either reduced form statistical models (see Hausman et al. 1998, among others) or structural economic models (see Kean and Wolpin, 2001, Sullivan, 2009, Roys and Taber, 2017, among others).

Our classification error model builds on Abowd and Zellner (1985) and Poterba and Summer (1986) in that it uses a garbling matrix that captures the errors made in classifying workers. These authors investigate the misclassification of individuals' employment status in the CPS and use reported employment status at the original interview date, at the re-interviewing date (which occurred one week after the original interview) and the reconciliation information provided by the CPS to directly observe the garbling matrix. Their key assumption is that the reconcile information provides the true individuals' labor market status. In contrast, we do not have auxiliary information that allows us to directly recover the garbling matrix. Our challenge is to estimate the garbling matrix.

Our classification error model also relates to Sullivan (2009) in that both approaches provide estimates of an occupation garbling matrix. In contrast, our approach does not rely on the observed optimal choices of economic agents for identification. Rather, miscoding can be estimated using the occupational transitions alone, without requiring further information on wages which themselves can be subject to measurement error. Therefore, it is much simpler to implement relative to Sullivan's (2009) approach which relies on time costly simulation maximum likelihood methods that end up restricting the size of the estimated garbling matrix. This is important as it reduces the overall computational burden when estimating our economic model.

To identify and estimate $\mathbf{\Gamma}$ we make three assumptions:

(\emph{A1}) \emph{Independent classification errors}: conditional on the true occupation, the realization of an occupational code does not depend on workers' labor market histories, demographic characteristics or the time it occurred in our sample. This assumption is also present in Poterba and Summers (1986) and Abowd and Zellner (1985) and is consistent with independent interviewing. In the standard practise of independent interviewing, professional coders base their coding on the verbatim description of the reported work activities without taking into account the respondents' demographic characteristics or earlier work history.\footnote{For example, during the 1980s and 1990s independent occupational coding in the PSID was done without reference to respondents' characteristics or their work history. However, this information was used in the retrospective coding exercise done to the 1970s occupational codes.} Errors introduced by the respondents, however, could be correlated with their characteristics. Assumption $\emph{A1}$ implies that errors in the individuals' verbatim responses are fully captured by the nature of the job these individuals are performing and hence only depend on their \emph{true} occupation. Another implication of assumption $\emph{A1}$ is that $\mathbf{\Gamma}$ is time-invariant. This is important as we will apply our correction method across all years in our sample and one could be concern whether the coding errors estimated in the 1980's are similar to those founds 20 years later. In Section 4 (below) we investigate further these implications.

(\emph{A2}) \emph{``Detailed balance'' in miscoding}: $\mathbf{diag(c)\Gamma}$ is symmetric, where $\mathbf{c}$ is a $O$x$1$ vector that describes the distribution of workers across occupations and $\mathbf{diag(c)}$ is the diagonal matrix of $\mathbf{c}$. This assumption is known as ``detailed balance''. It implies that the number of workers whose true occupation $i$ gets mistakenly coded as $j$ is the same as the number of workers whose true occupation $j$ gets mistakenly coded as $i$, such that the overall size of occupations do not change with coding error. This assumption allow us to invert $\Gamma$ and hence aids our identification arguments. Although undeniably strong, this assumption is a weaker version of the one proposed by Keane and Wolpin (2001) and subsequently used Roys and Taber (2017) to correct of occupation miscoding. We return to discuss \emph{A2} in Section 4.

(\emph{A3}) \emph{Strict diagonal dominance}: $\mathbf{\Gamma}$ is strictly diagonally dominant in that $\gamma_{ii}>0.5$ for all $i=1,2,...,O$. This assumption is also present in Hausman et al., (1998) and implies that it is more likely to correctly code a given occupation $i$ than to miscoded it. The converse would imply occupational mobility rates that are of a magnitude inconsistent with our data. Indeed, we derive an upper bound on code error directly from the data and find that \emph{A3} is verified in our data.

To estimate $\mathbf{\Gamma}$ we exploit the change of survey design between the 1985 and 1986 SIPP panels. Until the 1985 panel the SIPP used independent interviewing for all workers: in each wave all workers were asked to describe their job anew, without reference to answers given at an earlier date. Subsequently, a coder would consider that wave's verbatim descriptions and allocate occupational codes. This practise is known to generate occupational coding errors. In the 1986 panel, instead, the practise changed to one of dependent interviewing (see Lynn and Sala, 2007, J\"ackle, 2008, and J\"ackle and Eckman, 2020). Respondents were only asked ``independently'' to describe their occupation if they reported a change in employer or if they reported a change in their main activities without an employer change within the last 8 months. If respondents declared no change in employer \emph{and} in their main activities, the occupational code assigned to the respondent in the previous wave is carried forward.

It is important to note that during February 1986 to April 1987, the 1985 and 1986 panels overlap, representing the \emph{same} population under different survey designs. The identification theory we develop in the next section refers to this population. We then show how to consistently estimate $\mathbf{\Gamma}$ using the population samples.

\subsection{Identification of $\mathbf{\Gamma}$}

Consider the population represented by 1985/86 panels during the overlapping period and divide it into two groups of individuals across consecutive interviews by whether or not they changed employer or activity. Label those workers who stayed with their employers in both interviews and did not change activity as ``employer/activity stayers''. By design this group \emph{only} contains true occupational stayers. Similarly, label those workers who changed employers or changed activity within their employers as ``employer/activity changers''. By design this group contains all true occupational movers and the set of true occupational stayers who changed employers.

Suppose that we were to subject the employer/activity stayers in this population to dependent interviewing as applied in the 1986 panel. Let $\mathbf{c_{s}}$ denote the $O$x$1$ vector that describes their \emph{true} distribution across occupations and let $\mathbf{M_{s}}=\mathbf{diag(c_{s})}$. In what follows we will use the convention that the $(i,j)$'th element of an $M$ matrix indicates the flow from occupation $i$ to $j$. Let $\mathbf{c^{D}_{s}}$ denote the $O$x$1$ vector that describes their \emph{observed} distribution across occupations under dependent interviewing and let $\mathbf{M^{D}_{s}}=\mathbf{diag(c^{D}_{s})}$. Note that $\mathbf{c^{D}_{s}}=\mathbf{\Gamma^{'} M_{s} \overrightarrow{\bf{1}}}$, where $\overrightarrow{\bf{1}}$ describes a vector of ones. $\mathbf{M_{s}}$ is pre-multiplied by $\mathbf{\Gamma^{'}}$ as true occupations would have been miscoded in the first of the two consecutive interviews. ``Overall balance'', a weaker version of \emph{A2}, implies that $\mathbf{c^{D}_{s}}=\mathbf{diag(c_{s})\Gamma \overrightarrow{\bf{1}}}=\mathbf{c_{s}}$ and hence $\mathbf{M^{D}_{s}=M_{s}}$.\footnote{Overall balance only requires that classification errors do not change the overall occupational distribution rather than the bilateral flows between occupations as also required by detailed balance in \emph{A2}.}

Next suppose that instead we were to subject the employer/activity stayers in this population to independent interviewing as applied in the 1985 panel. Let $\mathbf{M^{I}_{s}}$ denote the matrix that contains these workers' \emph{observed} occupational transition \emph{flows} under independent interviewing. In this case $\mathbf{M^{I}_{s}}=\mathbf{\Gamma^{\prime} M_{s} \Gamma}$. Here $\mathbf{M_{s}}$ is pre-multiplied by $\mathbf{\Gamma^{'}}$ and post-multiplied by $\mathbf{\Gamma}$ to take into account that the observed occupations of origin and destination would be subject to coding error.

Let $\mathbf{M_{m}}$ denote the matrix that contains the \emph{true} occupational transition \emph{flows} of employer/activity changers in this population. The diagonal of $\mathbf{M_{m}}$ describes the distribution of true occupational stayers across occupations among employer/activity changers. The off-diagonal elements contain the flows of all true occupational movers. Under independent interviewing we observe $\mathbf{M^{I}_{m}}=\mathbf{\Gamma^{\prime} M_{m} \Gamma}$. Once again $\mathbf{M_{m}}$ is pre-multiplied by $\mathbf{\Gamma^{'}}$ and post-multiplied by $\mathbf{\Gamma}$ as the observed occupations of origin and destination would be subject to coding error.

Letting $\mathbf{M^{I}}=\mathbf{M^{I}_{m}+M^{I}_{s}}$ denote the matrix that contains the aggregate occupational transition flows across two interview dates under independent interviewing, it follows that $\mathbf{M^{I}_{s}}=\mathbf{M^{I}-M^{I}_{m}}=\mathbf{\Gamma^{\prime} M_{s} \Gamma}$. By virtue of the symmetry of $\mathbf{M_{s}}$ and ``detailed balance'' (\emph{A2}), $\mathbf{M_{s} \Gamma=\Gamma^{\prime} {M_{s}}^{\prime}=\Gamma^{\prime} M_{s}}$. Substituting back yields $\mathbf{M^{I}_{s}}=\mathbf{M_{s}\Gamma \Gamma}$. Next note that $\mathbf{M^{I}_{s}=M_{s}T^{I}_{s}}$, where $\mathbf{T^{I}_{s}}$ is the occupational transition probability matrix of the employer/activity stayers in this population \emph{observed} under independent interviewing. Substitution yields $\mathbf{M_{s} T^{I}_{s}}=\mathbf{M_{s} \Gamma \Gamma}$. Multiply both sides by $\mathbf{M_{s}^{-1}}$, which exists as long as all the diagonal elements of $\mathbf{M_{s}}$ are non-zero, yields the key relationship we exploit to estimate $\mathbf{\Gamma}$,

\begin{equation}\label{eq:est_gamma_2}
\mathbf{T^{I}_{s}}=\mathbf{\Gamma \Gamma}.
\end{equation}

To use this equation we first need to show that it implies a unique solution for $\mathbf{\Gamma}$. Towards this result, we now establish that $\mathbf{\Gamma}$ and $\mathbf{T^{I}_{s}}$ are diagonalizable. For the latter it is useful to interpret the coding error process described above as a Markov chain such that $\mathbf{\Gamma}$ is the one-step probability matrix associated with this process.

\vspace{3 mm}
\noindent{\bf{Lemma A.1:}} \emph{Assumptions A2 and A3 imply that $\mathbf{\Gamma}$ and $\mathbf{T^{I}_{s}}$ are diagonalizable.}

\begin{proof}
First note that without loss of generality we can consider the one-step probability matrix $\mathbf{\Gamma}$ to be irreducible. To show this suppose that $\mathbf{\Gamma}$ was not irreducible, we can (without loss of generality) apply a permutation matrix to re-order occupations in $\mathbf{\Gamma}$ and create a block-diagonal $\mathbf{\Gamma'}$, where each block is irreducible and can be considered in isolation. Given \emph{A3}, it follows directly that $\mathbf{\Gamma}$ is aperiodic. Further, assumption \emph{A2} implies that $\mathbf{c_{s}}$ is a stationary distribution of $\mathbf{\Gamma}$. The fundamental theorem of Markov chains then implies that $\mathbf{c_{s}}$ is the \emph{unique} stationary distribution of $\mathbf{\Gamma}$. Assumption \emph{A2} also implies that the Markov chain characterised by $\mathbf{\Gamma}$ is reversible with respect to $\mathbf{c_s}$. This means that $\mathbf{\Gamma}$ is similar (in matrix sense) to a symmetric matrix $\mathbf{G}$ such that $\mathbf{G}=\mathbf{diag(\sqrt{\mathbf{c_s}})} \ \mathbf{\Gamma} \ \mathbf{diag(\sqrt{\mathbf{c_s}})^{-1}}$. By symmetry, $\mathbf{G}$ is orthogonally diagonalizable by $\mathbf{Q \Delta Q^{-1}}$, where the diagonal matrix $\mathbf{\Delta}$ contains the associated (real) eigenvalues and $\mathbf{Q}$ is the orthogonal matrix of associated (normalized) eigenvectors. It then follows that $\mathbf{\Gamma}$ is diagonalizable as well. Further, $\mathbf{T^{I}_{s}}= \mathbf{diag(\sqrt{\mathbf{c_s}})^{-1}\  G \ G \ diag(\sqrt{\mathbf{c_s}})}=\mathbf{diag(\sqrt{\mathbf{c_s}})^{-1}\ Q \ \Delta^2 \ Q^{-1} \ diag(\sqrt{\mathbf{c_s}})}$, and hence $\mathbf{T^{I}_{s}}$ is also orthogonally diagonalizable, with a root of $\mathbf{P \Lambda^{0.5} P^{-1}}$, where $\mathbf{\Lambda}$ is the diagonal matrix of eigenvalues of $\mathbf{T^{I}_{s}}$, and $\mathbf{P}$ the associated orthogonal matrix with eigenvectors of $\mathbf{T^{I}_{s}}$.
 \end{proof}

In general one cannot guarantee the uniqueness, or even existence, of a transition matrix that is the $(nth)$ root of another transition matrix (see Higham and Lin, 2011). Here, however, existence is obtained by construction: $\mathbf{T_s}$ is constructed from $\mathbf{\Gamma}$, and in reverse, we can find its roots. The next result shows that $\mathbf{T_s}$ has a unique root satisfying assumptions \emph{A2} and \emph{A3}.

\vspace{3 mm}
\noindent{\bf{Proposition A.1:}} $\mathbf{\Gamma}$ \emph{is the unique solution to $\mathbf{T^{I}_{s}=\Gamma \ \Gamma}$ that satisfies assumptions A2 and A3. It is given by $\mathbf{P \Lambda^{0.5} P^{-1}}$, where $\mathbf{\Lambda}$ is the diagonal matrix with eigenvalues of $\mathbf{T^{I}_{s}}$, $0<\lambda_i\leq 1$, and $\mathbf{P}$ is the orthogonal matrix with the associated (normalized) eigenvectors.}

\begin{proof}
Following from the proof of Lemma A.1, a root of $\mathbf{T^{I}_{s}}$ is given by $\mathbf{P \Lambda^{0.5} P^{-1}}$, where $\mathbf{\Lambda}$ is the diagonal matrix with eigenvalues of $\mathbf{T^{I}_{s}}$ and $\mathbf{P}$ is the orthogonal matrix with the associated (normalized) eigenvectors. Since \emph{A3} implies $\mathbf{\Gamma}$ is strictly diagonally dominant, it follows that the determinant of all its leading principal minors are positive. Moreover, under the similarity transform by pre-/post-multiplication with the diagonal matrices $\mathbf{diag(\sqrt{c_s}), diag(\sqrt{c_s})^{-1}}$, the determinant of all principals minors of the symmetric matrix $\mathbf{G}=\mathbf{diag(\sqrt{\mathbf{c_s}})} \ \mathbf{\Gamma} \ \mathbf{diag(\sqrt{\mathbf{c_s}})^{-1}}$ are positive as well. Hence $\mathbf{G}$ is a symmetric positive definite matrix (with all eigenvalues between $0$ and $1$, as has $\mathbf{\Gamma}$). It follows that $\mathbf{G \ G}=\mathbf{S}$ is also positive definite, and $\mathbf{T^{I}_{s}}=\mathbf{diag(\sqrt{\mathbf{c_s}})^{-1} \ S \ diag(\sqrt{\mathbf{c_s}})}$ is positive definite in the sense that $\mathbf{v'\ T^{I}_{s}\ v}>0$ for all $\mathbf{v}\neq \mathbf{0}$, while also all eigenvalues of $\mathbf{T^{I}_{s}}$ will be between $0$ and $1$.

To show the uniqueness of the root of $\mathbf{T^{I}_{s}}$ suppose (towards a contradiction) that there exists two different roots $\mathbf{\Gamma}$ and $\mathbf{\Upsilon}$ such that each are similar (in matrix sense), with the same transform involving $\mathbf{diag(\sqrt{c_s})}$, to different symmetric positive definite matrices $\mathbf{G}$ and $\mathbf{Y}$, where $\mathbf{G\ G}= \mathbf{S}$ and $\mathbf{Y\ Y} = \mathbf{S}$. Both $\mathbf{G}$ and $\mathbf{Y}$ are diagonalizable, and have the square roots of the eigenvalues of $\mathbf{S}$ on the diagonal. Given that the squares of the eigenvalues need to coincide with the eigenvalues of $\mathbf{S}$ and assumptions \emph{A2} and \emph{A3} imply that all eigenvalues must be between $0$ and $1$, without loss of generality we can consider both diagonalizations to have the same diagonal matrix $\mathbf{\Delta}$, where $\mathbf{\Delta}$ is the diagonal matrix of eigenvalues of $\mathbf{T^{I}_{s}}$ and these eigenvalues are ordered using a permutation-similarity transform with the appropriate permutation matrices. Let $\mathbf{G}=\mathbf{H \Delta  H^{-1}}$ and $\mathbf{Y}=\mathbf{K \Delta K^{-1}}$. Then, it follows that $\mathbf{K^{-1}H \Delta^2 H^{-1} K = \Delta^2}$ and since $\mathbf{K^{-1} H}$ and $\mathbf{\Delta^2}$ commute, implies that $\mathbf{K^{-1} H}$ is a block-diagonal matrix with the size of the blocks corresponding to the multiplicity of squared eigenvalues. Again, since all eigenvalues of $\mathbf{\Delta}$ are positive, this equals the multiplicity of the eigenvalues $\delta_i$ itself. But then it must be true that $\mathbf{K^{-1} H \Delta H^{-1} K = \Delta}$. Then, $\mathbf{G}=\mathbf{H\ \Delta\  H^{-1}}=\mathbf{K K^{-1} H\ \Delta\  H^{-1} K K^{-1}}=\mathbf{K\ \Delta\  K^{-1}}=\mathbf{Y}$ which leads to a contradiction.
\end{proof}

The above results imply that under assumptions \emph{A2} and \emph{A3}, $\mathbf{\Gamma}$ is uniquely identified from the transition matrix of true occupational stayers under independent interviewing, $\mathbf{T^{I}_{s}}$.

\subsection{Estimation of $\mathbf{\Gamma}$}

The next lemma provides an intermediate step towards estimating $\mathbf{\Gamma}$. For this purpose let $PDT(.)$ denote the space of transition matrices that are similar, in the matrix sense, to positive definite matrices.

\vspace{3 mm}

\noindent{\bf{Lemma A.2:}} \emph{The function} $f:PDT(\mathbb{R}^{O\times O})\to PDT(\mathbb{R}^{O\times O})$ \emph{given by} $f(\mathbf{T})=\mathbf{T^{0.5}}$ \emph{exists and is continuous with} $f(\mathbf{T^{I}_{s}})=\mathbf{\Gamma}$ in the spectral matrix norm.

\begin{proof}
Existence follows from Lemma A.1 and Proposition A.1. To establish continuity of the mapping, we follow Horn and Johnson (1990). Let $\mathbf{T_{1}}$ and $\mathbf{T_{2}}$ be any two transition matrices in $PDT$ and let $\mathbf{U_{1}}$ and $\mathbf{U_{2}}$ be two symmetric positive definite matrices constructed as $\mathbf{U_{1}}=\mathbf{diag(\sqrt{\mathbf{c_1}})\ T_{1} \ diag(\sqrt{\mathbf{c_1}})^{-1}}$ and $\mathbf{U_{2}}=\mathbf{diag(\sqrt{\mathbf{c_2}}) \ T_{2} \ diag(\sqrt{\mathbf{c_2}})^{-1}}$, where $\mathbf{c_{1}}$ and $\mathbf{c_{2}}$ are the unique stationary distributions associated with $\mathbf{T_{1}}$ and $\mathbf{T_{2}}$, respectively. We want to show that if $\mathbf{U_{1}}\to \mathbf{U_{2}}$, then $\mathbf{U_{1}^{0.5}}\to \mathbf{U_{2}^{0.5}}$. First, note that $\| \mathbf{U_{1}-U_{2}} \|_2=\| \mathbf{U_{1}^{0.5}(U_{1}^{0.5}-U_{2}^{0.5})}+\mathbf{(U_{1}^{0.5}-U_{2}^{0.5})U_{2}^{0.5}} \|_2\geq | \mathbf{x'U_{1}^{0.5}(U_{1}^{0.5}-U_{2}^{0.5})x}+\mathbf{x'(U_{1}^{0.5}-U_{2}^{0.5})U_{2}^{0.5}x }|$, where $\mathbf{x}$ is any normalised vector. Assumptions \emph{A2} and \emph{A3} imply $\mathbf{U_{1}^{0.5}-U_{2}^{0.5}}$ exists and is a symmetric matrix. Let $|\lambda|=\rho(\mathbf{U_{1}^{0.5}-U_{2}^{0.5}})$ be the absolute value of the largest eigenvalue of $\mathbf{U_{1}^{0.5}-U_{2}^{0.5}}$ and let $\mathbf{z}$ be the normalized eigenvector associated with $\lambda$. Note that $\|\mathbf{U_{1}-U_{2}}\|_2=|\lambda|$ and $(\mathbf{U_{1}^{0.5}-U_{2}^{0.5}})\mathbf{z}=\lambda\mathbf{z}$. Then $| \mathbf{z'U_{1}^{0.5}(U_{1}^{0.5}-U_{2}^{0.5})z}+\mathbf{z'(U_{1}^{0.5}-U_{2}^{0.5})U_{2}^{0.5}z }|\geq |\lambda| \ |\lambda_{min}^{0.5}(\mathbf{U_{1}})+\lambda_{min}^{0.5}(\mathbf{U_{2}})| = \|\mathbf{U_{1}^{0.5}-U_{2}^{0.5}}\|_2 \ (\lambda_{min}^{0.5}(\mathbf{U_{1}})+\lambda_{min}^{0.5}(\mathbf{U_{2}})) $, where $\lambda_{min}(\mathbf{U_{1}})$ denotes the smallest eigenvalue of $\mathbf{U_{2}}$, which is positive by virtue of assumptions \emph{A2} and \emph{A3}. Then choose a $\delta=\varepsilon \lambda_{min}^{0.5}(\mathbf{U_{1}})$. It follows that if $\|\mathbf{U_{1}-U_{2}}\|_2<\delta$, then $\|\mathbf{U_{1}^{0.5}-U_{2}^{0.5}}\|_2 $ $\times$ $\ \frac{(\lambda_{min}^{0.5}(\mathbf{U_{1}})+\lambda_{min}^{0.5}(\mathbf{U_{2}}))}{\lambda_{min}^{0.5}(\mathbf{U_{1}})}$ $<\varepsilon$,  and therefore $\|\mathbf{U_{1}^{0.5}-U_{2}^{0.5}}\|_2<\varepsilon$, which establishes the desired continuity. From the fact that $\mathbf{U_{1} }\to \mathbf{U_{2}}$ implies $\mathbf{U_{1}^{0.5}}\to \mathbf{U_{2}^{0.5}}$, it then also follows that $f(\mathbf{T})$ is continuous.
\end{proof}

Let $\mathbf{\hat{T}^{I}_{s}}$ denote the sample estimate of $\mathbf{T^{I}_{s}}$ and let $\mathbf{\hat{\Gamma}}$ be estimated by the root $\mathbf{(\hat{T}^{I}_{s})}^{0.5} \in PDT(\mathbb{R}^{O \times O})$ such that $\mathbf{\hat{\Gamma}}=\mathbf{(\hat{T}^{I}_{s})^{0.5}}=\mathbf{\hat{P} \hat{\Lambda}^{0.5} \hat{P}^{-1}}$, where $\mathbf{\hat{\Lambda}}$ is the diagonal matrix with eigenvalues of $\mathbf{\hat{T}^{I}_{s}}$, $0<\hat{\lambda_i}^{0.5}\leq 1$ and $\mathbf{\hat{P}}$ the orthogonal matrix with the associated (normalized) eigenvectors. We then have the following result.

\vspace{3 mm}

\noindent{\bf{Proposition A.2:}} \emph{$\mathbf{\Gamma}$ is consistently estimated from $\mathbf{(\hat{T}^{I}_{s})}^{0.5} \in PDT(\mathbb{R}^{O \times O})$ such that $\mathbf{\hat{\Gamma}}=\mathbf{(\hat{T}^{I}_{s})^{0.5}}=\mathbf{\hat{P} \hat{\Lambda}^{0.5} \hat{P}^{-1}}$. That is, $plim_{n \to \infty} \mathbf{\hat{\Gamma}}=\mathbf{\Gamma}$.}
\begin{proof}
From Lemma A.1 and Proposition A.1 it follows that if we know $\mathbf{T^{I}_{s}}$, then we can find the unique $\mathbf{\Gamma}$ that underlies it, constructing it from the eigenvalues and eigenvectors of $\mathbf{T^{I}_{s}}$. To estimate $\mathbf{T^{I}_{s}}$ one can use the sample proportion $\mathbf{\hat{M}_{ij}}/\sum_{k=1}^O \mathbf{\hat{M}_{ik}}$ and note that this converges in probability to $\mathbf{(T^{I}_{s})_{ij}}$ (see Anderson and Goodman, 1957; Billingsley 1961, thm 1.1-3.) for all occupations, given assumptions \emph{A2} and \emph{A3}. Hence, $plim_{n\to \infty} \mathbf{\hat{T}^{I}_{s}} = \mathbf{T^{I}_{s}}$. Then, we derive $\mathbf{\hat{\Gamma}}$ from $\mathbf{\hat{T^{I}_{s}}}$ according to $\mathbf{\hat{P}\hat{\Lambda}^{0.5}\hat{P}^{-1}}$, per Proposition A.1. By continuity of the mapping in Lemma A.2, it follows that $plim_{n\to \infty} \mathbf{\hat{\Gamma}} = \mathbf{\Gamma}$, and our estimator is consistent.
\end{proof}

Note that to identify and estimate $\mathbf{\Gamma}$ in the SIPP it is not sufficient to directly compare the aggregate occupational transition flows under independent interviewing with the aggregate occupational transition flows under dependent interviewing. To show this let $\mathbf{M^{D}}=\mathbf{M^{I}_{m}+M^{D}_{s}}$ denote the matrix that contains the aggregate occupational transition flows across two interview dates under dependent interviewing for employer/activity stayers and under independent interviewing for employer/activity movers. Subtracting $\mathbf{M^{I}}=\mathbf{M^{I}_{m}+M^{I}_{s}}$ from $\mathbf{M^{D}}$ yields $\mathbf{M^{D}_{s}-M^{I}_{s}}=\mathbf{M_{s} -\Gamma' M_{s} \Gamma}$. Given the symmetry assumed in \emph{A2}, the latter expression has $0.5n(n-1)$ exogenous variables on the LHS and $0.5n(n+1)$ unknowns (endogenous variables) on the RHS, leaving $\mathbf{\Gamma}$ (and $\mathbf{M_s}$) unidentified.

In addition to $\mathbf{M^{D}-M^{I}}=\mathbf{M_{s} -\Gamma' M_{s} \Gamma}$ one can use $\mathbf{M^D}=\mathbf{\Gamma^{'} M_{m} \Gamma + M_{s}}$, which contains the remainder information. When $\mathbf{M_{m}}$ has mass on its diagonal, however, this additional system of equations has $n^{2}$ exogenous variables on the LHS and $n^{2}$ unknowns (arising from $\mathbf{M_m}$) on the RHS. This implies that with the $n$ unknowns remaining from $\mathbf{M^{D}-M^{I}}=\mathbf{M_{s} -\Gamma' M_{s} \Gamma}$, one is still unable to identify $\mathbf{\Gamma}$ and $\mathbf{M_s}$.

\vspace{3 mm}

\noindent{\bf{Corollary A.1:}}
\emph{If $\mathbf{M_{m}}$ has mass on its diagonal, $\mathbf{\Gamma}$ cannot be identified from $\mathbf{M^I}$ and $\mathbf{M^D}$ alone.}

\vspace{3 mm}

The intuition behind this result is that by comparing aggregate occupational transition flows under dependent and independent interviewing, it is unclear how many workers are `responsible' for the change in occupational mobility between $\mathbf{M^D}$ and $\mathbf{M^I}$. Only when the diagonal of $\mathbf{M_{m}}$ contains exclusively zeros, identification could be resolved and one can recover $\mathbf{M_s}$, $\mathbf{\Gamma}$ and $\mathbf{M_m}$ as the number of equations equals the number of unknowns.\footnote{However, in the SIPP this case is empirically unreasonable as it requires that all employer/activity changers be true occupational movers.} An implication of the above corollary is that interrupted time-series analysis that is based on the difference in occupational mobility at the time of a switch from independent to dependent interviewing, does not identify the precise extent of the average coding error, but provides a downwards biased estimate.

To identify $\mathbf{\Gamma}$, however, Proposition A.2 implies that one can use the observed occupational transition flows of a sample of \emph{true} occupational stayers that are subject to two rounds of independent interviewing. Some of these workers will appear as occupational stayers and some of them as occupational movers. Ideally, such a sample of workers should be isolated directly from the 1985 panel. Unfortunately, the questions on whether the individual changed activity or employer were only introduced in the 1986 panel, as a part of the switch to dependent interviewing. As a result, the 1985 panel by itself does not provide sufficient information to separate employer/activity stayers from employer/activity movers. Instead we use 1986 panel to estimate $\mathbf{\hat{M}^I_{m}}$. We can infer $\mathbf{M^I_{s}}$ indirectly by subtracting the observed occupational transition flow matrix $\mathbf{\hat{M}^I_m}$ in the 1986 panel from the observed occupational transition flow matrix $\mathbf{\hat{M}^I}$ in the 1985 panel. This is possible as the 1986 panel refers to the same underlying population as the 1985 panel and separates the employer/activity changers, who are independently interviewed.

\vspace{3 mm}

\noindent{\bf{Corollary A.2:}}
\emph{$\mathbf{\hat{\Gamma}}$ is consistently estimated from $\mathbf{\hat{T}_s^I}$ when the latter is estimated from $\mathbf{\hat{M}^{I}}-\mathbf{\hat{M}^{I}_{m}}$}
\vspace{3 mm}

This result is important to implement our approach. It follows as the population proportions underlying each cell of $\mathbf{\hat{M}_{s}}$, the sample estimate of $\mathbf{M_{s}}$, are consistently estimated. In turn, the latter follows from the standard central limit theorem for estimating proportions, which applies to $\mathbf{\hat{M}^{I}}$, $\mathbf{\hat{M}^{I}_{m}}$ and its difference. Proposition A.2 then implies that $\mathbf{\hat{\Gamma}}$ is consistently estimated.

\subsection{Implementation}

To implement our correction method we take the overlapping period of the 1985/86 panels. To increase the sample size we also use observations from the 1987 panel for the period between February 1987 and April 1987.\footnote{To avoid seasonality effects we re-weight all observations such that each observation in a given month has the same weight as another observation in any other month.} This panel has an identical setup to that of the 1986 panel (dependent interviewing and other relevant aspects) and is likewise representative of the population during the period of study.

Interviews throughout the SIPP are conducted every four months and collect information pertaining to the last four months, where these four months are considered to be a wave. We compare the reported occupational code of a worker in a given interview with the reported occupational code of that worker in the subsequent interview. An observation is therefore a pair of occupational codes, a reported `source' and (potentially identical) `destination' occupation. To keep comparability across interviews as clean as possible, and to focus on measuring occupations in the primary job, we only consider those workers who throughout the two waves stayed in full-time employment and who reported having only one employer at any point in time. These restrictions imply that in the estimation of $\mathbf{\Gamma}$ we do not include non-temporary laid off workers who experienced a short unemployment episodes and returned to their same jobs and employers. We also restrict attention to those workers who do not have imputed occupations, were not enrolled in school and were between 19-66 years old. These restrictions yield 28,302 wave/individual observations for the 1985 panel, 27,801 wave/individual observations for the 1986 panel and 5,922 wave/individual observations for the 1987 panel.

\begin{table}[!ht]
 \caption{Demographic characteristics - February 1986 to April 1987}\label{t:sumstats1}
  \centering
  {\small
     \begin{tabular}{lcccc}
     \hline
       & SIPP 1985 & SIPP 1986 & SIPP 1987 & p-value  \\
       &          &           &         & (no difference) \\ \hline \hline
       \bf{Education} &   &   &   &  \\
less than high school &    14.72&    15.27&    14.55&    0.386\\
high school grad &    38.10&    37.36&    36.80&    0.361 \\
some college &    24.51&    24.94&    24.71&    0.546 \\
college degree &    22.67&    22.43&    23.95&    0.746 \\
\bf{Age category} &   &   &   &  \\
19-24 & 12.29	& 12.62	& 12.90 & 0.458  \\
25-29  & 16.72	& 16.15 & 	 16.36 & 0.357 \\
 30-34 & 15.84	& 15.40	& 16.00 &	0.512 \\
35-39 & 15.02 & 15.30 &	13.99 &	0.806 \\
40-44  & 11.65	&  11.50	& 11.80 &	0.804 \\
45-49 & 9.10 &	8.87 & 9.22 &	0.667 \\
50-54 & 7.72	& 7.90 &	8.25	&  0.600 \\
55-59 & 7.07 &	7.31	&  7.01 & 0.600 \\
60-64  & 4.20 &	 4.64  & 4.05 &	0.221 \\
\bf{Ethicity}&   &   &   &  \\
white &    86.29 &  86.57 & 86.25 &	0.729 \\
black &    10.62 &  10.81 & 10.73 &	0.782 \\
american indian, eskimo &     0.49	& 0.62  &	0.42 & 0.401 \\
asian or pacific islander &     2.60 &	2.01 & 2.60 &	0.090 \\
\bf{Other}  &   &   &   &  \\
men &    54.20&    55.27&    53.93&    0.112 \\
married&    65.51 &	66.15  &	64.54  & 0.550 \\
living in metro area &    76.33 &  76.24 &	75.97 &   0.885\\ \hline
\multicolumn{5}{p{0.8\textwidth}}{\tiny{Workers aged 19-66, not enrolled in school, in two adjacent waves, measured in the first month of the current wave, with employment in one firm only in the previous wave, and employment in one (but possibly different) employer only in the wave that follows, without any self-employment, with un-imputed occupations reported in both waves. Person weights are used to scale observations per month within panel group (1985 versus 1986+1987).}}
   \end{tabular}
   }
\end{table}

Tables \ref{t:sumstats1} and \ref{t:sumstats2} show the demographic and occupational characteristics (based on the major occupations of the 1990 SOC), respectively, of the samples across the three panels. In the last column of each table we test characteristic-by-characteristic whether the proportion of workers with a given characteristic in the 1985 sample is statistically indistinguishable from the proportion of workers with the same characteristic in the pooled 1986/87 sample. Across all the characteristics analysed, we cannot reject at a 5\% level that the proportions in the 1985 sample and the corresponding proportions in the 1986-87 sample are the same. With the exception of the proportion of Asian Americans and the proportion of workers whose source occupation is management, similarity cannot be rejected even at a 10\% level. Although not shown here, we also cannot rejected at a 10\% level that the proportion of workers across source and destination industries are the same when comparing the 1985 and 1986/87 samples. This analysis thus confirm that the observations used for our exercise are taken from the same underlying population.\footnote{To further rule out any meaningful impact from the observed differences in occupational distributions, we re-calculated all statistics after re-balancing the weights on the source/destination occupations to create identical occupational distributions. This exercise yields minimal effects on our statistics. For example, the observed occupational mobility changes by 0.01 percentage point at most. The reason for this small change is because there is the large proportion of the implied true stayers in the sample, which means that $\mathbf{\hat{T}^I_s}$ is not very sensitive to proportional changes in $\mathbf{\hat{M}_m^I}$.}

\begin{table}[ht!]
 \caption{Distribution of workers across occupations - February 1986 to April 1987}\label{t:sumstats2}
  \centering
  {\small

  \begin{tabular}{lcccc}
     \hline

\multicolumn{5}{c}{Distribution across source occupations (occupation code) (\%)} \\ \hline
     & SIPP 1985 & SIPP 1986 & SIPP 1987 & p-value (no difference) \\ \hline \hline
managing occupations&    12.31 & 13.11 &  13.99 &	0.064 \\
professional speciality&    13.40 & 12.92 & 13.18  &	0.380 \\
technicians and related support&     3.85 & 3.89 & 4.08 &	0.829 \\
sales occ.&     9.74 & 9.98 & 9.86 &	0.599\\
admin support&    18.67 &	 18.17 &	18.26 &	0.366 \\
services&    10.96 &	11.66 & 11.19 & 0.166 \\
farming/fish/logging&     1.09 & 1.08 & 1.03 & 0.940\\
mechanics and repairers&     4.87 &	4.47 & 4.69 & 0.231 \\
construction and extractive&     3.47 & 3.65 & 3.60 &	 0.503 \\
precision production&     4.01 &	 4.19 & 3.75 &	0.643 \\
machine operators/assemblers&     9.27 &  8.89 & 8.63 &	0.355 \\
transportation and materials moving&     4.63 &	 4.34 & 4.31 & 0.340 \\
laborers&     3.73 &	3.66 & 3.42 &	0.714 \\ \hline

\multicolumn{5}{c}{Distribution across destination occupations (occupation code) (\%)} \\ \hline
managing occupations&    12.58 & 13.27 & 14.18 &	0.103  \\
professional speciality&    13.31 & 12.90 & 13.10 &	0.451 \\
technicians and related support&     3.82 & 3.92 &  4.01 &	0.693 \\
sales occ.&     9.76 & 9.89 & 9.77 &	0.797 \\
admin support&    18.53 &	 18.20 & 18.20 & 0.546 \\
services&    10.96 & 11.57 & 11.19 &	0.229 \\
farming/fish/logging&    1.08 &	1.07 & 1.03 &	0.924 \\
mechanics and repairers&     4.87 &	4.46 & 4.73 &	0.222 \\
construction and extractive&     3.59 & 3.60 & 3.56 &	0.954 \\
precision production&     4.01 &  4.26 & 3.77 &	0.480 \\
machine operators/assemblers&    9.24 &	8.93 & 8.77 &	0.455 \\
transportation and materials moving&    4.62 &	4.31 & 4.33 &	0.310 \\
laborers&     3.62 &	3.62 & 3.36 &	0.906 \\ \hline

\multicolumn{5}{p{0.8\textwidth}}{\tiny{Workers aged 19-66, not enrolled in school, in two adjacent waves, measured in the first month of the current wave, with employment in one firm only in the previous wave, and employment in one (but possibly different) firm only in the wave that follows, without any self-employment, with un-imputed occupations reported in both waves. Person weights are used to scale observations per month within panel group (1985 versus 1986+1987).}}
   \end{tabular}
   }
\end{table}

We estimate the occupational flows of employer/activity stayers by $\mathbf{\hat{M}^{I,85}}-\mathbf{\hat{M}^{I, 86/87}_m}=\mathbf{\hat{M}^{I}_s}$, where $x\!=\!85$ ($x\!=86/87$) in $\mathbf{\hat{M}_i^{I,x}}$ refers to the 1985 sample (1986/87 sample). In the 1986/87 sample, where we can observe employer/activity changers directly, we find that in 2.31\% of (weighted) observations workers changed employers and in 4.65\% workers report an activity change within their employers. This implies that more than 93\% of the 1985 sample should be made up of employer/activity stayers.

A potential concern from using this survey design change could be the reliability in the 1986 implementation of dependent interviewing and its comparison with the data collected in the overlapping period of the 1985 panel. For example, any trail/learning period in the implementation of dependent interviewing in the 1986 panel could affect our results. To evaluate whether any trial/learning period in the implementation of dependent interviewing affected the occupational mobility rates we compare the occupational mobility rates obtained from the 1986 panel to those obtained from subsequent ones. Assuming improvements made after 1986 affected the measurement of occupational mobility, we should observe a meaningful change in these rates when using subsequent panels. Figure \ref{fig:SIPP_occmob_estayers} (see below for a detail explanation of the graph) suggests that any trail/learning period in the implementation of dependent interviewing did not have a major effect on average occupational mobility.\footnote{This is consistent with Hill (1994), who instead points to a potential higher level of attrition in the ``older'' 1985 panel relative to the ``newer'' 1986 panel as a concern. He argues that using the ``1986 final panel weights, taken from the 1985 and 1986 Full Panel Longitudinal Research Files'', while not perfect, seem to be the best available solution to tackle this problem. We follow the same practice in our analysis.}

\section{Results and Discussion}

Table 3 shows the occupational transition matrix $\mathbf{\hat{T}^{I}_{s}}$ for true occupational stayers derived from $\mathbf{\hat{M}^{I}_s}$ using the 1985 sample based on the major occupations of the 1990 SOC. The $(i,j)$'th element of $\mathbf{\hat{T}^{I}_{s}}$ indicates the transition probability from occupation $i$ to $j$. This matrix implies that 18.46\% of true occupational stayers get classified as occupational movers.\footnote{This value lies within the expected bounds. To construct an upper bound consider the 1985 sample and assume that all observed occupational transitions are spurious. In this case we can expect that at most 19.71\% of the observations would be miscoded. To construct a lower bound consider the 1986/87 sample and calculate the number of observations in which an occupational move is reported among the employer/activity changers. These observations account for 2.60\% of all observations in the 1986/87 sample. Assuming that the effect of miscoding is to generate a net increase in the number of occupational changes, we can expect that at least 19.71\%-2.60\%$=$17.11\% of the observations would be miscoded.} Although not shown here, a similar conclusion arises when we calculate the same matrix based on the 2000 SOC, which we use for our results in the
main text.

\begin{table}[htbp]
\caption{Observed occupational transition matrix of true occupational stayers, SOC 1990, $\mathbf{\hat{T}^{I}_{s}}$, (\%)} }\label{t:tranmatrix_85}
\vspace{1mm}
\centering
{\small
\setlength{\tabcolsep}{3pt}
\renewcommand\arraystretch{0.8}
\begin{tabular}{lcccccccccccccc}\hline
\textbf{OCCUPATIONS} & (1)  & (2)  & (3)  & (4)  & (5)  & (6)  & (7)  & (8)  & (9)  & (10)  & (11)  & (12)  & (13)  \\ \hline \hline
(1) Managing Occ. &  \bf{75.7} & 3.7 & 1.0 & 5.1 & 8.8 & 1.5 & 0.1 & 0.7 & 0.7 & 1.5 & 0.5 & 0.4 & 0.3 \\
(2) Professional Spec. & 3.3 & \bf{87.9} & 2.9 & 0.4 & 2.2 & 1.5 & 0.1 & 0.5 & 0.2 & 0.6 & 0.2 & 0.1 & 0.1 \\
(3) Technicians   & 3.1 & 10.1 & \bf{69.5} & 0.5 & 5.9 & 3.8 & 0.1 & 2.7 & 0.6 & 1.3 & 2.1 & 0.2 & 0.3  \\
(4) Sales Occ. & 6.5 & 0.5 & 0.2 & \bf{83.3} & 3.8 & 1.5 & 0.1 & 0.7 & 0.2 & 0.5 & 0.2 & 0.8 & 1.8  \\
(5) Admin. Support  &  5.8 & 1.6 & 1.2 & 2.0 & \bf{84.5} & 1.0 & 0.1 & 0.3 & 0.1 & 0.5 & 0.7 & 0.6 & 1.6  \\
(6) Services   & 1.7 & 1.8 & 1.3 & 1.3 & 1.8 & \bf{87.2} & 0.2 & 1.3 & 0.7 & 0.5 & 0.6 & 0.6 & 0.9  \\
(7) Farm/Fish/Logging & 1.3 & 1.7 & 0.5 & 1.3 & 1.1 & 2.0 & \bf{84.7} & 0.6 & 0.6 & 0.2 & 1.0 & 3.1 & 2.0 \\
(8) Mechanics & 1.8 & 1.3 & 2.1 & 1.3 & 1.1 & 2.9 & 0.1 & \bf{79.2} & 2.1 & 2.6 & 3.0 & 0.8 & 1.6 \\
(9) Construction  & 2.4 & 0.7 & 0.6 & 0.5 & 0.4 & 2.2 & 0.2 & 3.0 & \bf{77.7} & 1.7 & 1.9 & 1.2 & 7.6 \\
(10) Precision Prod. & 4.5 & 2.0 & 1.2 & 1.1 & 2.4 & 1.4 & 0.0 & 3.2 & 1.5 & \bf{69.1} & 11.1 & 0.2 & 2.3 \\
(11) Mach. Operators &  0.7 & 0.3 & 0.9 & 0.2 & 1.4 & 0.7 & 0.1 & 1.6 & 0.7 & 4.8 & \bf{84.1} & 0.7 & 3.8 \\
(12) Transport & 1.1 & 0.4 & 0.2 & 1.6 & 2.2 & 1.4 & 0.7 & 0.8 & 0.9 & 0.2 & 1.5 & \bf{85.1} & 3.9\\
(13) Laborers &  1.0 & 0.4 & 0.3 & 4.6 & 8.2 & 2.7 & 0.6 & 2.1 & 7.3 & 2.5 & 9.7 & 5.0 & \bf{55.7} \\ \hline
  \end{tabular}
\end{table}

\subsection{The estimate of $\mathbf{\Gamma}$}

Following Proposition \emph{A.1} we can then recover the garbling matrix $\mathbf{\Gamma}$ by using $\mathbf{\hat{T}^{I}_{s}}$ and equation \eqref{eq:est_gamma_2}. Table 4 shows the estimated $\mathbf{\hat{\Gamma}}$ based on the major occupations of the 1990 SOC, while Table \ref{t:garblinggamma1} shows the estimated $\mathbf{\hat{\Gamma}}$ based on the major occupations of the 2000 SOC. These estimates imply that on average the incorrect occupational code is assigned in around 10\% of the cases. Since a spurious transition is likely to be created when either the source or destination occupation is miscoded, the probability of observing a spurious transition for a true occupational stayer is nearly twice as large. Our methodology then suggests that coding error is indeed substantial under independent interviewing. Its magnitude is of similar order as found in other studies analysing the extent of errors in occupational coding (see Campanelli et al., 1997, Sullivan, 2009, Roys and Taber, 2017, and vom Lehn et al., 2021).

\begin{table}[!th]
\caption{Estimate of the garbling matrix, SOC 1990, $\mathbf{\hat{\Gamma}}$, (\%)}}
\label{t:garblinggamma2}
\vspace{1mm}
\centering
{\small
\setlength{\tabcolsep}{3pt}
\renewcommand\arraystretch{0.8}
\begin{tabular}{lcccccccccccccc}\hline
\textbf{OCCUPATIONS} & (1)  & (2)  & (3)  & (4)  & (5)  & (6)  & (7)  & (8)  & (9)  & (10)  & (11)  & (12)  & (13)  \\ \hline \hline
(1) Managing Occ.  &  \textbf{86.8} & 2.0 & 0.5 & 2.8 & 4.9 & 0.8 & 0.1 & 0.4 & 0.4 & 0.8 & 0.2 & 0.2 & 0.1 \\
(2) Professional Spec.  &  1.8 & \textbf{93.7} & 1.6 & 0.2 & 1.1 & 0.8 & 0.1 & 0.2 & 0.1 & 0.3 & 0.1 & 0.1 & 0.0  \\
(3) Technicians      &  1.7 & 5.6 & \textbf{83.3} & 0.2 & 3.3 & 2.1 & 0.1 & 1.5 & 0.3 & 0.7 & 1.1 & 0.1 & 0.1 \\
(4) Sales Occ.     & 3.6 & 0.2 & 0.1 & \textbf{91.2} & 1.9 & 0.8 & 0.1 & 0.3 & 0.1 & 0.2 & 0.0 & 0.4 & 1.0  \\
(5) Admin. Support   & 3.2 & 0.8 & 0.7 & 1.0 & \textbf{91.8} & 0.5 & 0.0 & 0.1 & 0.0 & 0.3 & 0.3 & 0.3 & 1.0  \\
(6) Services        &  0.9 & 0.9 & 0.7 & 0.7 & 0.9 & \textbf{93.3} & 0.1 & 0.7 & 0.4 & 0.3 & 0.3 & 0.3 & 0.5 \\
(7) Farm/Fish/Logging  &  0.7 & 0.9 & 0.3 & 0.7 & 0.5 & 1.0 & \textbf{92.0} & 0.3 & 0.3 & 0.1 & 0.5 & 1.7 & 1.1 \\
(8) Mechanics  & 1.0 & 0.6 & 1.2 & 0.7 & 0.5 & 1.5 & 0.1 & \textbf{88.9} & 1.2 & 1.5 & 1.6 & 0.4 & 0.9 \\
(9) Construction & 1.3 & 0.3 & 0.3 & 0.2 & 0.0 & 1.1 & 0.1 & 1.6 & \textbf{88.0} & 0.9 & 0.8 & 0.6 & 4.7 \\
(10) Precision Prod.  & 2.6 & 1.0 & 0.7 & 0.6 & 1.2 & 0.7 & 0.0 & 1.8 & 0.8 & \textbf{83.0} & 6.3 & 0.1 & 1.3 \\
(11) Mach. Operators  & 0.3 & 0.1 & 0.5 & 0.0 & 0.7 & 0.4 & 0.1 & 0.8 & 0.3 & 2.7 & \textbf{91.5} & 0.4 & 2.3 \\
(12) Transport  & 0.5 & 0.2 & 0.1 & 0.8 & 1.1 & 0.7 & 0.4 & 0.4 & 0.4 & 0.0 & 0.7 & \textbf{92.2} & 2.3 \\
(13) Laborers   & 0.3 & 0.1 & 0.1 & 2.7 & 4.8 & 1.5 & 0.3 & 1.2 & 4.5 & 1.4 & 5.7 & 3.0 & \textbf{74.3} \\ \hline
\end{tabular}
\end{table}

\begin{sidewaystable}
  \centering
  \caption{Estimate of the garbling matrix, SOC 2000, $\mathbf{\hat{\Gamma}}$, (\%)}
  \label{t:garblinggamma1}
\resizebox{1.0\textwidth}{!}{\addtolength{\tabcolsep}{-3pt}%
    \begin{tabular}{|l|cccccccccccccccccccccc|}
\textbf{OCCUPATIONS} & (11)  & (13)  & (15)  & (17)  & (19)  & (21)  & (23)  & (25)  & (27)  & (29)  & (31)  & (33)  & (35)  & (37) & (39) & (41) & (43) & (45) & (47) & (49) & (51) & (53) \\ \hline \hline
(11) Management Occ.   &        {\bf{84.2}}  &  2.3  &  0.3  &  0.8  &  0.3  &  0.3  &  0.1  & 0.3  & 0.1  & 0.3  & 0.1  & 0.1  & 0.4  & 0.1  &  0.4  &  3.3  &  3.5  &  0.5  &  0.7  &   0.5  &  1.2  &  0.3  \\

(13) Business \& Finance Oper. &  4.7  &  {\bf{82.8}}  &   0.9  &  0.5  &  0.2  &  0.2  &  0.2  &  0.3  & 0.0  &  0.0  & 0.0  & 0.1  & 0.0  & 0.0 &  0.0  & 1.6  & 7.6  & 0.0  & 0.1  & 0.1  & 0.5  & 0.1  \\

 (15) Computer \& Math. Occ. & 2.6  &   3.7  &  {\bf{85.7}}  &  1.6  & 0.4  & 0.0  & 0.0 & 0.4  & 0.5  & 0.0 & 0.1  & 0.1  & 0.0  & 0.0  & 0.0  & 0.0  &      4.4  & 0.0  & 0.1  &  0.7  & 0.2  &  0.2  \\

 (17) Architect \& Eng. Occ.  & 2.2  &  0.7  & 0.5  &  {\bf{87.0}}  & 1.3  & 0.0  & 0.0 & 0.0  & 1.2  & 0.3  & 0.0  & 0.0 & 0.0 & 0.1  & 0.0 & 0.1  &         1.1  & 0.0 & 0.9  & 2.4  & 2.2  & 0.1  \\

  (19) Life, Phys, and Soc. Sci. Occ. &  2.4  & 0.6  & 0.4  & 3.3  & {\bf{82.9}}  & 0.0  & 0.2  & 1.1  & 0.4  & 2.2  & 0.5  & 0.0 & 0.0 &0.5  & 0.0 &         0.3  & 1.7  & 0.7  & 0.5  & 1.1  &1.0  & 0.3  \\

 (21) Comm \& Soc. Service Occ.  & 2.8  & 0.7  & 0.0 & 0.0 & 0.0 &  {\bf{89.8}}  & 0.0 & 1.0  & 0.4  & 0.5  & 0.7  & 0.3  & 0.2  & 0.1  & 0.4  &         0.0  & 2.6  & 0.0 & 0.2  & 0.1  & 0.1  & 0.3  \\

 (23) Legal  & 1.3  & 1.6  & 0.0 & 0.0 & 0.5  & 0.0 & {\bf{93.0}}  & 0.0  & 0.0  & 0.0 & 0.0  & 0.5  & 0.0  & 0.0 &  0.1  & 0.1  & 3.5  & 0.0  & 0.0  &    0.0  & 0.0  & 0.0  \\

 (25) Educ., Training \& Library  &  0.5  & 0.3  & 0.1  &  0.0  & 0.3  & 0.2  & 0.0 & {\bf{97.3}}  &  0.1  & 0.2  & 0.1  & 0.0 & 0.0 & 0.0  & 0.3  &  0.1  &        0.6  & 0.0  & 0.0 & 0.0 & 0.0 & 0.0 \\

 (27) Arts, Dsgn, Ent., Sports \& Media  &  1.0  & 0.2  & 0.6  & 3.8  &   0.5  &  0.4  &  0.0 &  0.3  & {\bf{89.2}}  & 0.0 & 0.0 & 0.0 & 0.0 & 0.0  &         0.0  & 0.7  & 1.3  & 0.0 & 0.2  & 0.2  & 1.7  &  0.4  \\

 (29) Healthcare Pract. \& Tech. Occ.  &  0.7  & 0.0  &  0.0 & 0.2  & 0.7  & 0.1  & 0.0  & 0.3  & 0.0 & {\bf{92.7}}  & 3.3  &   0.0 & 0.1  & 0.1  &   0.1  &  0.1  & 1.0  & 0.0 & 0.0 & 0.1  & 0.2  & 0.1  \\

 (31) Healthcare Support & 0.5  &  0.0  &  0.0 & 0.0  & 0.3  &  0.5  & 0.0 &  0.2  & 0.0 & 6.8  &  {\bf{88.5}}  & 0.0  & 0.6  & 0.4  &  0.3  & 0.0  &     1.6  & 0.0 & 0.0 & 0.1  &  0.3  & 0.0  \\

  (33) Protective Service  & 0.7  & 0.3  & 0.1  & 0.0  & 0.0  & 0.2  & 0.2  & 0.1  & 0.1  & 0.1  & 0.0  &  {\bf{95.0}} &  0.5  &  0.1  & 0.2  & 0.1  & 1.8  &         0.3  & 0.0 & 0.1  & 0.4  & 0.5  \\

 (35) Food Prep/Serving \& Rel.  & 0.9  & 0.0  & 0.0 & 0.0  & 0.0  & 0.0  & 0.0  & 0.0 & 0.0  &  0.1  & 0.3  & 0.2  &  {\bf{95.3}}  &  0.5  & 0.0  & 1.5  & 0.4  & 0.0  & 0.0  & 0.0 & 0.4  &  0.2  \\

 (37) Building/grounds Clean. \& Maint.  &  0.3  & 0.0  & 0.0  & 0.1  & 0.2  & 0.1  & 0.0  & 0.0  & 0.0  & 0.1  & 0.2  & 0.0 & 0.6  & {\bf{90.3}}  &         0.1  &  0.4  & 0.2  & 0.3  & 1.7  & 2.5  & 1.1  & 1.8  \\

 (39) Personal Care \& Service Occ.  & 3.3  & 0.0  & 0.0  & 0.0 & 0.0 & 0.4  & 0.1  & 1.6  & 0.0 & 0.5  &  0.6  &  0.4  & 0.0 &  0.3  & {\bf{91.4}}  &         0.9  &  0.6  & 0.0  & 0.0 & 0.1  & 0.0 & 0.2  \\

 (41) Sales \& Rel. Occ. &  3.0  & 0.7  & 0.0 &  0.0 & 0.0  & 0.0 & 0.0  & 0.0 &  0.1  & 0.0 & 0.0  &  0.0  & 0.5  & 0.1  & 0.1  & {\bf{91.7}}  &  1.7  &         0.1  & 0.1  & 0.4  & 0.3  & 1.3  \\

  (43) Office \& Admin. Support & 1.7  &  1.8  &  0.3  & 0.2  & 0.1  & 0.2  & 0.1  & 0.2  &  0.1  &  0.2  &  0.2  & 0.2  & 0.1  & 0.0 & 0.0 & 0.9  &         {\bf{91.6}}  &  0.0  & 0.0 & 0.1  & 0.6  & 1.3  \\

(45) Farm, Fish. \& Forestry & 4.5  & 0.2  & 0.0  & 0.0 &  1.0  & 0.1  & 0.0  & 0.1  & 0.0 & 0.0 & 0.0 & 0.5  & 0.1  & 0.9  & 0.1  & 0.7  & 0.3  &         {\bf{87.5}}  &  0.1  & 0.3  & 0.8  &   3.4  \\

 (47) Construction \& Extraction & 1.3  & 0.1  &  0.0  & 0.6  & 0.1  &  0.0  & 0.0  & 0.0 & 0.0 & 0.0 & 0.0 & 0.0 & 0.0 & 1.1  & 0.0 & 0.2  & 0.1  &         0.0 & {\bf{90.8}}  & 1.6  &  1.9  & 2.1  \\

 (49) Install., Maint. \& Repair Occ.   &  0.8  & 0.1  & 0.1  & 1.4  &  0.2  & 0.0  & 0.0 & 0.0  & 0.0 & 0.1  & 0.0 &  0.0 & 0.0 & 1.5  & 0.0 & 0.6  &         0.4  &  0.1  &  1.4  & {\bf{89.0}}  & 3.0  &  1.1  \\

  (51) Production Occ. &  0.8  &  0.2  & 0.0 & 0.5  &  0.1  &  0.0  & 0.0 & 0.0 & 0.1  & 0.1  & 0.0 & 0.0 & 0.1  & 0.3  & 0.0  & 0.2  &  0.8  &         0.0  & 0.6  &  1.1  &  {\bf{93.1}}  &  1.9  \\

 (53) Transportation \& Mater. Moving & 0.3  & 0.1  & 0.0 & 0.1  & 0.0  &  0.0  & 0.0  & 0.0  & 0.1  & 0.1  &  0.0  &  0.1  & 0.1  & 0.7  & 0.0  &  1.5  &         2.9  & 0.4  &  1.2  &  0.7  &  3.4  &  {\bf{88.2}}  \\

\end{tabular}}
\end{sidewaystable}

Two additional messages come out of Tables 4 and \ref{t:garblinggamma1}. (i) Different occupations have very different propensities to be assigned a wrong code. For example, when using the 1990 SOC we find that individuals whose true occupation is ``laborers'' have a 74\% probability of being coded correctly, while individuals whose true occupation is ``professional speciality'' have a 94\% probability of being coded correctly. (ii) Given a true occupation, some coding mistakes are much more likely than others. For example, workers whose true occupation is ``laborers'' have a much larger probability to be miscoded as ``machine operators'' (5.7\%), ``construction'' (4.5\%) or ``admin. support'' (4.8\%) than as ``managers'' (0.3\%) or ``professionals'' (0.1\%). Our methodology enable us to take these differences into account by correcting observed occupational flows by source-destination occupation pair. This provides cleaner net mobility estimates, where the identity of the origin and destination occupation matters.

\subsection{The effect of $\mathbf{\hat{\Gamma}}$ on the gross occupational mobility rate} 
\begin{figure}[!ht]
\centering
\includegraphics[width=4.0in]{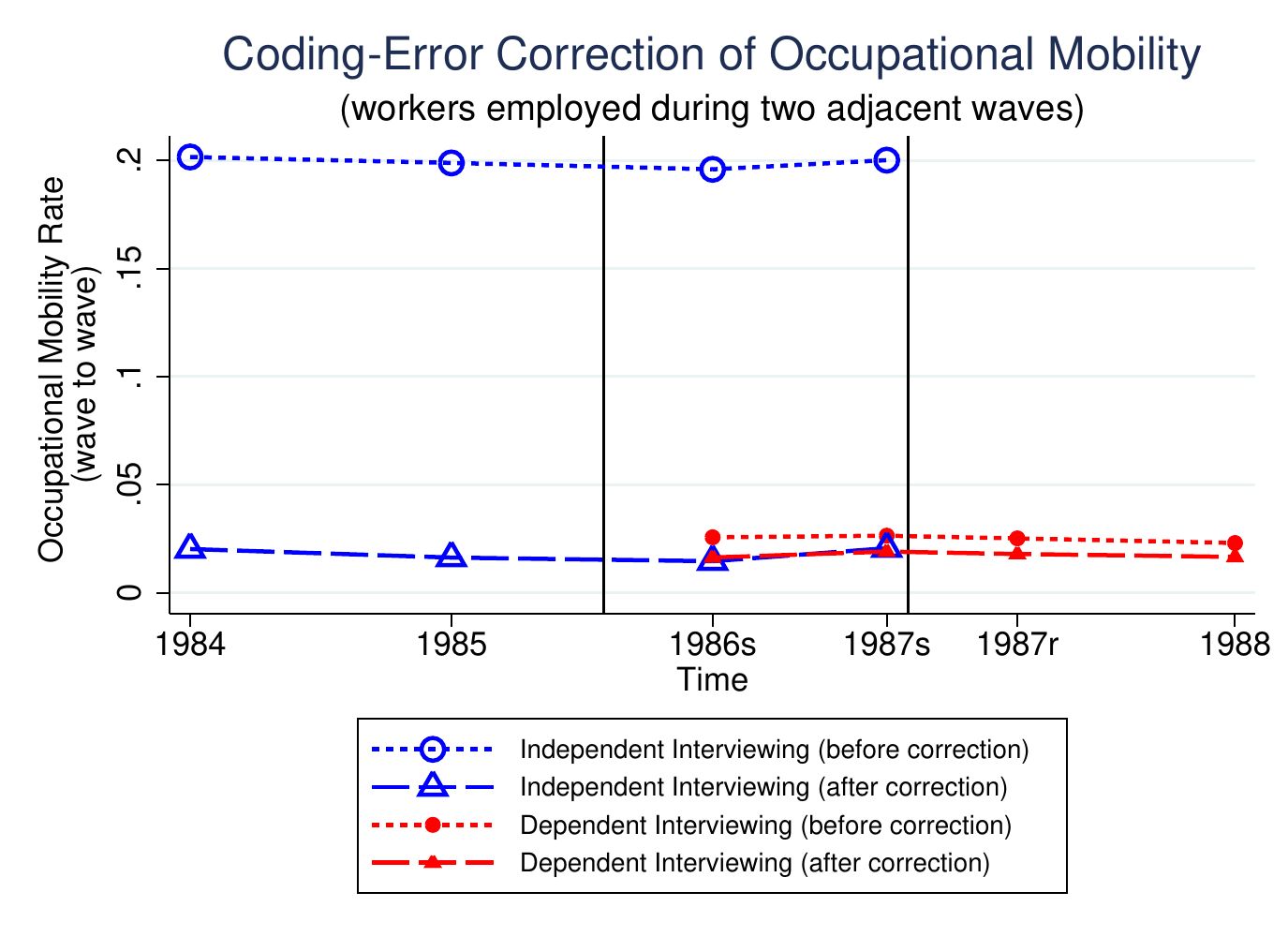}
\caption{Correcting occupational mobility rates for workers employed in two subsequent waves}
\label{fig:SIPP_occmob_estayers}
\end{figure}

Figure \ref{fig:SIPP_occmob_estayers} depicts the effect of $\mathbf{\hat{\Gamma}}$ when computing \emph{wave-to-wave} occupational mobility rates using the major occupational groups of the 2000 SOC. For this exercise we augment the 1985/86/87 samples with workers from the 1984 and 1988 panels, which satisfy the same sample restrictions as before. Figure \ref{fig:SIPP_occmob_estayers} depicts the average wave-to-wave occupational mobility rate obtained during each year. For the year 1986 we only use the observations that cover the February to December period obtained from the 1986 sample. We label these observations ``1986s'' as they are the ones we use in our original sample to estimate $\mathbf{\hat{\Gamma}}$. In the case of the year 1987, we present the average occupational mobility rate obtained for the January to April period (labelled ``1987s'') separately from the average occupational mobility rate obtained from the remaining months (labelled ``1987r''). The two vertical lines mark the time period in which dependent and independent interviewing overlap.

The short-dashed line with hollow circular markers depicts the observed occupational mobility rates obtained from pooling together employer/activity stayers and changers using the 1984/85 samples, where all respondents were subject to independent interviewing. This pooled sample yields occupational mobility rates that lie between 19.6\%-20\% and average 19.7\% between the vertical lines. Under independent interviewing, $\mathbf{M_m}$ is garbled both at the source and destination occupations and hence is observed as $\mathbf{\Gamma'} \mathbf{M_m }\mathbf{\Gamma}$. Pre- and post-multiplying the latter by the respective inverses $\mathbf{(\Gamma')^{-1}}, \mathbf{\Gamma^{-1}}$ recovers $\mathbf{(\Gamma')^{-1}\Gamma' M_m \Gamma \Gamma^{-1}}=\mathbf{M_m}$. Applying this procedure to the 1984/85 samples yields the long-dashed line with hollow triangular markers. The result is a drop in the occupational mobility rate to about 1.6\%.

Next consider the 1986/88 samples. Here independent interviewing is only applied to employer/activity changers. Therefore, the observed overall occupational mobility rate for these samples is based on $\mathbf{M^{D,86/87}_s} + \mathbf{M^{I, 86/87}_m}$. The matrix $\mathbf{M^{D,86/87}_s}$ does not contribute to any occupational transitions while matrix $\mathbf{M^{I, 86/87}_m}$ does. This difference implies a much lower observed aggregate mobility rate. The latter is depicted by the short-dashed line with filled circular markers in Figure \ref{fig:SIPP_occmob_estayers} and averages 2.6\%. Correcting the occupational flows of the 1986-88 samples using $\mathbf{\hat{M}^{D,86/87}_s} + \mathbf{(\hat{\Gamma}')^{-1}}\mathbf{\hat{M}^{I, 86/87}_m}\mathbf{\hat{\Gamma}^{-1}}$ yields the series depicted by the long-dashed line with solid triangular markers. The result is a drop in the occupational mobility rate to about 1.7\%, which is very close to the $\mathbf{\Gamma}$-corrected occupational mobility rate from the 1984/85 samples. Indeed, in Figure \ref{fig:SIPP_occmob_estayers} the blue (independent-interviewing) and red (dependent-interviewing) long-dashed lines of the $\mathbf{\Gamma}$-corrected measures nearly coincide between the two vertical lines.\footnote{Note that in the estimation of $\mathbf{\Gamma}$ we used $\mathbf{\hat{M}^{I,86/87}_m}$, but we did not impose the additional restriction on $\mathbf{(\hat{\Gamma}')^{-1}}\mathbf{\hat{M}^{I,86/87}_m }\mathbf{\hat{\Gamma}^{-1}}$ to equal $\mathbf{(\hat{\Gamma}')^{-1}}\mathbf{\hat{M}^{I,85}}\mathbf{\hat{\Gamma}^{-1}}$.}

The fact that we obtain very similar corrected mobility rates after using the same underlying $\mathbf{\Gamma}$-correction matrix in two different survey designs, suggests that our methodology captures the extent of coding error quite well. Our methodology also seems to work well in other dimensions. We find that the occupational mobility rates in the years before 1986 are adjusted downwards to numbers that are very similar to the numbers obtain during the 1986s-1987s window. For example, the $\mathbf{\Gamma}$-corrected occupational mobility in 1985 is 1.63\%, while during the 1986s-1987s window we obtain 1.62\%. For the years after the 1986s-1987s window, we find that the $\mathbf{\Gamma}$-corrected occupational mobility rate is within 0.02\% of the one obtained during this window. Further, Figure \ref{fig:SIPP_occmob_estayers} shows that any changes in the level of the $\mathbf{\Gamma}$-corrected occupational mobility rate series appear to track changes in the uncorrected series. This also suggests that our correction method does not seem to introduce additional randomness into the occupational mobility process.

We apply our $\mathbf{\Gamma}$-correction method to those who changed employers with an intervening spell of unemployment or non-employment. We then compute the host of statistics shown in the main text and Supplementary Appendix B. We show that occupational miscoding increases observed gross mobility. In the raw data we compute an occupational mobility rate at re-employment of 53.1\% based on the 2000 SOC. After applying our $\mathbf{\Gamma}$-correction, we obtain an occupational mobility rate at re-employment of 44.4\%. We also find that coding errors makes occupational mobility appear less responsive to unemployment duration. Since in short unemployment spells true occupational staying is more common, miscoding creates relatively more spurious mobility and therefore our method corrects more the short spells, leading to a steeper relationship between occupational mobility and unemployment duration (see Table 1 and Figure 1 in Supplementary Appendix B). Miscoding also reduces the degree of procyclicality of gross mobility. This arises as coding errors will generate more spurious mobility in times were there are more true stayers (see Table 6 and Figure 11 in Supplementary Appendix B).

In contrast, we find that miscoding reduces the contribution of net occupational mobility (see also Kambourov and Manovskii, 2013). The average net mobility rate (as defined in the main text) increases from 3.6\% (uncorrected) to 4.2\% (corrected), a nearly 15\% increase. To understand why this arises, consider the true net mobility transition flow matrix, $\mathbf{M_{net}}$, and note that this matrix does not have mass on its diagonal. Under independent interviewing coding errors imply that the true net mobility matrix would be observed as $\mathbf{M^{I}_{net}}=\mathbf{\Gamma'M_{net}\Gamma}$, which could have mass on the diagonal and hence biasing downwards net mobility flows. That is, coding errors mistakenly convert some true mobility flows into occupational stays, while miscoding for stayers is completely symmetric with respect to origin and destination occupations, and therefore should not give rise to spurious net mobility.

\subsection{The differential impact of coding error on employer/activity stayers and movers} Note that our $\mathbf{\Gamma}$-corrected occupational mobility rate for those workers who changed employers through a spell of unemployment is 16.4\% lower than the raw one (44.4\% vs 53.1\%, 2000 SOC). This adjustment is much smaller in relative terms than the one suggested by Kambourov and Manovskii (2008). They argue that on average around 50\% of \emph{all} year-to-year observed occupational mobility in the raw PSID data is due to coding error. Indeed when constructing the year-to-year occupational mobility rate for our pooled sample of employer/activity stayers and movers in the SIPP, we find that only 39.3\% of observed occupational changes are genuine (an uncorrected rate of 26.7\% vs a $\mathbf{\Gamma}$-corrected rate of 10.5\%).

These findings are not mutually inconsistent. The key to this difference is that the \emph{relative} importance of coding error varies greatly with the true propensity of an occupational change among employer stayers and among employer changers. True occupational changes are more likely to be accompanied by changes in employers (for examples of this argument see Hill, 1994, Moscarini and Thomsson, 2007, and Kambourov and Manovskii, 2009) and, vice versa, employer changes are more likely to be accompanied by occupational changes. As such, the \emph{relative} adjustment that Kambourov and Manovskii (2008) find does not automatically carry over to different subsets of workers or flows measured at a different frequency.

As an example, consider an individual who is observed moving from ``managers'' to ``laborers''. Suppose ``managers'' was this individual's true source occupation, but the observed destination occupation was the result of coding error. If this individual is a true occupational stayer, the observed transition will wrongly tell us that the individual \emph{stopped} being a manager and will generate a false occupational move. Instead, if this individual is a true occupational mover, the observed transition, although wrongly coded, will still capture the fact that the individual stopped being a manager and hence will capture a true occupational move. Given that true occupational changes are more likely to occur along side employer changes, there will be more workers among employer changers (relative to employer stayers) whose categorization as an observed occupational mover will not change after using the $\mathbf{\Gamma}$-correction. This implies that the measured occupational mobility rate of employer changers would have a relative smaller adjustment than the measured occupational mobility rate of employer stayers. Kambourov and Manovskii (2008) pooled together employer changers and stayers. Since the latter group represent the vast majority of workers in their sample (as well as in our sample), the relative adjustment proposed by these authors is naturally much larger.

Consider the following iterative back-of-the-envelope approximation to understand why there must be a larger proportion of true occupational movers among those who changed employers than among those who did not change employers. Recall that the $\mathbf{\Gamma}$-correction method implies that true occupational stayers will be coded as movers in about 20\% of the times. This happens irrespectively of whether the worker changed employers or not. If we were to suppose that all of the unemployed who regain employment were true occupational stayers, the difference between their observed mobility rate (53\%) and coding error (20\%) would immediately imply that among the unemployed there must be true occupational movers and these true movers would represent at least 33\% of the unemployed. This result then shrinks the population of occupational stayers (the ``population at risk'') among the unemployed to at most 67\%, which (proceeding iteratively) implies the maximum extent of spurious flows produced under the same miscoding propensity is reduced to $0.2 \times 67\%$. In turn, the latter implies an updated lower bound on the percentage of true occupational movers among the unemployed of 39.6\%. Proceeding iteratively, one arrives to a lower bound on the percentage of true occupational movers among the unemployed of 41.25\%, which is close to the gross mobility we obtained after applying the $\mathbf{\Gamma}$-correction. A similar procedure but applied to employer stayers shows a much lower `true' mobility rate among this group.

\begin{table}[ht!]

{\small
  \centering
   \caption{Inferred Coding Error Probabilities and Observed vs. Underlying Occupational Mobility}
\vspace{1mm}

     \begin{tabular}{lccccc}
    Classification & \multicolumn{1}{c}{$\mathbb{P}(\tilde{M} |S)$}  & \multicolumn{1}{c}{ \ \ \ \  $\mathbb{P}(\tilde{M}|U)$} & \multicolumn{1}{c}{\ \ \ \ \ \ \ $\mathbb{P}(M|U)$} & \multicolumn{1}{c}{\ \ \ \ $\frac{\mathbb{P}(M|U)}{\mathbb{P}(\tilde{M}|U)} - 1$ } & \multicolumn{1}{c}{$\mathbb{P}(\tilde{o}\neq o | U)$}  \\ \hline \hline
 2000 SOC (22 cat) & 0.178 &  \ \ \ \ \ 0.531 & \ \ \ \ \ \  0.444 & \ \ \ \  -0.164 & 0.095   \\
1990 SOC (13 cat) & 0.197 & \ \ \ \ \ 0.507 &\ \ \ \ \ \  0.401 & \ \ \ \   -0.209  & 0.105 \\
1990 SOC (6 cat) & 0.148  & \ \ \ \ \ 0.402 &\ \ \ \ \ \  0.317 & \ \ \ \  -0.213  & 0.077 \\
 NR/R Cognitive, NR/R Manual  (4 cat) & 0.110  & \ \ \ \ \ 0.332 &\ \ \ \ \ \ \  0.263 & \ \ \ \   -0.208 & 0.058 \\
 Cognitve, R Manual / NR Manual (3 cat) & 0.083  & \ \ \ \ \ 0.273 &\ \ \ \ \ \ \  0.218 & \ \ \ \   -0.199 & 0.043 \\
Major industry groups (15 cat) & 0.101  &  \ \ \ \ \ 0.523 &\ \ \ \ \ \  0.477 & \ \ \    -0.088  & 0.055 \\
    \hline
            \multicolumn{6}{p{1\textwidth}}{\tiny{Sample: unemployed between 1983-2013, in 1984-2008 SIPP panels, subject to conditions explained in data construction appendix (most importantly: unimputed occupations (resp. industries), with restrictions to avoid right and left censoring issues.) $\mathbb{P}(\tilde{M} | S)$: probability that the wrong code is assigned to a true stayer; $\mathbb{P}(\tilde{o}\neq o | U)$}: probability that the wrong code is assigned to an unemployed worker; $\mathbb{P}(\tilde{M})$: observed occupational mobility among the unemployed; $\mathbb{P}(M)$: inferred underlying true mobility (proportion of unemployed). \emph{NR/R} refers to routine vs. non-routine. Further details on the classifications are explained in the data construction appendix.}
    \end{tabular}%
  \label{tab:catcomparison}%
}
\end{table}%

Table \ref{tab:catcomparison} shows that the differential impact of coding error on employer changers and stayers is present when considering several alternative occupational classifications as well as mobility across industries. In all these cases we use wave-to-wave mobility rates. The first column, $\mathbb{P}(\tilde{M} |S)$, presents the probability that a true employer/activity stayer in the 1985 sample is assigned the wrong occupational code and hence is observed as an occupational mover $\tilde{M}$. This probability is 17.8\% when using the 2000 SOC. This implies that under independent interviewing we will observe an occupational mobility rate of 17.8\% among the employer/activity stayers. This rate increases slightly when using the 1990 SOC, 19.7\%, and remains high even when we aggregate occupations into six categories, 14.8\%.\footnote{The six groups are: (1) managers/professional speciality; (2) tech support/admin support/sales; (3) services; (4) farm/forest/fisheries; (5) precision production/craft/repair; and (6) operators, fabricators and laborers. These correspond to the summary occupational group of the 1990 SOC.} Aggregating occupations into four tasked based categories (routine vs. non-routine and manual vs. cognitive) only brings down the observed mobility rate of true occupational stayers to 11\%. As many other studies, we also find the probability that a true stayer is observed as a mover is lower when considering industries instead of occupations.

The second and third columns show the observed, $\mathbb{P}(\tilde{M}|U)$, and $\mathbf{\Gamma}$-corrected, $\mathbb{P}(M|U)$, occupational mobility rate of those workers who changed employers through unemployment. These are obtained using the probability that an unemployed worker in the 1984-2008 SIPP panels is assigned the wrong occupational code at re-employment, $\mathbb{P}(\tilde{o}\neq o | U)$. We observe that across all classifications the (relative) difference between the observed and the $\mathbf{\Gamma}$-corrected occupational mobility rates of the unemployed, $\mathbb{P}(\tilde{M}|U) - \mathbb{P}(M|U)$, is about half the size of $\mathbb{P}(\tilde{M} |S)$. The fourth column shows this difference in relative terms. It is clear that across all classifications the same coding error generates a much larger difference (in absolute and relative terms) between the observed and $\mathbf{\Gamma}$-corrected occupational mobility rates of employer stayers than among employer changers.

\section{Measuring coding error in the PSID}

We now broaden the above analysis and use probabilistic models based on the PSID (as in Kambourov and Manovskii, 2008) to assess the impact of coding errors on the probability of an occupational change. The advantage of the $\mathbf{\Gamma}$-correction method is that it captures all sources of error in assigning occupation codes. It delivers an identification procedure that is not subject to the issue highlighted in Corollary A.1 and that recovers the extent to which coding errors arise at the level of each occupation. Further, the SIPP provides a large sample size in which we can apply our correction method. The advantage of the PSID is that retrospective coding affected directly the way occupational transitions among employer changers were measured. We exploit this feature and assess the impact of coding error on employer movers and employer stayers separately and compare the results with the ones obtained from the SIPP using our $\mathbf{\Gamma}$-correction method. We find a very consistent picture across the two data sets.

To assess the impact of retrospective coding in reducing coding errors, we use the PSID retrospective occupation-industry supplementary data files, which contain the re-coding the PSID staff performed on the occupational mobility records obtained during the 1968-1980 period. Since the 1981-1997 records were not re-coded and collected under independent interviewing, the earlier period can be used to construct ``clean'' occupational mobility rates and to analyse the effect of measurement error at the coding stage. In constructing our sample we closely follow Kambourov and Manovskii (2008, 2009). The details of this sample are described in the Supplementary Appendix B.7.

\subsection{Gross occupational mobility rates}

As in Kambourov and Manovskii (2008) we define the \emph{overall} occupational mobility rate as the fraction of employed workers whose occupational code differs between years $t$ and $t+1$ divided by the number of workers who were employed in year $t$. As these authors we also consider those workers who were employed at the time of the interview in year $t-1$, unemployed in $t$ and employed at the time of the interview in year $t+1$. Further, we define the \emph{within-employer} occupational mobility rate as the fraction of workers employed who did not change employers but exhibit a different occupational code between years $t$ and $t+1$, divided by the number of employed workers who did not change employers between years $t$ and $t+1$. Similarly, we define the \emph{across-employer} occupational mobility rate be the fraction of employed workers who's occupational code differs between years $t$ and $t+1$ and reported an employer change between these years, divided by the number of employed workers in year $t$ who have reported an employer change between years $t$ and $t+1$. To identify employer changes we follow the procedure detailed in Kambourov and Manovskii (2009), Appendix A1.

\begin{figure}[!ht]
\centering
\includegraphics[width=5.0in]{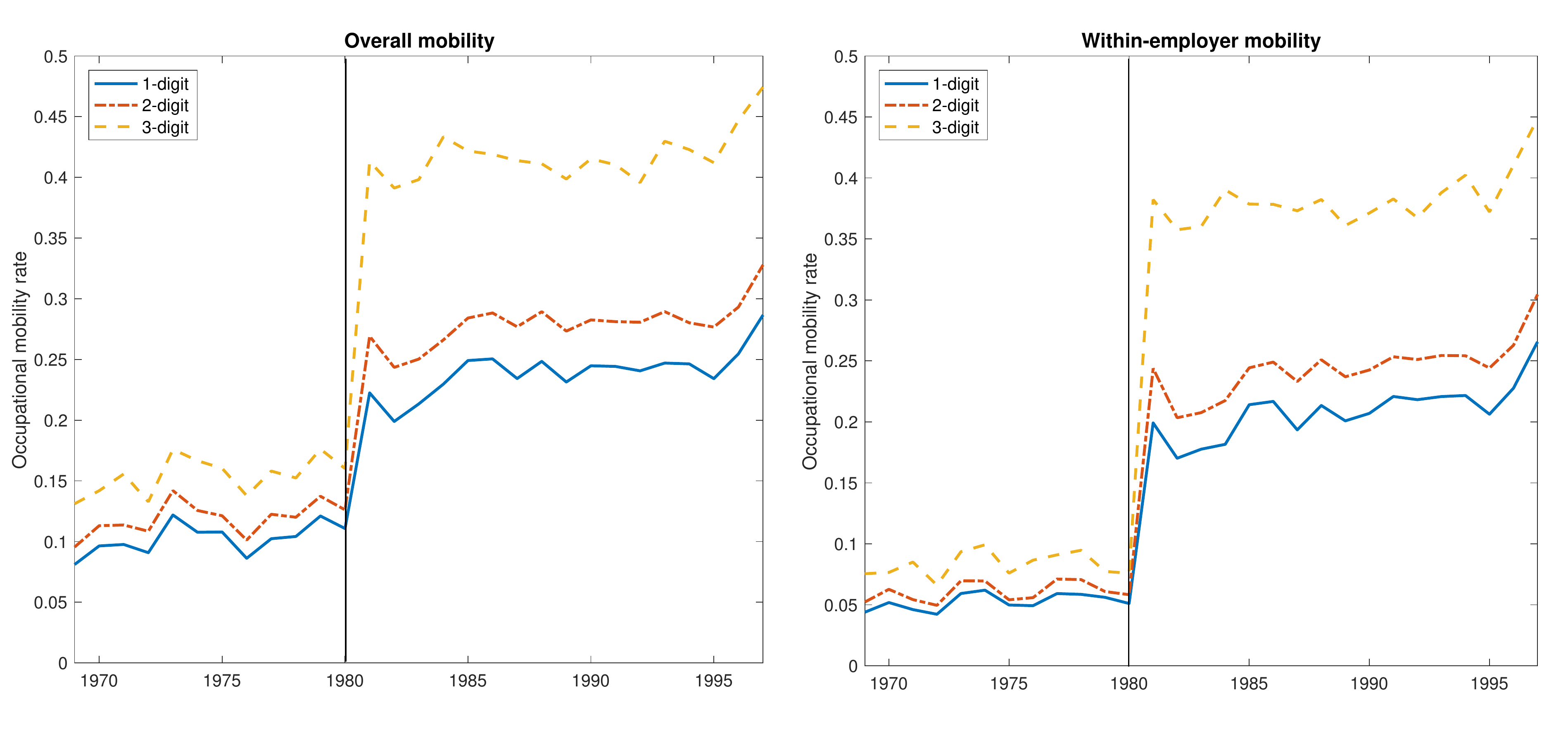}
\caption{Overall and within-employer occupational mobility rates}
\label{fig:KM_occ_mob}
\end{figure}

The left panel of Figure \ref{fig:KM_occ_mob} depicts the yearly overall occupational mobility rate at a one-, two- and three-digit level of aggregation, using the 1970 SOC. When retrospective re-coding was used, the overall occupational mobility rate experienced a large downward shift, ranging between 10 to 25 percentage points, depending on the level of aggregation of the occupational codes. These drops suggest that only between 38\% to 45\% of all occupational moves are genuine. This is very similar to the conclusion reached by Kambourov and Manovskii (2008).

\begin{figure}[!ht]
\centering
\includegraphics[width=5.0in]{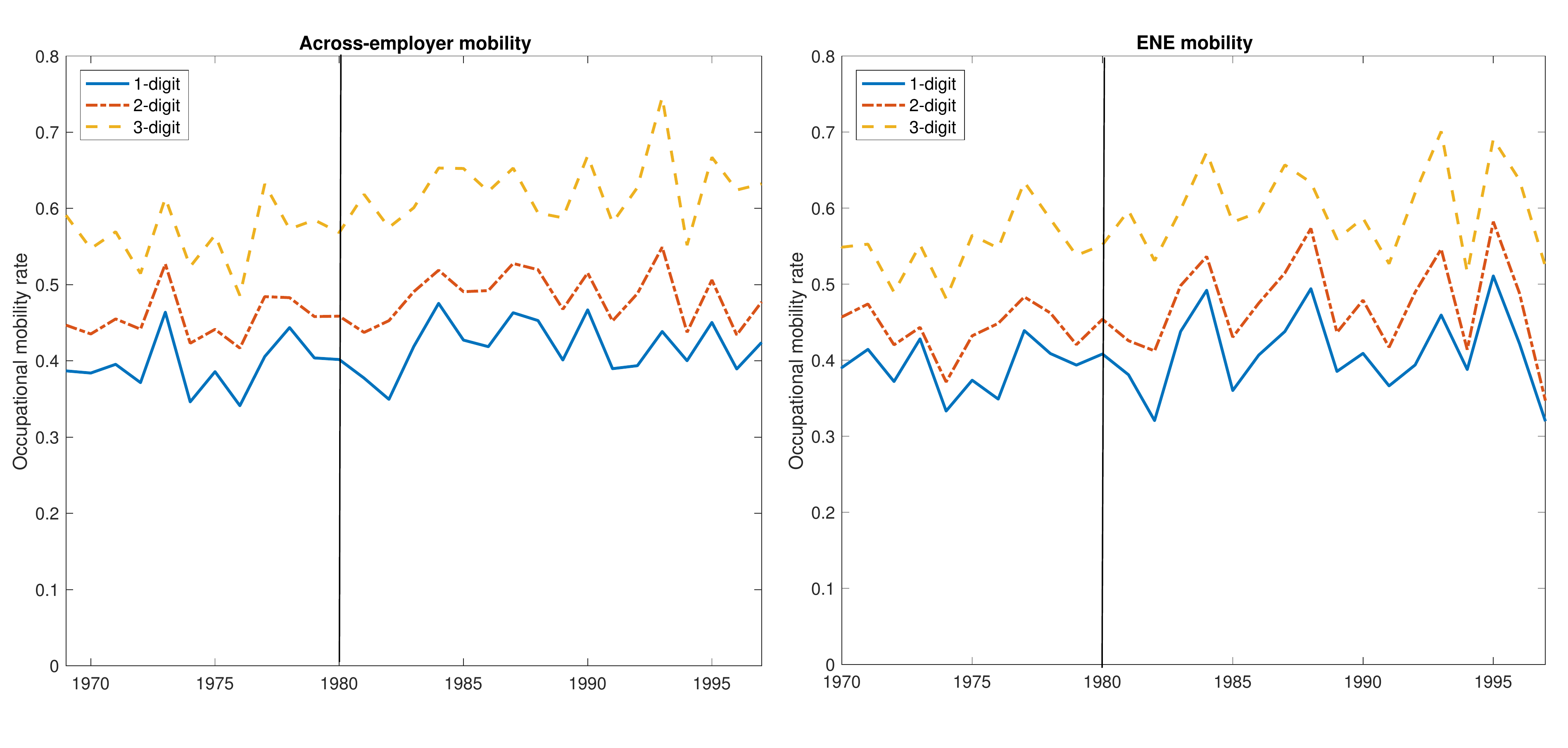}
\caption{Across-employer occupational mobility rates}
\label{fig:Occ_emp_mob}
\end{figure}

The right panel of Figure \ref{fig:KM_occ_mob} shows that the \emph{within-employer} occupational mobility rates experienced even stronger drops than the overall ones under retrospective coding. In contrast, the left panel of Figure \ref{fig:Occ_emp_mob} shows that the impact of coding error in the \emph{across-employer} occupational mobility rates is much more moderate and hardly visible when aggregating occupations at a one-digit level. These results then suggest that the impact of coding error on the overall mobility rates mainly arises from those workers who did not change employers, where employer stayers account on average for 87.1\% of all employed workers in a given year, while those who changed employers account for the remainder 12.9\%. We find a similar conclusion based on the SIPP data.

Next consider the effects of coding error on the occupational mobility rate of only those workers who changed employers through a spell of non-employment $(ENE)$. To construct the $ENE$ occupational mobility rate we consider (i) those workers who were employed at the interview date in year $t-1$, non-employed at the interview date in year $t$ and once again employed at the interview date in year $t+1$; and (ii) those workers employed at the interview dates in years $t$ and $t+1$, but who declared that they experienced an \emph{involuntary} employer change between these two interviews. An involuntary change is defined as those cases were the worker declared a job separation due to ``business or plant closing'', due to ``being laid off or were fired'' or their ``temporary job ended'' (see Supplementary Appendix B.7 for details). We divide these flows by the number of workers who changed employers through a spell of non-employment during the corresponding years. The right panel of Figure \ref{fig:Occ_emp_mob} shows that coder error once again seems to have a small effect on the occupational mobility rates of these workers.

\subsection{Probabilistic models}

The visual impressions given by the above figures on the effects of coding error are confirmed when estimating the effects of retrospective coding using a probit or a linear probability model. In these regressions the dependent variable takes the value of one if the worker changed occupation and zero otherwise. We include the indicator variable ``break'' which takes the value of one during the years in which the PSID used retrospective coding. In addition we control for age, education, full or part-time work, occupation of origin, region of residence, aggregate and regional unemployment rates, a quadratic time trend and number of children.\footnote{As in Kambourov and Manovskii (2008), the education indicator variable takes the value of one when the worker has more than 12 years of education and zero otherwise. This is to avoid small sample problems if we were to divide educational attainment in more categories. The regional unemployment rates are computed using US states unemployment rates.}

Table \ref{t:Probit1} shows the marginal effects for the probit regressions.\footnote{These estimates are obtained using the personal weights provided by each survey, but similar results are obtained when using the unweighted data. We also obtained very similar results when using the linear probability model on weighted and unweighted data and when using robust standard errors and clustering standard errors at a yearly level.} It shows that retrospective coding had a large and significant effect on reducing the probability of changing occupations when all workers were included in the sample. Furthermore, the values of the marginal effects of the ``break'' indicator are very close to the amount by which the overall occupational mobility series shifted when retrospective coding was used, as depicted in Figure \ref{fig:KM_occ_mob}.

\begin{table}[ht!]
  \centering
  \caption{The effect of measurement error on the PSID (probit marginal effects)}\vspace{3mm}
  {\footnotesize
    \begin{tabular}{l|lll|llllll}
    \hline
          & \multicolumn{3}{c|}{{\bf{All workers}}} & \multicolumn{3}{c|}{{\bf{Across employer}}} & \multicolumn{3}{c}{{\bf{ENE}}} \\
          & \multicolumn{1}{c}{1-digit} & \multicolumn{1}{c}{2-digits} & \multicolumn{1}{c|}{3-digits} & \multicolumn{1}{c}{1-digit} & \multicolumn{1}{c}{2-digits} & \multicolumn{1}{c|}{3-digits} & \multicolumn{1}{c}{1-digit} & \multicolumn{1}{c}{2-digits} & \multicolumn{1}{c}{3-digits} \\
\hline
    Unemp rate & \multicolumn{1}{c}{-0.002} & \multicolumn{1}{c}{-0.004} & \multicolumn{1}{c|}{0.001} & \multicolumn{1}{c}{-0.026$^{**}$} & \multicolumn{1}{c}{-0.025$^{**}$} & \multicolumn{1}{c|}{-0.021$^{**}$} &  \multicolumn{1}{c}{-0.058$^{***}$} & \multicolumn{1}{c}{-0.057$^{***}$} & \multicolumn{1}{c}{-0.032$^{*}$} \\

    Reg unemp rate & \multicolumn{1}{c}{-0.003} & \multicolumn{1}{c}{-0.003} & \multicolumn{1}{c|}{-0.003} & \multicolumn{1}{c}{0.015$^{**}$} & \multicolumn{1}{c}{0.014$^{*}$} & \multicolumn{1}{c|}{0.013$^{*}$} & \multicolumn{1}{c}{0.046$^{***}$} & \multicolumn{1}{c}{0.039$^{***}$} & \multicolumn{1}{c}{0.022$^{*}$} \\

    Age   & \multicolumn{1}{c}{-0.008$^{***}$} & \multicolumn{1}{c}{-0.008$^{***}$} & \multicolumn{1}{c|}{-0.011$^{***}$} &\multicolumn{1}{c}{-0.007} & \multicolumn{1}{c}{-0.008} & \multicolumn{1}{c|}{-0.007} & \multicolumn{1}{c}{-0.025$^{*}$} & \multicolumn{1}{c}{-0.029$^{**}$} & \multicolumn{1}{c}{-0.022$^{*}$} \\

    Age squared & \multicolumn{1}{c}{0.6 e-4$^{***}$} & \multicolumn{1}{c}{0.6 e-4$^{**}$} & \multicolumn{1}{c|}{0.9 e-4$^{***}$} &  \multicolumn{1}{c}{0.4 e-4} & \multicolumn{1}{c}{0.3 e-4} & \multicolumn{1}{c|}{0.2 e-4} & \multicolumn{1}{c}{0.26 e-4} & \multicolumn{1}{c}{0.29 e-4$^{*}$} & \multicolumn{1}{c}{0.19 e-3} \\

    Education & \multicolumn{1}{c}{0.015$^{***}$} & \multicolumn{1}{c}{0.016$^{***}$} & \multicolumn{1}{c|}{0.007} & \multicolumn{1}{c}{0.026$^{*}$} & \multicolumn{1}{c}{0.026$^{*}$} & \multicolumn{1}{c|}{0.030$^{*}$} & \multicolumn{1}{c}{0.024} & \multicolumn{1}{c}{0.034} & \multicolumn{1}{c}{0.069$^{**}$} \\

    Break & \multicolumn{1}{c}{-0.133$^{***}$} & \multicolumn{1}{c}{-0.165$^{***}$} & \multicolumn{1}{c|}{-0.260$^{***}$} & \multicolumn{1}{c}{-0.040} & \multicolumn{1}{c}{-0.069$^{**}$} & \multicolumn{1}{c|}{-0.065$^{**}$} & \multicolumn{1}{c}{-0.029} & \multicolumn{1}{c}{-0.077} & \multicolumn{1}{c}{-0.065} \\

    Full-time & \multicolumn{1}{c}{0.034$^{***}$} & \multicolumn{1}{c}{0.017} & \multicolumn{1}{c|}{-0.001} & \multicolumn{1}{c}{-0.019} & \multicolumn{1}{c}{-0.055} & \multicolumn{1}{c|}{-0.101$^{***}$} & \multicolumn{1}{c}{0.051} & \multicolumn{1}{c}{0.043} & \multicolumn{1}{c}{-0.065} \\

    \hline
    \emph{N Obs} & \multicolumn{1}{c}{39,047} & \multicolumn{1}{c}{39,047} & \multicolumn{1}{c|}{38,841} &\multicolumn{1}{c}{4,962} & \multicolumn{1}{c}{4,935} & \multicolumn{1}{c|}{4,656} & \multicolumn{1}{c}{1,792} & \multicolumn{1}{c}{1,782} & \multicolumn{1}{c}{1,576} \\
    \emph{R$^2$}    & \multicolumn{1}{c}{0.010} & \multicolumn{1}{c}{0.119} & \multicolumn{1}{c|}{0.184} & \multicolumn{1}{c}{0.040} & \multicolumn{1}{c}{0.067} & \multicolumn{1}{c|}{0.134}  & \multicolumn{1}{c}{0.064} & \multicolumn{1}{c}{0.081} & \multicolumn{1}{c}{0.175} \\
    \hline
    \multicolumn{10}{l}{{\footnotesize{Levels of significance: $^{*} p<0.1$, $^{**} p<0.05$, $^{***} p<0.01$}}}
    \end{tabular}
    }
\label{t:Probit1}
\end{table}%

Our estimates also show that the effect of retrospective coding is much more moderate when we condition the sample on workers who changed employers and when we consider the $ENE$ sample.\footnote{We do not include the within-employer occupational mobility in Table \ref{t:Probit1} because, as suggested by the graphical analysis, the results are very similar to the ones obtained with the full sample.}

\subsection{A comparison of coding errors across the PSID and SIPP}

The point estimates obtained in Table \ref{t:Probit1} suggest that the probability of an occupational change for those who changed employers and for those who changed employer through non-employment spells, should be lowered on average by 3 percentage points at a one-digit level, 8 percentage points at a two-digit level and 7 percentage points at a three-digit level to capture the effect of coding error. To compare these estimates to the ones obtained from the SIPP, first note that retrospective coding is done using the same descriptions of the ``kind of work'' individuals gave in past interviews and hence captures coding errors introduced at the coding stage (see Sullivan, 2010). In addition to this coder error, our $\mathbf{\Gamma}$-correction method also takes into account that the source of code disagreement can originate in different descriptions of the same work (respondent error). Hence we would expect a higher correction when using the $\mathbf{\Gamma}$-correction than when using retrospective coding. Taking this feature into account and noting that coder error is expected to be the most important source error (see Mathiowetz, 1992), the PSID estimates compare very well with the adjustments implied by the $\mathbf{\Gamma}$-correction method.

In particular, when aggregating occupations into major categories (2000 SOC or 1990 SOC) and using the $\mathbf{\Gamma}$-correction, the corrected average occupational mobility rate for the non-employed was approximately 11 percentage points lower than the one obtained using the raw SIPP data. Noting that the occupational aggregations used in the SIPP and the two-digit aggregation used in the PSID lead to very similar $ENE$ occupational mobility rates, the 11 percentage point adjustment obtained from the SIPP is thus close to the 8 percentage points suggested by Table \ref{t:Probit1}.\footnote{For the 1985-1995 period, during which the PSID and SIPP overlap, the average ``year-to-year'' occupational mobility rate of the non-employed in the PSID and the SIPP was both around 53.1\% and 47.6\% when using the major occupational categories of the 1990 SOC.} The difference between the adjustments obtained from the SIPP and the PSID (3 percentage points) then provides a rough estimate of the impact of the respondent error. This estimate then implies that the importance of coder error is about 2.6 times larger than the importance of the respondent error. This is remarkably consistent with Mathiowetz (1992), who shows that the importance of coder error is two times larger than the importance of the respondent error when aggregating codes at a one-digit level and five times larger than the importance of the respondent error when aggregating codes at a three-digit level.

Furthermore, both the SIPP and the PSID data sets strongly suggest that the percentage reduction of occupational mobility due to coding error varies substantially between employer stayers and employer movers. In the case of employer stayers, a large percentage of transitions are implied to be spurious. In the PSID we find that at a two-digit level 45\% of yearly transitions are spurious, while in the SIPP we find that 40\% of the yearly transitions of employer stayers are spurious. In the case of employer movers, occupational mobility is reduced by about 10 percentage points, but high occupational mobility remains (around 40\% comparing before and after an employer changers), after applying retrospective coding or after using the $\mathbf{\Gamma}$-correction. As discussed earlier, this difference arises as among employer stayers the proportion of true occupational stayers is high and coding errors translate into a large amount of spurious mobility. Among those who changed employers through non-employment there is a much smaller proportion of true occupational stayers and hence the ``population at risk'' to be assigned a spurious occupational change is smaller.\footnote{vom Lehm et al. (2021) analyses the impact of coding errors on occupational mobility in the CPS when pooling together movers and stayers and find similar results to the ones we derive in the SIPP for the same pooled sample.}

\section{Discussion of assumptions $A1$ and $A2$}

We now turn to discuss the two assumptions that appear the most restrictive in our analysis. As mentioned earlier, assumption $A3$ is verified in our data.

\subsection{Assumption $A1$}

This assumption requires that the realization of an occupational code does not depend on workers' labor market histories, demographic characteristics or the time it occurred in our sample. Therefore it implies that errors in the individuals' verbatim responses are fully captured by the nature of their job and hence only depend on their \emph{true} occupation. It also implies that $\mathbf{\Gamma}$ is time-invariant. Since we use these implications extensively in the implementation of our correction method, we now investigate them further to help us evaluate the strength of $A1$. Our main conclusion is that our $\mathbf{\Gamma}$ correction method appears to pick up heterogeneity in miscoding across occupations that is robust to estimations on subsamples of the population. Likewise, we find evidence that even though the $\mathbf{\Gamma}$ matrix was estimated using 1985-1986 data, it captures well miscoding observed in recent years.

\subsubsection{Worker heterogeneity}

We investigate two aspects of worker heterogeneity that could deliver very different estimates of $\mathbf{\Gamma}$. We first consider whether the accuracy of the answer to the occupation question is affected by differences in workers' education attainment. A concern would be that more educated workers can better explain the type of job they are performing and hence coding errors would be less severe among these workers than in those with lower education. We then consider whether the accuracy of the occupation information is affected by a worker's interview status. In the SIPP either the worker reports his/her occupation him/herself to the interviewer (self-report) or another person of the same household reports his/her occupation (proxy). One would be concern that proxies answers are more prone to coding error.

Our analysis relies on re-estimating $\mathbf{\Gamma}$ using the 1985-1986 sample but on subsamples based on different educational attainment categories and interview status. As discussed above, the estimation process recovers a transition matrix of purely spurious mobility, $\mathbf{\hat{T}^{I}_{s}}$, (as shown in Table 3 for the 1990 SOC). We can then compare the estimated matrix of each subsample with the one computed in our main analysis. This comparison allows us to gauge whether the aforementioned characteristics lead to different miscoding even when conditioning on occupation.

\paragraph{Education differences}

To study the effect of education differences, we divide the 1985-1986 sample into two groups: (i) high skilled workers which captures all those individuals with at least some college and (ii) low skilled workers which capture all those individuals with at most a high school degree. About one-third of the sample covers high skilled group. Note that occupations and education are highly correlated in the data. For example, very few non-college-educated workers can be found in architecture and engineering occupations, while very few college-educated workers can be found in production and some service occupations. Hence, part of the impact of education would already be captured by conditioning on occupations. However, the aim of this exercise is to evaluate the impact of the within-occupation variation in education on the level of miscoding.

\begin{figure}
    \centering%
\resizebox{0.90\textwidth}{!}{
      \subfloat[High skilled - All cells]{
          \includegraphics[width=\textwidth]{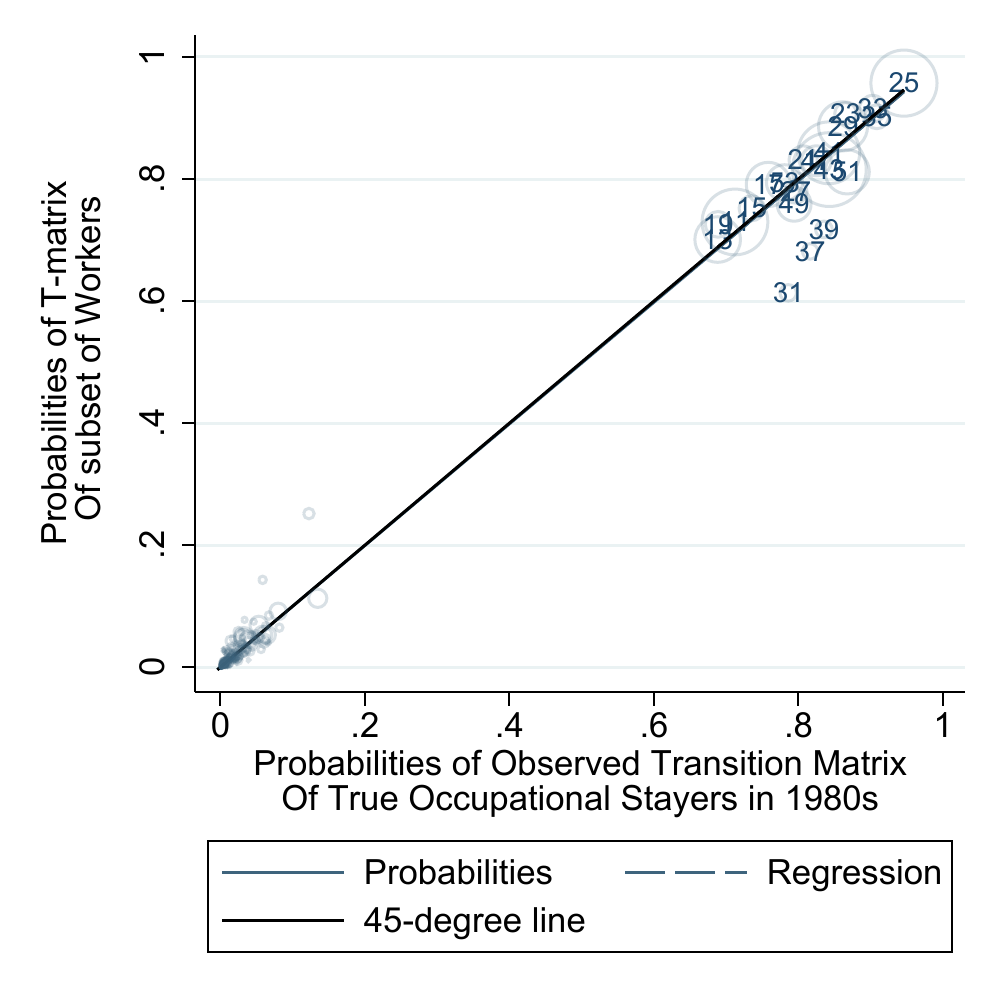}}
      \subfloat[High skilled - Diagonal]{
       \includegraphics[width=\textwidth]{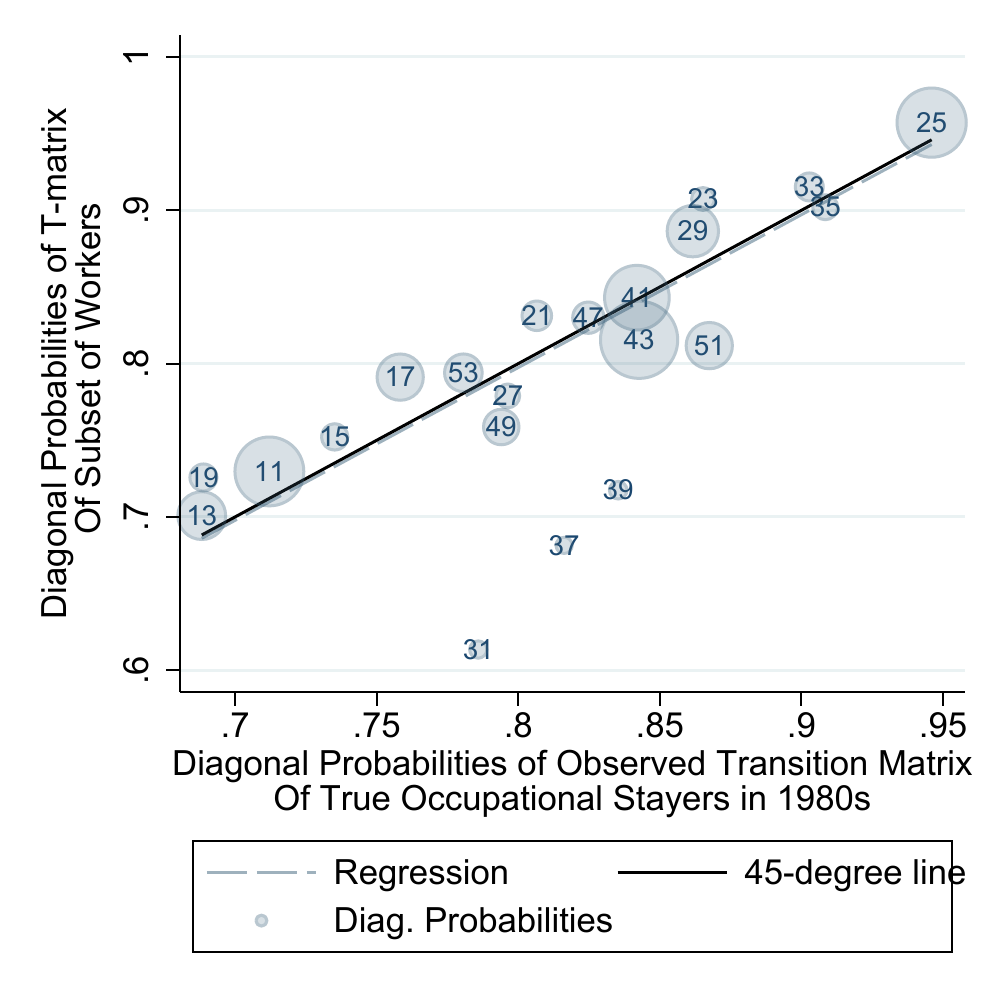}}
          }

\resizebox{0.90\textwidth}{!}{
    \subfloat[Low skilled - All cells]{
          \includegraphics[width=\textwidth]{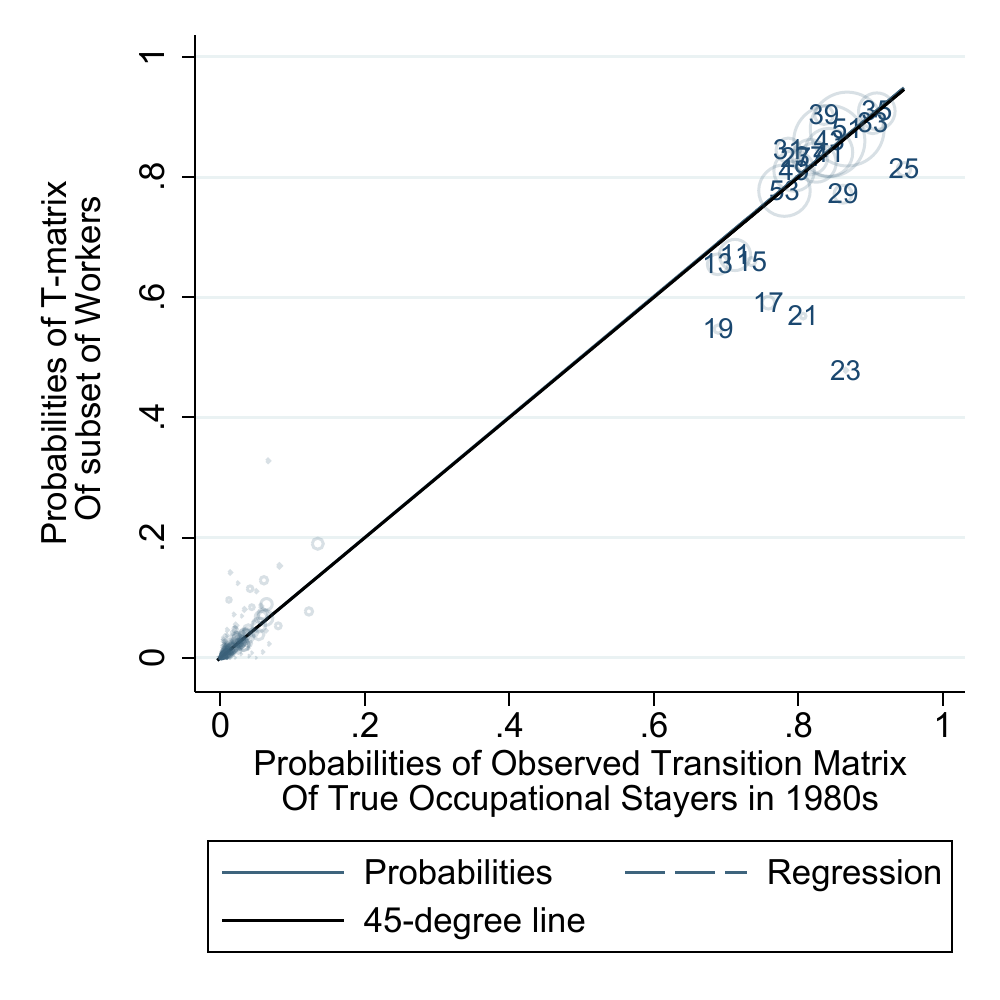}}
      \subfloat[Low skilled - Diagonal]{
          \includegraphics[width=\textwidth]{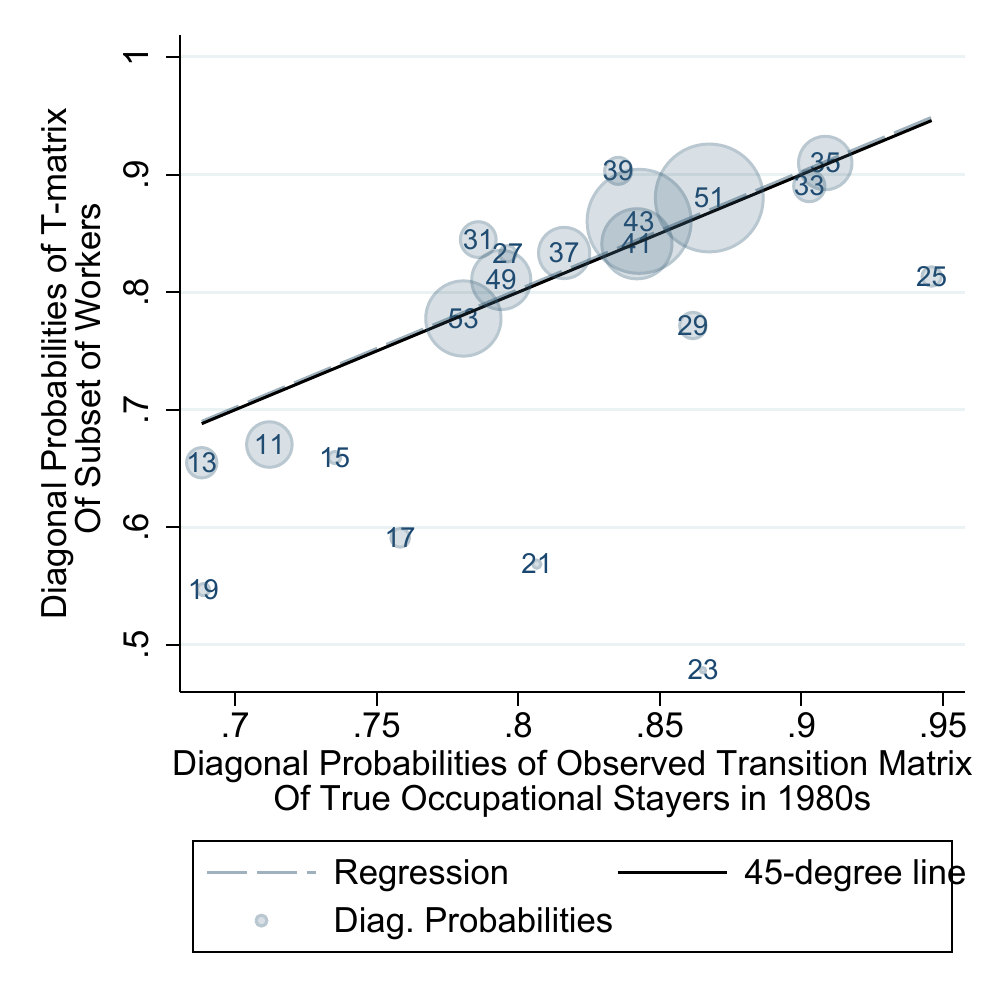}}
          }

\resizebox{0.90\textwidth}{!}{
      \subfloat[Dominant education - All cells]{
          \includegraphics[width=\textwidth]{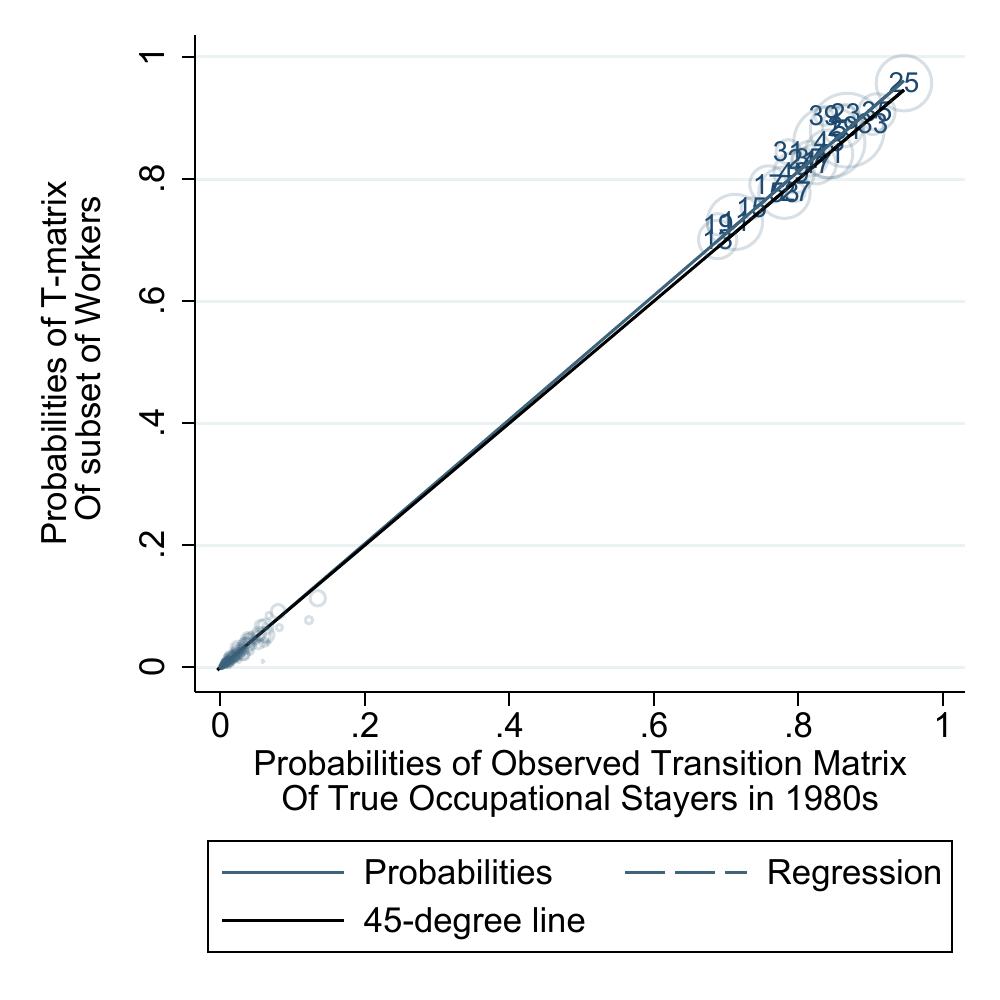}}
      \subfloat[Dominant education - Diagonal]{
          \includegraphics[width=\textwidth]{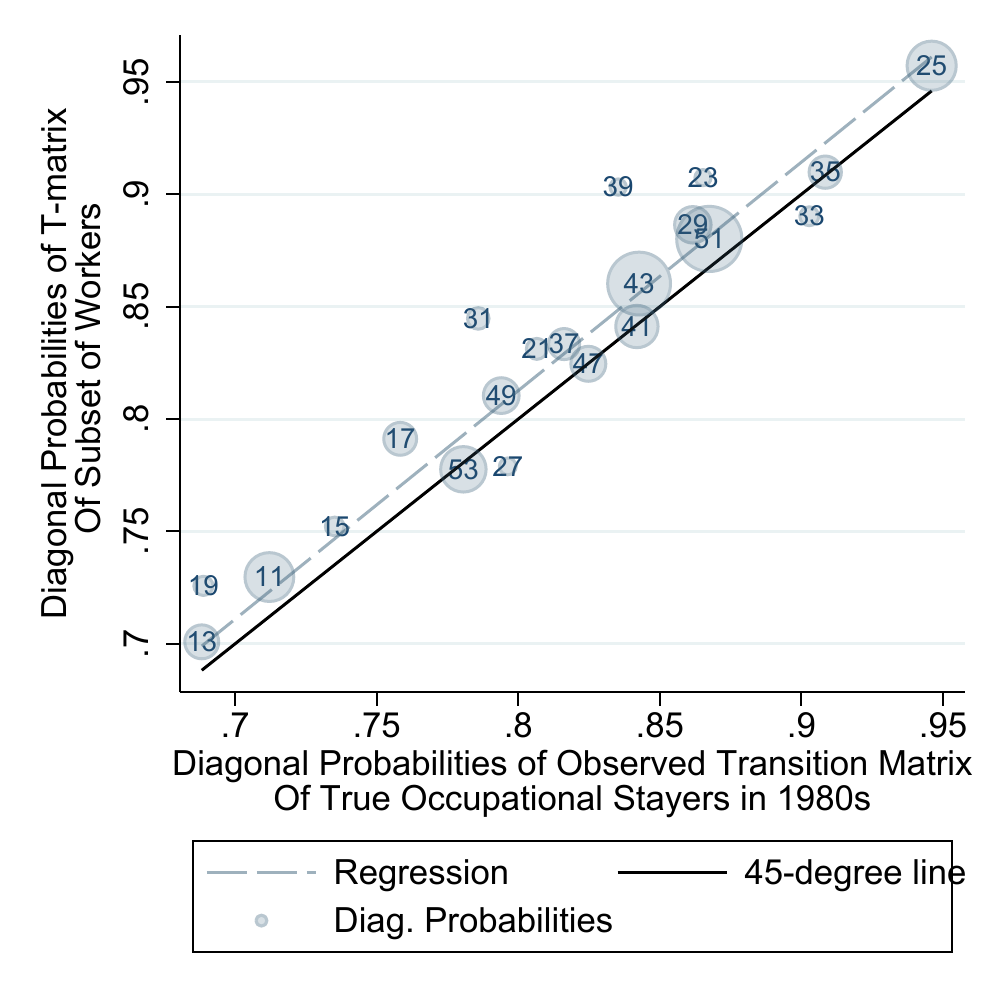}}
         }
          \caption{Comparing the estimated spurious transition matrices by education}\label{f:education}
\end{figure}

Each panels (a)-(d) of Figure \ref{f:education} depicts a scatter plot in which on the $y$-axis are the elements of the transition matrix $\mathbf{\hat{T}^{I}_{s}}$ estimated for either the high skilled subsample (panels (a)-(b)) or the low skilled subsample (panels (c)-(d)) and on the $x$-axis are the elements of $\mathbf{\hat{T}^{I}_{s}}$ estimated in our main analysis. We use the major occupations categories of the 2000 SOC and depict the same numbering of occupations as in Table \ref{t:garblinggamma1} for the diagonal elements. Panels (a) and (c) show all the elements of the corresponding matrices, while panels (b) and (d) focus only on the diagonal elements. It is in the latter where the $\mathbf{\Gamma}$ matrix captures the heterogeneity in the probability of being coded correctly. Although the correction method formally involves applying the inverse of $\mathbf{\Gamma}$, there is an intuitive relation between the diagonal elements and the probability of being miscoded. The higher a diagonal element, the lower miscoding tends to be. Further note that as most of the mass in these transitions matrices lies on the diagonal elements they will be observed on the top right of the graph, while the off-diagonal elements will be closer to the origin. For the former we also show circles around them representing the relative size of the occupations.

If educational differences across workers created very different coding errors, we would observe large deviations from the 45-degree line. The latter would suggest important differences between the coding errors obtained in our main analysis and the ones obtained when taking into account differences in workers' education. Instead, panels (a)-(d) shows that this is not the case. The correlation of coding errors is very close to one. Using OLS to fit a regression line among the observations, we observe that the regression line lies nearly on top of the 45-degree line.

Panels (e)-(f) of Figure \ref{f:education} present a different approach to evaluate the effect of education differences on coding errors. In this case we only consider, for each occupation, the observations of those workers with the most dominant education (more than 50\%) in that occupation. That is, we counterfactually impose the coding error attributed to the dominant education group on all workers in such an occupation. As before, if educational differences have meaningful effects on coding errors, controlling for occupations, then we would observe the regression line diverge significantly from the 45-degree line. Instead, we once again observe that these lie very close to each other, with only a slight deviation due to the diagonal elements (see also Table \ref{tab:selfproxycorr} below).

\begin{figure}[!ht]
    \centering
    \resizebox{0.90\textwidth}{!}{
    \subfloat[Self-interview - All cells]{
          \includegraphics[width=\textwidth]{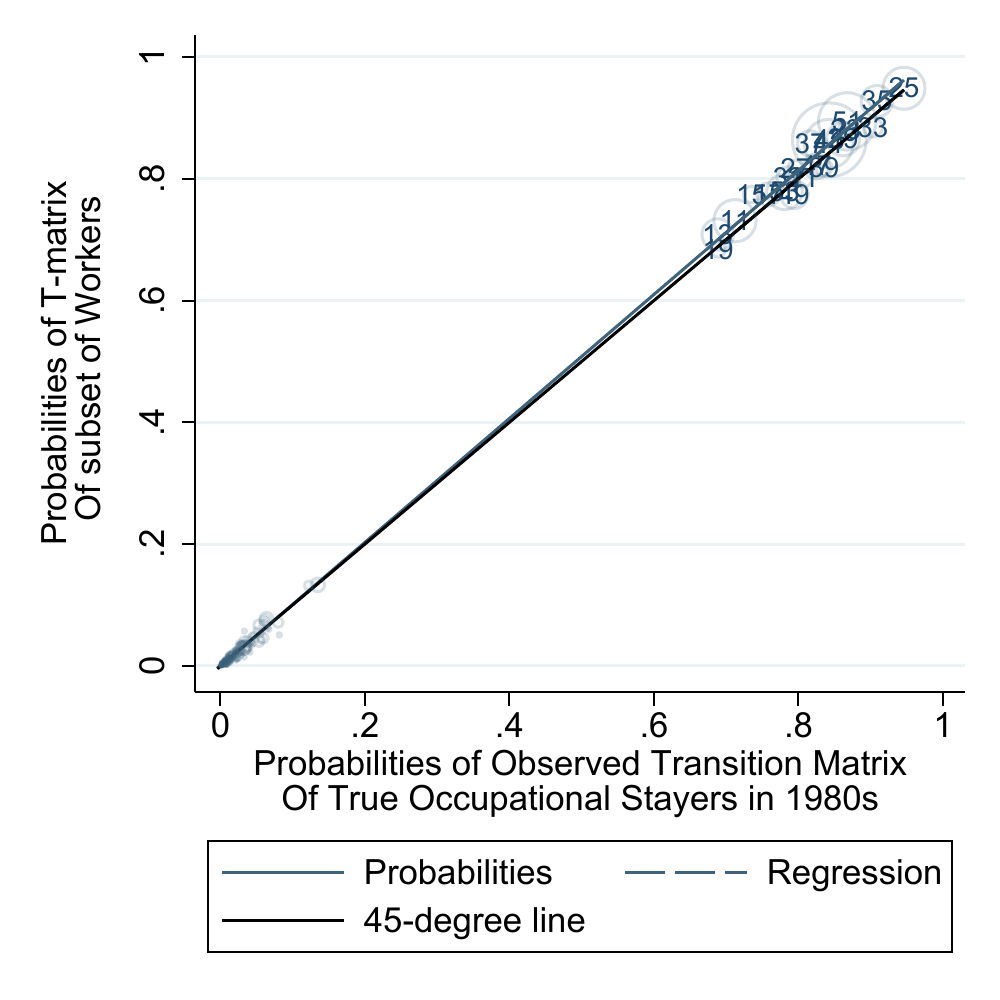}}
    \subfloat[Self-interview - Diagonal]{
          \includegraphics[width=\textwidth]{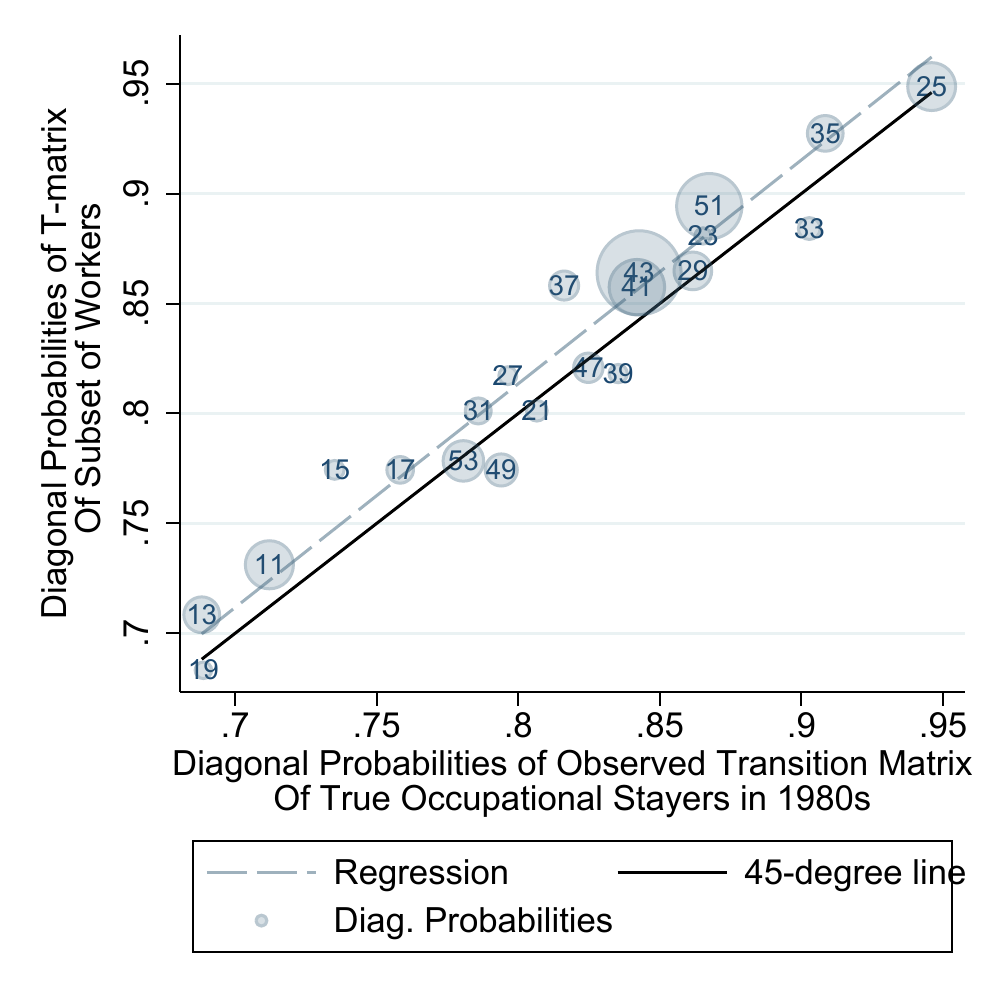}}
      }

\resizebox{0.90\textwidth}{!}{
      \subfloat[Proxy-interviewed - All cells]{
          \includegraphics[width=\textwidth]{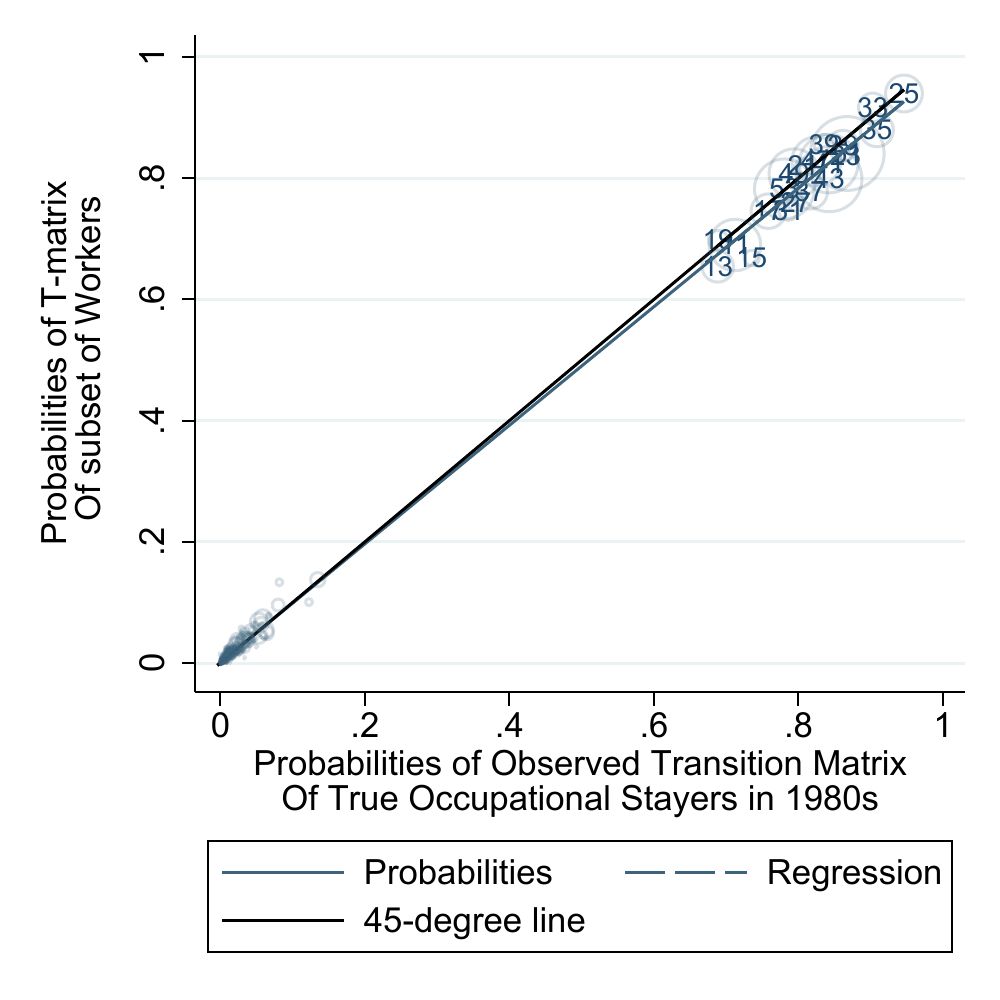}}
      \subfloat[Proxy-interview - Diagonal]{
          \includegraphics[width=\textwidth]{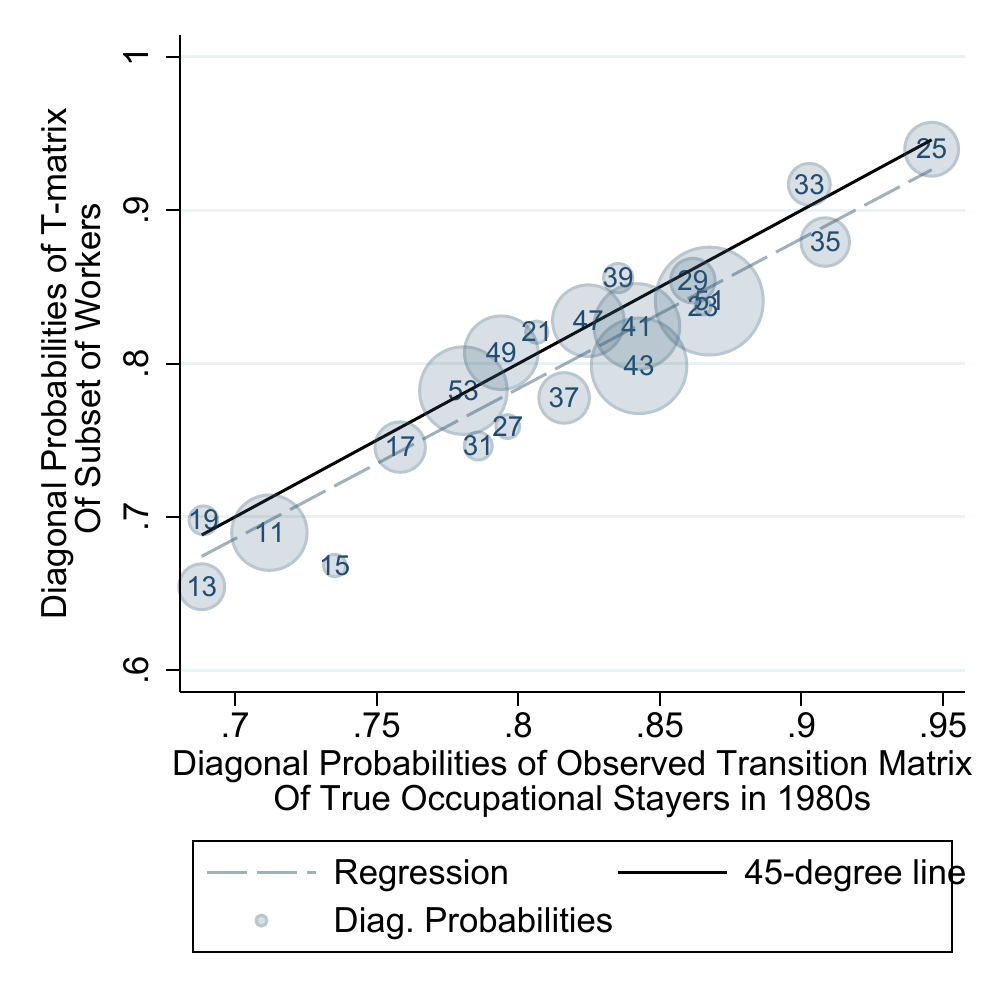}}
          }
          \caption{Comparing the estimated spurious transition matrices by interview status}\label{f:intvwstatus}
\end{figure}

\paragraph{Interview status differences}

To investigate the impact of differences in the interview status of a worker on miscoding, we divide the 1985-1986 sample into those who were interviewed in person (self-interviewed) in two consecutive waves and those who had their information given by a proxy (proxy interview) at least in one of the waves. This divides the sample roughly in half: 55\% were interviewed in person and 45\% involve a proxy interview.

Figure \ref{f:intvwstatus} presents the same scatter plot exercises as describe above, but this time using the interview status instead of education attainment. We can observe a very high correlation between the elements of the transition matrix $\mathbf{\hat{T}^{I}_{s}}$ for each of these subsamples and of the transition matrix $\mathbf{\hat{T}^{I}_{s}}$ estimated in our main analysis. In this case the regression line is also nearly on top of the 45-degree line, with a very slight deviation at the diagonal elements. This deviation implies that for those interviewed in person the slope is a little steeper than one, while it is a little lower than one for those interviewed by proxy. This captures that the self-interviewed appear slightly more accurate, with slightly less spurious mobility, than the proxy interviewed. Nevertheless, in both cases we obtained a very similar conclusion. Occupations like managers, business and financial operators, computer and mathematical occupations and physical scientists are  more prone to miscoding; while occupations like education, training and library occupations, food services and protective services are less prone to miscoding.

One could further subdivide the analysis using the interaction between education and interview status categories to gain a further insight. However, the above analysis suggests that we would find once again that subdividing the sample into these categories would not meaningfully change our estimated $\mathbf{\hat{T}^{I}_{s}}$ and hence $\mathbf{\Gamma}$ matrix.

\paragraph{Implications for measured occupational mobility}

Table \ref{tab:selfproxycorr} summarises the above results, showing that the overall level of occupational staying among true stayers estimated across all subsamples is very similar to the one estimated in our main analysis. The correlations when using all elements of the matrices are nearly one. As mentioned above these correlations drop but still remain very high when only considering the diagonal elements. That is, the $\mathbf{\Gamma}$ matrix capture miscoding differences across occupation that is present (i) whether the individual is the one being interviewed or a proxy provides his/her information and (ii) across education categories.\footnote{Interestingly, high skilled workers are a bit more likely to be miscoded than low skilled workers. This could arise as the occupation typically performed by the high skilled are more specialized than those performed by the low skilled, making miscoding more likely in the former. Indeed the increased miscoding reflects mostly the occupations that typically performed by the high skilled (11 to 29  in the 2000 SOC). These are associated with a 78.1\% of occupational stayers in the estimated $\mathbf{\hat{T}^{I}_{s}}$, while the proportion of occupational staying in the occupations typically performed by low skilled workers (31 to 53  in the 2000 SOC) is 83.7\%. If we were only to consider the high skilled in all occupations we find 79.9\% of occupational staying, while if we only consider the low skilled in all occupations we obtain 82.8\% of occupational staying.}

\begin{table}[ht!]
  \centering
  \caption{Observed Occupation Staying of True Stayers, Correlations across $\mathbf{\hat{T}^{I}_{s}}$ Estimates. }\label{tab:selfproxycorr}%
  {\small
    \begin{tabular}{lccrlrrr}
    \hline
              & level & \multicolumn{2}{c}{corr. w/ baseline} &       & \multicolumn{1}{c}{level} & \multicolumn{2}{c}{corr. w/ baseline}   \\
          & \multicolumn{1}{l}{occ stay} & all   & \multicolumn{1}{c}{diag} &       & \multicolumn{1}{c}{occ. stay} & \multicolumn{1}{c}{all} & \multicolumn{1}{c}{diag} \\
          \hline
    Baseline & 0.822 & 1.000 & 1.000     & \multicolumn{4}{c}{\textbf{Education}} \\ \hline
    \multicolumn{4}{c}{\textbf{By interview status}} & Dominant educ. & 0.832 & 1.000 & 0.982 \\ 
    Self-interviewed  & 0.838 & 1.000 & 0.979 & High skilled & 0.799 & 0.998 & 0.900 \\
    Proxy interview & 0.799 & 0.999 &  0.953 & Low skilled & 0.828 & 0.999 & 0.867 \\
    \hline
    \end{tabular}%
    }
\end{table}

\begin{figure}[!ht]
    \centering
    \resizebox{1.0\textwidth}{!}{
    \subfloat[Educational attainment]{
          \includegraphics[width=\textwidth]{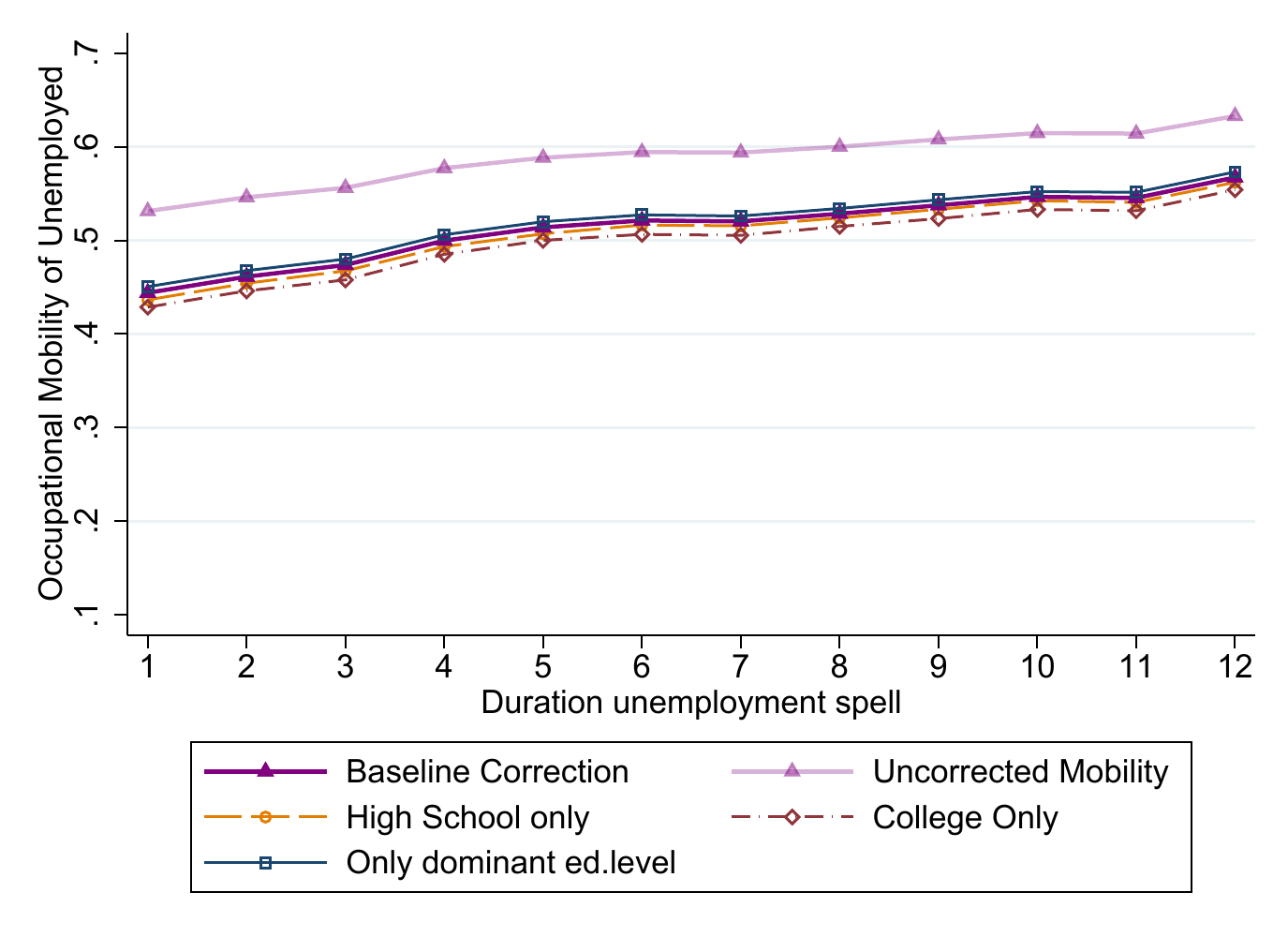}
          \label{fig:mobdur_misceduc}}
    \subfloat[Interview status]{
          \includegraphics[width=\textwidth]{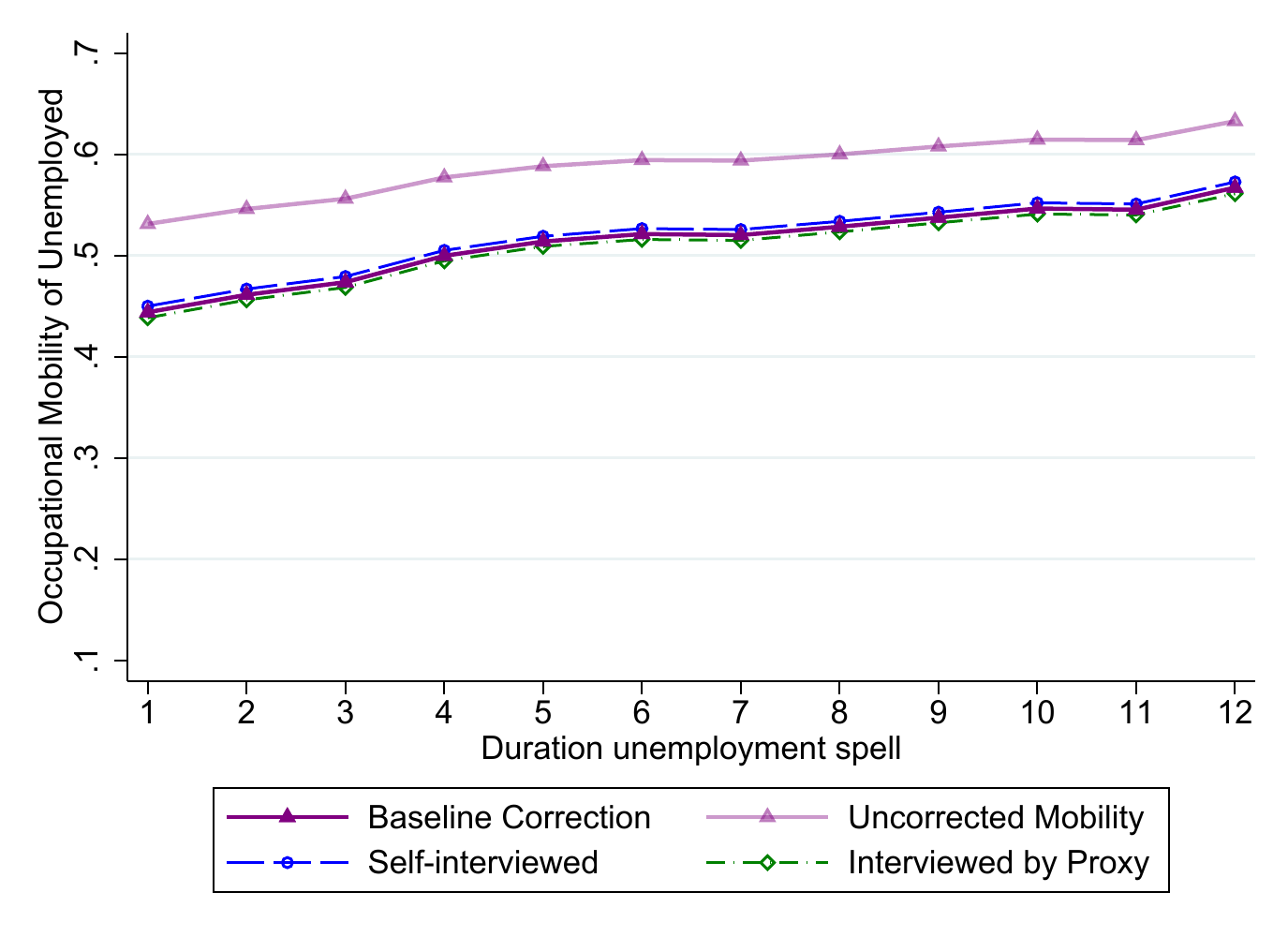}
          \label{fig:mobdur_selfproxy}}
        }
        \caption{Corrected Mobility-Duration Profile from $\mathbf{\Gamma}$ estimated on subsamples}
\end{figure}

Next we apply the implied $\mathbf{\Gamma}$ obtained for each of the above subgroup of workers to correct the mobility-duration profile documented in Section 2.2 of the main text. Figure \ref{fig:mobdur_misceduc} considers the mobility-duration profiles when using the $\mathbf{\Gamma}$ matrices obtained from high skilled and low skilled workers. Figure \ref{fig:mobdur_selfproxy} considers the same mobility-duration profile but this time using the $\mathbf{\Gamma}$ matrices obtained from the self and proxy interviewed. In both graphs we depict the uncorrected mobility duration profile (without any smoothing), and the one corrected with our baseline $\mathbf{\Gamma}$.

We can observe that when using the $\mathbf{\Gamma}$ matrices from the education attainment subsamples the implied mobility-duration profiles are very similar to the obtained in our main analysis. When we use high skilled sample to measure miscoding in all occupations, including those that have very few college workers in them, the associated profile is only by 1-1.5 percentage point different from the baseline one. We also obtain a very similar conclusion when using the low skilled sample across all occupations. The mobility-duration profile corrected by using only information on those with the dominant education in a given occupation also track the original mobility duration profile closely. Moreover, when using the $\mathbf{\Gamma}$ matrices implied by the self-interviewed or proxy subsamples, we once again obtain mobility-duration profiles that hardly differ from the one obtained in our main analysis. This evidence thus suggests that differences in education attainment or interview status (or their interaction) do not affect the conclusion obtained from our main correction analysis: occupational mobility is high, a little over 40\%, and increases moderately with unemployment duration.

\subsubsection{Time invariance of $\mathbf{\Gamma}$}

As previously discussed we estimate the $\mathbf{\Gamma}$ correction matrix using the 1985-1986 SIPP panels and then apply it to occupational mobility data up to 2014. An important concern that arises from this application is whether the estimated coding errors remain relevant in the later years of our sample as implied by assumption $A1$. To directly evaluate whether this is the case we would need to re-apply our correction method at a later date and compare the more recent coding error correction matrix to our baseline one. To perform this exercise we will require a US data set that switched from dependent to independent interviewing with respect to occupations in the 2000s. However, we are not aware of any data set in which this re-design took place. Due to this barrier we instead take a different approach. We consider a group of workers who are likely occupation stayers, but coded independently such that some of them would be observed as occupational movers. We then evaluate the extent to which our $\mathbf{\Gamma}$ matrix can predict these workers' observed occupational mobility, particularly in more recent years. A high predictive power would suggest that coding errors estimated using data from 1985-1986 remain relevant throughout our sample.

Motivated by Fujita and Moscarini (2017) we use temporary laid-off workers to approximate this group of occupational stayers. These authors' empirical work suggests that workers in temporary layoff have a very low chance of an occupation switch once recalled by their previous employers. Therefore it is not unreasonable to assume that this set of workers are \textit{largely} made up of true occupation stayers and can provide a good approximation to the latter group. Instead of using the SIPP to measure temporary layoffs (as Fujita and Moscarini, 2017), however, we use the Current Population Survey (CPS). The main reason for this choice is that dependent interviewing in the CPS only applies when a person is employed both in the current month and the month before (see e.g. the CPS interviewing manual 2015). This implies that workers who are \textit{unemployed on temporary layoff} will have their occupations coded independently. In contrast, as interviews in the SIPP are conducted every four months, one cannot guarantee that temporary layoffs with spells of unemployment of at most 4 months will have their occupations independently coded at re-employment.

\begin{table}[ht!]
  \centering
    \begin{tabular}{lrccccc}
          \hline \hline
          &       \multicolumn{6}{c}{(Cell-by-Cell) Correlation across Transition Matrices} \\ \hline
          &       & \multicolumn{2}{c}{All, <13 weeks} &       & \multicolumn{2}{c}{All 3-dgt Industry Stayers} \\
          \hline 
CPS years          &       & All cells & Diagonal &       & All cells & Diagonal \\
          \hline
    1994-2021 &       & 0.991 & 0.844 &       & 0.993 & 0.805 \\
    1994-2004 &       & 0.986 & 0.816 &       & 0.990 & 0.823 \\
    2004-2014 &       & 0.989 & 0.790 &       & 0.992 & 0.742 \\
    2014-2021 &       & 0.989 & 0.812 &       & 0.990 & 0.710 \\ \hline \hline
    \end{tabular}%
  \caption{Correlations Observed Transition Matrix of Temporary Layoffs with Spurious Transitions (from $\mathbf{\Gamma}$)}\label{tab:corr_templayoffs}%
\end{table}%

Therefore, if the vast majority of temporary layoffs are independently-coded true occupation stayers \textit{and} the $\mathbf{\Gamma}$ coding errors persist over time, we would observe a positive correlation between the elements of the transition matrix $\mathbf{\hat{T}^{I}_{s}}$ estimated in our main analysis and the elements of the observed transition matrix of those workers returning to work out of a temporary layoff, even if we consider temporary layoffs three decades later. Table \ref{tab:corr_templayoffs} presents the results of such an exercise using several time periods post the CPS 1994-redesign. The first two columns refer to all those workers who were in temporary layoff for less than 13 weeks before re-employment; while the second two columns refers to the subset of temporary layoffs with less than 13 weeks in unemployment what were also observed as industry stayers at re-employment when considering a 3-digit industry aggregation. Although the latter group reduces the sample size, it is more likely to contain true occupational stayers. This occurs as occupation and industry mobility tends to go hand in hand. We return to this point below.\footnote{The post-1994 sample size is about 45,000 for all temporary layoffs with duration less than 13 weeks and about 30,000 for temporary layoffs who are industry stayers. The decadal samples (1994-2004, 2004-2014, 2014-2021) are about 1/3 of these numbers.}

Across both samples of temporary layoffs Table \ref{tab:corr_templayoffs} shows the correlation between their observed occupational transition matrix and $\mathbf{\hat{T}^{I}_{s}}$, obtained from the 1985-1986 SIPP data, when using the 2000 SOC. One can immediately see the very high correlations in the observed occupational mobility patterns among those in temporary layoff and the one implied by the $\mathbf{\Gamma}$ matrix. Moreover, the value of the correlation is nearly unchanged over time even when focusing on the years 2014-2021, the period after our SIPP analysis ended. In particular, the correlations are nearly one when taking all elements of the matrices (or, very similar but not shown here, all cells with positive probabilities). Note, however, that some degree of positive correlation may not be unexpected for this exercise, as the diagonal of both matrices will unsurprisingly consists of numbers closer to one, while some other cells will naturally be closer or equal to zero. Therefore, a more stringent test is to consider the correlation only of the diagonal elements. Here we once again observe high correlations ($\rho >0.7$ across all periods). Indeed coding errors according to the $\mathbf{\Gamma}$ matrix explain about two-thirds of the variance of heterogeneity on the diagonal of occupational transition of temporary layoffs in the second half of the CPS sample ($\rho=0.80$).

\begin{figure}[!ht]
    \centering
        \resizebox{0.65\textwidth}{!}{
    \subfloat[Temp Layoffs, Occ. Mobility, 1994+]{
          \includegraphics[width=\textwidth]{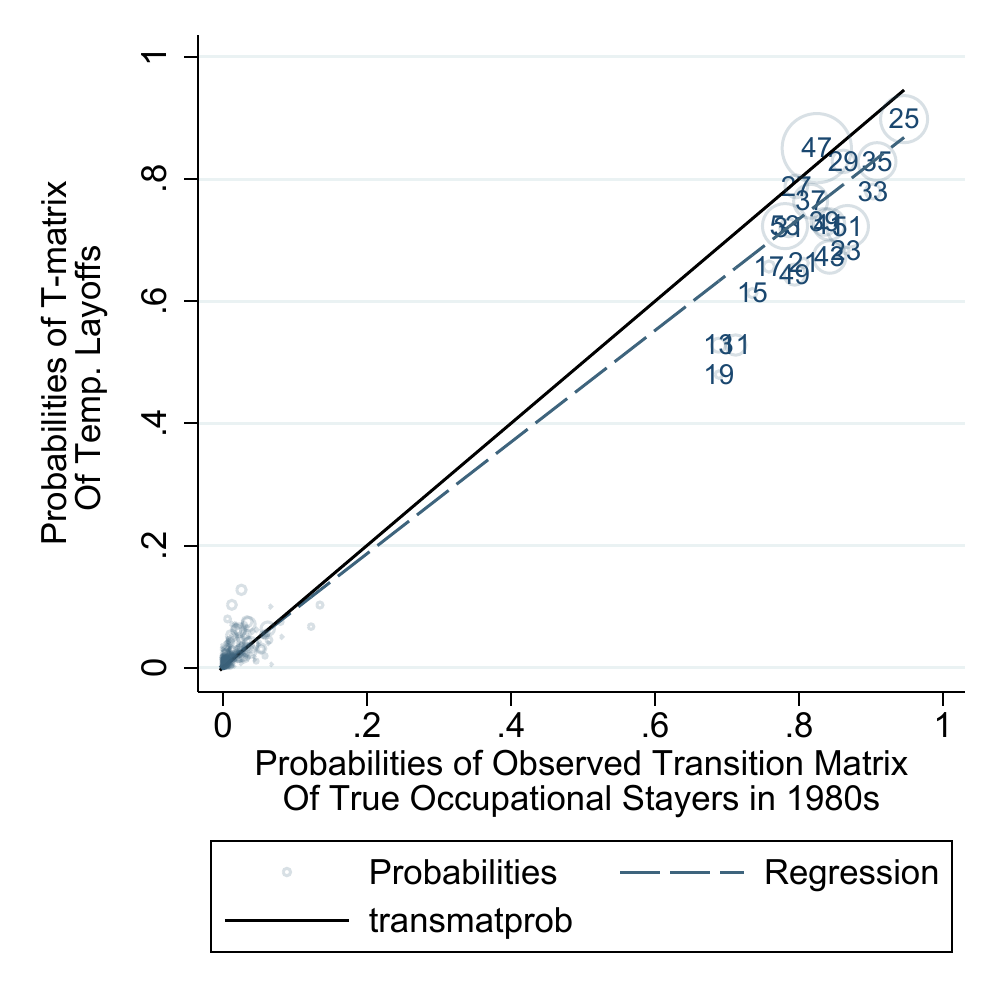}}
      \subfloat[Temp Layoffs, Occ. Mobility, 2014-21]{
            \includegraphics[width=\textwidth]{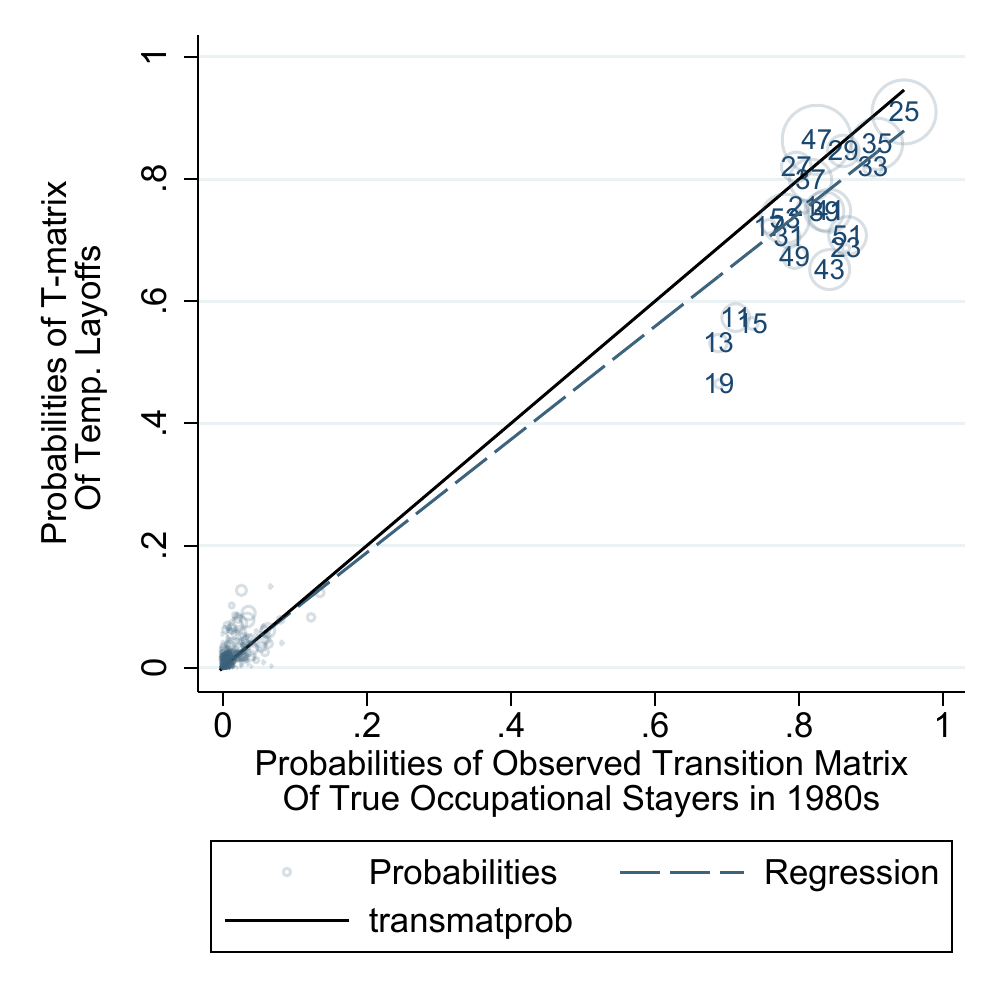}}
            }

    \resizebox{0.65\textwidth}{!}{
    \subfloat[Temp. Layoffs, Ind. Stayers, 1994+ ]{
          \includegraphics[width=\textwidth]{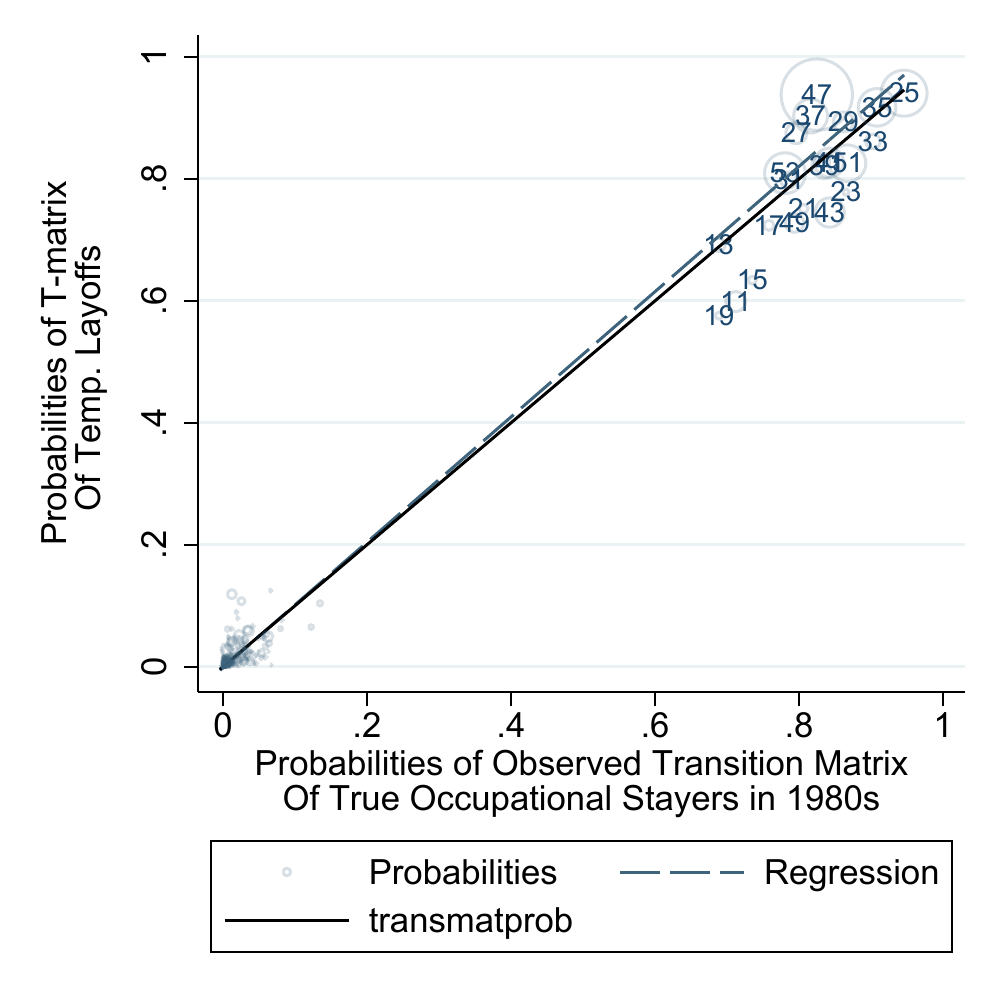}}
    \subfloat[Temp. Layoffs, Ind. Stayers, 2014-21]{
          \includegraphics[width=\textwidth]{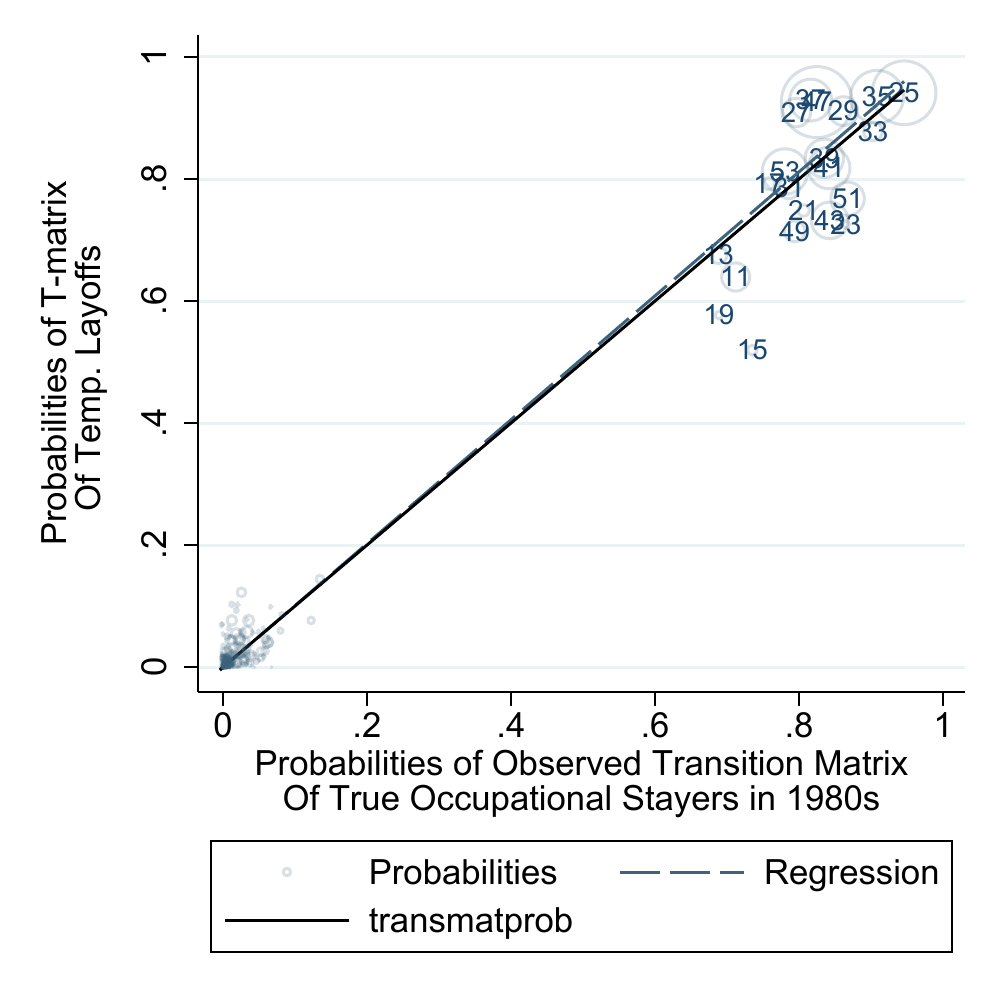}}
      }
      \caption{Comparing temporary layoffs in the CPS with the spurious transition matrix from the SIPP}\label{f:tlayoffs}
\end{figure}

Figure \ref{f:tlayoffs} displays the same exercise as in the previous section, where we now depict the values of each element of the occupation transition matrix of temporary layoffs on the $y$-axis relative to the associate value of the transition matrix implied by $\mathbf{\Gamma}$. The top row considers all temporary layoffs (unemployed for less than 13 weeks), while the bottom row considers the subgroup of temporary layoffs that are also industry stayers. We observe that the same occupations most prone to miscoding according to $\mathbf{\Gamma}$ are also the ones observed among temporary layoffs: managers, business and financials operations, computer and math occupations, and physical sciences. Conversely, many of the occupations that are least likely to be miscoded according to $\mathbf{\Gamma}$ are also the ones among temporary layoffs: healthcare practitioners, protective services, food preparation services and above all educators. Figures \ref{f:tlayoffs}.b and Figure \ref{f:tlayoffs}.d consider temporary layoffs for the more recent period 2014-2021 and show that these patterns are largely maintained over time (if a bit more noisy).

\paragraph{Some caveats}

Even if coding mistakes were completely persistent, the comparison between the observed transition matrix of temporary layoffs and our correction matrix may be affected by other factors not considered in the previous section.

First, and perhaps most importantly, the comparison of the occupational transitions behind $\mathbf{\Gamma}$ with those of temporary layoffs relies on the assumption that the latter set of workers captures both the whole set of occupation stayers (and its miscoding), but simultaneously no other type of worker, i.e. true movers. Deviations from this requirement will likely lower the observed correlation. One way to investigate the presence of true movers among temporary layoffs is to consider their amount of net mobility -- as miscoding inflates excess mobility rather than net mobility. When considering all temporary layoffs, the average net occupational mobility rate (as defined in Section 2.3 of the paper) is significantly lower for these workers than for all the unemployed, only about 1/3 of the latter. If we assume that true gross mobility and net mobility move roughly in proportion, this would suggest that a much larger share of temporary layoffs are true stayers. However, since net mobility does not drop to zero, some true mobility must remain. The latter seems a likely explanation of why, even though the top row of Figure \ref{f:tlayoffs} shows a high correlation between the two transition matrices, the slope of the regression line lies a bit further from the 45-degree line relative to the cases studied in Section 4.1.1. However, if we focus on the subgroup of temporary layoffs that are also industry stayers we find that their average net occupational mobility rate drops by another third relative to all temporary layoffs.\footnote{Note that in theory, it could be the case that a spurious occupation change correlates with the \textit{realization} of a spurious industry change, something not ruled by assumption $A1$.} This strongly suggests that among this subgroup of temporary layoff workers the vast majority are indeed occupational stayers and hence provide a more accurate way to evaluate the persistence of the coding errors in $\mathbf{\Gamma}$. Consistent with this conclusion, we observe in the bottom row of Figure \ref{f:tlayoffs} that the regression line of temporary layoffs who are also three-digit industry stayers lies much closer to the 45-degree line.\footnote{As noted before, to the extent that some realizations of spurious industry mobility correlate with realizations of spurious occupation mobility (which is not ruled out by assumption $A1$), we may have restricted our sample too much to fully satisfy the condition that a representative set of true occupational stayers are included. However, the difference between the regression line and the 45-degree line remains small.}

Second, there may be some differences between the CPS and SIPP that affect the occupational information and its coding, e.g. (slight) differences in the questions about occupations.\footnote{For example, in case the interviewee reports an occupation description that the CPS considers to be too general, the interviewer follows up by providing a list of more specific occupations from which the interviewee should chose from (Appendix 2 of the CPS Manual, 2015). Standard SIPP documentation does not report a similar routing.} Given many similarities in terms of occupations across both surveys (and across time), we expect this factor not to be of great importance. Third, there may be finite sample considerations. The sample sizes of the 10-year CPS windows are lower than the sample on which the $\mathbf{\Gamma}$ is estimated in the SIPP (see footnote 15 and Section 1.3, respectively). We observe some indication that for the smaller CPS 10-year windows, the sample size may leave some role for finite sample noise. In particular, the overall correlation of the temporary layoff transition matrix over 1994-2020 is higher (at 0.843) than the correlation with respect to the underlying subsets 1994-2004, 2004-2014, 2014-2020 (around 0.80).

Having highlighted these factors, it is noteworthy that the coding error correction matrix estimated in the 1985-1986 SIPP panels can explain between half and two-thirds of the \textit{variance} along the diagonal of the transition matrix of temporary layoffs, even in the last decade. These high correlations occur even though the set of workers in the SIPP behind $\mathbf{\Gamma}$ and the temporary layoffs in the CPS are in a very different labor market situation (and formally even in a different employment status). Overall, this suggests that when using the $\mathbf{\Gamma}$-implied miscoding correction we gain a much better sense of miscoding, beyond a simple uniform level adjustment applied evenly across all occupation (which by construction has zero correlation with heterogeneity on the diagonal of spurious occupational mobility transition matrix).

\begin{figure}[ht!]
\centering
        \resizebox{0.45\textwidth}{!}{
          \includegraphics[width=\textwidth]{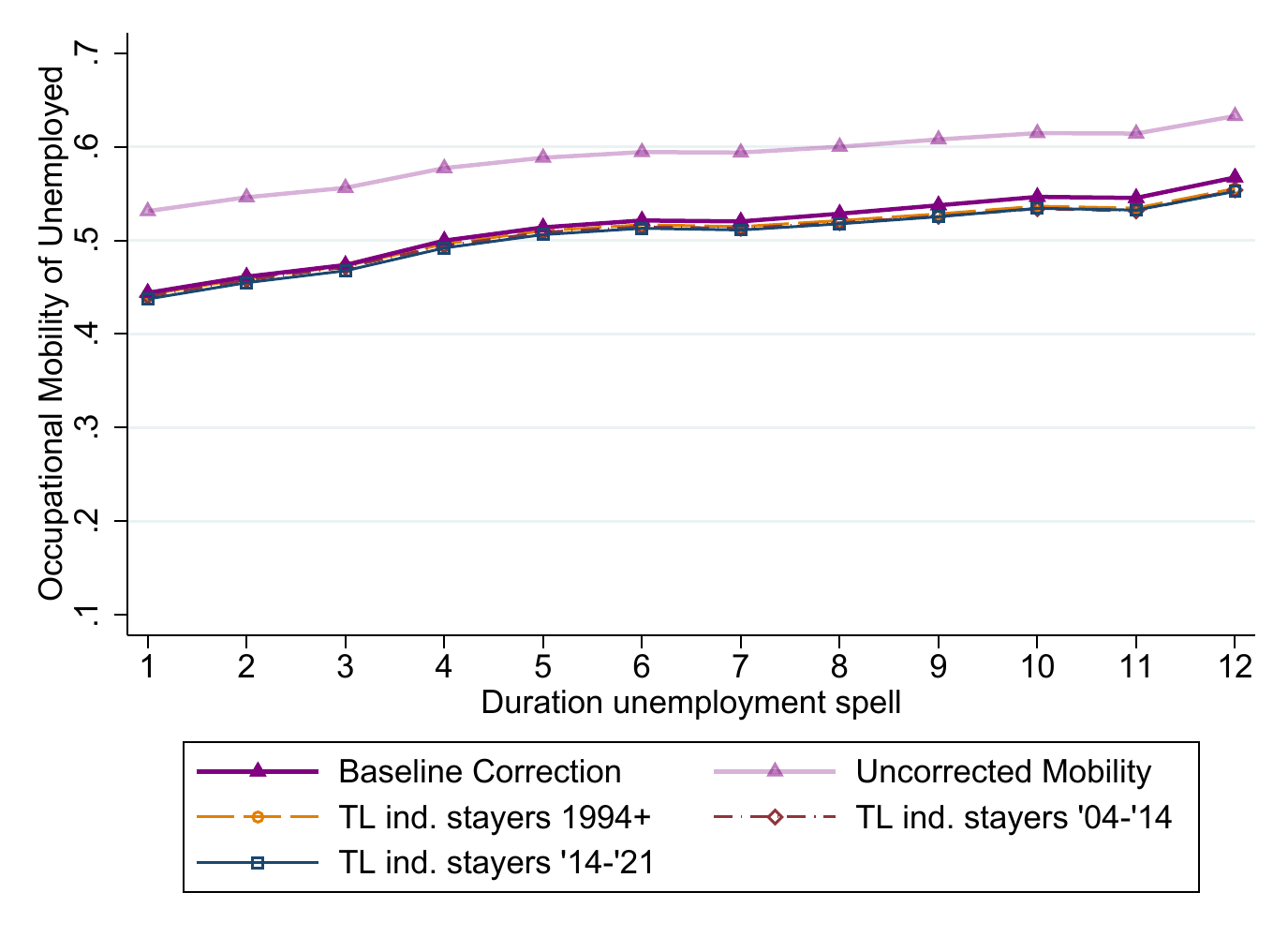}}
\caption{Corrected Mobility-Duration Profile from Temp Layoff and $\mathbf{\Gamma}$}\label{fig:mobdur_templayoff}
\end{figure}

\paragraph{Implications for measured occupational mobility}

As a final exercise we use the observed occupational transition matrix of the subset of workers who were on temporary layoff and observed as industry stayers instead of the $\mathbf{\Gamma}$ matrix in order to correct the occupational mobility-duration profile of the unemployed in our SIPP sample. We consider this subset of workers as they have the largest proportion of true occupational stayers and provide a more accurate comparison with the mobility-duration profile derived from the $\mathbf{\Gamma}$ matrix. Figure \ref{fig:mobdur_templayoff} shows that the two corrected profiles are nearly indistinguishable, particularly during the first six month of the unemployment spell. This occurs irrespectively of whether we consider all industry-staying temporary layoffs as the base for the correction matrix, or focus on the subperiods 2004-2014, or even 2014-2021. In all these cases we find a high occupational mobility rate that increases moderately with unemployment duration as we did in our main analysis.

\bigskip

Taken together, the above evidence strongly suggests that the $\mathbf{\Gamma}$ coding errors do remain relevant even in the 2010s.

\subsection{Assumption $A2$}

This assumption requires that the number of workers whose true occupation $i$ gets mistakenly coded as $j$ and the number of workers whose true occupation $j$ gets mistakenly coded as $i$ and is needed to derive equation \eqref{eq:est_gamma_2} and solve for $\mathbf{\Gamma}$. Although this assumption is clearly strong, it is important to note it is a weaker version of the one proposed by Keane and Wolpin (2011) and subsequently used in the structural estimation of discrete choice models. In particular, Roys and Taber (2017) used this assumption to correct for occupation classification error. They require that the number of workers whose true occupation $i$ gets mistakenly coded as occupation $j$ is independent of $i$ and $j$ for all $i\neq j$ and given by a constant. This implies that in their ``garbling'' matrix all off-diagonal elements are the same. In contrast, our approach allows (and recovers) different off-diagonal elements in $\mathbf{\Gamma}$, showing the large heterogeneity in bilateral occupation miscoding (see also Sullivan, 2009).

\begin{table}[ht!]
  \centering
  \caption{Occupational mobility and unemployment duration}
  \resizebox{0.9\textwidth}{!}{
    \begin{tabular}{lccccc}
    \hline
	&Health prof, 	& Account, archit, legal	& Managers, officials, &	Clerical 	& Retail and sales \\
	&School/college teachers &	tech., others &proprietors	&workers	&workers  \\
	\hline									
Break &	-0.0002   &	0.0040	&-0.0014    &	0.0034	&0.0044	  \\
	&{\footnotesize{(0.0027)}}	&{\footnotesize{(0.0039)}}	&{\footnotesize{(0.0043)}}	&{\footnotesize{(0.0033)}}	&{\footnotesize{( 0.0030)}}	 \\
Constant &	0.0458$^{***}$	& 0.0774$^{***}$	&0.1459$^{***}$	&0.0966$^{***}$	&0.0510$^{***}$	\\
	&{\footnotesize{(0.0040)}}	&{\footnotesize{(0.0100)}}	&{\footnotesize{(0.0064)}}	&{\footnotesize{(0.0134)}}&{\footnotesize{(0.0121)}}	\\
Time trend &	X	&X	&X	&X	&X \\
\hline
$R^{2}$	& 0.8482	&0.9619	&0.8552	&0.8294	&0.8793 \\
\hline
\hline
	& Foremen	& Craftsmen	 &Operatives 	& Laborer	& Other service \\
	&	workers	&kindred workers	&kindred workers	&workers	& workers \\
	\hline									
Break &	-0.0004  	&0.0012	&0.0031	&-0.0085$^{***}$   &	-0.0031   \\
	&{\footnotesize{(0.0023)}}	&{\footnotesize{(0.0029)}}	&{\footnotesize{(0.0044)}}	&{\footnotesize{(0.0017)}}	& {\footnotesize{(0.0023)}}\\
Constant &	0.0179$^{***}$&	0.0494$^{***}$&	0.1717$^{***}$	&0.0524$^{***}$	& 0.0907$^{***}$\\
	&{\footnotesize{(0.0034)}}	& {\footnotesize{(0.0043)}} &	{\footnotesize{(0.0066)}}	&{\footnotesize{(0.0015)}}	&{\footnotesize{(0.0027)}} \\
Time trend &	X	&X	&X	&X	&X \\
\hline
$R^{2}$	& 0.6108	&0.7911	&0.9110&	0.8437	&0.4639 \\
\hline
\hline
\multicolumn{6}{l}{{\footnotesize{Levels of significance: $^{*} p<0.1$, $^{**} p<0.05$, $^{***} p<0.01$}}}
    \end{tabular}}
  \label{tab:occ_dist_psid}
\end{table}

Unfortunately, we cannot directly test assumption $A2$ in our SIPP data. Instead, we can investigate whether there is evidence of an important implication of $A2$: coding errors do not change the distribution of occupations across workers. The PSID retrospective coding exercise is suitable for this exercise as it provides cleaned occupations codes for workers during the period 1968-1980, but does not correct for coding errors in the occupations assigned to the same workers during the period 1981-1997. Assumption $A2$ would imply that we should not observe any significant change in the contribution of each individual occupation across these two periods after controlling for time trends.

Table \ref{tab:occ_dist_psid} reports the results of regressing the contribution of a given occupation across the period 1968-1997 on the ``break'' indicator and cubic polynomial time trend.\footnote{Using higher order polynomials generate similar results. We also exclude the categories referring to armed forces and agricultural occupations as these are not included in our main analysis.} We can observe that classification errors across occupational coding do not seem to meaningfully change the distribution of occupation across our PSID sample. We observe that the break indicator is nearly zero and not statistically significant, except for the unskilled laborer occupation. However, even in this case the effect of coding error is to decrease the contribution by an average of only 0.85\%. Considering that this is a small occupation, contributing 4.20\% in 1980 and 3.60\% in 1981, the effect of coding error in this respect does not seem to be of first order importance.

\clearpage


\noindent {\Large \bf{Supplementary Appendix B: Occupational Mobility in the Data}}
\renewcommand{\thesection}{B.\arabic{section}}
\setcounter{section}{0}
\label{suppappx_B_firstpage}
\vspace{0.75cm}

In this Appendix we investigate in detail the occupational mobility patterns of those workers who changed employers through spells of unemployment, complementing the empirical patterns documented in the paper. Section 1 documents the gross occupational mobility - unemployment duration profile. We show that gross occupational mobility is high and increases moderately with spell duration when considering different occupation classifications and types of non-employment spells. We also show that this profile is widely shared among demographic groups and individual occupations. The main message of Section 1 is that the aggregate mobility-duration profile documented in the main text is \emph{not} driven by composition effects whereby some demographic groups and/or individual occupations are characterised by high mobility rates and short unemployment spells while, simultaneously, others are characterised by low mobility rates and longer unemployment spells. This finding motivates the assumption in our theoretical model of a common idiosyncratic ($z$-productivity) process shared across different occupations.

Section 2 investigates the extent to which excess and net mobility drive gross occupational mobility, once again using different alternative classifications and non-employment spells. The main message here is that excess occupational mobility is the main force behind the above gross occupational mobility - unemployment duration profile. Nevertheless, we find a clear pattern in the net mobility flows that is consistent with the job polarisation literature. These findings motivate the way we model workers' occupational choice in our theoretical model.

Section 3 investigates the cyclical patterns of occupational mobility among unemployed workers. We document that gross occupational mobility is procyclical. This pattern is present not only in the average mobility rate but also along the entire occupational mobility - unemployment duration profile. Here we show that the cyclicality of gross occupational mobility does not depend on occupational classifications or the inclusion of non-employment spells and that it is also present when controlling for demographic characteristics and individual occupation fixed effects. We also show that the procyclicality of gross occupational mobility is driven by excess mobility. In contrast, we document that net occupational mobility is countercyclical, once again showing a cyclical pattern consistent with the job polarisation literature.

Section 4 investigates the aggregate hazard functions of the unemployed and the non-employed in our sample. Our definition of unemployment implies that these hazard functions exhibit negative duration dependence but the degree of duration dependence is small. We then show that occupational movers take longer to leave unemployment than occupational stayers and that this difference increases in recessions and decreases in expansions. We argue that the difference in unemployment durations between movers and stayers is related to the process of mobility itself and not due to occupational or demographic composition effects.

In Section 5 we use the CPS and PSID to further investigate the extent and cyclicality of gross occupational mobility among the non-employed. Section 6 constructs occupational mobility rate using self-reported occupational and job tenure data obtained from the topical modules of the SIPP. The advantage here is that the information on occupational change does not use occupational codes, but relies on the workers own judgement of `occupation', `kind of work' or `line of work'. Once again we find that gross occupational mobility among the unemployed is high and procyclical, increasing moderately with non-employment duration. Section 7 present the details of the data construction.

\section{The occupational mobility - unemployment duration profile}

Here we analyse the long-run relationship between occupational mobility and unemployment duration observed during the 1983-2013 period. We refer to this relation as the occupational mobility - unemployment duration profile or the mobility-duration profile for short. We provide further support to one of our main empirical findings: the extent of occupational mobility among those workers who changed employers through spells of unemployed is \emph{high} and increases \emph{moderately} with the duration of the unemployment spell. We use `moderately' to indicate that at least 40\% of long-duration unemployed workers return to their previous major occupation upon re-employment. We show that this pattern holds under alternative occupational classifications and when aggregating major occupations into task-based categories (see Acemoglu and Autor, 2011). This pattern also holds when considering non-employment spells instead of just unemployment spells, and can be found within gender, education and age groups as well as at the level of individual occupations. These findings are important as they counter concerns that the aggregate mobility-duration profile is driven by differences in the demographic or occupational composition of the outflows from unemployment at different durations. This suggests that there is a common underlying process shared across these subgroups which leads long-duration unemployed workers to be still significantly attached to their previous occupation.

Table \ref{tab:basic_demog} reports the estimated mobility-duration profiles across occupational categories and demographic groups. In particular, panel A reports the $\Gamma$-corrected mobility-duration profile, where the first row shows the average occupational mobility rates over all unemployment (non-employment) spells and the second row reports the OLS estimate of the linear relation between completed unemployment duration and occupational mobility, $\beta_{\text{dur}}$. To estimate the latter we compute the raw occupation transition matrix of all workers with the same unemployment duration. We then apply to this matrix our $\Gamma$-correction and compute the average corrected occupational mobility rate. This is done for each month of completed unemployment duration between 1 and 14 months. We estimate $\beta_{\text{dur}}$ by regressing the set of average mobility rates on completed duration, weighing each monthly observation by the number of workers in the corresponding duration group (which itself involves summing the person weights by month).

Panel B reports the uncorrected mobility-duration profile in the same fashion as in panel A. However, since in the uncorrected data we can use individual workers' unemployment spells to estimate $\beta_{\text{dur}}$, we regress the following linear probability model (probits give nearly identical results)
\begin{align}
 \mathbf{1}_{\text{occmob}} &= \beta_0 + \beta_{\text{dur}} \ \text{duration of U (or N) spell} + \varepsilon, \tag{R1} \label{e:app_basic}
\end{align}
where $\mathbf{1}_{\text{occmob}}$ is a binary indicator that takes the value of one (zero) if a worker changed (did not change) occupation at the end of his/her unemployment (non-employment) spell, ``duration of U (or N) spell'' is the individual's \emph{completed} unemployment (non-employment) spell and $\varepsilon$ is the error term. The advantage of the individual-level uncorrected data is that we can take into account worker's characteristics and evaluate their role in shaping the mobility-duration profile. This is done in panels C, D and E.

Column (1) of Table \ref{tab:basic_demog} presents our benchmark result which is based on major occupational groups of the 2000 Census Classification of Occupations (or 2000 SOC). To consistently use the 2000 SOC throughout the period of study, we applied the IPUMS homogenisation procedure to convert previous classifications used in the SIPP. The 2000 SOC has 23 major occupational groups as its most aggregated classification, which we further reduce to 21 as we leave ``army'' and ``agricultural occupations'' out of our sample.

Figure \ref{2000} depicts the mobility-duration profile using this occupational classification, but based on an alternative formulation. Namely, for a given unemployment duration $x$, the figure depicts gross occupational mobility as the fraction of workers who had at least $x$ months in unemployment and changed occupation at re-employment among those workers who had at least $x$ months in unemployment before regaining employment. This way of presenting the profile allows to read directly from the graph the average occupational mobility rate of the sample of workers who had completed unemployment durations of at least $x$ months. In what follows every time we graph the mobility-duration profile, we will be using this alternative formulation.

In both the table and figure the average ($\Gamma$-corrected and uncorrected) probability of an occupational change is high and increases moderately with unemployment duration. Further, note that the slope of the mobility-duration profile is slightly steeper with the corrected measures. This arises as miscoding (and hence spurious mobility) is more common among observations with short unemployment durations. The $\Gamma$-correction will then adjust more the shorter than the longer duration spells and yield a steeper profile relative to the uncorrected data. Also with considerably more short than long spells in the sample, the regression coefficient on duration is more representative of the slope at short completed durations.

The rest of the columns of Table \ref{tab:basic_demog} present the results for a number of different ways of classifying occupational mobility. Column (2) uses instead the major occupational groups of the 1990 Census Classification of Occupations (or 1990 SOC). Columns (3) to (5) aggregate occupations into task-based categories. Column (6) considers the case of simultaneous occupational and industry mobility as an alternative way to minimise the effects of coding errors in the occupational mobility rates. Column (7) considers industry mobility, as a comparison to occupational mobility. Column (8) uses the major groups of the 2000 SOC but instead considers non-employment spells that contain at least one month of unemployment (`NUN'-spells). We now discuss each one in turn. We then discuss the effects of worker demographics characteristic and the role of individual occupations.

\begin{table}[ht!]
  \centering
  \caption{The occupational mobility - unemployment duration profile}
   \resizebox{1\textwidth}{!}{
  \normalsize
    \begin{tabular}{lcccccccc}
    \toprule
    \toprule
          & 2000 SOC    & 1990 SOC    & NR/R-M/C & NR/R-M/C* & C/NRM/RM & OCC*IND & IND   & 2000 SOC-NUN \\
          & (1)   & (2)   & (3)   & (4)   & (5)   & (6)   & (7)   & (8) \\
    \bottomrule

    no. obs. & \multicolumn{1}{c}{19,115} & \multicolumn{1}{c}{19,051} & \multicolumn{1}{c}{24,815} & \multicolumn{1}{c}{18,527} & \multicolumn{1}{c}{18,604} & \multicolumn{1}{c}{19,054} & \multicolumn{1}{c}{19,055} & \multicolumn{1}{c}{19,386} \\
    \midrule
    \multicolumn{9}{c}{Panel A: miscoding corrected mobility, no demographic characteristics, no time, no occ/ind controls} \\
    \midrule
    av occmob & 0.444*** & 0.419*** & 0.271*** & 0.296*** & 0.227*** & 0.3838*** & 0.485*** & 0.469*** \\
    (s.e.) & (0.0043) & (0.0043) & (0.0037) & (0.0034) & (0.0035) & (0.0053) & (0.0039) & (0.0060) \\
    dur coef & 0.0173*** & 0.0197*** & 0.0111*** & 0.0123*** & 0.0098*** & 0.0116*** & 0.0146*** & 0.0184*** \\
    (s.e.) & (0.0017) & (0.0023) & (0.0018) & (0.0022) & (0.0023) & (0.0016) & (0.0012) & (0.0016) \\
    \midrule
    \multicolumn{9}{c}{Panel B: uncorrected, no demog, no time, no occ/ind controls } \\
    \midrule
    av occmob & 0.5312*** & 0.5221*** & 0.3391*** & 0.3659*** & 0.2809*** & 0.3838*** & 0.5285*** & 0.5504*** \\
    (s.e.) & (0.0054) & (0.0054) & (0.0052) & (0.0052) & (0.0049) & (0.0053) & (0.0055) & (0.0044) \\
    dur coef & 0.0142*** & 0.0149*** & 0.0093*** & 0.0103*** & 0.0086*** & 0.0116*** & 0.0131*** & 0.0149*** \\
    (s.e.) & (0.0015) & (0.0015) & (0.0015) & (0.0015) & (0.0015) & (0.0016) & (0.0015) & (0.0010) \\
    \midrule
    \multicolumn{9}{c}{Panel C: uncorrected, with demog, time controls, no occ/ind controls} \\
    \midrule
    dur coef & 0.0150*** & 0.0156*** & 0.0103*** & 0.0112*** & 0.0085*** & 0.0123*** & 0.0137*** & 0.0145*** \\
    (s.e.) & (0.0015) & (0.0015) & (0.0015) & (0.0015) & (0.0015) & (0.0016) & (0.0016) & (0.0010) \\
    female & 0.0208** & -0.0283*** & 0.0713*** & 0.0292*** & -0.0273*** & 0.0012 & 0.0154* & 0.0165** \\
    (s.e.) & (0.0083) & (0.0083) & (0.0080) & (0.0081) & (0.0074) & (0.0082) & (0.0084) & (0.0069) \\
    hs drop & -0.0421*** & -0.0406*** & -0.0545*** & -0.0504*** & -0.0379*** & -0.0212* & -0.0309** & -0.0436*** \\
    (s.e.) & (0.0118) & (0.0118) & (0.0106) & (0.0110) & (0.0104) & (0.0116) & (0.0120) & (0.0099) \\
    some col & 0.0223** & 0.0099 & 0.0364*** & 0.0312*** & -0.0206** & 0.0125 & 0.0287*** & 0.0235** \\
    (s.e.) & (0.0108) & (0.0108) & (0.0104) & (0.0106) & (0.0101) & (0.0107) & (0.0109) & (0.0092) \\
    col grad & 0.0377*** & -0.0151 & 0.0181 & -0.0090 & -0.1391*** & 0.0014 & 0.0047 & 0.0244** \\
    (s.e.) & (0.0122) & (0.0122) & (0.0117) & (0.0117) & (0.0101) & (0.0119) & (0.0124) & (0.0102) \\
    black & 0.0315** & -0.0000 & 0.0265** & 0.0159 & 0.0468*** & 0.0315** & 0.0189 & 0.0247** \\
    (s.e.) & (0.0124) & (0.0122) & (0.0119) & (0.0120) & (0.0116) & (0.0125) & (0.0126) & (0.0103) \\
    time trend & 0.0013*** & 0.0012*** & 0.0010*** & 0.0009*** & 0.0009*** & 0.0010*** & 0.0012*** & 0.0011*** \\
    (s.e.) & (0.0002) & (0.0002) & (0.0002) & (0.0002) & (0.0002) & (0.0002) & (0.0002) & (0.0002) \\
    \midrule
    \multicolumn{9}{c}{Panel D: uncorrected, additionally interactions of demog. with duration; demog. \& time ctrls} \\
    \midrule
    female*dur & -0.0014 & -0.0013 & 0.0048 & 0.0034 & 0.0018 & 0.0008 & 0.0001 & -0.0025 \\
    (s.e.) & (0.0030) & (0.0031) & (0.0031) & (0.0031) & (0.0029) & (0.0032) & (0.0031) & (0.0021) \\
    hs drop*dur & -0.0022 & -0.0051 & -0.0000 & -0.0023 & -0.0037 & 0.0021 & -0.0006 & -0.0007 \\
    (s.e.) & (0.0043) & (0.0042) & (0.0041) & (0.0042) & (0.0040) & (0.0044) & (0.0044) & (0.0030) \\
    (some col)*dur & 0.0048 & 0.0019 & -0.0011 & -0.0001 & 0.0022 & 0.0074* & 0.0004 & 0.0025 \\
    (s.e.) & (0.0037) & (0.0037) & (0.0039) & (0.0039) & (0.0038) & (0.0039) & (0.0039) & (0.0026) \\
    (col grad)*dur & 0.0045 & 0.0005 & 0.0005 & -0.0017 & -0.0032 & 0.0081* & 0.0054 & 0.0047 \\
    (s.e.) & (0.0043) & (0.0043) & (0.0045) & (0.0045) & (0.0041) & (0.0045) & (0.0045) & (0.0030) \\
    black*dur & 0.0020 & 0.0005 & 0.0022 & 0.0027 & 0.0064 & 0.0002 & -0.0012 & 0.0010 \\
    (s.e.) & (0.0043) & (0.0043) & (0.0044) & (0.0045) & (0.0045) & (0.0045) & (0.0044) & (0.0030) \\
    \midrule
    \multicolumn{9}{c}{Panel E: F-test equality duration coefficient across gender, educ and race} \\
    \midrule
    p-value & 0.756 & 0.904 & 0.852 & 0.7   & 0.382 & 0.621 & 0.906 & 0.453 \\
    \bottomrule
    \bottomrule
    \multicolumn{9}{c}{{\scriptsize *$\ p<0.1$;\ **$\ p<0.05$;\ ***$\ p<0.01$}}
    \end{tabular}}
  \label{tab:basic_demog}%
\end{table}%

\subsection{Gross occupational mobility under alternative occupational classifications}

\paragraph{1990 Census Classification of Occupations} The 2000 SOC provided a major revision of the 1990 SOC as a response to the changing structure of jobs in the US. The major difference between these two classifications relies on the former grouping occupations based on the concept of ``job families'' by placing individuals who work together in the same occupational group. The result was that some occupations that belonged to different groups in the 1990 SOC were pulled together in the 2000 SOC revision. This led to an increase in the size of occupations like ``professional'', ``technical'', ``management'' and ``services''.

Influential work, however, relies on the 1990 SOC to understand occupational change in the US (see Autor and Dorn, 2013, among others). To verify that our conclusions are not affected by the type of classification used, we compute the mobility-duration profile based on the 1990 SOC. In this case we apply the homogenisation procedure proposed by Autor and Dorn (2013). The 1990 SOC provides 13 major occupational groups from which we aggregate the services related occupations (``protective services'', ``private households'' and ``others'') into one single major group. We do this as the ``protective services'' and ``private households'' occupations are very small in size. At the same time we expand the ``precision, production, craft and repair'' into three new major groups: ``precision production'', ``mechanics and repair'' and ``construction trade occupations''. This allows us to evaluate ``construction'' as a separate group.

\begin{figure}[ht!]
\centering
\subfloat[2000 SOC]{\label{2000} \includegraphics [width=0.5 \textwidth] {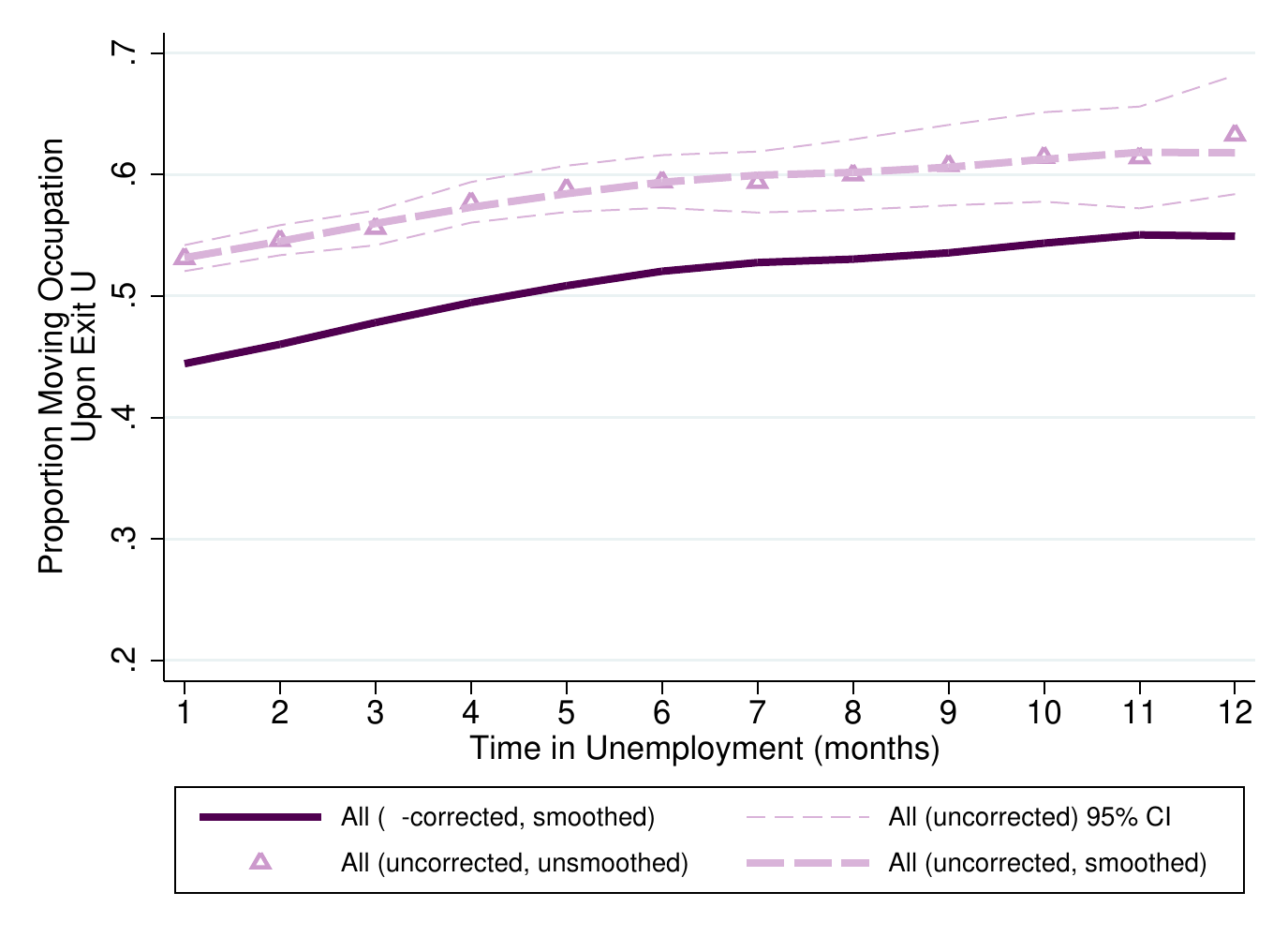}}
\subfloat[1990 SOC]{\label{1990} \includegraphics [width=0.5 \textwidth] {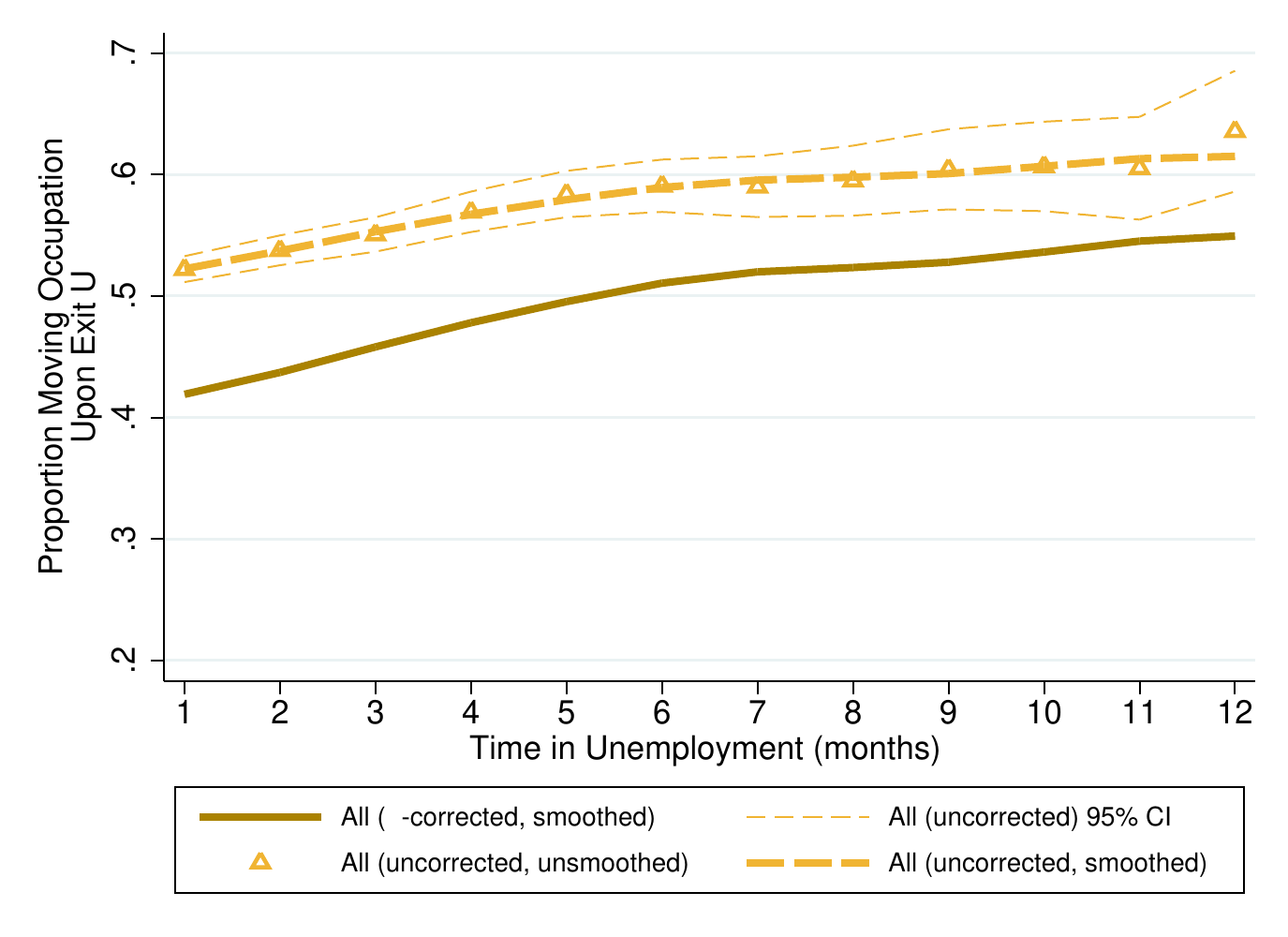}}
\caption{Extent of mobility by unemployment duration - Major occupational groups}
\label{f:2000_1990}
\end{figure}

Column (2) of panel B in Table \ref{tab:basic_demog} shows that in the uncorrected data we obtain nearly identical results when using the 2000 SOC or the 1990 SOC. Applying the $\Gamma$-correction yields a slightly lower average mobility rate and a slightly larger duration coefficient when using the 1990 SOC. Despite these differences, Figure \ref{f:2000_1990} shows the mobility-duration profiles and the associate 95\% confidence intervals on uncorrected data. It shows that long-duration unemployed with completed spells of at least 9 months change occupations in 53\% of cases in both the 1990 and 2000 classifications.

\paragraph{Routine vs Non-Routine, Cognitive vs Manual Occupational Categorization} A related set of important work has documented a pattern of job polarization during our period of study (1983-2013). This pattern is characterised by a decline in the employment shares and levels of occupations that have a high content of routine tasks (see Autor et al., 2003, Goos and Manning, 2007, and Acemoglu and Autor, 2011, among others). Although at the heart of this evidence lies the changing direction of net flows across occupations (a topic we address in Section 2 below), it is important for our study to first investigate whether the mobility-duration profile documented above is also observed when aggregating the major occupations into the task-based categories suggested by the job polarization literature. In particular, we aggregate the 13 major occupational groups of the 1990 SOC into routine manual (RM), non-routine manual (NRM), routine cognitive (RC) and non-routine cognitive (NRC).

Column (3) of Table \ref{tab:basic_demog} shows the estimates based on the RM/NRM/RC/NRC grouping. In this case we follow Cortes et al. (2016) and define the RM occupations to be (i) ``precision production'', (ii) ``machine operators and assemblers'', (iii) ``mechanics and repairers'', (iv) ``laborers and helpers'', (v) ``transportation and material moving'' and (vi) ``construction and extractive''. The NRM occupations to be the ``service occupations''. The RC occupations to be (i) ``sales occupations'' and (ii) ``administrative support and clerical occupations''. The NRC occupations to be (i) ``management occupations'', (ii) ``professional specialties'' and (iii) ``technicians and related support occupations''.

Column (4) considers an alternative classification in which we move the ``transportation and material moving'' occupation to the NRM group (see Autor et al., 2003). This is done mainly because this occupation exhibits a low routine-intensity score. Table \ref{tab:rtclass} illustrates this feature by reporting the average routine task-intensity (together with the abstract and manual task intensity) of each major occupation of the 2000 SOC, using the occupation-task intensity crosswalk of Autor and Dorn (2013). Its third column reports the average routine task intensity of the pre-separation three-digit occupation held by unemployed workers. It is clear that ``transportation and material moving'' has one of the lowest routine task-intensity scores.\footnote{A similar ranking occurs when using the 1990 SOC. We use the 2000 SOC in Table \ref{tab:rtclass} to refer back to the results presented in the main text, which are based in the latter classification.} Additional reasons to include ``transportation and material moving'' into the NRM groups are that this occupation exhibits net inflows, while the other conventional RM occupations are net losers of workers; and that it also behaves cyclically similar to other NRM occupations.

\begin{figure}[ht!]
\centering
\subfloat[RM/NRM/RC/NRC (1990 SOC-based, Transport in RM)]{\label{4cat_hs_cat} \includegraphics [width=0.5 \textwidth] {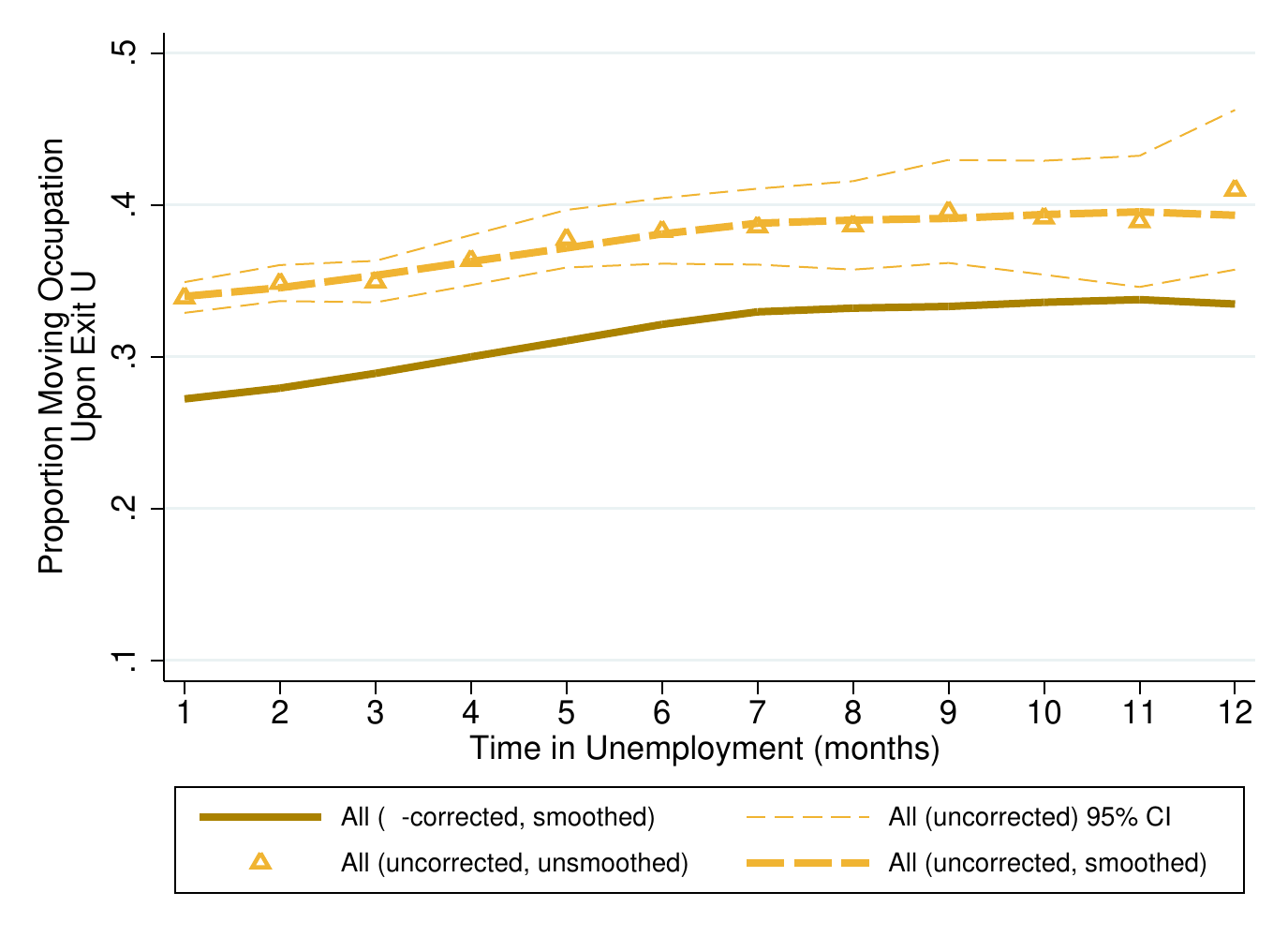}}
\subfloat[RM/NRM/RC/NRC (1990 SOC-based, Transport in NRM)]{\label{4cat_ths_cat} \includegraphics [width=0.5 \textwidth] {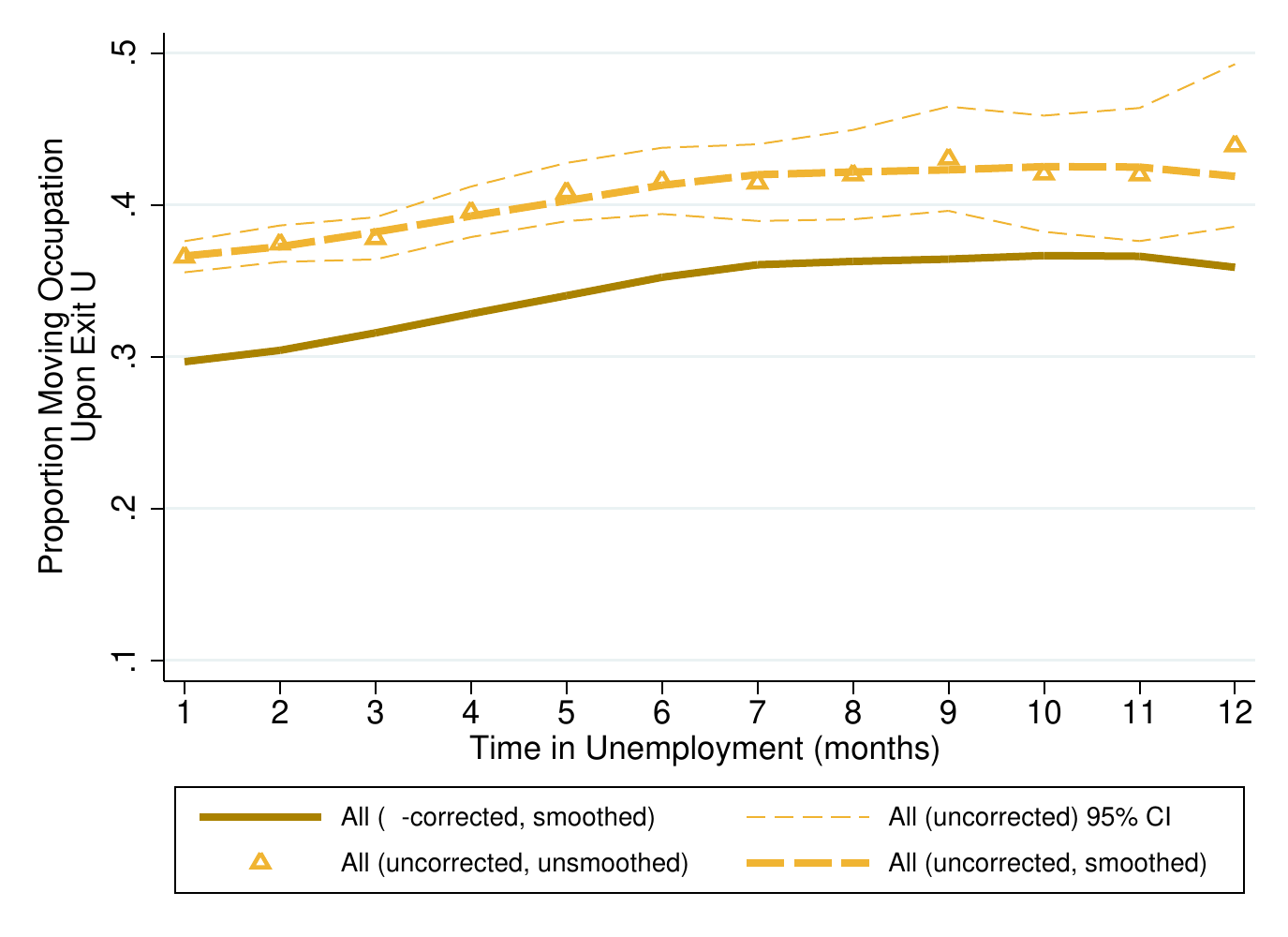}}

\subfloat[RM/NRM/RC/NRC (2000 SOC-based, Transport in NRM)]{\label{4_cat} \includegraphics [width=0.5 \textwidth] {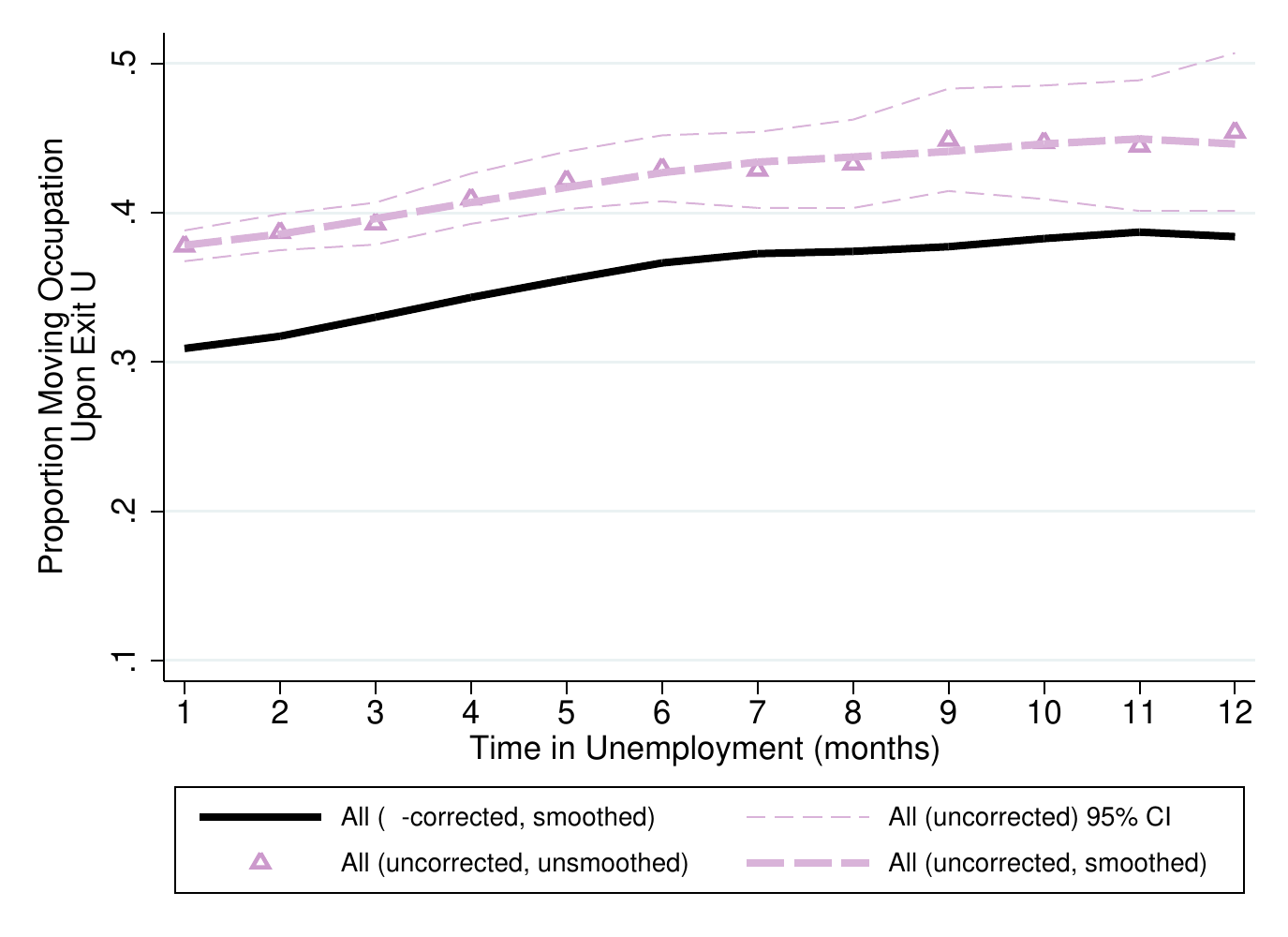}}
\subfloat[RM/NRM/C (1990 SOC-based, Transport in NRM)]{\label{3_cat} \includegraphics [width=0.5 \textwidth] {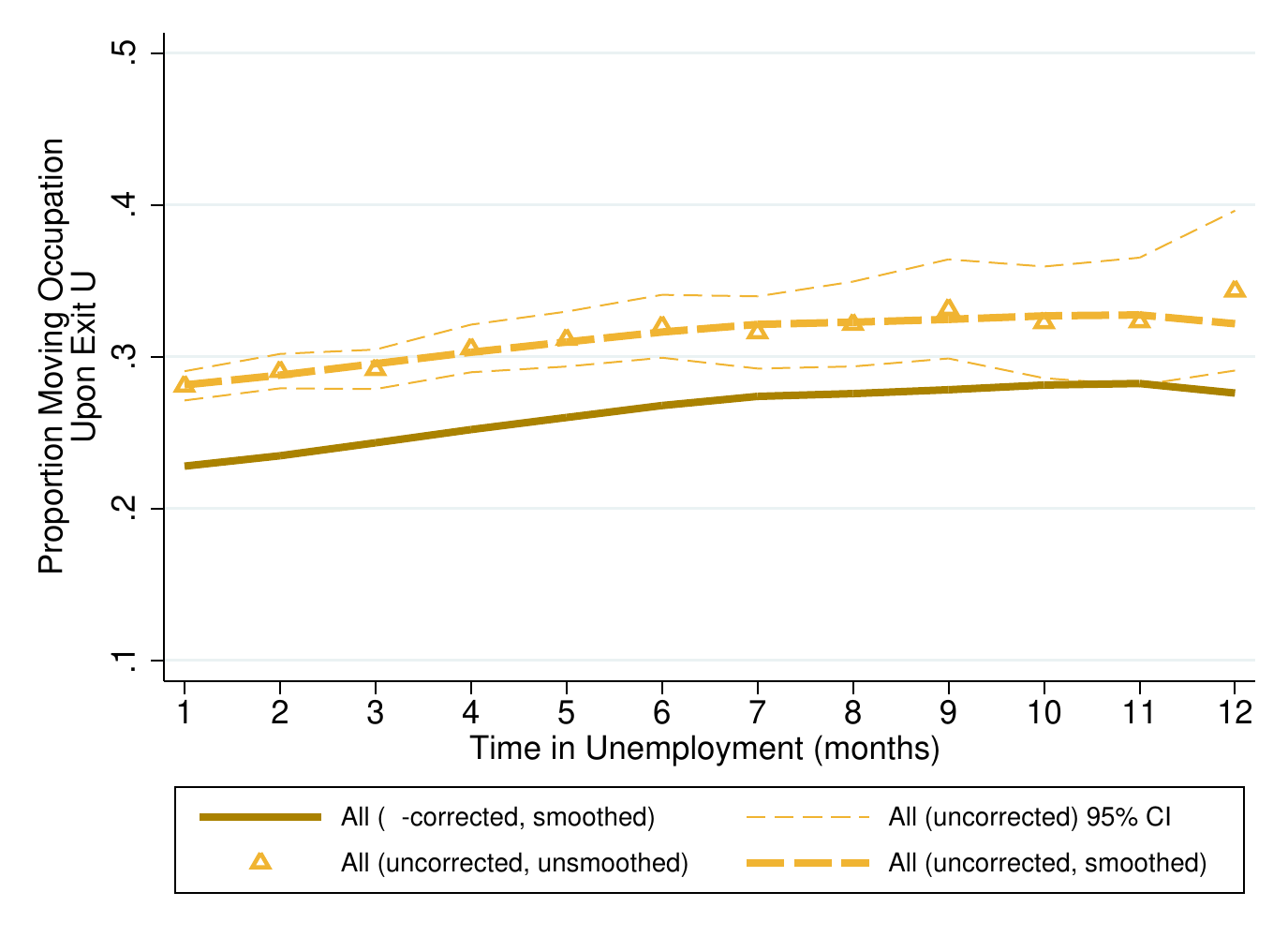}}
\caption{Extent of mobility by unemployment duration - Task based categorisation}
\label{f:task_based}
\end{figure}

Table \ref{tab:basic_demog} shows that under both task-based categorizations the average occupational mobility rates is high: 26\% (33\%) and 29\% (35\%) of all unemployment spells in the $\Gamma$-corrected (uncorrected) data involved workers changing task-based occupational categories. It also shows that the duration coefficient implies a modestly increasing profile. Figures \ref{4cat_hs_cat} and \ref{4cat_ths_cat} present the mobility-duration profiles and the associate confidence intervals. It shows that workers who had unemployment spells of at least 9 months experience about a 34\% and 37\% probability of changing task-based groups. Figure \ref{4_cat} shows the mobility-duration profile by assigning ``transportation and material moving'' into the NRM category, using the 2000 SOC. Once again we observe that choosing the 2000 SOC or the 1990 SOC to classify occupations makes little difference.

\begin{table}[htbp]
\small
  \centering
  \caption{Routine (Abstract, Manual) task intensity for source occupations of U inflow}
    \begin{tabular}{lrrrrl}
\toprule
\toprule  Occupation & \multicolumn{1}{l}{U inflow distr} & \multicolumn{1}{l}{Routine Int.} & \multicolumn{1}{l}{Abstract Int.} & \multicolumn{1}{l}{Manual Int.} & Class. \\
\cmidrule{1-6}    Protective service & 1.30  & \cellcolor[rgb]{ .816,  .925,  .843} 1.52 & 0.87  & 0.83  & NRM \\
    Management & 6.22  & \cellcolor[rgb]{ .847,  .933,  .843} 1.91 & 6.95  & 0.29  & NRC \\
    Educ, training, and library & 2.84  & \cellcolor[rgb]{ .871,  .937,  .843} 2.17 & 3.97  & 1.22  & NRC \\
    Personal care/Service & 1.51  & \cellcolor[rgb]{ .875,  .937,  .843} 2.21 & 1.54  & 0.97  & NRM \\
    Transportation \& Mat moving & 11.58 & \cellcolor[rgb]{ .886,  .941,  .843} 2.35 & 0.84  & 2.92  & NRM \\
    Comm \& Social service & 0.64  & \cellcolor[rgb]{ .91,  .949,  .839} 2.60 & 4.89  & 0.17  & NRC \\
    Building/Grounds clean \& maint. & 4.97  & \cellcolor[rgb]{ .941,  .957,  .839} 2.97 & 0.93  & 2.30  & NRM \\
    Food prep/Serving \& rel. & 6.85  & \cellcolor[rgb]{ .945,  .957,  .839} 3.00 & 1.38  & 1.01  & NRM \\
    Legal & 0.31  & \cellcolor[rgb]{ .976,  .965,  .839} 3.39 & 3.23  & 0.28  & NRC \\
    Healthcare support & 2.08  & \cellcolor[rgb]{ .988,  .969,  .839} 3.53 & 1.70  & 1.67  & NRM \\
    Computer and Math. occ & 1.12  & \cellcolor[rgb]{ 1,  .973,  .835} 3.62 & 5.78  & 1.22  & NRC \\
    Arts/Dsgn/Entrtmnt/Sports/Media & 1.21  & \cellcolor[rgb]{ 1,  .973,  .835} 3.69 & 3.61  & 0.97  & NRC \\
    Sales \& related occ & 10.42 & \cellcolor[rgb]{ 1,  .973,  .839} 3.79 & 2.92  & 0.65  & RC \\
    Buss \& Financial operations & 2.67  & \cellcolor[rgb]{ 1,  .973,  .839} 3.86 & 6.50  & 0.43  & NRC \\
    Life, phys, and social science & 0.68  & \cellcolor[rgb]{ 1,  .957,  .867} 4.74 & 4.67  & 0.84  & NRC \\
    Architect \& Eng.  & 1.62  & \cellcolor[rgb]{ 1,  .937,  .902} 5.96 & 6.52  & 1.29  & NRC \\
    Production & 12.88 & \cellcolor[rgb]{ 1,  .937,  .906} 6.03 & 1.40  & 1.20  & RM \\
    Office/Admin Support & 13.04 & \cellcolor[rgb]{ 1,  .933,  .91} 6.16 & 2.03  & 0.30  & RC \\
    Construction/Extraction & 12.53 & \cellcolor[rgb]{ 1,  .933,  .914} 6.37 & 1.76  & 3.02  & RM \\
    Healthcare pract \& Tech  & 1.72  & \cellcolor[rgb]{ 1,  .929,  .922} 6.64 & 3.17  & 1.22  & NRC \\
    Install/Maint/Repair & 3.80  & \cellcolor[rgb]{ .996,  .925,  .925} 6.65 & 2.05  & 1.79  & RM \\
    \bottomrule
    \bottomrule
    \end{tabular}%
  \label{tab:rtclass}%
\end{table}%

Column (5) of Table \ref{tab:basic_demog} presents the results for another task-based classification. In this case we merge the routine cognitive and non-routine cognitive categories together, to focus on transitions between routine manual and non-routine manual, relative to all other (cognitive) occupations. This aggregation is motivated by the observed direction of the net flows across the task-based occupations, which we discuss in detail in Section 2. It aims to highlight the disappearance of routine manual jobs and the rise of non-routine manual jobs as a key feature of unemployed workers' occupational mobility, and reduce the role of ``management occupations'' which accounts for the vast majority of net outflows from non-routine cognitive occupations. Nevertheless, we again observe that, even with this very coarse subdivision, there remains substantial occupational mobility (over 20\%) which modestly increases with unemployment duration. Figure \ref{3_cat} presents graphically the mobility-duration profile and the associate confidence intervals.

\subsection{Industry Mobility, and Simultaneous Occupation and Industry Mobility}

Next we consider the case of simultaneous occupational and industry mobility as an alternative way to minimise the effects of coding errors in the occupational mobility rates in the raw data. This approach follows Neal (1999), Moscarini and Thomsson (2007), Kambourov and Manovskii (2008) and Pavan (2011), and it is based on the assumption that workers who are observed changing occupations, are more likely to be true movers when they are  also observed changing employers and industries. Column (4) of Table \ref{tab:basic_demog} shows the average mobility rate and duration coefficient for the sample of workers for which we have valid (and non imputed) occupation and industry information. In this case we use the major occupational groups of the 2000 SOC and the 15 major industry groups of the 1990 Census Bureau industrial classification system. Since simultaneous occupational and industry mobility can be taken as an alternative measure of true occupational mobility (see references above), we report this measure both in panels A and B without any further adjustment. Once again we observe that the extent of occupational mobility among those who changed employers through spells of unemployed is high and increases moderately with unemployment duration.

\begin{figure}[ht!]
\centering
\subfloat[Simultaneous Occ and Ind. mob]{\label{OxI} \includegraphics [width=0.5 \textwidth] {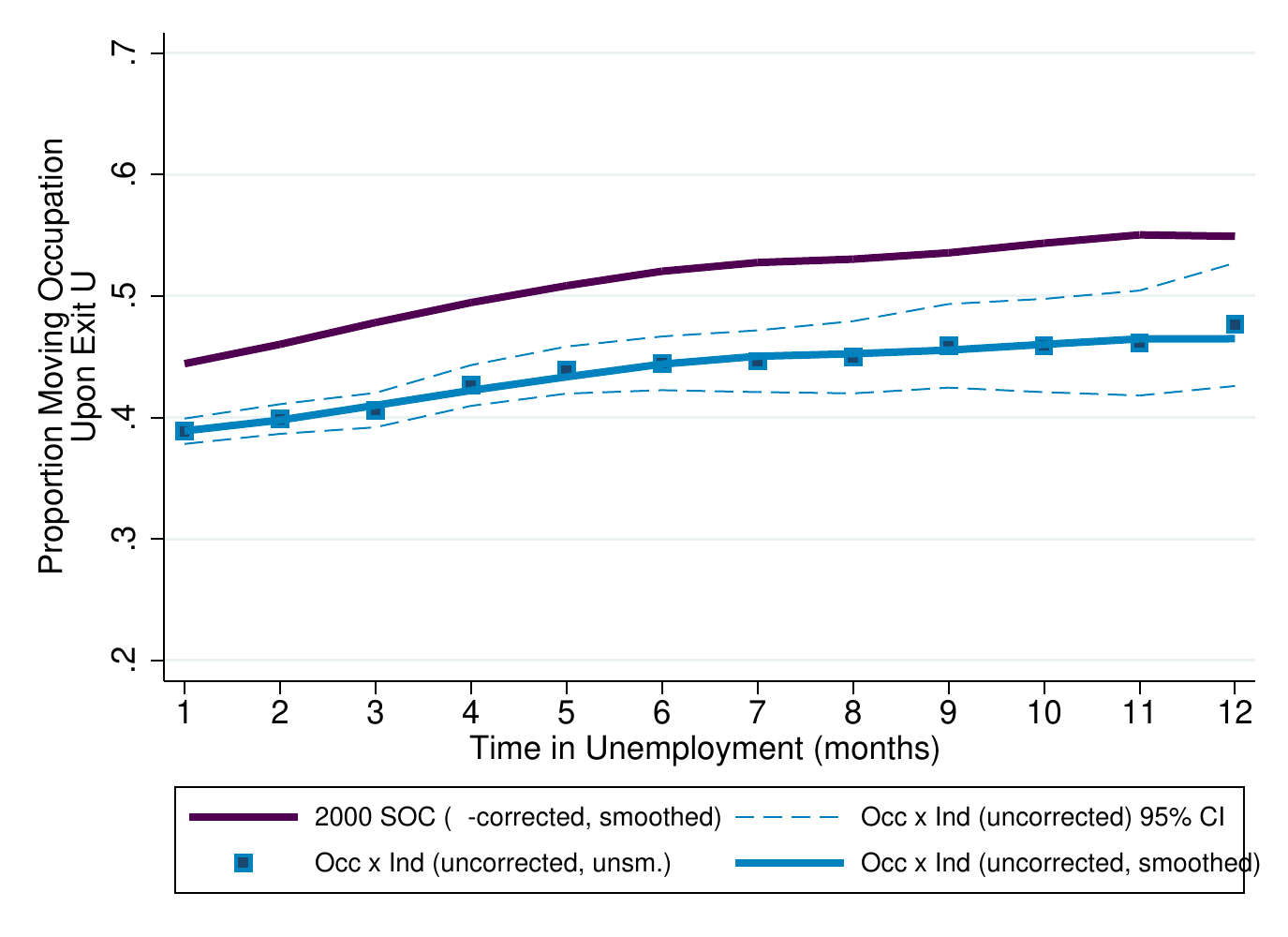}}
\subfloat[Industry mobility]{\label{Industries} \includegraphics [width=0.5 \textwidth] {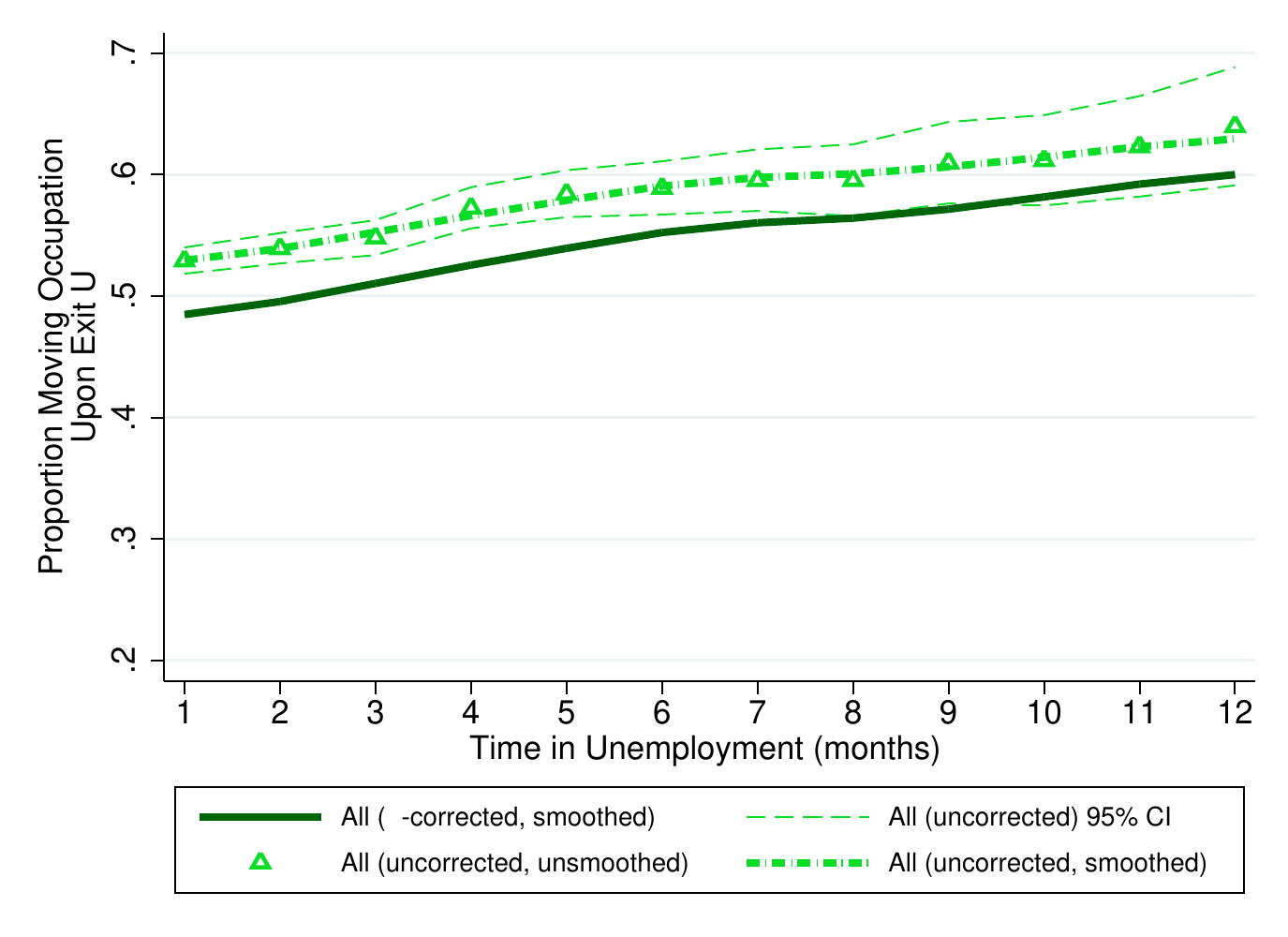}}
\caption{Extent of mobility by unemployment duration - Relation with industry mobility}
\label{f:Industry}
\end{figure}

The mobility-duration profile obtained from simultaneous occupational and industry mobility is depicted in Figure \ref{OxI}, together with the $\Gamma$-corrected mobility-duration profile depicted in Figure \ref{2000} for comparison and the associate confidence intervals. The former shows that about 38\% of workers who had unemployment spells of at least one month changed major occupations and industries at re-employment, while about 46\% of workers who had unemployment spells of at least 9 months changed major occupations and industries at re-employment. These rates are around 5 percentage points lower but not too dissimilar from the ones obtained from the $\Gamma$-corrected profile for occupational mobility. This suggests that conditioning on simultaneous industry and occupational transitions can provide a useful alternative to gauge the level of gross occupational mobility rates.

Caution is advised, however, when constructing statistics that lean more heavily on occupational identities or capture measurement of change in occupational mobility rates. When using the occupation/industry cross-product to inform true occupational mobility, two mistakes will be made. (i) True occupational movers who are true industry stayers will be mistakenly left aside. This issue might be more prevalent in some occupations and less so in others, and hence can \emph{unevenly} affect the measurement of net mobility. For this reason, in what follows we will use simultaneous industry and occupation transitions to construct gross mobility rates and not net mobility measures. (ii) For true industry movers who are also true occupational stayers, there will be a (on average) 20\% probability of mistakenly considering them as simultaneous occupation and industry movers. It is not clear whether the size of these two types of errors will stay constant when, for example, measuring differences between subsamples with different true mobility rates.\footnote{It is relatively straightforward to verify that the (uncorrected) slope of the mobility-duration profile can suffer from attenuation bias arising from miscoding. While our $\Gamma$-corrections pushes against this bias, this does not seem to be the case when using the simultaneous mobility measure. In this case we observe that the mobility-duration profile has a somewhat lower slope than both the raw occupational and industrial mobility-duration profiles, and therefore even lower than the $\Gamma$-corrected slopes.} Nevertheless, it is evident that the simultaneous mobility measure also indicates a high level of occupational mobility that moderately increases with unemployment duration.

Column (5) of Table \ref{tab:basic_demog} report the $\Gamma$-corrected and the uncorrected average mobility rates and the duration coefficients of workers who changed employers through unemployment, using only the 15 major industry groups of the 1990 Census Bureau industrial classification system. Figure \ref{Industries} also presents the $\Gamma$-corrected and the uncorrected mobility-duration profiles. This evidence confirms that many unemployed workers also changed industries at re-employment. In the uncorrected data we observe that around 53\% of workers who had unemployment spells of at least a month changed industries at re-employment, while about 60\% of workers who had unemployment spells of at least 9 months changed industries at re-employment. These rates are remarkably close to the raw occupational mobility rates reported in columns (1) and (2). The $\Gamma$-correction mobility-duration profile of industry mobility, however, drops by less than the occupation profiles, with an average mobility of 48.5\%. This is consistent with the well-known fact that industry mobility rates are less prone to measurement error than occupational mobility rates (as also reported in Table 5 of Supplementary Appendix A).

\subsection{Occupational mobility through non-employment spells}

The above analysis restricts attention to non-employment spell in which workers where unemployed every month, categorised as ``no job/business - looking for work or on layoff'' in the SIPP. We now consider the case in which workers spend part of their non-employment spell outside of the labor force, categorised as ``no job/business - not looking for work and not on layoff'' in the SIPP. This allows us to investigate whether the mobility-duration patterns documented above are also present when we include joblessness periods in which workers reported no active job search. This case also allows us to investigate whether the mobility-duration profile based on ``pure'' unemployment spells is driven by a composition effect based on workers' differential propensities to drop out of the labor force. In particular, if workers had ex-ante different propensities to change occupations and if these propensities were negatively associated with the probability of dropping out of the labor force after long periods of joblessness (for example, due to discouragement effects), a restriction to ``pure'' unemployment spells could be effectively selecting occupational movers at higher unemployment durations. This would then lead to a different interpretation of the mobility-duration profile than the one proposed by our theoretical framework.

To address this concern we compute the mobility-duration profile for workers with different degrees of labor market attachment: (1) `U' spells, our baseline, where the individual is ``looking for work'' every month of the non-employment spell. (2) `UNU' spells, where the worker starts the non-employment spell ``looking for a work'' and ends it also ``looking for a work'', but can have intervening months outside the labor force. (3) `UN' spells, where the worker starts the non-employment spell ``looking for work'' and this can be followed by periods out of the labor force, before regaining employment (not necessarily reporting unemployment just before re-employment). (4) `NU' spells, where the worker might or might not be ``looking for work'' after separation, but eventually ``looks for work'' and finds one shortly thereafter. (5) `N*' spells, where the worker loses his job for economic reasons tied in with the job (as indicated by the `reason for ending previous job', or starts looking for a job when he becomes non-employed.\footnote{The reasons we consider in this case are: (i) employer bankrupt, (ii) employer sold business; (iii) job was temporary and ended, (iv) slack work or business conditions, (v) unsatisfactory work arrangements (hours, pay, etc), (vi) quit for some other reason. We do not consider: quit to take another job, retirement or old age, childcare problems, other family/personal obligations, own illness, own injury, school/training.} (6) `NUN' spells, where the worker ``looks for work'' at least one month during the non-employment spell. (7) `N' spells, covering all non-employment spells in the sample that are completed within 18 months.

\begin{table}[ht!]
  \centering
    \caption{Non-employment spells - basic statistics}
  {\small
  \begin{tabular}{lccc}
     & Num. obs & Occ. mobility (\%) & Job finding rate (\%) \\
     \hline
     (1) U  & 12,278 & 44.4 & 23.1 \\
     (2) UNU  & 14,861 & 45.3 & 19.7 \\
     (3) UN & 16,106 & 45.9 & 18.1 \\
     (4) NU & 17,579 & 46.4 &17.2\\
     (5) N* & 18,559 & 46.0 & 17.5 \\
     (6) NUN & 19,060 & 47.0  & 15.9\\
     (7) N & 27,931 & 43.2 & 16.7 \\
     \hline
   \end{tabular}}
  \label{app:nonemployment:t1}
\end{table}

Column (8) of Table \ref{tab:basic_demog} reports the average mobility rate and duration coefficient for the `NUN' case using the major occupation categories of the 2000 SOC. It shows that the mobility-duration profile is very similar to the one obtained when considering only unemployment spells in between jobs. Table \ref{app:nonemployment:t1} then compares the `NUN' case with the rest of the cases described above. The first column shows the amount of eligible spells for the 1983-2013 period, where we have arranged the number of spells in ascending order. The number of eligible spells takes into account that we do not want to create a bias because of left-censoring. Therefore we only count those spells that end after more than 16 months into the sample. At the same time we want to ensure that the relevant observations are not too close to the end of the panel, at least 1 year away from it, to avoid biases due to right-censoring.\footnote{When controlling for non-/unemployment duration in the regressions, censoring is less of an issue. In that case we use a less stringent selection criterion, and as a result we have more observations, as can be seen in the first line of Table \ref{tab:basic_demog}.} Analogously to our treatment of unemployment spells, we consider workers to enter non-employment only if they have not been employed for more than a month.

\begin{figure}[ht!]
\centering
\subfloat[U, NUN, UN, NU]{\label{Nonemp1} \includegraphics [width=0.5 \textwidth] {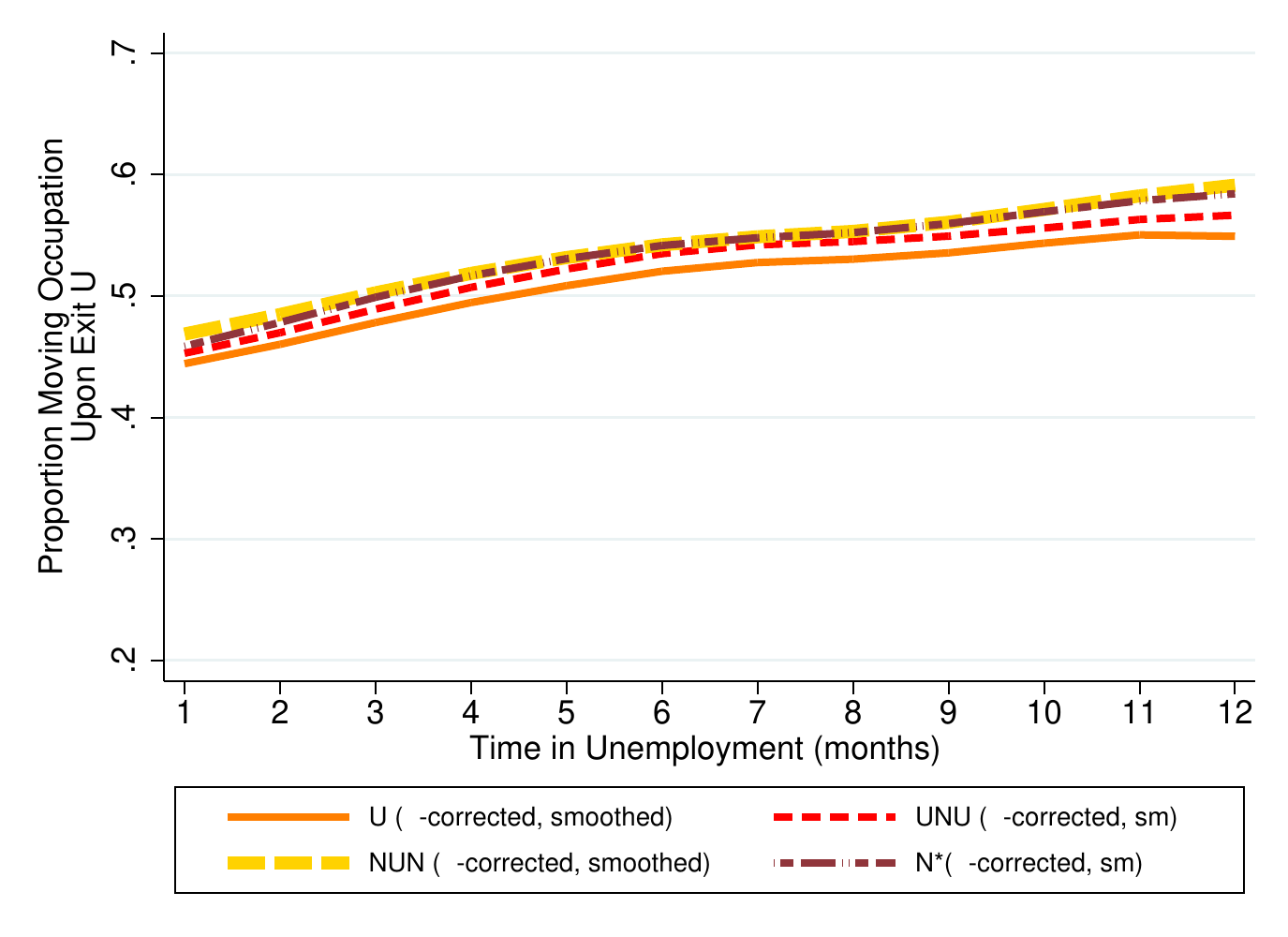}}
\subfloat[UNU, N*, N]{\label{Nonemp2} \includegraphics [width=0.5 \textwidth] {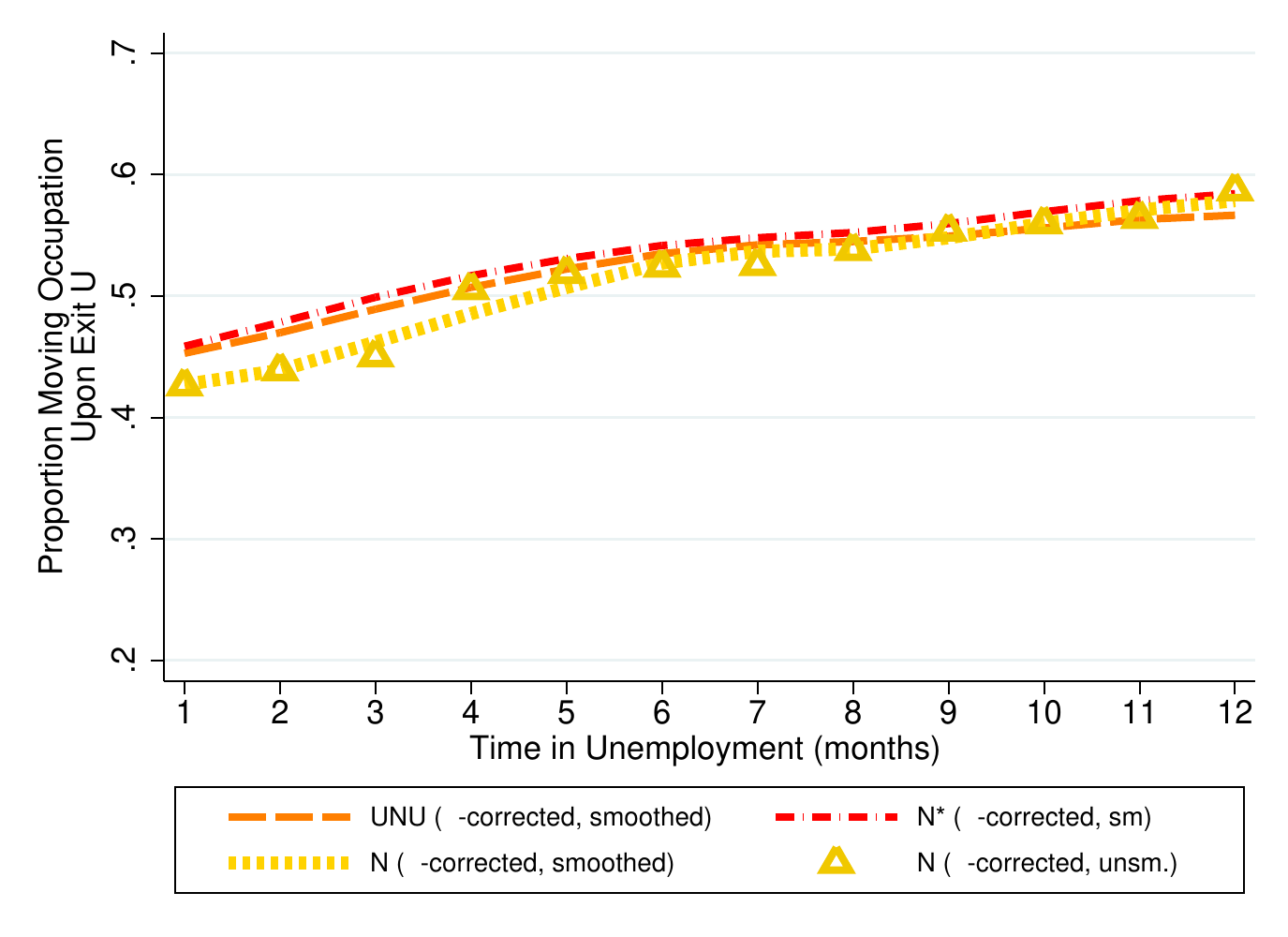}}
\caption{Extent of mobility by non-employment duration - $\Gamma$ corrected}
\label{f:occ_non_emp1}
\end{figure}

The second column of Table \ref{app:nonemployment:t1} shows the $\Gamma$-corrected average occupational mobility rates of all those workers who had at least one month in non-employment before regaining employment. We can readily observe very similar and high average occupational mobility rates across different degrees of labor market attachment. Figure \ref{f:occ_non_emp1} shows the $\Gamma$-corrected mobility-duration profiles for all these cases. It is immediate that these profiles are also very similar to each other. Overall, this means that the concern that our baseline restriction to (pure) unemployment is selecting occupational movers at long unemployment durations appears not to be supported by the data. Workers who spend part of their non-employment spell outside the labor force have very similar mobility-duration profiles as those who spend all their non-employment spell actively looking for work.\footnote{Note, however, that including non-employment spells that are spent wholly outside of the labor force (`N' spells), leads to a somewhat steeper mobility-duration profile. This is visible especially when we consider the non-smoothed observations in Figure \ref{Nonemp2}. This difference originates entirely by the set of short completed spells ($\leq$ three months) in which workers were exclusively out of the labor force and exhibited a significantly lower probability of changing occupation at re-employment. These short `pure N'-spells may reflect employer-to-employer moves with a delayed start. After three months the`N' mobility-duration profile then follows the other profiles.}

Although not shown in the figure, we also observe that when using non-employment spells the $\Gamma$-corrected and uncorrected mobility-duration profiles relate to each other in a similar way as documented when using pure unemployment spells. For example, in the case of the `UNU' spells we find a 7.1 percentage points average difference between the corrected and uncorrected profiles, while for the `NUN' spells the average difference is of 6.7 percentage points.

\subsection{Demographics}

We now investigate the extent to which different demographic groups have different propensities to change occupations and how these propensities change with unemployment (non-employment) duration. In particular, we want to know whether there is evidence of a demographic composition effect driving the aggregate mobility-duration profile. To analyse the impact of demographic characteristics we use the uncorrected, individual-based, data and augment equation \eqref{e:app_basic} by including dummies for gender, race and education, a quartic in age, a linear time trend, and dummies for the classification used to report occupations (industries) in each panel. The resulting regression is then given by
\begin{align}
 \mathbf{1}_{\text{occmob}} &= \beta_0 + \beta_{\text{dur}} \ \text{duration of U (or N) spell} + \beta_{\text{educ}} \text{dum}_{\text{Education}} + \beta_{\text{race}} \text{dum}_{\text{Race}} \nonumber \\
 &\qquad + \beta_{\text{sex}} \text{dum}_{\text{Gender}} +\beta_{\text{time}} \text{Quarter}+ \beta_{cls} \text{dum}_{\text{classification}} +\beta_{\text{age}}(\text{Quartic in Age}) + \varepsilon, \tag{R2} \label{e:app_demctrls}
\end{align}
where as a baseline we chose white high school educated male individuals. The demographic dummies attempt to capture any fixed characteristics that differentiate the average probability of an occupational change across these groups. Further, if some demographic characteristics are associated with higher (lower) propensities to move occupations at re-employment and if these propensities are positively associated with longer (shorter) unemployment durations, the inclusion of the demographic dummies will also capture any composition effects that could be driving the slope of the mobility-duration profile. Hence significant changes to the estimated value of the duration coefficient, $\beta_{\text{dur}}$, when using equation \eqref{e:app_demctrls} instead of \eqref{e:app_basic} would suggest that demographic composition effects are at work.

\paragraph{Gender and Education}

The first row of panel C of Table \ref{tab:basic_demog} shows that the inclusion of the gender and education dummies do not meaningfully affect the coefficient on unemployment duration. Across all columns, the point estimates of the duration coefficient are marginally higher relative to the point estimates in panel B and their differences are not statistically significant. Therefore we do not find evidence that the increase in occupational mobility with unemployment duration is a result of different demographic composition among those that re-gain employment at different unemployment durations. The rest of the estimated coefficients in panel C show that the average probability of an occupational change only differs a few percentage points across gender and education groups. For example, using the 2000 SOC the average occupational mobility rate of females is 2.1 percentage points higher than for males; while the average occupational mobility rate of a high school graduate is around 4.2 percentage points higher than for high school dropouts and 3.4 percentage points lower than college graduates.\footnote{The exception is the college graduates group in the 3-category task-based classification (see column (8)). This arises because in this case we have aggregated all cognitive occupations, which represent the bulk of occupations chosen by college graduates, into one task-based group. The remaining mobility of these workers is therefore mostly into (or from) manual occupations. Around 10\% of college graduates ($\Gamma$-corrected measure) move across these task-based groups rising to about 15\% for long-term unemployed workers.}

Panel D shows the estimates of augmenting equation \eqref{e:app_demctrls} by interactions between the completed unemployment duration, gender and education. This is done to investigate whether the different demographic groups exhibit different slopes in their respective mobility-duration profiles. The lack of statistical significance of the interaction terms then suggests that the mobility-duration profiles specific to the gender and educational subgroups exhibit similar slopes. This is further confirmed in panel E, where we test the joint equality of the duration coefficients across these demographic groups. We cannot reject that the duration coefficients are equal, even at higher p-values.

\begin{figure}[ht!]
\centering
\subfloat[Gender]{\label{f:gender} \includegraphics [width=0.5 \textwidth] {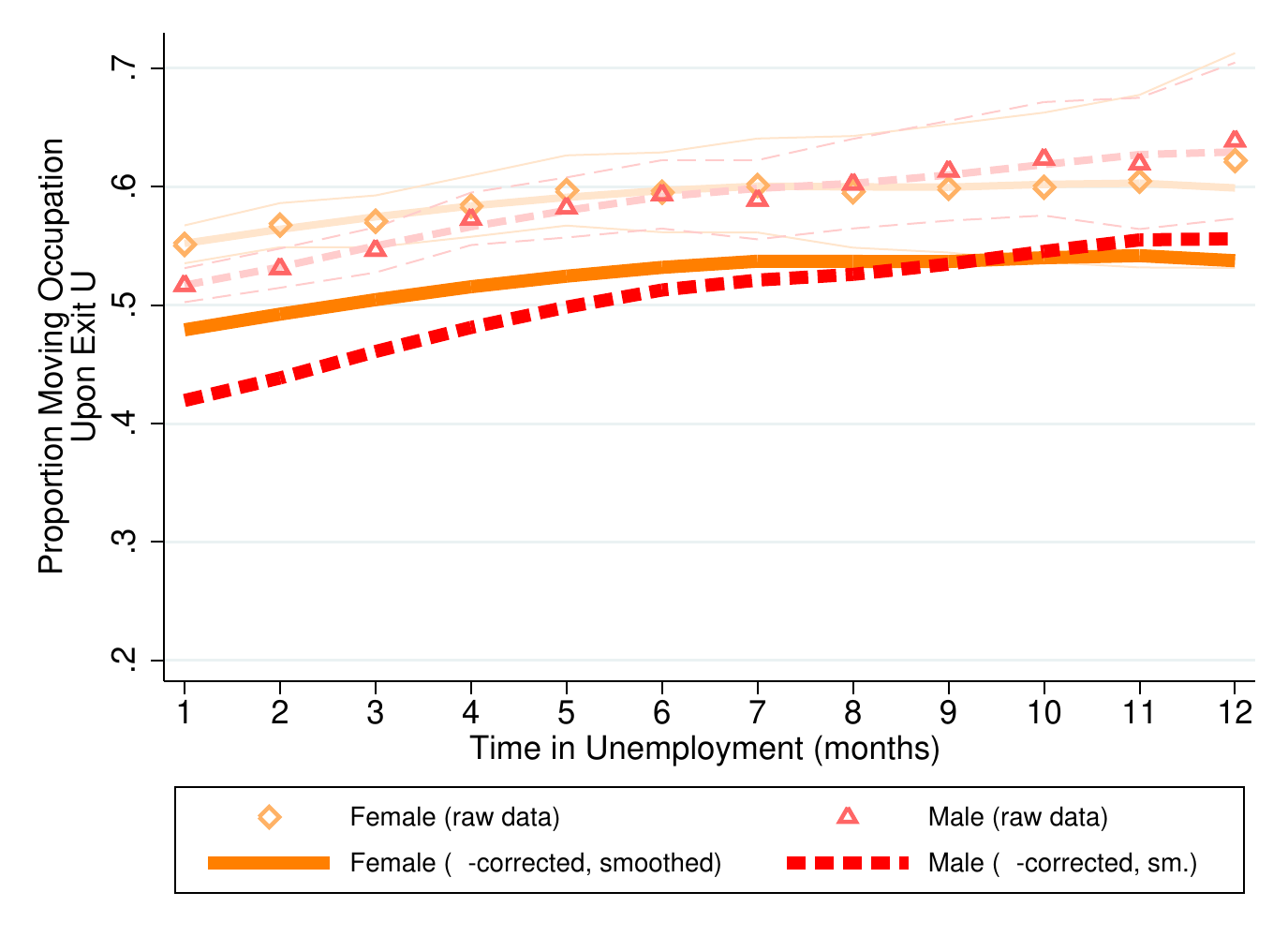}}
\subfloat[High school graduates and college graduates]{\label{f:education} \includegraphics [width=0.5 \textwidth] {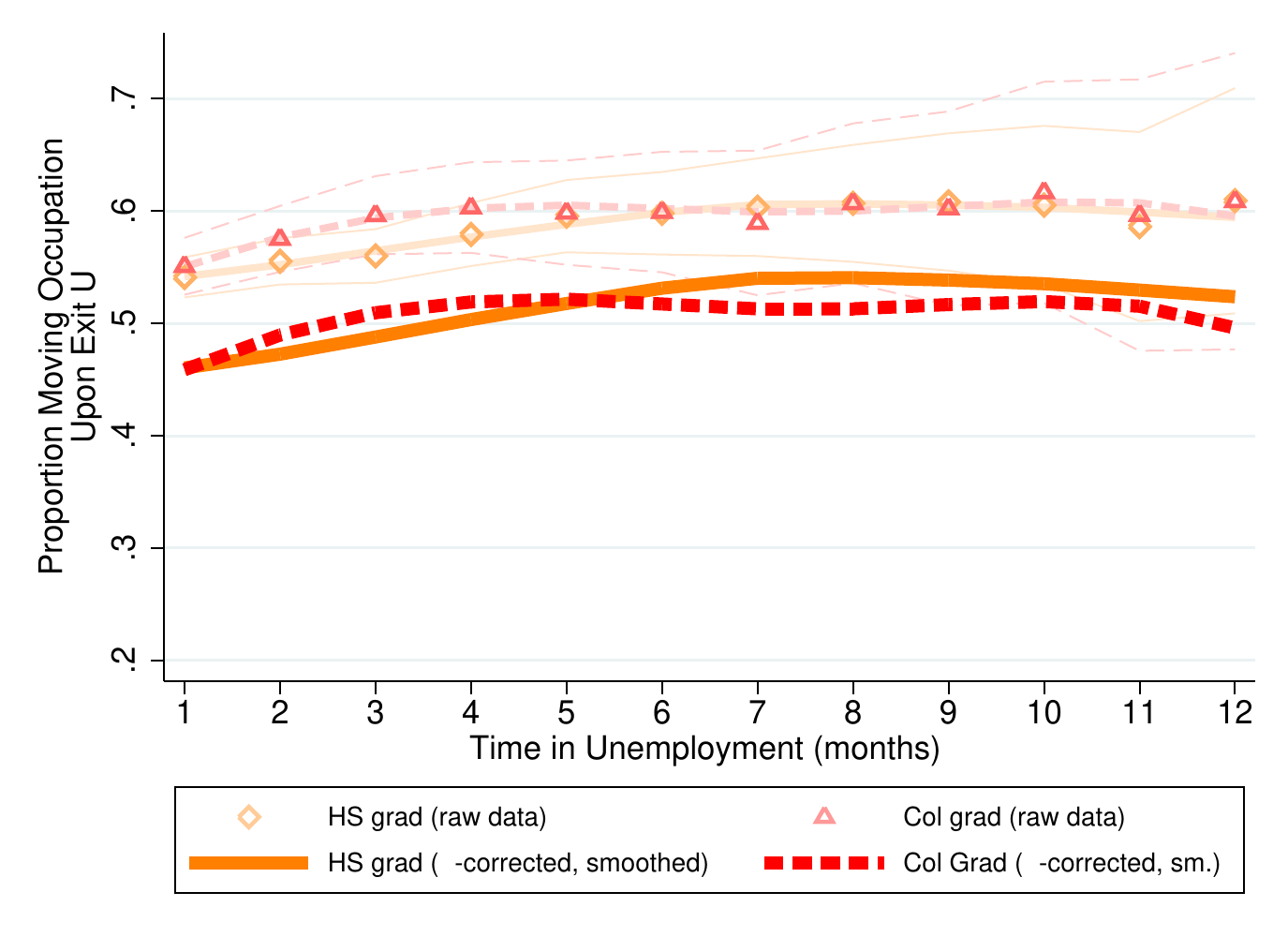}}
\caption{The mobility-duration profile by gender and education groups}
\label{f:main_duration_gender_educ}
\end{figure}

Figure \ref{f:main_duration_gender_educ} depicts the mobility-duration profiles by gender and two education subgroups. Figure \ref{f:gender} depicts the occupational mobility-duration profile by gender, while Figure \ref{f:education} depicts the profiles for high school graduates and college graduates. We observe that across these demographics occupational mobility is high, above 40\%. Moreover, although the slope is somewhat stronger for males, the increase of occupational mobility with unemployment duration is moderate, in the sense longer-term unemployed more often change occupations, yet between 45-50\% will still return to the previous occupation, across gender and education.

\paragraph{Linear Time Trend}

The last two rows of panel C of Table \ref{tab:basic_demog} shows that the average occupational mobility rate of the unemployed has been increasing over time. The estimated coefficients are statistically significant and economically meaningful. This is consistent with the rise in \emph{overall} occupational mobility documented by Kambourov and Manovskii (2008). As the main focus of our paper is on the cyclical patterns of occupational mobility, we leave the investigation of these long-run increase for future research. However, we note that controlling for a linear time trend does not have a major impact on the documented behavior of the mobility-duration profile and its cyclical responsiveness.

\paragraph{Age groups}

\begin{figure}[ht!]
\centering
\subfloat[Young workers]{\label{f:young} \includegraphics [width=0.5 \textwidth] {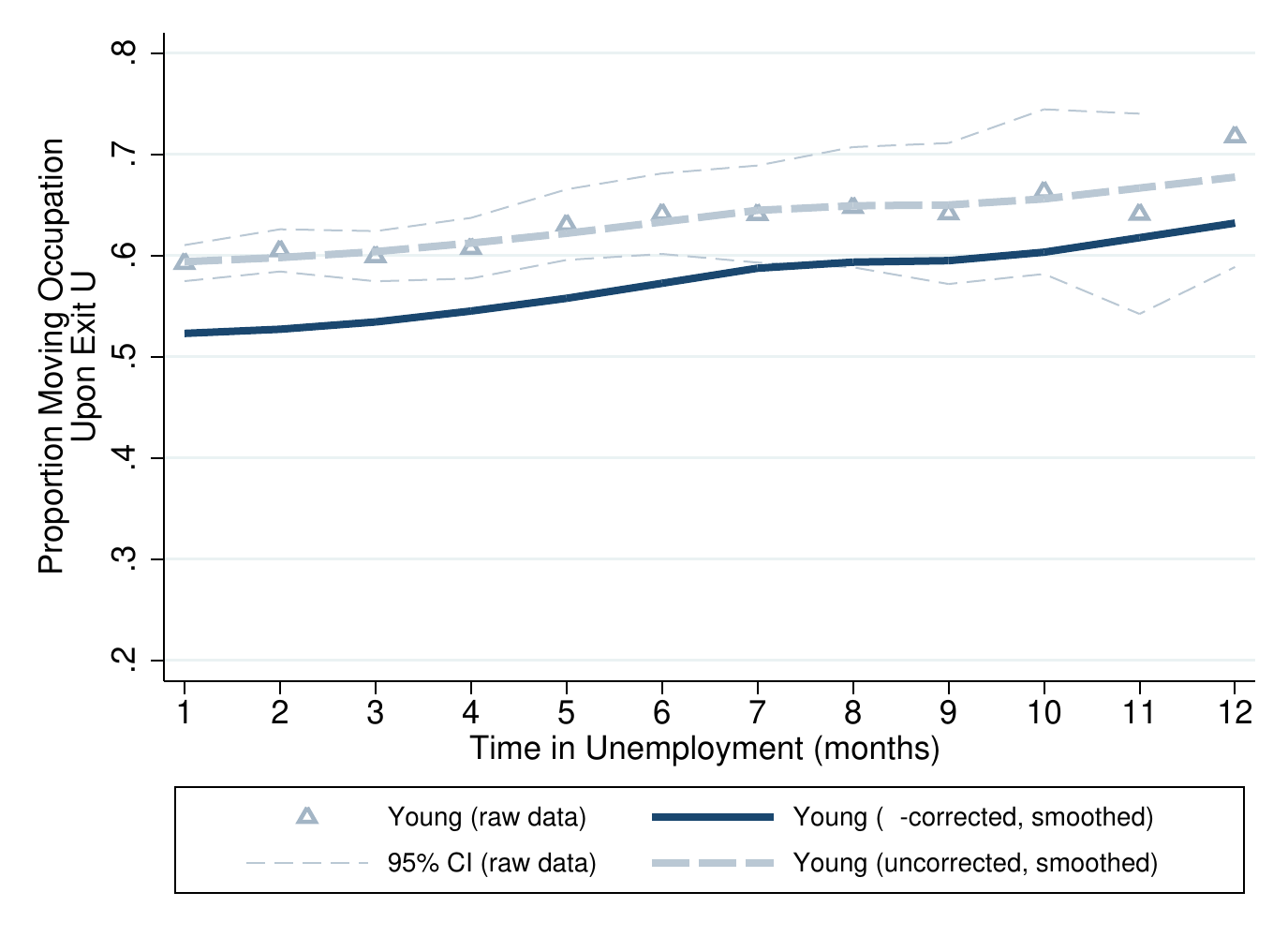}}
\subfloat[Prime-aged workers]{\label{f:prime} \includegraphics [width=0.5 \textwidth] {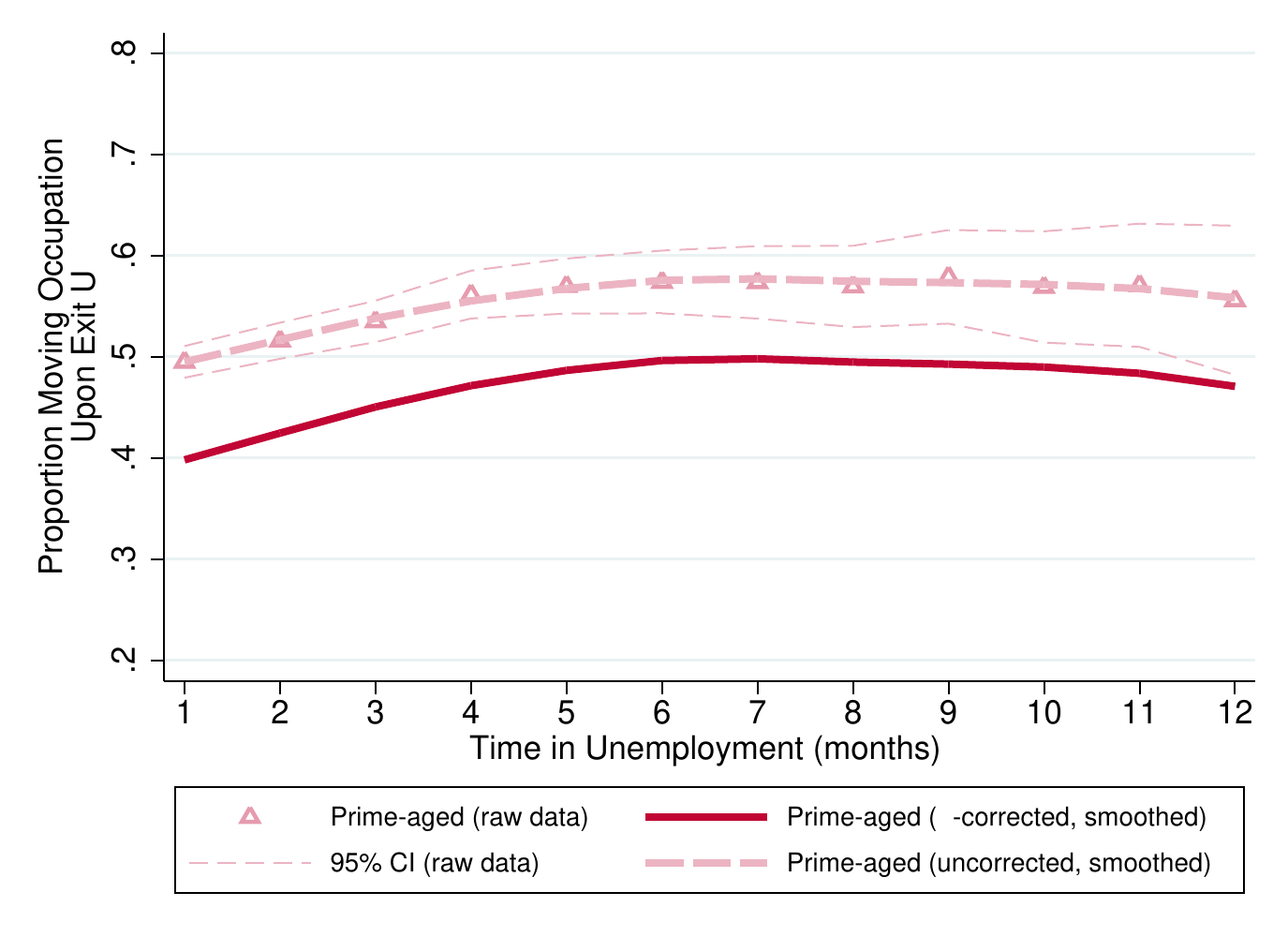}}
\caption{The mobility-duration profile by age groups}
\label{f:age}
\end{figure}

The most significant difference across all the demographic groups considered is between young and prime-aged workers. We define young workers as those who left education (and hence fully entered the labor market) and are between 20-30 years. Prime-aged workers are those workers who are between 35 and 55 years of age. Figure \ref{f:age} depicts the uncorrected and $\Gamma$-corrected mobility-duration profiles of these workers.

Panel A of Table \ref{tab:occmob_age} reports the average occupational mobility rate of young and prime-aged workers, after correcting for miscoding. The first two rows show that across classifications the difference between young and prime-aged workers' average occupational mobility rates is about 10 percentage points. The second two rows show that this difference diminishes somewhat with unemployment duration. The last two rows summarise all this information by showing the ratio between the occupational mobility rate of those long-term unemployed (with spells of at least 9 months) and the occupational mobility rate of all those unemployed of the same age group. In particular, we observe that across the different classifications young workers with unemployment spells of at least 9 months have an occupational mobility rate that is about 15\% higher than the average mobility rate of all young unemployed workers. In the case of prime-aged workers, this ratio is about 20\%. Panel B shows very similar conclusions for the uncorrected data.

\begin{table}[ht!]
  \footnotesize
  \centering
  \caption{Occupational mobility of young and prime-aged }
   \resizebox{1.0\textwidth}{!}{
        \begin{tabular}{lcccccccc}
    \toprule
    \toprule
    & 2000 SOC    & 1990 SOC    & NR/R-M/C & NR/R-M/C* & C/NRM/RM & OCC*IND & IND   & 2000 SOC-NUN \\
          & (1)   & (2)   & (3)   & (4)   & (5)   & (6)   & (7)   & (8) \\
    \midrule
    \midrule
    \multicolumn{9}{c}{Panel A: Overall mobility of different age groups (corrected)} \\
    \midrule
    young -all & 0.521 & 0.489 & 0.335 & 0.365 & 0.305 & 0.443 & 0.533 & 0.528 \\
    prime -all & 0.398 & 0.375 & 0.229 & 0.250 & 0.179 & 0.344 & 0.447 & 0.428 \\
    young -9+ months & 0.585 & 0.582 & 0.383 & 0.439 & 0.354 & 0.511 & 0.636 & 0.598 \\
    prime -9+ months & 0.499 & 0.489 & 0.300 & 0.331 & 0.255 & 0.435 & 0.562 & 0.531 \\
    \midrule
    \multicolumn{9}{c}{relative mobility increase Unemp 9mth+ / All Unemp} \\
    \midrule
    \multicolumn{1}{c}{young} & 0.114 & 0.173 & 0.136 & 0.185 & 0.148 & 0.143 & 0.176 & 0.124 \\
    \multicolumn{1}{c}{prime-aged} & 0.225 & 0.265 & 0.270 & 0.278 & 0.349 & 0.235 & 0.229 & 0.214 \\
\midrule    \multicolumn{9}{c}{Panel B: Overall mobility of different age groups (uncorrected for miscoding)} \\
    \midrule
    young -all & 0.593 & 0.583 & 0.598 & 0.443 & 0.571 & 0.393 & 0.424 & 0.350 \\
    prime -all & 0.495 & 0.485 & 0.518 & 0.344 & 0.497 & 0.303 & 0.327 & 0.238 \\
    young -9+ months & 0.642 & 0.647 & 0.655 & 0.511 & 0.663 & 0.432 & 0.485 & 0.393 \\
    prime -9+ months & 0.579 & 0.569 & 0.600 & 0.435 & 0.597 & 0.363 & 0.396 & 0.304 \\
    \midrule
    \multicolumn{9}{c}{Panel C: Regression, uncorrected, no demog, no time controls, no occ/ind controls} \\
    \midrule
    prime-aged dum & -0.1032*** & -0.1031*** & -0.0938*** & -0.1006*** & -0.0714*** & -0.0891*** & -0.0944*** & -0.1087*** \\
    (s.e.) & (0.0092) & (0.0091) & (0.0077) & (0.0091) & (0.0094) & (0.0088) & (0.0089) & (0.0084) \\
    \midrule
    \multicolumn{9}{c}{Panel D: Regression, uncorrected, demog and time controls, no occ/ind controls} \\
    \midrule
    prime-aged dum & -0.1121*** & -0.1077*** & -0.1004*** & -0.1038*** & -0.0783*** & -0.1001*** & -0.1014*** & -0.1002*** \\
    (s.e.) & (0.0093) & (0.0093) & (0.0078) & (0.0093) & (0.0095) & (0.0089) & (0.0090) & (0.0085) \\
    \midrule
    \multicolumn{9}{c}{Panel E: Regression, uncorrected, interactions with demog, time controls, and occ/ind controls} \\
    \midrule
    prime-aged dum & -0.1359*** & -0.1188*** & -0.1212*** & -0.1454*** & -0.1138*** & -0.0846*** & -0.0989*** & -0.1043*** \\
    (s.e.) & (0.0164) & (0.0168) & (0.0143) & (0.0163) & (0.0165) & (0.0153) & (0.0159) & (0.0155) \\
    gender*prm age & 0.0252 & 0.0187 & 0.0354** & 0.0356* & 0.0420** & -0.0055 & -0.0015 & -0.0126 \\
    (s.e.) & (0.0182) & (0.0183) & (0.0153) & (0.0184) & (0.0182) & (0.0178) & (0.0181) & (0.0168) \\
    hs drop*prm age & -0.0150 & -0.0322 & -0.0112 & 0.0180 & -0.0046 & -0.0251 & -0.0209 & -0.0145 \\
    (s.e.) & (0.0254) & (0.0254) & (0.0213) & (0.0256) & (0.0255) & (0.0225) & (0.0240) & (0.0229) \\
    some col*prm age & 0.0256 & -0.0061 & 0.0200 & 0.0342 & 0.0281 & 0.0003 & -0.0003 & 0.0171 \\
    (s.e.) & (0.0231) & (0.0232) & (0.0196) & (0.0236) & (0.0237) & (0.0223) & (0.0230) & (0.0225) \\
    col grad*prm age & 0.0267 & 0.0075 & -0.0085 & 0.0348 & 0.0164 & -0.0094 & -0.0029 & 0.0668*** \\
    (s.e.) & (0.0266) & (0.0268) & (0.0220) & (0.0268) & (0.0266) & (0.0267) & (0.0269) & (0.0221) \\
    black*prmage  & 0.0176 & 0.0091 & 0.0114 & 0.0412 & 0.0285 & -0.0269 & -0.0118 & 0.0012 \\
    (s.e.) & (0.0266) & (0.0261) & (0.0221) & (0.0273) & (0.0269) & (0.0254) & (0.0262) & (0.0253) \\
    \midrule
    \multicolumn{9}{c}{Panel F: Regression, uncorrected, no demog, no time  and no occ/ind controls, interaction coeff dur and age} \\
    \midrule
    dur*prm. age & 0.0060* & 0.0039 & 0.0073*** & 0.0076** & 0.0056 & 0.0053 & 0.0049 & 0.0060* \\
    (s.e.) & (0.0034) & (0.0034) & (0.0023) & (0.0035) & (0.0035) & (0.0035) & (0.0035) & (0.0033) \\
    \midrule
    \multicolumn{9}{c}{Panel G: Regression, uncorrected, with demog, time  and occ/ind controls, interaction coeff dur and age} \\
    \midrule
    dur*prm. age & 0.0048 & 0.0032 & 0.0066*** & 0.0071** & 0.0045 & 0.0051 & 0.0047 & 0.0073** \\
    (s.e.) & (0.0033) & (0.0033) & (0.0022) & (0.0035) & (0.0034) & (0.0034) & (0.0034) & (0.0033) \\
    \bottomrule
    \bottomrule
    \multicolumn{9}{c}{{\scriptsize *$\ p<0.1$;\ **$\ p<0.05$;\ ***$\ p<0.01$}}
    \end{tabular}}
  \label{tab:occmob_age}%
\end{table}%

Panel C reports the estimated difference between young and prime-aged workers' average occupational mobility rates obtained from regressing $\mathbf{1}_{\text{occmob}}$ on a constant and a dummy that takes the value one if the worker belongs to the prime-age group and zero otherwise. In this case we restrict to the sample that only contains either young or prime-aged workers. As in panel A, we find that the difference between the occupational mobility rates of young and prime-aged workers is of about 10 percentage points. Panel D shows the estimated coefficient of the prime-age dummy when augmenting this regression with controls for demographic characteristics (gender, race, education) and a linear time trend. In this case we find that adding these additional regressors does not meaningfully affect the estimated difference between the average occupational mobility rates of young and prime-aged workers.

The first row of panel E reports the estimated coefficient on the prime-age dummy when augmenting the regression underlying panel D with occupational/industry controls and interactions between gender, education and race with the prime-age dummy. In this case we find a slightly larger difference between the occupational mobility rates of young and prime-aged workers, with the exception of the task-based classification. The remainder rows of panel E show the estimated coefficients of the interaction terms. The interaction terms allow us to investigate whether the lower occupational mobility rate of prime-aged workers can be explain by life-cycle shifts towards certain occupations, or is concentrated in certain demographic groups. Given the lack of statistical significance of most of the coefficients, we do not find strong evidence for either explanation.\footnote{An exception being for gender in the cases of  `NUN' spells, industry mobility and simultaneous occupation and industry mobility. Here we find a drop of 2 to 3 percentage points. Another exception is for college graduates in the case of the 3 task-based category classification. Here all cognitive occupations are merged into one task-based group, limiting overall occupational mobility and its responsiveness of college graduates.}

Panels F and G consider the interaction between age and unemployment duration. In panel F (G) we add an interaction term between the prime-aged dummy and unemployment duration to the regression underlying the estimates in panel C (E). In both cases, we observe that the point estimates indicate a steeper slope for prime-aged workers, as suggested by the last two rows of panel A. However, the estimated coefficients are small and are not always statistically significant.

\subsection{Occupation identities}

We now turn to investigate whether the aggregate mobility-duration profile is driven by composition effects at the level of individual occupations. In particular, we want to know to what extent a subset of occupations is associated with high occupational mobility rates and longer non-employment durations, while another subset is associated with lower mobility rates and shorter non-employment durations. If this were to be the case, one could potentially explain the aggregate mobility-duration profile as a result of selection effects across occupations.

Figures \ref{f:netgross_mmo_dd_ind} and \ref{f:netgross_4cat} (in Section 2) show the average gross occupational mobility rates by major occupations and task-based occupations together with the corresponding overall average occupational mobility rate. The height of each light colored bar corresponds to the average gross occupation(industry)-specific mobility rate, while the width of each bar corresponds to the proportion of the inflow into unemployment that originate from a given occupation (industry). The light colored horizontal line depicts the $\Gamma$-corrected occupational (industry) mobility rate.

The graphs in Figure \ref{f:netgross_mmo_dd_ind} show that across the vast majority of occupations (and industries) the extent of gross mobility is high. The $\Gamma$-corrected occupation-specific mobility rate in nearly all occupations is either close or above 40\%, covering over 80\% of all unemployment spells. Figures \ref{sf:occdistr22occ} and \ref{sf:22occ-nun} use the 2000 SOC and show that this feature is robust to whether we consider pure unemployment spells or `NUN' spells. Figure \ref{sf:occdistr13occ} shows that this feature is also robust to using the 1990 SOC instead of the 2000 SOC. Figure \ref{sf:inddistr13occ} shows that this feature is also found across major industries. In both industry and occupations, however, the main exception is ``construction'', which exhibits a mobility rate of around 25\%. Figure \ref{f:netgross_4cat} shows that gross occupational mobility is also high and nearly identical across all task-based occupational categories.

To analyse the impact of occupational identities on the slope of the mobility-duration profile, we estimate regressions of the general form:
\begin{align}
 \mathbf{1}_{\text{occmob}} = \beta_0 + \beta_{\text{dur}} \ \text{u.duration} + \beta_{\text{occ}}\text{occ.dum} + \beta_{\text{age}}(\text{Quartic in Age}) + \beta_{\text{dm}} \ \text{demog.ctrls} + \varepsilon, \tag{R3} \label{eq:appocc1}
\end{align}
where ``occ.dum'' denotes occupation identity dummies and the demographic controls include dummies for gender, education and race. If a subset of occupations are associated with higher (lower) mobility probabilities and if these probabilities are positively associated with longer (shorter) unemployment durations, the inclusion of occupation identity dummies will capture composition effects that could be driving the slope of the aggregate mobility-duration profile. We consider two cases when evaluating these occupation dummies: (i) source and (ii) destination occupations. The former are the occupations that were performed by workers immediately before becoming unemployed (non-employed), while the latter are the occupations to which workers got re-employed into. For comparability, panel A of Table \ref{tab:occspec} reports the estimated duration coefficient based on regression \eqref{e:app_demctrls} without occupation identity dummies, as reported in the first row of panel C of Table \ref{tab:basic_demog}. If significant changes to the estimated value of the duration coefficient, $\beta_{\text{dur}}$, were observed when adding occupation identity dummies, this would suggest the presence of composition effects across occupations.

\begin{table}[ht!]
  \centering
    \footnotesize
  \caption{The role of individual occupations}
     \resizebox{1.0\textwidth}{!}{
    \begin{tabular}{lcccccccc}
\toprule
    \toprule
    & 2000 SOC    & 1990 SOC    & NR/R-M/C & NR/R-M/C* & C/NRM/RM & OCC*IND & IND   & 2000 SOC-NUN \\
          & (1)   & (2)   & (3)   & (4)   & (5)   & (6)   & (7)   & (8) \\
    \midrule
    \midrule
    \multicolumn{9}{c}{Panel A: baseline regression, with demog, time controls, but no occ/ind controls} \\
    \bottomrule
        dur coef & 0.0150*** & 0.0156*** & 0.0103*** & 0.0112*** & 0.0085*** & 0.0123*** & 0.0137*** & 0.0145*** \\
    (s.e.) & (0.0015) & (0.0015) & (0.0015) & (0.0015) & (0.0015) & (0.0016) & (0.0016) & (0.0010) \\
    \midrule
    \multicolumn{9}{c}{Panel B: uncorrected, source occupation controls, time and demographic controls} \\
    \midrule
    dur coef & 0.0136*** & 0.0145*** & 0.0106*** & 0.0113*** & 0.0088*** & 0.0109*** & 0.0116*** & 0.0136*** \\
    (s.e.) & (0.0015) & (0.0015) & (0.0015) & (0.0015) & (0.0015) & (0.0016) & (0.0015) & (0.0010) \\
    female & -0.0144 & -0.0315*** & 0.0042 & -0.0059 & -0.0286*** & -0.0166* & 0.0097 & -0.0073 \\
    (s.e.) & (0.0094) & (0.0092) & (0.0091) & (0.0088) & (0.0081) & (0.0098) & (0.0091) & (0.0079) \\
    hs drop & -0.0311*** & -0.0306*** & -0.0386*** & -0.0389*** & -0.0426*** & -0.0160 & -0.0268** & -0.0346*** \\
    (s.e.) & (0.0116) & (0.0117) & (0.0103) & (0.0109) & (0.0105) & (0.0115) & (0.0116) & (0.0098) \\
    some col & 0.0047 & 0.0014 & 0.0056 & 0.0097 & -0.0159 & 0.0088 & 0.0314*** & 0.0104 \\
    (s.e.) & (0.0107) & (0.0107) & (0.0103) & (0.0106) & (0.0102) & (0.0107) & (0.0106) & (0.0091) \\
    col grad & -0.0200 & -0.0391*** & -0.0692*** & -0.0743*** & -0.1219*** & -0.0147 & 0.0078 & -0.0166 \\
    (s.e.) & (0.0137) & (0.0138) & (0.0139) & (0.0138) & (0.0116) & (0.0138) & (0.0123) & (0.0113) \\
    black & 0.0299** & 0.0116 & 0.0241** & 0.0169 & 0.0362*** & 0.0291** & 0.0128 & 0.0240** \\
    (s.e.) & (0.0123) & (0.0120) & (0.0118) & (0.0120) & (0.0116) & (0.0124) & (0.0123) & (0.0102) \\
    time (qtr) & 0.0011*** & 0.0012*** & 0.0008*** & 0.0008*** & 0.0009*** & 0.0010*** & 0.0012*** & 0.0010*** \\
    (s.e.) & (0.0002) & (0.0002) & (0.0002) & (0.0002) & (0.0002) & (0.0002) & (0.0002) & (0.0002) \\
    \midrule
    \multicolumn{9}{c}{Panel C: uncorrected, destination occupation controls, time and demographic controls} \\
    \midrule
    dur coef & 0.0137*** & 0.0144*** & 0.0103*** & 0.0112*** & 0.0088*** & 0.0112*** & 0.0117*** & 0.0132*** \\
    (s.e.) & (0.0015) & (0.0015) & (0.0015) & (0.0015) & (0.0014) & (0.0016) & (0.0015) & (0.0010) \\
    \midrule
    \multicolumn{9}{c}{Panel D: F-test source occupation-specific duration slopes (demog \&time \& source occ controls)} \\
    \midrule
    p-value & 0.545 & 0.77  & 0.494 & 0.679 & 0.85  & 0.913 & 0.808 & 0.188 \\
    \bottomrule
    \bottomrule
    \multicolumn{9}{c}{{\scriptsize *$\ p<0.1$;\ **$\ p<0.05$;\ ***$\ p<0.01$}}
    \end{tabular}}
  \label{tab:occspec}%
\end{table}%

Panel B reports the results from regression \eqref{eq:appocc1} based on workers' \emph{source} occupations. By comparing the duration coefficients of panels A and B, it is immediate that across all occupational classifications the slopes of the mobility-duration profile are hardly affected when adding source occupation fixed effects. The rest of the coefficients in panel B show a similar pattern as the one reported in Table \ref{tab:basic_demog}. Panel C reports the results from regression \eqref{eq:appocc1} based on workers' \emph{destination} occupations. Note that the duration coefficients remain virtually identical to the case in which we use source occupation instead. This suggests that the aggregate mobility-duration profile does not seem to be a result of selection, whereby those occupations with high gross occupational flows are also associated with long unemployment durations.

Panel D further investigates whether the slopes of the mobility-duration profiles are different across source occupations. It reports the results of testing whether the implied slopes of the mobility-duration profiles across source occupations are equal. Using an F-test, we cannot reject the hypothesis that all slopes are equal, with high p-values in many cases.\footnote{The somewhat lower p-value for the `NUN' measure (0.25) is driven by women in the education/library occupational category. Excluding this occupation yields a p-value for equality of slopes in all remaining occupations of 0.66.} Although not shown here, we also tested whether the semi-elasticity of the \emph{relative} increase in occupational mobility associated with a one-month higher duration is equal across occupations. Once again we find that we cannot reject the hypothesis that they are equal across occupations and this result holds for all classifications and samples used in Table \ref{tab:occspec}.

\section{Excess and net occupational mobility}

In this section we show that excess occupational mobility accounts for the vast majority of the gross mobility documented in Section 1 and is the main driver of the mobility-duration profile. We also show that although the extent of net mobility is small compared to the extent of gross mobility or the overall amount of unemployment spells, it exhibits a well defined pattern. Consistent with the job polarization literature we observe that during the 1983-2013 period routine manual occupations have experienced net outflows, while non-routine manual occupations have experienced net inflows. At the same time we find that routine cognitive occupations have experienced net inflows, while non-routine cognitive occupations have experienced net outflows. We document the importance of ``management'' occupations in driving the net mobility patterns within the set of cognitive occupations.

\subsection{Net and gross flows per occupation}

\begin{figure}[hp!]
\centering
\subfloat[Major Occupational Groups (2000 SOC) - Unemployed]{\label{sf:occdistr22occ} \includegraphics[width=0.65\textwidth] {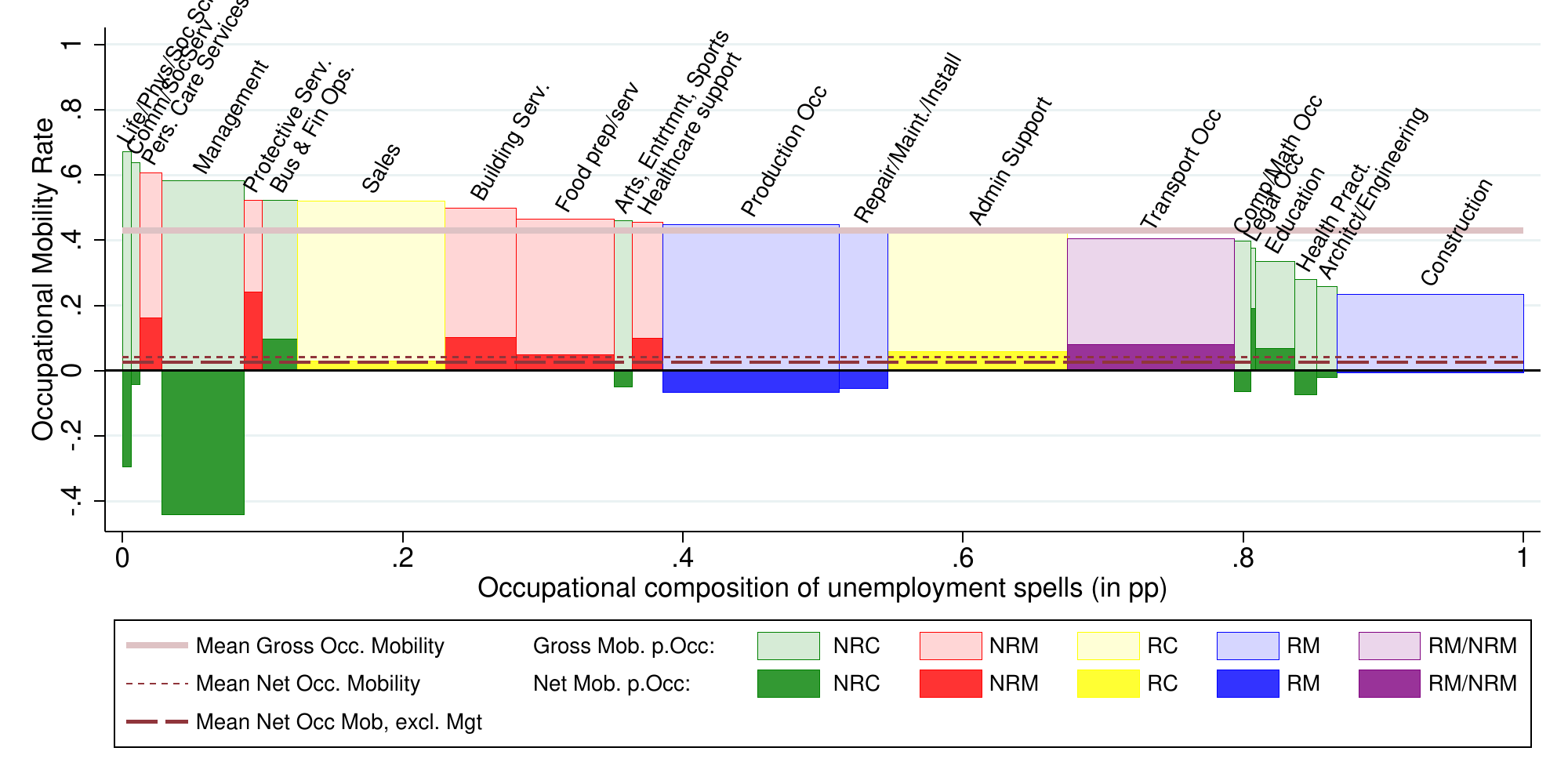}}

\subfloat[Major Occupational Groups (2000 SOC) - `NUN']{\label{sf:22occ-nun} \includegraphics[width=0.65\textwidth] {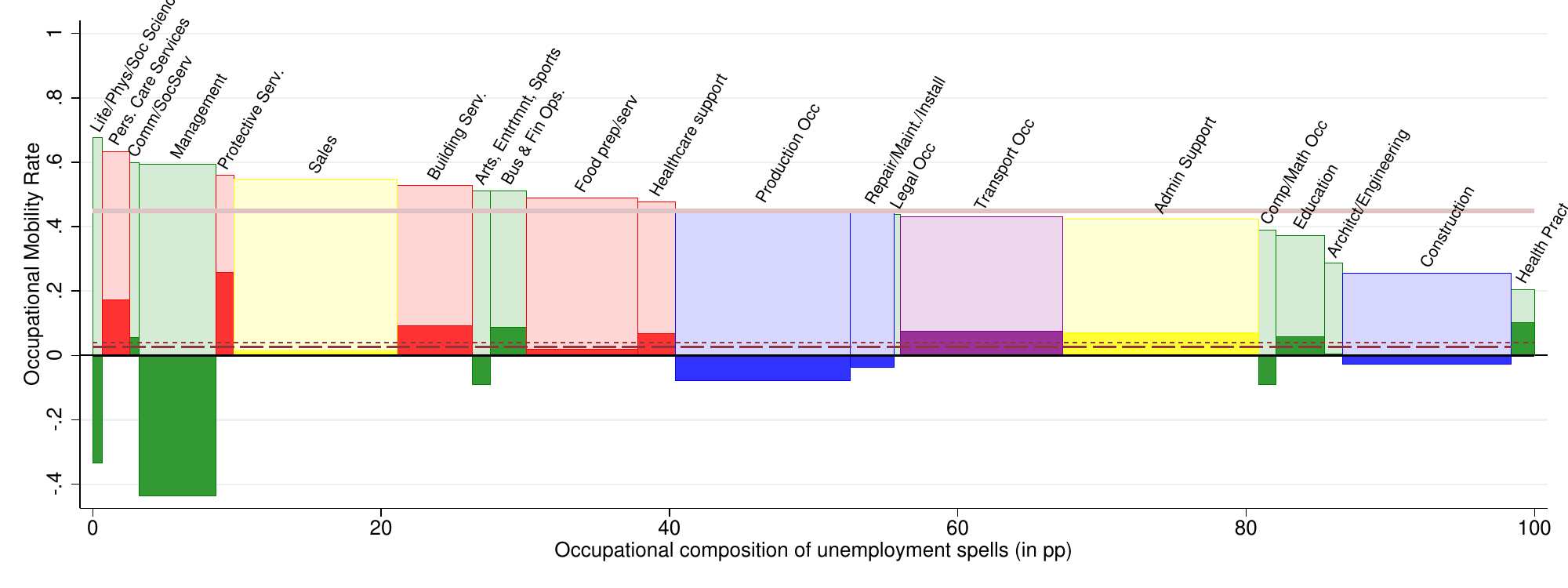}}

\subfloat[Major Occupational Groups (1990 SOC) - Unemployed]{\label{sf:occdistr13occ} \includegraphics[width=0.65\textwidth] {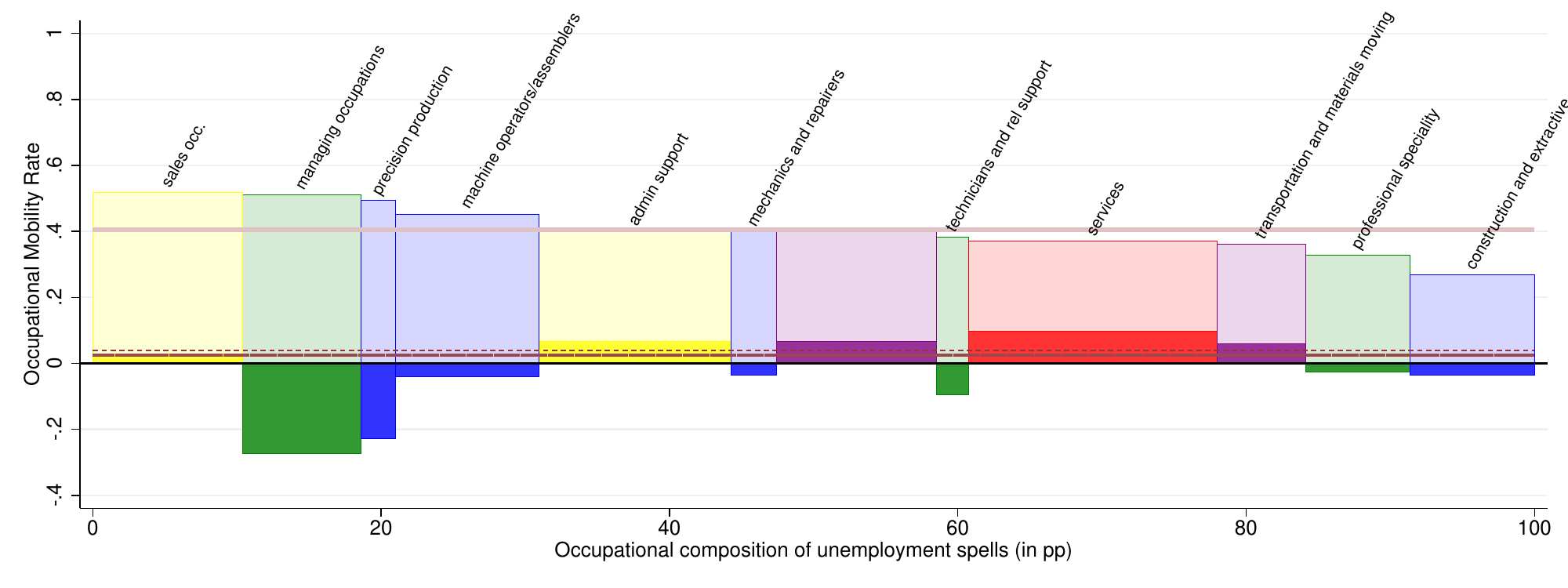}}

\subfloat[Major Industry Groups  - Unemployed]{\label{sf:inddistr13occ} \includegraphics[width=0.65\textwidth] {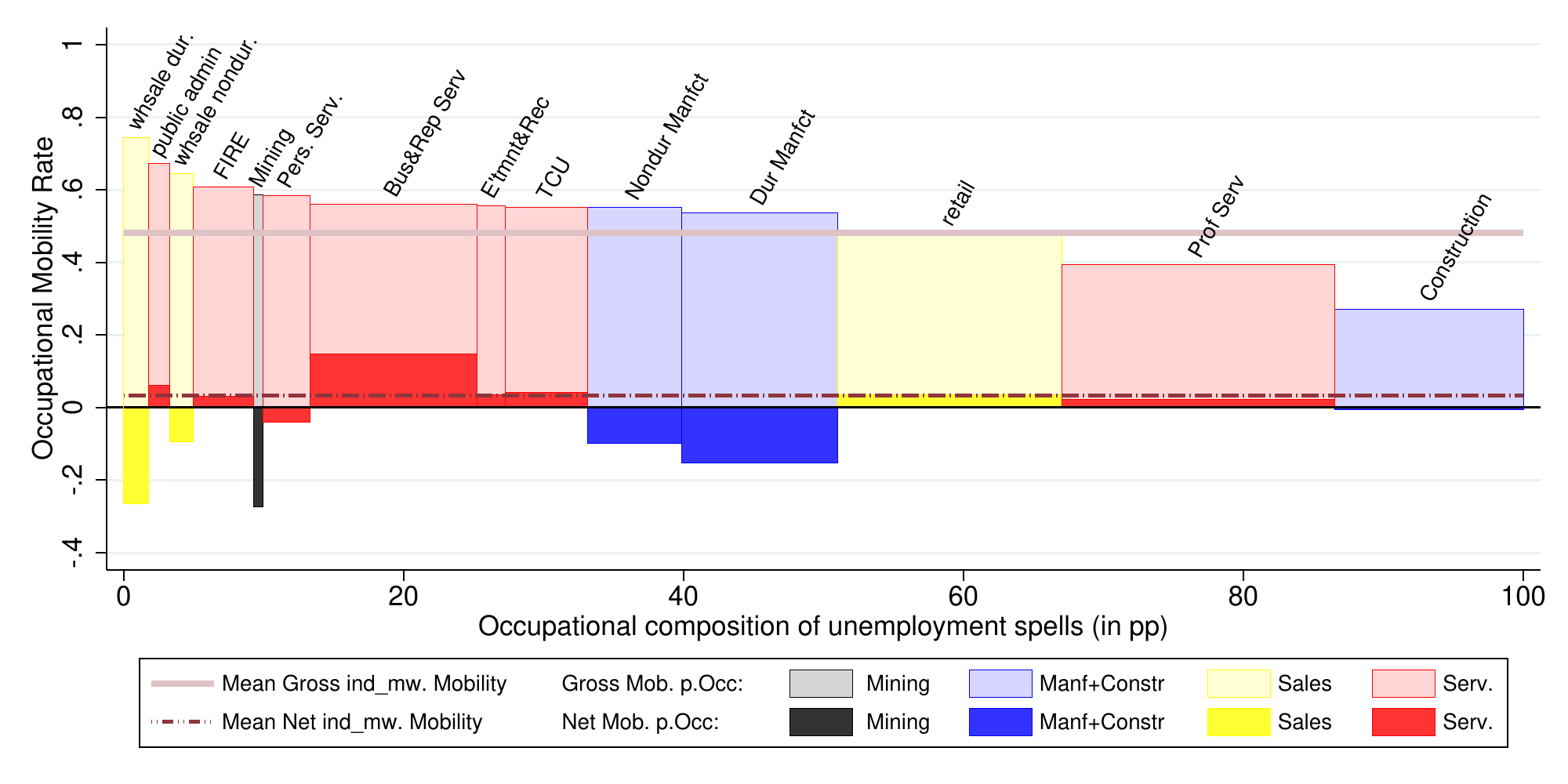}}
\caption{Net and Gross Mobility per Occupation (Industry)}
\label{f:netgross_mmo_dd_ind}
\end{figure}

Figure \ref{f:netgross_mmo_dd_ind} depicts the $\Gamma$-corrected gross and net occupational mobility per individual major occupation and industry. The width of each bar corresponds to the proportion of workers' unemployment spells that originate from a given occupation (industry) among all workers' unemployment spells in our sample. The height of each \emph{light} colored bar corresponds to the proportion of workers' unemployment spells that originate from a given occupation and that end with an occupational change. That is, the height of each light colored bar measures the occupation-specific gross mobility rate. On the other hand, the height of each \emph{dark} colored bar corresponds to the proportion of the workers' unemployment spells that originate from a given occupation that cover the total net flows from that occupation. A positive value for the height of a dark colored bar refers to inflows, while a negative value refers to outflows. The area of each light (dark) colored bar then gives the occupation-specific gross (net) flows as a proportion of all workers unemployment spells. It is important to note that a net flow appears twice on the graph, once as an outflow and then as an inflow. It is also important to note that because the SIPP has a longitudinal dimension a workers may have more than one unemployment spell in which he ends up changing occupation. This is the reason why we describe our measures of mobility based workers' unemployment spells. Total net flows are then obtained as the absolute value of the sum of all occupation-specific outflows, or the sum of all occupation-specific inflows, or the absolute value of the sum of both divided by two. Total gross flows are obtained as the sum of all occupation-specific gross flows. The dashed lines depict total net flows as a proportion of all unemployment spells. The light colored line depicts the average gross occupational mobility rate.

It is evident that total gross flows are much larger than total net flows. Using the 2000 SOC, for example, the proportion of all workers unemployment spells in our sample that cover total net flows is just 4.2\%, while the proportion of all workers' unemployment spells in our sample that cover total gross flows is 44.4\% (see Table \ref{tab:basic_demog}). The proportion of all gross occupational flows that are necessary to generate the observe net flows between major occupations is then 9.5\%. When considering ``NUN'' spells the proportion of all workers' unemployment spells that cover total net flows is 4.1\%, while the proportion of total gross flows is 9.1\%. Using the 1990 SOC and considering unemployment spells we find that these proportions hardly change: 4\% and 9.9\%, respectively. We find that when considering industry mobility these proportions are even smaller: 3.4\% and 7\%, respectively. Taken together this evidence implies that about 95\% of workers' unemployment spells and about 90\% of all gross occupational flows are driven by excess mobility.\footnote{The overwhelming importance of excess relative to net mobility we document is consistent with the results of Murphy and Topel (1987), Jovanovic and Moffitt (1990) and Kambourov and Manovskii (2008), who obtained large differences between excess and net mobility on pooled samples of employer movers and stayers.}

It is also evident that the importance of excess relative to net mobility occurs across nearly all major occupations (industries). The main exception to this pattern is ``management''. This is clearest when using the 2000 SOC. Figures \ref{sf:occdistr22occ} and \ref{sf:22occ-nun} show that the proportion of workers' unemployment spells that originate from ``management'' that cover the total net flows of that occupation is around 40\%. The latter reflects two underlying forces. (i) A high outflow rate: around 62\% of workers who lose their jobs as managers change occupation after their unemployment (non-employment) spell, where the majority of the ``management'' outflows end up in ``sales and related occupations'' (22.1\%), ``office and admin support'' (19.9\%), ``business and financial operators'' (13\%) and ``food preparation/serving'' (12.7\%). (ii) A very small inflow rate: very few unemployed workers from other major occupations obtain jobs as managers at re-employment. We find that less than 1\% of all unemployment spells end up in non-managers becoming managers. Excluding all flows involving ``management'' (7\% of all workers' unemployment spells) implies that now 2.6\% (instead of 4.2\%) of all unemployment spells and 6.1\% (instead of 9.5\%) of all gross occupational flows are needed to generate the observed net flows among the remainder occupations.\footnote{Using the 1990 SOC we obtain a lower net mobility rate because in this categorisation ``management'' includes the 2000 SOC ``management'' and ``business and financial operators'' occupations. Also note that in the 1990 SOC both ``transportation'' and ``helpers/laborers'' are in purple, reflecting that the average routine intensity score of ``helpers/laborers'' is also low (see Section A.1).}

\begin{figure}[ht!]
\centering
\subfloat[RM/NRM/RC/NRC incl. Management]{\label{sf:rtmm} \includegraphics[width=0.45\textwidth] {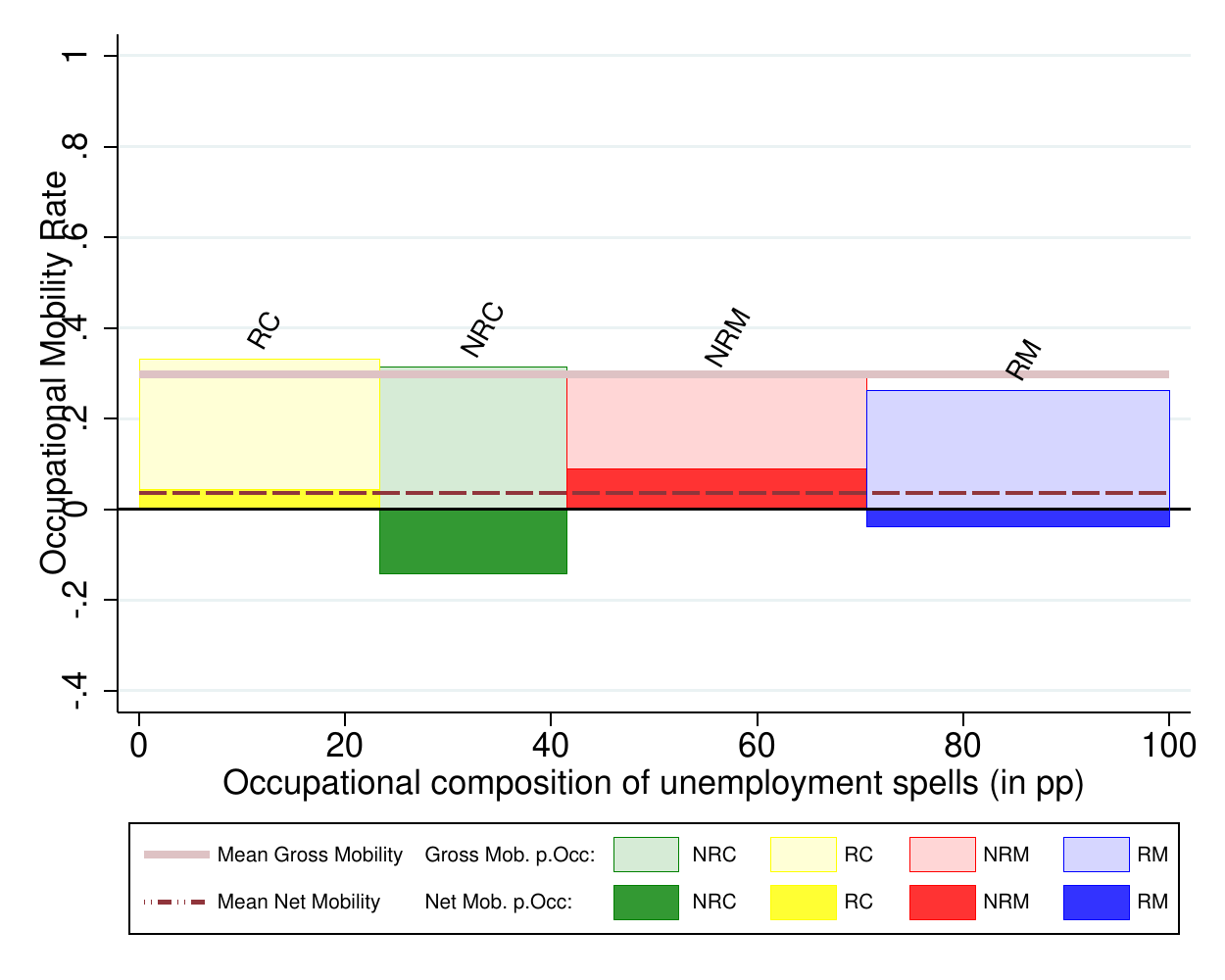}}
\subfloat[RM/NRM/RC/NRC excl. Management]{\label{sf:rtmm_nomgt} \includegraphics[width=0.45\textwidth] {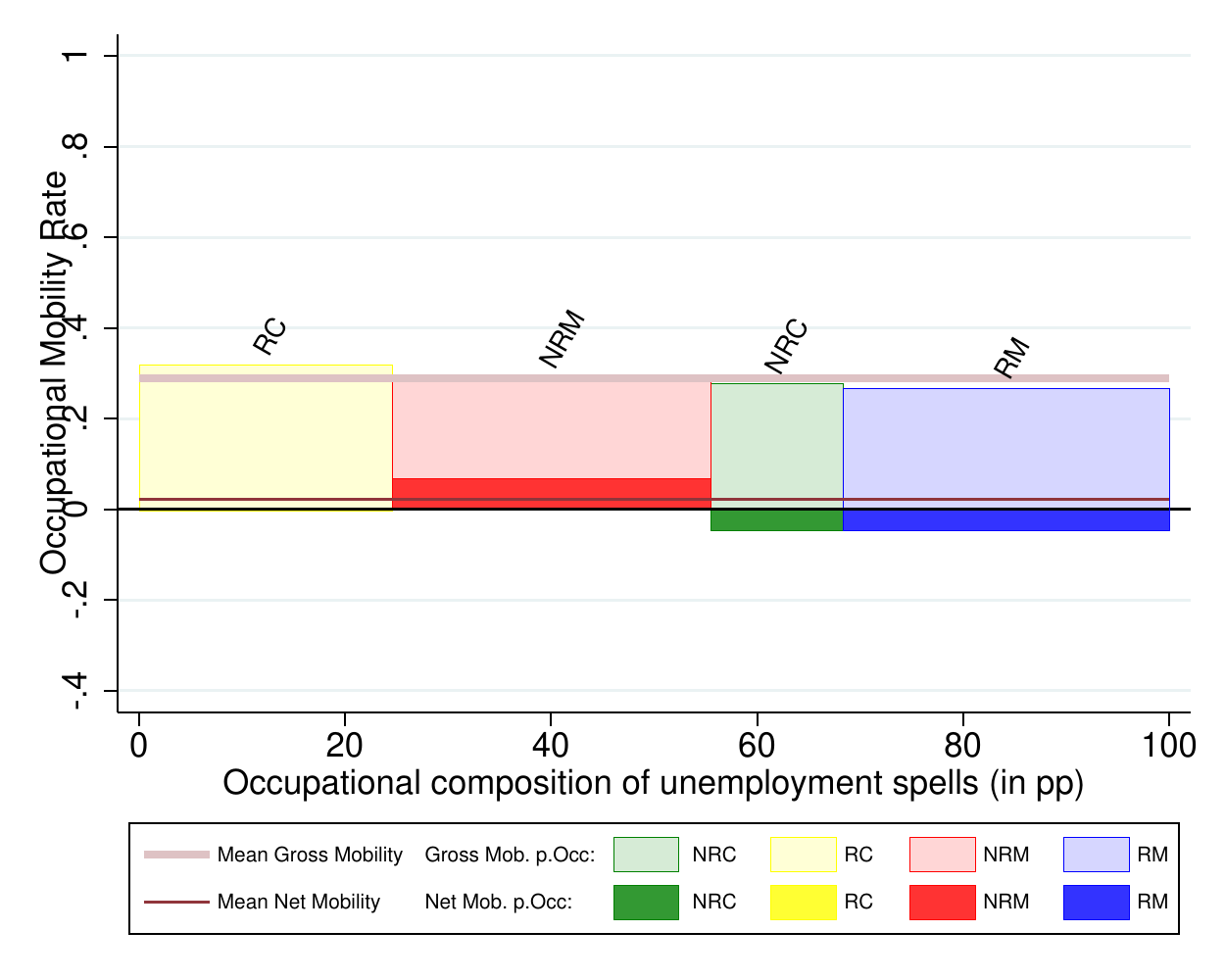}}
\caption{Net and Gross Mobility per Occupation Task-based Categories}
\label{f:netgross_4cat}
\end{figure}

Figure \ref{f:netgross_4cat} presents the same information as Figure \ref{f:netgross_mmo_dd_ind}, but when aggregating the major occupational groups into task-based categories. The left panel includes ``management'', while the right panel excludes it. Once again we find an overwhelming importance of excess mobility relative to net mobility. When including ``management'' net mobility can be covered by 3.7\% of all unemployment spells and 12.4\% of all gross flows across the four task-based occupational groups. When excluding workers with employment in ``management'' before or after unemployment, net mobility can be covered by 2.1\% of all remaining unemployment spells, which is 7.4\% of all grows flows in this set.

Even though net mobility is small, we find clear patterns among the net flows across these task-based categories. In particular, Figure \ref{f:netgross_4cat} shows that during the 1983-2013 period more workers left jobs in routine manual occupations than took up jobs in these occupations. It also shows that more workers took up jobs in non-routine manual occupations than left these occupations. There is also a clear pattern in the net flows of non-routine cognitive and routine cognitive occupations. Figure \ref{sf:rtmm} shows that the non-routine cognitive occupations experienced net outflows. It is immediate from comparing Figures \ref{sf:rtmm} and \ref{sf:rtmm_nomgt} that the vast majority of the net outflows from the non-routine cognitive occupations come from ``management''. At the same time, Figure \ref{sf:rtmm} shows that the routine cognitive occupations experience net inflows. As suggested by Figure \ref{sf:rtmm_nomgt} by far the main contributor to these net inflows is ``management''. We observe that by excluding the ``management'' flows the net inflow into the routine cognitive category basically disappears.

Taken together this evidence suggests that the net mobility that occurs through unemployment or non-employment spells across task-based categories is best understood through the manual versus cognitive dimensions. Within the manual set of occupations there is clear evidence of job polarization: routine jobs are disappearing while non-routine jobs are on the rise. Within the cognitive set of occupations job polarization is not so evident. Instead we observe that much of the net flows that arise between non-routine cognitive and routine cognitive occupations take the form of managers losing their jobs and then re-gaining employment as sales or office/administrative support workers. This type of mobility does not seem much related to structural change, but suggests a picture in which workers in higher skilled jobs move down their career ladder to perform less skilled jobs after experiencing job loss.

\begin{figure}[ht!]
\centering
\subfloat[Major Occupational Groups (2000 SOC) - Unemployed]{\label{2000SOC} \includegraphics [width=0.5 \textwidth] {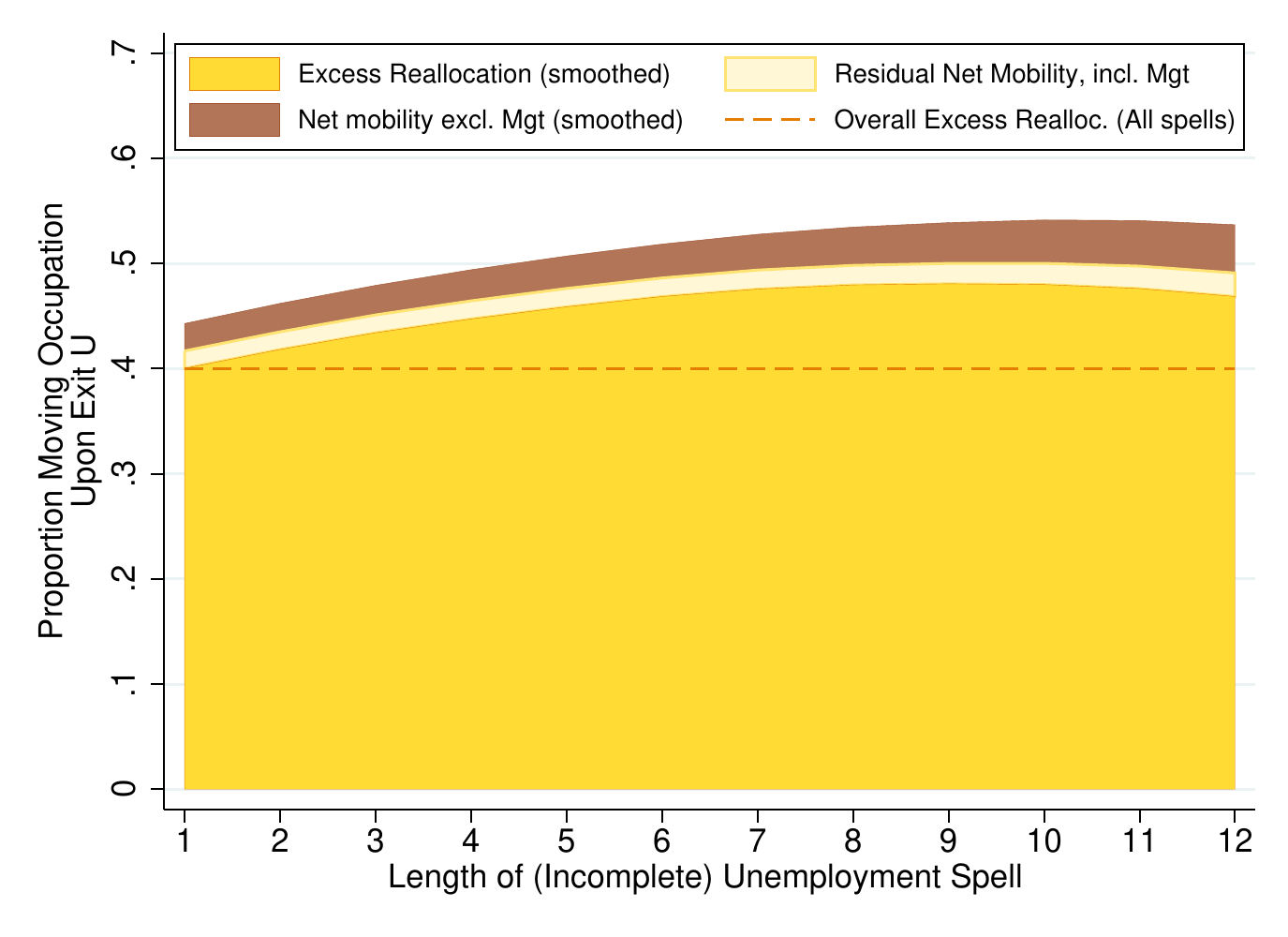}}
\subfloat[Major Occupational Groups (2000 SOC) - `NUN']{\label{Nonemp1} \includegraphics [width=0.5 \textwidth] {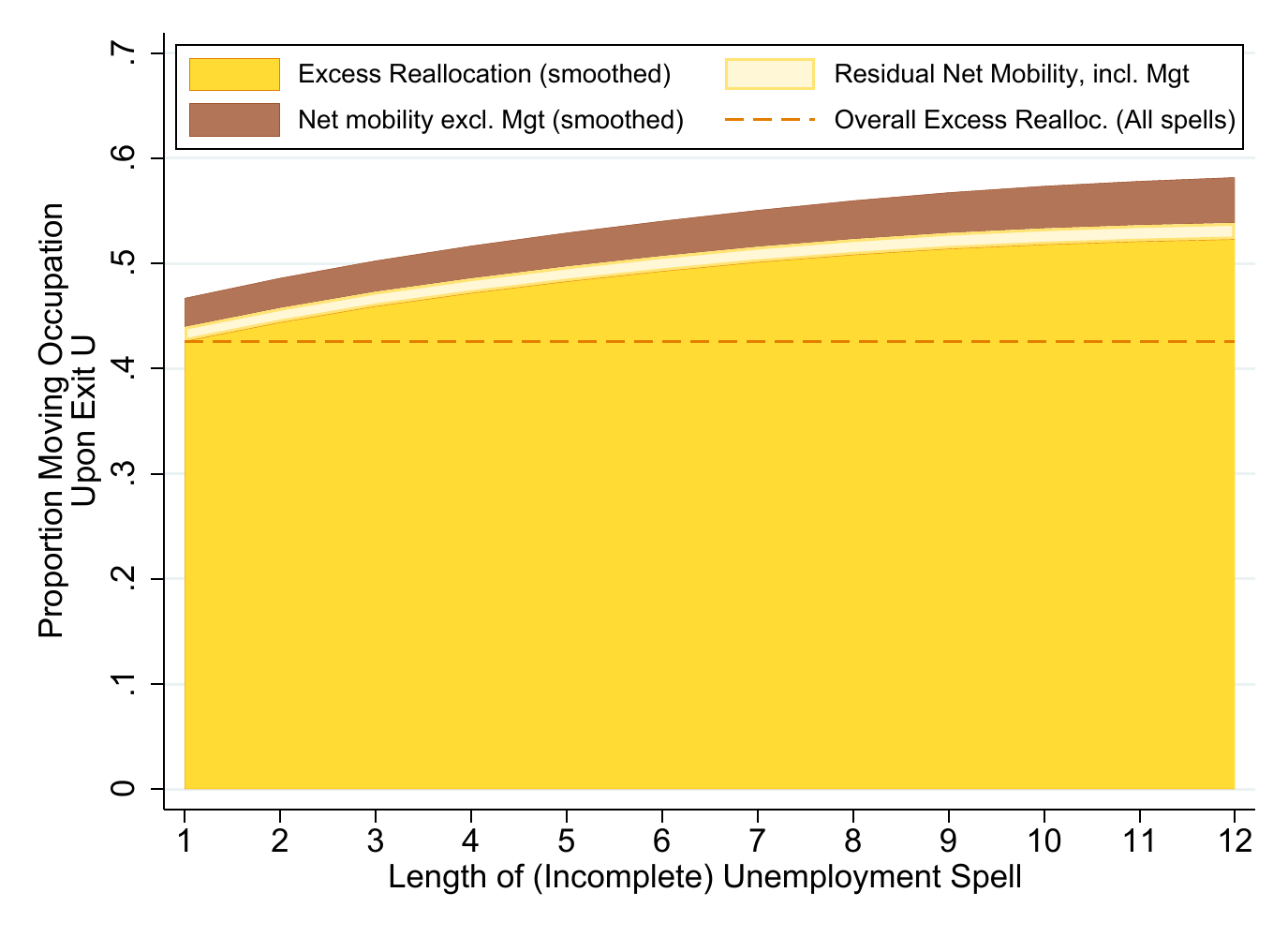}}

\subfloat[Major Occupational Groups (1990 SOC) - Unemployed]{\label{1990SOC} \includegraphics [width=0.5 \textwidth] {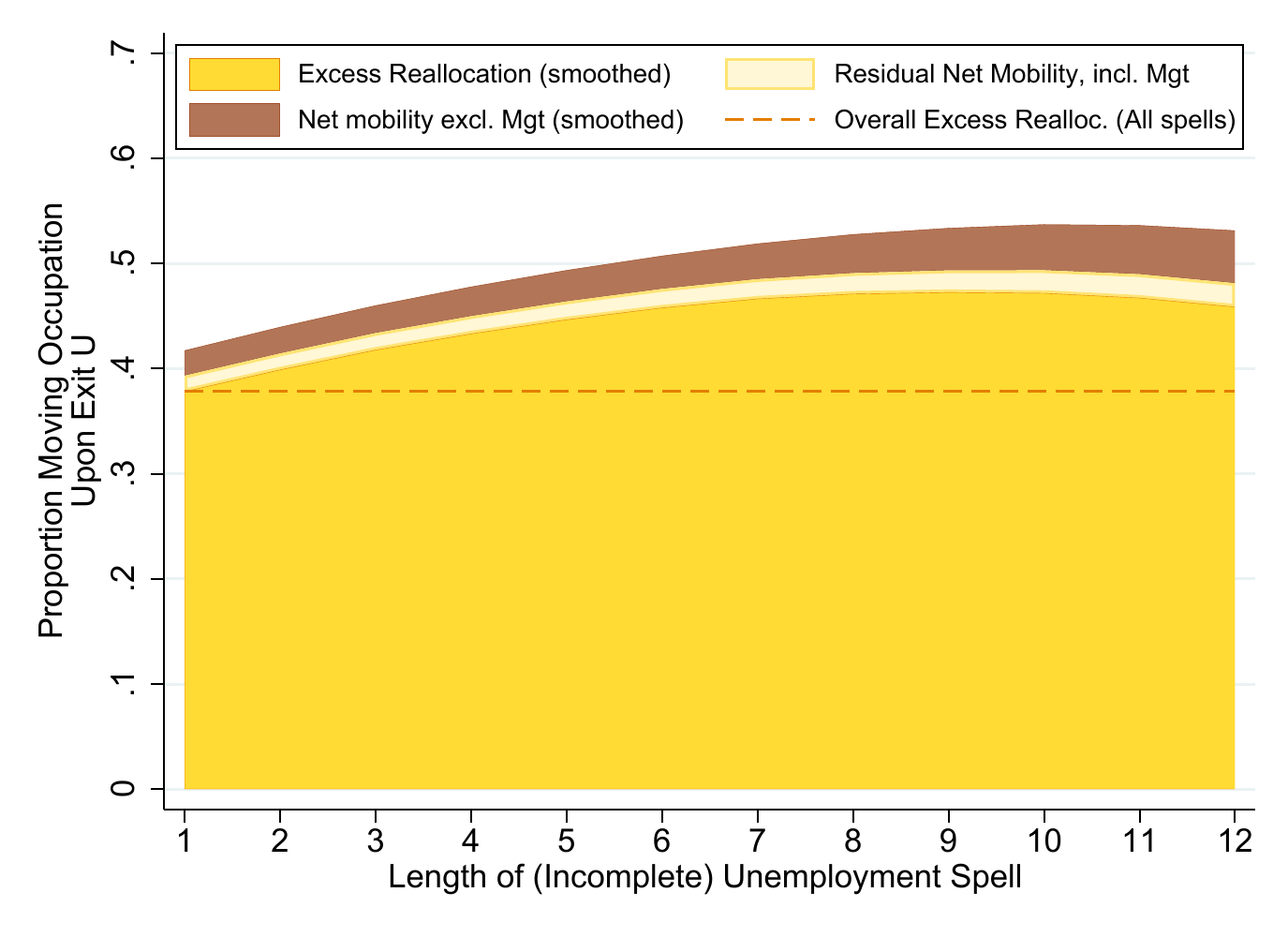}}
\subfloat[Major Industry Groups  - Unemployed]{\label{Ind} \includegraphics [width=0.5 \textwidth] {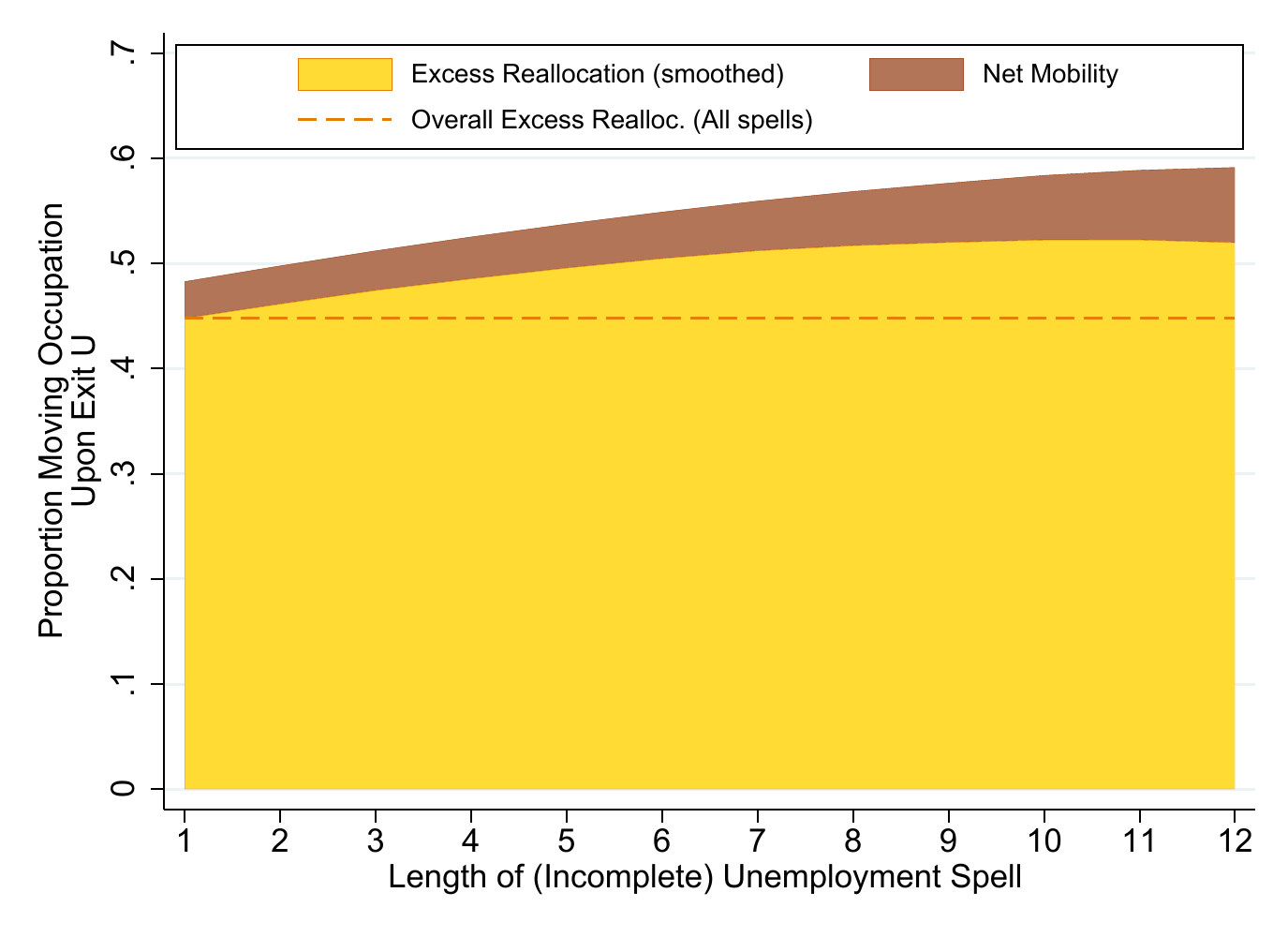}}
\caption{The importance of excess and net mobility}
\label{f:exc_net}
\end{figure}

\subsection{Mobility - duration profile}

We now turn to show that excess mobility is also the main driver of the mobility-duration profile. Figure \ref{2000SOC} depicts the $\Gamma$-corrected mobility-duration profile based on the major occupational groups of the 2000 SOC (as in Figure \ref{2000} in Section A) and subdivides the area below it into the contribution of net and excess mobility. Recall that for a given unemployment duration $x$, the profile shows the gross occupational mobility rate as the fraction of workers who had at least $x$ months in unemployment and changed occupation at re-employment among those workers who had at least $x$ months in unemployment before regaining employment. To compute the net and excess mobility rate for a given unemployment duration, we only use those employment-unemployment-employment spells that had at least $x$ months in unemployment.

The area immediately below the profile show the contribution of total net mobility at each unemployment duration. This area is further subdivided into the contribution that corresponds to the ``management'' flows and to the contribution that corresponds to the flows of the remainder occupations. The area below net mobility show the contribution of excess mobility at each unemployment duration. The horizontal dashed line crossing this area shows the average excess occupational mobility rate. The rest of the graphs in Figure \ref{f:exc_net} depicts the same information, but instead consider occupational mobility through non-employment spells with at least one period of unemployment ('NUN' spells), occupational mobility using the 1990 SOC and industry mobility. It is immediate from these graphs that excess mobility is the largest component of gross mobility at all unemployment (non-employment) durations. Further, the importance of excess mobility increases with unemployment (non-employment) duration. At the same time we observe that the importance of net mobility also increases with unemployment (non-employment) duration, particularly the net mobility flows that corresponds to the non-``management'' occupations.

\begin{figure}[ht!]
\centering
\subfloat[RM/NRM/RC/NRC]{\label{4groups} \includegraphics [width=0.5 \textwidth] {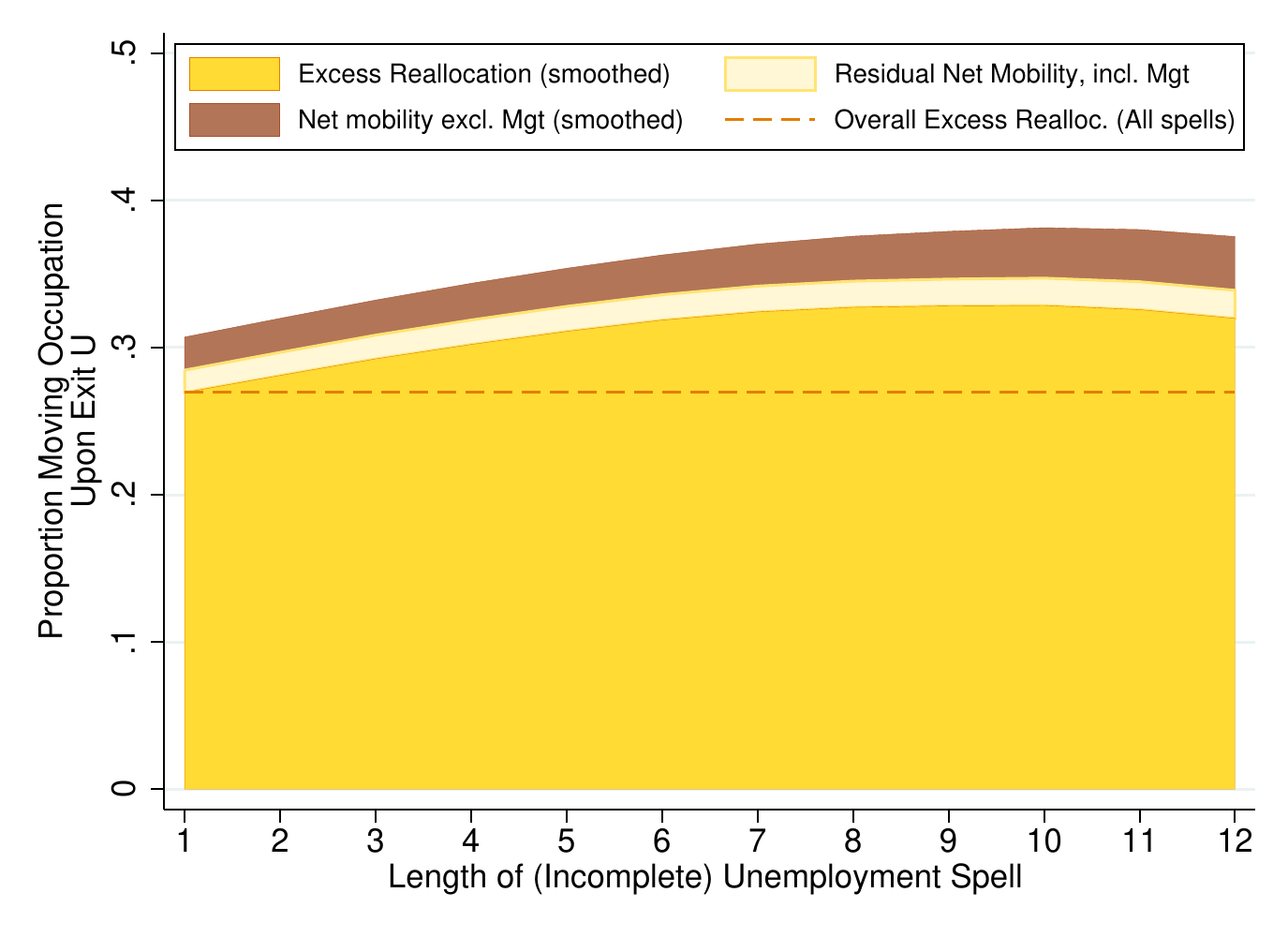}}
\subfloat[RM/NRM/C]{\label{3groups} \includegraphics [width=0.5\textwidth] {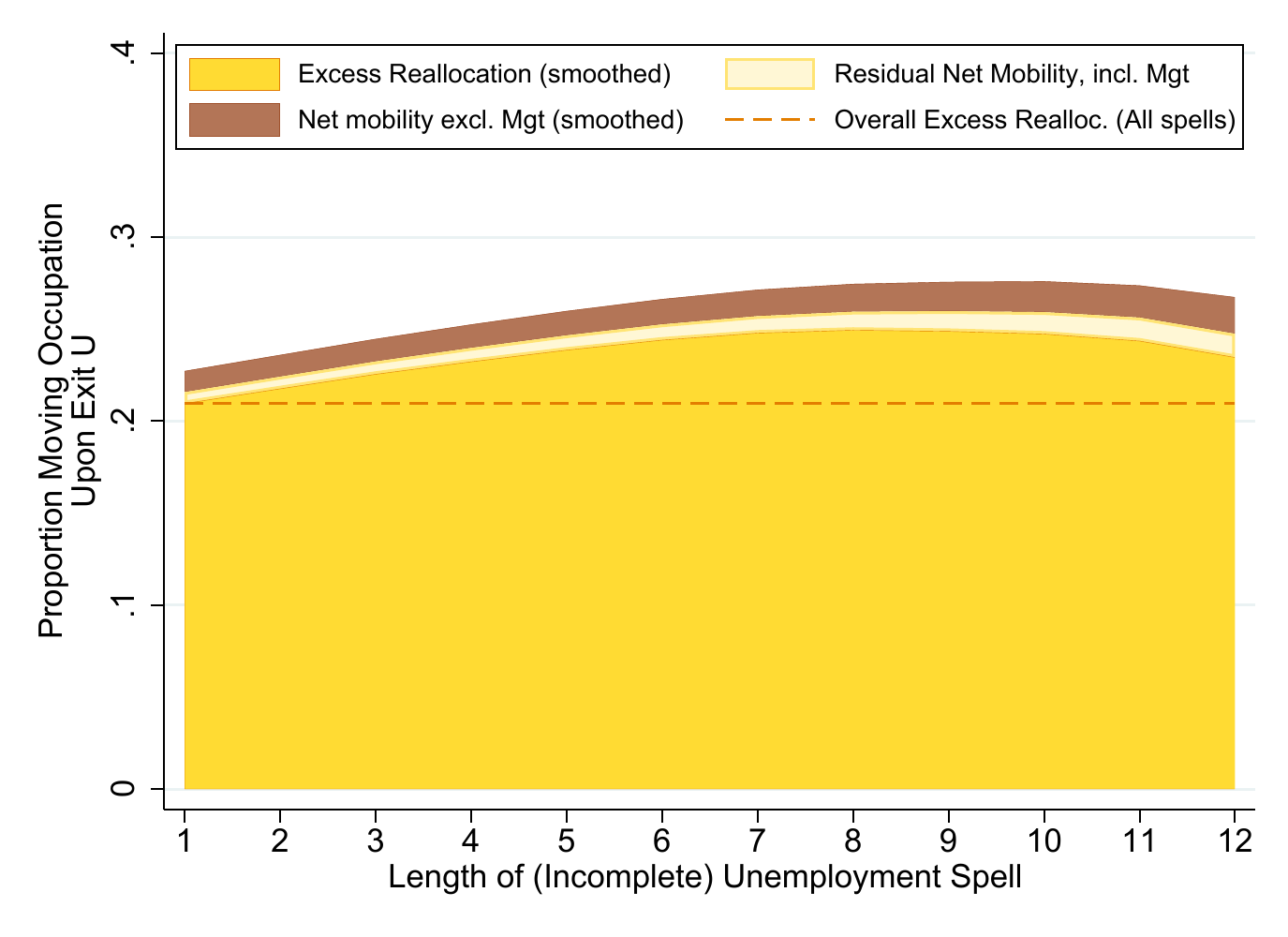}}
\caption{The importance of excess and net mobility - Task based categorisation}
\label{f:occ_Cog_Routine}
\end{figure}

Figure \ref{4groups} present the same information as above but now aggregating occupations using a task-based classification. Figure \ref{3groups} presents a very similar pattern as in Figure \ref{4groups}, but now aggregating non-routine cognitive and routine cognitive into one category to subsume the ``management'' flows into one (larger) cognitive category. Once again we observe the importance of excess mobility in shaping the mobility-duration profile.

\section{Cyclical patterns of occupational mobility}

In Section 2 of the paper we document the cyclical patterns of occupational mobility for those workers who changed employers through intervening unemployment (non-employment) spells. Here we investigate in more detail these cyclical patterns. Below we return to these patterns using the CPS and PSID.

\subsection{Cyclical responsiveness of gross occupational mobility}

\begin{figure}[ht!]
\centering
\subfloat[2000 SOC ]{\label{f:cycl4_mm} \includegraphics [width=0.5 \textwidth]{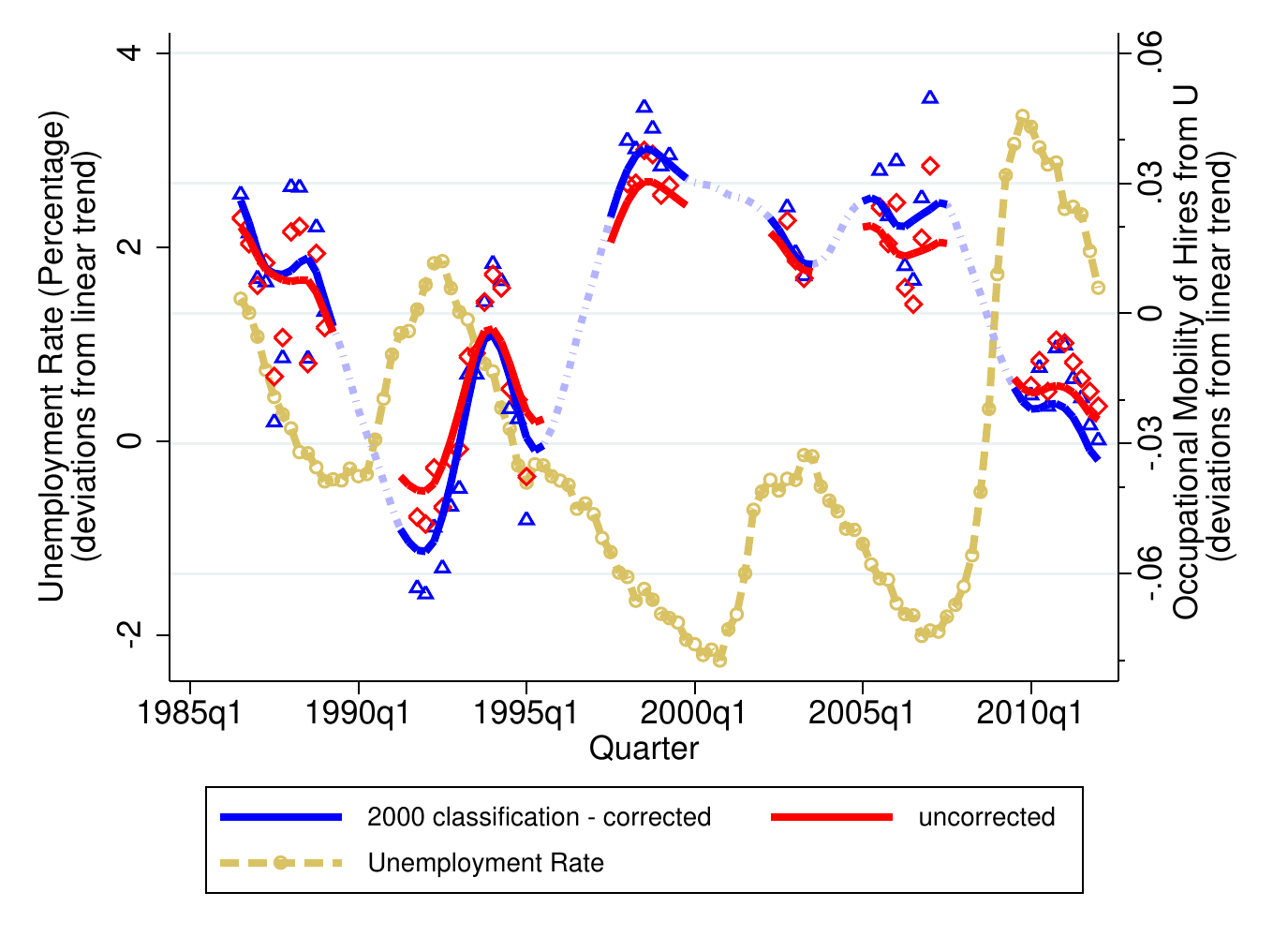}}
\subfloat[4 task-based categories]{\label{f:cycl4_rtmm} \includegraphics [width=0.5 \textwidth]{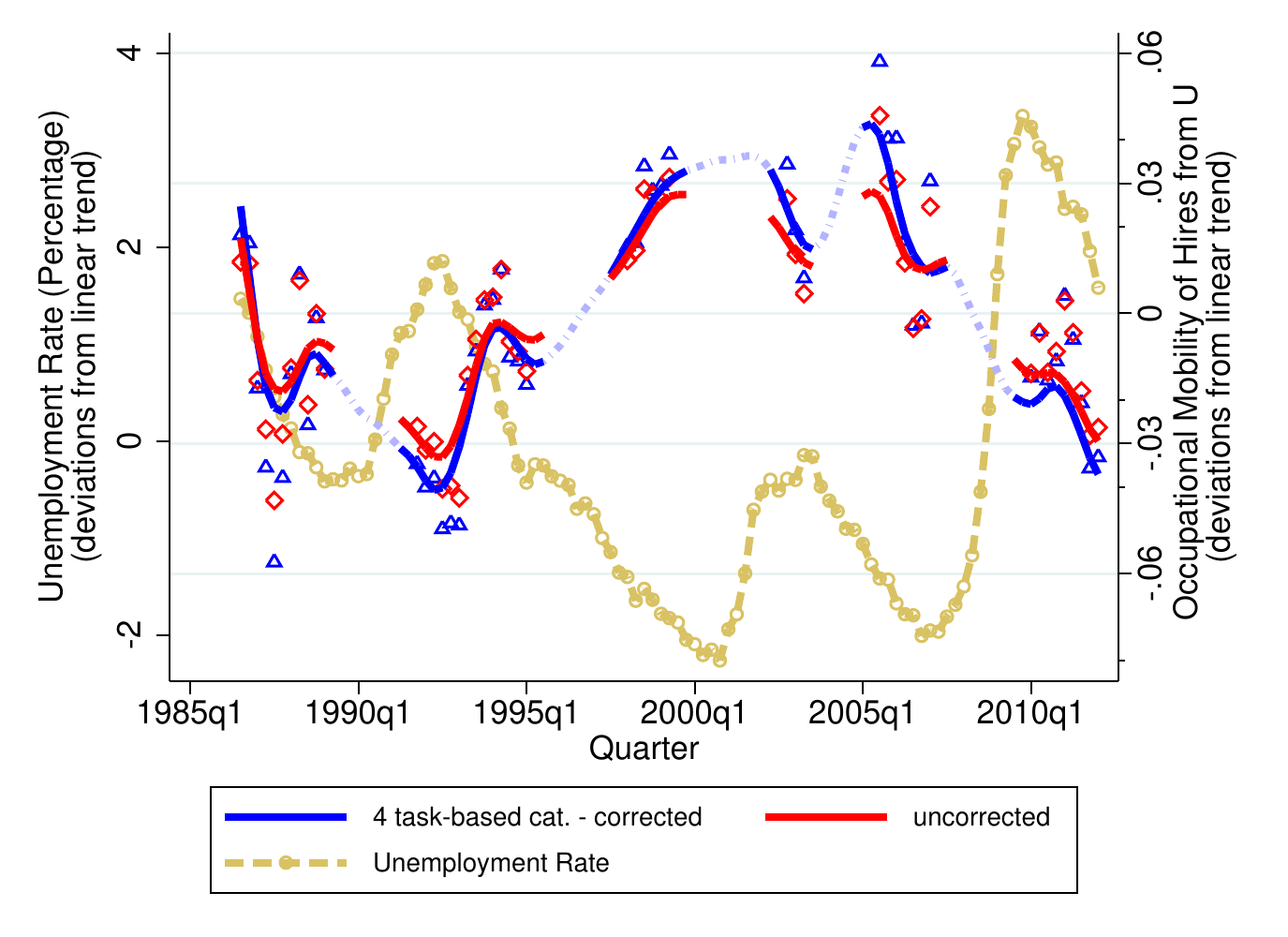}}

\subfloat[1990 SOC]{\label{f:cycl4_dd} \includegraphics [width=0.5 \textwidth]{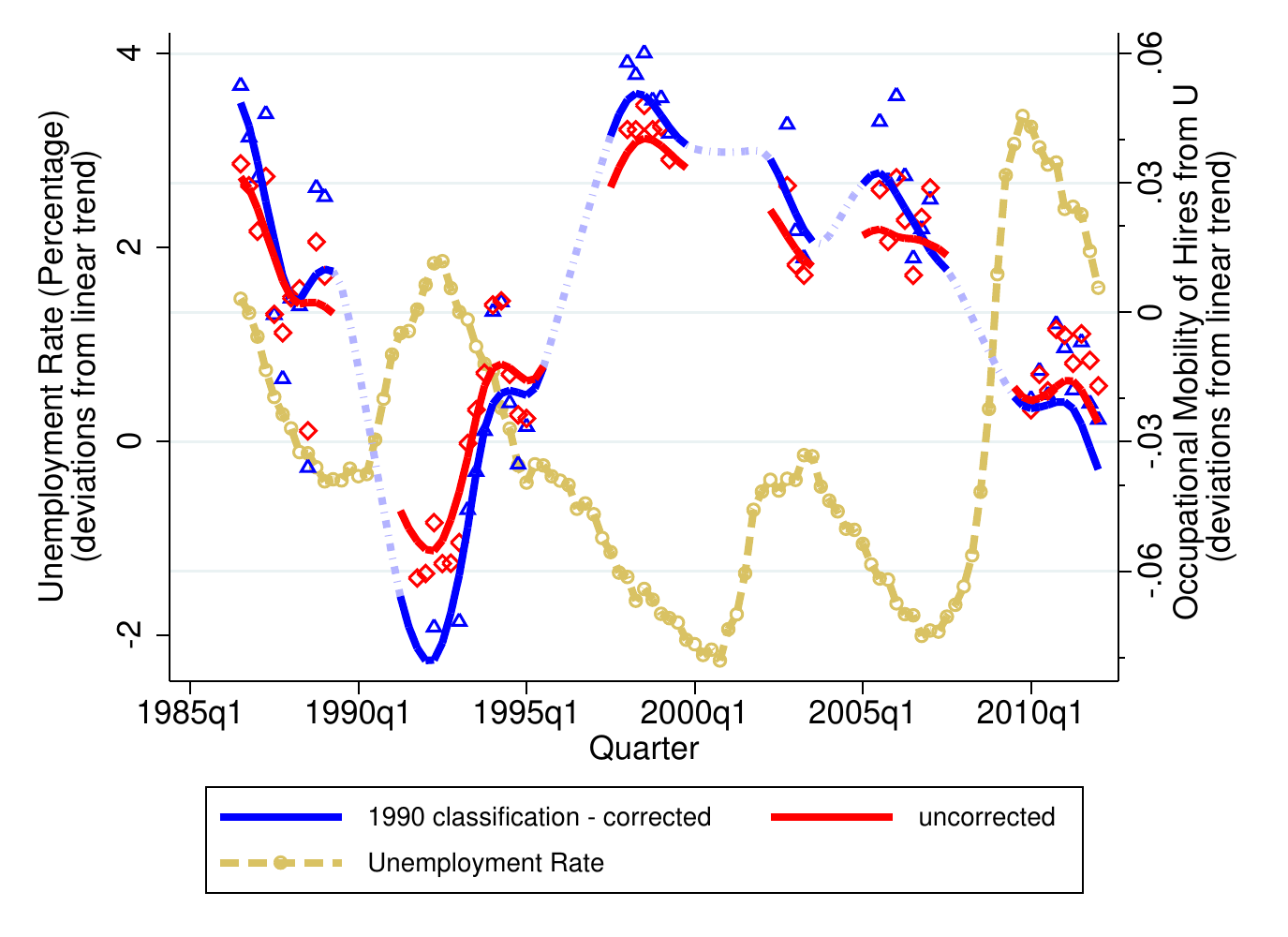}}
\subfloat[Industry (1990 SIC)]{\label{f:cycl4_ind} \includegraphics [width=0.5 \textwidth]{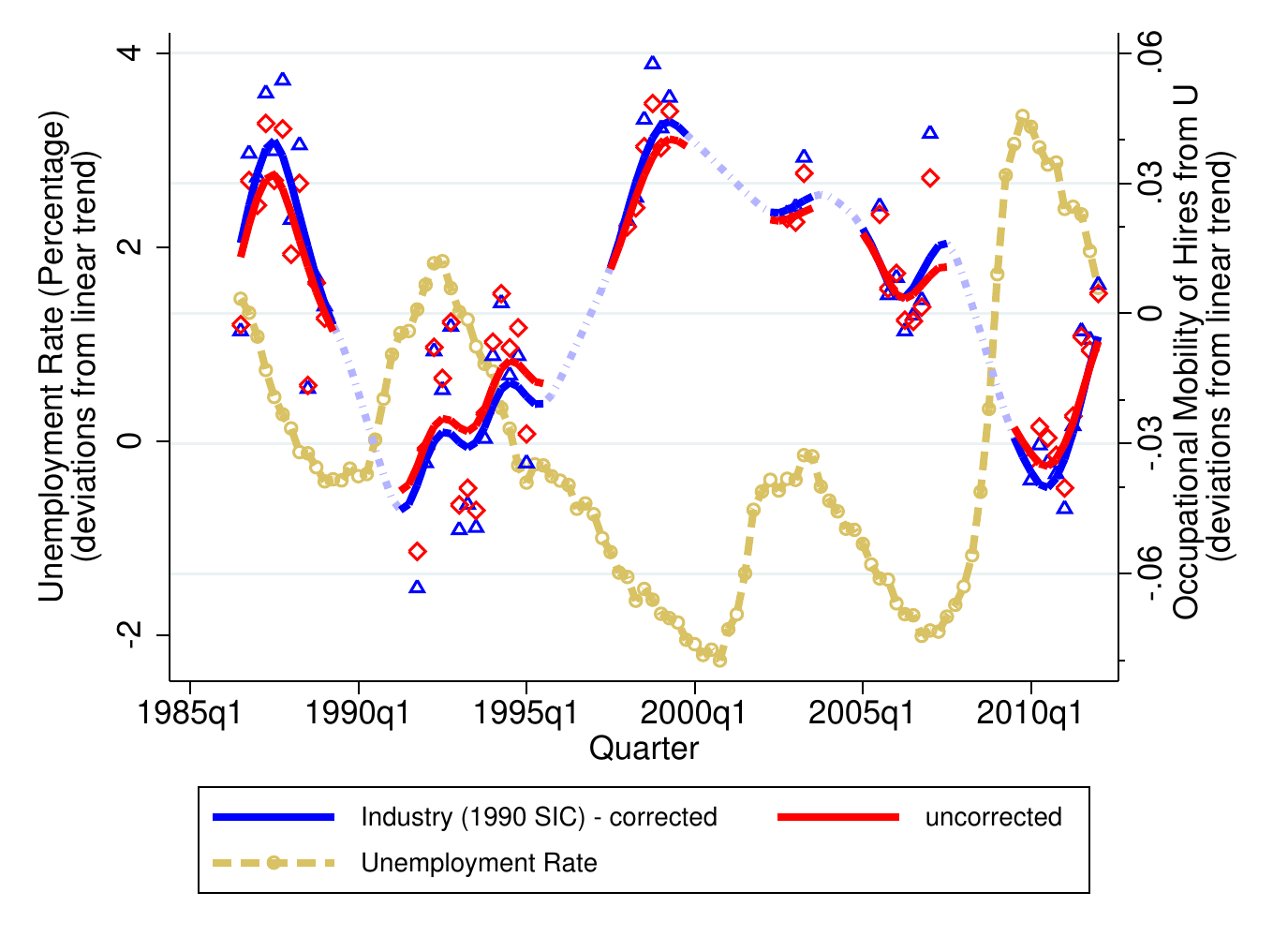}}
\caption{Occupational Mobility of the Unemployed\label{f:cycl_bench1}}
\end{figure}

Figure \ref{f:cycl_bench1} displays the time series of the $\Gamma$-corrected and uncorrected occupational (industry) mobility rates of unemployed workers, together with that of the aggregate unemployment rate, in level deviations from their respective linear trends to first investigate the cyclical behaviour of the series with any formal filtering method. Figure \ref{f:cycl_bench1}a displays these patterns using the major occupational groups of the 2000 SOC, while Figure \ref{f:cycl_bench1}b displays these patterns using the 4 task-based categories: Non-routine cognitive, routine cognitive, non-routine manual and routine manual. The markers in the graphs depict the 5-quarter centered moving averages of the time series, while the curves around these markers smooth the averages locally. The key message from both graphs is that gross occupational mobility among unemployed workers is \emph{procyclical}. When comparing the linearly de-trended time series of the centered 5-quarter moving average of the (log) occupation mobility rate with that of the (log) unemployment rate, we find correlations of -0.62 and -0.47 for the 2000 SOC and the 4 task-based categories classification, respectively.\footnote{Observations at the end of de-trended time series are estimated more imprecisely than those in the middle. For this reason, bandpass filtering typically excludes those observations. If we restrict our attention to the time series window for which we would have bandpass de-trended observations, 1988q3 - 2010q1, the aforementioned correlations rise to (in absolute value) -0.82 for the case of the 2000 SOC and -0.80 for the 4 task-based categories. Restricting the sample to this window does not change any of our conclusions. In particular, all empirical mobility elasticities from the linearly de-trended series stay significant at the 1\%. For the HP filtered series, conclusions also carry over, with the restriction sharpening the pro-cyclicality of occupational mobility across the 4 task-based groups, while blunting somewhat the cyclicality of industry mobility (see below). An exercise in which we directly use each individual quarterly observation, which are much noisier (for example, there are quarters in the data in which relatively few unemployed workers are hired), also yields broadly similar results in these restricted window. While in this case correlations drop to around (-0.50,-0.30) when linearly de-trended and around (-0.30,-0.10) when HP-filtered, all linear de-trended series discussed in this section stay statistically significantly procyclical (with respect to unemployment) at the 1\%. Further, the elasticities of the HP-filtered procyclical series of occupational mobility with respect to HP-filtered unemployment show procyclicality and still reach significance at the 5\% level (1990 and 2000 SOC), and 10\% (4 task-based categories), within the 1988q3-2010q1 window.} Figure \ref{f:cycl_bench1}c shows that the procyclicality of occupational mobility among the unemployed is also present when using the 1990 SOC. Differences in the cyclical patterns between the 2000 SOC and 1990 SOC appear to be of second order. If anything, occupational mobility seems somewhat more cyclically responsive when using the 1990 SOC. Figure \ref{f:cycl_bench1}d further shows the cyclical patterns of industry mobility. It shows that the gross industry mobility rate among the unemployed is also procyclical, exhibiting a correlation of -0.56 between the linear de-trended industry mobility and unemployment rates.

\begin{table}[h]
\small
  \centering
  \caption{Cyclicality Occupational Mobility Using Unemployment Spells}
     \resizebox{1.0\textwidth}{!}{
    \begin{tabular}{lllllllll}

    \toprule
    \toprule
          & \multicolumn{4}{c}{5Q  MA  $\Gamma$-CORRECTED} & \multicolumn{4}{c}{5Q MA UNCORRECTED} \\
\cmidrule{2-9}          & \multicolumn{2}{c}{linear de-trend} & \multicolumn{2}{c}{HP 1600} & \multicolumn{2}{c}{linear de-trend} & \multicolumn{2}{c}{HP 1600} \\
\cmidrule{2-9}    \multicolumn{1}{l}{Category} & \multicolumn{1}{l}{elasticity} & \multicolumn{1}{c}{$\rho$} & \multicolumn{1}{l}{elasticity } & \multicolumn{1}{c}{$\rho$} & \multicolumn{1}{l}{elasticity} & \multicolumn{1}{c}{$\rho$} & \multicolumn{1}{l}{elasticity } & \multicolumn{1}{c}{$\rho$} \\
    \midrule \hline
    \multicolumn{9}{c}{\textbf{Panel A. Mobility wrt Unemployment }} \\

        2000 SOC  & -0.19*** & -0.62 & -0.17*** & -0.38 & -0.12*** & -0.63 & -0.10*** & -0.36 \\
          & (0.03)  & & (0.06)  &  & (0.02)  &  & (0.03)  & \\
    1990 SOC  & -0.24*** & -0.51 & -0.19*** & -0.34 & -0.15*** & -0.58 & -0.12*** & -0.40 \\
          & (0.06)  &  & (0.07)  &  & (0.03)  & & (0.04)  & \\
    4 task-based categories & -0.20*** & -0.47 & -0.08 & -0.11 & -0.14*** & -0.49 & -0.05 & -0.12 \\
          & (0.05)  &  & (0.09)  &       & (0.03)  & & (0.06)  &  \\
    4 task-based cat. (Excl. Manag.) & -0.23*** & -0.50 & -0.21** & -0.29 & -0.16*** & -0.52 & -0.13** & -0.29 \\
          & (0.05)  & & (0.09)  &  & (0.03)  & & (0.06)  & \\
    Industries (1990 SIC) & -0.16*** & -0.56 & -0.15** & -0.33 & -0.13*** & -0.58 & -0.12** & -0.32 \\
          & (0.03)  &  & (0.06)  &  & (0.02)  &  & (0.05)  & \\ \hline \hline
    \multicolumn{9}{c}{\textbf{Panel B. Mobility wrt Productivity}} \\
    2000 SOC  & 3.08***  & 0.73  & 2.20**  & 0.28  & 1.86***  & 0.72  & 1.20*  & 0.25 \\
          & (0.38)  & & (1.01)  &  & (0.24)  &  & (0.63)  & \\
    1990 SOC  & 4.81***  & 0.73  & 1.45  & 0.15  & 2.64***  & 0.75  & 0.89  & 0.17 \\
          & (0.60)  & & (1.27)  &  & (0.32)  &  & (0.68)  &  \\
    4 task-based categories  & 4.06***  & 0.68  & 3.85**  & 0.33  & 2.55***  & 0.67  & 2.25**  & 0.29 \\
          & (0.58)  & & (1.48)  &       & (0.38)  & & (0.98)  & \\
    4 task-based cat. (Excl. Manag.)  & 4.37***  & 0.68  & 4.38***  & 0.35  & 2.85***  & 0.69  & 2.66**  & 0.33 \\
		  & (0.63)  &  & (1.58)  &  & (0.40)  & & (1.00)  & \\
    Industries (1990 SIC) & 1.73***  & 0.46  & -0.31 & -0.04 & 1.36***  & 0.46  & -0.27 & -0.04 \\
          & (0.45)  &       & (1.08)  &       & (0.35)  &  & (0.84)  &  \\ \hline \hline
    \multicolumn{9}{c}{\textbf{Panel C. Unemployment wrt Productivity}} \\
     Unemployment     & -8.72*** & -0.64 & -4.44* & -0.25 & -8.72*** & -0.64 & -4.44* & -0.25 \\
          & (1.41)  & & (2.25)  &  & (1.41)  & & (2.25)  & \\ \hline
    N(quarters)     & \phantom{x}58    & & \phantom{x}58    & & \phantom{x}58    &    & \phantom{x}58    & \\
    \bottomrule
    \bottomrule
    \multicolumn{9}{c}{{\scriptsize *$\ p<0.1$;\ **$\ p<0.05$;\ ***$\ p<0.01$}}
    \end{tabular}}
  \label{tab:cycl_occmob_u}%
\end{table}%

Panel A of Table \ref{tab:cycl_occmob_u} reports the OLS elasticities and correlation coefficients of the 5-quarter centered moving average of the (log) occupational mobility rate relative to the 5-quarter centered moving average of the (log) unemployment rate. Panel B present the same statistics using instead (log) productivity. In both cases we perform the analysis using either the corresponding linearly de-trended or HP-filtered series for the 1985q3-2013q1 period. Given the gaps in the SIPP before implementing the HP-filter we interpolate the data, but then drop the interpolated observations from the resulting series to estimate the regressions.\footnote{While we do not use any interpolated quarters for the calculation of correlations and estimation of elasticities, the HP-trend can only be obtained using interpolated time series. This likely introduces some further noise for those quarters adjacent to interpolated quarters. On the other hand, including interpolated quarters in the HP-filtered time series allows some further information that is contained in the linearly de-trended time series to weigh in the HP-filtered series as well. Indeed, including interpolated quarters in our calculation appears to strengthen the statistical significance of our coefficients. To conservatively minimize the impact of interpolation, we focus on the statistics excluding any interpolated quarters. In contrast, the linearly de-trended series are derived without restoring to interpolation.}  In the main text we have also shown evidence suggesting that the gaps in the SIPP series no do meaningfully affect the degree of procyclicality. This is done using the CPS (see Section 5, below) as an alternative data source and comparing the SIPP and CPS uncorrected series and estimated elasticities with respect to unemployment. The first four columns report the results for the $\Gamma$-corrected mobility series and the last four columns for the uncorrected series. The results in panel A confirm the conclusion obtained from Figure \ref{f:cycl_bench1}: gross occupational mobility of the unemployed is procyclical. For all linearly de-trended series, the correlations are substantially negative and the elasticities are statistically significantly negative at a 1\% level. HP-filtering leads to lower correlations because unemployment and productivity are less aligned (see panel C of the table) and the relative impact of noise is higher after HP-filtering. Nevertheless, using the HP-filtered series yields elasticities with respect to the unemployment rate that are significant at the 5\% level, with the exception of the 4 task-based categories. The latter appears to be driven in part by the mobility patterns of those workers in managerial occupations. As we highlighted before, management occupations behave differently throughout the entire sample period, with consistently large relative net outflow. Excluding those unemployment spells that are related to managerial occupations results once again in a statistically significant procyclical series when using the 4 task-based categories.

Panel B shows that gross occupational mobility through unemployment is also procyclical when using (log) productivity instead of (log) unemployment rate. For the linearly de-trended series, the correlations are again high (around 0.70) and the empirical elasticities are positive and significant at a 1\% level. Although the HP-filtered series are more noisy, they still yield statistically significantly positive elasticities at a 5\% level for the 2000 SOC and 4 task-based categories series; and at a 1\% level when considering the 4 task-based categories series without managerial occupations. In the case of mobility across industries, we also find positive and statistically significant elasticities with respect to labor productivity. In terms of the correlation coefficients we find that these are positive with respect to the linearly de-trended productivity, but is nearly zero in the HP-filtered series.\footnote{One reason is that the HP-filtered productivity series peaks around 2005, at which time industry mobility of the unemployed is high compared to the average level, but at the same time somewhat lower than in the earlier 2000s. The HP-filtered series weighs the latter more heavily. Further, while industry mobility subsequently meaningfully dropped in the Great Recession, it recovered at a rate mirroring the recovery of unemployment. In contrast, HP-filtered productivity dropped sharply at the beginning of the Great Recession, but also recovered relatively sharply before the beginning of 2010. This behavior is partly behind the relatively low correlation between productivity and unemployment as well. Interestingly, the correlations between mobility and unemployment, and between mobility and output typically are at the same level, and at times even stronger, than the correlation between output and unemployment for the main occupational mobility series.}

Note that across panels A and B all elasticities are higher for the $\Gamma$-corrected series than for the uncorrected series, illustrating that miscoding reduces the cyclical response of gross occupational mobility.\footnote{A further exercise, in which we reduced the impact of noise (and very high frequency movements), by isolating (using TRAMO/SEATS) the trend-cycle component, yields similar results. As a consequence of the reduction in noise, standard errors are much smaller, and more of the elasticities are significant at the 1\% level.}

\paragraph{Cyclical responsiveness using non-employment spells}

\begin{table}[h]
\small
  \centering
  \caption{Cyclicality Occupational Mobility Using NUN Spells}
    \begin{tabular}{lllllllll}

    \toprule
    \toprule
          & \multicolumn{4}{c}{5Q  MA  $\Gamma$-CORRECTED} & \multicolumn{4}{c}{5Q MA UNCORRECTED} \\
\cmidrule{2-9}          & \multicolumn{2}{c}{linearly de-trend} & \multicolumn{2}{c}{HP 1600} & \multicolumn{2}{c}{linearly de-trend} & \multicolumn{2}{c}{HP 1600} \\
\cmidrule{2-9}    \multicolumn{1}{l}{Category} & \multicolumn{1}{l}{elasticity} & \multicolumn{1}{c}{$\rho$} & \multicolumn{1}{l}{elasticity } & \multicolumn{1}{c}{$\rho$} & \multicolumn{1}{l}{elasticity} & \multicolumn{1}{c}{$\rho$} & \multicolumn{1}{l}{elasticity } & \multicolumn{1}{c}{$\rho$} \\
    \midrule
    \multicolumn{9}{c}{\textit{\textbf{Mobility across NUN-spells, wrt Unemployment Rate}}}\\
    2000 SOC  & -0.14*** & -0.73 & -0.13*** & -0.43 & -0.09*** & -0.73 & -0.08*** & -0.42 \\
          & (0.02) &       & (0.04) &       & (0.01) &       & (0.02) &  \\
    1990 SOC  & -0.17*** & -0.60 & -0.16*** & -0.35 & -0.11*** & -0.67 & -0.11*** & -0.42 \\
          & (0.03) &       & (0.06) &       & (0.02) &       & (0.03) &  \\
    4 task-based categories  & -0.11*** & -0.38 & -0.00 & -0.01 & -0.08*** & -0.43 & -0.02 & -0.07 \\
          & (0.04) &       & (0.07) &       & (0.02) &       & (0.04) &  \\
    4 task-based cat. (Excl. Manag.) & -0.14*** & -0.48 & -0.08 & -0.21 & -0.10*** & -0.53 & -0.06* & -0.23 \\
          & (0.03) &       & (0.05) &       & (0.02) &       & (0.04) &  \\
    Industries (1990 SIC) & -0.16*** & -0.72 & -0.21*** & -0.55 & -0.13*** & -0.72 & -0.16*** & -0.53 \\
          & (0.02) &       & (0.04) &       & (0.02) &       & (0.03) &  \\ \hline
    N(quarters)     & \phantom{x}58    & & \phantom{x}58    & & \phantom{x}58    &    & \phantom{x}58    & \\

    \bottomrule
    \bottomrule
    \multicolumn{9}{c}{{\scriptsize *$\ p<0.1$;\ **$\ p<0.05$;\ ***$\ p<0.01$}}
    \end{tabular}%
  \label{tab:cycl_occmob_nun}%
\end{table}%

To complement the above analysis, Table \ref{tab:cycl_occmob_nun} considers the cyclical behavior of occupational mobility among those workers who mixed periods of unemployment with periods of out-of-labor-force during their non-employment spells (NUN-spells). In this case we also find that occupational and industry mobility are procyclical. Point estimates of the elasticities of occupational mobility with respect to unemployment are somewhat lower when using NUN spells rather than pure unemployment spells, but generally these differences are not statistically significant. Point estimates of the procyclical responsiveness of industrial mobility are instead higher. The linearly de-trended data once again yield statistically significant procyclicality at the 1\% level across all mobility measures. The HP-filtered series also yields similar results when considering mobility across major occupational groups (both 1990 and 2000 classification) and industry mobility. However, the pattern is statistically weaker for mobility across the 4 task-based categories, but still marginally significant (at the 10\% level) in the uncorrected data when managers are excluded.

\paragraph{Cyclical responsiveness at different moments of the unemployment spell}

\begin{table}[!t]
\small
  \centering
  \caption{Cyclicality of Mobility at Different Moments of the Unemployment Spell}
     \resizebox{1.0\textwidth}{!}{
    \begin{tabular}{llllllll}
    \toprule
    \toprule
          & 2000 SOC    & 1990 SOC    & 2000 SOC-NUN & OCC*IND & IND   & NR/R-M/C  & C/NRM/RM  \\
          & (1)   & (2)   & (3)   & (4)   & (5)   & (6)   & (7) \\
    \toprule
    \multicolumn{8}{c}{\textbf{Panel 1. Regression of Individual Mobility on Linearly De-trended Unemployment Rate}} \\
    \bottomrule
    \multicolumn{8}{c}{1(a) unemployment in quarter of hiring} \\
    \hline
    U hiring & -0.0788*** & -0.0890*** & -0.0583*** & -0.0606*** & -0.0952*** & -0.0575*** & -0.0589*** \\
    (s.e.) & (0.0180) & (0.0211) & (0.0153) & (0.0192) & (0.0219) & (0.0192) & (0.0191) \\
    \hline
    \multicolumn{8}{c}{1(b) unemployment in quarter of separation} \\
    \hline
    U sep & -0.0628*** & -0.0691*** & -0.0536** & -0.0420* & -0.0586** & -0.0700*** & -0.0659*** \\
    (s.e.) & (0.0211) & (0.0239) & (0.0206) & (0.0249) & (0.0256) & (0.0211) & (0.0202) \\
    \hline
    \multicolumn{8}{c}{1(c) unemployment at hiring and at separation averaged} \\
    \hline
    U ave & -0.0868*** & -0.0969*** & -0.0704*** & -0.0630*** & -0.0946*** & -0.0776*** & -0.0758*** \\
    (s.e.) & (0.0218) & (0.0259) & (0.0204) & (0.0239) & (0.0247) & (0.0231) & (0.0222) \\
    \hline
    \multicolumn{8}{c}{1(d) unemployment at moment of hiring \& separation} \\
    \hline
    U. hiring & -0.0676*** & -0.0782*** & -0.0429** & -0.0583** & -0.0996*** & -0.0241 & -0.0306 \\
    (s.e.) & (0.0236) & (0.0253) & (0.0204) & (0.0265) & (0.0316) & (0.0242) & (0.0256) \\
    U sep. & -0.0182 & -0.0175 & -0.0271 & -0.0037 & 0.0071 & -0.0540** & -0.0456* \\
    (s.e.) & (0.0277) & (0.0293) & (0.0268) & (0.0334) & (0.0352) & (0.0261) & (0.0267) \\
    \toprule
    \multicolumn{8}{c}{\textbf{Panel 2. Regression of Individual Mobility on HP-filtered Unemployment Rate}} \\
    \bottomrule
    \multicolumn{8}{c}{2(a) unemployment in quarter of hiring} \\
    \hline
    U hiring & -0.1594*** & -0.1768*** & -0.1182*** & -0.1198** & -0.2092*** & -0.0840** & -0.0959** \\
    (s.e.) & (0.0420) & (0.0449) & (0.0366) & (0.0458) & (0.0543) & (0.0408) & (0.0419) \\
    \hline
    \multicolumn{8}{c}{2(b) unemployment in quarter of separation} \\
    \hline
    U sep & -0.0916* & -0.0943* & -0.1017** & -0.0628 & -0.1015 & -0.1292*** & -0.1238*** \\
    (s.e.) & (0.0479) & (0.0505) & (0.0455) & (0.0535) & (0.0633) & (0.0402) & (0.0406) \\
    \hline
    \multicolumn{8}{c}{2(c) unemployment at hiring and at separation averaged} \\
    \hline
    U ave & -0.2187*** & -0.2358*** & -0.2078*** & -0.1588** & -0.2697*** & -0.1895*** & -0.1938*** \\
    (s.e.) & (0.0590) & (0.0677) & (0.0601) & (0.0635) & (0.0685) & (0.0563) & (0.0547) \\
    \hline
    \multicolumn{8}{c}{2(d) unemployment at moment of hiring \& separation} \\
    \hline
    U. hiring & -0.1492*** & -0.1666*** & -0.1127*** & -0.1130** & -0.1985*** & -0.0671 & -0.0794* \\
    (s.e.) & (0.0437) & (0.0460) & (0.0387) & (0.0483) & (0.0579) & (0.0439) & (0.0451) \\
    U sep. & -0.0723 & -0.0727 & -0.0960** & -0.0482 & -0.0756 & -0.1205*** & -0.1132*** \\
    (s.e.) & (0.0467) & (0.0503) & (0.0458) & (0.0521) & (0.0575) & (0.0400) & (0.0406) \\
    \bottomrule
    \bottomrule
        \multicolumn{8}{c}{{\scriptsize *$\ p<0.1$;\ **$\ p<0.05$;\ ***$\ p<0.01$}}
    \end{tabular}}
  \label{tab:4_cycl_timing}%
\end{table}

The preceding analysis documents the procyclicality of occupational (industry) mobility, measuring the state of the economy at the end of the unemployment spell; i.e. at the time of job finding. Table \ref{tab:4_cycl_timing} investigates whether this result survives when instead we measure the unemployment rate at the beginning of the worker's unemployment spell (at the time of job separation) or use the average of the unemployment rates at job separation or job finding. The first panel considers the case in which we use the linearly de-trended (log) unemployment rate as our cyclical measure and reports estimates of a regression incorporating a linear time trend and a dummy variable for the change of occupational classification at the 2004 panel. The main message from the first two rows is that measuring the unemployment rate at the moment of job separation or job finding does not affect our conclusions. In both cases we find that the occupational (industry) mobility of unemployed workers is procyclical. Note that the point estimates obtained in the first five columns decrease when measuring the unemployment rate at the moment of job separation. However this decrease is not meaningful and the point estimates remain statistically significant. Taking the average of the unemployment rates at job separation and job finding yields the highest empirical responsiveness in the point estimates, with significance at the 1\% across all columns. Further, including both the unemployment rate at job separation and at job finding, the point estimates reveal an additional negative impact on mobility when unemployment rates are high both at the time of job separation and job finding (and, in practice, in between), but this impact is not statistically significant.

In the second panel of Table \ref{tab:4_cycl_timing} we relate occupational (industry) mobility to the HP-filtered (log) unemployment rate, measured again at the time of job separation or of job finding. Here we also observe that occupational (industry) mobility remains procyclical. However, we observe a meaningful drop in the responsiveness of occupational mobility with respect to the unemployment rate when using the 2000 SOC and 1990 SOC and in the case of industry mobility. As in the previous case, the clearest response is obtained when we use the average HP-filtered unemployment at the begin and end of the spell, with statistically significant coefficients at the 1\% level. When including both HP-filtered unemployment rates, although measured imprecisely, the impact of higher unemployment rate at the end of the spell, ceteris paribus, is additionally negative (and not economically insignificant) for mobility. For mobility across the 4 task-based categories, this impact is at least marginally significant at the 10\% level, while for those NUN spells it is significant at the 5\% level. %

The above results then suggest that our benchmark analysis is conservative in that we have not selected the timing of unemployment to maximize the cyclical responsiveness of the occupational mobility rate among unemployed workers.

\paragraph{Cyclical responsiveness when controlling for demographic characteristics and occupational identities}

Next we investigate whether the cyclical behavior of occupational mobility among the unemployed is affected by demographic characteristics and/or the possible effects of source or destination occupations. Panels A and B of Table \ref{tab:3} consider regressions of the form
of the form
\begin{align}
 \mathbf{1}_{\text{occmob}} &= \beta_0 + \beta_{1}\text{Cyclical Variable} +\text{Controls} + \varepsilon, \tag{R1}\label{e:cycl_app_controls}
\end{align}
where as the cyclical variable we use either the linearly de-trended (log) unemployment rate and on the HP-filtered (log) unemployment rate. Each panel is divided into a number of sub-panels that present different set of controls. In the first sub-panel we consider a regression relating the uncorrected mobility to the relevant cyclical variable, a linear time trend, and a dummy for 2000 SOC (implemented in the 2004 and 2008 panels).\footnote{We further have investigated the cyclical patterns during 1985-2003 period in isolation. This analysis gives very similar results, although they produce less precise estimates.} In the second sub-panel we add demographic controls to the previous regression, while in the third sub-panel we further add source occupation dummy variables. In the fourth sub-panel we further control for destination occupation dummy variables instead of controlling for source occupation. Column (i) considers the 2000 SOC, column (ii) the 1990 SOC, while column (iv) considers simultaneous occupation and industry mobility and column (v) considers industries based on the 1990 Census classification. We analyse the 4 task-based categories including managerial occupations in column (vi) and excluding managerial occupations in column (vii). In addition, in column (iii) we consider NUN spells instead of just pure unemployment spells. Note that all of these regression extend the corresponding regressions reported in Section 1 of this appendix by including the relevant cyclical variable. Standard errors are derived by clustering at the quarter level.

\begin{table}[ht!]
\small
  \centering
  \caption{Cyclicality of Mobility Controlling for Demographics and Occupation Identities}
     \resizebox{1.0\textwidth}{!}{
    \begin{tabular}{llllllll}
        \toprule
    \toprule
          & 2000 SOC    & 1990 SOC    & 2000 SOC-NUN & OCC*IND & IND   & NR/R-M/C  & C/NRM/RM  \\
          & (1)   & (2)   & (3)   & (4)   & (5)   & (6)   & (7) \\
    \toprule

    \multicolumn{8}{c}{\textbf{Panel A. Regression of Individual Mobility on Linearly De-trended Unemployment Rate}} \\
    \midrule

    \multicolumn{8}{c}{A1: uncorrected, time controls, no demog, no occ/ind controls} \\
    \hline
    U.rate & -0.0788*** & -0.0890*** & -0.0583*** & -0.0606*** & -0.0952*** & -0.0575*** & -0.0589*** \\
    (s.e.) & (0.0180) & (0.0211) & (0.0153) & (0.0192) & (0.0219) & (0.0192) & (0.0191) \\
    \hline
    \multicolumn{8}{c}{A2: uncorrected, time and demog. controls, no occ/ind controls} \\
    \hline
    U.rate & -0.0763*** & -0.0910*** & -0.0552*** & -0.0579*** & -0.0904*** & -0.0542*** & -0.0547*** \\
    (s.e.) & (0.0176) & (0.0213) & (0.0151) & (0.0199) & (0.0217) & (0.0198) & (0.0198) \\
    \hline
    \multicolumn{8}{c}{A3: uncorrected, time, demog. \& source occ. Controls} \\
    \hline
    U.rate & -0.0707*** & -0.0810*** & -0.0521*** & -0.0546*** & -0.0783*** & -0.0560*** & -0.0578*** \\
    (s.e.) & (0.0176) & (0.0217) & (0.0148) & (0.0200) & (0.0217) & (0.0193) & (0.0192) \\
    \hline
    \multicolumn{8}{c}{A4: uncorrected, time, demog. Controls and dest. Occ controls} \\
    \hline
    U.rate & -0.0766*** & -0.0831*** & -0.0564*** & -0.0568*** & -0.0817*** & -0.0545*** & -0.0568*** \\
    (s.e.) & (0.0176) & (0.0214) & (0.0146) & (0.0197) & (0.0218) & (0.0197) & (0.0197) \\
    \midrule
    \multicolumn{8}{c}{\textbf{Panel B. Regression of Individual Mobility on HP-filtered Unemployment Rate}} \\
    \midrule

    \multicolumn{8}{c}{B1: uncorrected, time controls, no demog, no occ/ind controls} \\
    \hline
    HP U.rate & -0.1594*** & -0.1768*** & -0.1182*** & -0.1198** & -0.2092*** & -0.0840** & -0.0959** \\
    (s.e.) & (0.0420) & (0.0449) & (0.0366) & (0.0458) & (0.0543) & (0.0408) & (0.0419) \\
    \hline
    \multicolumn{8}{c}{B2: uncorrected, time and demog. controls, no occ/ind controls} \\
    \hline
    HP U.rate & -0.1520*** & -0.1782*** & -0.1088*** & -0.1137** & -0.2012*** & -0.0751* & -0.0829* \\
    (s.e.) & (0.0418) & (0.0457) & (0.0367) & (0.0472) & (0.0536) & (0.0421) & (0.0438) \\
    \hline
    \multicolumn{8}{c}{B3: uncorrected, time, demog. \& source occ. Controls} \\
    \hline
    HP U.rate & -0.1409*** & -0.1623*** & -0.1045*** & -0.1090** & -0.1825*** & -0.0782* & -0.0884** \\
    (s.e.) & (0.0396) & (0.0451) & (0.0347) & (0.0472) & (0.0542) & (0.0413) & (0.0427) \\
    \hline
    \multicolumn{8}{c}{B4: uncorrected, time, demog. Controls and dest. Occ controls} \\
    \hline
    HP U.rate & -0.1476*** & -0.1577*** & -0.1172*** & -0.1115** & -0.1697*** & -0.0770 & -0.0906* \\
    (s.e.) & (0.0427) & (0.0495) & (0.0364) & (0.0479) & (0.0542) & (0.0466) & (0.0474) \\
    \hline
    obs   & 12639 & 12591 & 16574 & 12260 & 12309 & 12639 & 11506 \\
    \bottomrule
    \bottomrule
        \multicolumn{8}{c}{{\scriptsize *$\ p<0.1$;\ **$\ p<0.05$;\ ***$\ p<0.01$}}
    \end{tabular}}
  \label{tab:3}%
\end{table}

The main message from this table is that including demographic controls, or occupation (industry) fixed effects do not seem to meaningful change the measured responsiveness of mobility with respect to the business cycle. Our evidence therefore strongly suggests that the procyclicality of  occupational/industry mobility among the unemployed is not driven by compositional shifts, whereby in expansions (recessions) we observe more (less) workers who are linked to occupation/industries or demographic characteristics that are associated with higher mobility.

\paragraph{Cyclical responsiveness when controlling for unemployment duration}

We now investigate the role of unemployment (non-employment) duration in the cyclicality of gross occupational mobility. To do so, we estimate regressions of the form
\begin{align}
 \mathbf{1}_{\text{occmob}} &= \beta_0 + \beta_{1}\text{Cyclical Variable} +\beta_{2}\ \text{(u. duration)} + \text{Controls} + \varepsilon. \tag{R-dur}\label{e:cycl_app_dur}
\end{align}
Panel A show the estimated coefficients of the HP-filtered (log) unemployment rate together and spell duration when using the uncorrected mobility data (see the main text for the $\Gamma$-corrected estimates. Panel B presents these estimates using the uncorrected data by further adding a time trend and controlling for classification changes. Panels C, D and E progressively add demographic controls (gender, education, race, age), source occupation fixed effects or destination occupation fixed effects, respectively. Panel F considers the estimated duration coefficient with added controls over the same uncorrected data, but without a cyclical regressor. Note that because of we use only those quarters with an uncensored duration distribution in the cyclical analysis (and in panel F), these data is a subset of the one used to estimate the overall duration profile in the main text and the previous sections of this appendix.

\begin{table}[ht!]
\small
  \centering
  \caption{Cyclicality of Mobility Controlling for Unemployment Duration}
     \resizebox{1.0\textwidth}{!}{
    \begin{tabular}{lccccccl}
    \toprule
    \toprule
          & 2000 SOC    & 1990 SOC    & 2000 SOC-NUN   & OCC*IND & IND   & NR/R-M/C  & C/NRM/RM  \\\\
          & \multicolumn{1}{c}{(1)} & \multicolumn{1}{c}{(2)} & \multicolumn{1}{c}{(3)} & \multicolumn{1}{c}{(4)} & \multicolumn{1}{c}{(5)} & \multicolumn{1}{c}{(6)} & \multicolumn{1}{c}{(7)} \\
    \midrule
    \midrule
    \multicolumn{8}{c}{Panel A: uncorrected, no demog, no time, no occ/ind controls } \\ \hline
    HP U.rate & -0.1655*** & -0.1831*** & -0.1126*** & -0.1251*** & -0.2068*** & -0.0834* & -0.0974** \\
    (s.e.) & (0.0475) & (0.0518) & (0.0385) & (0.047) & (0.0554) & (0.0449) & (0.0444) \\
    Duration & 0.0139*** & 0.0148*** & 0.0139*** & 0.0110*** & 0.0115*** & 0.0093*** & 0.0093*** \\
    (s.e.) & (0.0019) & (0.0017) & (0.0017) & (0.0019) & (0.0018) & (0.0015) & (0.0016) \\ \hline
    \multicolumn{8}{c}{Panel B: uncorrected, with time and classification controls} \\ \hline
    HP U.rate & -0.2037*** & -0.2237*** & -0.1494*** & -0.1549*** & -0.2458*** & -0.1137*** & -0.1240*** \\
    (s.e.) & (0.0421) & (0.0453) & (0.0360) & (0.0459) & (0.0540) & (0.0418) & (0.0422) \\
    Duration & 0.0142*** & 0.0151*** & 0.0141*** & 0.0112*** & 0.0117*** & 0.0095*** & 0.0095*** \\
    (s.e.) & (0.0019) & (0.0017) & (0.0017) & (0.0019) & (0.0018) & (0.0015) & (0.0016) \\ \hline
    \multicolumn{8}{c}{Panel C: uncorrected, with demog, time controls, no occ/ind controls} \\ \hline
    HP U.rate & -0.2019*** & -0.2297*** & -0.1419*** & -0.1527*** & -0.2418*** & -0.1085** & -0.1138** \\
    (s.e.) & (0.0417) & (0.0461) & (0.0361) & (0.0471) & (0.0530) & (0.0432) & (0.0444) \\
    Duration & 0.0160*** & 0.0166*** & 0.0143*** & 0.0125*** & 0.0130*** & 0.0107*** & 0.0106*** \\
    (s.e.) & (0.0019) & (0.0018) & (0.0016) & (0.0018) & (0.0017) & (0.0015) & (0.0016) \\ \hline
    \multicolumn{8}{c}{Panel D: uncorrected, source occupation controls, time and demographic controls} \\ \hline
    HP U.rate & -0.1863*** & -0.2106*** & -0.1357*** & -0.1436*** & -0.2178*** & -0.1123*** & -0.1206*** \\
    (s.e.) & (0.0397) & (0.0457) & (0.0340) & (0.0471) & (0.0534) & (0.0423) & (0.0432) \\
    Duration   & 0.0148*** & 0.0155*** & 0.0134*** & 0.0112*** & 0.0112*** & 0.0109*** & 0.0109*** \\
    (s.e.) & (0.0019) & (0.0017) & (0.0016) & (0.0019) & (0.0018) & (0.0015) & (0.0016) \\ \hline
    \multicolumn{8}{c}{Panel E: uncorrected, destination occupation controls, time and demographic controls} \\ \hline
    HP U.rate  & -0.2036*** & -0.2085*** & -0.1462*** & -0.1545*** & -0.2259*** & -0.1081** & -0.1188*** \\
    (s.e.) & (0.0397) & (0.0457) & (0.0332) & (0.0466) & (0.0521) & (0.0426) & (0.0433) \\
    Duration   & 0.0152*** & 0.0156*** & 0.0133*** & 0.0117*** & 0.0113*** & 0.0109*** & 0.0109*** \\
    (s.e.) & (0.0019) & (0.0017) & (0.0016) & (0.0019) & (0.0018) & (0.0015) & (0.0016) \\ \hline
    \multicolumn{8}{c}{Panel F: comparison (duration profile, no cycle, with time + demog controls)} \\ \hline
    Duration  & 0.0149*** & 0.0153*** & 0.0136*** & 0.0117*** & 0.0119*** & 0.0102*** & 0.0101*** \\
    (s.e.) & (0.0019) & (0.0018) & (0.0016) & (0.0018) & (0.0018) & (0.0015) & (0.0016) \\
    \bottomrule \bottomrule
    \multicolumn{8}{c}{{\scriptsize *$\ p<0.1$;\ **$\ p<0.05$;\ ***$\ p<0.01$}}
    \end{tabular}}
  \label{tab:regress_duration}%
\end{table}%

Relative to the results in Table \ref{tab:3} (panel B), we observe that controlling for the duration of the unemployment (non-employment) spell increases the cyclical responsiveness of occupational mobility. This arises as the estimated unemployment coefficient conditional on duration captures the vertical shift of the mobility-duration profile. Without controlling for duration, the unemployment coefficient captures both the vertical shift of the profile and the rightward shift of the unemployment duration distribution that one observes in recessions. The inclusion of further controls does not meaningfully change either the slope of the profile or the cyclical responsiveness of occupational mobility. This suggests that the scope to link compositional shifts across source occupations (or demographics) to the observed cyclicality of mobility \emph{conditional on completed unemployment duration} is limited. Further, it also runs counter to a role of occupational or demographical shifts in our finding that recessions also exhibit a moderately increasing mobility-duration profile for the unemployed
.
\begin{figure}[h!t]
\centering
\subfloat[2000 SOC - Aggregate mobility-duration profile]{\label{f:cycl4_mm} \includegraphics [width=0.5 \textwidth]{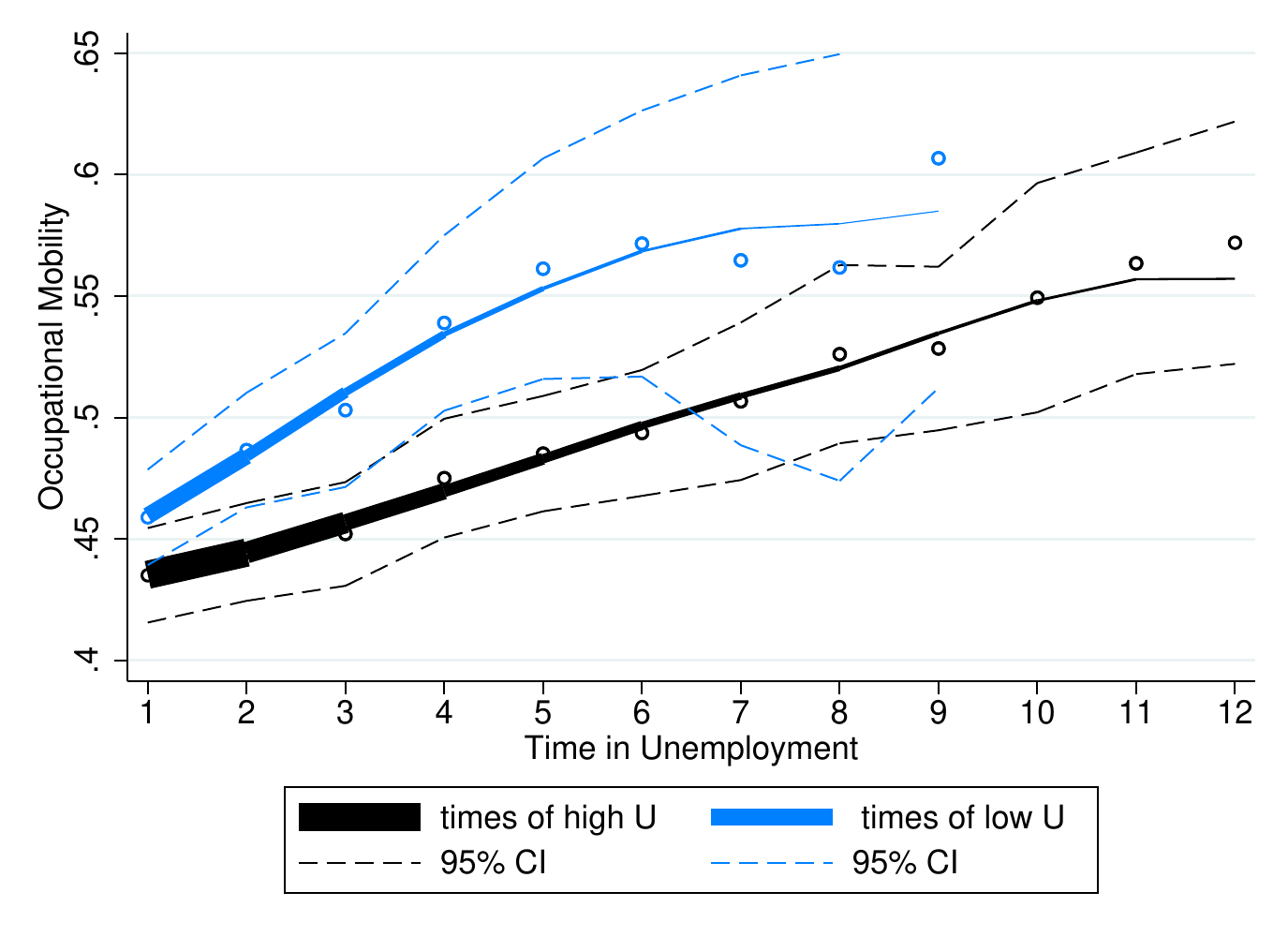}}
\subfloat[2000 SOC - Marginal mobility-duration profile]{\label{f:cycl4_mm_distr} \includegraphics [width=0.5 \textwidth]{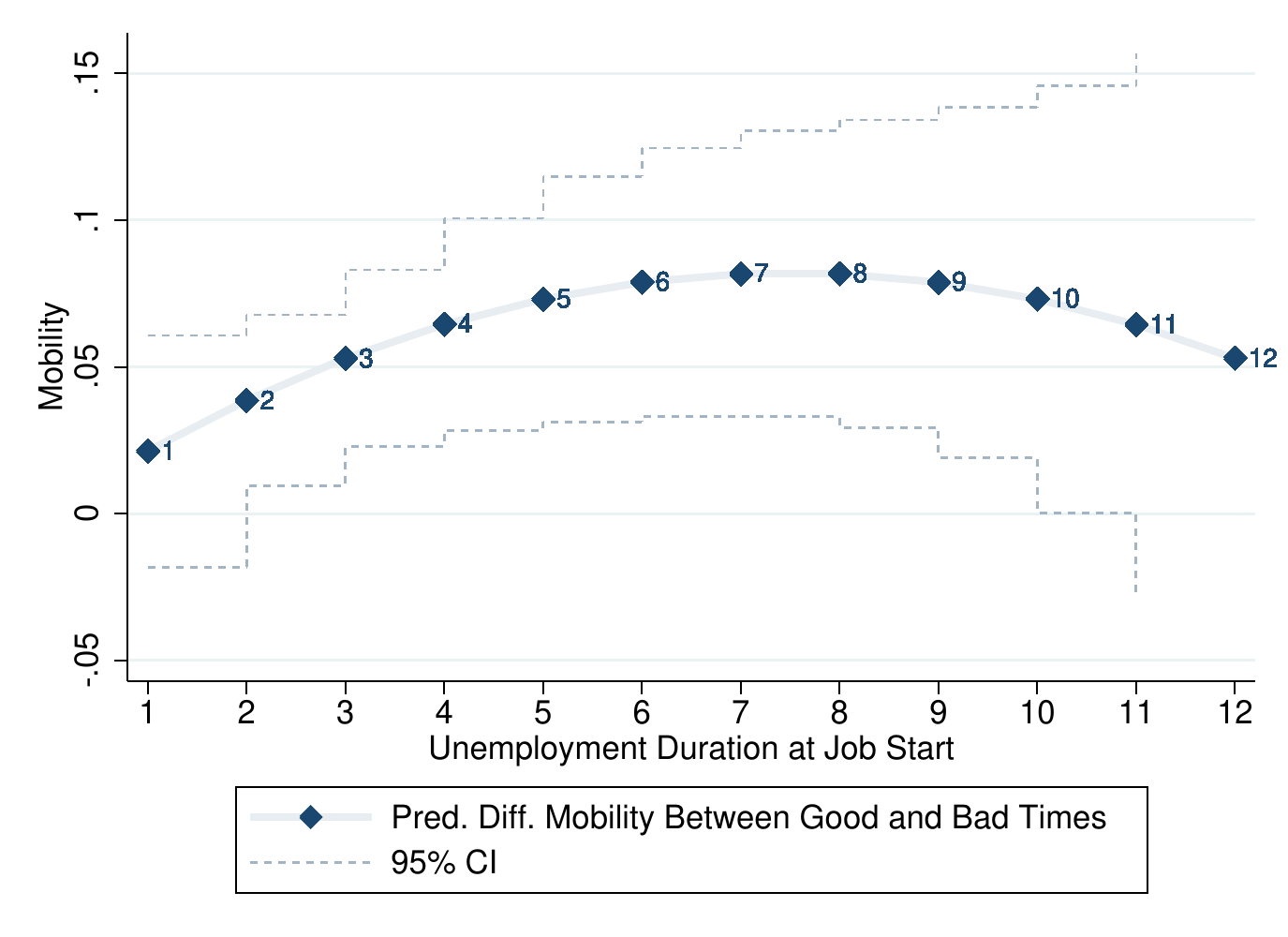}}

\subfloat[4RTMC - Aggregate mobility-duration profile]{\label{f:cycl4_mm} \includegraphics [width=0.5 \textwidth]{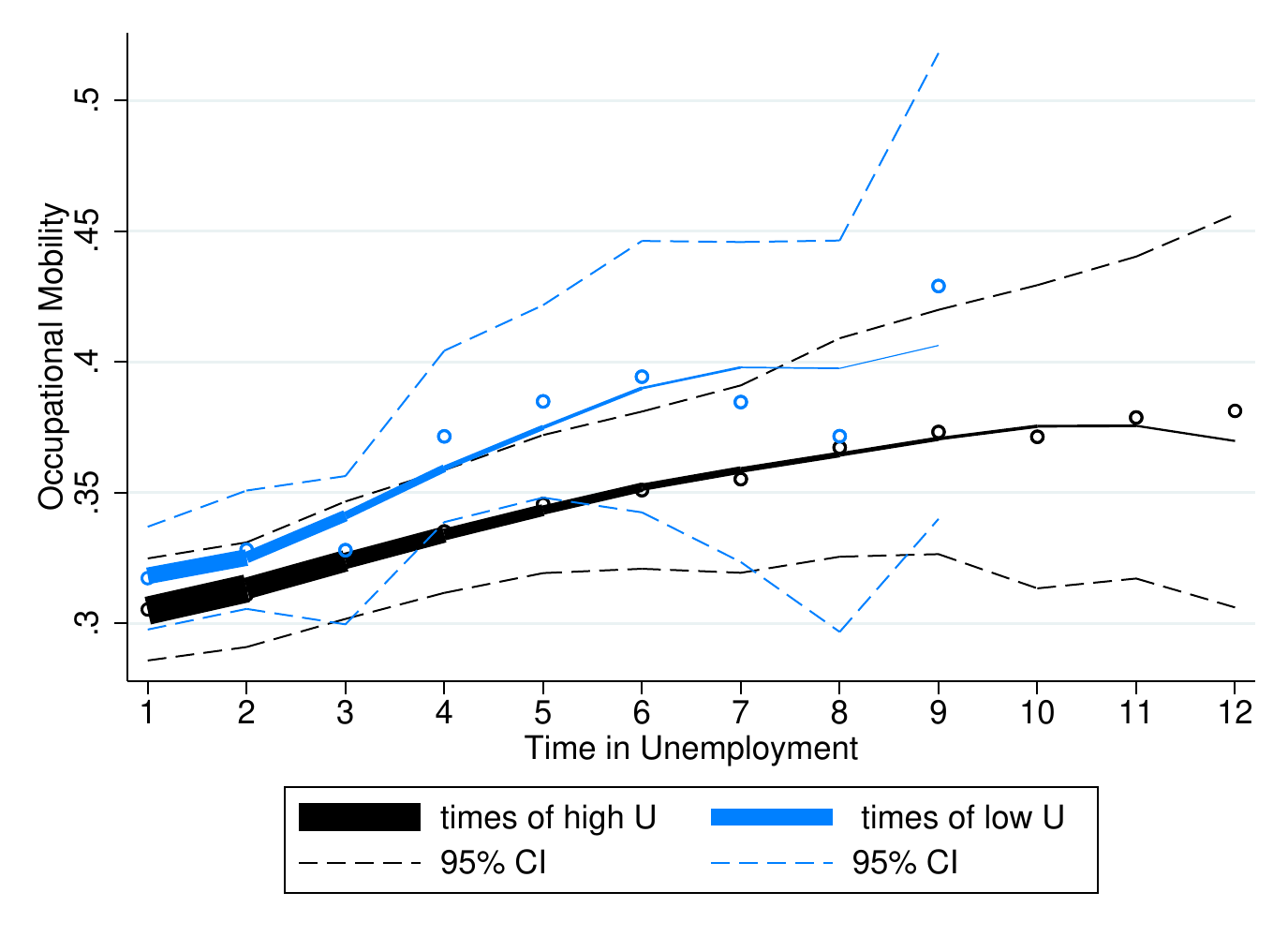}}
\subfloat[4RTMC - Marginal mobility-duration profile]{\label{f:cycl4_mm_distr} \includegraphics [width=0.5 \textwidth]{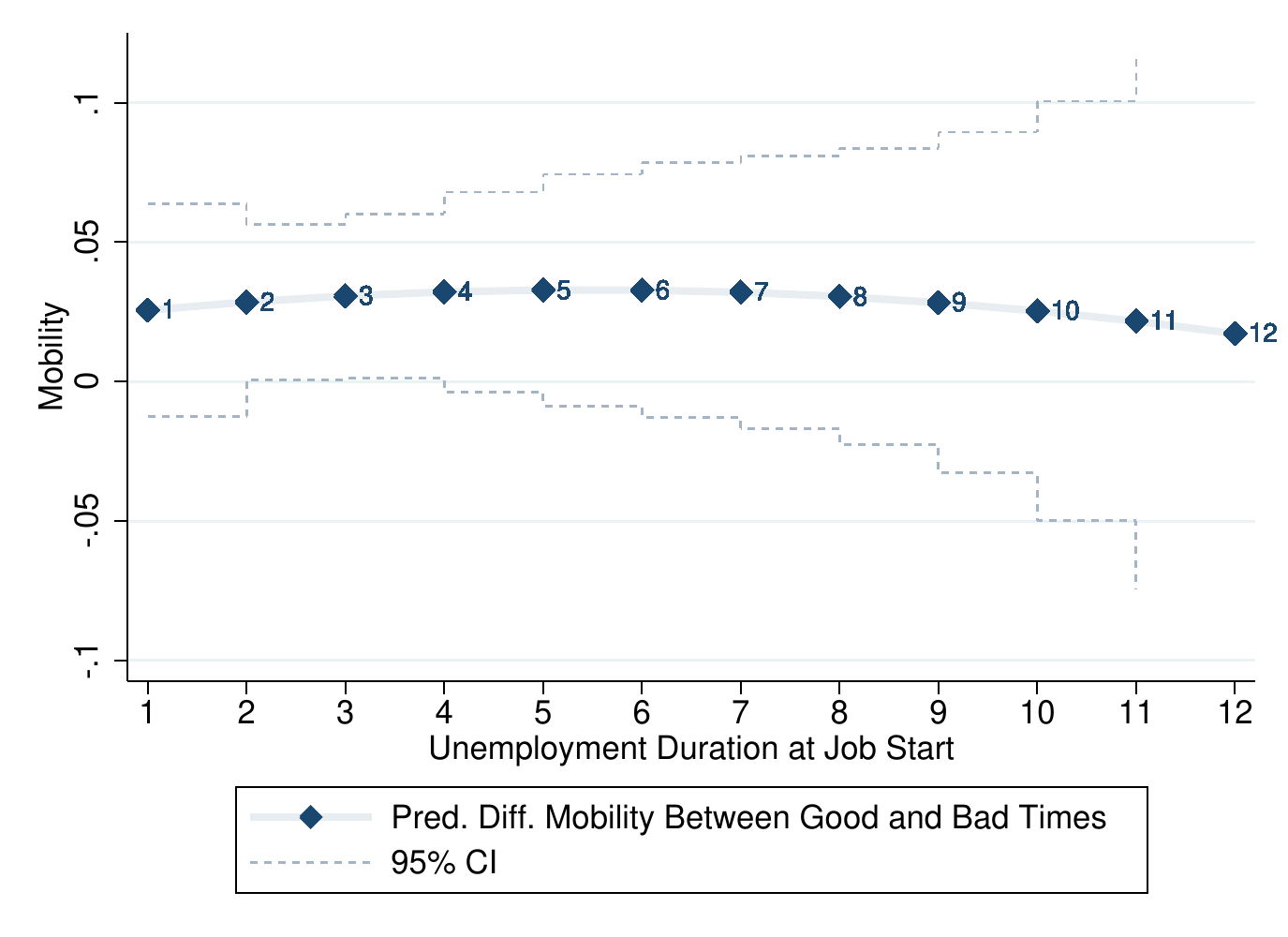}}

\subfloat[4RTMC, excl. Mgt - Aggregate mobility-duration profile]{\label{f:cycl4_mm} \includegraphics [width=0.5 \textwidth]{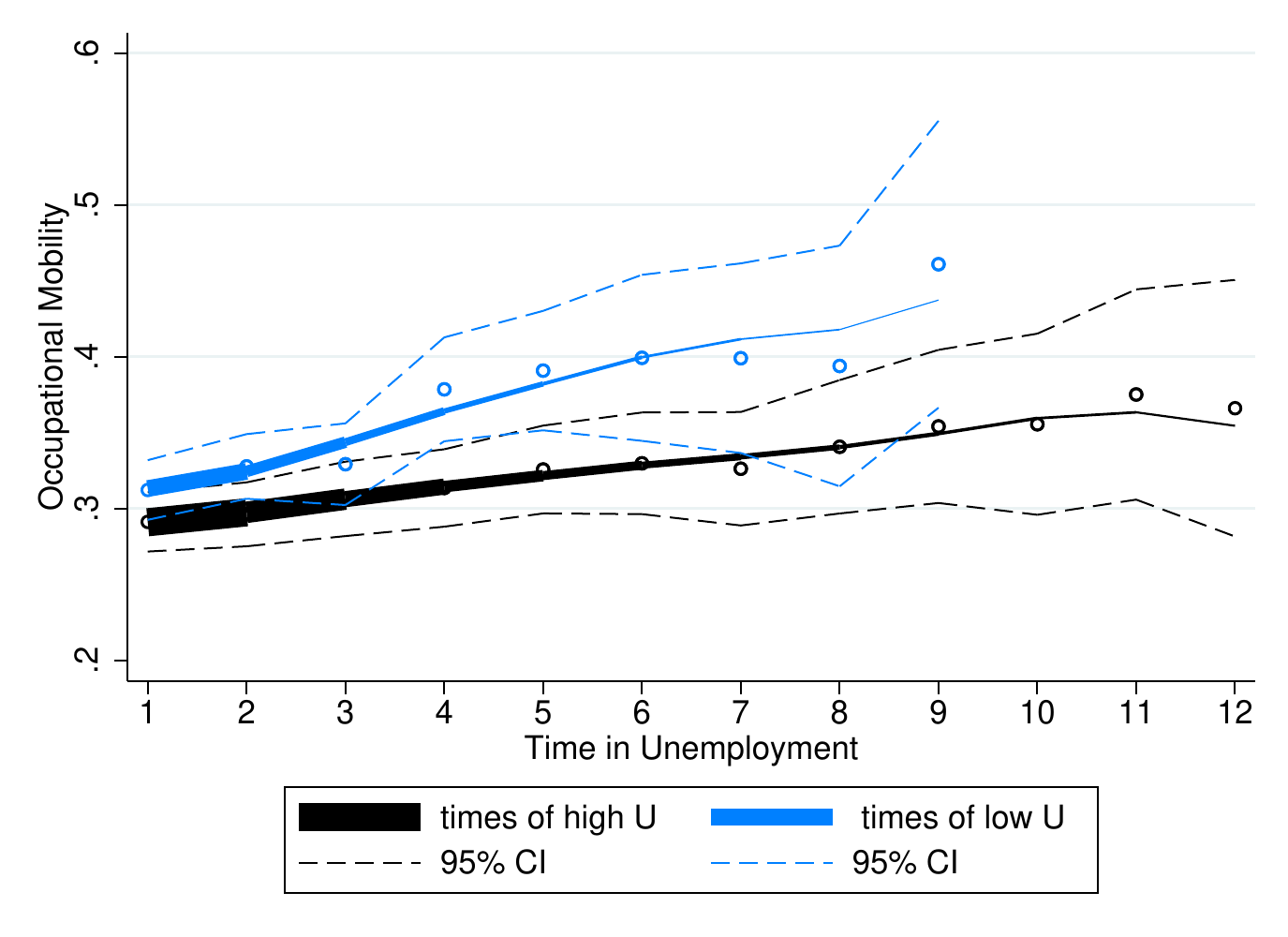}}
\subfloat[4RTMC, excl. Mgt - Marginal mobility-duration profile]{\label{f:cycl4_mm_distr} \includegraphics [width=0.5 \textwidth]{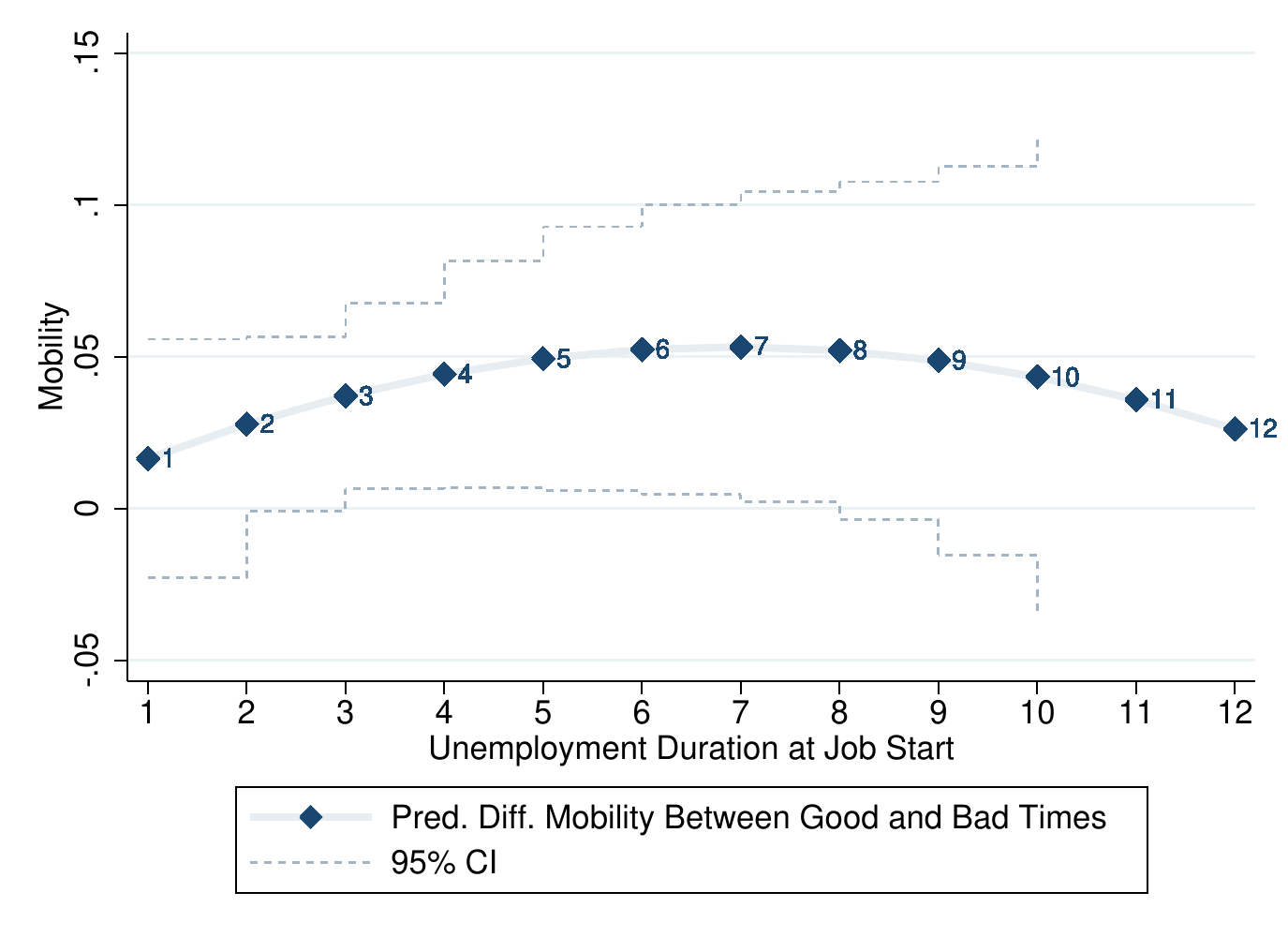}}
\caption{Cyclical responsiveness of the mobility-duration profile\label{f:cycl_90occmog}}
\end{figure}

\subsection{Cyclical responsiveness of the mobility-duration profile}

We now turn to investigate the cyclicality of the mobility-duration profile. Figures \ref{f:cycl_90occmog}a, c and e depict the mobility-duration profile in periods of high and low unemployment, using the major occupations of the 2000 SOC and the 4 task-based categories. Periods of high unemployment are defined as those in which the HP-filtered (log) unemployment rate lies within the top third of the distribution of HP-filtered (log) unemployment rates; while periods of low unemployment are defined as those in which the HP-filtered (log) unemployment rate lies within the bottom third of the distribution. The thickness of each profile reflects the number of unemployed workers with an unemployment spell of at least $x$ months of duration. It is readily seen that in periods of high unemployment there are both more unemployed workers and longer unemployment spells. The main conclusion from these graphs is that occupational mobility is higher in periods of low unemployment than in periods of high unemployment \emph{at all unemployment durations}. Figure \ref{f:cycl_90occmog}a shows that when considering major occupations the two mobility-duration profiles are statistically different from each other for those unemployment durations of up to 6 months. Beyond this point, the mobility-duration profile associated with periods of low unemployment becomes thinner and its confidence bands become wider.\footnote{We do not plot the duration profile when the associated confidence interval becomes wider than 0.2.} Figure \ref{f:cycl_90occmog}e shows that when considering the 4 task-based occupations without the managerial occupations, the two mobility-duration profiles are also statistically different from each at shorter unemployment durations. Including the managerial occupations decreases the precision of the estimates, generating wider confidence intervals.

As a complementary way to investigate the shift of the mobility-duration profile between periods of high and low unemployment, Figures \ref{f:cycl_90occmog}b, d and f depict the \emph{marginal} mobility-duration profile. The latter measures the change in occupational mobility at the same \emph{completed} unemployment duration between periods of high and low unemployment. This is in contrast to the mobility-duration profiles considered in Figures \ref{f:cycl_90occmog}a, c and e where, for example, a substantial part of those in unemployment at 4 months will still be in unemployment at 5 months, and thus contributing to the average occupational mobility at 4 and 5 months. Since the construction of the marginal mobility-duration profile relies on a much lower number of observations at each duration, we make a functional form assumption on the shift of this profile over the cycle. Specifically, we estimate the probability that a worker changed occupation (industry) at a given unemployment duration as
\begin{align}
 \mathbf{1}_{\text{occmob}} &= \beta_0 + \beta_{1}\mathbf{1}_{\text{Cycl}} +\sum_{n=1}^{12}\beta_{2n} \ \text{(u. duration dummy)} + \nonumber \\
 & \qquad \qquad  +\beta_{3}\left(\mathbf{1}_{\text{Cycl}}\times \text{u. duration}\right) + \beta_{4}\left(\mathbf{1}_{\text{Cycl}}\times \text{(u. duration)}^2\right) + Controls + \varepsilon, \tag{R-X}\label{e:cycl_app_r1}
\end{align}
where $\mathbf{1}_{\text{Cycl}}$ is the cyclical indicator (0 for periods of high unemployment and 1 for periods of low unemployment), and unemployment duration refers to completed spell duration. We estimate this equation on the uncorrected (from miscoding) SIPP data, controlling for a linear time trend and classification changes. Note that equation \ref{e:cycl_app_r1} allows us to shift the marginal mobility-duration profile differently at different durations following a quadratic relationship.\footnote{We have also experimented with a cubic and quartic relationship. These alternatives at times make the response stronger for durations between 6-10 months, while declining beyond 10 months. However, beyond 10 months, the confidence intervals in all cases become rather wide. Using an alternative specification, where instead of duration dummies for the profile, we use a linear baseline relation between mobility and duration does not change the presented relationship meaningfully. Using a cubic or quartic formulation does not meaningfully change the direction of the shifts or its statistical significance, throughout this section.} Figures \ref{f:cycl_90occmog}b, d and f show that in times of high unemployment, workers who end their unemployment spells at any duration have a lower probability of changing occupation. The differences in probability of an occupational change is statistically significant for durations between 2 and 10 months when using the 2000 SOC and between 2 and 7 months when using the 4 task-based categories, excluding managerial occupations.

\begin{figure}[ht!]
\centering
\subfloat[Major Occupation Groups (1990 SOC) ]{\label{f:cycl4_mm} \includegraphics [width=0.5 \textwidth]{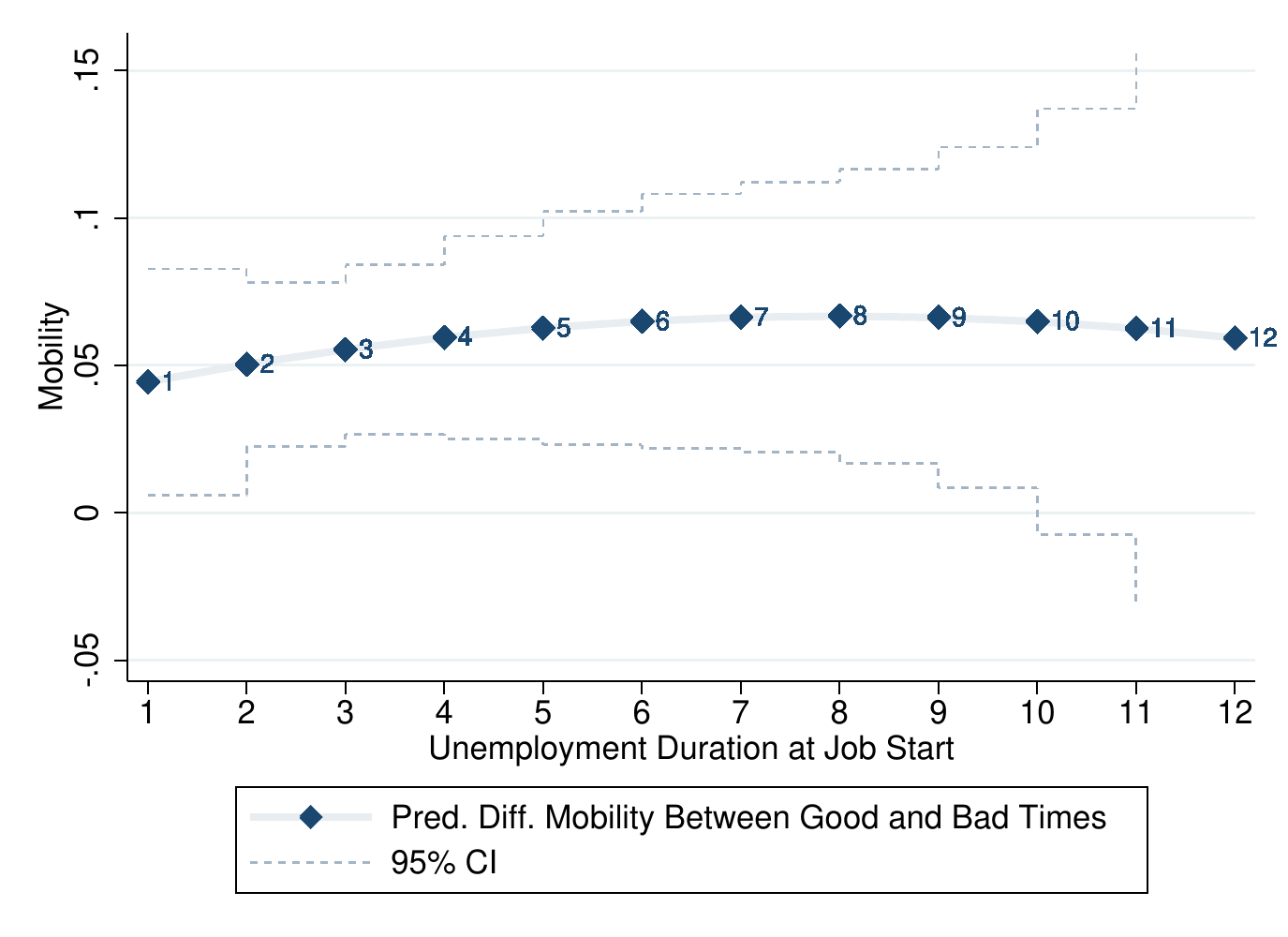}}
\subfloat[NRMC ]{\label{f:cycl4_rtmm_} \includegraphics [width=0.5 \textwidth]{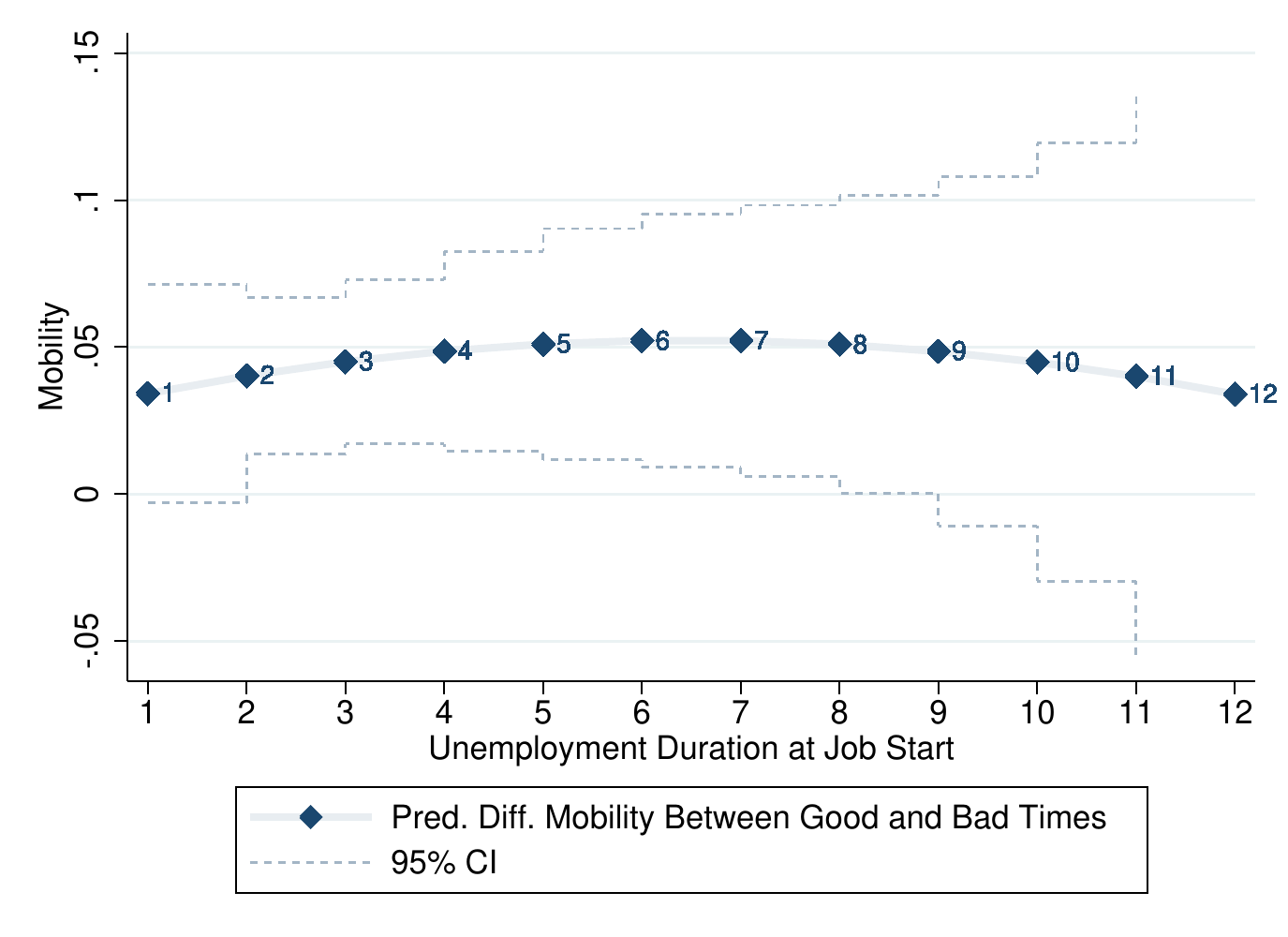}}
\caption{Cyclical Occupational Mobility Shift of the Unemployed\label{f:cycl_lindt}}
\end{figure}

Figure \ref{f:cycl_lindt} shows the marginal mobility-duration profile by estimating regression \ref{e:cycl_app_r1} using instead the linearly de-trended unemployment rate as the cyclical indicator. In this case we observe an even stronger difference in the occupational mobility rates of unemployed workers at any completed duration during periods of high and low unemployment rates. In particular, for the 4 task-based categories, including managers, we now see that in periods of high unemployment the entire profile shifts down, and statistically significantly so between 2 and 8 months.

\begin{figure}[ht!]
\centering
\subfloat[2000 SOC - Corrected Mobility in Surviving Spells]{\label{f:cycl4_mm} \includegraphics [width=0.45 \textwidth]{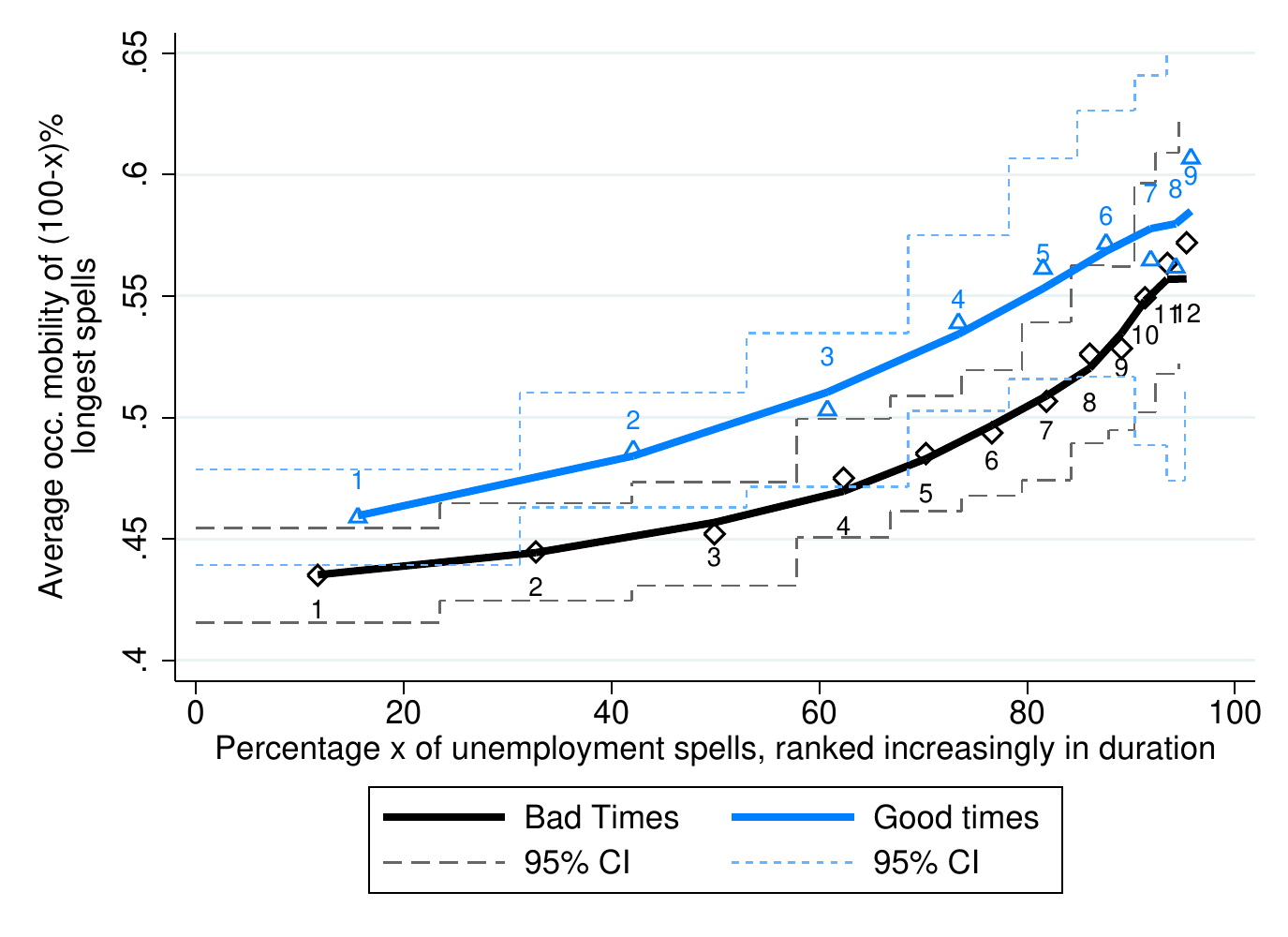}}
 \quad  \subfloat[2000 SOC - Cyclical Shift of Mobility across Percentiles of Completed Spell Distribution]{\label{f:cycl4_mm_distr} \includegraphics [width=0.45 \textwidth]{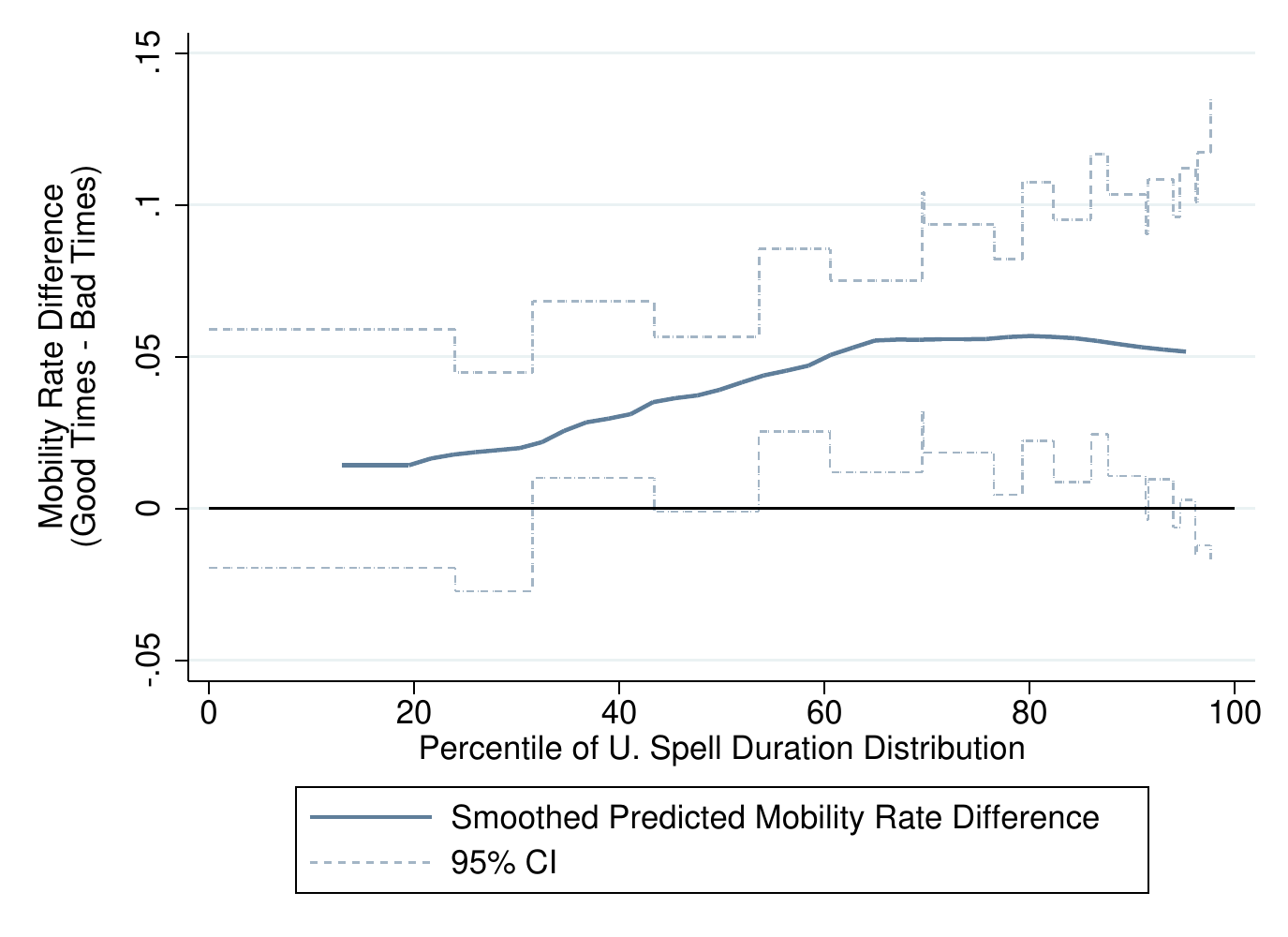}}

\subfloat[NRMC excl. Mgt, Code-Error Corr in Surviving Spells ]{\label{f:cycl4_rtnm} \includegraphics [width=0.45 \textwidth]{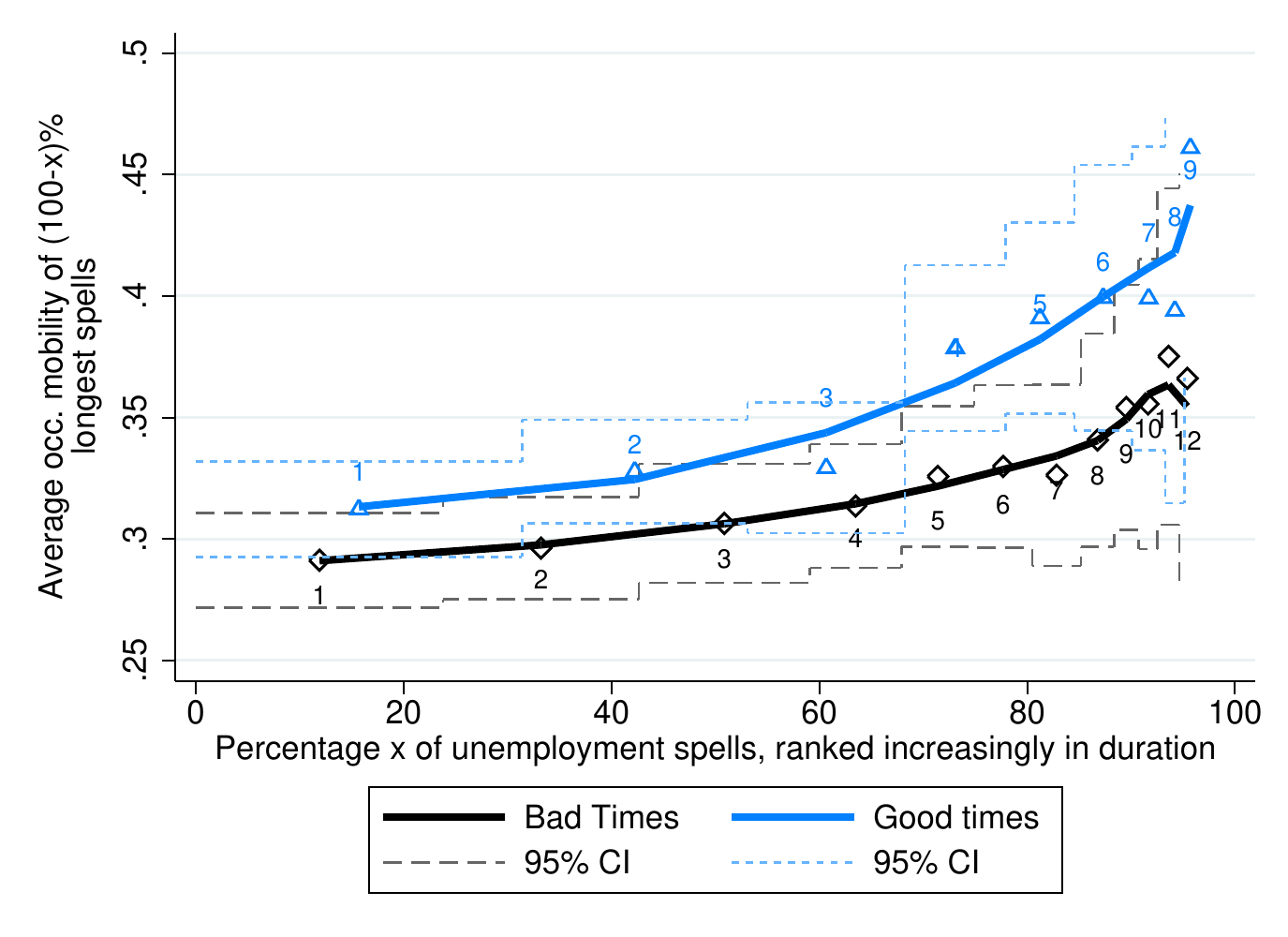}}
\quad \subfloat[NRMC excl. Mgt, Cyclical Shift of Mobility across Percentiles of Completed Spell Distribution]{\label{f:cycl4_rtnm_distr} \includegraphics [width=0.45 \textwidth]{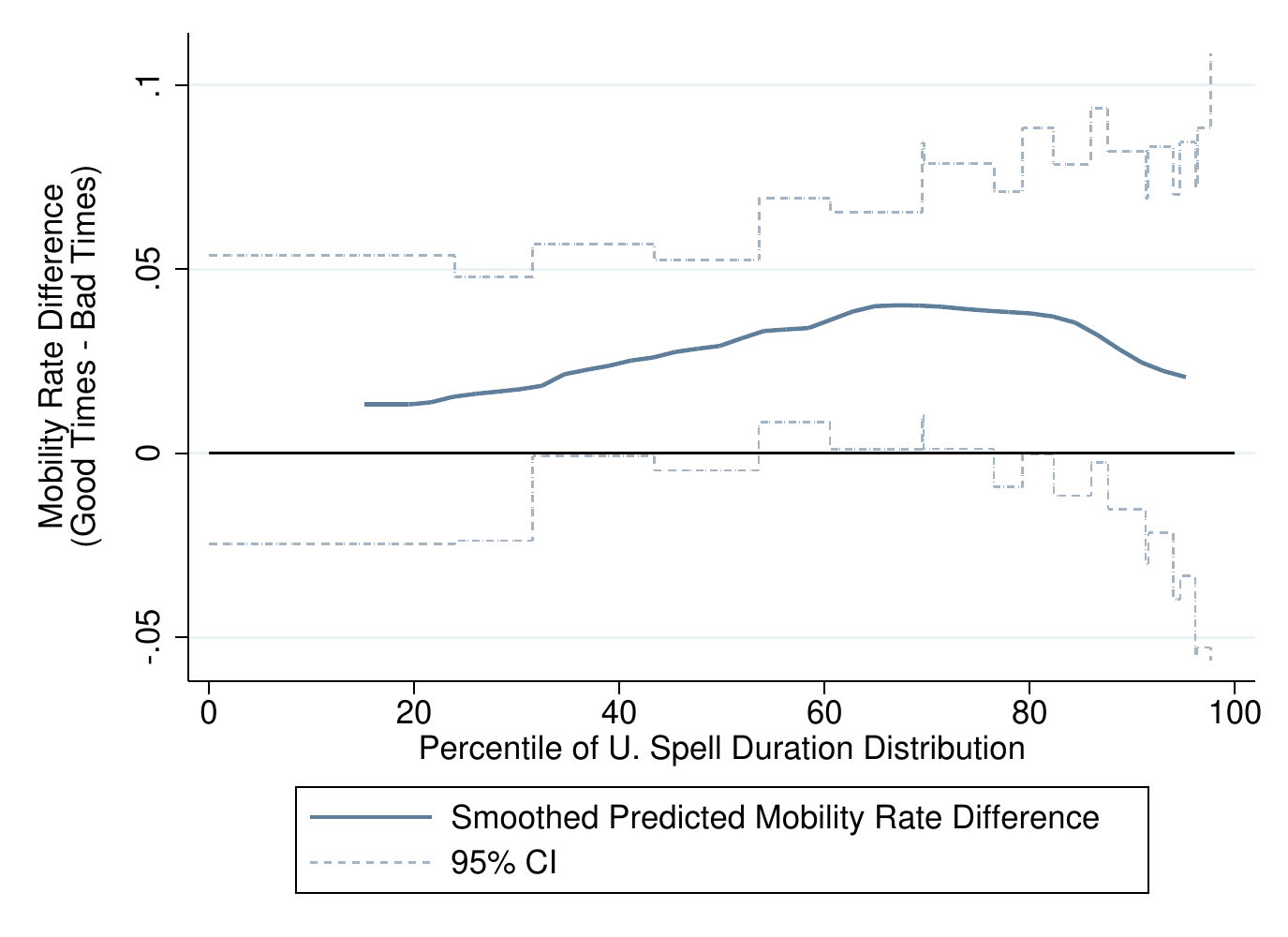}}
            \caption{Cyclical change in Mobility, and Rank in Unemployment Duration Distribution \label{f:cycl_durdistr_shift_mm_rtnm}}
\end{figure}

\paragraph{Occupational Mobility and Rank in the Duration Distribution over the Business Cycle}

To investigate to what extend the rightward shift in the mobility-duration profile observed in Figures \ref{f:cycl_90occmog}a, c and e is due to a rightward shift in the unemployment duration distribution, we derive workers' occupational mobility as a function of the rank of their unemployment spell in the duration distribution. We do this both in times of high and low unemployment, as defined above. Figures \ref{f:cycl_durdistr_shift_mm_rtnm}a and c, depict the relation between the average occupational mobility rate of the 100-x\% of longest unemployment spells with the rank of these unemployment spells in the unemployment duration distribution. For both the major occupational groups of the 2000 SOC and the 4 task-based categories we observe that at \emph{any} given rank, occupational mobility is lower in periods of high unemployment and this difference appears statistically significant for a wide interval of ranks around the median. This implies that the downward shift of the mobility-duration profile in times of high unemployment goes beyond the rightward shift of the duration distribution associated with recessions. In particular, we do not observe that the business cycle shifts mobility at lower quantiles of the duration distribution in a different direction than it does at the higher quantiles. Therefore, it is not the case that \emph{relatively} shorter unemployment spells display more occupational attachment in a recession, while simultaneously the relatively longer spells display less occupational attachment.

Figures \ref{f:cycl_durdistr_shift_mm_rtnm}b and d, investigate this issue further and depict the estimated change in occupational mobility between period of high and low unemployment on individual panel (uncorrected) data, controlling for a time-trend and classification effects. We analyse occupational mobility \emph{at} a given percentile of the (completed) duration distribution, rather than the average occupational mobility of the subset of all spells including \emph{and above} that percentile, as in Figures \ref{f:cycl_durdistr_shift_mm_rtnm}a and c. Once again since this exercise involves a lower number of observations we make a functional form assumption on the relationship between unemployment duration and occupational mobility. In particular, we estimate a variant of \ref{e:cycl_app_r1} as
\begin{align}
 \mathbf{1}_{\text{occmob}} &= \beta_0 + \beta_{1}\mathbf{1}_{\text{Cycl}} +\beta_{2,1}\ \text{(u. duration)} + \beta_{2,2}\ \text{(u. duration)}^2 \nonumber \\
 & \qquad \qquad +\beta_{3}\left(\mathbf{1}_{\text{Cycl}}\times \text{u. duration}\right) + \beta_{4}\left(\mathbf{1}_{\text{Cycl}}\times \text{(u. duration)}^2\right) + \text{Controls} + \varepsilon, \tag{R-XX}\label{e:cycl_app_r2}
\end{align}
where $\mathbf{1}_{\text{Cycl}}$ is the cyclical indicator (0 for periods of high unemployment and 1 for periods of low unemployment) and instead of unemployment duration dummies we have a smooth quadratic duration profile in recessions, and a (potentially) different quadratic duration profile in booms.\footnote{We also have experimented with linear, cubic and quartic specifications of the duration profile, which do not lead to different conclusions for the first-order patterns discussed here, unless stated.}

\begin{figure}[th!]
\centering
\subfloat[21 MOG (2000) Mobility in Outflow from U ]{\label{f:cycl4_rtmm_distr_lindt} \includegraphics [width=0.5 \textwidth]{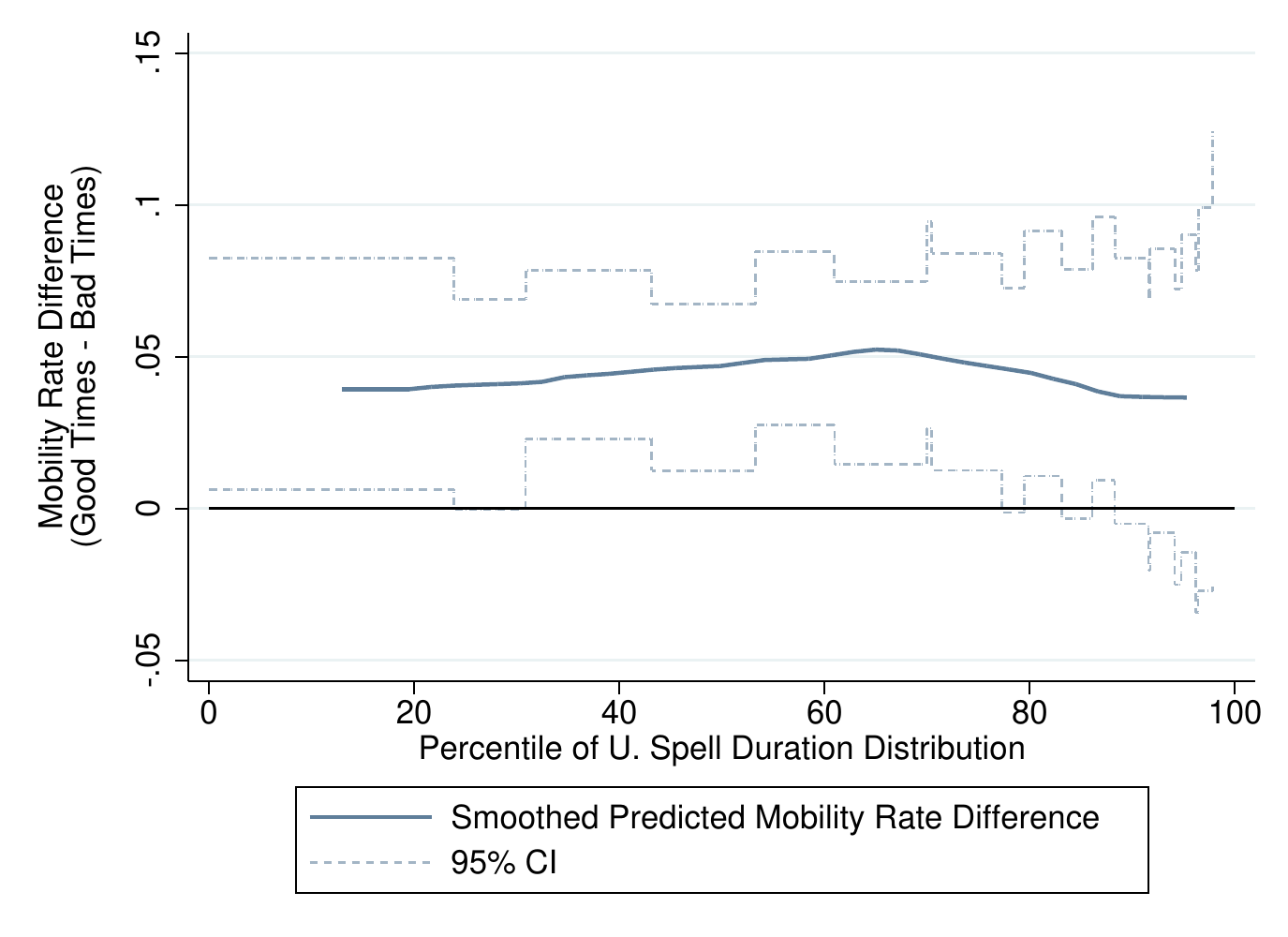}}
\subfloat[4 NRMC excl. Mgt Mobility in Outflow from U ]{\label{f:cycl4_rtnm_distr_lindt} \includegraphics [width=0.5 \textwidth]{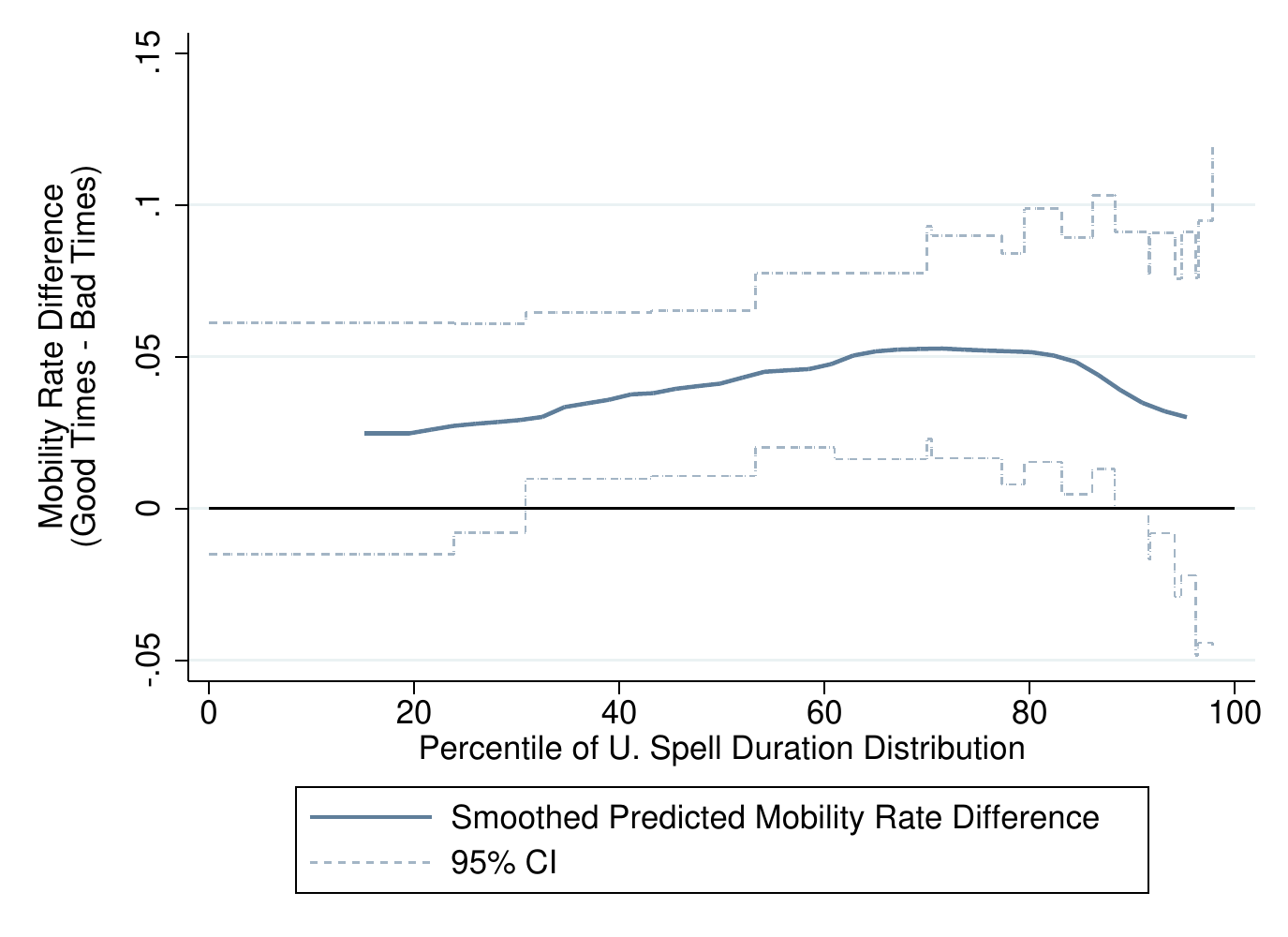}}
\caption{Good and Bad Times according to Linearly-Detrended Unemployment Series\label{f:cycl_durdistr_shift_mm_rtnm_lindt}}
\end{figure}

For mobility across major occupational groups we observe that at any given percentile of the completed unemployment duration distribution, mobility is higher in periods of low unemployment relative to periods of high unemployment. This difference is statistically significant for the vast majority of spells between the 35th and 90th percentile. This pattern is very similar for mobility across the 4 task-based categories, albeit statistically weaker. Thus recessions appear to reduce occupational mobility \emph{across} a wide range of the percentiles of the unemployment spell distribution. Figure \ref{f:cycl_durdistr_shift_mm_rtnm_lindt} shows the estimates for the same exercises as before, but using the linearly de-trended unemployment rates as a cyclical indicator. Here we also observe a cyclical shift in mobility across the whole distribution of unemployment spells. In this case, the difference between periods of high and low unemployment is statistically significant at almost all quantiles of the distribution up to the 80-90th percentile (2000 SOC) and between the 35th and 95th percentile (4 task-based caterories).

\begin{figure}[h!]
\centering
\subfloat[Major Occ. Groups (1990 SOC) ]{\label{f:cycl4_dd} \includegraphics [width=0.5 \textwidth]{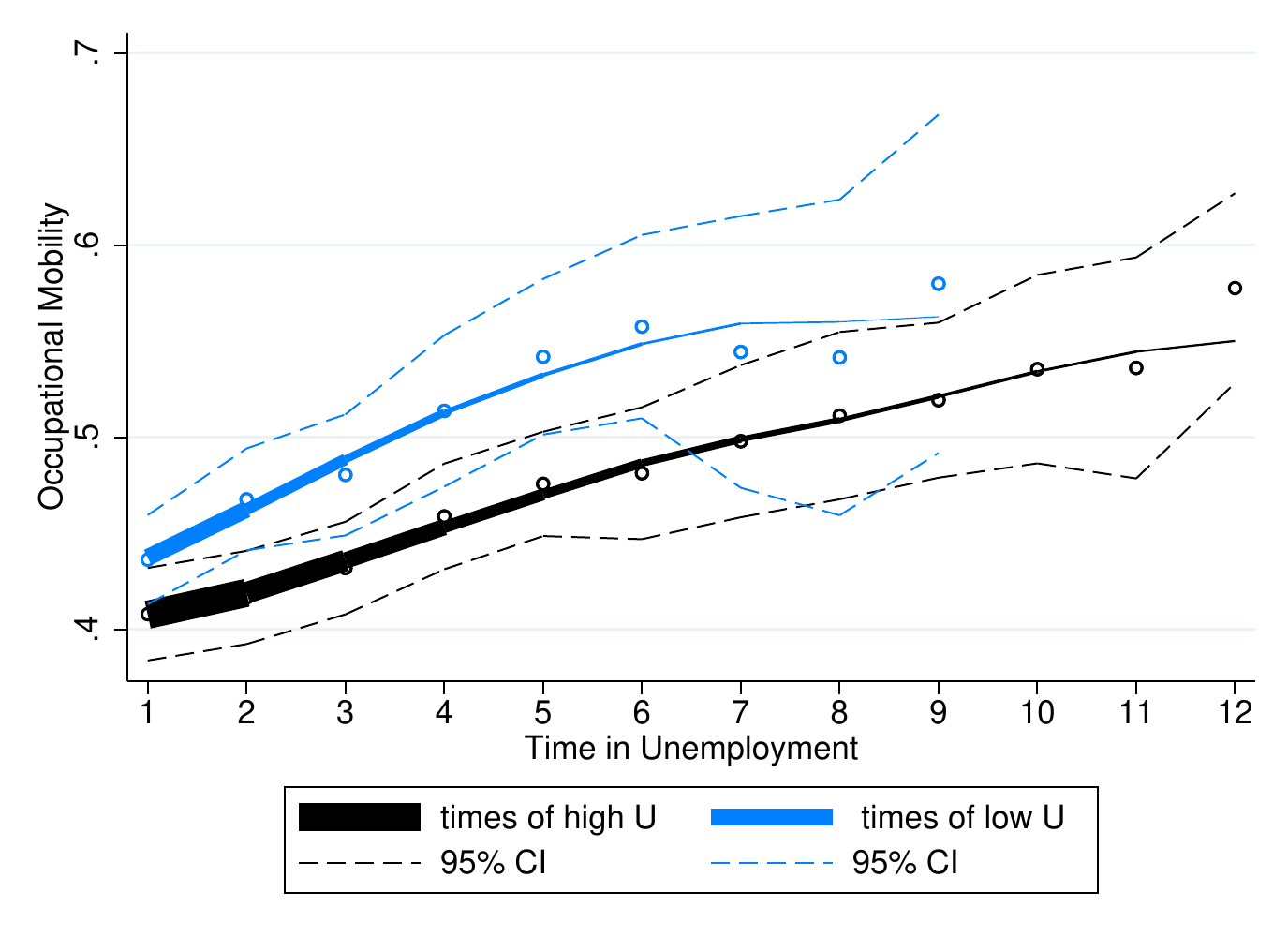}}
\subfloat[Major Occ. Groups (1990 SOC)]{\label{f:cycl4_dd_distr} \includegraphics [width=0.5 \textwidth]{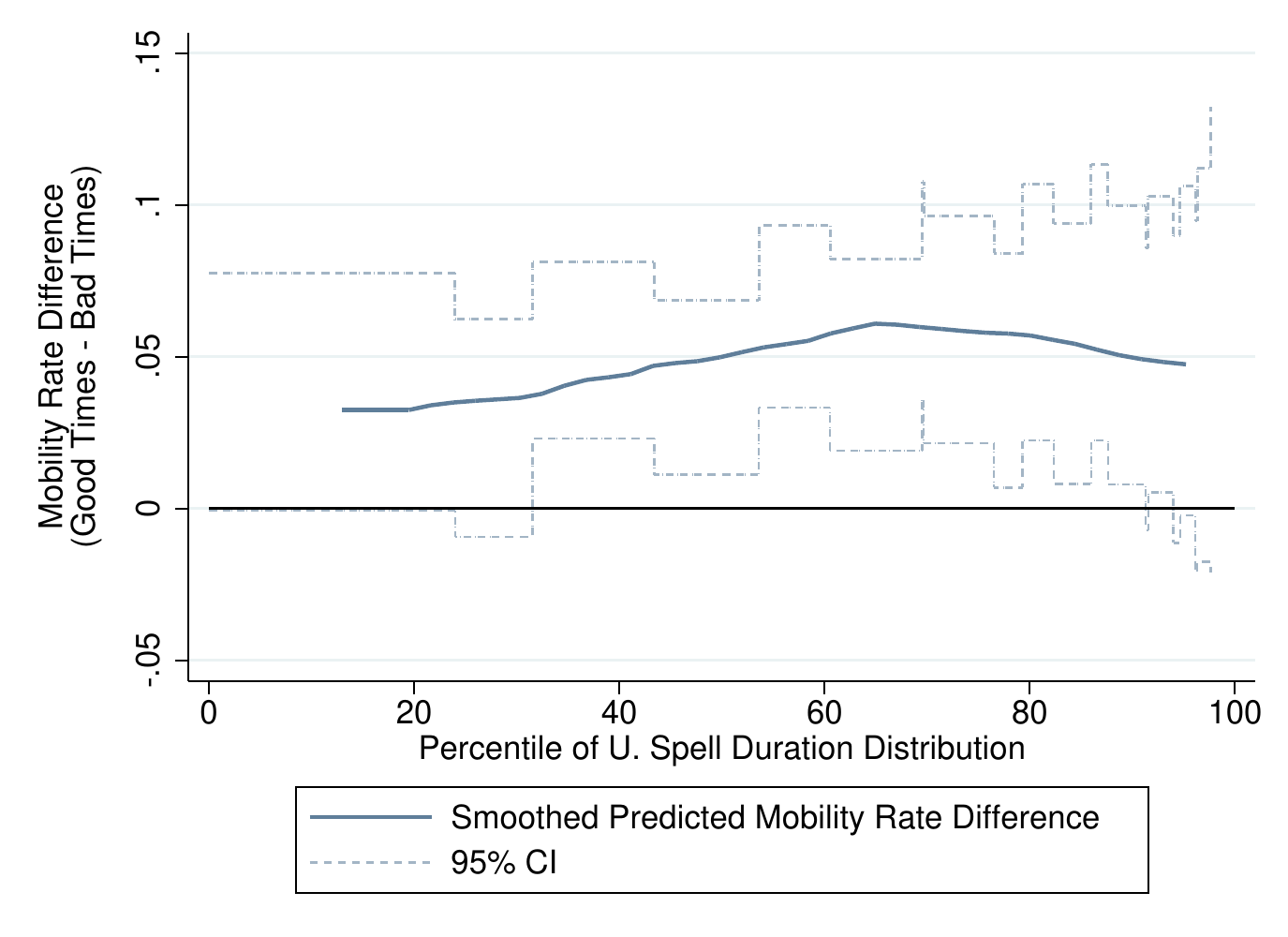}}

\subfloat[Major Industry Groups (1990 Census Ind. Cl.) ]{\label{f:cycl4_ind} \includegraphics [width=0.5 \textwidth]{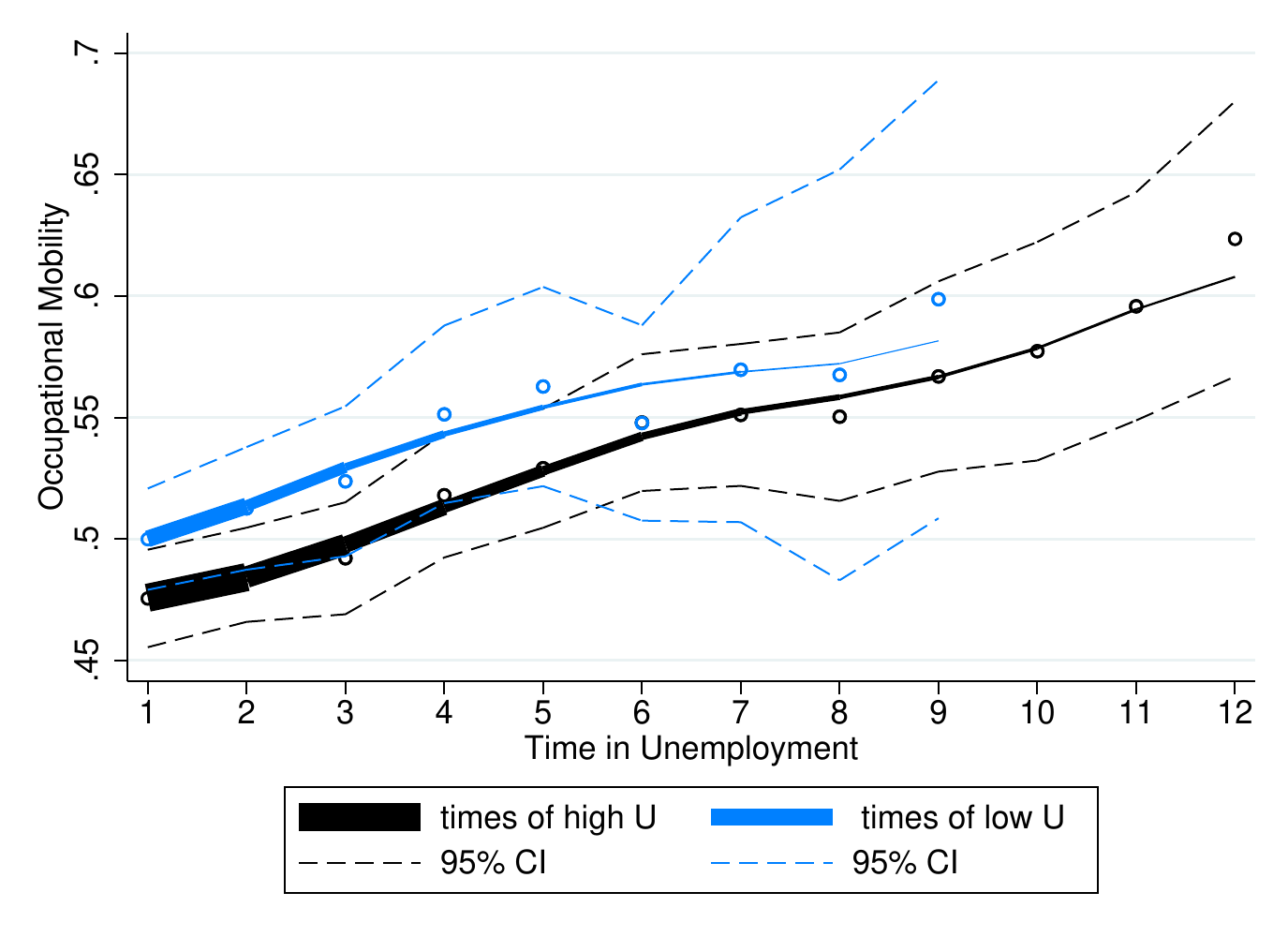}}
\subfloat[Major Industry Groups (1990 Census Ind. Cl.)]{\label{f:cycl4_ind_distr} \includegraphics [width=0.5 \textwidth]{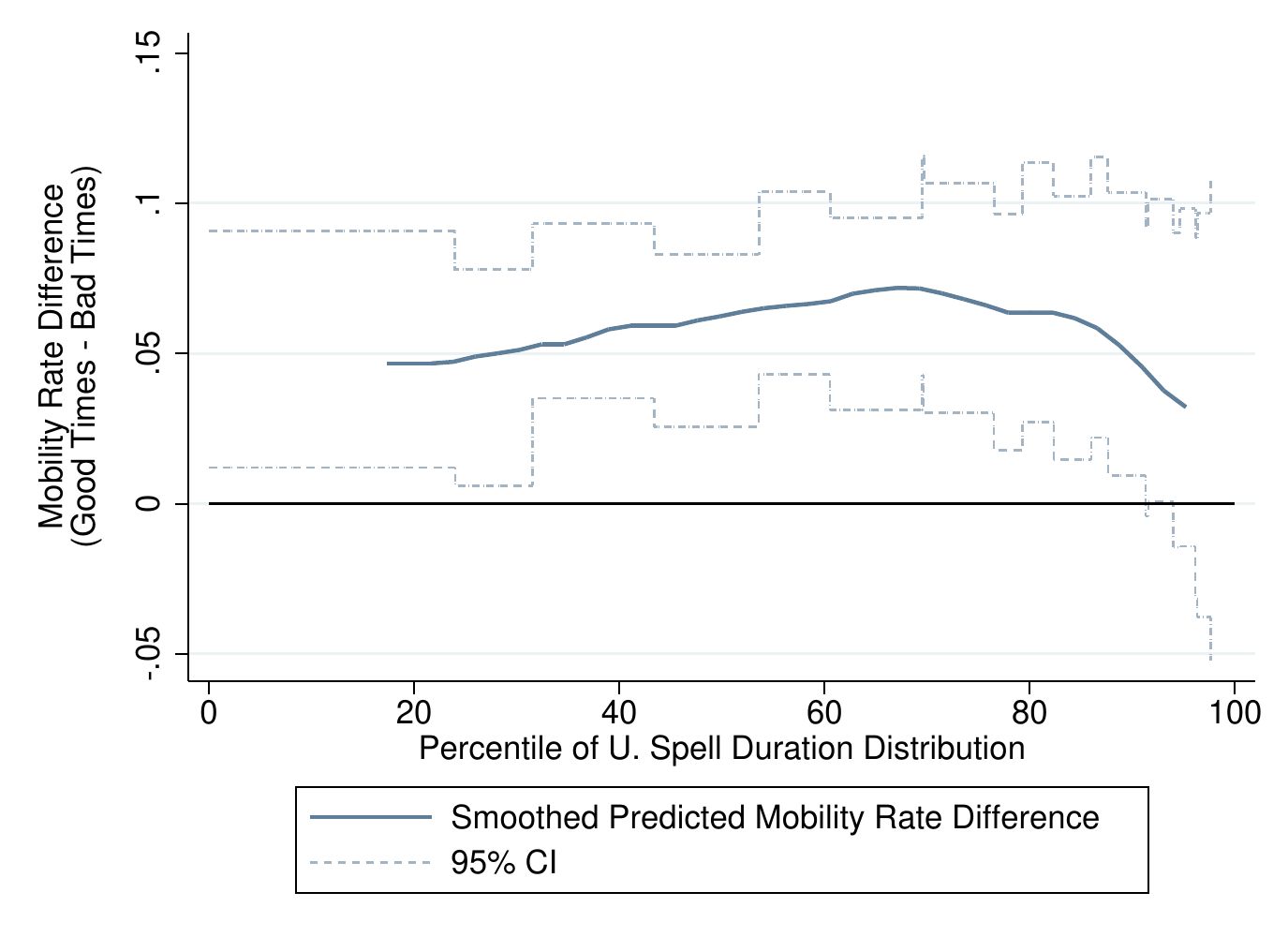}}
\caption{Occupation and Industry Mobility of the Unemployed (1990 Census Classifications) \label{f:cycl_ind_dd}}
\end{figure}

\paragraph{Conclusion}  In recessions, unemployment spells are longer and unemployed workers are less mobile. This pattern is shared across many subsets of the population, when dividing by gender, education, age, occupations and industries. Within a recession as within an expansion, longer-unemployed workers are relatively more mobile. We show that the mobility-duration profile appears to shift down in recessions. However, the cyclical shift down of the duration profile in recession goes beyond a slowdown in the job finding rate. Throughout the quantiles of the unemployment spell distribution, occupational mobility goes down in recessions (statistically so for a substantial part of the distribution). This pattern appears a general feature of occupational and industrial mobility of the unemployed, and is likewise present  for the 1990 SOC and for 1990 Industry Census Classification, in Figure \ref{f:cycl_ind_dd}.

\subsection{The cyclical behaviour of net occupational mobility}

To study the cyclical behavior of net mobility we split the sample of unemployment spells into two groups. The first group is composed by those spells that ended in quarters with the \emph{highest} $(100-\underline{x}_b)\%$ of HP filtered (log) unemployment rates. We label these quarters as ``downturns''. The second group is composed by those unemployment spells that ended in quarters with the \emph{lowest} $ \overline{x}_g\%$ of HP filtered (log) unemployment rates. We label these quarters as ``expansions''. To proceed we need to choose values for $\overline{x}_g$ and $\underline{x}_b$. The choice of these percentile cutoffs, however, presents the following trade-off. On the one hand, a low $\overline{x}_g$ (and/or a high $\underline{x}_b$) implies a relatively small sample of unemployment spells. This could be problematic, as the law of large numbers is not necessarily strong enough in small samples to mitigate randomness in the direction of observed occupational flows.\footnote{Consider for example a set of gross flows obtained from an underlying distribution of flows were net mobility is zero. If this set contains only one observation of occupational mobility, this observation will be categories as a net flow and one would need 100\% of gross flows to cover the net flow.} On the other hand, it could be the case that for structural reasons the response of net mobility is larger when we restrict to more extreme cyclical periods. The latter will imply that a low $\overline{x}_g$ (and/or a high $\underline{x}_b$) by itself might go hand-in-hand with more net mobility, because it concentrates on the most responsive periods.

\begin{figure}
  \centering
\subfloat[Net Mobility (and no. observations)] {\includegraphics [width=0.5 \textwidth]{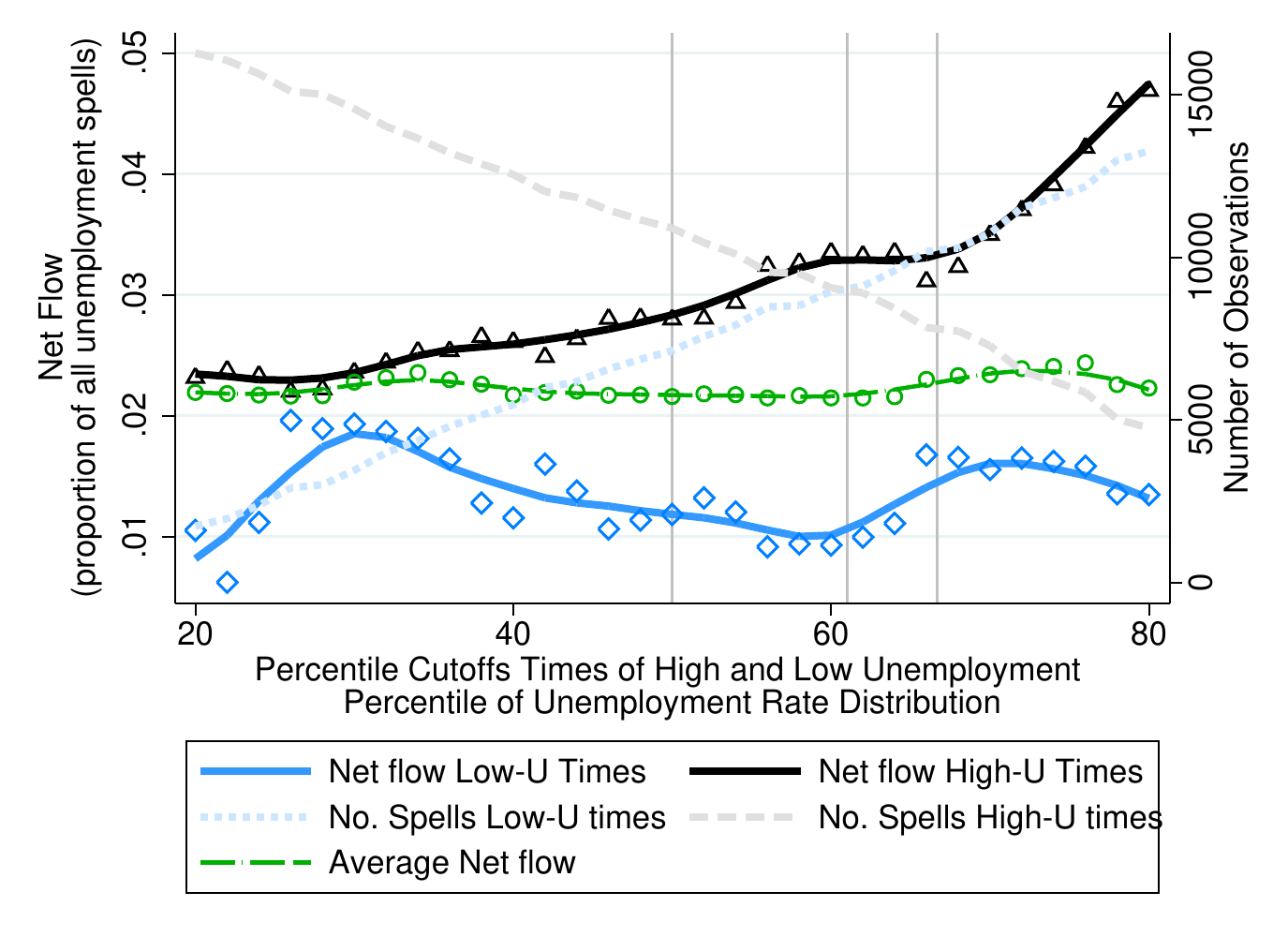}}
\subfloat[High-U times (above 67th pctile), Low-U (below 50th pctile)] {\includegraphics [width=0.5 \textwidth]{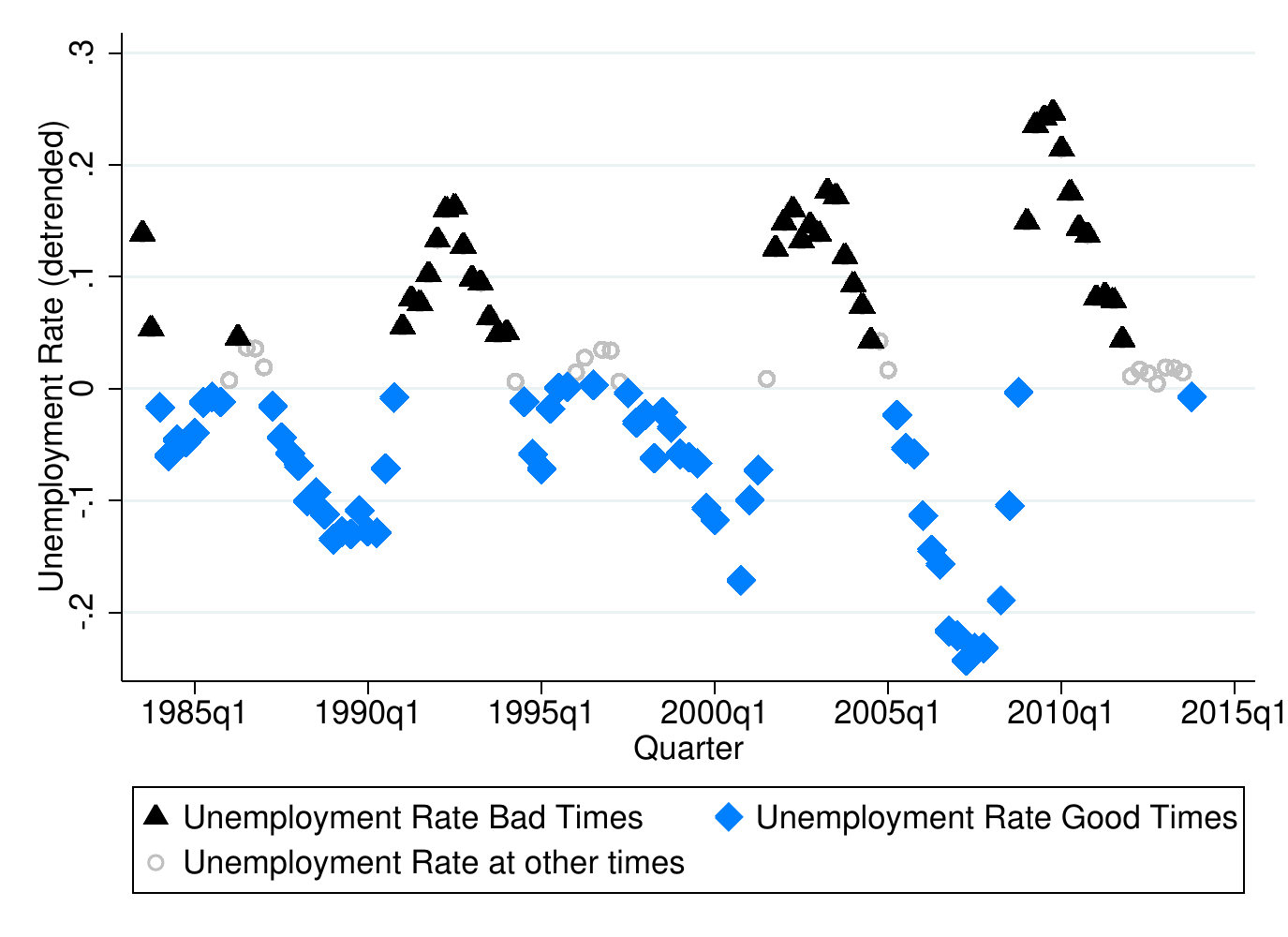} \label{f:times_high_low_u}}
\caption{Net Mobility and Definition of High-U and Low-U times }\label{f:netmob_definition_Utimes}
\end{figure}

\begin{figure}[!ht]
  \centering
\subfloat[Times of High U] {\includegraphics [width=0.5 \textwidth]{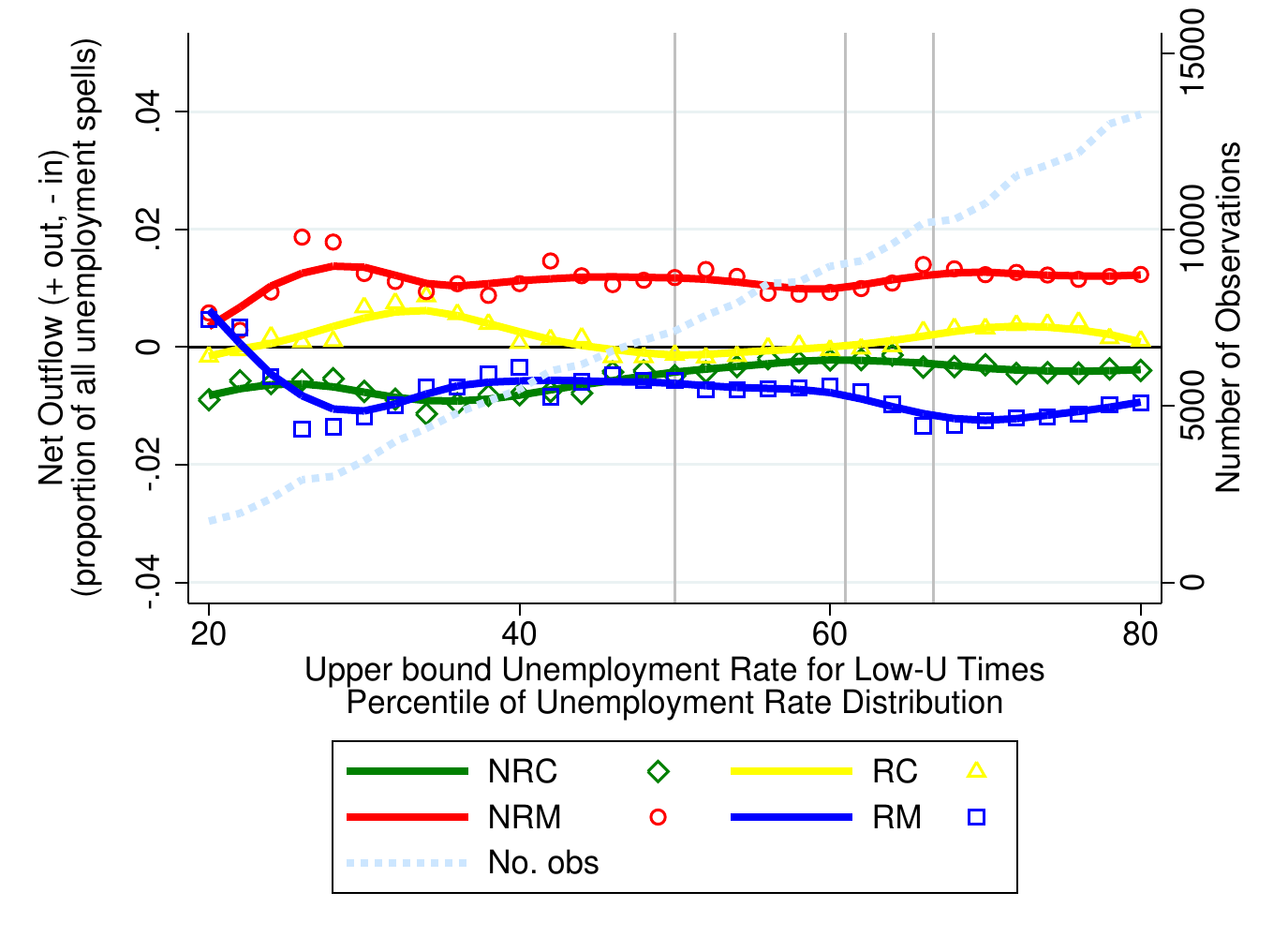}}
\subfloat[Times of Low U] {\includegraphics [width=0.5 \textwidth]{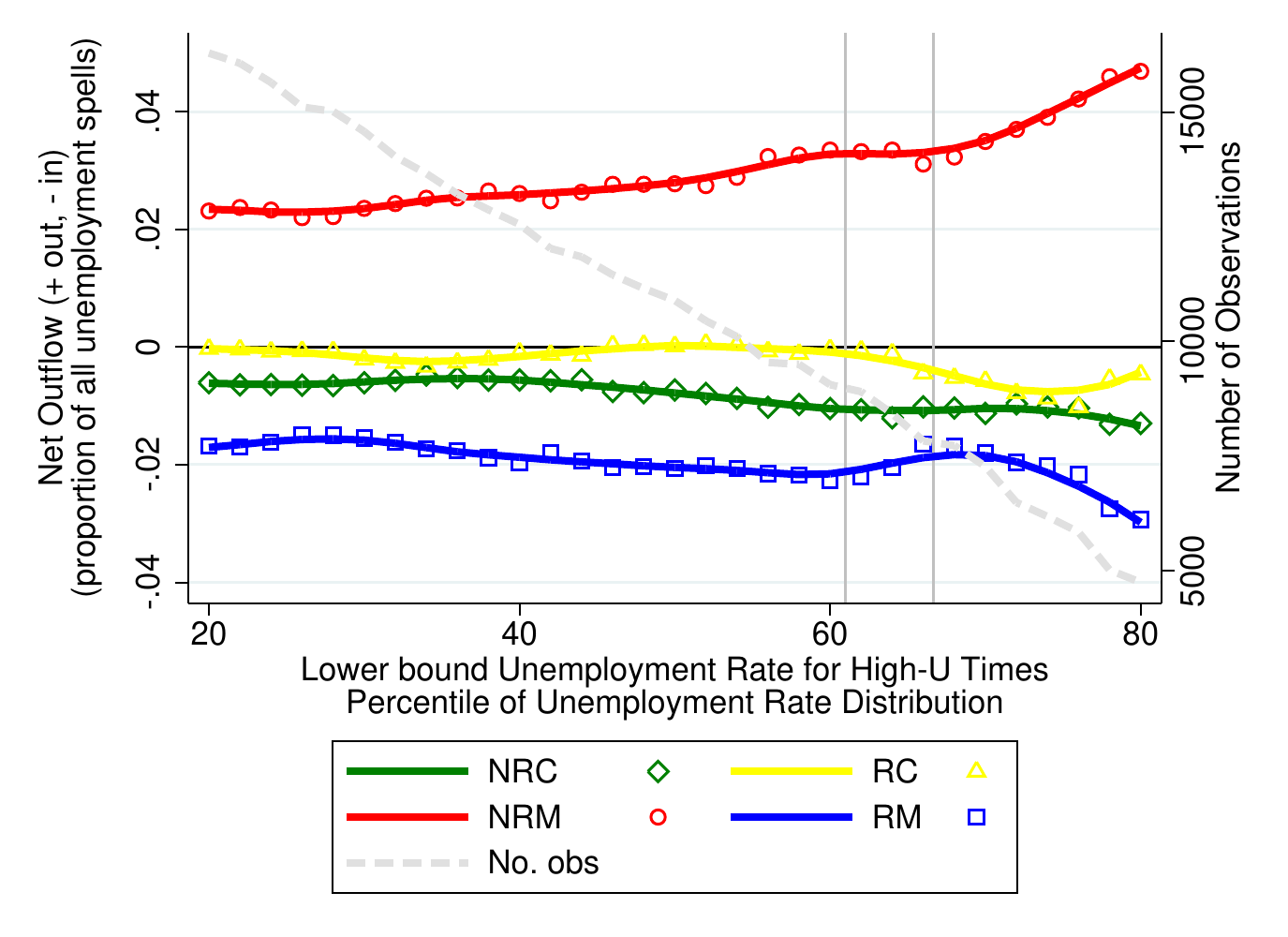}}
  \caption{Net Mobility per (NRMC) Occupation and Definition of High-U and Low-U times }\label{f:netmob_definition_Utimes_indiv_occs}
\end{figure}

To investigate whether this trade-off is important for the cyclicality of net mobility, Figure \ref{f:netmob_definition_Utimes} shows the net flows across the 4 task-based categories as a function of $\underline{x}_b , \overline{x}_g \in[0.2,0.8]$. As described in the main paper these flows are normalised by the number of employment-unemployment-employment spells observed during either expansion and recessions as defined above. The blue curve depicts net flows in expansions and the black curve depicts net flows in downturns. To compare net mobility between expansions and recessions, defined for example as periods with the lowest and highest 33\% of HP filtered (log) unemployment rates, one needs to compare the value of the blue curve at the x-coordinate 0.33 with the value of the black curve at the x-coordinate 0.67. The green dashed line (with circles) denotes the \emph{average} net mobility obtained from calculating net mobility in expansions and in downturns and then averaging over these, weighting them by the number of underlying unemployment spells. In this case, the percentile on the x-axis represents the upper bound of unemployment rates to define expansions, and simultaneously the lower bound to define downturns. In all these cases the net flows are calculated from the implied occupational transition matrix of all unemployment spells that ended in an expansion or a downturn. We then applied the $\Gamma$-correction to this matrix.

\begin{figure}[!ht]
\centering
\subfloat[NRC] {\includegraphics [width=0.5 \textwidth]{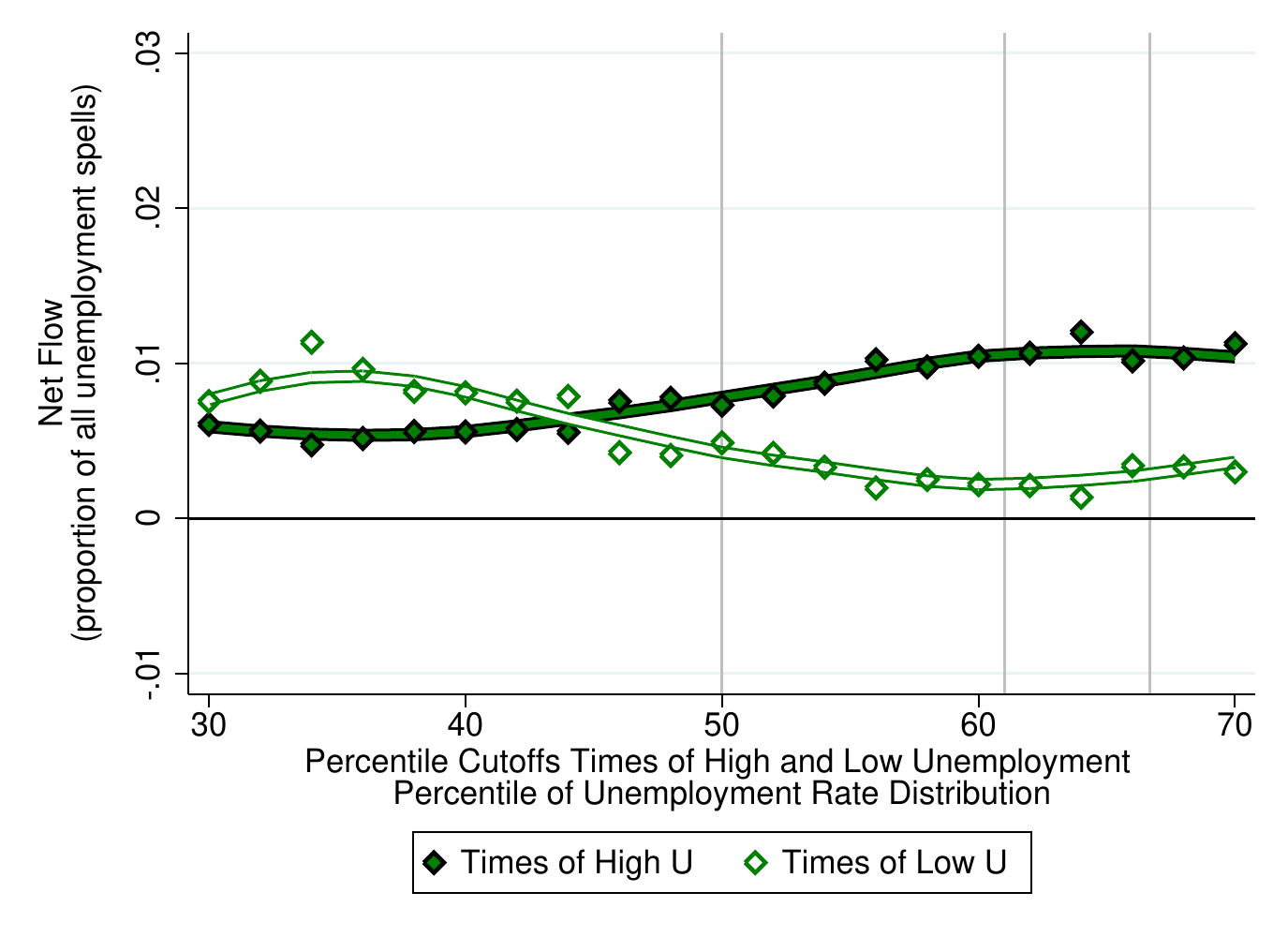}}
\subfloat[RC] {\includegraphics [width=0.5 \textwidth]{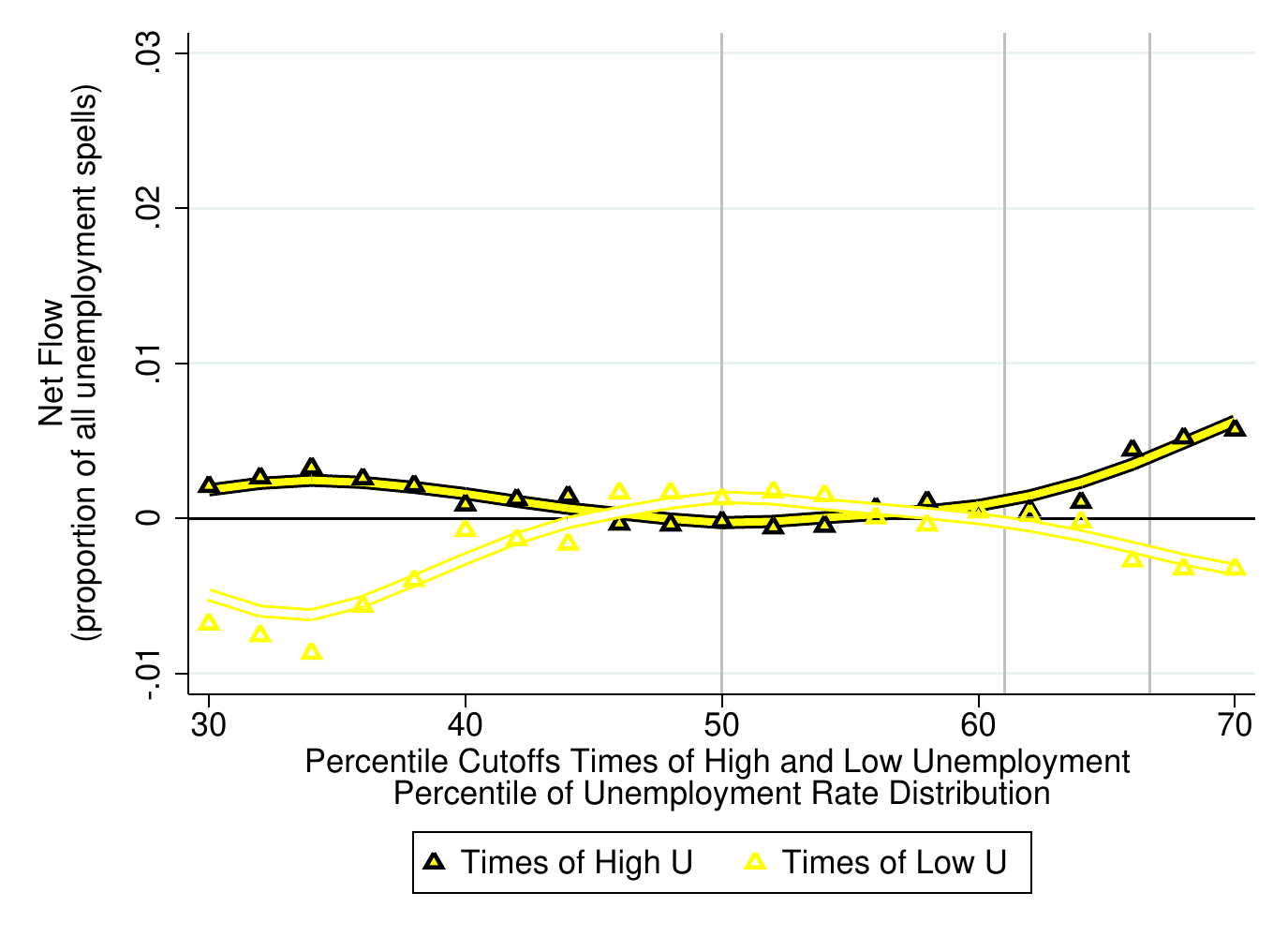}}

\subfloat[NRM] {\includegraphics [width=0.5 \textwidth]{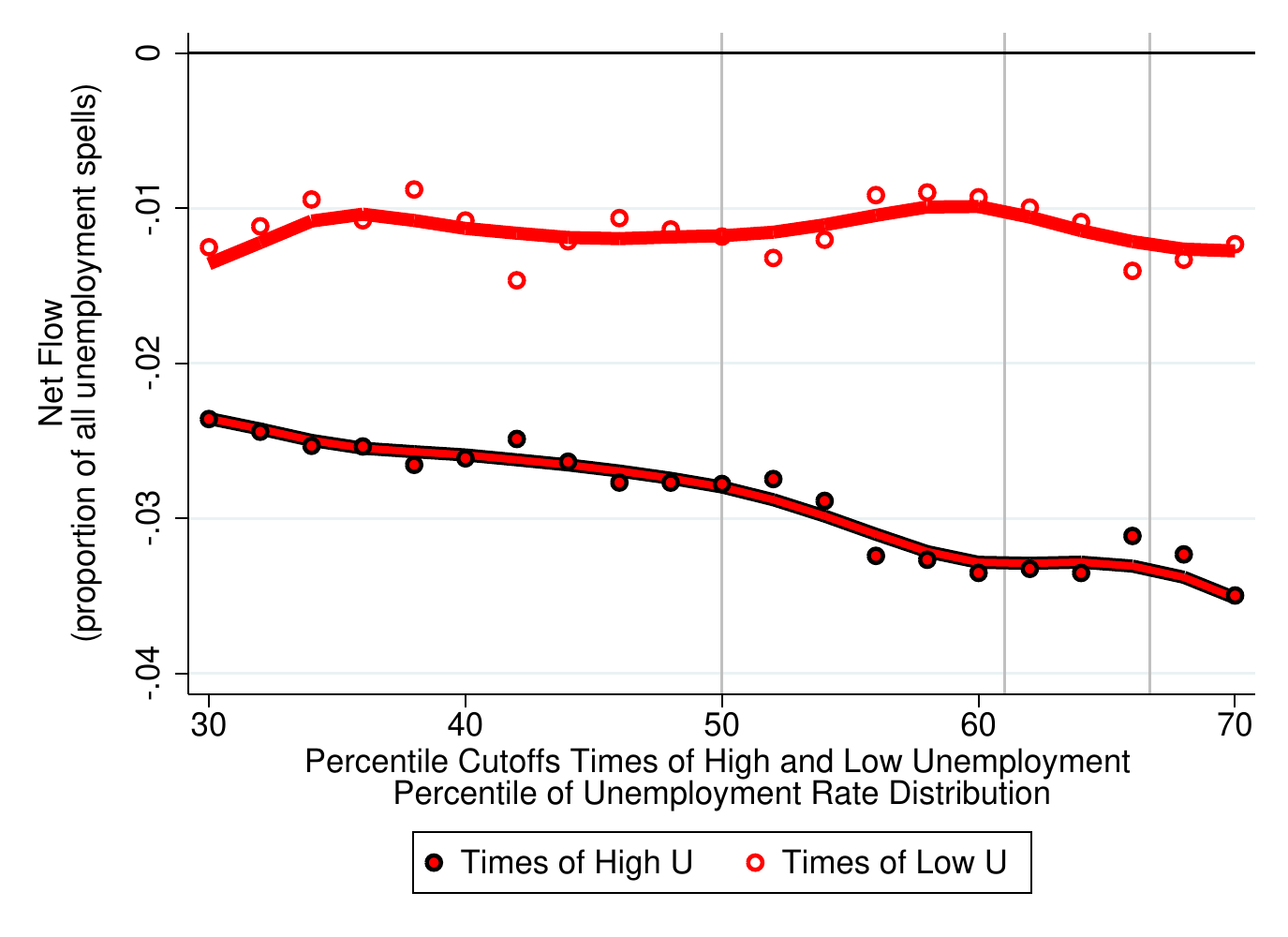}}
\subfloat[RM] {\includegraphics [width=0.5 \textwidth]{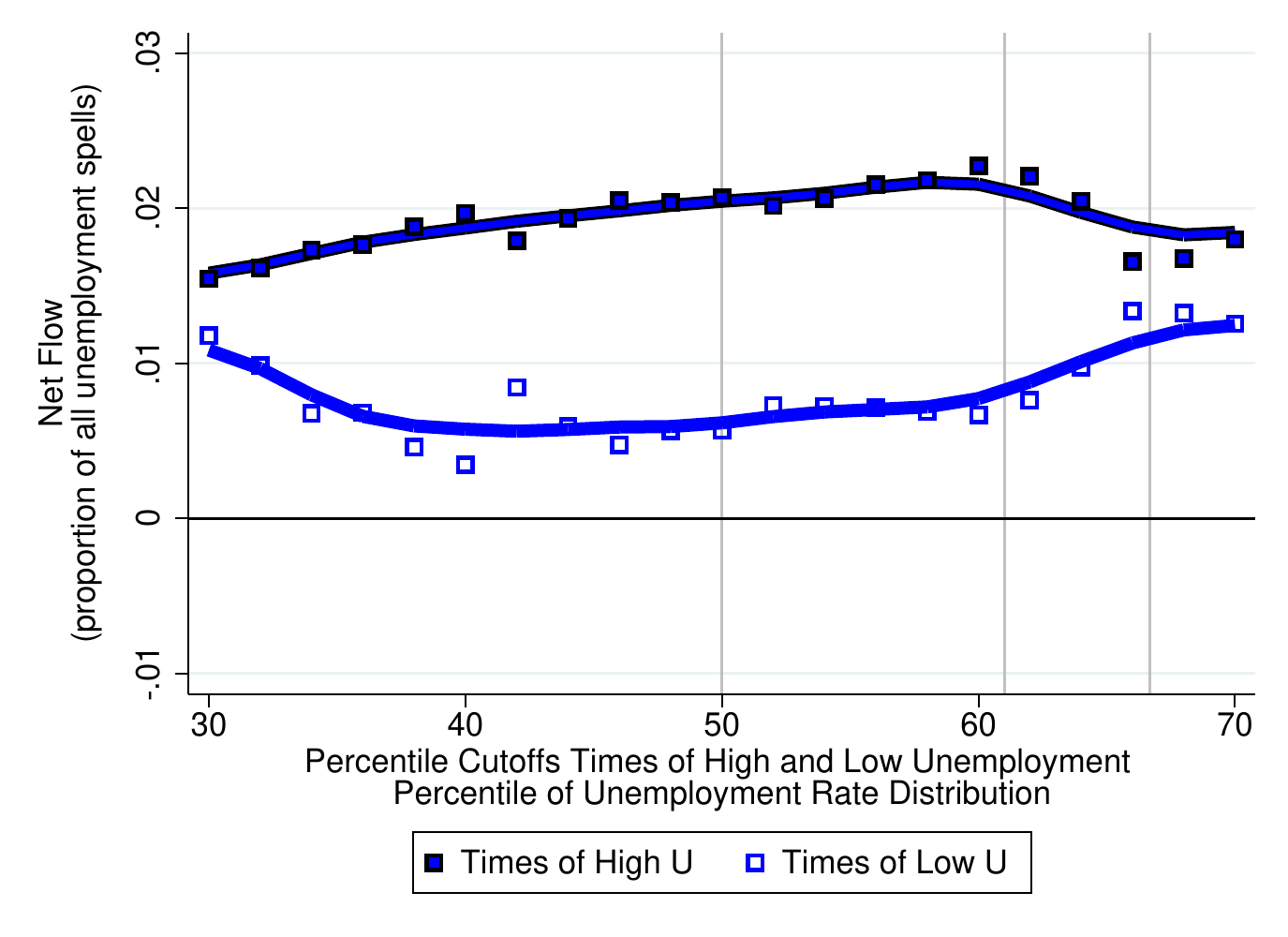}}
\caption{Net Mobility Outflows (Inflows) per Occupations, and definition of High-U and Low-U times}
\label{f:netmob_robust_pctile_change}
\end{figure}

The main message that comes out of Figure \ref{f:netmob_definition_Utimes} is that overall net mobility across task-based categories is \emph{countercyclical} for any $\underline{x}_b , \overline{x}_g \in[0.2,0.8]$. Thus, the variations in the size of our samples do not seem to affect our finding that net mobility is countercyclical. Figure \ref{f:netmob_definition_Utimes_indiv_occs} presents the same exercise but focusing on individual task-based categories. Figure \ref{f:netmob_definition_Utimes_indiv_occs}a shows net mobility in expansions as a function of $\overline{x}_g$; while Figure \ref{f:netmob_definition_Utimes_indiv_occs}b shows net mobility in downturns as a function of $\underline{x}_b$. Here we observe that the net flows of the routine manual and non-routine manual categories are larger in downturns than in expansions, implying that the net flows of these categories are countercyclical for any $\underline{x}_b , \overline{x}_g \in[0.2,0.8]$. Further, both in expansions and downturns the non-routine manual category exhibits \emph{net inflows} while the routine manual category exhibits \emph{net outflows}, consistent with the job polarization literature. The patterns for the cognitive categories, however, are not as clear. For example, we find that the net flows of the non-routine cognitive category are only higher in downturns when considering $\underline{x}_b , \overline{x}_g \in[0.45,0.8]$. Figure \ref{f:netmob_robust_pctile_change} shows all these net mobility flows in more detail, depicting the pairwise expansion and downturn comparison of net flows for each category separately.

In light of the above, when analysing net mobility flows we will take expansions to represent quarters with below median HP filtered (log) unemployment rates and downturns to represent quarters with the 33\% highest HP filtered (log) unemployment rates. The benefit of defining the business cycle in this way is that the aforementioned small sample issue seems to be less important in this definition of an expansion, where unemployment spells are less frequently observed. Indeed, the blue curve in Figure \ref{f:netmob_definition_Utimes}a appears well behaved around the median.\footnote{Also note the there is relatively little variation in the average net mobility between the 40th and 65th percentile. As we restrict sample sizes when moving across these different cutoffs, noisier observations of net flows in the smaller set can drive up the average net mobility rate. This does not seem to be the case for our measures.} Figure \ref{f:times_high_low_u} also shows that the standard NBER recessions and their immediate aftermath closely correspond to this definition of downturns, while the second half of the 1990s, late 1980s, and mid-2000s till the beginning of the Great Recession closely correspond to this definition of expansions.

\begin{figure}[!h]
  \centering
\subfloat[Cyclical Net Mobility (NRMC excl. Management)] {\includegraphics[width=0.5\textwidth]{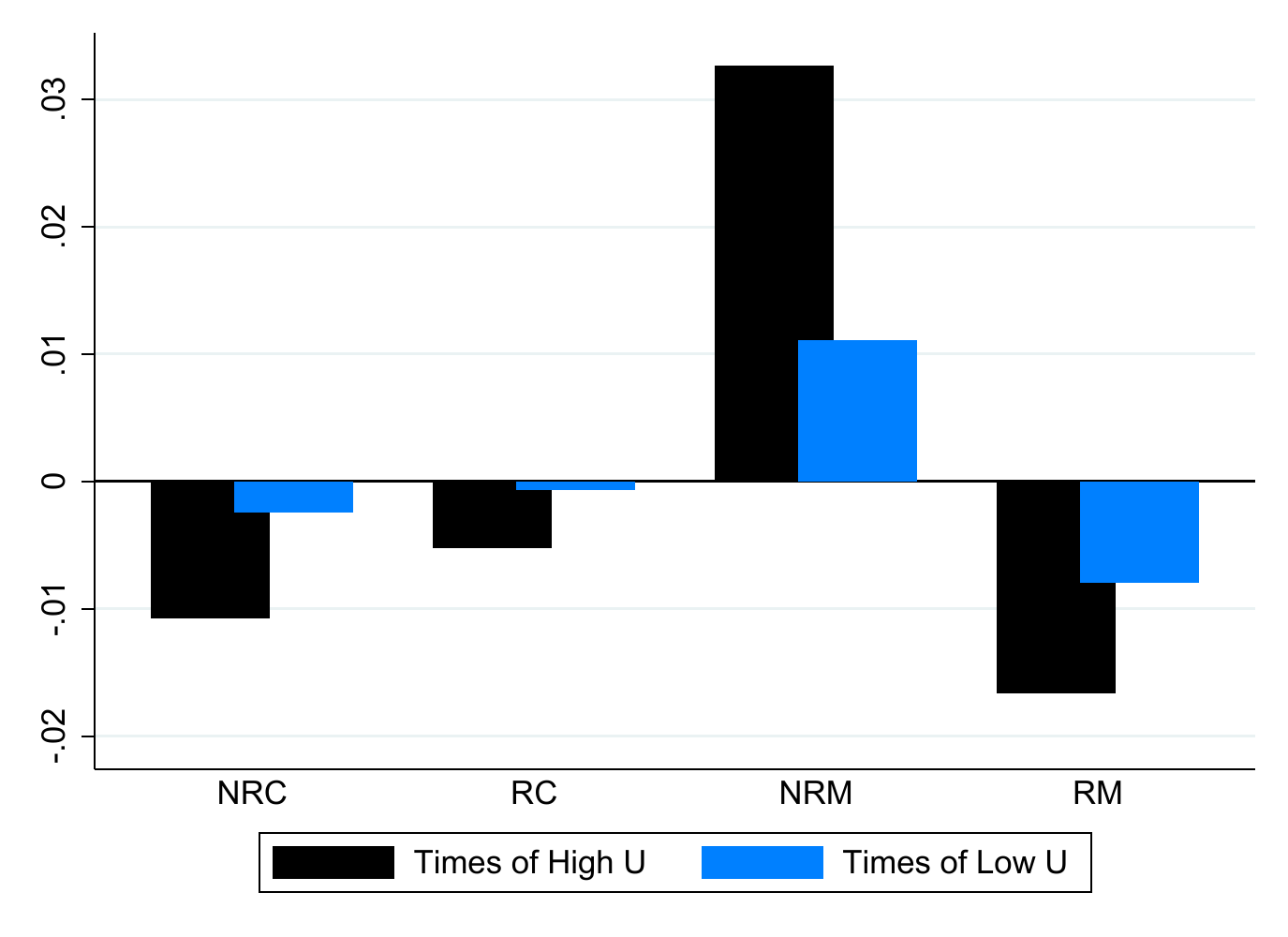}
  \label{f:supp_app_netmob_baseline}}
\subfloat[Cyclical Net Mobility (NRMC incl. Management)] {\includegraphics[width=0.5\textwidth]{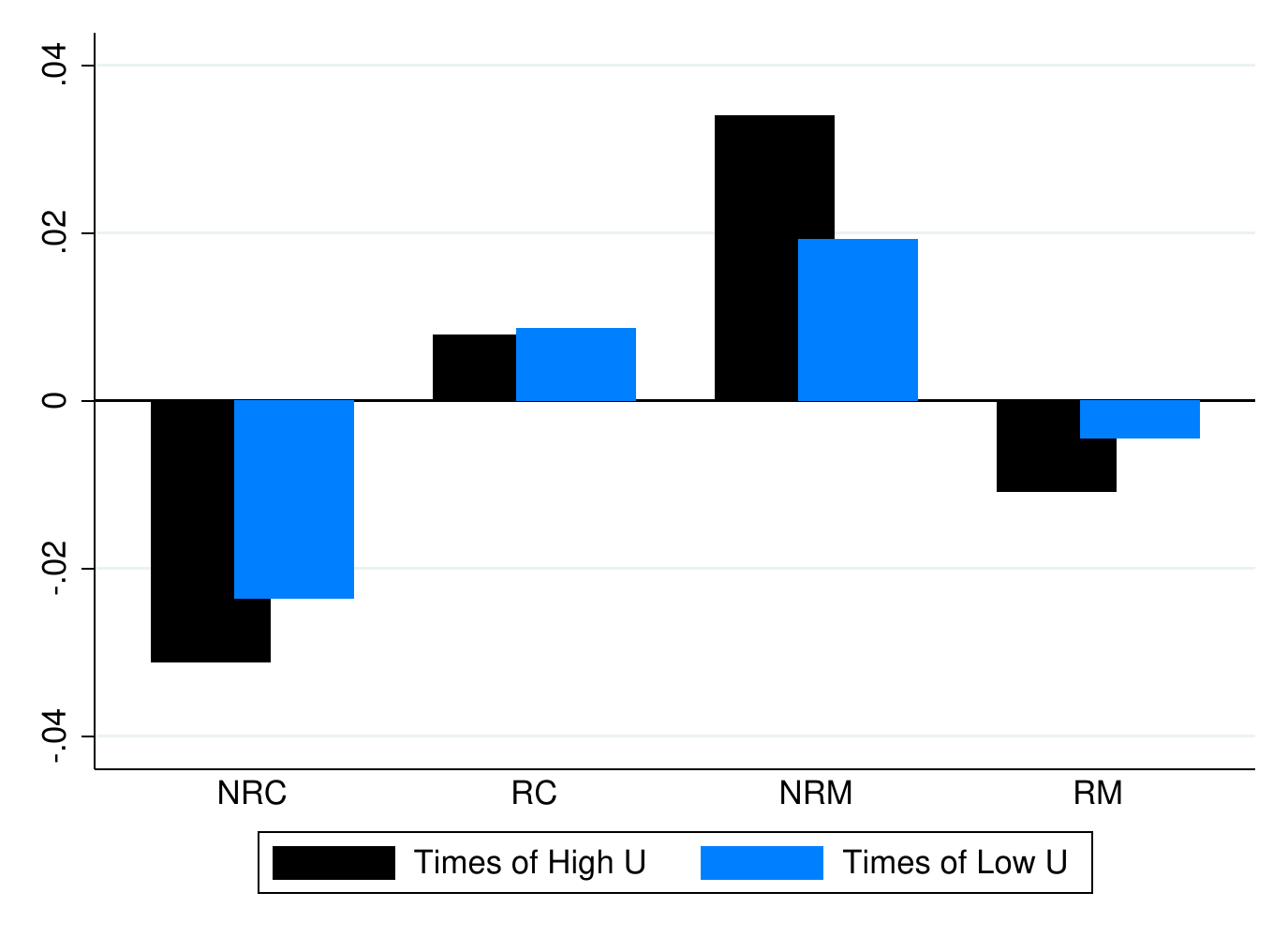}
   \label{f:supp_app_netmob_NRMC_mgt}}\caption{Net Mobility by Task-based categories}
\end{figure}

Analysing the business cycles in such a way implies that overall net mobility accounts for 1.1\% of workers unemployment spells in expansions, while it accounts for 3.2\% of workers unemployment spells in downturns. Figure \ref{f:netmob_definition_Utimes} depicts these values as the intersection between the left-most vertical line with the blue curve and the intersection between the right-most vertical line with the black curve, respectively.\footnote{As an alternative to the above benchmark we also split the sample into two parts with an equal number of observations. In this case expansions correspond to the lowest 61\% of HP filtered (log) unemployment rates, and recessions correspond to the highest 39\% of HP filtered (log) unemployment rates. This alternative generates a similar cyclical change in net mobility, whereby net mobility accounts for 1.0\% of unemployment spells in expansions and accounts for 3.3\% of unemployment spells in recessions.} Figure \ref{f:supp_app_netmob_baseline} displays the net mobility patterns for each of the 4 task-based categories also using this business cycle definition. As suggested by Figures \ref{f:netmob_definition_Utimes_indiv_occs} and \ref{f:netmob_robust_pctile_change}, the non-routine manual category exhibits a countercyclical increase in net inflows: in downturns 3.2\% of workers' unemployment spells cover the net mobility of workers into non-routine manual occupations, while only 1.1\% of spells in expansions. In contrast, the routine manual category exhibits a countercyclical increase in net outflows: in downturns 1.7\% of workers' unemployment spells are needed to cover the net mobility of workers out of routine manual occupations, while only 0.7\% of spells in expansions.

Figure \ref{f:supp_app_netmob_NRMC_mgt} display the net mobility across the four task-based categories, but now including managerial occupations in the non-routine cognitive category. In this case net flows of over 2\% of all unemployment spells now originate from the non-routine cognitive category, while the routine cognitive category now experiences a net inflow as a result of former managers taking up administrative or sales jobs. Further, the inflow from management mutes somewhat the outflow from the routine manual category. Nevertheless, the same cyclical patterns regarding non-routine and routine manual categories emerge. Net inflows into the former and net outflows from the latter are larger in absolute value during downturns.

\begin{figure}[ht!]
  \centering
\subfloat[Baseline episodes] {\includegraphics[width=0.50\textwidth]{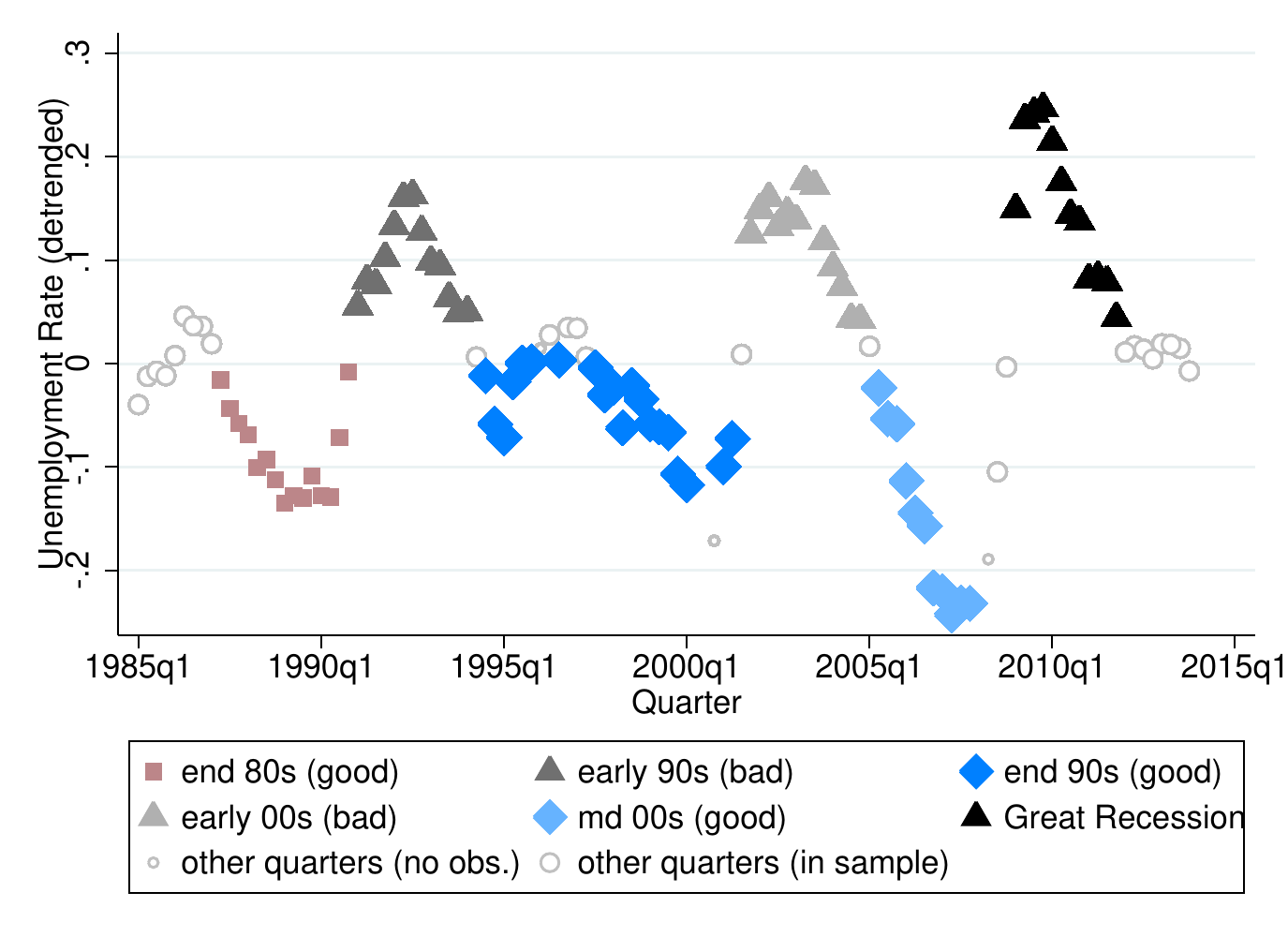}
  \label{f:episodes_50_67}}
 \subfloat[Alternative (NBER) episodes] {\includegraphics[width=0.50\textwidth]{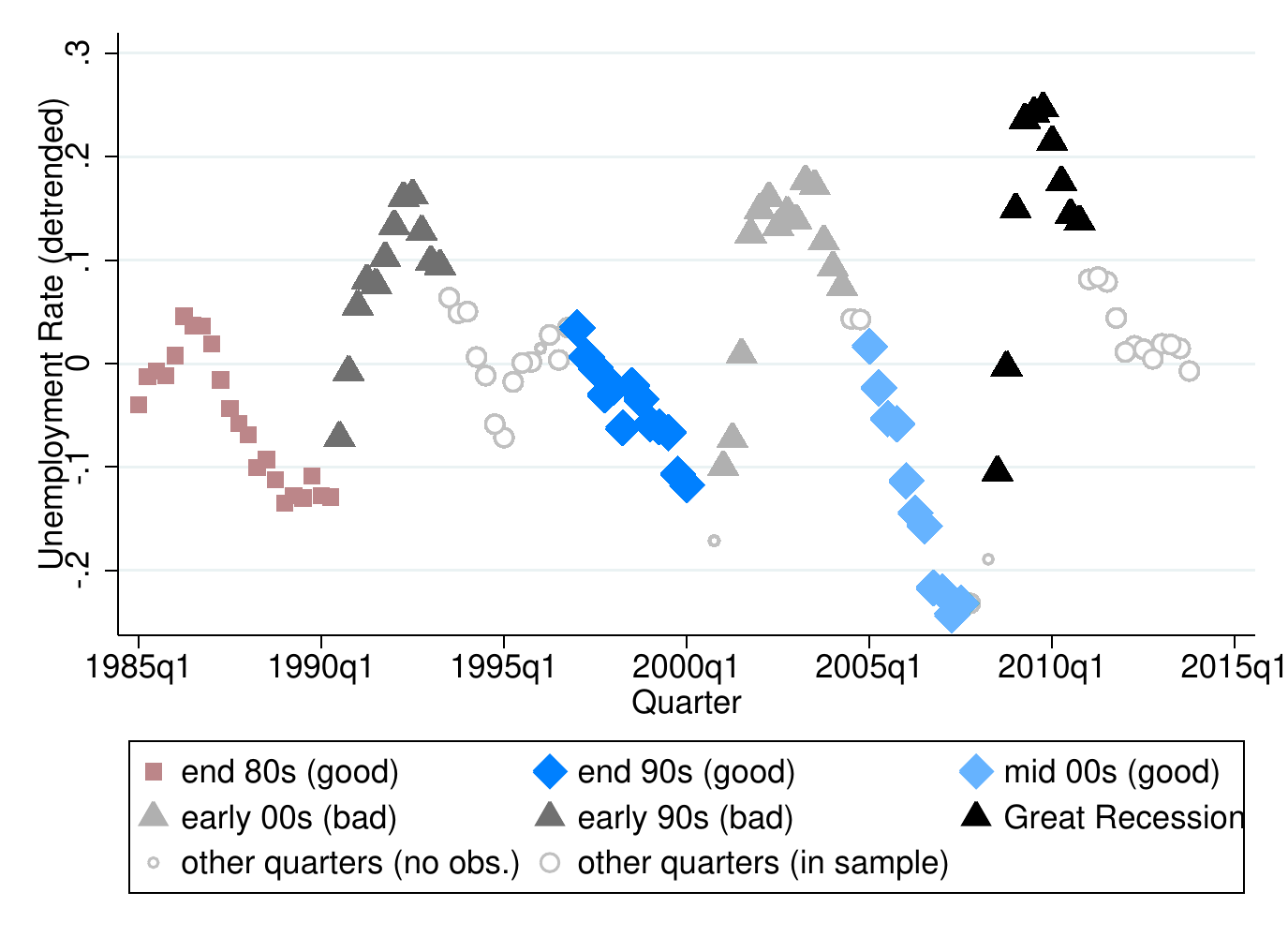}
  \label{f:episodes_1yr}} \caption{Expansion and downturn episodes: 1986-2013}
  \end{figure}

We now investigate whether the above net mobility patterns appear common across the various expansion and downturn episodes we observe during our sample period. Using the above definition of the business cycle, Figure \ref{f:episodes_50_67} distinguishes three expansion and three downturn episodes, where the quarters of our sample are divided into largely connected (continuous) episodes.\footnote{We take those quarters with below-median unemployment rates in the second part of the 90s as one period, and exclude the 2008 quarters in the Great Recession that still have below-median unemployment rates from the expansion.} To check the robustness of our results, Figure \ref{f:episodes_1yr} considers an alternative definition of the business cycle that is closer to the NBER one. In this case we label as downturns the set of quarters that starting with an NBER recession go until one year after peak unemployment is reached. Expansions are the set of quarters preceding these downturns, going backwards until a previous downturn is reached (with a two quarter gap) or until the sample sizes are about balanced. This alternative definition differs from the baseline one by including early NBER recession quarters in which HP-filtered (log) unemployment rates were low but rising fast, and by limiting the 1990s expansion to the 1997-2000 period. Given that the overall number of net mobility flows is small (see Section 2 of this appendix), note that the main caveat of these exercises is that each episode ends up containing an even smaller number of net flows. As discussed earlier this can generate nosier and potentially (upward) biased estimates.

\begin{figure}[!th]
  \centering
\subfloat[Cyclical Net Mobility (Baseline episodes)] {\includegraphics[width=0.7\textwidth]{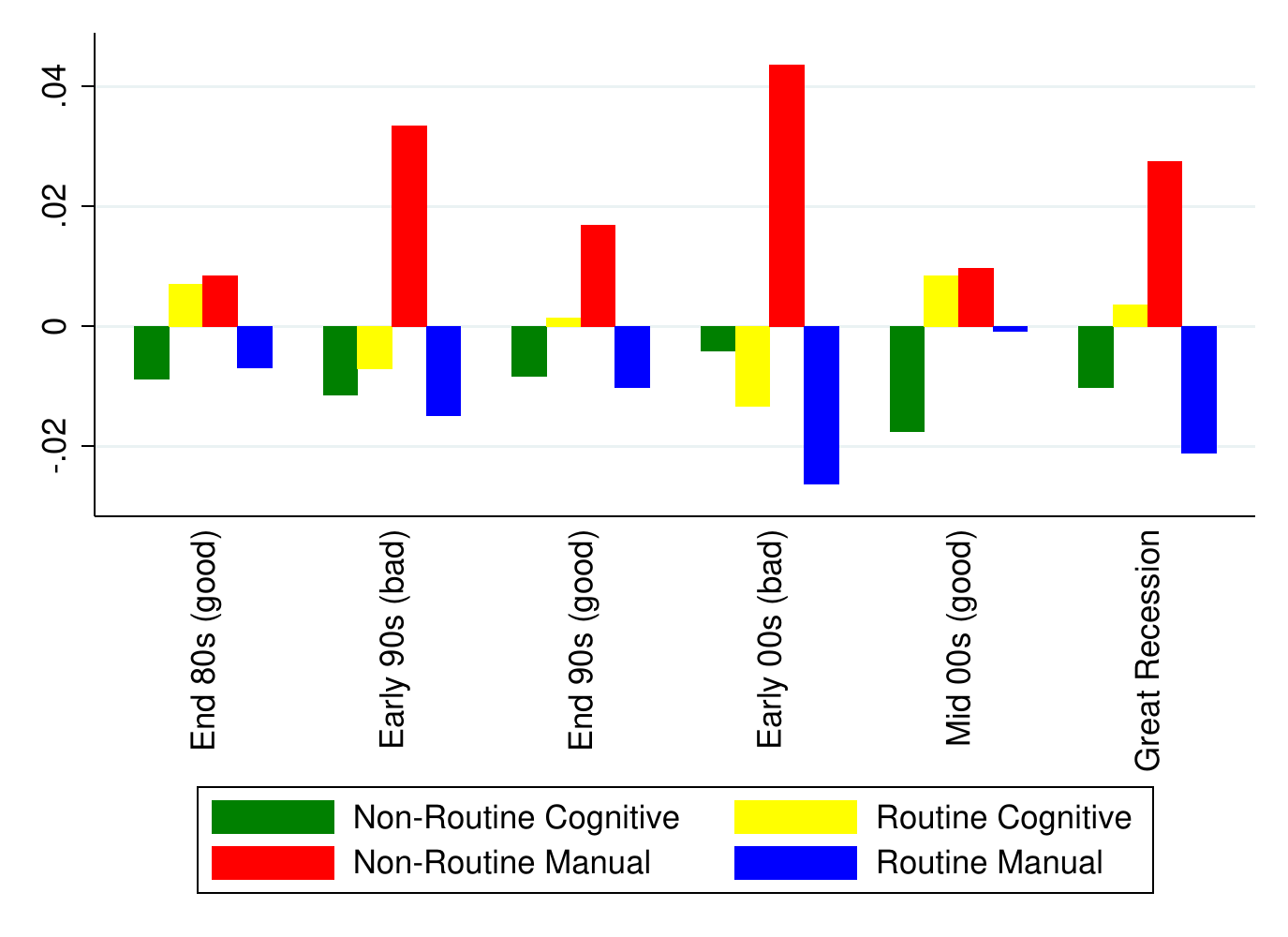}\label{f:supp_app_netmob_eps_baseline}}

\subfloat[Cyclical Net Mobility (Alternative (NBER) episodes)] {\includegraphics[width=0.7\textwidth]{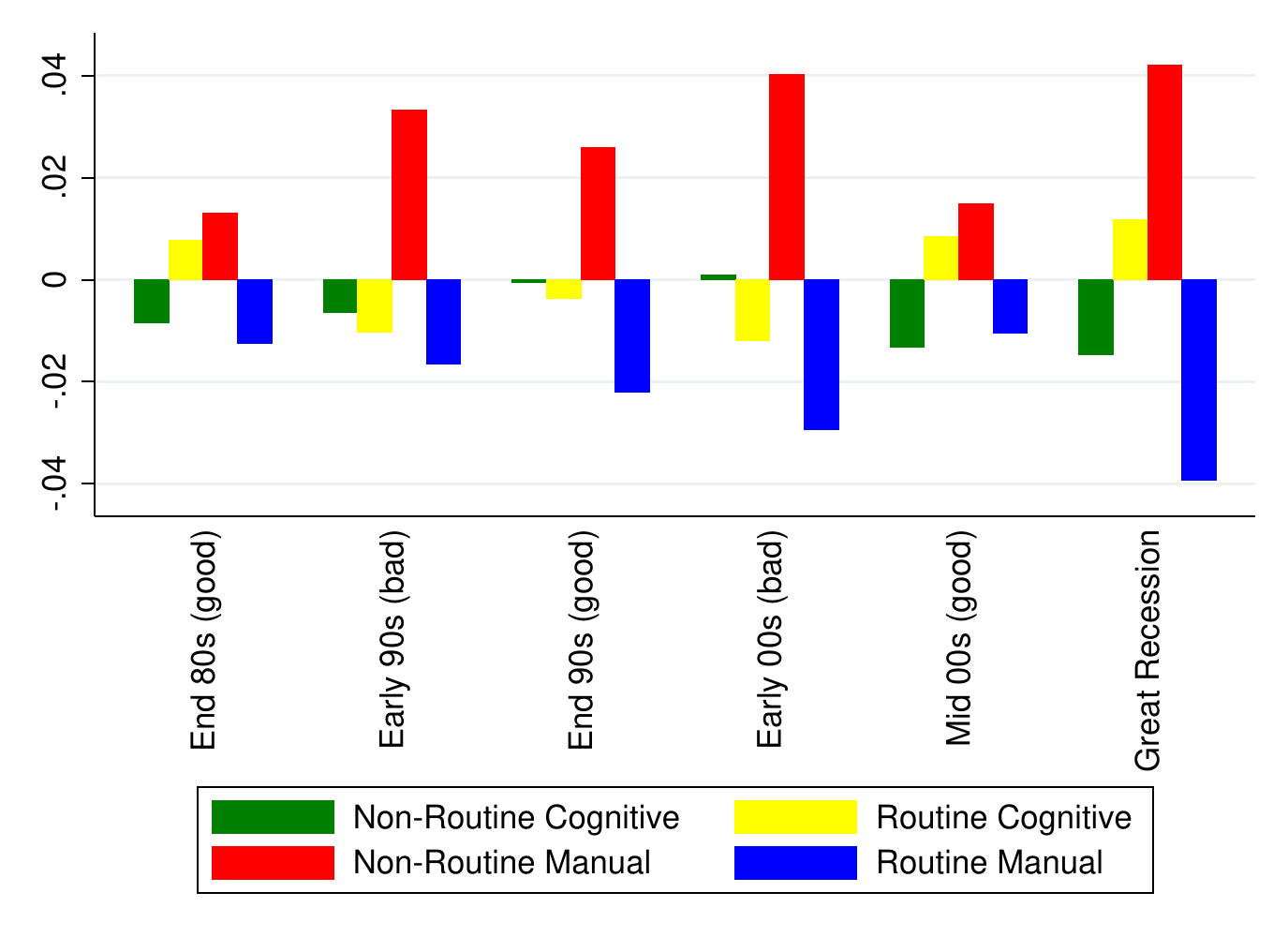}\label{f:supp_app_netmob_eps_nber}}\caption{Net mobility at different expansions and downturns}
\end{figure}

The resulting net mobility patterns are displaying in Figures \ref{f:supp_app_netmob_eps_baseline} and \ref{f:supp_app_netmob_eps_nber}. We observe that the overall patterns described in Figure \ref{f:supp_app_netmob_baseline} seem to be common across all expansions and downturns. In particular, we observe that across all episodes the non-routine manual category exhibit net inflows, while the routine manual category exhibits net outflows. The non-routine cognitive category also exhibits net outflows in most episodes across the two business cycle definitions. The routine cognitive net flows, however, are close to zero but change direction across episodes. This means that the overall low net mobility rate over the entire sample period does not mask meaningful reversals of direction or more substantial net mobility over time and this seems to be consistent across different definitions of the business cycle.

We also observe that each expansion episode is associated with less net mobility than in the downturn episodes. Moreover, across both business cycle definitions the net inflows into the non-routine manual category are larger in downturns than in expansions. A difference, however, is that the net inflows during the Great Recession are more pronounced using the business cycle definition that is close to the NBER one. This appears to reflect that in our data net mobility into the non-routine manual category is noticeably higher during the NBER recession quarters of the Great Recession, and less so in the aftermath. Since Figure \ref{f:supp_app_netmob_eps_nber} includes more of the aforementioned NBER recession quarters, it naturally observe a stronger responds. Note also the cyclical pattern in the net outflows from the routine manual category. These net outflows appear typically larger in downturns than in expansions. This pattern is strongest when using our baseline definition of the business cycle, but when using the definition closer to the NBER one we observe an increase in the net outflows during the late 1990s expansion episode.

\section{Job Finding Hazards and Spell Duration with Occupational Mobility}

In this section we first investigate the re-employment hazard functions of those workers who changed employers through spells of non-employment, differentiating between these workers' degrees of labor market attachments. This connects to our calibration targets. We then analyse the differences in unemployment durations between those workers who changed occupations and those who did not, and how these durations respond over the cycle. This further connects to the model and calibration, including the calibration outcomes. In particular, we document that in cyclical downturns, it is the unemployment duration of movers that lengthens more. We show that this result is robust even when controlling for destination and/or source destinations.

\subsection{Job Finding Hazard}

\begin{figure}[ht!]
\centering
\subfloat[Survival in Nonemployment with Duration]{\label{Hazard1} \includegraphics [width=0.5 \textwidth] {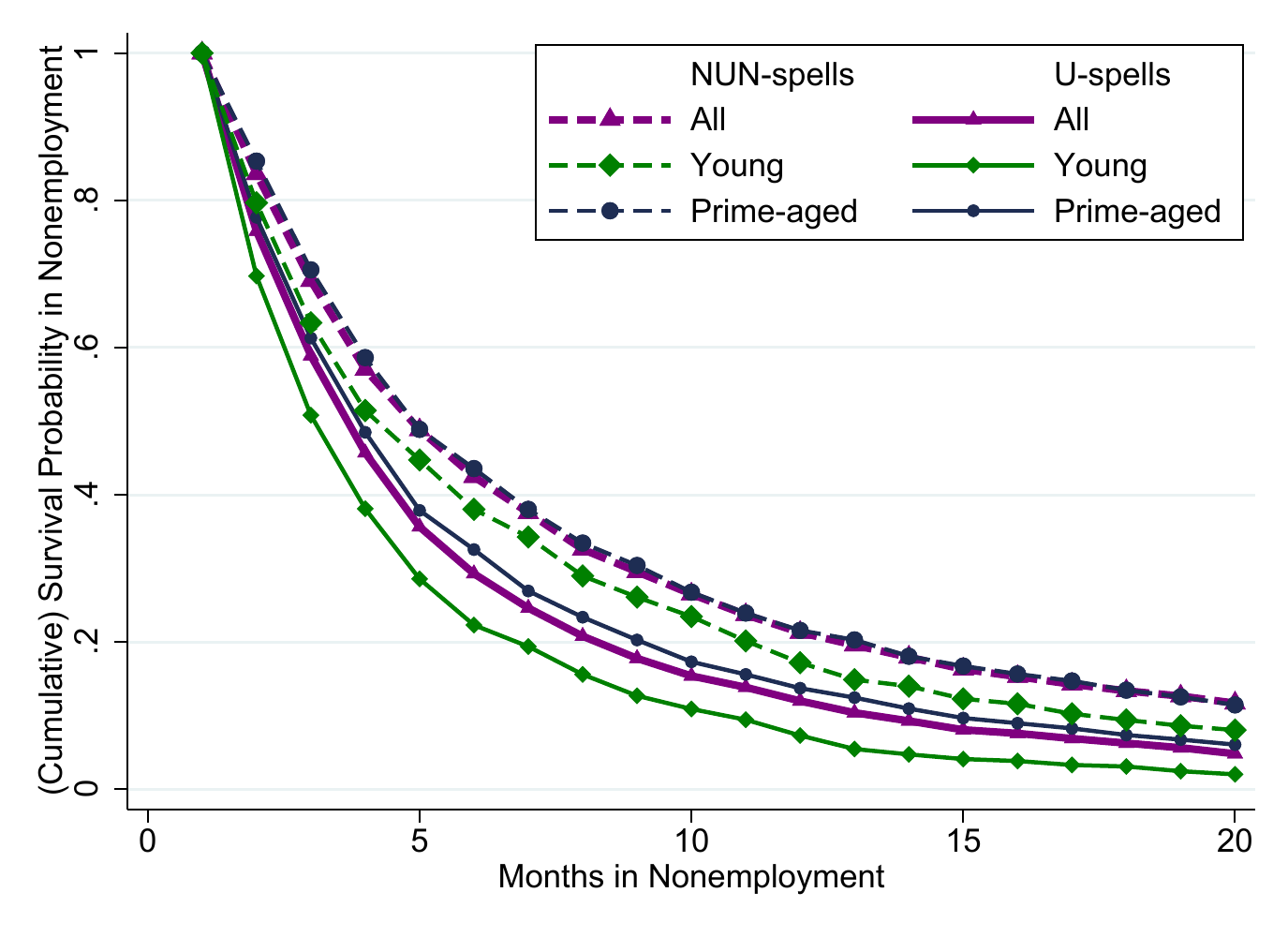}}
\subfloat[Job Finding Rate with Duration]{\label{Hazard2} \includegraphics [width=0.5 \textwidth] {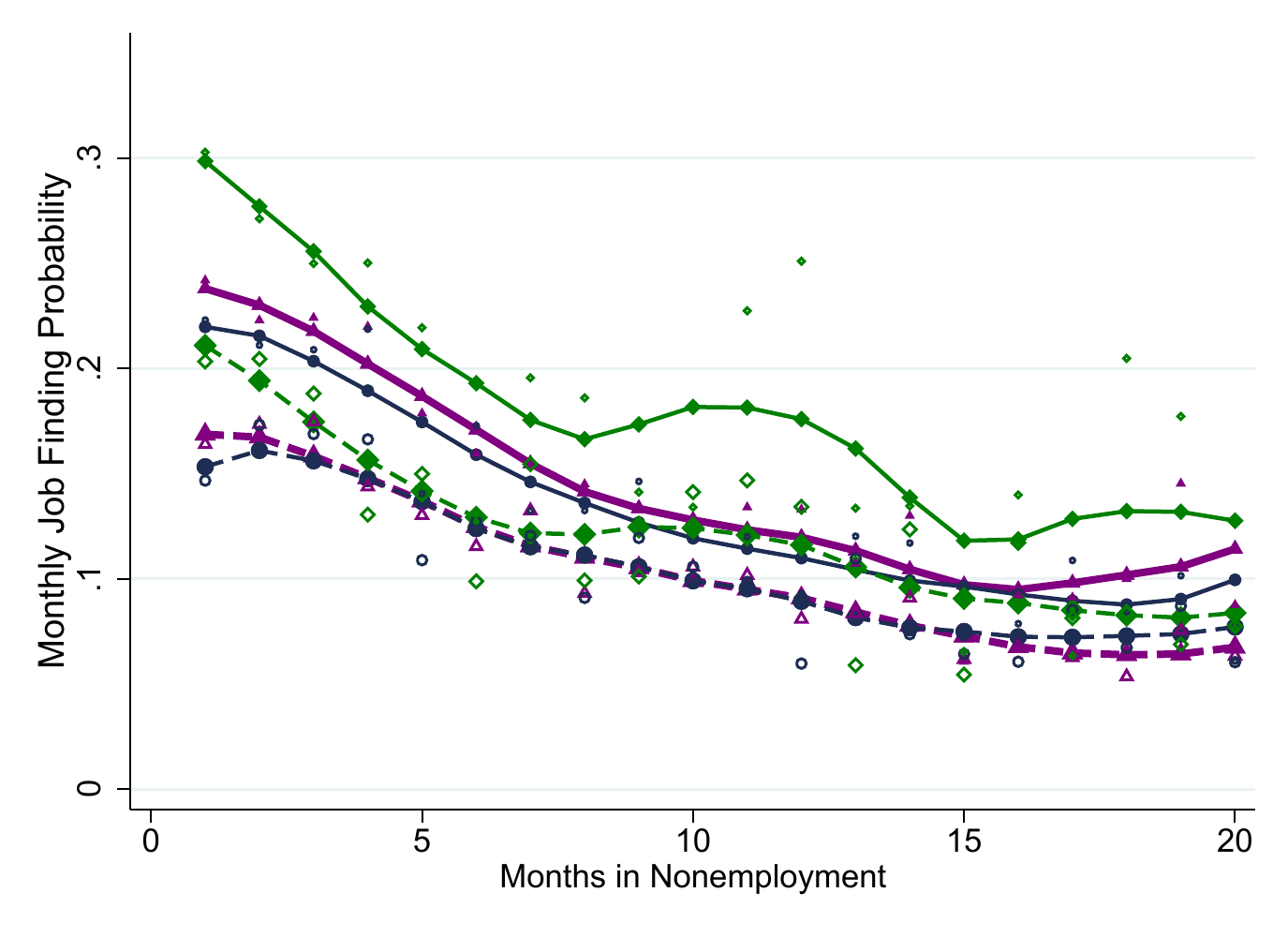}}
\caption{Job finding out of non-employment}
\label{f:hazard1}
\end{figure}

The right-hand panel of Figure \ref{f:hazard1} shows the probability that a worker is still without a job as a function of the number of months that have passed since losing his previous job. The solid lines capture the survival functions for the set of workers who have experienced \emph{un}employed for all months since losing their jobs, while the dashed lines refer to male workers who have been unemployed for at least one month since losing their job, but not necessarily all months, i.e. have a non-employment spell that mixes out of the labour force with unemployment (labelled a ``NUN-spell''). We restrict our attention to males here because females have interestingly different patterns (with higher survival in non-employment at longer duration), which nevertheless might be driven by different motives than the conditions in their labour market alone. Males which have been unemployed for every month since losing their job are not displayed separately, because the associated graph stays very close to the depicted ``U-spell'' category for both males and females.

Within these categories, we can separately investigate the subset of young workers (between 20-30 years old) and prime-aged (between 35-55 years old), drawn in green (with diamond-shaped markers) and dark-blue (with circle markers) respectively. We observe that survival in non-employment is indeed higher for those workers who are not searching for a job in every month, but the general shape and age differences seem preserved across both groups.

The left-hand panel of Figure \ref{f:hazard1} depicts the implied monthly job finding probabilities, with Epanechnikov-kernel weighted local polynomial smoothing (bandwidth 2.5). First note the extent of duration dependence in job finding in the first 10 months, with a job finding rate that falls about 1 percentage point per month for all unemployed workers. This ``moderate'' duration dependence is in part because we focus on those workers who have become more strongly detached from employment. By construction, the workers we consider are without work for at least a whole month (hence, the job finding rate in the first month refers to the probability of finding a job within the next month, after having been unemployed for a month). We also exclude workers in temporary layoff, who have a higher job finding rate.

As a result of these restrictions, our hazard functions do not exhibit the steep drop typically observed during the first month in unemployment (which can be seen e.g. in Farber and Valletta, 2015, Figure 3, based on the CPS). As argued in the main text, our restrictions are motivated by the finding in the literature that entrants into unemployment can be separated roughly into two different groups: a set of workers who has high job finding rates and behaves differently over the cycle, with most of the cyclical movement in the unemployment rate due to those who are in the second, slower job-finding group. We want to focus on the latter group. Fujita and Moscarini (2017) highlight the roll of recalls for the first group and, importantly also for our paper, highlight the different job finding expectations of the first group. This motivates our exclusion from unemployment/non-employment of those who are classified in the SIPP as ``with a firm, on layoff''. Ahn and Hamilton (2019) similarly argue that two different sets of workers enter unemployment, with cyclical movements largely driven by the group which exhibits in comparison less propensity to return to employment.

\begin{figure}[ht!]
  \centering
  \includegraphics[width=3in]{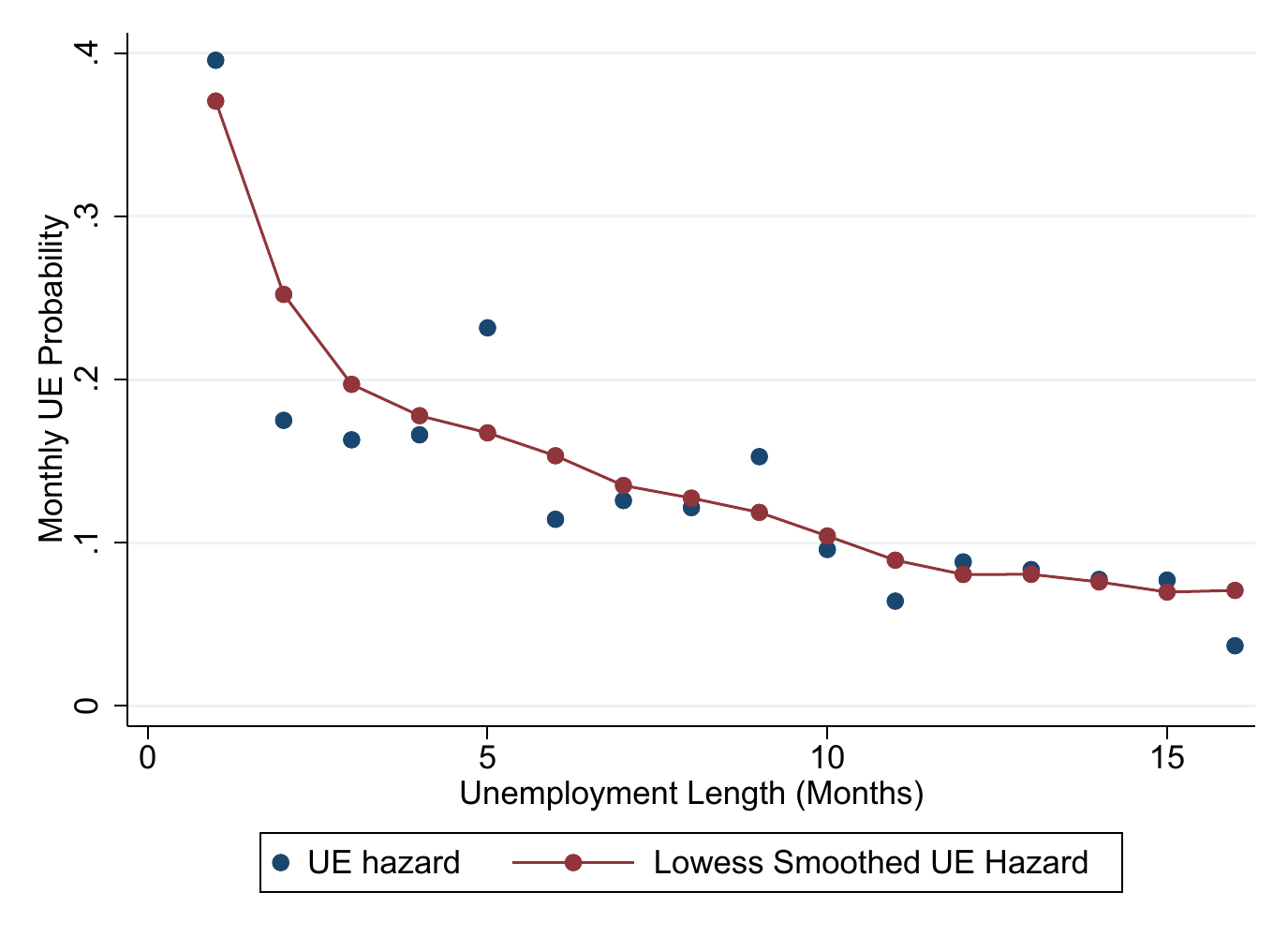}\\
  \caption{Aggregate Job Hazard - unconstrained unemployment definition}\label{f:job_haz_unc}
\end{figure}

Figure \ref{f:job_haz_unc} depicts the hazard function of those workers who reported \emph{conventional} unemployment the month before the interview (to `mimic' the CPS), after dropping the aforementioned restrictions on entering unemployment and non-employment. In this case we indeed observe a much stronger duration dependence, where there is a large drop in the hazard function during the first month (see Fujita and Moscarini, 2017, for a similar result also using the SIPP). Thus, the negative duration dependence among the unemployed (non-employed) we consider in this paper is indeed relatively weaker than when considering the full set of conventionally-unemployed workers in the same data. Aside from this, the \emph{seam effect} on the job finding rate is an additional issue that is clearly visible in Figure \ref{f:job_haz_unc}, at four month intervals. Although with our restrictions the seam effect is weaker in Figure \ref{f:hazard1}, we will attempt to minimize its impact in the calibration by using survival rates at four month intervals (in addition to the first-month job finding rate).

Note that the extent of negative duration dependence exhibited by each group-specific hazard function depicted in Figure \ref{Hazard2} is not too dissimilar from each other. Workers in the `U' group exhibit the higher absolute change in the job finding rate with duration and workers in the `NUN' group exhibit the lower absolute change, but with similar relative changes of the job finding rate with duration. The main difference between these hazard functions seem to be more in levels, where workers in the `U' group exhibit the highest hazard rates and workers in the `NUN' groups exhibit the lowest hazard rates across all durations. The relative reduction in job finding of prime-aged workers in ``pure'' unemployment spells (when compared to younger workers) seems to be similarly present in the ``mixed'' spells of non-employment with some unemployment. Interesting, although not further emphasized in our paper is the non-monotonicity in the job finding rate with duration of the young around 12 months, which is suggestive of a sensitivity to benefit exhaustion that is particularly strong for this group. We abstract from this feature in the paper, but it is worth noting that our calibration targets, which consider survival probabilities in four months intervals, smooth this non-monotonicity out.

\subsection{Job finding differences among occupational movers and stayers}

\subsubsection{Long-run patterns}

The positive slope of the mobility-duration profile implies that occupational movers take longer to find jobs than occupational stayers. We now investigate whether this difference is still present after controlling for demographic characteristics and occupational identities. To do this we report several estimates of the difference between the unemployment \emph{durations} of occupational stayers and movers based on the uncorrected data. These estimates are obtained from regressions of the form
\begin{align}
 \text{Duration of U} = \beta_0 + \beta_{\text{occmob}} \mathbf{1}_{\text{occmob}} + \beta_{\text{dm}} \ \text{demog.ctrls} + \beta_{\text{occ}}\text{occ.dum} + \varepsilon, \tag{R4} \label{eq:udur_reg}
\end{align}
where ``Duration of U spell'' is the individual's \emph{completed} unemployment spell, $\mathbf{1}_{\text{occmob}}$ is a binary indicator that takes the value of one (zero) if a worker changed (did not change) occupation at the end of his/her unemployment (non-employment) spell, the demographic controls include dummies for gender, education, marital status, race and age, ``occ.dum'' denotes occupation identity dummies and $\varepsilon$ is the error term. Table \ref{tab:Uduroccmob} present the results of these regressions by progressively adding demographic and occupation identity controls. Columns (2) and (3) use the sample of completed durations among all unemployed workers, while Columns (4)-(8) restricts the sample to the completed durations of young and prime-aged unemployed workers.

Column (2) shows the results from estimating \eqref{eq:udur_reg} without any controls other than the occupational mobility dummy. It shows that occupational movers take on average an additional 0.5 months to find a job. This difference arises as the average unemployment duration of occupational stayers is 3.6 months, while the average unemployment duration of occupational movers is 4.1 months. This difference is also economically significant: it represents nearly half of the differences between the average unemployment spell duration in times of low unemployment (expansions) and times of high unemployment (recessions).\footnote{We consider times of high (low) unemployment as those quarters in which the HP de-trended (log) unemployment rate lies within the 33\% highest (lowest) HP de-trended unemployment rates. The average unemployment length in those quarters with high unemployment is 4.4 months, whereas the average unemployment spell lasts 3.3 months in the quarters with the low unemployment.}

The rest of the columns show that the estimated difference between the average unemployment durations of occupational movers and stayers does not seem to be driven by composition effects. Column (3) reports the results from estimating \eqref{eq:udur_reg} when adding worker demographic characteristics. It shows that the estimated coefficient of the occupational mobility dummy hardly changes. Column (4) shows that the estimated coefficient of the occupational mobility dummy also hardly changes when restricting the sample to young and prime-aged workers and adding demographic characteristics. In columns (5)-(7) we further add source and destination occupational identity dummies. Here we observe a drop of up to 5 percentage points in the difference between average unemployment durations of occupational movers and stayer. However, note that a difference of 0.45 months still remains economically significant. This suggests that the difference in the average duration of the unemployment spell between occupation movers and stayers is not a result of workers moving out of (or into) occupations in which typically all workers take longer to find jobs. The increased unemployment duration of occupational movers thus seems to be associated with the act of moving itself. Therefore this evidence does not seem to support theories that are based on workers moving into a particular subset of occupations in which newcomers need to spend relatively more time to find jobs because of, for example, re-training.

\begin{table}[!t]
  \centering
  \footnotesize
  \caption{Unemployment Duration and Occupational Mobility}
     \resizebox{1.0\textwidth}{!}{
    \begin{tabular}{lcccccccc}
    \toprule
    \toprule
          & (1)   & (2)   & (3)   & (4)   & (5)   & (6)   & (7)   & (8) \\
          & all U & all U & all U & U yng+prm  & U yng+prm  & U yng+prm  & U yng+prm  & U yng+prm  \\

    \midrule
    {\bf{Average Unemp. Duration}} &   &       &       &       &       &       &       &  \\
     All & 3.91  &       &       &       &       &       &       &  \\
          & (.033) &       &       &       &       &       &       &  \\
    Occ. Stayers &       & 3.646 & 3.646 & 3.627 & 3.627 & 3.627 & 3.627 & 3.627 \\
          &       & (.048) & (.047) & (.052) & (.051) & (.052) & (.051) & (.056) \\
    Occ. Movers &       & 4.146 & 4.146 & 4.063 & 4.063 & 4.063 & 4.063 & 4.063 \\
          &       & (.045) & (.045) & (.048) & (.048) & (.048) & (.048) & (.055) \\

      \midrule
      {\bf{Regression Coefficients}} &   &       &       &       &       &       &       &  \\

     Coeff. Occ. Mob Dummy &       & 0.500 & .507  & .499  & .462  & .472  & .451  & .262 \\
          &       & (.065) & (.065) & (.071) & (.072) & (.072) & (.073) & (.114) \\
    Occ. Mob x Prime-Age Dum. &       &       &       &       &       &       &       & .336 \\
          &       &       &       &       &       &       &       & (.157) \\
    \midrule
    Worker's Characteristics &       &       & X     & X     & X     & X     & X     & X \\
    Age (prime-age dummy) &       &       &       & X     & X     & X     & X     & X \\
    Source Occupation Dummies &       &       &       &       & X     &       & X     & X \\
    Dest. Occupation Dummies &       &       &       &       &       & X     & X     & X \\
    \midrule
    \multicolumn{9}{c}{F-test Interactions (p-value) } \\
    \midrule
    Age x Occ Mob  &       &       &       & 0.008 & 0.012 &       &       & 0.024 \\
    Worker Char. x Occ Mob. &       &       & 0.029 & 0.118 & 0.194 & 0.616 & 0.630 & 0.607 \\
    Occupation x Occ Mob.  &       &       &       &       & 0.806 & 0.463 & 0.931 & 0.927 \\
    \midrule
    Num. of Observations & 10886 & 10886 & 10886 & 8887  & 8887  & 8887  & 8887  & 8887 \\
    \bottomrule
    \bottomrule
    \end{tabular}}
  \label{tab:Uduroccmob}%
\end{table}%

Table \ref{tab:Uduroccmob}, however, does show important differences across age groups. Column (8) reports that prime-aged workers take 0.33 months longer to find a job when they changed occupations than young occupational movers. In this case the role of other demographic characteristics factors appears more limited, once we control for age and occupational identities. The bottom panel of Table \ref{tab:Uduroccmob} present several F-tests evaluating the equality of the occupational mobility dummies specific to demographic characteristics and occupational identities. These F-test highlight two important results. There is a statistically significant interaction between age and the additional unemployment duration of occupational movers. There is not a statistically significant interaction between the rest of the demographic or occupational identity dummies and the additional unemployment duration of occupational movers.

\subsubsection{Business cycle patterns}

Table \ref{tab:dur_movers_bc} extends the previous analysis and considers the cyclicality of the difference between the completed unemployment durations of occupational movers and stayers. In Section 3.2 we documented that the mobility-duration profile is procyclical: at any duration the occupational mobility rate is higher in expansions than in recessions. We now show that ties in with a countercyclical distance between the unemployment durations of movers versus stayers. In particular, we estimate
\begin{align}
 \text{Duration of U} = \beta_0 + \beta_{\text{occmob}} \mathbf{1}_{\text{occmob}} + \beta_{\text{occ.un}} \mathbf{1}_{\text{occmob}} \times urate+ \beta_{\text{dm}} \ \text{demog.ctrls} + \beta_{\text{occ}}\text{occ.dum} + \varepsilon, \tag{R5} \label{eq:udur_reg_bc}
\end{align}
where the new explanatory variable is relative to equation \eqref{eq:udur_reg} is the interaction between occupational mobility and the unemployment rate. The latter appears as ``diff resp. (responsiveness between) movers vs stayers'' in Table \ref{tab:dur_movers_bc}.

Panel A of Table \ref{tab:dur_movers_bc} considers as the measure of the business cycle the log of the aggregate unemployment rate, controlling for a linear time trend. The results presented in the first set of four columns use as the dependent variable spells in which the worker was classified unemployed since loosing his/her job until finding a new one (U-spells). The second set of four columns use as the dependent variable spells in which the worker was classified at least one month as unemployed and the rest as out of the labour force since loosing his/her job until finding a new one (NUN-spells). In each case, the first three columns consider all workers, while the last column restricts the sample to young and prime-aged workers.

\begin{table}[!t]
  \centering
  \caption{Unemployment Duration and Occupational Mobility over the Business Cycle}
  \resizebox{1.0\textwidth}{!}{\addtolength{\tabcolsep}{-3pt}
        \begin{tabular}{lcccccccccc}
\toprule
\toprule
    \multicolumn{11}{c}{Panel A: Semi-Elasticity Un-/Nonemployment Duration with Log linearly detrended Unemployment rate} \\ \midrule
          &       & \multicolumn{4}{c}{Unemployment Duration} &       & \multicolumn{4}{c}{NUN-spell duration} \\ \cmidrule(lr){3-6} \cmidrule(lr){8-11}
    Coefficient      &       &  (1)   & (2)   & (3)   & (4)   &       & (1)   & (2)   & (3)   & (4) \\ \midrule
    Occupation Stayers x U. rate &       & 1.65***  & 1.64***  & 1.63***  & 1.64***  &       & 1.07***  & 1.15***  & 1.13***  & 1.14*** \\
    (s.e) &       & (.18) & (.18) & (.18) & (.18) &       & (.25) & (.25) & (.25) & (.25) \\
    Occupation Movers x U. rate &       & 2.04***  & 2.03***  & 2.04***  & 2.03***  &       & 1.72***  & 1.81***  & 1.81***  & 1.81*** \\
    (s.e) &       & (.17) & (.17) & (.17) & (.17) &       & (.23) & (.23) & (.23) & (.23) \\
    difference resp. mover-stayer &       & 0.40**  & 0.40**  & 0.41**  & 0.40**  &       & 0.65**  & 0.66**  & 0.68**  & 0.67** \\
    (s.e) &       & (.18) & (.18) & (.18) & (.18) &       & (.31) & (.29) & (.30) & (.29) \\ \hline
    Worker's Characteristics &       &       & X     & X     & X     &       &       & X     & X     & X \\
    Source Occupation &       &       & X     &       & X     &       &       & X     &       & X \\
    Destination Occupation &       &       &       & X     & X     &       &       &       & X     & X \\ \toprule
    \multicolumn{11}{c}{Panel B: Semi-Elasticity Un-/Nonemployment Duration with HP-Filtered Log Unemployment rate} \\ \midrule
    &       & \multicolumn{4}{c}{Unemployment Duration} &       & \multicolumn{4}{c}{NUN-spell duration} \\ \cmidrule(lr){3-6} \cmidrule(lr){8-11}
    Coefficient      &       & (1)   & (2)   & (3)   & (4)   &       & (1)   & (2)   & (3)   & (4) \\ \midrule
    Occupation Stayers x U. rate &       &           2.47***  &           2.47***  &           2.49***  &           2.48***  &       & 1.345*** & 1.591*** & 1.608*** & 1.616*** \\
    (s.e) &       &  (.46)  &  (.43)  &  (.43)  &  (.43)  &       & (.62) & (.60) & (.60) & (.60) \\
    Occupation Movers x U. rate &       &           3.20***  &           3.14***  &           3.21***  &           3.16***  &       & 3.516*** & 3.67***  & 3.705*** & 3.691*** \\
    (s.e) &       &  (0.56)  &  (.54)  &  (.54)  &  (.54)  &       & (.60) & (.60) & (.60) & (.59) \\
    difference resp. mover-stayer &       &           0.73*  &           0.67*  &           0.72*  &           0.68*  &       & 2.17***  & 2.08***  & 2.098*** & 2.074*** \\
    (s.e) &       &  (.42)  &  (.40)  &  (.39)  &  (.40)  &       & (.76) & (.71) & (.72) & (.72) \\ \hline
    Worker's Characteristics &       &       & X     & X     & X     &       &       & X     & X     & X \\
    Source Occupation &       &       & X     &       & X     &       &       & X     &       & X \\
    Destination Occupation &       &       &       & X     & X     &       &       &       & X     & X \\
    \midrule
    Number of observations  &       & 9840  & 9840  & 9840  & 9840  &       & 15506 & 15506 & 15506 & 15506 \\
    \bottomrule
    \bottomrule
    \end{tabular}%
}
\emph{\small{}{}Notes: }{\small{}{}Standard errors in parenthesis. $^{***}$ significant at a 1\% level; $^{**}$ significant at a 5\% level; $^{*}$ significant at a 10\% level. Controls: gender, age, age squared, number of years of education, family status.}
  \label{tab:dur_movers_bc}%
\end{table}%

For both types of non-employment spells, we observe that when the aggregate unemployment rate increases the differences between the completed spell duration of occupational movers and stayers increases. That is, during recession (times when unemployment is high) occupational movers experience even stronger increases in unemployment duration than occupational stayers do. As shown across the columns, this result also holds (with rather stable coefficients for U-spells) when controlling for demographic characteristics (including a quadratic in age) and dummies for the occupation of origin and destination.\footnote{Kroft et al. (2016) find that compositional shifts matter little for the increase in long-term unemployment in recessions; our results are related in the sense that we find that compositional shifts matter little for the increase in the duration difference between occupational movers and stayers in recessions. This increase is proportionally stronger than the drop in occupational mobility of the unemployed that is broadly shared across occupations. Hence the lengthening of spells of occupational movers contributes to the lengthening of unemployment and, in particular, long-term unemployment across occupations in recessions.} Our results also show that the responsiveness of the difference in unemployment durations to the unemployment rate is stronger when considering NUN-spells.

Panel B of Table \ref{tab:dur_movers_bc} instead considers as the measure of the business cycle the cyclical component of the log of aggregate unemployment rate, where the cyclical component is obtained through an HP filter. It is immediate from the table that the conclusions obtained from using this measure are the same as the ones obtained when using the linearly de-trended logged unemployment rate.

\subsubsection{Summary} In Section 1 we have documented that at any unemployment duration, occupational movers largely move across all occupations. Here we have shown that conditional on a given occupation, an occupational mover takes longer to find a job than an occupational stayer and that this difference increases in recessions. It is important to highlight that this does not mean that different occupations exhibit the same average unemployment duration. Indeed, we do find differences in the estimated coefficients of the source and destination occupation dummies (see Section 1.5), suggesting that some occupations do lead \emph{all workers} (occupational movers and stayers) to find jobs fasters than workers in other occupations (in line with Wiczer, 2015). What our results suggest is that even though a worker might have experienced job loss in an occupation that exhibits short or long overall unemployment durations, this worker will take on average longer to find a job if he/she is to change occupations at re-employment and even longer if he/she changes in a recession.

\section{Occupational mobility in the PSID and the CPS}

We now turn to investigate some of our main findings using alternative data sources: the Current Population Survey (CPS) and the Panel Study of Income Dynamics (PSID). Analysing occupational mobility through the CPS is helpful even though these data is not corrected for measurement error. In particular, the CPS has the advantage of providing the longest, uninterrupted series of occupational mobility, even spanning into 2021. This allows us to evaluate whether the breaks in the SIPP time series have a meaningful effect on the extent and cyclicality of gross occupational mobility. For this purpose and because the individual-worker panel dimension of the CPS is much shorter relative to the SIPP, we only use these data to investigate the average gross mobility rate. In Section 2 of the main paper we conclude that the CPS and the (uncorrected) SIPP gross mobility series have very similar degrees of procyclicality. Another advantage of using the CPS is that it is easily accessible. We use the CPS data available via IPUMS.\footnote{Our STATA do-file is available upon request, while the underlying data can be download from IPUMS, at \texttt{https://cps.ipums.org/cps/}}

The PSID is also useful for several reasons: (i) It provides a longer panel dimension than the SIPP. (ii) We can compare our main results with the analysis of Kambourov and Manovskii (2008), who use this data set to provide a highly influential analysis of the occupational mobility patterns found in the US labor market. Therefore, in constructing our sample we closely follow Kambourov and Manovskii (2008, 2009). The details of this sample are described in the ``Data Construction'' section of this appendix. (iii) It allows us to evaluate the extend of coding error using retrospective coding as an alternative method. In particular, we use the PSID retrospective occupation-industry supplementary data files, which contain the re-coding the PSID staff performed on the occupational mobility records obtained during the 1968-1980 period. Since the 1981-1997 records were not re-coded and collected under independent interviewing, the earlier period can be used to construct ``clean'' occupational mobility rates and to analyse the effect of measurement error at the coding stage.

\subsection{Occupational mobility in the CPS}

\subsubsection{Extent of occupational mobility}

To derive the average gross occupational mobility rate among the unemployed in the CPS we compare the occupation coded immediately before the worker became unemployed with the occupational code at re-employment. From this set of workers, the gross mobility rate is then computed as the proportion of occupational movers among all those who went through unemployment and subsequently found a job. We focus on the 22 major occupational groups of the 2010 SOC, which are provided homogenized for the entire 1976-2021 period by IPUMS. Using the full extent of our sample, we obtain an average occupational mobility rate of 47.5\%. One concern could be that since in 1994 the CPS underwent a significant re-design, the occupational mobility rate should be affected. However, computing this rate for the period 1994-2021 yields 46.6\%, hardly changing the extent of mobility.

A perhaps more important concern is that in this sample we are including workers in temporary layoffs, who are very likely to return to their previous occupations and employers (see Fujita and Moscarini, 2017). This would bias downward the occupational mobility rate. To leave out temporary layoffs, we drop those unemployed classified as ``job loser/on layoff'' in the reported reasons for unemployment. This category corresponds to individuals who are on \textit{temporary layoff}, with the expectation to be recalled with a specific date or within 6 months. Note that in this case we only use data from 1994 onwards as after the 1994 redesign we have available a homogenous series for ``reasons for unemployment'', without discrete jumps associated with shifting definitions/survey questions. Once we drop these temporary layoff workers, the average occupational mobility rate indeed increases to 56.9\%. This value is very similar to the 55.0\% we obtained from the uncorrected SIPP when using the 21 major occupational groups of the 2000 SOC and dropping the temporary layoffs (See Table \ref{tab:basic_demog} in Section 1 of this Appendix).

\begin{figure}[ht!]
\centering
\subfloat[Occup. mobility - smoothed series]{\label{fig:bc1} \includegraphics [width=0.5 \textwidth] {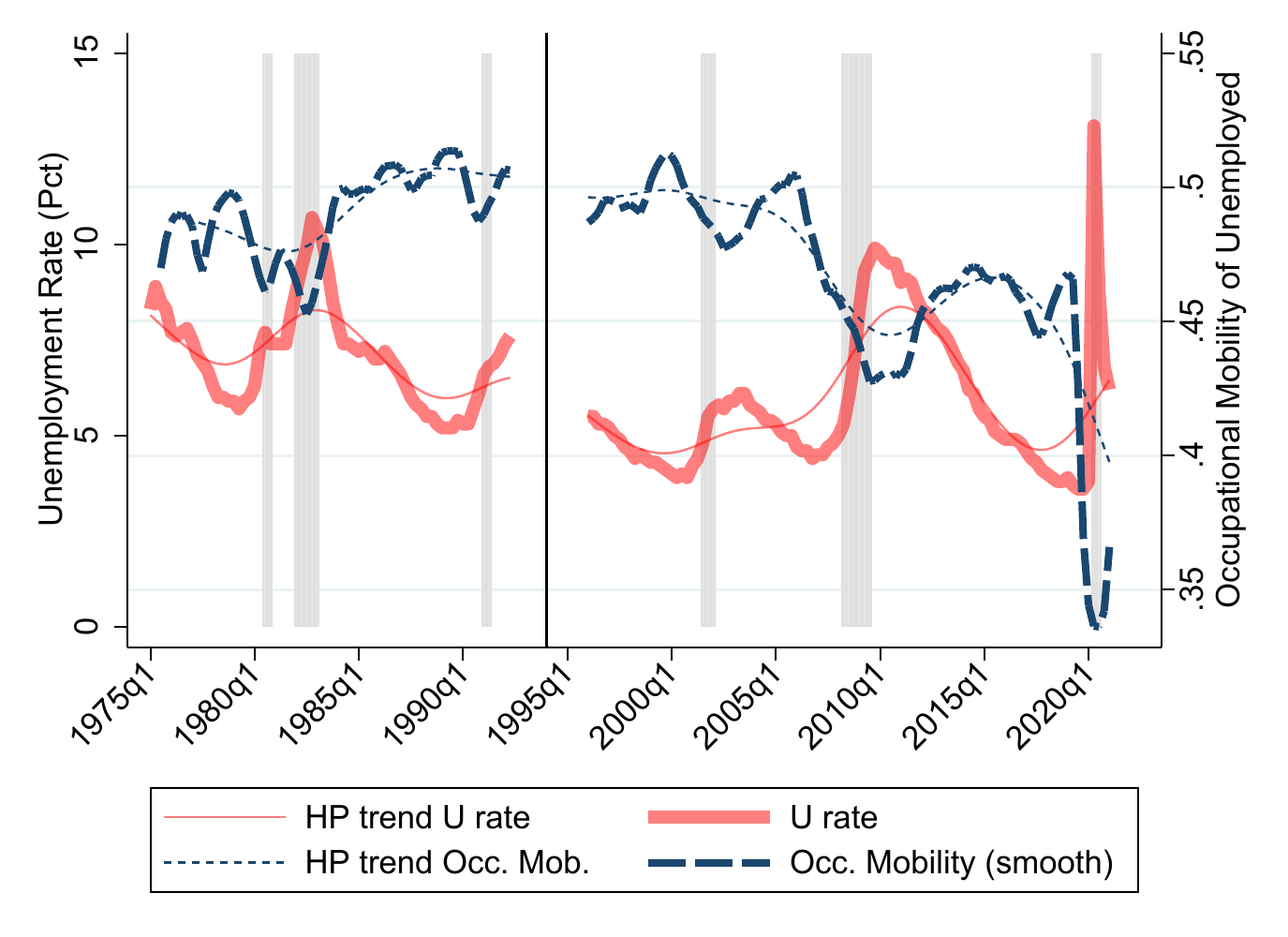}}
\subfloat[Occup. mobility - deviations from HP trend]{\label{fig:bc2_allu} \includegraphics [width=0.5 \textwidth] {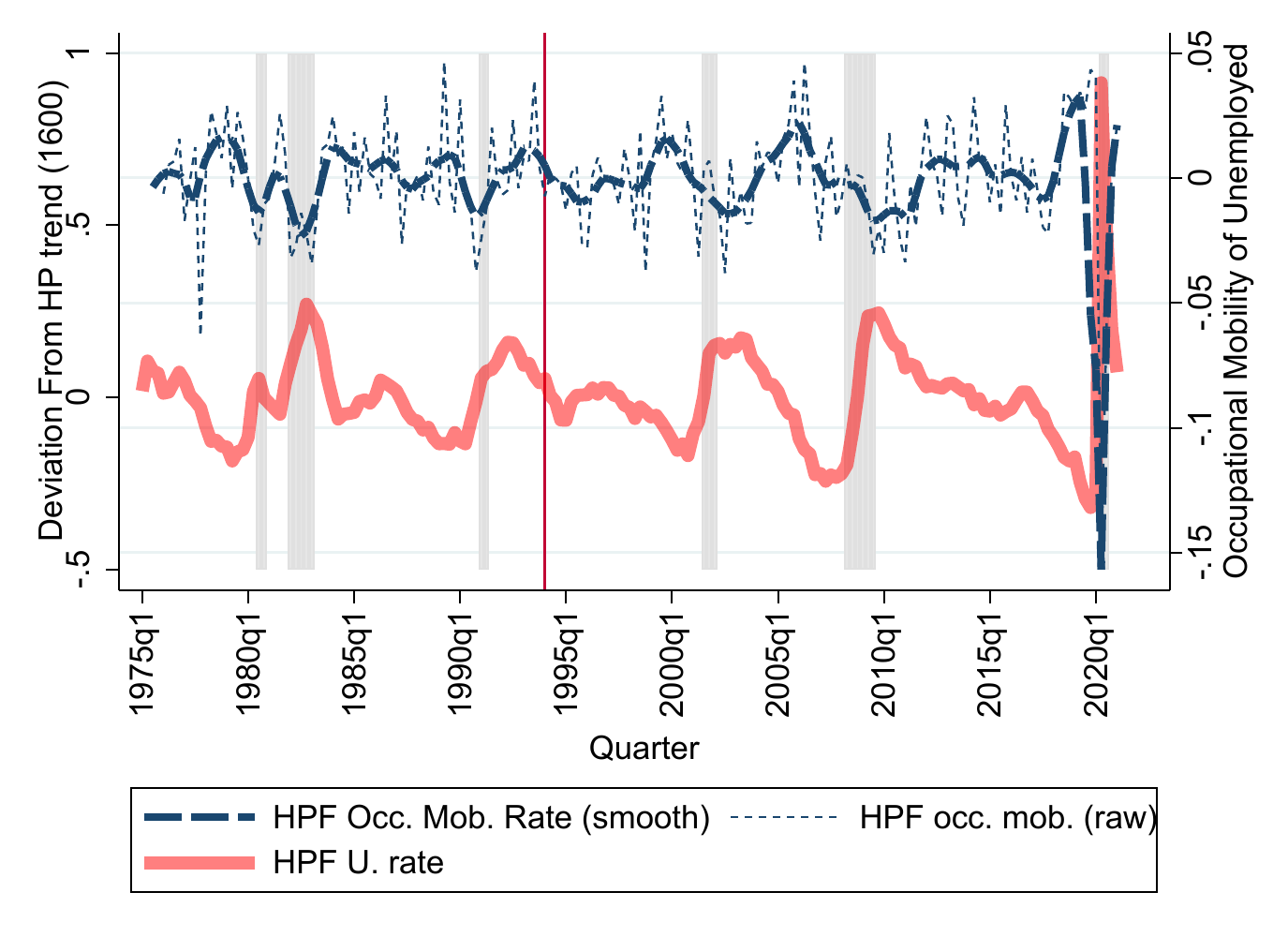}}

\subfloat[Occup. mobility - no temp layoffs]{\label{fig:bc2_notemplayoff} \includegraphics [width=0.5 \textwidth] {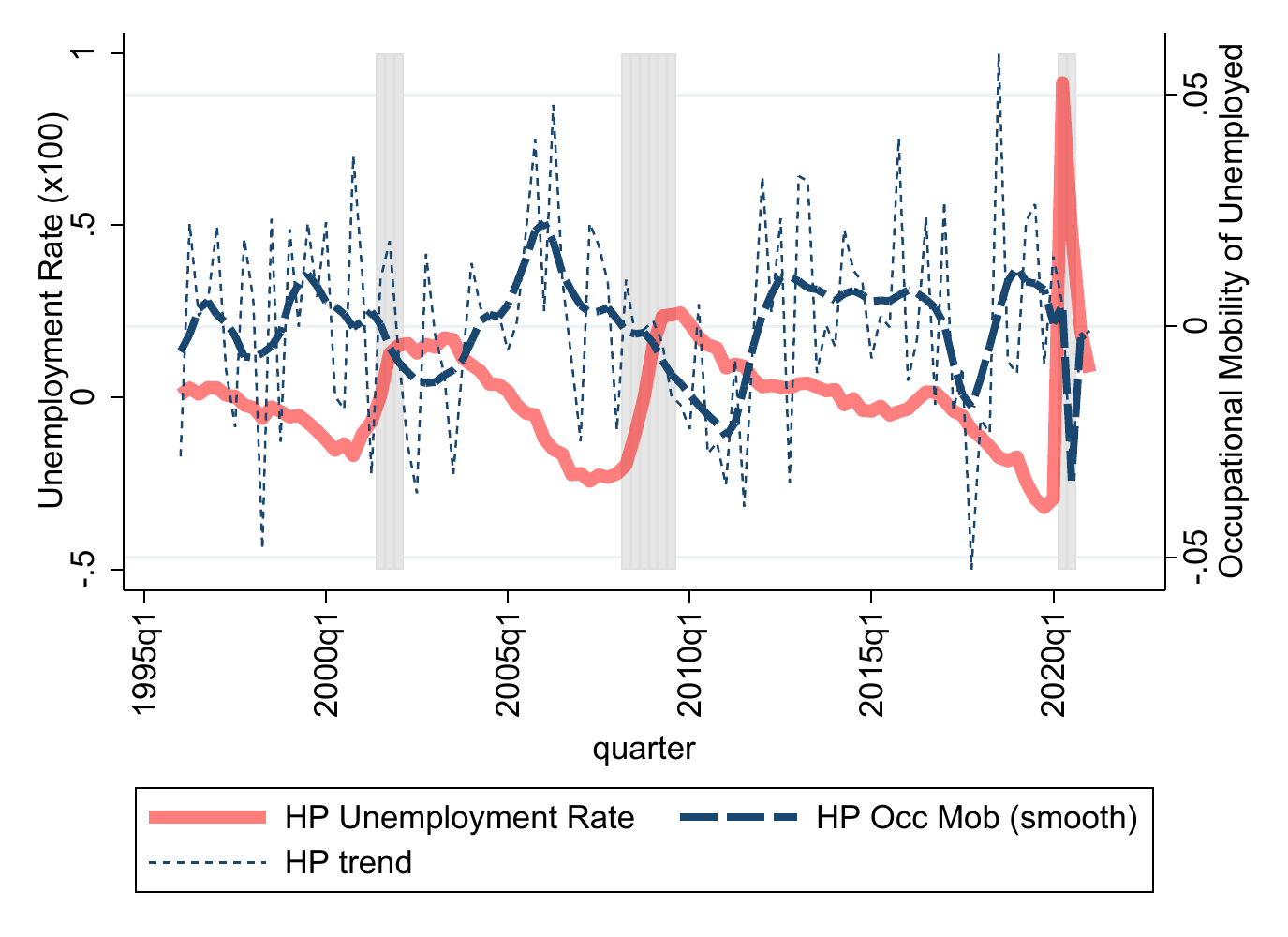}}
\caption{Occupational mobility of the unemployed - CPS}
\end{figure}

\subsubsection{Cyclicality of occupational mobility}

Figure \ref{fig:bc1} plots the time series of the average gross occupational mobility rate of workers hired from unemployment. To smooth out some of the noise in the quarterly observations, we present a lowess smoothed version in such a way that business cycle patterns remain visible. We also depict this series HP trend (filtered with parameter 1600). The redesign of the CPS in the mid-1990s means that the pre-1994 series might not be fully comparable to the post-1994 series. Following a conservative approach, we have HP filtered the entire 1976-2021 series but then removed observations for the period 1992-1995 to stay clear of the 1994 design break. To capture business cycle conditions, Figure \ref{fig:bc1} depicts the unemployment rate (and its HP trend with parameter 1600).

A comparison of these series reveals a procyclical pattern in occupational mobility. When the occupational mobility rate is above its trend, unemployment is typically below its trend. This is clearly visible around the last three recessions (including the Covid recession), but also in the double-dip recession in early 1980s. Figure \ref{fig:bc2_allu} shows a similar conclusion using the HP-filtered unemployment and occupational mobility rates for the full sample.  A reasonable concern is that the observed occupational mobility in the previous graphs is affected by an increased importance of temporary layoffs in recessions. Figure \ref{fig:bc2_notemplayoff} considers the series without temporary layoffs and once again shows a procyclical occupational mobility rate.

\begin{figure}[ht!]
\centering
\subfloat[HP-filter: Occup. mobility - up to 2019q4]{\label{fig:robust3_1} \includegraphics [width=0.5 \textwidth] {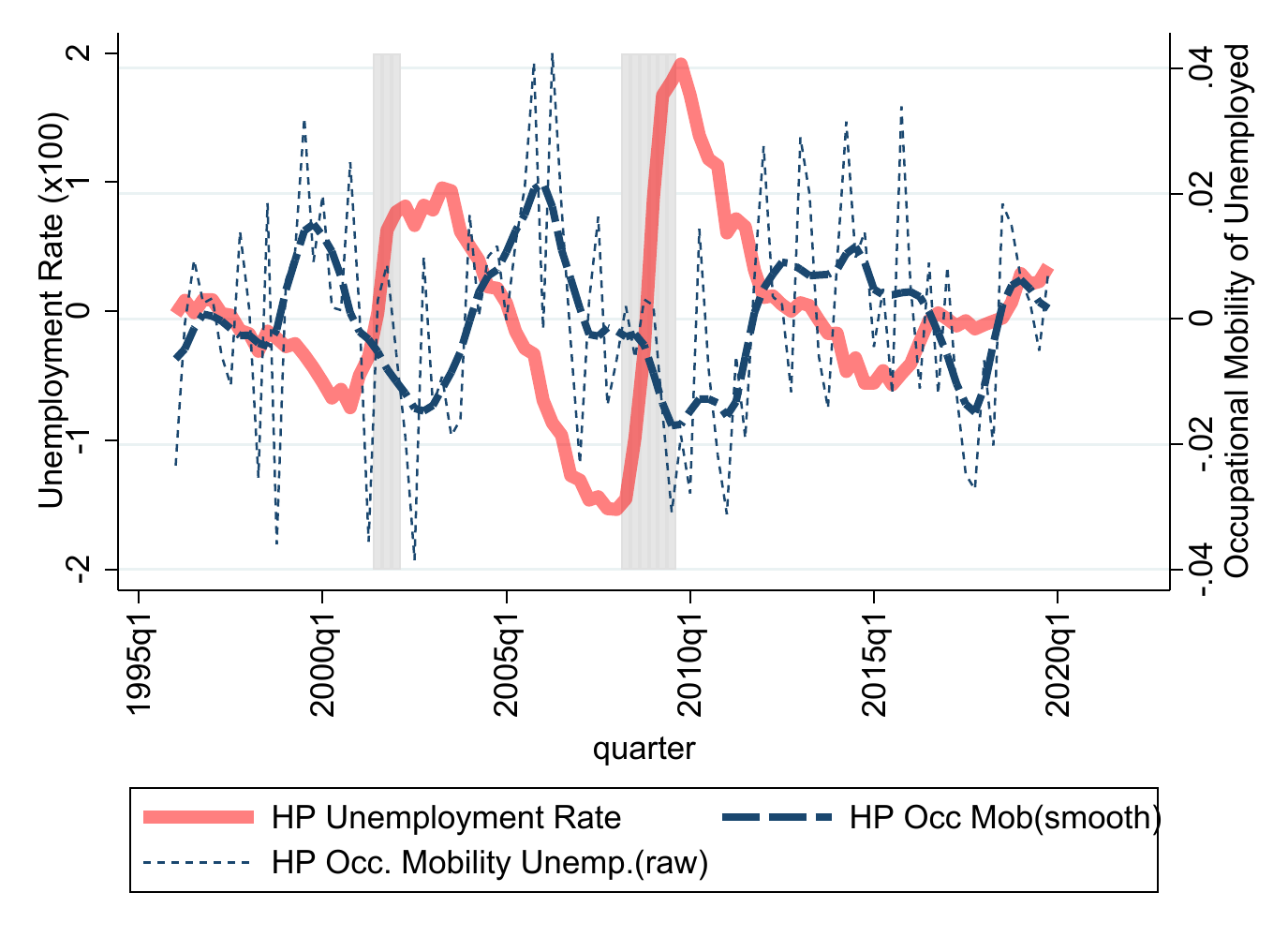}}
\subfloat[HP-filter:  Occup. mobility - up to 2019q4, no temp layoffs]{\label{fig:robust3_2} \includegraphics [width=0.5 \textwidth] {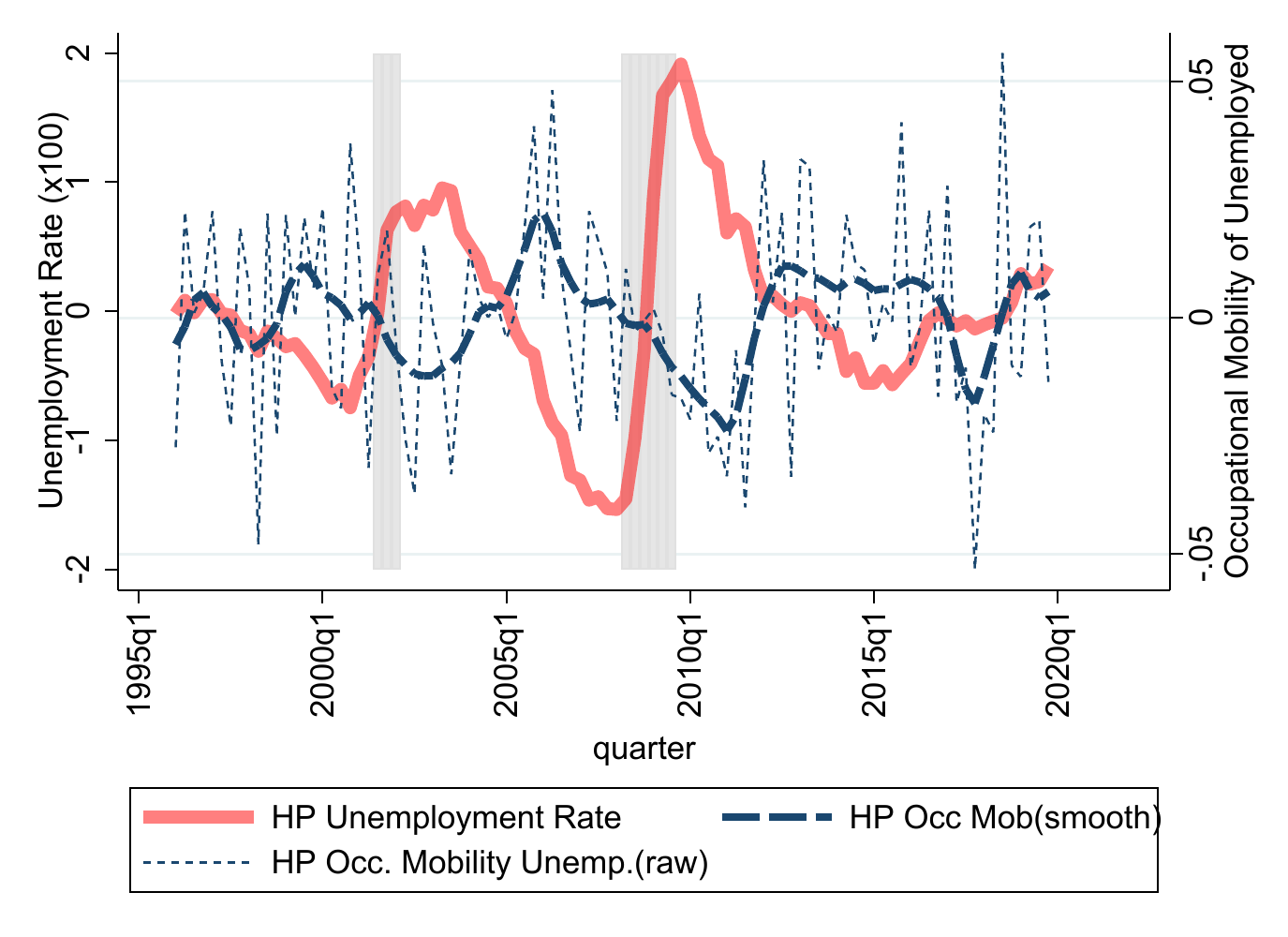}}

\subfloat[Bandpass filter: Occup. mobility - up to 2019q4]{\label{fig:robust3_3} \includegraphics [width=0.5 \textwidth] {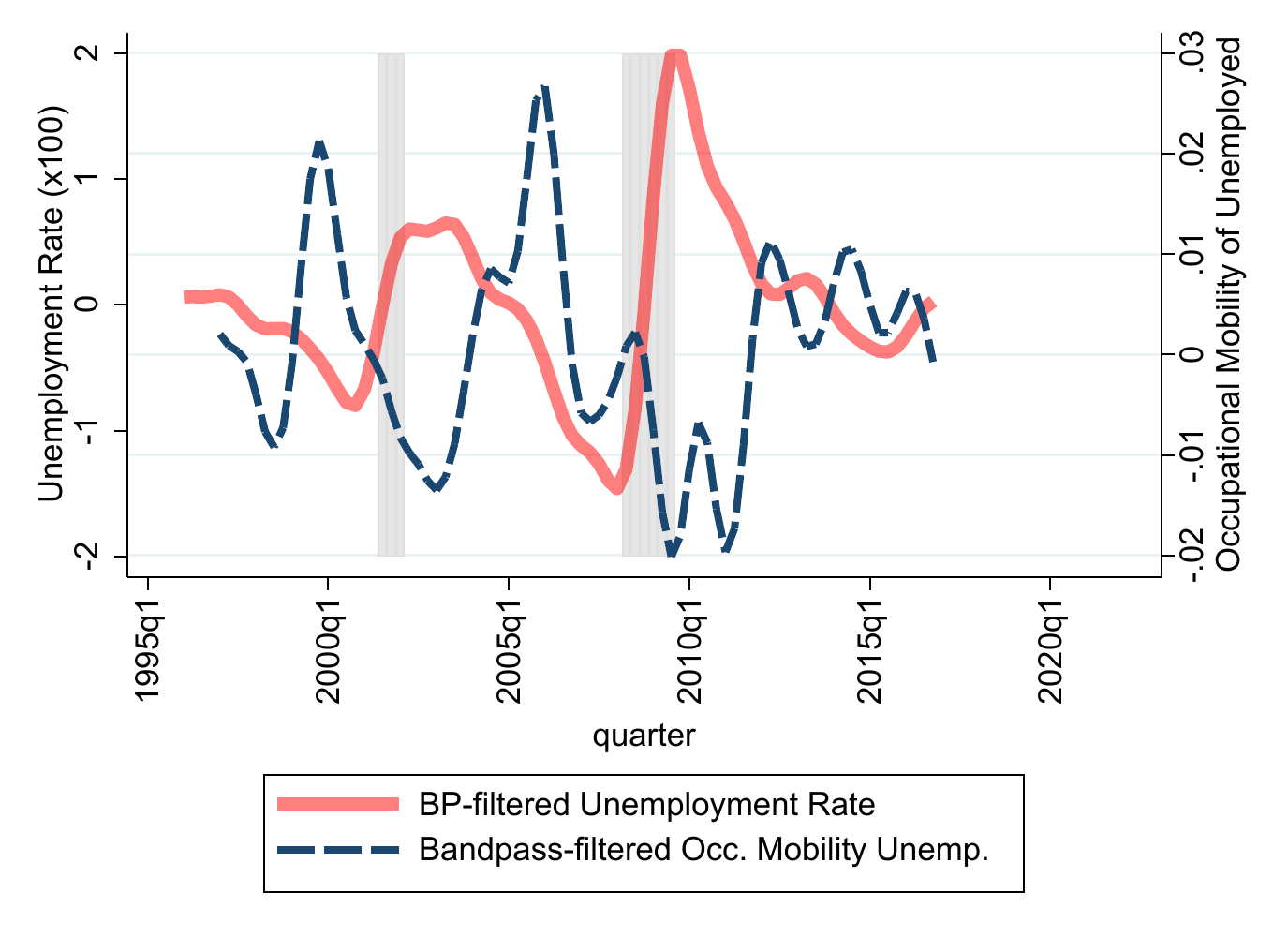}}
\subfloat[Bandpass filter: Occup. mobility - up to 2019q4, no temp layoffs]{\label{fig:robust3_4} \includegraphics [width=0.5 \textwidth] {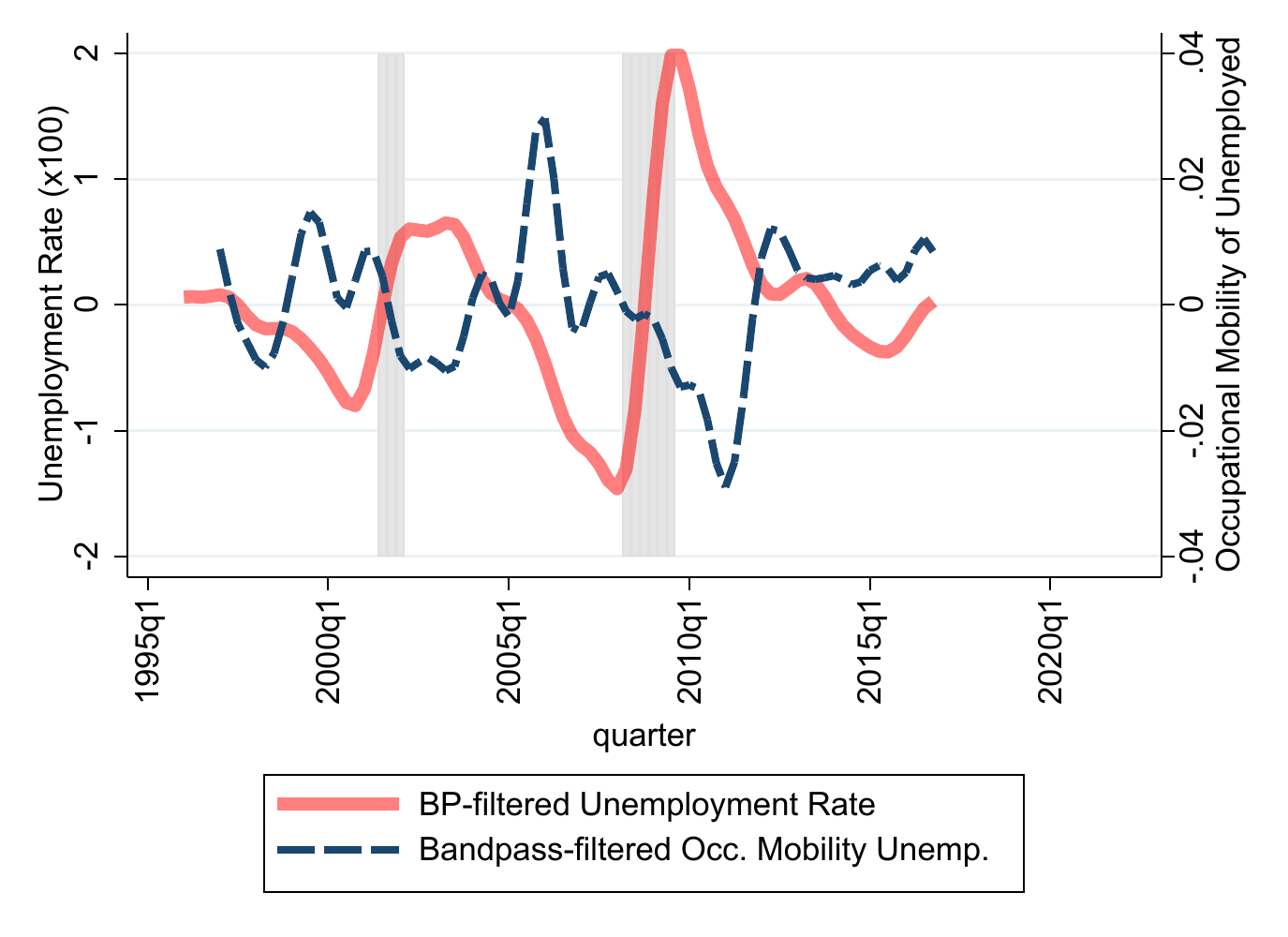}}

\caption{Occupational Mobility under Alternative Filtering}\label{fig:robustoccmobfilt}
\end{figure}

To further investigates the robustness of our findings, in Figures \ref{fig:robust3_1} and \ref{fig:robust3_2} we only consider the time series from 1994 up to the fourth quarter of 2019. In this way we focus on the period after the 1994 re-design and drop the large decrease in occupational mobility observed during the Covid recession. While this somewhat changes the de-trended behavior in the last five years of the series, it does not seem to meaningfully affect its procyclicality. In Figures \ref{fig:robust3_3} and \ref{fig:robust3_4} we re-do this exercise but now using bandpass filtering instead of the HP-filter to better deal with the noise in the raw data. If anything, the bandpass filtered series depicts even stronger procyclicality in the occupational mobility series, with clearly visible drops in occupational mobility during and in the aftermath of the 2001 and 2008 recessions.

Finally, Table \ref{tab:bc1} investigates the cyclicality of HP and bandpass filtered observations without additional smoothing. In particular, we regress the quarterly observations of the \textit{de-trended log gross occupational mobility rate} on a constant and the de-trended log unemployment rate. We report the coefficient and standard error associated with the latter. We observe that, after taking out the HP-trend (with parameter 1600), occupational mobility decreases in times of higher unemployment. This relationship is statistically significant when using all the sample period or after the 1994 CPS re-design (with and without temporary layoffs). Using the bandpass filtered series confirms these results.

\begin{table}[t]
  \centering
  \caption{OLS regression - Cycalility of occupational mobility}
    \begin{tabular}{rlllll}
    \toprule
    \multicolumn{6}{c}{Level and Responsiveness of Mobility to the Unemployment Rate} \\
    \midrule
    & Level & \multicolumn{4}{c}{Cyclical responsiveness} \\ \cmidrule(lr){2-2} \cmidrule(lr){3-6}
         & Ave. & \multicolumn{2}{c}{HP Filtered (1600)} & \multicolumn{2}{c}{Band-Pass Filtered} \\
         \cmidrule(lr){3-4} \cmidrule(lr){5-6}
&    Mob.       & \multicolumn{1}{l}{up to '21} & \multicolumn{1}{l}{up to `19} & \multicolumn{1}{l}{up to '21} & \multicolumn{1}{l}{up to '19} \\
    \midrule
    \multicolumn{6}{c}{Quarterly Occupational Mobility Rate, Hires from U (2010 MOG)} \\
    \midrule
    \multicolumn{1}{l}{All unemployed, 1976-} & 0.475 & -0.223$^{\star\star\star}$ & -0.105$^{\star\star}$ & -0.105   $^{\star\star\star}$ & -0.104$^{\star\star\star}$ \\
          & & (0.030) & (0.039) & (0.016) & (0.017) \\
    \multicolumn{1}{l}{All unemployed, 1994-} & 0.464 & -0.254$^{\star\star\star}$ & -0.103$^{\star\star}$ & -0.096$^{\star\star\star}$ & -0.104$^{\star\star\star}$ \\
          & & (0.036) & (0.050) & (0.018) & (0.019) \\
    \multicolumn{1}{l}{Unemployed, not on temporary layoffs, 1994-} & 0.569 & -0.044$^{\diamond}$ & -0.091$^{\star}$ & -0.063$^{\star\star}$ & -0.085$^{\star\star\star}$ \\
          & & (0.024) & (0.037) & (0.019) & (0.017) \\
    \bottomrule
    \bottomrule
    \multicolumn{5}{l}{\footnotesize $^{\star\star\star}$ p-value<0.001; $^{\star\star}$ p-value<0.01; $^{\star}$ p-value<0.05; $^{\diamond}$ p-value<0.1.}
    \end{tabular}%
  \label{tab:bc1}%
\end{table}%

\subsection{Occupational mobility in the PSID}

\subsubsection{Extent and cyclicality}

Using these data, Supplementary Appendix A.3 documents that when conditioning the sample only on those who changed employers through non-employment ($ENE$), the gross occupational mobility rates are high across different levels of aggregation. In particular, the raw data shows that the average occupational mobility rates among the non-employed are 39.6\%, 44.7\% and 56.2\% at a one-, two- and three-digit levels, respectively, for the period 1970-1980. The average occupational mobility rates for these workers for the period 1981-1997 are 41.1\%, 47.4\% and 60.1\% at a one-, two- and three-digit levels, respectively.

To compare these mobility rates with the occupational mobility rate among all workers (employer movers and stayers pooled together), we divide the numerator of the $ENE$ rate by the total number of employed workers at year $t$ (the denominator of our overall rate). In this case we find that the occupational mobility rates through non-employment are 2.5\%, 2.8\% and 3.5\% for the period 1970-1980 and 3.2\%, 3.7\% and 4.7\% for the period 1981-1997, at a one-, two- and three-digit levels, respectively. These rates are consistent with the results of Kambourov and Manovskii (2008), who show that unemployed workers contribute 2.5\% to the year-to-year occupational mobility rate of a pooled sample of employer movers and stayers based on a two-digit level. It is important to note, however, that the information presented in Kambourov and Manovskii (2008) \emph{does not} allow us to infer that their 2.5\% estimate is consistent with mobility rates among the non-employed of over 40\%. To arrive to this conclusion one has to re-do Kambourov and Manovskii's (2008) analysis as done here.

Table 6 in Supplementary Appendix A.3 shows the marginal effects obtained from probit regressions to further investigate the cyclicality of gross occupational mobility.\footnote{These estimates are obtained using the personal weights provided by each survey, but similar results are obtained when using the unweighted data. We also obtained very similar results when using the linear probability model on weighted and unweighted data and when using robust standard errors and clustering standard errors at a yearly level.} In these regressions the dependent variable takes the value of one if the worker changed occupation and zero otherwise. We control for age, education, full or part-time work, occupation of origin, region of residence, aggregate and regional unemployment rates, a quadratic time trend, number of children and the impact of retrospective coding.\footnote{As in Kambourov and Manovskii (2008), the education indicator variable takes the value of one when the worker has more than 12 years of education and zero otherwise. This is to avoid small sample problems if we were to divide educational attainment in more categories. The regional unemployment rates are computed using US states unemployment rates.} Our results show that gross occupational mobility of the non-employed is also procyclical.

These results corroborate the ones obtained using the SIPP. Hence, we find that across data sets the probability of an occupational transition among the unemployed (or non-employed) is high and increases in expansions and decreases in recessions. The procyclicality of gross occupational mobility among the unemployed complements the result found by Kambourov and Manovskii (2008), who show that the year-to-year occupational mobility rate of a pooled sample of employer movers and stayers is procyclical. In principle there is no reason to expect that the procyclical pattern these authors find in the overall occupational mobility rate would translate to the occupational mobility rate for the unemployed. Indeed, as discussed above, the latter contributes a small proportion to the overall rate.

\subsubsection{Repeat mobility}

In the SIPP analysis we found that a large proportion of workers who experienced an occupational change in their previous non-employment spell also experienced an occupational change at the end of their current non-employment spell. Similarly, we found that a large proportion of workers who did not change occupation in their previous non-employment spell experienced an occupational change at the end of their current non-employment spell.

A potential concern with the SIPP structure is that it does not follow workers for a long enough period. While in the calibrated model, we deal with this by imposing an analog restriction on the time window of measurement, such that the data and model statistics capture exactly the same type of repeat mobility, one would still want to extrapolate repeat mobility patterns to apply more generally (to longer periods of workers' labour market history). This extrapolation might generate a bias in the repeat occupational mobility statistics as the statistics in the SIPP will disproportionally capture: (i) those workers with shorter non-employment spells, even though we leave enough time towards the end of the SIPP panel to try to capture longer non-employment spells; and (ii) those workers with short employment durations between consecutive non-employment spells. To analyse the extent of this bias we re-compute the repeat mobility statistics using the PSID based on the sample used to construct the $ENE$ occupational mobility rate depicted in Figure 3.b in Supplementary Appendix A.3. Since these data sets allow us to follow the same workers for a longer period of time, we expect (i) and (ii) to have a much smaller impact.

\begin{table}[htbp]
  \centering
  \caption{Repeat Mobility - PSID 1968-1997}\vspace{3mm}
     \resizebox{1.0\textwidth}{!}{
  {\footnotesize
    \begin{tabular}{l|cc|cc}
    \hline
          & \multicolumn{2}{c|}{All workers } & \multicolumn{2}{c}{Male workers}  \\
    \hline
          & \multicolumn{1}{c}{Occ. Mobility } & \multicolumn{1}{c|}{Occ.+Ind. Mobility} &  \multicolumn{1}{c}{Occ. Mobility} & \multicolumn{1}{c}{Occ.+Ind. Mobility}  \\
           \hline \hline
    \emph{1-digit}         &          &  & & \\
    Stayer - Stayer & \multicolumn{1}{c}{67.3} & \multicolumn{1}{c|}{77.2} & \multicolumn{1}{c}{67.4} & \multicolumn{1}{c}{77.0}  \\
    Stayer - Mover & \multicolumn{1}{c}{32.7} & \multicolumn{1}{c|}{22.8} & \multicolumn{1}{c}{32.6} & \multicolumn{1}{c}{23.0}  \\
    \hline
    Mover - Stayer & \multicolumn{1}{c}{46.9} & \multicolumn{1}{c|}{58.8} & \multicolumn{1}{c}{45.8} & \multicolumn{1}{c}{57.8}  \\
    Mover - Mover & \multicolumn{1}{c}{53.1} & \multicolumn{1}{c|}{41.2} & \multicolumn{1}{c}{54.2} & \multicolumn{1}{c}{42.2} \\
    \hline \hline
   \emph{2-digits}    &       &       &          &   \\
    Stayer - Stayer & \multicolumn{1}{c}{61.9} & \multicolumn{1}{c|}{69.9} & \multicolumn{1}{c}{62.8} & \multicolumn{1}{c}{71.0} \\
    Stayer - Mover & \multicolumn{1}{c}{38.1} & \multicolumn{1}{c|}{30.1} & \multicolumn{1}{c}{37.2} & \multicolumn{1}{c}{29.0}   \\
    \hline
    Mover - Stayer & \multicolumn{1}{c}{40.6} & \multicolumn{1}{c|}{49.7} & \multicolumn{1}{c}{40.5} & \multicolumn{1}{c}{48.8} \\
    Mover - Mover & \multicolumn{1}{c}{59.4} & \multicolumn{1}{c|}{50.2}  & \multicolumn{1}{c}{59.5} & \multicolumn{1}{c}{51.2} \\
    \hline \hline
   \emph{3-digits}    &       &       &   & \\
    Stayer - Stayer & \multicolumn{1}{c}{54.3} & \multicolumn{1}{c|}{61.2} & \multicolumn{1}{c}{57.8} & \multicolumn{1}{c}{64.1} \\
    Stayer - Mover & \multicolumn{1}{c}{45.7} & \multicolumn{1}{c|}{38.8} & \multicolumn{1}{c}{42.2} & \multicolumn{1}{c}{35.9} \\
    \hline
    Mover - Stayer & \multicolumn{1}{c}{25.9} & \multicolumn{1}{c|}{33.2}  & \multicolumn{1}{c}{27.5} & \multicolumn{1}{c}{34.8} \\
    Mover - Mover & \multicolumn{1}{c}{74.1} & \multicolumn{1}{c|}{66.8}  & \multicolumn{1}{c}{72.5} & \multicolumn{1}{c}{65.2} \\
    \hline
    \multicolumn{5}{l}{\footnotesize{Note: Total number of observations among all workers (male) = 3,261 (2,467).}}

    \end{tabular}}
    }
  \label{tab:Repeat_mobility_PSID}%
\end{table}%

Table \ref{tab:Repeat_mobility_PSID} shows the repeat mobility statistics at a one-, two- and three-digit level of aggregation, using the 1970 SOC. The proportions are based on weighted data, but similar proportions are obtained using unweighted data. For each level of aggregation we divide the sample by whether a worker was an occupational stayer or an occupational mover after the first non-employment spell. We then compute the proportion of stayers (movers) who, after a subsequent non-employment spell, did not change occupation and the proportion who changed occupation. For each level of aggregation, the first two rows show the proportions for stayer- stayer and stayer-mover. These proportions add up to one. Similarly, the second two rows show the proportions for mover-stayer and mover-mover. Further, the columns labelled ``Occ. Mobility'' consider workers who only changed occupations. Since these statistics are based on raw (uncorrected) data, the propensity to change occupations would be biased upward. As a way to deal with the latter, we consider simultaneous occupation and industry mobility and re-compute the repeat mobility statistics. These are presented in the columns labelled ``Occ. + Ind. Mobility''. As discussed in Section 1 of this appendix, conditioning on simultaneous occupation and industry mobility provides an alternative way to correct for coding errors. In particular, we consider a worker to be an occupational mover if and only if he/she reported a change in occupation and a simultaneous change in industry at the same level of aggregation, where industry changes are based on the 1970 census industries codes.\footnote{These repeat mobility statistics are calculated based on up to five consecutive non-employment cycles, where a cycle is constructed as non-employment, employment, non-employment, employment sequence. The number of observations in our repeat mobility sample is then the product of the number of cycles per individual. The majority of workers, however, experience only one cycle, which make up for 80\% of the total number of observations. We find that our results do not change if we were to compute these same statistics based only on workers' first cycle.}

The PSID shows a very similar picture to the one obtained from the SIPP. There is a high proportion of workers who changed occupation after a non-employment spell and once again changed occupations after a subsequent non-employment spell. Out of all those workers who were occupational movers after a non-employment spell in the raw data, between 53\% (at a one-digit level) and 74\% (at a three-digit level) moved occupations once again after a subsequent non-employment spell. There is also an important proportion of workers who did not change occupation after a non-employment spell, but did change occupations after a subsequent non-employment spell. Out of all those workers who were stayers after a non-employment spell, between 33\% (at a one-digit level) to 48\% (at a three-digit level) moved occupations after the subsequent non-employment spell.

Conditioning simultaneous occupation and industry changes does not drastically change these results. In this case, out of all those workers who were occupational movers after a non-employment spell, between 41\% (at a one-digit level) and 67\% (at a three-digit level) moved occupations once again after a subsequent non-employment spell. Out of all those workers who were stayers after a non-employment spell, between 23\% (at a one-digit level) to 33\% (at a three-digit level) moved occupations after the subsequent non-employment spell. That is, conditioning on simultaneous occupation and industry changes, decreases by about 10 percentage points the occupational mobility rates, but still shows a high propensity for repeat mobility among the non-employed.

\section{Self-reported retrospective occupational mobility}

In this section we propose an alternative way to analyse the occupational mobility patterns of the non-employed that is not subject to coding errors. In particular, we use information on workers' self-reported employer and occupational tenure obtained from the SIPP core panels and topical modules. For the majority of SIPP panels the first topical module asks ``for how many [months/years] has [the worker] done the kind of work [he] does in this [current] job'', or a variation of it.\footnote{This question is asked in topical module in the 1984, 1987, 1990, 1991, 1992 and 1993 panels. In the 1996, 2001, 2004 and 2008, the question is in the core waves and makes explicit reference to the entire working life, while invoking  `occupation'/`line of work' rather than `kind of work': ``Considering  entire working life, how many years has [the worker] been in this occupation or line of work?''.} In addition, it records the start date with the \emph{current} employer \textit{and} the start and end dates of the most recently finished employer spell previous to the panel.\footnote{The question referring to the current employer is ``when did [the worker] start working for [explicit employer name]''. The wording means that employer recalls are not (or, at least, should not be) recorded as starts of employment. In the 1996-2008 panels, the question also refer to ``... the month/year when [worker's name] began employment with [employer name]''.}

Using this information we restrict attention to those workers whose completed non-employment spells lasted between one and twelve months. This restriction helps us capture a group of workers who have not lost their attachment to the labour market even though they might be categorised as non-participants at some point during the spell\footnote{For most of the panels we obtain the duration of the non-employment spell by subtracting the date in which the job with the previous employer ended from the day the re-employment job started with the current employer. In the 1984 panel the duration of the non-employment spell is asked directly. In the 2004 and 2008 panels we can only observe a lower bound of the non-employment spell as we only have information on the year the job with the previous employer ended. For this reason some of our analysis (see below) will not use the 2004 and 2008 panels. Note that the mobility reported in the 1984 panel lies } Below, we also focus on subsets of this group that e.g. are observed with periods of unemployment. To determine whether or not one of these workers changed occupations, we compare the employer tenure with the occupational tenure information. In particular, out of all those workers who found their current job out of non-employment, we label a worker as an occupational mover when his/her occupational tenure equals (or is very close to) his/her employer tenure and as an occupational stayer when his/her occupational tenure is (sufficiently) greater than his/her employer tenure.\footnote{When occupational tenure does not exactly match the employer tenure due to, potentially, recall errors, we consider several plausible adjustments to identify occupational movers and stayers. We discuss these in the next section.} We then compute the occupational mobility rate by dividing the number of non-employed workers who re-gained employment and changed occupation over the number of non-employed workers who re-gained employment.

\subsection{The extent of self-reported occupational mobility}

Table \ref{table_retro_1} shows the extent of occupational mobility of the non-employed using four different samples. The first column (overall) considers all employed workers who (i) went through a spell of non-employment before starting with the current employer, provided that the non-employment spell occurred during the last ten years; (ii) have spent more than two years in the labor force, (iii) have finished their previous job at least one year after entering the labor market and (iv) have held their previous job for at least twelve months.\footnote{We have excluded workers who have imputed start and dates of the previous job in the data, or an imputed start date of the current job. In case ambiguity remains about occupational changing, we have excluded these workers as well. If we take a strict approach to assignment of mobility within this subset of workers, mobility is closer to 50\% than in the unambiguous group. Hence, we consider the exclusion of ambiguous observations to yield conservative levels of self-reported occupational mobility.} These restrictions are made to focus on those who have had meaningful employment before any spell of non-employment, so we are considering changes of occupation, rather than a start within an occupation at the beginning of working life. These restrictions also help with inferring occupational mobility from occupational tenure.

The second column (attached workers) restricts the ``overall'' sample to those spells in which the worker entered non-employment because they were laid off, discharged, a temporary job ended, indicated dissatisfaction with the previous job, or because of a non-family/non-personal (``other'') reason.\footnote{Including ``Other'' (but excl.: ``other, family reasons'') does not change our conclusions.} We consider these workers to be attached to the labor market as they did not leave their jobs for reasons that indicate leaving the labor force.\footnote{For the ``overall'' and ``attached'' samples, we only count an occupational move at the end of the non-employment spell when the current job started within six months of the start of the occupation and the previous job lasted for at least one year. For a worker to be considered an occupational mover in these samples, the current job starting date must be within a six months interval around the implied start of the occupation and the previous job must have lasted at least a year. Further, given that occupational tenures above twelve months are reported in full years it was difficult to assess an occupational change when a relatively long tenure in the current job is preceded by a short tenure in the previous job. We take a conservative approach and categorise the latter cases as a employer move without an occupational change.} Given that the above restrictions could not be applied across all panels, these two samples only rely on the topical modules of the 1984, 1987-88 and 1990-1993 SIPP panels.

\begin{table}[!ht]
  \caption{Self-reported occupational mobility rates}\label{table_retro_1} \vspace{3mm}
  \centering
  {\small
  \begin{tabular}{lcccc}
    \hline
    &  (a) &  (b) &  (c) &  (d) \\
     & Overall  & Attached workers  & Recent hires  & Hires after U \\
     & 1984-1993 & 1984-1993  & 1984-2001 & 1984-2001 \\ \hline \hline \\
    All workers         & 0.446     & 0.434     & 0.372     &  0.397\\ \vspace{1mm}
                        & (0.007)   & (0.008)   & (0.013)   & (0.019) \\ \hline \\
    Males               & 0.467     & 0.437     & 0.393     & 0.439 \\ \vspace{1mm}
                        & (0.010)   & (0.012)   & (0.018)   & (0.026) \\
    Females             & 0.429     & 0.433     & 0.352     & 0.343 \\ \vspace{1mm}
                        & (0.010)   & (0.012)   & (0.017)   & (0.026) \\ \hline \\
    HS dropouts         & 0.476     & 0.476     & 0.409     & 0.507 \\ \vspace{1mm}
                        & (0.021)   & (0.025)   & (0.032)   & (0.046) \\
    HS grads            & 0.478     & 0.486     & 0.405     & 0.405 \\ \vspace{1mm}
                        & (0.013)   & (0.015)   & (0.021)   & (0.031) \\
    Some college        & 0.437     & 0.413     & 0.342     & 0.347 \\ \vspace{1mm}
                        & (0.014)   & (0.017)   & (0.024)   & (0.037) \\
    College grads       & 0.412     & 0.380     & 0.330     & 0.350 \\ \vspace{1mm}
                        & (0.013)   & (0.015)   & (0.026)   & (0.042) \\ \hline \\
    20-30 years old     & 0.555     & 0.548     & 0.441     & 0.476 \\ \vspace{1mm}
                        & (0.015)   & (0.018)   & (0.025)   & (0.039) \\
    35-55 years old     & 0.387     & 0.385     & 0.323     & 0.347 \\ \vspace{1mm}
                        & (0.010)   & (0.012)   & (0.018)   & (0.027) \\
    \hline
        \multicolumn{5}{p{0.7\textwidth}}{\footnotesize{Standard Errors in parentheses. Sample weights: using person weights of SIPP panels within panel, normalized such that average weight of observations across all panels is one, while cross-sectionally weights make the sample representative).}}
  \end{tabular}
    }
\end{table}

The third column (recent hires) considers those workers whose re-employment job started within six months prior to the time of interview and had a job tenure of at least 12 months before transiting into non-employment. The last restriction enable us to more confidently distinguish the current employer's tenure from the sum of the current and previous employer's tenure, which reduces ambiguous cases when inferring occupational mobility from self-reported occupational tenure. The cost, obviously, is a smaller sample of workers, as logically we can include only workers with a low job tenure at the moment that the retrospective occupational questions are asked. Note also that this sample is selected differently, columns (c) naturally samples more those in the population who are more likely to experience non-employment spells. The fourth column (``Hires after U'') uses an even more restricted sample of ``recent hires'' for whom the employment status is observed within the core waves and who were unemployed for at least one month prior to re-gaining employment. Since the retrospective question is asked in the early waves of a panel, this leaves relatively little room to observe the unemployment status of those workers in the core wave dataset (which is a requirement to be included in this group). For these samples we use the 1984 up to 2001 panels.\footnote{For the most recent two panels, 2004 and 2008, we can find a lower bound and an upper bound on the non-employment spell, but typically not a precise monthly duration. When restricting the 1996 and 2001 panels to provide the same amount of information as the 2004 and 2008 panels (by ignoring start and end months of the previous jobs), we find that the loss of information appears to be small.}

For a worker to be considered as an occupational mover, the current job starting date must be within a one month interval around the implied start of the occupation, or before.\footnote{Here we use a one month `margin' to reflect one month of rounding error. The impact of including an additional margin of one month is small, ranging from 0 to 3 percentage points change in the occupational mobility rate at most. Below we use an alternative measure to check the robustness of our measure.} Across all these sample we once again find that non-employed workers' self-reported occupational mobility is high. Combining retrospective employer and occupational tenure information implies that roughly about 40\% of non-employed workers accept jobs in a new line of work. Average mobility of recently hired workers in columns (c) and (d), which uses recall of more recent non-employment spells, and allows us to distinguish tenures more clearly, is broadly in line with columns (a) and (b), where the former have a reported mobility in 35-42\% range, versus 42-46\% for all workers in columns (a) and (b). Retrospective measures of occupational mobility do subtly differ from the ones derived by comparing occupational codes, first because they reflect the workers' own assessment of 'line of work'/occupations, rather than the Census', but also because the occupational tenure in question can reasonably refer to jobs even before the previous job. Nevertheless, we find that the extent of occupational mobility obtained from individuals reporting a ``different kind/line of work'' aligns very well with the adjusted rate of occupational mobility for the non-employed obtained by comparing the major occupations of the 2000 SOC or the 1990 SOC described in Section 1.1 of this appendix. This confirms that, after a gap in their employment history, about two out of five workers ends up in a different line of work than they had before.\footnote{The retrospective question, is consistent with the interpretation of occupational mobility in the main paper, where an occupational change is a start of a new career. In the calibration, this occurs in 44\% of unemployment spells, according to the retrospective question, workers start something they have not done before in about 40\% of the cases.}

Table \ref{table_retro_1} shows that the retrospective occupational mobility rate is also high across several demographic groups. In particular, both male and female workers have mobility rates around 40\%, with an average mobility rate across the four samples that is higher for males. Interestingly, in some measures (for example in column (d)) this gap appears meaningful, but on the other hand is smaller when considering women who were classified as ``attached'' (in column (b)). Occupational mobility upon re-employment is substantial for workers from all education groups, though again we can discern an education gradient, with college workers more often taking re-employment in a line of work they held before. Finally, we find that the perhaps clearest differences in the occupational mobility rates by demographic characteristics is across age groups, where the mobility rate decreases significantly as workers get older. Nevertheless, the mobility rate of prime-aged workers also remains high, between 32.5 and 39\%.\footnote{The relative drop across age groups is somewhat stronger in our retrospective measure than in the occupational mobility based on comparing codes, falling between 12 to 16 percentage points from young to prime-aged workers. A potential explanation is the difference between comparing occupational codes after an unemployment spell with the occupational code just before, rather than asking whether, at some point in their entire working life, workers have had a line of work similar to the current one. The relevance of this difference, intuitively, seems larger for older workers.}

It is important to note some differences across panels. For example, how the duration of non-employment was solicited: directly as a duration or implied by the starting and end dates of a job. In the 1984 panel individuals are asked explicitly for the duration of the non-employment spell and the tenure on the job, rather than reporting start and end months of employment, as done in more recent panels. Another difference is that the labor market topical module is asked in the interview of wave 2 between 1987 and 1991 panels, but from the 1992 panel onwards in the interview of wave 1. The SIPP was redesigned in 1996 panel and carried over to 2001 panel. Generally, a large part the new design of 1996 is carried forward till the 2008 panel. However, in the labor market topical module there are changes that lead the 2004 and 2008 panels to differ from the 1996 and 2001 panels and make it much harder to determine the combination of non-employment duration and occupational mobility. For this reason these former panels are omitted from the analysis. The setup of the labor market history questionnaire is perhaps the most distinct in the 1984 panel; however, it also contains the most direct question on the employment gap. Implied occupational mobility in this panel is somewhat higher than in other panels, while its sample size is relatively large relative to the other panels in the 1980s. Excluding it from the analysis means that the average implied occupational mobility across the remaining panels is lower, typically, by 1-4 percentage points. For example, mobility of all workers (pooled) in the first column, drops to 42.9\%. Nevertheless, even excluding this panel, mobility is high, and broadly shared across many demographic groups.

\subsection{Self-reported occupational mobility and non-employment spell duration}

We now turn to analyse the mobility-duration profile that arises from the retrospective measure of occupational mobility. Table \ref{table_retro_3} reports the results of linear probability models based on the implied occupational mobility rates obtained from self-reported occupational tenure as a function of the time in non-employment between jobs and a set of controls. These include an indicator variable for gender, a quadratic in age and categorical variables for the completed duration of the non-employment spell (where we take 1-4 months as the baseline category) and for workers' educational levels (where we take high school graduate as the baseline category). We also add controls for previous and current occupational tenure and occupation of origin and destination, include a linear time trend and grouped-panel fixed effects to control for differences between SIPP panel setups.
\begin{table}[!ht]
  \caption{Self-reported Occupational Mobility and Non-employment Duration}\label{table_retro_3} \vspace{3mm}
  \centering
  {\small

  \begin{tabular}{llllll}
  \hline
  \multicolumn{6}{c}{ \textbf{Panel A. Retrospective Occ. Mobility of All Workers (1984-1993 panels)} }\\ \hline
            & (i)   & (ii)  & (iii) & (iv)  & (v) \\
          & Overall & Overall & Overall & Attached & Attached \\ \hline \hline
  no obs. & 5219  & 5219  & 5170  & 3656  & 3622 \\ \hline
  NE duration & 0.0109*** & 0.0112*** & 0.0096*** & 0.0095*** & 0.0094*** \\ \vspace{1mm}
  (s.e.) & (0.0024) & (0.0024) & (0.0023) & (0.0028) & (0.0028) \\ \hline
      linear time trend          & \phantom{XX}X  & \phantom{XX}X & \phantom{XX}X & \phantom{XX}X  & \phantom{XX}X\\
    panel FEs  &   & \phantom{XX}X &  \phantom{XX}X&  & \phantom{XX}X \\
    demogr. controls & &\phantom{XX}X &\phantom{XX}X &  & \phantom{XX}X\\
    source occup. &       &       &   \phantom{XX}X    &       &  \phantom{XX}X\\ \hline
          &       &       &       &       &  \\ \hline
  \multicolumn{6}{c}{ \textbf{Panel B. Retrospective Occ. Mobility of Recent Hires (1984-2001 panels) }}  \\ \hline
          & (vi)  & (vii) & (viii) & (ix) & (x) \\
          & All Hires & All Hires & Hires after U & Hires After U & Hires After U$^\star$ \\ \hline \hline
 no. obs & 1549  & 1549  & 692   & 692   & 679 \\ \hline
 NE duration & 0.0062 & 0.0068* & 0.0110* & 0.0107* & 0.0114* \\ \vspace{1mm}
 (s.e.) & (0.0039) & (0.0039) & (0.0062) & (0.0061) & (0.0063) \\ \hline
        linear time trend          &  \phantom{XX}X & \phantom{XX}X &  \phantom{XX}X& \phantom{XX}X & \phantom{XX}X\\
    panel FEs  &   & \phantom{XX}X &  & \phantom{XX}X & \\
    demogr. controls & &\phantom{XX}X & & \phantom{XX}X & \\
    \hline
        \multicolumn{6}{p{0.8\textwidth}}{\footnotesize{Levels of significance: $^{*} p<0.1$, $^{**} p<0.05$, $^{***} p<0.01$. Observation weighted by person weights within panel, by number of total observations per panel across panels. ``Recent Hires'' means hired within 6 months of interview. ``Hires After U$^\star$'' refers to an alternative measure of occupational mobility, where everyone with occupational tenure weakly less than current job tenure (allowing for one month of ambiguity) is a mover, while everyone with an occupational tenure larger than current job tenure + 4 months is a stayer.}}
  \end{tabular}
  }
\end{table}

In panel A we consider the retrospective information on the (most recent) spell of non-employment of all workers, if they have gone through non-employment after entering the labor force. We observe that self-reported occupational mobility increases with non-employment duration, roughly at a rate of one percentage point per month. These results are well in line with the results presented in the main text, for occupational mobility according to changes in the occupational code.

In the first column we relate the probability of self-reported change in occupation to the reported time between employment. We only allow for a linear time trend in occupational mobility. In the second column, we add controls for gender, race, a quartic in age and marital status, and grouped panel fixed effects. This hardly affects the slope of the mobility-duration profile. In the third column, we include the occupational code on the job previous to the non-employment spell. Interestingly, the earlier panels of the SIPP, up to 1993, include both a question on occupational tenure and the occupation code of this previous job. Controlling for the `source' occupation reduces slightly the point estimate of the mobility - duration profile, but not in a significant way. In the fourth and fifth columns we repeat these regressions for the smaller subset of workers who do not separate for ``personal reasons'', i.e. our ``attached'' sample discussed above. Again, we see that a broadly similar picture appearing, again without much difference when controlling for demographics, grouped panel fixed effects and source occupations.

In panel B, we concentrate on recent hires, i.e. those who started a job, after non-employment, within 6 months of the interview on their labor market history. We apply the same sample selection criteria as discussed above. The smaller sample size means that the gradients of mobility with duration are more imprecisely estimated. Nevertheless, a similar broad (but less precise) conclusion can be drawn. There is a positive slope of mobility with duration that is not too steep, with point estimates well in line with panel A and the occupational code-driven mobility in the main text. Although care has to be taken, given the size of the standard errors, we observe that the point estimates of the slope of the mobility-duration profile of all workers hired after non-employment is somewhat lower, while for those who have been unemployed the point estimates of the slope is rather close to panel A of Table \ref{table_retro_3}, and the main text. Although not shown here, the coefficients on education, age, gender paint a similar relative picture as Table \ref{table_retro_1}.

The above results show that occupational mobility is high at any non-employment duration and exhibits a moderate increase with the duration of the non-employment spell, where between 50\% and 60\% of workers with at least 9 months of non-employment duration return to previous occupations at re-employment. These are very similar characteristics as the ones found in our main analysis based on the comparison of occupational codes. Once again we find a prominent decline in the probability of an occupational change with age. We also find small differences in the probability of an occupational change between males and females and across education levels, with the exception of the group of recently hired from unemployment (as in Table \ref{table_retro_1}). Note that including indicator variables for the occupations of origin and destination (as done in 1b) does not alter greatly the outcomes for non-employment spell durations and other coefficients, with the exception of the point estimates for the education coefficients. This suggests that specific occupational identities are not the main drivers of both non-employment duration and occupational mobility of workers. This is once again in line with the conclusions of Section 1.5 in this appendix.

\subsection{The cyclicality of self-reported occupational mobility for the non-employed}

\begin{figure}[th!]
\centering
\includegraphics[width=0.8 \textwidth]{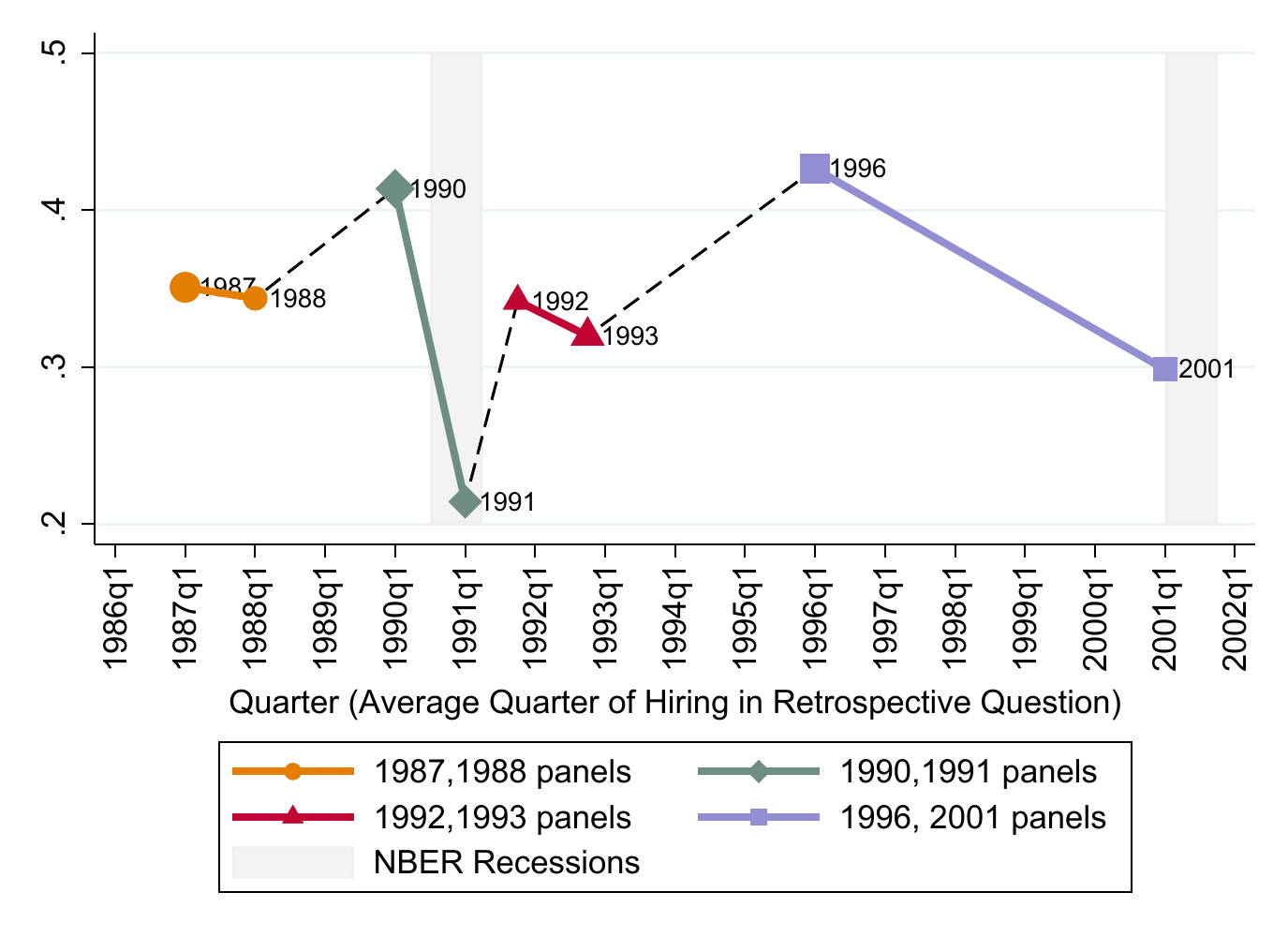}
\caption{Retrospective self-reported occupational mobility of recent hires (after u), by panel}
\label{f:retro_ts}
\end{figure}

We now turn to investigate the cyclicality of our occupational mobility measure. When comparing the occupational mobility of workers recently hired from non-employment across different panels, Figure \ref{f:retro_ts} suggests a procyclical pattern. This figure depicts, for each panel, the occupational mobility implied by the answer to the occupational tenure question of those who were recently hired after a period of unemployment. Note that the retrospective work history questions are only asked at the beginning of a panel, so if we were to focus on recent hires we would not have a complete quarterly time series. Rather, we measure this mobility at various points over the business cycle. In the graph, we show the observations in the quarter in which the ``recent hiring" typically took place for each panel. One can observe that, with the switch from overlapping panels to sequential panels in 1996, the length between the points for which we have observations widens significantly over time.

We focus on those panels which share their survey design with at least one other panel, so that we can also compare within survey design. In particular, the 1987-1993 panels share the same retrospective questioning, but we separate the 1990-1993 panels because we can use the re-coded firm identities to reduce measurement error in these panels. However, this has only a very minor impact on the recent hires, as firm identifiers do not play a large role in extracting the implied occupational mobility from self-reported occupational tenure. Potentially more important is that the timing of the retrospective question changes from wave 2 to wave 1 between the 1990-91 panels and the 1992-93 panels. This means that we observe less of the non-employment spell inside the core waves (and consequently have less monthly observations in which we could distinguish unemployment), which may have relevance for our ``hires from unemployment'' measure. As noted, from the 1996 panel onwards, the SIPP was redesigned such that we can only use the occupational tenure question to gauge occupational mobility of those who are recently hired.\footnote{The 1984 has a different design that is not shared with any other panel, and is omitted from the picture but is included in the regressions below.}

The first observation from Figure \ref{f:retro_ts} is the procyclicality of occupational mobility over the entire time series. In times of low unemployment such as 1996 but also late 1989 to early 1990, self-reported occupational mobility is high, while in times of recession (high, or at least rather rapidly increasing unemployment) as in late 1990 to early 1991 and 2001, mobility is lower than in preceding ``good times''. Second, to assuage concerns about changes in survey design, we observe that the aforementioned observation occurs precisely across panels that share the same survey design, i.e. the 1990-1991 panels, and the 1996-2001 panels.

\begin{table}[!ht]
  \caption{The cyclicality of the probability of an occupational change}\label{table_retro_4} \vspace{3mm}

  \centering
  {\footnotesize
       \resizebox{1.0\textwidth}{!}{
  \begin{tabular}{lllllllll} \hline
    \multicolumn{9}{c}{\textbf{Panel A: Recent Hires from Nonemployment (1984-2001)}} \\ \hline
        & (i)   & (ii)  & (iii) & (iv)  & (v)   & (vi)  & (vii) & (viii) \\
          &   occ.move    & occ.move      &  occ.move     & occ.move      &  occ.move      & occ.move$^\star$  & occ.move$^\star$  & occ.move$^\star$ \\\hline \hline
    no obs. & 1546  & 1546  & 1546  & 1546  & 1546  & 1507  & 1507  & 1507 \\ \hline
    HP-filt.log(U) & -0.288*** & -0.315*** & -0.291** & -0.323*** & -0.587*** & -0.310*** & -0.383*** & -0.648** \\ \vspace{1mm}
    (s.e.) & (0.103) & (0.085) & (0.112) & (0.098) & (0.209) & (0.108) & (0.101) & (0.215) \\ \hline
    panel FE v1 &       &   X   &       &   X   &       &       &   X   &     \\
    panel FE v2 &       &       &       &       &   X   &       &       &   X \\
    demog. ctrls &       &       &   X   &   X   &   X   &       &   X   &   X \\ \hline
          &       &       &       &       &       &       &       &  \\ \hline
    \multicolumn{9}{c}{\textbf{Panel B: Recent Hires after \emph{Unemployment} (1984-2001)}} \\  \hline
          & (ix)  & (x)   & (xi)  & (xii) & (xiii) & (xiv) & (xv)  & (xvi) \\
     &   occ.move    & occ.move      &  occ.move     & occ.move      &  occ.move      & occ.move$^\star$  & occ.move$^\star$  & occ.move$^\star$ \\\hline \hline
    no obs. & 689   & 689   & 689   & 689   & 689 & 676   & 676   & 676 \\ \hline
    HP-filt.log(U) & -0.360* & -0.270 & -0.342* & -0.246 & -0.765*** & -0.418** & -0.325 & -0.828*** \\ \vspace{1mm}
    (s.e.) & (0.184) & (0.164) & (0.189) & (0.183) & (0.176) & (0.184) & (0.196) & (0.188) \\ \hline
    panel FE v1 &       &   X   &       &   X   &       &       &   X   &     \\
    panel FE v2 &       &       &       &       &   X   &       &       &   X \\
    demog. ctrls &       &       &   X   &   X   &   X   &       &   X   &   X \\ \hline
    \multicolumn{9}{p{0.9\textwidth}}{\footnotesize{Levels of significance: $^{*} p<0.1$, $^{**} p<0.05$, $^{***} p<0.01$. Dependent variable occ.move$^\star$ is our alternative mobility indicator that excludes cases where occupational between 1-4 months longer than current job tenure, while previous job tenure is larger than 12 months. All regressions include a linear time trend. Demographic controls are a quartic in age, a gender dummy, race and education dummies. \textit{panel FE v1} has dummies for the following groups of panels, \{1984\}, \{1987, 1988\}, \{1990, 1991, 1992, 1993 \}, \{1996, 2001\}; \textit{panel FE v2} for \{1984\}, \{1987, 1988\}, \{1990, 1991\}, \{1992, 1993 \}, \{1996, 2001\} Observation weighted by person weights within panel, by number of total observations per panel across panels. Standard errors clustered by quarter.}}
  \end{tabular}}
    }
\end{table}

We confirm the procyclicality of occupational mobility using regression analysis. Table \ref{table_retro_4} shows the estimates of a linear probability model, were the dependent variable again takes the value of one if the worker reported a new line of work and zero otherwise. Our baseline is again occupational mobility as inferred from the occupational tenure question, where all those who have an occupational tenure higher than the current job tenure (allowing for a month of ambiguity) are considered stayers. In our alternative measure (denoted with $\star$, in columns (vi-viii) and (xiv-xvi)), we drop those observations who were coded as stayers but reported an occupational tenure within four months of the current job tenure.

When only controlling for a linear time trend, in columns (i) and (ix), we observe that times of high unemployment, as captured by positive deviations from HP-filtered log unemployment trend, are times of lower self-reported occupational mobility. This is also observed for the alternative measure of self-reported occupational mobility, in columns (vi) and (xiv). In panel A, we take all recent hires, while in panel B we consider all recent hires for whom we observed a period of unemployment. Naturally, the number of observations in panel B is lower, which makes our inference harder. Nevertheless, we also observe a procyclical response of self-reported occupational mobility, statistically significant at the 10\%, having clustered standard errors at the quarter level. Restricting stayers to have an occupational tenure more than four months longer than current job tenure leads to a slightly stronger observed procyclical responsiveness.

As noted above, the SIPP's survey design changed over time, which may affect our results. To address this issue, we consider two variants of dummies for sets of panels that share the same survey design. As noted above, the 1990 to 1993 panel share the same survey design apart from a change of the timing of the retrospective question, from wave 2 to wave 1. This can be relevant for observing unemployment during the preceding non-employment spell. Therefore, in the first version (v1), we add four dummies for grouped panels, one for the 1984 panel, one for both 1987 and 1988 panel, one for the 1990-1993 panels, and one for the 1996-2001 panels. In the second version (v2), we add two separate dummies for the 1990-1991 panels and the 1992-1993 panels.

We observe that the responsiveness to the unemployment is much stronger in (v2). This responsiveness is driven by the variation within pairs of two panels, which puts a lot of weight on the behavior of the 1990 vs 1991 panel (and also on the 1996 vs 2001) panel, and less on the lower-frequency (but still business-cycle) behavior of the time series of observations in the subsequent recovery of the 1991 recession \emph{as measured relative} to the pre-recession 1990 panel and full-recession 1991 panel. While these responses are statistically significant in both panels, we find it conservative to prefer \emph{v1} and consider results in these two panels on a similar base to those in 1992 and 1993 panels, and estimate the responsiveness of mobility also taking into account that response to unemployment in the recovery of the 1991 recession.

Overall, controlling for grouped panel effects (v1) does not change the estimated empirical responses much relative to e.g. columns (i) and (ix), while grouped panel effect (v1) leads to stronger procyclical responses. Comparing the estimates in panel B to panel A we do not observe meaningful differences in the point estimates, even though the underlying samples are smaller, with this naturally affecting the precision of our inference. Controlling for demographic characteristics does not change the empirical responsiveness to cyclical unemployment, fully in line with the results of the occupational code-based analysis. Taken together, it appears clear that occupational mobility of recent hires from non-employment, inferred from retrospective questions on occupational tenure and job history, is procyclical. For those hired after unemployment, the evidence points in the same direction, though tends to be statistically weaker as a result of a low numbers.

\begin{table}[!ht]
  \caption{The cyclicality of the probability of an occupational change}\label{table_retro_5} \vspace{3mm}

  \centering
  {\footnotesize
       \resizebox{1.0\textwidth}{!}{
  \begin{tabular}{lllllllll} \hline
    \multicolumn{9}{c}{\textbf{Panel A: Recent Hires from Nonemployment (1984-2001)}} \\ \hline
        & (i)   & (ii)  & (iii) & (iv)  & (v)   & (vi)  & (vii) & (viii) \\
          &   occ.move    & occ.move      &  occ.move     & occ.move      &  occ.move      & occ.move$^\star$  & occ.move$^\star$  & occ.move$^\star$ \\\hline \hline
    no obs. & 1546  & 1546  & 1546  & 1546  & 1546  & 1507  & 1507  & 1507 \\ \hline
    HP-filt.log(U) & -0.270** & -0.304*** & -0.271** & -0.311*** & -0.583*** & -0.291** & -0.370*** & -0.660*** \\
    (s.e.) & (0.104) & (0.082) & (0.113) & (0.095) & (0.201) & (0.108) & (0.100) & (0.215) \\
    NE duration & 0.006** & 0.006** & 0.007** & 0.007** & 0.007** & 0.006** & 0.007** & 0.007** \\ \vspace{1mm}
    (s.e.) & (0.003) & (0.003) & (0.003) & (0.003) & (0.003) & (0.003) & (0.003) & (0.003) \\ \hline
    panel FE v1 &       &   X   &       &   X   &       &       &   X   &     \\
    panel FE v2 &       &       &       &       &   X   &       &       &   X \\
    demog. ctrls &       &       &   X   &   X   &   X   &       &   X   &   X \\ \hline
          &       &       &       &       &       &       &       &  \\ \hline
    \multicolumn{9}{c}{\textbf{Panel B: Recent Hires after \emph{Unemployment} (1984-2001)}} \\  \hline
          & (ix)  & (x)   & (xi)  & (xii) & (xiii) & (xiv) & (xv)  & (xvi) \\
     &   occ.move    & occ.move      &  occ.move     & occ.move      &  occ.move      & occ.move$^\star$  & occ.move$^\star$  & occ.move$^\star$ \\\hline \hline
    no obs. & 689   & 689   & 689   & 689   & 689 & 676   & 676   & 676 \\ \hline
  HP-filt.log(U) & -0.349* & -0.271 & -0.333* & -0.250 & -0.877*** & -0.405** & -0.327 & -0.890*** \\
    (s.e.) & (0.187) & (0.175) & (0.189) & (0.192) & (0.175) & (0.188) & (0.206) & (0.188) \\
    NE duration & 0.013** & 0.013** & 0.014** & 0.013** & 0.015** & 0.014** & 0.013** & 0.014** \\ \vspace{1mm}
    (s.e.) & (0.006) & (0.006) & (0.006) & (0.006) & (0.006) & (0.006) & (0.006) & (0.006) \\ \hline
    panel FE v1 &       &   X   &       &   X   &       &       &   X   &     \\
    panel FE v2 &       &       &       &       &   X   &       &       &   X \\
    demog. ctrls &       &       &   X   &   X   &   X   &       &   X   &   X \\ \hline
    \multicolumn{9}{p{0.9\textwidth}}{\footnotesize{Levels of significance: $^{*} p<0.1$, $^{**} p<0.05$, $^{***} p<0.01$. Dependent variable occ.move$^\star$ is our alternative mobility indicator that excludes cases where occupational between 1-4 months longer than current job tenure, while previous job tenure is larger than 12 months. All regressions include a linear time trend. Demographic controls are a quartic in age, a gender dummy, race and education dummies. \textit{panel FE v1} has dummies for the following groups of panels, \{1984\}, \{1987, 1988\}, \{1990, 1991, 1992, 1993 \}, \{1996, 2001\}; \textit{panel FE v2} for \{1984\}, \{1987, 1988\}, \{1990, 1991\}, \{1992, 1993 \}, \{1996, 2001\} Observation weighted by person weights within panel, by number of total observations per panel across panels. Standard errors clustered by quarter.}}
  \end{tabular}}
    }
\end{table}

\paragraph{Cyclical Shift of the Mobility-duration profile} Finally, we investigate whether the self-reported occupational mobility profile with non-employment duration shifts down with recessions and whether on average exhibits a similar slope as in the pooled cross-sectional sample. Table \ref{table_retro_5} shows that this is indeed the case. We observe that the modest positive slope on the non-employment duration is preserved when including the HP-filtered log of the unemployment rate. Controlling for non-employment duration largely leaves the empirical cyclical responsiveness unaffected. This result again mimics the result for the code-based mobility measures as they change with duration and the cycle.

\subsection{Retrospective Self-reported Occupational Mobility -- Conclusion} In this section, we have investigate occupational mobility after non-employment spells using very different data than in the main text. None of the occupational information used in the occupational code-based measures of the main text has been used in the measures in this section. Further, in the code-based measures of the main text, census coders are the judge of occupational change, while in the retrospective occupational tenure, this judgement is made by the interview subject. It is then comforting to observe that both measures line up well. First, according to workers, they are starting in a new line of work after nonemployment in about 40\% cases. Second, with a longer nonemployment duration comes an increased tendency to change one's line of work. However, this tendency is modest, around 6 p.p. higher after 6 months of unemployment, fully consistent with the occupational-code results in the main text and in Section 1 of this appendix. Third, when cyclical unemployment is high, self-reported occupational mobility tends to be cyclically low. Fourth, controlling for the business cycle, we still observe a modest increase of self-reported mobility with nonemployment duration, again keeping with the pattern of code-based occupational mobility documented in the main text.

Once again our estimates show that the probability of an occupational change for the non-employed is procyclical. This holds true for both samples and with or without controlling for the duration of the non-employment spell. In particular, for the sample of recent hires that were unemployed for at least one month before re-employment, we still observe the moderate increase in the probability of an occupational mobility with the duration of non-employment. Likewise, controlling for destination occupations, we also find strong procyclicality in the probability of an occupational change. This suggests that the procyclicality is not driven by shifts in the occupational destinations over the business cycle. This is once again inline with the results of Section 4 of this appendix.

\section{Data Construction}

\subsection{Survey of Income and Program Participation}

The Survey of Income and Programme Participation (SIPP) is a longitudinal data set based on a representative sample of the US civilian non-institutionalized population. It is divided into multi-year panels. Each panel comprise a new sample of individuals and is subdivided into four rotation groups. Individuals in a given rotation group are interviewed every four months such that information for each rotation group is collected for each month. At each interview individuals are asked, among other things, about their employment status as well as their occupations and industrial sectors during employment in the last four months.\footnote{See http://www.census.gov/sipp/ for a detailed description of the data set.}

The SIPP offers a high frequency interview schedule and aims explicitly at collecting information on worker turnover. Further, its panel dimension allows us to follow workers over time and construct uninterrupted spells of unemployment (or non-employment) that started with an employment to unemployment transitions and ended in a transition to employment. Its panel dimension also allows us to analyse these workers' occupational mobility patterns conditional on unemployment (or non-employment) duration and their post occupational mobility outcomes as outlined in Section 2 in the main text.

\paragraph{Survey design and use of data.} We consider the period 1983 - 2013. To cover this period we use 13 panels in total: the 1984-1988, 1990-1993, 1996, 2001, 2004 and 2008 panels. For the 1984-1988 and 1990-1993 panels we have used the Full Panel files as the basic data sets, but appended the monthly weights obtained from the individual waves (sometimes referred to as core wave data). Until the 1993 panel we use the occupational information from the core waves. We do this for two reasons: (i) the full panel files do not always have an imputation flag for occupations; and (ii) between the 1990 and 1993 panels firm identities were retrospectively recoded, based on core wave firm identifiers. For our study it is important to be clear to which firm the occupation belongs. We exclude the 1989 SIPP panel because the US Census Bureau does not provide the Full Panel file for the 1989 data set and this panel was discontinued after only three waves (12 months). Since we want to be conservative regarding censoring, we opted for not using this data set. This is at a minor cost as the 1988 panel covers up to September 1989 and the 1990 panel collects data as from October 1989. For the 1996, 2001, 2004 and 2008 panels there is no longer a Full Panel file nor a need for one. One can simply append the individual wave information using the individual identifier ``lgtkey'' and merge in the person weights of those workers for whom we have information from the entire panel (or an entire year). In this case, the job identifier information is also clearly specified. Two important differences between the post and pre-1996 panels are worth noting. The pre-1996 panels have an overlapping structure and a smaller sample size. Starting with the 1996 panel the sample size of each panel doubled in size and the overlapping structure was dropped. We have constructed our pre-1996 indicators by obtaining the average value of the indicators obtained from each of the overlapping panels.

The SIPP's sample design implies that in \emph{all} panels the first and last three months have less than four rotation groups and hence a smaller sample size. For this reason, in our time series analysis, we only consider months that have information for all four rotation groups. For statistics for which the distribution of unemployment duration matters we require that workers have at least 14 months of labor market history at the moment of re-employment in their corresponding SIPP panel. If necessary (and discussed in detail below), we impose further restrictions to deal with censored spells in order to generate a representative distribution of unemployment spells for at least up to one year. This restriction addresses that e.g. short completed unemployment spells typically have lower mobility, while in the first waves of a panel spells started and completed within the panel are necessarily of short duration. This is even more important when constructing the job finding rates, especially when we want to focus on the job finding rates in completed spells for which we know the occupational mobility outcomes. For the cumulative survival profile in unemployment, we consider all spells that at their start have at least 32 months of subsequent continuous presence in the sample and restrict these observations to be in the first 4 waves of the panel. For job finding rates in the time series, which are based on incomplete spells, we require that all workers whose job finding rate is measured at duration $x$ remain continuously in the sample for at least 19-$x$ more months.

The data also shows the presence of seams effects between waves, where transitions are more likely to occur at a seam (i.e. between waves, and therefore at 4,8, 12\ldots months) than based on other characteristics, e.g. duration. When we consider time series and given the above restrictions, there is always one rotation at the seam in every month we consider which effectively smoothes out the clustering at the seam. In the case of the duration statistics for which the seam effect matters, we either consider observations in 4 months bins (e.g. survival at 4, 8, 12, 16 months of unemployment) or use the standard methods to reducing the seam bias by smoothing out the survival rates or considering ``cumulative'' mobility rates with duration (see e.g. for the mobility-duration profile reported in Section 2.2 of the main text).

We use the person weights per wave (``wpfinwgt'', and equivalent), but normalize these such that the average weight within a panel is equal to one. This is done because the size of panels is not constant, and we do not want to weigh panels with less observations more heavily as within a wave of a panel ``wpfinwgt'' adds up to population totals and thus is higher on average when sample size is smaller. We think of our normalization as a reasonably agnostic approach that keeps the relative weights within a panel intact, but also takes into account the number of available observations.

\paragraph{Sample selection and labor market status.} For the 1984-2008 panels, we consider all workers between 18 and 65 years of age who are not in self-employment or in the armed forces nor in the agricultural occupations.\footnote{As agricultural occupations could be miscoded nonagricultural occupations and vice versa, in our code-error corrected measure, we take agricultural workers according to reported occupational codes into account, apply our correction method, and remove agricultural workers \emph{after correction} from our sample and associated statistics.} We measure an individual's monthly labor force status in the SIPP using two sources of information. The first one relies on the labor force status reported at the second week of each month. The second relies on the monthly employment status recode. Using these two sources (and using the SIPP 2001 wording as an example) we consider a worker to be employed during a month if the individual reported in the second week of that month that he/she was ``with job/business - working'', ``with job/business - not on layoff, absent without pay'' or ``with job/business - on layoff, absent without pay''. The category ``with a job'' (Census 2008) is assigned if the person either (a) worked as paid employees (or worked in their own business or profession or on their own farm or worked without pay in a family business or farm) or (b) \emph{``were \textbf{temporarily} [emphasis added] absent from work either with or without pay.''} Thus, the employment status recode category ``with job/business - on layoff, absent without pay'' appears to capture temporary layoffs. Note that as a result our definition of employment differs from the CPS definition of employment. As the SIPP documentation points out: `` ``With a job'' includes those who were temporarily absent from a job because of layoff and those waiting to begin a new job In 30 days; in the CPS these persons are not considered employed.''

We consider a worker to be \emph{employed} if the individual reported in the monthly employment status recode variable that he/she was ``with a job entire month, worked all weeks", but also when ``with a job all month, absent from work without pay 1+ weeks, absences not due to layoff'', or ``with a job all month, absent from work without pay 1+ weeks, absences due to layoff''. If workers have spent part of the month in employment and part of the month in unemployment, workers are nonemployed only if they are nonemployed in week 2 \emph{and} have been nonemployed for at least four weeks in total. That is, those who have less than a month of nonemployment in week 2 are still counted as employed. If the worker is ``no job/business - looking for work or on layoff'' during one of the weeks in nonemployment (i.e. in the "no job/business") state, we consider the worker to be unemployed. We have chosen this classification, because we want entry into unemployment to capture the serious weakening of the link with the previous firm of employment, rather than to be a definite period of nonproduction after which the worker would return to the previous employer. The restriction of nonemployment for at least four weeks is meant to further limit the role of short-term absences from the same firm and temporary layoffs. This is motivated by the analysis of Fujita and Moscarini (2017), who document that many workers with very short unemployment spells return to their previous employer. We want to focus on those unemployed who at least \emph{consider} employment in other firms and possibly other occupations.

\begin{figure}[th!]
\centering
\subfloat[Sample: quarters after seam-correction SIPP]{\label{sf:occdistr22occ} \includegraphics[width=0.5\textwidth] {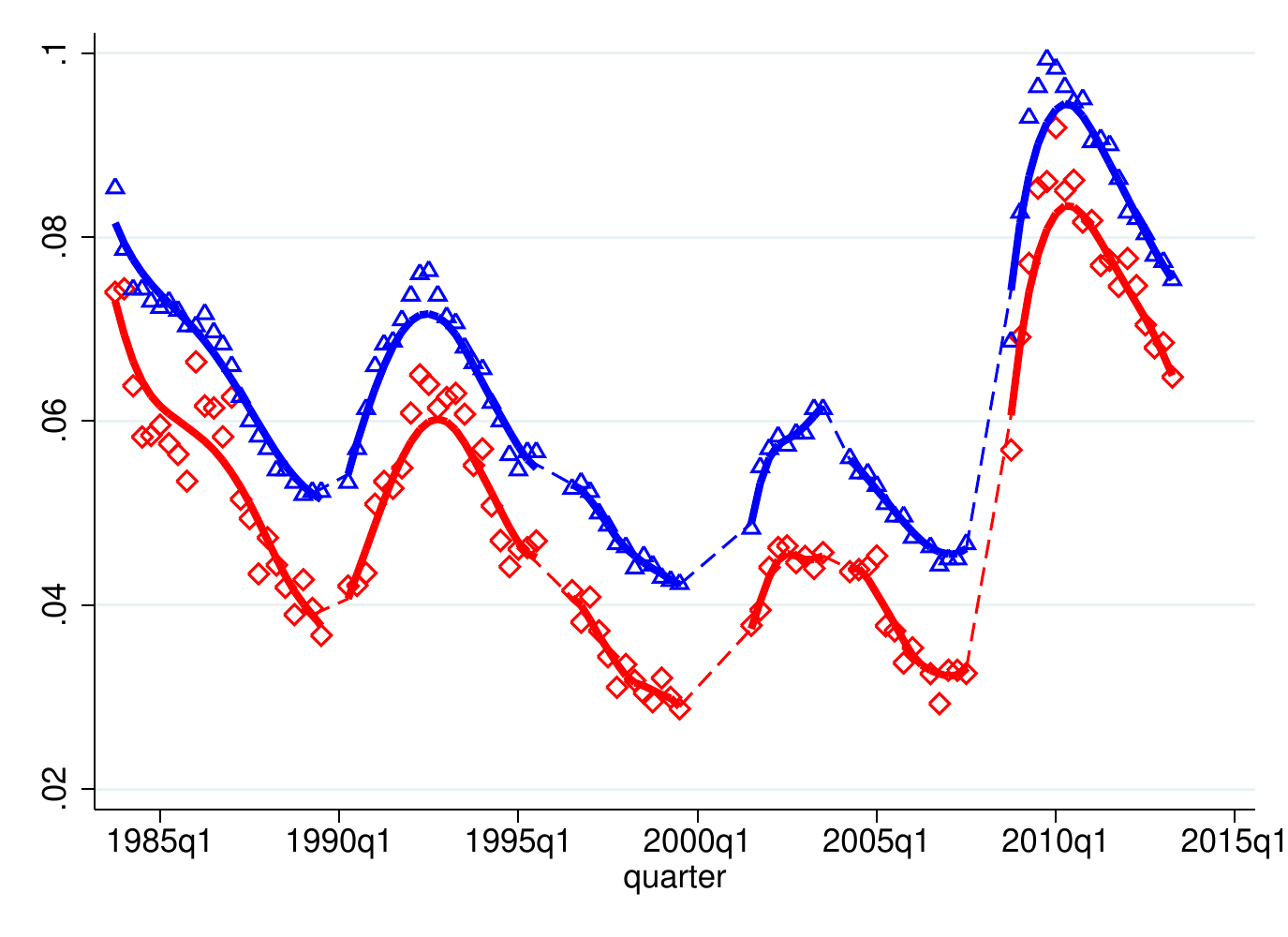}}
\subfloat[Sample: quarters with workers in sample >14months ]{\label{sf:22occ-nun} \includegraphics[width=0.5\textwidth] {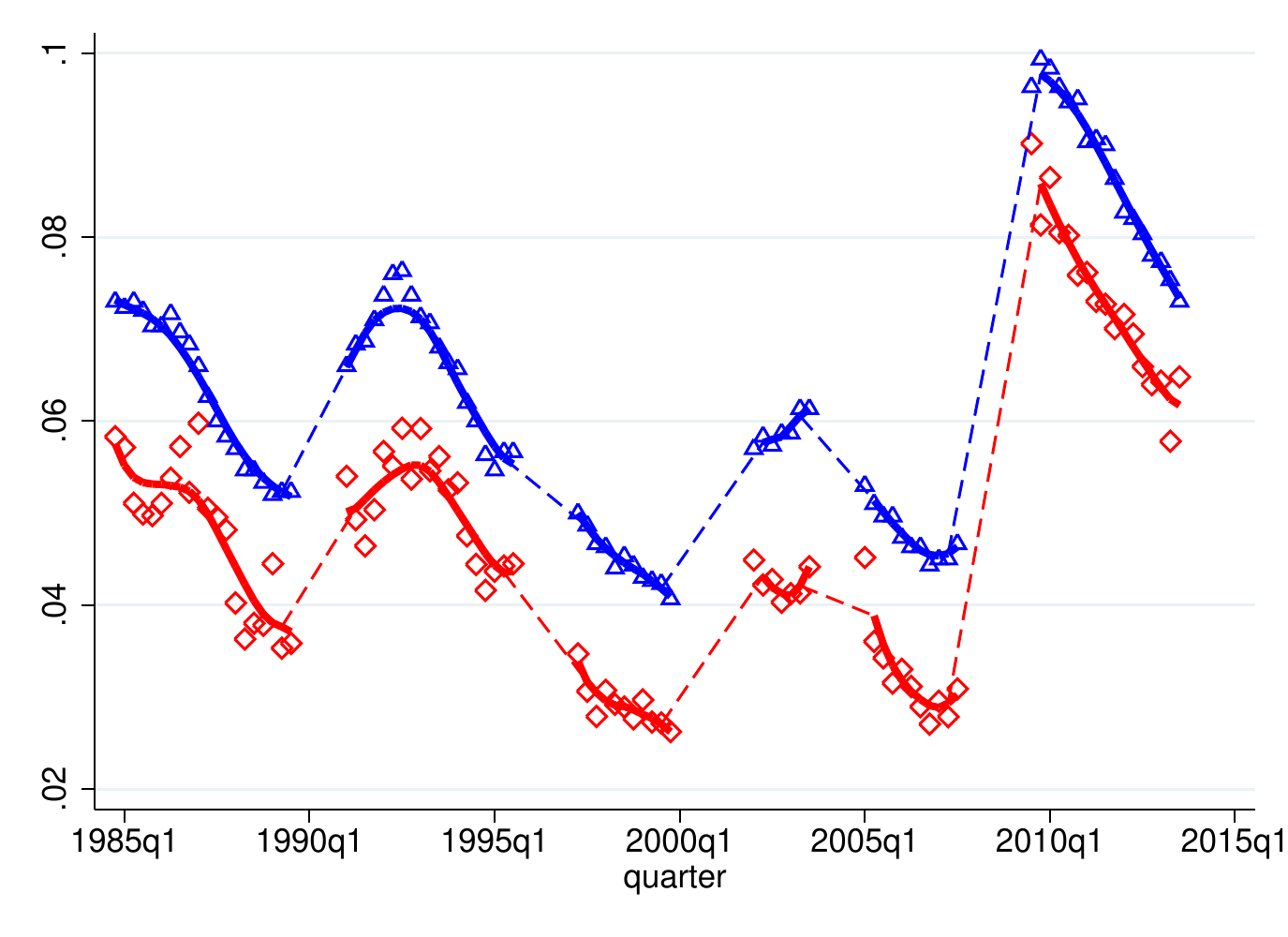}}
\caption{Unemployment Rate SIPP (red, our measures) and BLS-CPS (blue, official)}
\label{f:u_comparison}
\end{figure}

With these restrictions, our measured unemployment rate is somewhat lower than the official unemployment rate. In Figure \ref{f:u_comparison}, we plot the unemployment rate in our SIPP sample, constructed according to our definition, next to the standard (CPS-based) unemployment rate from the BLS. The left panel displays the unemployment rate in the SIPP when excluding those quarters in which panels phase in or out. In this case all quarters considered are symmetric with respect to seams between waves: each month there is at least one rotation group which switches interview wave. The right panel considers only observations that have been in the sample for at least 14 months. This necessarily means bigger gaps between panels, but addresses issues of left-censoring when considering complete unemployment spells for which we can observe the occupational mobility. This is important when e.g. considering the distribution of unemployment durations of completed spells as at the beginning of the panel the only completed unemployment spells are necessarily those of short duration, which have a lower occupational mobility rate. Not taking this into account would lead to mistaken time series patterns of occupational mobility of the unemployed.

Both panels of Figure \ref{f:u_comparison} show that our restrictions lower the unemployment rate by about 1.5pp. This level effect is nearly uniform over time and over the cycle. That is, the time-series evolution of our measure of unemployment follows BLS unemployment closely, while the level effect is nearly unaffected by the business cycle. If anything, our measure responds slightly stronger to the business cycle (a 1pp rise in the official rate raises our unemployment measure by 1.04pp, this is small but statistically significant). The correlation between our unemployment measure and the BLS measure (for those months for which we have our SIPP measure of unemployment) is 98-99\% for both unemployment series in Figure \ref{f:u_comparison}. This implies that the cyclical pattern in unemployment, to a very large extent, originates in the set of unemployed we consider in this paper (see also Hornstein, 2013, and Ahn and Hamilton, 2019).

We further check that these properties also hold when we restrict unemployment to those who \emph{start} and \emph{end} unemployment within the SIPP sample, taking care to address both the left- and right-censoring involved in this measure. The correlation of this unemployment rate (smoothed, because of the smaller set of observations) with the BLS unemployment rate is over 94\% when considering nonemployment spells that include months of unemployment, and 93\% when considering pure unemployment spells, where workers are unemployed in every month of nonemployment.

\paragraph{Assigning ``source''/``destination'' - occupations to unemployed workers.} The SIPP collects information on a maximum of two jobs an individual might hold simultaneously. For each of these jobs we have information on, among other things, hours worked, total earnings, 3-digit occupation and 3-digit industry codes. We drop all observations with imputed occupations (and industries). If the individual held two jobs simultaneously, we consider the main job as the one in which the worker spent more hours. We break a possible tie in hours by using total earnings. The job with the highest total earnings will then be considered the main job. In most cases individuals report to work in one job at any given moment. In the vast majority of cases in which individuals report two jobs, the hours worked are sufficient to identify the main job. Once the main job is identified, the worker is assigned the corresponding two, three or four digit occupation.\footnote{For the 1990-1993 panels we correct the job identifier variable following the procedure suggested by Stinson (2003).}

Each unemployment spell that is started and finished inside the panel can be assigned a ``source''-occupation (main occupation right before the start of the unemployment spell), and a ``destination''-occupation (main occupation right after becoming employed again). If the occupation code is missing just before the unemployment spell (e.g. due to imputation) and an occupation code is reported in a previous wave, while employment is continuous from the time that the occupation was reported until the start of the unemployment spell under consideration, we carry the latter occupation forward as source occupation. A worker is an occupation mover if source and destination occupations do not coincide. We thus conservatively count the following situation also as an occupational stay: the worker is simultaneously employed in two firms at the moment the worker becomes unemployed, and finds a job afterwards in an occupation that matches the occupation in one of the two previous jobs, even when it matches the job with less hours. The effect on the occupational mobility statistics of counting as occupational stays the unemployment spells with two simultaneous jobs at either side is small.

We construct the occupational mobility statistics from transitions of the form: at least a month in employment (with a non-imputed occupational code), followed by an unemployment spell which has a duration of at least a month, followed by at least a month in employment (with a non-imputed occupational code). We label these transitions as EUE transitions. We also consider transitions of the form: at least a month in employment (with a non-imputed occupational code), followed by a non-employment spell which has a duration of at least a month and involved at least one month of unemployment. We call these E-NUN-E transitions, or NUN-spells of nonemployment. Further convexifying the space between EUE and E-NUN-E, we also consider spells that started with a EU transition, i.e. employment directly followed by unemployment (though later the worker can report to stop looking for work), and those that ended with UE transition. We label these transitions as E-UN-E, E-NU-E, and if both restrictions apply, E-UNU-E transitions. We also tried other versions of the latter in which the full jobless spell was non-employed (ENE).

\paragraph{Occupational Classifications.} The SIPP uses the Census of Population Occupational System, which relates closely to the Standard Occupational Code (SOC). The 1984-1991 panels use the 1980 Census Occupational classification, while the 1992-1996 and 2001 panels use the 1990 Census Occupational classifications. These two classifications differ only slightly between them. The 2004 and 2008 panels use the 2000 Census occupational classification, which differs more substantially from the previous classifications. We use David Dorn's recoding of the 1980 and 2000 Census Occupational Classification (Dorn, 2009, and Autor and Dorn, 2013) into the 1990 Census Occupational Classification to have a uniform coding system. In robustness exercises, we instead use the IPUMS crosswalk to map the 1980 and 1990 Census occupational system into the 2000 Census occupational system.

We aggregate the information on ``broad'' occupations  (3-digit occupations) provided by the SIPP into ``minor'' and ``major'' occupational categories.\footnote{In any of these classifications we have not included the Armed Forces. The 1980 and 1990 classifications can be found at https://www.census.gov/people/io/files/techpaper2000.pdf. The 2000 classification can be found in http://www.bls.gov/soc/socguide.htm. Additional information about these classifications can be found at http://www.census.gov/hhes/www/ioindex/faqs.html.} Measurement error in occupational codes might give rise to spurious transitions, as discussed for example in Kambourov and Manovskii (2008, 2009) and Moscarini and Thomsson (2007). We correct for coding error using the $\mathbf{\Gamma}$-correction method we propose in Supplementary Appendix A. The occupational categories of the classifications used can be found in Tables 4 and 5 of that appendix.

\paragraph{Time series construction.} We construct monthly time series for the unemployment rate, employment to unemployment transition rate (job separation rate), unemployment to employment transition rate (job finding rate), occupational mobility rates and the other measures described in the main text. Job finding rates are simply $UE_{t}/U_{t}$, the proportion of unemployed at time $t$ that moves to employment at time $t+1$. Similarly, the separation rate is $EU_{t}/E_{t}$, the proportion of employed at time $t$ that moves to unemployment at time $t+1$. We start measuring job finding and separation rates at the first month were we have information for all four rotation groups.

For construction of de-trended time series we have to address the issue of gaps (missing observations) in time series. There are a few quarters that are not covered at all by the 1984-2008 SIPP panels: 2000Q2-2000Q3, and 2008Q1. The gaps around these times can become larger, and new gaps can be created, when censoring issues cause us to drop further quarters from the analysis. We discuss this in more detail below.

To cover the missing observations we interpolate the series using the TRAMO (Time Series Regression with ARIMA Noise, Missing Observations and Outliers) procedure developed by Gomez and Maravall (1999), with interpolation of missing observations through regression (``Additive Outlier Approac'').\footnote{See also Fujita, et al. (2007) for a similar procedure using the SIPP. Tramo/Seats is a parametric, ARIMA model based method (AMB), that works in two steps. In the first step (TRAMO) the series is interpolates missing observations and deals with outliers. The second step (SEATS), among other things, decomposes the time series resulting from step 1 into e.g. a trend-cycle, an irregular, and a seasonal component. Seasonal adjustment is broadly similar to the Census Bureau's X12 procedure, and is used e.g. by Eurostat. (Hood et al. (Hood, Ashley, Finley, US Census Bureau: ``An Empirical Evaluation of the Performance of Tramo/Seats on Simulated Series''), who also argue that SEATS does better than X12-ARIMA with longer time series that have large irregular components.)} In our baseline cyclical timeseries, de-trended data series are produced with after HP-filtering the resulting (logged) time series, with smoothing parameter 1600.

Two aspects are especially important for the time series of the propensity of hires from U to start in a new occupation: (1) the code error correction, and (2) addressing censoring issues, to counter noise and bias due to shifts in the duration distributions between adjacent quarters that are orthogonal to the business cycle.

We select only observations of individuals who have been in sample for more than 14 interviews and are hired beyond wave 4. Given this, we correct for coding error at the level of quarter $\times$ (completed) spell duration. This means that, for all hires from unemployment with a given completed duration in a given quarter, we calculate the transition matrix of occupational mobility and correct it using the appropriate $\mathbf{\Gamma}$ (code-error) correction matrix. We then calculate the cumulative mobility of all hires with unemployment durations up to (and including) 12 months, and in an alternative measure, up to and including 14 months. We find little difference across these two measures, and hence report only the former measure. The same corrected observations at the level of duration$\times$quarter will subsequently be used to calculate the mobility-duration profile shift from times of low unemployment to times of high unemployment.

While time series gaps are small for many statistics, this issue is more important for the average occupational mobility of all hires from unemployment (of $\leq$ 12-14 months). The restriction to wave 5 and beyond means that, by default, we leave out those observed unemployment spells that arise from workers losing their jobs and being hired within the first 16 months of each panel. As argued above, this is to make sure that we capture the nearly full duration distribution of unemployment at each quarter considered in our analysis. This is especially important here as not doing so would generate a downward bias in the occupational mobility rates at the beginning of a panel, which is particularly strong in e.g. 2004. In addition, we take into account that, given the rotating survey design, some further quarters early in the panel do not have a seam in each month (this is also a relevant issue for the job finding and separation time series). We are also conservative on this issue and disregard those quarters as well.

We also note that censoring issues may arise at the very end of the panel, where again a quarter may contain varying proportion of monthly observations that are the final seam. We observe that in practice, some of these ``ending'' quarters have duration distributions that are very different from the duration distributions in previous quarters. As we show below, this variation is significantly larger and more abrupt than changes that appear to move with business cycle conditions. For this reason, we also exclude the last quarter of the 1988 panel and all panels from the 1993 panel onwards, with the exception of the 2001 panel. The quarters which are used for our analysis are then given in Table \ref{t:durdistr_selection}. These observations span a number of different moments during multiple business cycles and therefore the cyclical patterns can indeed be clear and significant, as is shown in Section 2.5 of the main text and this appendix.

\begin{table}
\centering
\begin{tabular}{lcc}
  \hline
  SIPP Panels & Qtr from & Qtr to (inclusive) \\ \hline \hline
  1984,85,86,87,1988 & 1985q1 & 1989q3 \\
  1990,91,92,1993 & 1991q2 & 1995q3 \\
  1996 & 1997q3 & 1999q4 \\
  2001 & 2002q2 & 2003q4 \\
  2004 & 2005q1 & 2007q3 \\
  2008  & 2009q3 & 2013q3 \\
  \hline
\end{tabular}\caption{Quarters included in time series for overall occupational mobility rate of all unemployed (with unemployment duration between 1-14 months }\label{t:durdistr_selection}
\end{table}

\begin{figure}[!ht]
\centering
\subfloat[1984-1988 SIPPs] {\includegraphics [width=0.5 \textwidth]{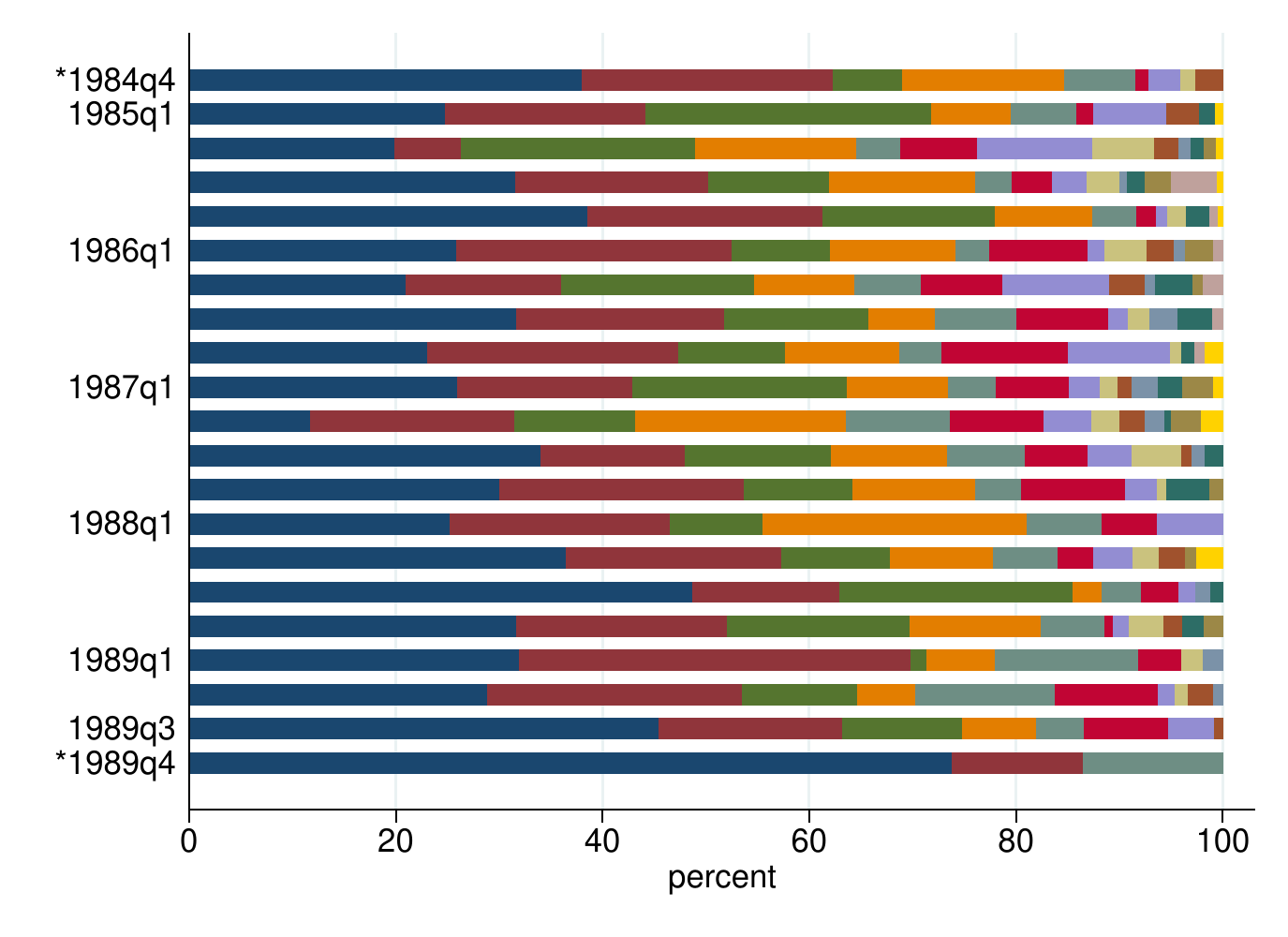}}
\subfloat[1990-1993 SIPPs] {\includegraphics [width=0.5 \textwidth]{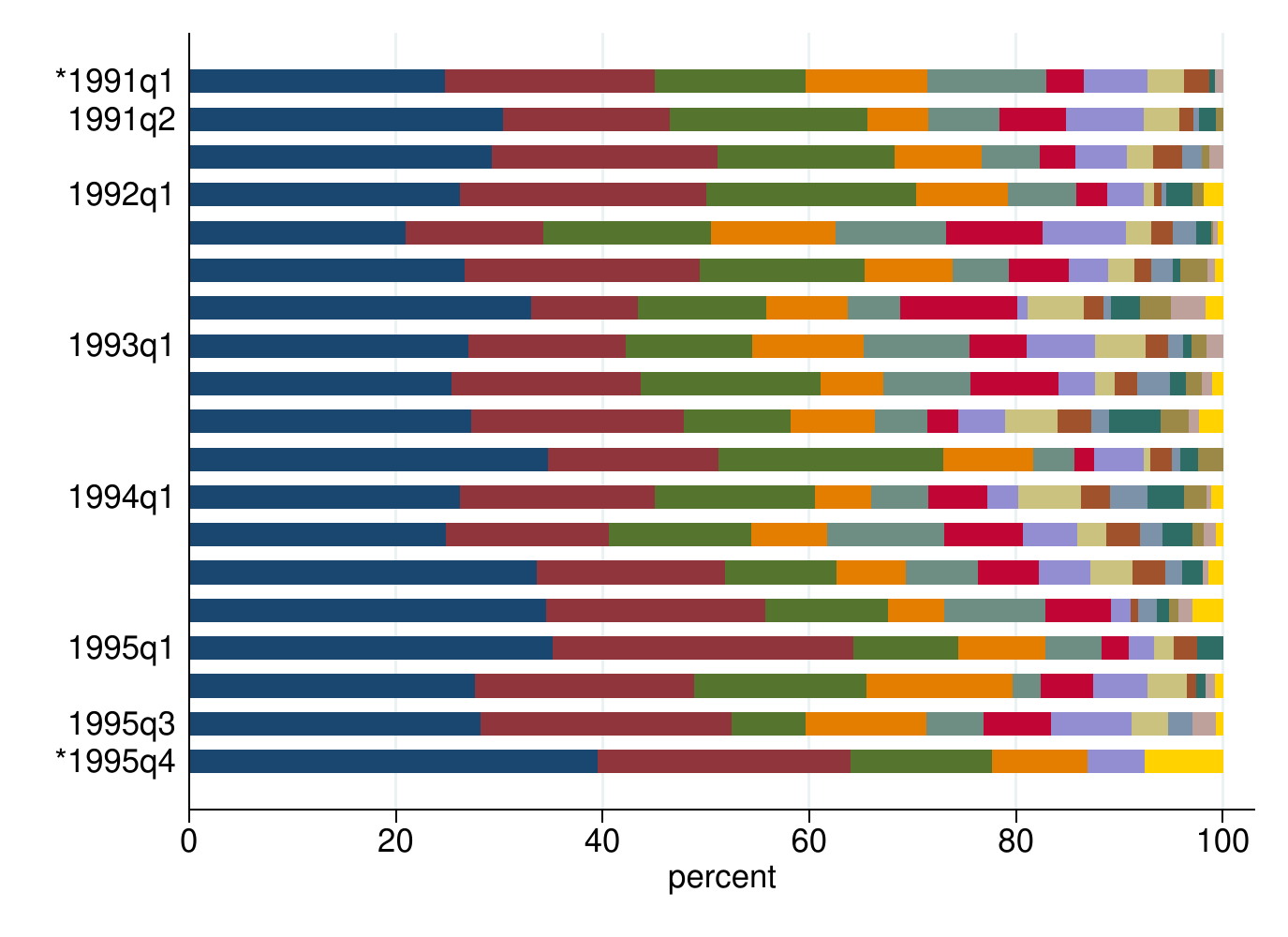}}

\subfloat[1996 SIPP] {\includegraphics [width=0.5 \textwidth]{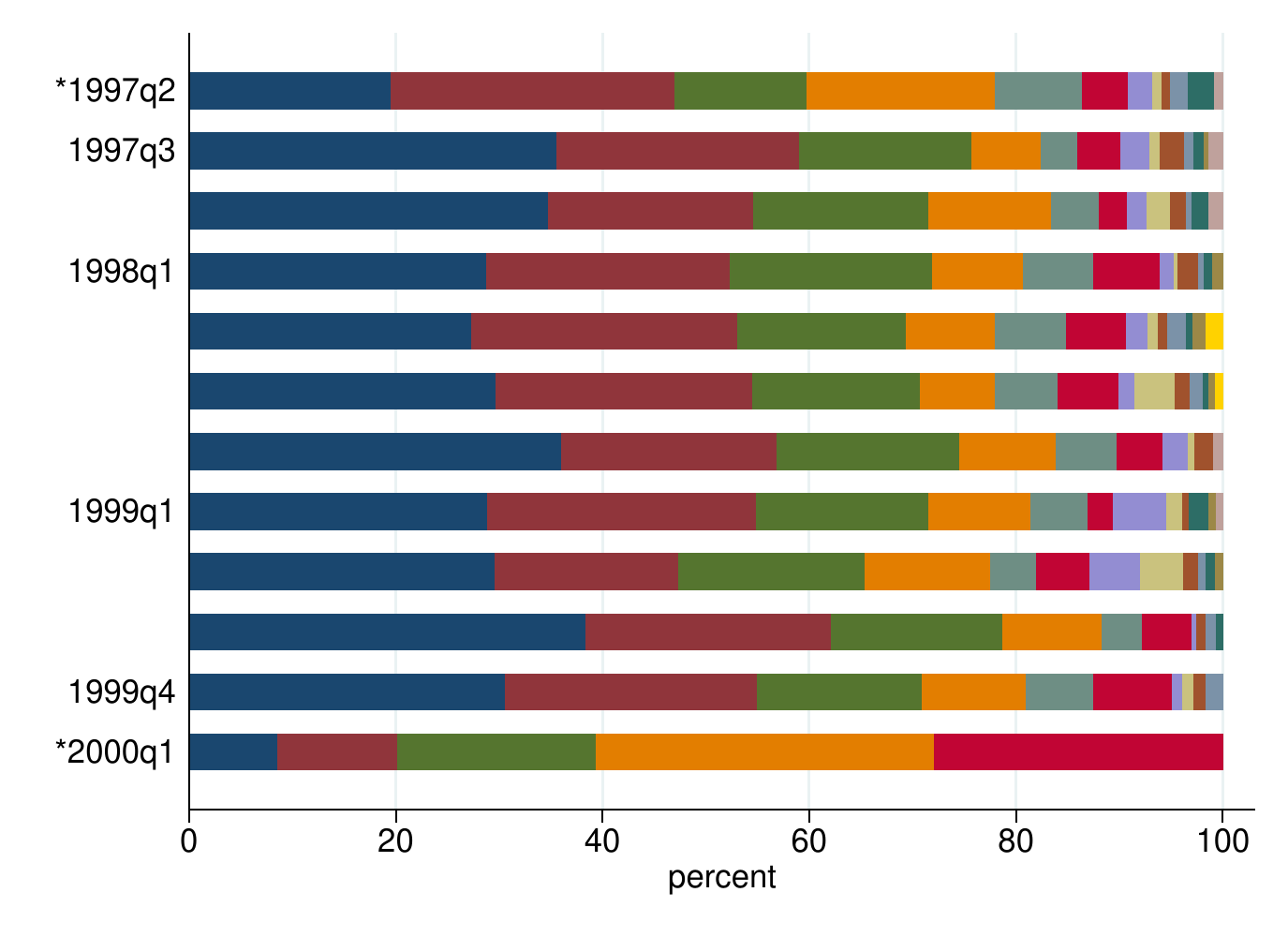}}
\subfloat[2001 SIPP] {\includegraphics [width=0.5 \textwidth]{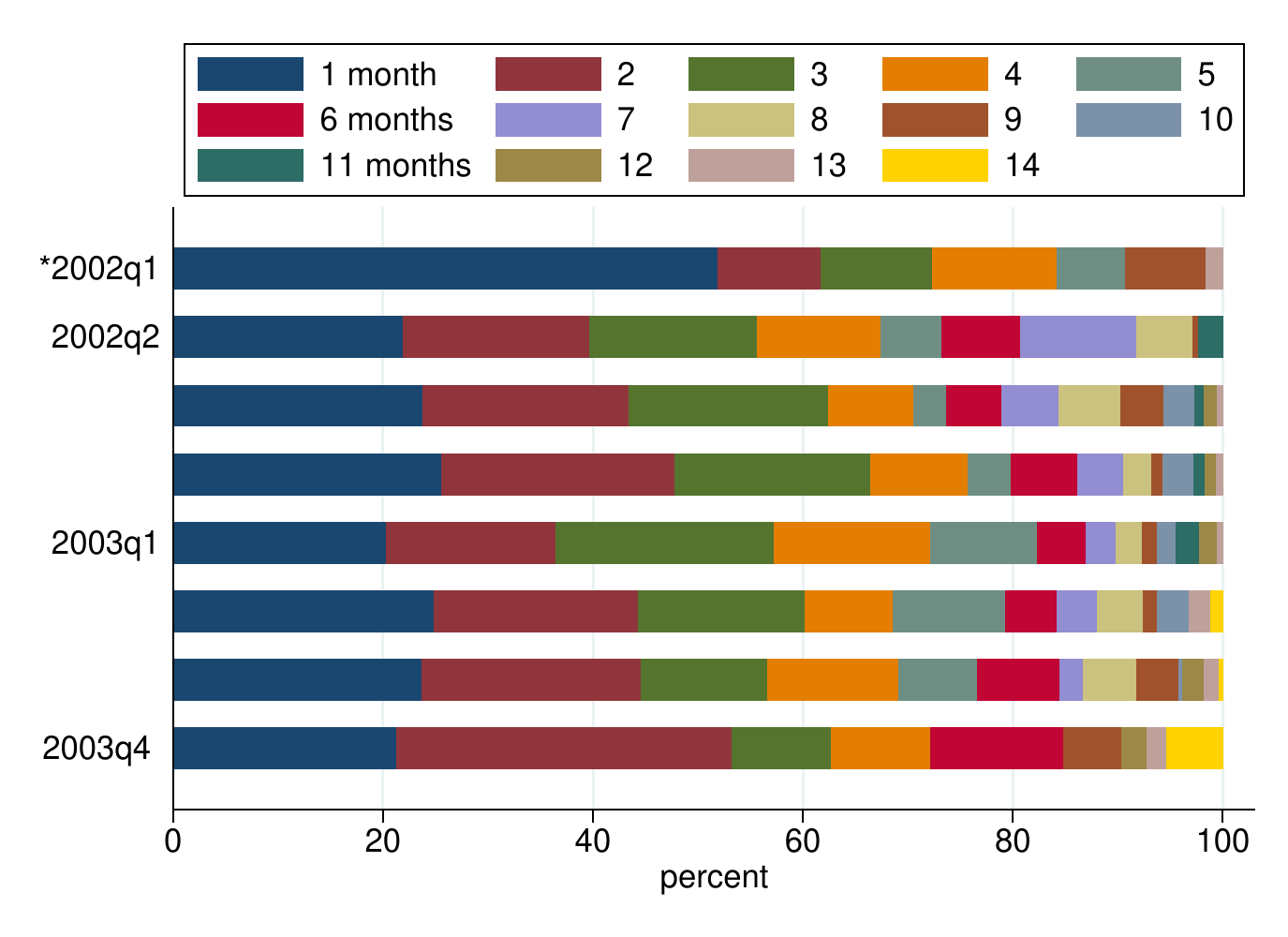}}

\subfloat[2004 SIPP] {\includegraphics [width=0.5 \textwidth]{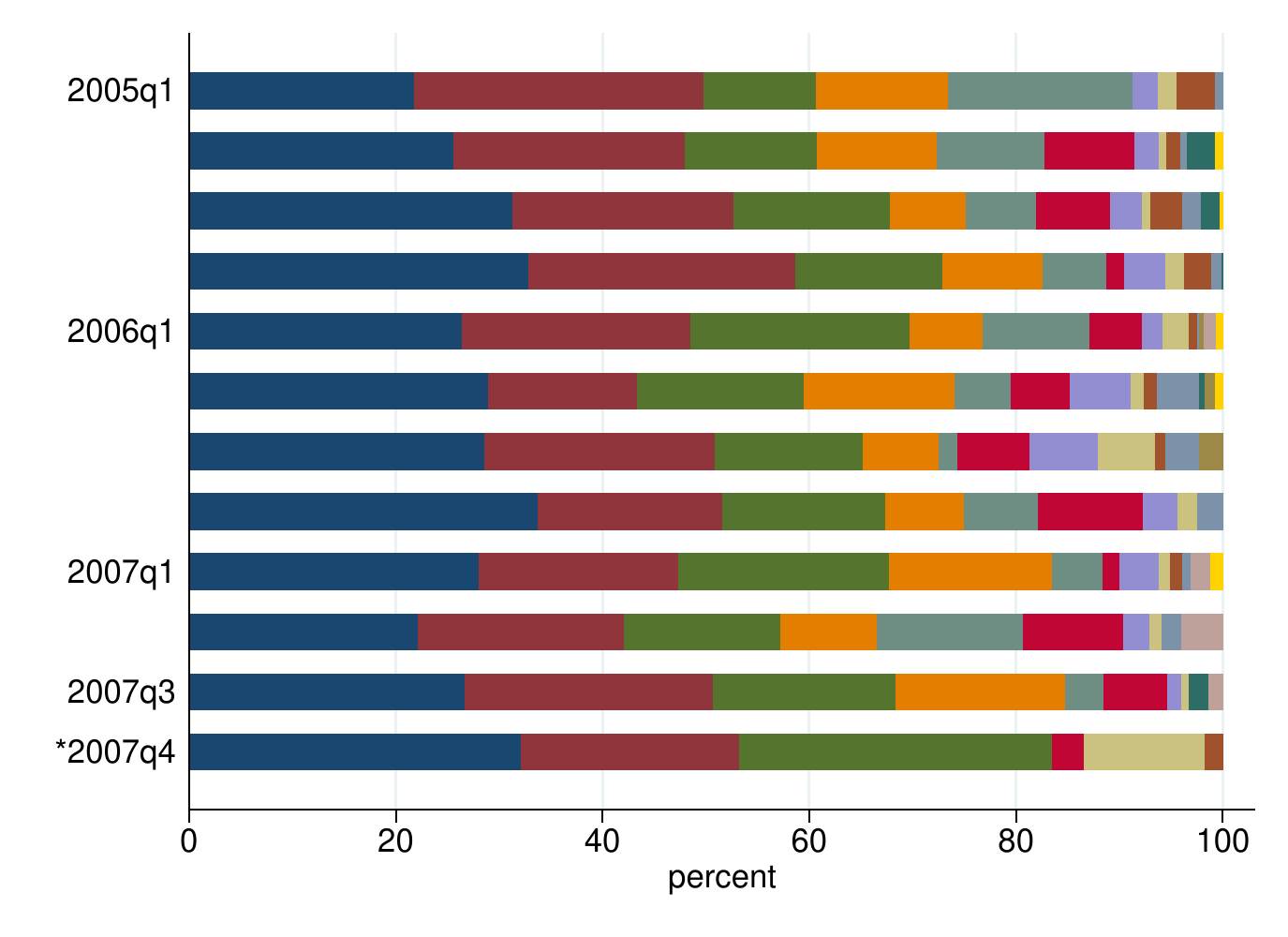}}
\subfloat[2008 SIPP] {\includegraphics [width=0.5 \textwidth]{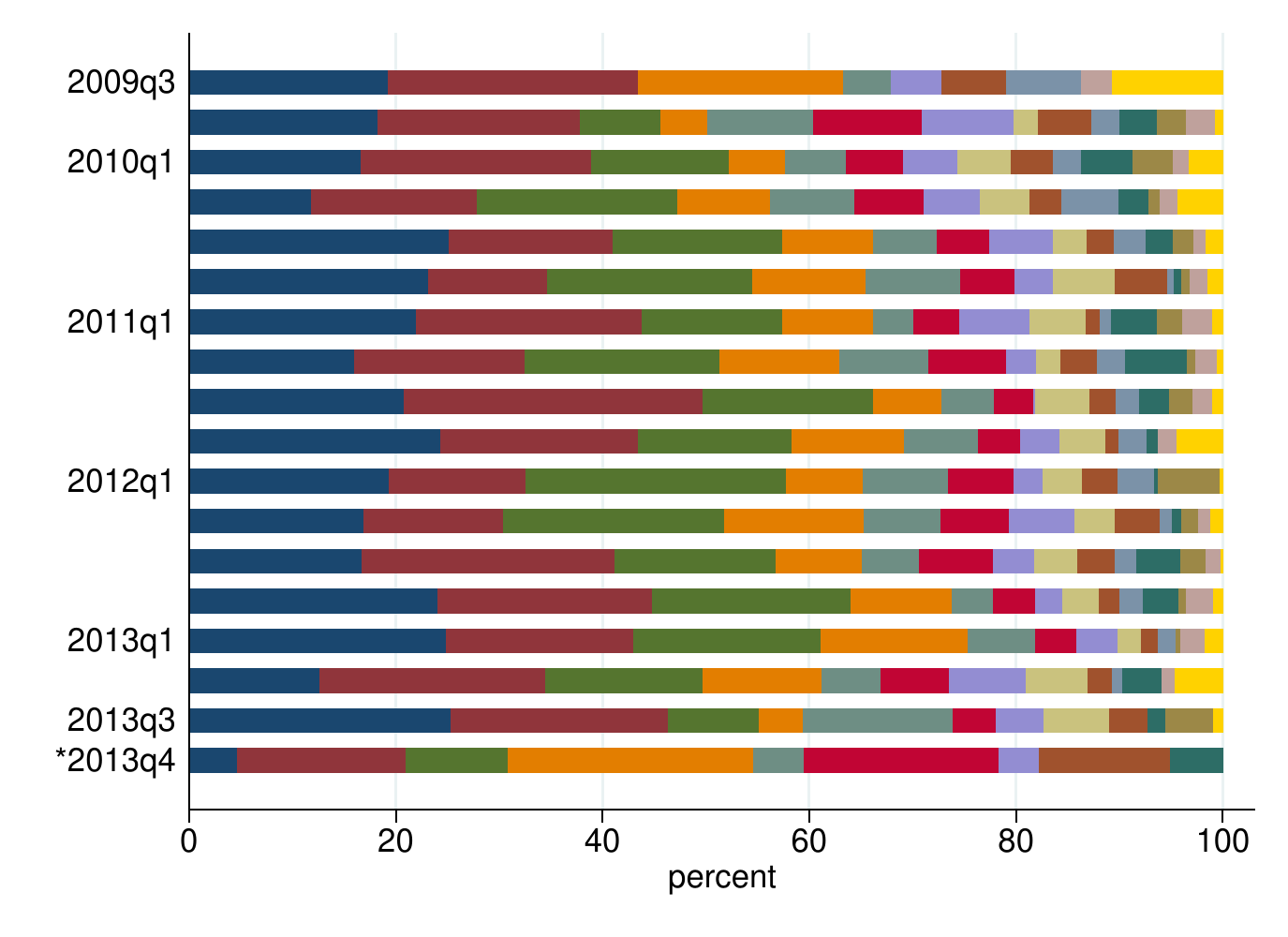}}

\caption{Completed Unemployment Duration of Hires from U, per quarter}
\label{f:durdistr_stability}
\end{figure}

To visualize the impact of censoring, Figure \ref{f:durdistr_stability} presents the duration distribution of hires from unemployment per quarter. Each horizontal bar specifies the unemployment durations of hires (from 1 month to 14 months, in order). On the y-axis are the quarters in which we have observations beyond wave. Those quarters that we exclude from our time series calculations, due to one of the reasons mentioned above, are prefaced by an $*$.\footnote{One can distinguish some cyclical patterns directly, clearest perhaps when comparing across panels, where the red bands (hires after 6 months of unemployment) in the early 1990s panels are to the left of those in the 1996 panel, and likewise for the 2001 SIPP vs the 1996 and 2004 SIPP, and the 2008 vs the 2004 SIPP, signifying more long-term unemployment in the ``recession'' 1990-1993, 2001, and 2008 SIPP panels.}

Here we can indeed observe the some of the last and first of the quarters considered are affected by these (censoring) issues and hence are excluded from our analysis. We also observe that the remaining duration distributions are relatively stable across individual quarters, with the amount of quarter-to-quarter variation dropping with the 1996 panel and thereafter. This occurs because the sample sizes per panel are larger, while the overlapping nature of panels beforehand imply that some additional noise is generated when panels are phased in and out, while other panels are ongoing. We also tried to further restrict the sample to exclude observations from these phase-in and -out, but the impact of this on time series patterns is minor.

\paragraph{Aggregate Series Productivity, Unemployment, Vacancy Series} For output per worker, we take nonfarm business output from the Major Sector Productivity and Costs section of Bureau of Labor Statistics (series id: PRS85006043), and divide this by the sum of employment of private industry wage and salary workers (not in agriculture) and employment of self-employed unincorporated workers in these sectors (series LNS12032189Q and lns12032192q, respectively) Unemployment rates in the cyclical analysis are BLS Unemployment rates (UNRATE in the St. Louis Fred data). We take the vacancy data from extended Help-Wanted Index constructed by Regis Barnichon.

\paragraph{Repeat Mobility} In our baseline repeat mobility measures, we concentrate on pure u-spells that follow pure u-spells within a SIPP panel, for workers that are at least 4 years in sample. To avoid including seasonal unemployment in our measures (which are not informative about attachment as interpreted in the model), we exclude construction and agricultural workers. We are left with 610 observations of repeat unemployment. For our NUN measures, we consider NUN-spell following NUN-spells with the same restrictions, leaving 1306 observations of repeat unemployment. We average the measures that consider the occupation of intermediate employment at the beginning and end of that spell. We correct coding errors using the procedure described in footnote 13 in the main text. Finally, to gauge occupational moving, we exclude any observations of return movers.

\subsection{Panel Survey of Income Dynamics}

Following Kambourov and Manovskii (2008) sample restrictions, we consider the 1968-1997 period as during these years the PSID interviews were carried out annually. We also consider males head of households between the ages of 23-61 years who were not self- or dual-employed and were not public sector works. This sample restriction then gives 1,643 employed individuals in the year 1968 and 2,502 employed individuals in 1997. In an alternative sample we also included women, younger workers and self or dual-employed workers with no meaningful change in our main results.

To construct the \emph{across-employer} occupational mobility rate we compute the fraction of employed workers who's occupational code differs between years $t$ and $t+1$ and have reported an employer change between these years, divided by the number of employed workers in year $t$ that have reported an employer change between years $t$ and $t+1$. As Kambourov and Manovskii (2009), to identify employer changes in the PSID we use Brown and Lights' (1992) Partition T method. As robustness we also use Brown and Lights' (1992) Partition 24T method. To identify employer changes using these two methods we followed exactly the same procedure as specified in Appendix A1 of Kambourov and Manovskii (2009). In addition, we used Hospicio's (2015) method to identified employer changes in the PSID, as described in Hospido (2015), Section 3.2. We find that our conclusions are not affected by the method used.

When constructing the $ENE$ occupational mobility rate we consider (i) those workers who were employed at the interview date in year $t-1$, non-employed at the interview date in year $t$ and once again employed at the interview date in year $t+1$; and (ii) those workers employed at the interview dates in years $t$ and $t+1$ but who declared that they experienced an involuntary employer change during these two interviews. In an alternative specification we also added those workers who were employed at the interview date in year $t-2$, non-employed at the interview date in years $t-1$ and $t$ and once again employed at the interview date in year $t+1$. Given the small number of workers in the latter category the results hardly change.

We follow Stevens (1997) and Hospido (2015) and classify an involuntary employer change as those cases were the worker declared a job separation due to business or plant closing, due to being laid off or were fired or their temporary job ended. We also added an ``other'' category as a reason why workers left their employers to increase the number of observations. This category encompasses other reasons such as military draft. Pooling together the ``involuntary'' and ``others'' categories and computing the $ENE$ occupational mobility rate gives very similar results. Further, in constructing the $ENE$ occupational mobility rates we were not able to eliminate those workers in temporary layoff. The analysis of Fujita and Moscarini (2016), however, suggests that unemployed workers in temporary layoff will bias downwards our $ENE$ occupational mobility rates as these workers have a very high probability of re-gaining employment in the same occupation.

\clearpage

\noindent {\LARGE \bf{Supplementary Appendix C: Theory}}
\label{suppappx_C_firstpage}
\renewcommand{\thesection}{C.\arabic{section}}
\setcounter{section}{0}

\bigskip

This appendix complements Section 3 of the paper. The first part investigates, using a simplified version of the model, (i) the conditions under which rest unemployment arises and (ii) the cyclical properties of workers' job separation and occupational mobility decisions. The second part presents the equations describing worker flows in a BRE. The third part shows that the sub-market structure we impose in paper endogenously arises from a competitive search model. The fourth part provides the definition of a BRE, the proof of Proposition 2 (in the main text), the proof of existence of the separation and reallocation cutoffs as well as of the results presented in Sections 1 and 3 of this appendix.

\section{Model Implications and Comparative Statics}

We start by exploring the main implications of our theory: the occurrence of rest unemployment and the cyclical properties of workers' decisions to search across occupations and to separate from jobs. To keep the intuition as clear as possible we study a simplified version of our model without occupational human capital accumulation, setting $x_{h}=1$ for all $h$, and without occupational-wide productivity shocks, setting $p_{o}=1$ for all $o$. These restrictions imply that all occupations are identical, worker mobility across occupations is fully undirected and purely driven by occupation-worker idiosyncratic shocks such that gross mobility equals excess mobility. Further, within an occupation workers differ only in their $z$-productivities and labor market segmentation is done along this dimension. Agents' value functions are still given by the Bellman equations described in Section 3.2 of the main text, but now with state space $(z,A)$ instead of $(z,x,o,A,p)$. The net value of searching across occupations then simplifies to
\begin{align}\label{a:r}
R(A) = -c + \int_{\underline{z}} ^{\overline{z}} W^U(A,\tilde{z})dF(\tilde{z}).
\end{align}

In Carrillo-Tudela and Visschers (2013) we show that in this setting and assuming $F(z'|z)<F(z'|\tilde{z})$ for all $z,z'$ if $z>\tilde{z}$, the value functions $W^U(A,z)$, $W^E(A,z)$, $J(A,z)$ and $M(A,z)$, exist, are unique and increase in $z$. This implies that $\theta(A,z)$ also exists, is unique and increases with $z$. Further we show that if $\delta +\lambda (\theta(A,z))<1$ for all $A,z$, in equilibrium there exists a unique cutoff function $z^s$ that depends only on $A$, such that $d(A,z)=\sigma(A,z)=1$ if and only if $z<z^s(A)$, and $d(A,z)=\sigma(A,z)=0$ otherwise. Since $R(A)$ is constant in $z$, $W^U(A,z)$ is increasing in $z$, and by the existence of a unique $z^s$ so is $\max\{\lambda(\theta(A,z^r(A)))(W^E(A,z^r(A))-W^U(A,z^r(A))),0\}$, there also exists a reallocation cutoff function $z^r(A)$ such that workers decide to search across occupations if and only if $z<z^r(A)$ for every $A$, where $z^r(A)$ satisfies
 \begin{equation}\label{e:reallocation}
R(A)=W^U(A,z^r(A))+\max\{\lambda(\theta(A,z^r(A)))(W^E(A,z^r(A))-W^U(A,z^r(A))),0\}. \\
\end{equation}

Using this simplified framework we first study the relative positions of the job separation and reallocation cut-off functions and hence gain insights on the conditions under which rest unemployment arises. We then study the slopes of these cut-off functions and gain insights into the cyclicality of separations and (excess) occupational mobility in our model. Using our calibrated model we have verified that the same properties as derived below apply to the more general setup considered there.

\subsection{The Occurrence of Rest Unemployment}\label{ss:occurrence of rest unemployment}

We first analyse how the value of waiting in unemployment and in employment changes with $c$, $b$ and the persistence of the $z$-productivity process, and how these changes determine the relative position of $z^r$ and $z^s$. The simplest setting that captures a motive for waiting is one in which the $z$-productivity is redrawn randomly with probability $0<(1-\gamma)<1$ each period from cdf $F(z)$ and $A$ is held constant. Time-variation in $z$ is essential here because a worker can decide to stay unemployed in his occupation, even though there are no jobs currently available for him, when there is a high enough probability that his $z$-productivity will become sufficiently high in the future. All other features of the model remain the same, with the exception that we do not consider human capital accumulation or occupational-wide productivity differences. In Section C.3.2 of this appendix we formulate the value functions for this stationary environment and provide the proofs of Lemmas 1 and 2, below.

In this stationary setting, the expected value of an unemployed worker with productivity $z$, measured at the production stage, is given by
\begin{eqnarray}\label{e:w^u iid}
W^U(A,z)&=&\gamma \Big(b + \beta \max \Big\{R(A), W^U(A,z)+ \max \{\lambda(\theta(A,z))(1-\eta)(M(A,z)-W^U(A,z)), 0\}\Big\}\Big)  \notag \\
 &+ &(1-\gamma)\Exp_{z}[W^U(A,z)].
\end{eqnarray}
Equation \eqref{e:w^u iid} shows that there are two ways in which an unemployed worker with a $z<z^s$ can return to production. \emph{Passively}, he can wait until his $z$-productivity increases exogenously. Or, \emph{actively}, by paying $c$ and sampling a new $z$ in a different occupation. In the case in which the worker prefers to wait, the inner $\max \{.\}=0$ as the workers is below the separation cutoff and the outer $\max \{.\}=W^U(A,z)=W^U(A,z^s(A))$, where the latter equality follows as in this simplified environment the $z$-productivity process is assumed to be iid. In the case in which the worker prefers to search across occupations the outer $\max \{.\}=R(A)=W^U(A,z^r(A))$.
The difference $W^U(A,z^s(A))-R(A)$, then captures the relative gain of waiting for one period over actively sampling a new $z$ immediately. If $W^U(A,z^s(A))-R(A)\geq0$, then $z^s\geq z^r$ and there is rest unemployment. If $W^U(A,z^s(A))-R(A)<0$, then $z^r>z^s$, and endogenously separated workers immediately search across occupations.

Changing $c$, $b$, or $\gamma$ will affect the relative gains of waiting, in employment and in unemployment. In the following lemma we derive the direction of the change in $W^U(A,z^s(A)) - R(A)$, where we take fully into account the feedback effect of changes in $c$, $b$, or $\gamma$ on the match surplus $M(A,z)-W^U(A,z)$ that arises due to the presence of search frictions as discussed in Section 3.3 in the main text.

\begin{lemma} \label{e: rest_unemp} Changes in $c$, $b$ or $\gamma$ imply
\[\frac{d(W^U(A,z^s(A))-R(A))}{dc}>0, \frac{d(W^U(A,z^s(A))-R(A))}{db}>0, \frac{d(W^U(A,z^s(A))-R(A))}{d\gamma}<0.\]
\end{lemma}
It is intuitive that raising $c$ directly increases the relative gains of waiting as it make occupational mobility more costly. An increase in $c$, however, also leads to a larger match surplus because it reduces $W^U(A,z)$, making employed workers less likely to separate and hence reducing rest unemployment. The lemma shows that, overall, the first effect dominates. A rise in $b$ lowers the effective cost of waiting, while at the same time decreasing the match surplus by increasing $W^U(A,z)$, pushing towards more rest unemployment. An increase in $\gamma$, decreases the gains of waiting because it decreases the probability of experiencing a $z$-shock without paying $c$ and hence increases the value of sampling a good $z$-productivity. In the proof of Lemma 1 we further show that an increase in $W^U(A,z^s(A))-R(A)$ leads to an increase in $z^s(A)-z^r(A)$. This implies that for a sufficiently large $c$, $b$ or $1-\gamma$ rest unemployment arises.

\paragraph{Rest Unemployment and Occupational Human Capital} Occupational human capital accumulation makes a worker more productive in his current occupation. This implies that workers are willing to stay longer unemployed in their occupations because, for a given $z$, they can find jobs faster and receive higher wages. At the same time, a higher $x$ makes the employed worker less likely to quit into unemployment, generating a force against rest unemployment. Taken together, however, the first effect dominates as the next result shows.

\begin{lemma} \label{l:ohc} Consider a setting where $A$ is fixed, $z$ redrawn with probability $(1-\gamma)$, and production is given by $y=Axz$. Consider an unexpected, one-time, permanent increase in occupation-specific human capital, $x$, from $x=1$. Then
\begin{align*}
\frac{d(W^U(A,z^{s}(A,x),x)-R(A))}{dx}>0.
\end{align*}
\end{lemma}
This result implies that the difference $z^s(A,x)-z^r(A,x)$ becomes larger when human capital increases. Thus, more occupational human capital leads to rest unemployment.

\subsection{The Cyclicality of Occupational Mobility and Job Separation Decisions}

In Section 2.5 of the main text we documented that occupational mobility through unemployment (non-employment) is procyclical, while it is well established that job separations into unemployment (non-employment) are countercyclical (see Section 5 in the main text). In the model, the cyclicalities of workers' occupational mobility and job separation decisions are characterised by the slopes of the cutoff functions $z^r$ and $z^s$ with respect to $A$. As discussed in Section 3.3 of the main text, occupational mobility decisions are procyclical and job separation decisions are countercyclical when $dz^r/dA>0$ and $dz^s/dA<0$, respectively. We now explore the conditions under which such slopes arise endogenously in our model.

\paragraph{Occupational Mobility Decisions} We start by investigating the impact of rest unemployment on the slope of $z^r$ using the simplified version of the model (i.e. $x_{h}=1$ for all $h$ and $p_{o}=1$ for all $o$). Using \eqref{e:reallocation} and noting that $R(A)-W^U(A, z^r(A))=0$, we obtain
 \begin{equation}\label{e: dz^r/dp general}
 \frac{dz^r}{dA}=\frac{\int_{z^r}^{\overline{z}} \left(\frac{\partial W^U(A,z)}{\partial A}-\frac{\partial W^U(A,z^r)}{\partial A}\right)dF(z) - \frac{\partial }{\partial A }\Big(\lambda(\theta(A,z^r))(W^E(A,z^r)-W^U(A,z^r))\Big)}{\frac{\partial W^U(A,z^r)}{\partial A}+ \frac{\partial}{\partial z^r }\Big(\lambda(\theta(A,z^r))(W^E(A,z^r)-W^U(A,z^r))\Big)}.
 \end{equation}
Since workers who decided to search across occupations must sit out one period unemployed, the term
\[\lambda(\theta(A,z^r))(W^E(A,z^r)-W^U(A,z^r))\] captures the expected loss associated with the time cost of this decision: by deciding not to search across occupations, the worker could match with vacancies this period. When $z^r>z^s$, $\lambda(\theta(A,z^r))(W^E(A,z^r)-W^U(A,z^r))>0$ and is increasing in $A$ and $z^r$. Therefore, an increase in $A$ in this case, increases the loss associated with the time cost of searching across occupations and decreases $dz^r/dA$. However, when rest unemployment occurs ($z^s>z^r$), $\lambda(\theta(A,z^r))(W^E(A,z^r)-W^U(A,z^r))=0$ and this effect disappears. This follows as during rest unemployment periods workers have a contemporaneous job finding rate of zero and hence by searching across occupations, the worker does not lose out on the possibility of matching this period. In this case, the cyclicality of occupational mobility decisions purely depends on the remaining terms, in particular $\int_{z^r}^{\overline{z}} \left(\frac{\partial W^U(A,z)}{\partial A}-\frac{\partial W^U(A,z^r)}{\partial A}\right)dF(z)$. Thus, the presence of rest unemployment adds a procyclical force to occupational mobility decisions.

Now consider the impact of search frictions on the slope of $z^r$. We focus on the more general case of $z^r>z^s$, which includes the additional countercyclical force discussed above. To gain analytical tractability we consider the stationary environment used in the previous subsection and set $\gamma=1$ such that both $A$ and workers' $z$-productivities are permanent. This allows us to link wages and labor tightness to $y(A,z)$ in closed form. We then analyse the effects of a one-time, unexpected, and permanent change in $A$ on $z^r$.\footnote{This approach follows Shimer (2005), Mortensen and Nagypal (2007), and Hagedorn and Manovskii (2008). Since the equilibrium value and policy functions only depend on $A$ and $z$, analysing the change in the expected value of unemployment and joint value of the match after a one-time productivity shock is equivalent to compare those values at the steady states associated with each productivity level. This is because in our model the value and policy functions jump immediately to their steady state level, while the distribution of unemployed and employed over occupations takes time to adjust.} To isolate the role of search frictions, we compare this case with one without search frictions in which workers (who are currently not changing occupations) can match instantaneously with firms and are paid $y(A,z)$. In both cases we keep in place the same reallocation frictions. Let $z^r_c$ denote the reallocation cutoff in the case without search frictions. The details of both cases, including the corresponding value functions and the proof of the following lemma can be found in Section C.4 of this appendix.

\begin{lemma} \label{l:reallocations}
\label{p realloc behavior} Consider a one-time, unexpected, permanent increase in $A$. With search frictions the cyclical response of the decision to search across occupations is given by
\begin{align}\label{dzr1}
\frac{d z^r}{d A}& = \frac{\beta \int_{z^r}^{\bar{z}}\Big(\big( \frac{C_s(A,z)}{C_s(A,z^r)}\big) y_A(A,z)- y_A(A,z^r)\Big)dF(z)-(1-\beta)y_A(A,z^r)} {(1-\beta F(z^r))y_z(A,z^r)},
\end{align}
where $y_i(A,z)=\partial y(A,z)/\partial i$ for $i=A,z$ and $C_s(A,z)=\frac{\beta \lambda(\theta(A,z))}{(1-\beta)(1-\beta+\beta \lambda(\theta(A,z)))}$.
Without search frictions the cyclical response is given by
\begin{align}\label{dzr2}
\frac{d z^r_c}{d A}& = \frac{\beta \int_{z^r_c}^{\bar{z}} (y_A(A,z)-y_A(A,z^r_c))dF(z)-(1-\beta)y_A(A, z^r_c)} {(1-\beta F(z^r_c))y_z(A,z^r_c)}.
\end{align}
\end{lemma}

The first term in the numerator of \eqref{dzr1} and \eqref{dzr2} relates to $\int_{z^r}^{\overline{z}} \left(\frac{\partial W^U(A,z)}{\partial A}-\frac{\partial W^U(A,z^r)}{\partial A}\right)dF(z)$ in equation \eqref{e: dz^r/dp general}, while the second term captures the opportunity cost of the reallocation time. The proof of Lemma \ref{l:reallocations} shows that $C_s(A,z)/C_{s}(A,z^r)$ is increasing in $z$ and $\frac{dz^r}{dA}>\frac{dz^r_c}{dA}$ at $z^r=z^r_c$. Thus, the presence of search frictions also adds procyclicality to the decision to search across occupations. Search frictions lead to a steeper reallocation cutoff function because, in this case, an increase in $A$ increases $W^U(A,z)$ through \emph{both} the wage and the job finding probability. In contrast, an increase in $A$ in the frictionless case only affects $W^U(A,z)$ through wages (with a proportionally smaller effect on $w-b$), as workers always find jobs with probability one. The fact that $C_s(A,z)/C_{s}(A,z^r)$ is increasing in $z$ for $z>z^r$, reflects that the impact of $y(A,z)$ on $W^U(A,z)$ is increasing in $z$. Indeed, from the proof of Lemma \ref{l:reallocations} one obtains $\frac{\partial W^U(A,z)}{\partial A}\Big/\frac{\partial W^U(A,z^r)}{\partial A}=\frac{C_s(A,z)y_A(A,z)}{C_s(A,z^r)y_A(A,z^r)}>\frac{y_A(A,z)}{y_A(A,z^r) }$ for $z>z^r$.

One can get further intuition by considering the planners' problem (for details see Carrillo-Tudela and Visschers, 2013). The envelope condition implies that the planner, at the optimum allocation, does not need to change labor market tightness at each $z$ for a infinitesimal change in $A$ to still obtain the maximum net increase in expected output. At a given $z$, this means that an increase of $dA$ creates \[dW^U(A,z)=\frac{\beta \lambda(\theta(A,z))}{(1-\beta)(1-\beta +\beta \lambda(\theta(A,z)))}y_A(A,z)dA=C_s(A,z)y_A(A,z)dA\] in additional life-time expected discounted output for the planner. Since our economy is constrained efficient, the change in $A$ also creates the same lifetime expected discounted income for an unemployed worker with such $z$.

In addition, \eqref{dzr1} and \eqref{dzr2} show that the cyclicality of $z^r$ in either case depends on the production function $y(A,z)$, in particular on the sign of $y_A(A,z)-y_A(A,z^r)$ for $z>z^r$. In the proof of Lemma \ref{l:reallocations} we show that with search frictions, the decisions to search across occupations will already be procyclical when the production function is modular and $z^r$ is sufficiently close to $z^s$. This follows because the opportunity cost of searching across occupations becomes smaller as $z^r$ approaches $z^s$ from above. With rest unemployment this opportunity cost is zero. If $z^r$ is substantially above $z^s$, we will need sufficient complementarities between $A$ and $z$ in the production function to obtain a procyclical $z^r$. Without search frictions, in contrast, a supermodular production function is only a necessary condition to generate procyclicality in $z^r$ .

\paragraph{Job Separations} The main aspect of having endogenous job separation and occupational mobility decisions is that the two can potentially interact. In particular, if $z^r>z^s$, workers separate endogenously to search across occupations and this could lead to $z^r$ and $z^s$ having the same cyclical behavior. For example, in the setting of Lemma \ref{l:reallocations} were both $A$ and workers' $z$-productivities are permanent, we show at the end of Section C.3.2 of this appendix that $dz^s/dA$ depends directly on $dz^r/dA$. Namely,
\begin{align}  \label{e: z^s lemma main eq}
\frac{dz^s(A)}{dA}=-\frac{y_A(A, z^s(A))}{y_z(A, z^s(A))}+\frac{\beta \lambda(\theta(A, z^r(A)))}{1-\beta(1-\delta)+\beta \lambda (\theta(A, z^r(A)))}\frac{y_A(A, z^r(A))}{y_z(A, z^s(A))}\left(1+\frac{y_z(A,z^r(A))}{y_A(A, z^r(A))}\frac{dz^r(A)}{dA}\right). \nonumber
\end{align}
The second term makes explicit the interaction between the decisions to separate from a job and to search across occupations when there is no rest unemployment. It captures the change in the gains of search across occupations, $dR(A)/dA$, and shows that $z^r$ and $z^s$ can have the same cyclicality. When instead $z^s>z^r$, workers endogenously separate into a period of rest unemployment. In this case, $R(A)$ has a smaller impact on the value of unemployment at the moment of separation. This is because searching across occupations would only occur further in the future, and then only if a worker's $z$-productivity would deteriorate below $z^r$. Thus, the presence of rest unemployment weakens any feedback of a procyclical $z^r$ onto $z^s$. Indeed, by setting $\lambda(.)=0$ in the above expression we get that $z^s$ is always countercyclical.

\section{Worker Flows} \label{flows}

In a BRE the evolution of the distribution $\mathcal{G}$ of employed and unemployed workers across labor markets $(z,x)$, occupations $o$ and employment status $es$ is a result of (i) optimal vacancy posting $\theta(.)$, job separation decisions $d(.)$ and occupational mobility decisions $\rho(.)$ and $\mathcal{S}(.)$, all depending on the state vector $\omega=(A,p,z,x,o)$; and (ii) the exogenous retiring probability $\mu$. To obtain the laws of motions of unemployed and employed workers it is then useful to derive the measure of unemployed and employed workers at each stage $j$ within a period, where $j=s,r,m,p$ represent separations, reallocations, search and matching and production as described in the main text. It is also useful to consider the following Markov Chain: in period $t$ an employed worker with human capital level $x_{h}$ increases his human capital to $x_{h+1}$ with probability $\chi^{e}(x_{h+1}|x_{h})$, where $\chi^{e}(x_{h+1}|x_{h})=1-\chi^{e}(x_{h}|x_{h})$, $x_{h}<x_{h+1}$, $h=1,\ldots,H$ and $x_{H}<\infty$. Human capital depreciation occurs during unemployment with probability $\chi^{u}(x_{h-1}|x_{h})$, where $\chi^{u}(x_{h-1}|x_{h})=1-\chi^{u}(x_{h}|x_{h})$, $x_{h}>x_{h-1}$, $h=1,\ldots,H$. Let $u_{t}^{j}(z,x_{h},o)$ denote the measure of unemployed workers in labor market $(z,x_{h})$ in occupation $o$ at the beginning of stage $j$ in period $t$. Similarly, let $e_{t}^{j}(z,x_{h},o)$ denote the measure of employed workers in labor market $(z,x_{h})$ in occupations $o$ at the beginning of stage $j$ in period $t$.

\subsection{Unemployed Workers}

Given the initial conditions $(A_0, p_{0}, \mathcal{G}^{p}_0)$, the measure of unemployed workers characterised by $(z,x_h)$ in occupation $o$ at the beginning of next period's separation stage is
\begin{eqnarray}
u_{t+1}^{s}(z,x_{h},o) dz & = & (1-\mu) \Bigg [ \chi^{u}(x_{h}|x_{h}) \int_{\underline{z}}^{\overline{z}}u_{t}^{p}(\tilde{z},x_{h},o)dF(z|\tilde{z})d\tilde{z} + \chi^{u}(x_{h}|x_{h+1}) \int_{\underline{z}}^{\overline{z}}u_{t}^{p}(\tilde{z},x_{h+1},o)dF(z|\tilde{z})d\tilde{z}  \Bigg] \notag \\
&+& \mu  \bigg [ \sum_{\tilde{o}=1}^{O} \sum_{\tilde{h}=1}^{H}  \int_{\underline{z}} ^ {\overline{z}} \Big [u_{t}^{p}(\tilde{z},x_{\tilde{h}},\tilde{o})  +  e_{t}^{p}(\tilde{z},x_{\tilde{h}},\tilde{o}) \Big] d\tilde{z}  \bigg]  \psi_{o}(\mathbf{1}_{h=1})dF(z). \label{uflow_s_t+1}
\end{eqnarray}
Conditional on not retiring from the labor market, the terms inside the first squared bracket show the probability that unemployed workers in labor markets $(\tilde{z},x_{h},o)$ and $(\tilde{z},x_{h+1},o)$ in the previous period's production stage will be in labor market $(z,x_{h},o)$ immediately after the $z$ and $x_{h}$ shocks occur. The term in the second squared bracket refers to the measure of new workers who entered the economy to replace those who left at the beginning of the period due to the $\mu$-shock. We assume that the population of workers is constant over time, making the inflow equal to the outflow of workers. New workers are assumed to enter unemployed with a $\tilde{z}$ randomly drawn from $F(.)$ and with the lowest human capital level $x_{1}$. The above equation considers the inflow who has been assigned productivity $z$ and occupation $o$, where $\psi_{o}$ denotes the probability that workers in the inflow are assigned occupation $o$ and $\mathbf{1}_{h=1}$ denotes an indicator function which takes the value of one when the labor market $(z,x_{h})$ is associated with $x_{1}$ and zero otherwise.

During the separation stage some employed workers will become unemployed. Since by assumption these newly unemployed workers do not participate in the current period's reallocation or search and matching stages, it is convenient to count them at the production stage. This implies that $u_{t+1}^{s}(z,x_{h},o)dz=u_{t+1}^{r}(z,x_{h},o)dz$. Similarly, we will count at the production stage those unemployed workers who arrived from other occupations during the reallocation stage, as they also do not participate in the current period's search and matching stage. This implies that the measure of unemployed workers characterised by $(z,x_h)$ in occupation $o$ at the beginning of the search and matching stage is given by \[u_{t+1}^{m}(z,x_{h},o)dz=(1-\rho(A,p,z,x_{h},o))u_{t+1}^{r}(z,x_{h},o)dz.\]

Noting that $(1-\lambda(\theta(A,p,z,x_{h},o)))u_{t+1}^{m}(z,x_{h},o)dz$ workers remain unemployed after the search and matching stage, the above assumptions on when do we count occupational movers and those who separated from their employers imply that the measure of unemployed workers characterised by $(z,x_h)$ in occupation $o$ during the production stage is given by
\begin{align}
u_{t+1}^{p}(z,x_{h},o) dz & = \!(1\!-\!\lambda(\theta(A,p,z,x_{h},o)))u_{t+1}^{m}(z,x_{h},o)dz + d(z,x_{h},o, A,p)e_{t+1}^{s}(z,x_{h},o)dz \label{uflow_p_t+1} \\
&+  (\mathbf{1}_{h=1})  \bigg [ \sum_{\tilde{o} \neq o} \sum_{\tilde{h}=1}^{H} \Big [ \int_{\underline{z}} ^ {\overline{z}} \!\!\!\!\rho(\tilde{z},\tilde{x}_{h},\tilde{o},A,p)\alpha(s_{o}(\tilde{z},\tilde{x}_{h},\tilde{o},A,p),\tilde{o}) u_{t+1}^{r}(\tilde{z},x_{\tilde{h}},\tilde{o}) d\tilde{z} \Big]  \bigg] dF(z) \Bigg]. \nonumber
 \end{align}

\subsection{Employed Workers}

Next we turn to describe the laws of motion for employed workers. Given the initial conditions $(A_0, p_{0}, \mathcal{G}^{p}_0)$, the measure of employed workers characterised by $(z,x_h)$ in occupation $o$ at the beginning of next period's separation stage is given by
\begin{eqnarray}
e_{t+1}^{s}(z,x_{h},o) dz & = & (1-\mu) \Bigg [ \chi^{e}(x_{h}|x_{h}) \int_{\underline{z}}^{\overline{z}}e_{t}^{p}(\tilde{z},x_{h},o)dF(z|\tilde{z})d\tilde{z} \notag \\
& +  &(\mathbf{1}_{h>1})\chi^{e}(x_{h}|x_{h-1}) \int_{\underline{z}}^{\overline{z}}e_{t}^{p}(\tilde{z},x_{h+1},o)dF(z|\tilde{z})d\tilde{z}  \Bigg] . \label{eflow_s_t+1}
\end{eqnarray}
Conditional on not retiring from the labor market, the terms inside the squared bracket show the probability that employed workers in labor markets $(\tilde{z},x_{h},o)$ and $(\tilde{z},x_{h-1},o)$ in the previous period's production stage will be in labor market $(z,x_{h},o)$ immediately after the $z$ and $x_{h}$ shocks occur. In this case, the indicator function $\mathbf{1}_{h>1}$ takes the value of one when the labor market $(z,x_{h})$ is associated with a value of $x_{h}>x_{1}$ and zero otherwise.

Since we count those employed workers who separated from their employers in the production stage and employed workers do not participate in the reallocation or the search and matching stages, it follows that $e_{t+1}^{s}(z,x_{h},o)dz=e_{t+1}^{r}(z,x_{h},o)dz=e_{t+1}^{m}(z,x_{h},o)dz$. This implies that the measure of employed workers characterised by $(z,x_h)$ in occupation $o$ during the production stage is given by
\begin{align}
e_{t+1}^{p}(z,x_{h},o) dz &= (1-d(z,x_{h},o, A,p))e_{t+1}^{s}(z,x_{h},o)dz + \lambda(\theta(\omega))u_{t+1}^{m}(z,x_{h},o)dz,\label{eflow_p_t+1}
\end{align}
where the last term describes those unemployed workers in labor market $(z,x_{h})$ who found a job in their same occupation $o$.

\section{Competitive Search Model}

In the model described in the main text we exogenously segment an occupation into many sub-markets, one for each pair $(z,x)$. We assumed that workers with current productivities $(z,x)$ in occupation $o$ only participate in the sub-market $(z,x)$ in such an occupation. We now show that this sub-market structure endogenously arises from a competitive search model in the spirit of Moen (1997) and Menzio and Shi (2010). To show this property in the simplest way, we focus on the case in which all occupations have the same productivities (only excess mobility) and workers only differ in their $z$-productivities within an occupation. This is the same simplification we used in Section C.1 of this appendix. Adding differences in occupation productivities and occupational human capital is a straightforward extension. The full theoretical and quantitative analysis of this competitive search model can be found in our earlier working paper Carrillo-Tudela and Visschers (2013).

\subsection{Basic Setup}

As in the main text, we look for an equilibrium in which the value functions and decisions of workers and firms in any occupation only depend on the productivities $\omega=(A,z)$ and workers' employment status. Following Menzio and Shi (2010) and Menzio, Telyukova and Visschers (2016) we divide the analysis in two steps. The first step shows that at most one sub-market is active for workers with current productivity $z$. The second step shows that in equilibrium firms will post wage contracts such that a worker with current productivity $z$ does not find it optimal to visit any other sub-market other than the one opened to target workers of this productivity.

Assume that in each occupation firms post wage contracts to which they are committed. For each value of $z$ in an occupation $o$ there is a continuum of sub-markets, one for each expected lifetime value $\tilde{W}$ that could potentially be offered by a vacant firm. After firms have posted a contract in the sub-market of their choice, unemployed workers with productivity $z$ (henceforth type $z$ workers) can choose which appropriate sub-market to visit. Once type $z$ workers visit their preferred sub-market $j$, workers and firms meet according to a constant returns to scale matching function $m(u_{j},v_{j})$, where $u_{j}$ is the measure of workers searching in sub-market $j$, and $v_{j}$ the measure of firms which have posted a contract in this sub-market. From the above matching function one can easily derived the workers' job finding rate $\lambda (\theta_{j})=m(\theta_{j})$ and the vacancy filling rate $q(\theta_{j} )=m(1/\theta_{j})$, where labor market tightness is given by $\theta_{j} =v_{j}/u_{j}$. The matching function and the job finding and vacancy filling rates are assumed to have the following properties: (i) they are twice-differentiable functions, (ii) non-negative on the relevant domain, (iii) $m(0,0)=0$, (iv) $q(\theta)$ is strictly decreasing, and (v) $\lambda(\theta)$ is strictly increasing and concave.

We impose two restrictions on beliefs off-the-equilibrium path. Workers believe that, if they go to a sub-market that is inactive on the equilibrium path, firms will show up in such measure to have zero profit in expectation. Firms believe that, if they post in an inactive sub-market, a measure of workers will show up, to make the measure of deviating firms indifferent between entering or not. We assume, for convenience, that the zero-profit condition also holds for deviations of a single agent: loosely, the number of vacancies or unemployed, and therefore the tightness will be believed to adjust to make the zero-profit equation hold.

\subsection{Agents' Problem}

\paragraph{Workers}

Consider the value function of an unemployed worker having productivity $z$ in occupation $o$ at the beginning of the production stage, $W^U(\omega)= b + \beta \mathbb{E}[ W^{R}(\omega^{\prime})]$. The value of unemployment consists of the flow benefit of unemployment $b$ this period, plus the discounted expected value of being unemployed at the beginning of next period's reallocation stage,
\begin{equation}
W^{R}(\omega^{\prime})=\max_{\rho(\omega^{\prime})} \{\rho(\omega^{\prime}) R(\omega^{\prime}) + (1-\rho(\omega^{\prime})) \mathbb{E}[S(\omega^{\prime})+W^U(\omega^{\prime})]\},\end{equation}
where $\rho(\omega^{\prime})$ takes the value of one when the worker decides to reallocate and take the value of zero otherwise and recall $R$ is the expected value of reallocation. In this case,
\[ R(\omega)= -c+\sum_{o'\neq o} \int W^{U}(A^{\prime },\tilde{z})\frac{dF(\tilde{z})}{O-1}. \] The worker's expected value of staying and searching in his current occupation is given by $\mathbb{E}[S(\omega^{\prime})+W^{U}(\omega^{\prime})]$. In this case, $W^{U}(\omega^{\prime})=\mathbb{E}[W^{U}(\omega^{\prime})]$ describes the expected value of not finding a job, while $S(\omega^{\prime})$ summarizes the expected value added of finding a new job. The reallocation decision is captured by the choice between $R(\omega^{\prime})$ and the expected payoff of search in the current occupation.

To derive $S(.)$ note that $\lambda (\theta (\omega,W_{f}))$ denotes the probability with which a type $z$ worker meets a firm $f$ in the sub-market associated with the promised value $W_{f}$ and tightness $\theta(\omega)$. Further, let $\alpha (W_{f})$ denote the probability of visiting such a sub-market. From the set $\mathcal{W}$ of promised values which are offered in equilibrium by firms for a given $z$, workers only visit with positive probability those sub-markets for which the associated $W_{f}$ satisfies
\begin{equation}  \label{e:application_max}
W_{f}\in \arg \max_{\mathcal{W}}\lambda (\theta (\omega^{\prime},W_{f}))(W_{f}-W^{U}(\omega^{\prime}))\equiv S(\omega^{\prime}).
\end{equation}
When the set $\mathcal{W}$ is empty, the expected value added of finding a job is zero and the worker is indifferent between visiting any sub-market.

Now consider the value function at the beginning of the production stage of an employed worker with productivity $z$ in a contract that currently has a value $\tilde{W}_{f}(\omega)$. Similar arguments as before imply that
\begin{align}  \label{employed}
\tilde{W}_{f}(\omega)=&w_{f} + \beta \mathbb{E}\bigg[ \max_{d(\omega^{\prime})} \{ (1-d(\omega^{\prime}))\tilde{W}_{f}(\omega^{\prime})+ d(\omega^{\prime})W^U(\omega^{\prime})\}\bigg],
\end{align}
where $d(\omega^{\prime})$ take the value of $\delta$ when $\tilde{W}_{f}(\omega^{\prime})\geq W^U(\omega^{\prime})$ and the value of one otherwise. In equation \eqref{employed}, the wage payment $w_{f}$ at firm $f$ is contingent on state $\omega$, while the second term describes the worker's option to quit into unemployment in the separation stage the next period. Note that $W^U(\omega^{\prime})= \mathbb{E}[W^U(\omega^{\prime})]$ as a worker who separates must stay unemployed for the rest of the period and $\tilde{W}_f(\omega^{\prime})=\mathbb{E}[\tilde{W}_f(\omega^{\prime})]$ as the match will be preserved after the separation stage.

\paragraph{Firms}

Consider a firm $f$ in occupation $o$, currently employing a worker with productivity $z$ who has been promised a value $\tilde{W}_{f}(\omega)\geq W^{U}(\omega)$. Noting that the state space for this firm is the same as for the worker and given by $\omega$, the expected lifetime discounted profit of the firm can be described recursively as
\begin{eqnarray}  \label{filledjob}
J(\omega;\tilde{W}_{f}(\omega)) =\max_{w_{f}, \tilde{W}_{f}(\omega^{\prime})} \bigg\{y(A,z)-w_{f}+\beta \mathbb{E}\Big[\max_{\sigma (\omega^{\prime})}\Big\{(1-\sigma (\omega^{\prime}))J(\omega^{\prime};\tilde{W}_{f}(\omega^{\prime}))  +\sigma (\omega^{\prime})\tilde{V}(\omega^{\prime})\Big\}\Big]\bigg\},
\end{eqnarray}
where $\sigma (\omega^{\prime})$ takes the value of $\delta$ when $J(\omega^{\prime};\tilde{W}_{f}(\omega^{\prime}))\geq \tilde{V}(\omega^{\prime})$ and the value of one otherwise, $\tilde{V}(\omega^{\prime})=\max \big\{\overline{V}(\omega^{\prime}),0\big\}$ and $\overline{V}(\omega^{\prime})$ denotes the maximum value of an unfilled vacancy in occupation $o$ at the beginning of next period. Hence (\ref{filledjob}) takes into account that the firm could decide to target its vacancy to workers of with a different productivity in the same occupation or withdraw the vacancy from the economy and obtain zero profits.

The first maximisation in (\ref{filledjob}) is over the wage payment $w_{f}$ and the promised lifetime utility to the worker $\tilde{W}_{f}(\omega^{\prime})$. The second maximisation refers to the firm's layoff decision. The solution to \eqref{filledjob} then gives the wage payments during the match (for each realisation of $\omega$ for all $t$). In turn these wages determine the expected lifetime profits at any moment during the relation, and importantly also at the start of the relationship, where the promised value to the worker is $\tilde{W}_{f}$.

Equation \eqref{filledjob} is subject to the restriction that the wage paid today and tomorrow's promised values have to add up to today's promised value $\tilde{W}_{f}(\omega)$, according to equation \eqref{employed}. Moreover, the workers' option to quit into unemployment, and the firm's option to lay off the worker imply the following participation constraints
\begin{equation}  \label{partconst}
\Big(J(\omega^{\prime};\tilde{W}_{f}(\omega^{\prime}))-\tilde{V}(\omega^{\prime})\Big)\geq 0 \text{\ \ \ and \ \ \ }  \Big(\tilde{W}_{f}(\omega^{\prime})-W^{U}(\omega^{\prime})\Big)\geq 0.
\end{equation}

Now consider a firm posting a vacancy in occupation $o$. Given cost $k$ and knowing $\omega$, a firm must choose which unemployed workers to target. In particular, for each $z$ a firm has to decide which $\tilde{W}_{f}$ to post given the associated job filling probability, $q(\theta (\omega,\tilde{W}_{f}))$. This probability summarises the pricing behaviour of other firms and the visiting strategies of workers. Along the same lines as above, the expected value of a vacancy targeting workers of productivity $z$ solves the Bellman equation
\begin{equation}  \label{unfilled job fe}
V(\omega)=-k+\max_{\tilde{W}_{f}}\Big\{q(\theta (\omega,\tilde{W}_{f}))J(\omega,\tilde{W}_{f})+(1-q(\theta (\omega,\tilde{W}_{f})))\tilde{V}(\omega)
\Big\}.
\end{equation}

We assume that there is free entry of firms posting vacancies within any occupation. This implies that $V(\omega)=0$ and $\tilde{W}_{f}$ that yield a $\theta(\omega,\tilde{W}_{f})>0$, and $V(\omega)\leq0$ for all those $\omega$ and $\tilde{W}_{f}$ that yield a $\theta(\omega,\tilde{W}_{f})\leq0$. In the former case, the free entry condition then simplifies (\ref{unfilled job fe}) to $k=\max_{\tilde{W}_{f}}q(\theta (\omega,\tilde{W}_{f}))J(\omega,\tilde{W}_{f})$.

\subsection{Endogenous Market Segmentation}

We now show that if there are positive gains to form a productive match with a worker of type $z$, firms offer a unique $\tilde{W}_f$ with associate tightness $\tilde{\theta}(A,z)$ in the matching stage. This implies that only one sub-market opens for a given value of $z$ and workers of productivity $z$ optimally chose to search in such a sub-market. Consider a value of $z$. For any promised value $W^E$, the joint value of the match is defined as $W^E+J(A,z,W^E)\equiv\tilde{M}(A,z,W^E)$. Lemma \ref{l:jointvalue} shows that under risk neutrality the value of a job match is constant in $W^E$ and $J$ decreases one-to-one with $W^E$.

\begin{lemma}\label{l:jointvalue}
The joint value $\tilde{M}(A,z, W^E)$ is constant in $W^E\geq W^U(A,z)$ and hence we can uniquely define $M(A,z)\overset{def}{=}\tilde{M}(A,z,W^E), \ \forall \ M(A,z)\geq W^E \geq W^U(A,z)$ on this domain. Further, $J_W(A,z,W^E)=-1, \ \forall \ M(A,z)>W^E>W^U(A,z)$.
\end{lemma}

The proof of Lemma \ref{l:jointvalue} is presented in Section C.4.4. It crucially relies on the firms' ability to offer workers inter-temporal wage transfers such that the value of the job match is not affected by the (initial) promised value. Note that Lemma \ref{l:jointvalue} implies that no firm will post vacancies for $z$ values such that $M(A,z)-W^U(A,z)\leq0$. Lemma \ref{l:application} now shows that for a given $z$, for which $M(A,z)-W^U(A,z)>0$, firms offer a unique $\tilde{W}_f$ in the matching stage and there is a unique $\theta$ associated with it.

\begin{lemma}\label{l:application}
If the elasticity of the vacancy filling rate is weakly negative in $\theta$, there exists a unique $\theta^*(A,z)$ and $W^*(A,z)$ that solve \eqref{e:application_max}, subject to \eqref{unfilled job fe}.
\end{lemma}

The proof of Lemma \ref{l:application} is also presented in Section C.4.4. The requirement that the elasticity of the job filling rate with respect to $\theta$ is non-positive is automatically satisfied when $q(\theta)$ is log concave, as is the case with the Cobb-Douglas and urn ball matching functions. Both matching functions imply that the job finding and vacancy filling rates have the properties described in Lemma \ref{l:application} and hence guarantee a unique pair $\tilde{W}_{f},\theta$. Consider a Cobb-Douglas matching function as it implies a constant $\varepsilon _{q,\theta }(\theta )$. Using $\eta$ to denote the (constant) elasticity of the job finding rate with respect to $\theta $, we find the well-known division of the surplus according to the Hosios' (1991) rule
\begin{equation}
\eta (W^{E}-W^{U}(A,z))-(1-\eta )J(A,z,W^{E})=0.  \label{hosios}
\end{equation}

Since for every value of $z$ there is at most one $\tilde{W}_{f}$ offered in the matching stage, the visiting strategy of an unemployed worker is to visit the sub-market associated with $\tilde{W}_{f}$ with probability one when $S(A,z)>0$ and to randomly visit any sub-market when $S(A,z)=0$ (or not visit any submarket at all). Let $\tilde{W}^{z}_{f}$ denote the unique expected value offered to workers of productivity $z$ in equilibrium.

The final step is to verify that if we allow firms to post a menu of contracts which specify for each $z$ a different expected value, then the above equilibrium allocation and payoffs can be sustained in this more general setting. To show this we closely follow Menzio and Shi (2010) with the addition of endogenous reallocation. Since the expected value of reallocation $R$ is a constant across $z$-productivities, it is easy to verify that the proof developed in Menzio and Shi (2010) applies here as well.

Intuitively, consider a BRE in which firms that want to only attract a worker of type $z$ offer a contract with expected value $\tilde{W}^{z}_{f}$, as derived above, to any worker of type $z$ or higher; and for lower worker types, the firm offers contracts with expected values strictly below these types' value of unemployment. Since $\tilde{W}^{z}_{f}$ and $\theta^{z}$ in this candidate equilibrium are increasing in $z$, this implies that only the type $z$ workers visit these firms. To show that this is indeed optimal for firms, first consider the deviation in which a firm opens a new sub-market that attracts workers of different types $z$ and $\hat{z}$ by offering expected values $W^{z}$ and $W^{\hat{z}}$. Without loss of generality assume that firms obtain higher profit from matching with a worker of type $z$ instead of type $\hat{z}$. Next consider an alternative deviation in which a firm opens a sub-market that only attract workers of type $z$ by offering $W^{z}$ and the value of unemployment to any other worker type. Given the off-equilibrium beliefs, a higher tightness is associated with this alternative deviation which makes it more attractive for type $z$ workers to visit relative to the original deviation. That is, if there is a strictly profitable deviation in which more than one type visits the same sub-market, there is always another strictly more profitable deviation in which a sub-market is visited by only one type. However, Lemma \ref{l:application} shows that the latter deviation cannot exist and therefore there also cannot be a profitable deviation in which more than type visits the same sub-market in equilibrium.

\section{Proofs}

\paragraph{Definition}
A Block Recursive Equilibrium (BRE) is a set of value functions $W^U(\omega)$, $W^E(\omega)$, $J(\omega)$, workers' policy functions $d(\omega)$, $\rho(\omega)$, $\mathcal{S}(\omega)$, firms' policy function $\sigma(\omega)$, tightness function $\theta(\omega)$, wages $w(\omega)$, laws of motion of $A$, $p$, $z$ and $x$ for all occupations, and laws of motion for the distribution of unemployed and employed workers over all occupations, such that: \textit{(i)} the value functions and decision rules follow from the firm's and worker's problems described in equations (1)-(5) in the main text; \textit{(ii)} labor market tightness $\theta(\omega)$ is consistent with free entry on each labor market, with zero expected profits determining $\theta(\omega)$ on labor markets at which positive ex-post profits exist; $\theta(\omega)=0$ otherwise; \textit{(iii)} wages solve equation (6) in the main text; \textit{(iv)} the worker flow equations map initial distributions of unemployed and employed workers (respectively) over labor markets and occupations into next period's distribution of unemployed and employed workers over labor markets and occupations, according to the above policy functions and exogenous separations.

\paragraph{Proposition 2}
Given $F(z'|z)<F(z'|\tilde{z})\ $for\ all $z,z'$ when $z>\tilde{z}$: (i) a BRE exists and it is the unique equilibrium, and (ii) the BRE is constrained efficient.

\bigskip

\subsection{Proof of Proposition 2}

We divide the proof into two parts. In the first part we show existence of equilibrium by deriving the operator $T$, showing it is a contraction and then verifying that the candidate equilibrium functions from the fixed point of $T$ satisfy all equilibrium conditions. The second part shows efficiency of equilibrium.

\subsubsection{Existence}

\paragraph{\emph{Step 1:}}

Let $M(\omega)\equiv W^E(\omega)+J(\omega)$ denote the value of the match. We want to show that the value functions $M(\omega)$, $W^{U}(\omega)$ and $R(\omega)$ exist. This leads to a three dimensional fixed point problem. It is then useful to define the operator $T$ that maps the value function $\Gamma(\omega,n)$ for $n=0,1,2$ into the same functional space, such that $\Gamma(\omega,0)=M(\omega),$ $\Gamma(\omega,1)=W^{U}(\omega)$, $\Gamma(\omega,2)=R(\omega)$ and
\begin{align*}
T(\Gamma(\omega,0))&=y(A,p_{o},z,x)+\beta \mathbb{E}_{\omega'}\Big[\max_{d^{T}}\{(1-d^{T})M(\omega')+d^{T}W^{U}(\omega')\}\Big], \non \\
T(\Gamma(\omega,1))&=b+\beta \mathbb{E}_{\omega'}\Big[\max_{\rho^{T}}\{\rho^{T} R(\omega')+(1-\rho ^{T})(D^{T}(\omega')+W^{U}(\omega'))\}\Big], \non \\
T(\Gamma(\omega,2))&=\max_{\mathcal{S}(\omega)}\Big( \sum_{\tilde{o} \in \vec{O}^{-}} \alpha(s^{T}_{\tilde{o}}) \int_{\underline{z}}^{\overline{z}}W^{U}(\tilde{z},x_{1},\tilde{o},A,p)dF(\tilde{z}) + (1-\sum_{\tilde{o} \in \vec{O}^{-}} \alpha(s^{T}_{\tilde{o}}))[b+ \beta \mathbb{E}_{\omega'} R(\omega')] - c \Big),
\end{align*}%
where the latter
maximization is subject to $s_{o}\in[0,1]$ and $\sum_{o \in \vec{O}^{-}} s_{o}=1$, and
\begin{align}\label{eq:Ddef}
D^{T}(\omega')\equiv \lambda (\theta (\omega'))(1-\eta) \Big(M(\omega')-W^{U}(\omega')\Big), \text{ with } \theta(\omega')= \left(\frac{\eta(M(\omega')-W^U(\omega'))}{k}\right)^{\frac{1}{1-\eta}}.
\end{align}%

\paragraph{Lemma A.1:} $T$ is (i) a well-defined operator mapping functions from the closed space of bounded continuous functions $\mathcal{E}$ into $\mathcal{E}$, and (ii) a contraction.

First we show that the operator $T$ maps bounded continuous functions into bounded continuous functions. Let $W^{U}(\omega)$, $R(\omega)$ and $M(\omega)$ be bounded continuous functions. It then follows that $\lambda (\theta(\omega) )$ and $D(\omega)$ are continuous functions. It also follows that $\max\{M(\omega),W^{U}(\omega)\}$ and $\max\{R(\omega),D(\omega)+W^{U}(\omega)\}$ are continuous. Further, since the constraint set for $\mathcal{S}$ is compact-valued and does not depend on $\omega$, functions $\alpha(.)$ are continuous, and the integral of continuous function $W^U$ is continuous, the theorem of the maximum then implies that the expression
\begin{equation}
\max_{\mathcal{S}(\omega)}\Big( \sum_{\tilde{o} \in \vec{O}^{-}} \alpha(s^{T}_{\tilde{o}}) \int_{\underline{z}}^{\overline{z}}W^{U}(\tilde{z},x_{1},\tilde{o},A,p)dF(\tilde{z}) + (1-\sum_{\tilde{o} \in \vec{O}^{-}} \alpha(s^{T}_{\tilde{o}})) [ b+ \beta \mathbb{E}_{\omega'} R(\omega') ] - c \Big)
\end{equation}
is continuous. Therefore $T$ maps continuous functions into continuous functions. Moreover, since the domain of $\omega$ is bounded, and $\alpha(.)$ and $\lambda(.)$ are bounded on bounded domains, $T$ maps the space of bounded continuous functions into itself.

Second we show that $T$ defines a contraction. Consider two functions $\Gamma,\Gamma^{\prime } \in \mathcal{E}$, such that $\Vert \Gamma-\Gamma^{\prime }\Vert_{\sup }<\varepsilon $. It then follows that $\Vert W^{U}-W^{U}{^{\prime }}\Vert _{\sup }<\varepsilon $, $\|R-R'\|_{sup}<\varepsilon$ and $\Vert M-M^{\prime} \Vert _{\sup }<\varepsilon $, where $W^{U}$, $R$ and $M$ are part of $\Gamma$ as defined above. We first establish that
\begin{align}\label{e:prop1_i_step1_eq1}
\|D+W^U-  D'-W^{U}{'}\|_{sup}<\varepsilon,
\end{align}
when function-tuples $D$ and $D'$ are derived from $(M, W^U)$, and $(M', W^U{'})$ respectively. With this aim consider, without loss of generality, the case in which $M(\omega)-W^U(\omega)>M^{' }(\omega)-W^U{' }(\omega)$ at a given $\omega$. Instead of $D(\omega)$, write $D(M(\omega)-W(\omega))$ to make explicit the dependence of $D$ on $M$ and $W$. To reduce notation we suppress the dependence on $\omega$ for this part of the proof and further condense $W^U=W$. From $M-W>M^{' }-W^{' }$, it follows that $\varepsilon >W^{' }-W\geq M^{' }-M>-\varepsilon $. Construct $M^{'' }=W^{' }+(M-W)>M^{' }$ and $W^{''}=M^{' }-(M-W)<W^{' }$. Equation \eqref{eq:Ddef} implies $D(M-W)+W$ is increasing in $M$ and in $W$. To verify this, hold $M$ constant and note that
\begin{align}\label{e:incr_z}
\frac{d(D(M-W)+W)}{dW}=-\lambda(\theta(.))+1\geq 0
 \end{align}
by virtue of $d(D(M-W))/d(M-W)=\lambda$ and that $\lambda\in[0,1]$, where the inequality in \eqref{e:incr_z} is strict when $\lambda\in[0,1)$ and weak when $\lambda=1$. Then, it follows that
\begin{align}
-\varepsilon <D(M^{' }-W^{''})+W^{'' }-D(M-W)-W &
\leq D(M^{' }-W^{' })+W^{' }-D(M-W)-W  \notag \\
& \leq D(M^{''}-W^{' })+W^{' }-D(M-W)-W<\varepsilon
\notag
\end{align}
where $D(M^{' }-W^{''})=D(M-W)=D(M^{''}-W^{' })$ by construction. Note that the outer inequalities follow because $M-M^{' }>-\varepsilon$ and $W^{' }-W<\varepsilon $. Given that $D$, $M$ and $W$ are bounded continuous functions on a compact domain and the above holds for every $\omega$, it then must be that $\Vert D+W^{U}-D^{' }-W^{U}{^{' }}\Vert_{\sup } <\varepsilon $. Since $\Vert \max \{a,b\}-\max \{a^{\prime},b^{\prime }\}\Vert <\max \{\Vert a-a^{\prime }\Vert ,\Vert b-b^{\prime}\Vert \}$, as long as the terms over which to maximize do not change by more than $\varepsilon$ in absolute value, the maximized value does not change by more $\varepsilon$, it then follows that $\Vert T(\Gamma(\omega, n))-T(\Gamma^{' }(\omega,n)\Vert <\beta \varepsilon $ for all $\omega$; $n=0,1$.

To show that $\Vert T(\Gamma(\omega, 3))-T(\Gamma^{' }(\omega,3)\Vert <\beta \varepsilon $ for all $\omega$, with a slight abuse of notation, define $TR(\omega, \mathcal{S})$ as
\begin{align}\label{eq:app_reall_s}
T R(\omega,\mathcal{S}) =&\sum_{\tilde{o} \in \vec{O}^{-}} \alpha(s_{\tilde{o}}(\omega)) \int_{\underline{z}}^{\overline{z}}\Big[b+\beta \mathbb{E}_{\omega'}\Big[\max_{\rho^{T}}\{\rho^{T} R(\omega')+(1-\rho ^{T})(D^{T}(\omega')+W^{U}(\omega'))\}\Big]\Big]dF(\tilde{z}) \\ \notag
+& (1-\sum_{\tilde{o} \in \vec{O}^{-}} \alpha(s_{\tilde{o}}(\omega)))[b+ \beta \mathbb{E}_{\omega'} R(\omega')] - c.
\end{align}
Let $\mathcal{S}_*(\omega)=\{s^*_1,\ldots, s^*_{o-1}, s^*_{o+1},\ldots, s^*_O\}$ be the maximizer of $TR$ at $\omega$ and $\mathcal{S}_*{'}(\omega)$ of $TR'$. Without loss of generality assume that $TR(\omega,\mathcal{S})> TR'(\omega,\mathcal{S}')$. Since our previous arguments imply that
\[
\Vert \Big[\max_{\rho}\{\rho R+(1-\rho)(D+W^{U})\}\Big] - \Big[\max_{\rho}\{\rho R'+(1-\rho)(D'+W')\}\Big] \|_{sup} <\beta \varepsilon
\]
it then follows that $\Vert TR(\omega,\mathcal{S}_*)-TR'(\omega,\mathcal{S}_*)\|_{sup}<\beta \varepsilon $. Further, since $TR(\omega,\mathcal{S}_*)> TR'(\omega,\mathcal{S}^{'}_*)> TR'(\omega,\mathcal{S}_*)$ we have that $0<TR(\omega,\mathcal{S}_*)-TR'(\omega,\mathcal{S}_*)<\beta \varepsilon$. Using the latter inequality to obtain $0< TR(\omega,\mathcal{S}_*)-TR'(\omega,\mathcal{S}^{'}_*)+TR'(\omega,\mathcal{S}^{'}_*)-TR'(\omega,\mathcal{S}_*)<\beta \varepsilon$ and noting that $TR'(\omega,\mathcal{S}^{'}_*)-TR'(\omega,\mathcal{S}_*) >0$, it follows that $0< TR(\omega,\mathcal{S}_*)-TR'(\omega,\mathcal{S}^{'}_*)<\beta \varepsilon$. This last step is valid for all $\omega$, then it implies that $\Vert T(\Gamma(\omega, 3))-T(\Gamma^{' }(\omega,3)\|_{sup}<\beta \varepsilon $ and hence the operator $T$ is a contraction and has a unique fixed point.

\paragraph{\emph{Step 2 (Linking the Mapping $T$ and BRE Objects):}}

From the fixed point functions $M(\omega)$, $W^{U}(\omega)$ and $R(\omega)$ define the function $J(\omega)=\max\{(1-\zeta)[M(\omega)-W^{U}(\omega)],0\}$, and the functions $\theta (\omega)$ and $V(\omega)$ from $0=V(\omega)=-k+q(\theta (\omega))J(\omega)$. Also define $W^{E}(\omega)=M(\omega)-J(\omega)$ if $M(\omega)>W^{U}(\omega)$, and $W^{E}(\omega)=M(\omega)$ if $M(\omega)\leq W^{U}(\omega)$. Finally, define $d(\omega)=d^{T}(\omega)$, $\sigma(\omega)=\sigma^{T}(\omega)$, $\rho (\omega)=\rho^{T}(\omega)$, $\mathcal{S}(\omega)=\mathcal{S}^{T}(\omega)$ and a $w(\omega)$ derived using the Nash bargaining equation in the main text given all other functions.

Given $1- \zeta=\eta$ and provided that the job separation decisions between workers and firms coincide, which they are as a match is broken up if and only if it is bilaterally efficient to do so according to $M(\omega)$ and $W^{U}(\omega)$, then equations (5) and (6) in the main text (describing $J(\omega)$ and surplus sharing) are satisfied. Further, equation (3) in the main text (describing $W^E(\omega)$) is satisfied by construction, $\theta(\omega)$ satisfies the free-entry condition and $w(\omega)$ satisfies equation (6) in the main text. Hence, the constructed value functions and decision rules satisfy all conditions of the equilibrium and the implied evolution of the distribution of employed and unemployed workers also satisfies the equilibrium conditions.

Uniqueness follows from the same procedure in the opposite direction and using a contradiction argument. Suppose the BRE is not unique. Then a second set of functions exists that satisfy all the equilibrium conditions. Construct $\hat{M}$, $\hat{W}^{U}$ and $\hat{R}$ from these conditions. Since in any equilibrium the job separation decisions have to be bilaterally efficient and the occupational mobility decisions ($\rho$ and $\mathcal{S}$) are captured in $T$, then $\hat{M}$, $\hat{W}^{U}$ and $\hat{R}$ must be a fixed point of $T$, contradicting the uniqueness of the fixed point established by Banach's Fixed Point Theorem. Hence, there is a unique BRE.

\subsubsection{Efficiency}

The social planner, currently in the production stage, solves the problem of maximizing total discounted output by choosing job separations decisions $d(.)$, occupational mobility decisions $\rho(.)$ and $\mathcal{S}(.)$, as well as vacancy creation decisions $v(.)$ for each pair $(z,x_{h})$ across all occupations $o \in O$ in any period $t$. The first key aspect of the planner's choices is that they could potentially depend on the entire state space $\Omega^j=\{z, x, o, A, p, \mathcal{G}^j\}$ for each of the four within period stages $j=s,r,m,p$ (separation, reallocation, search and matching, production) and workers' employment status, such that its maximization problem is given by
\begin{align}
\max_{\{d(\Omega^s),\rho(\Omega^r), \mathcal{S}(\Omega^r),v(\Omega^m)\}} \text {{\Large {$\mathbb{E}$}}}
 & \sum_{t=0} ^ {\infty} \bigg( \sum_{o=1} ^ {O} \sum_{h=1}^{H}\int_{\underline{z}} ^{\overline{z}} \beta ^{t} \left[ u_{t}^{p}(z,x_{h},o) b + e_{t}^{p}(z,x_{h},o) y(A_{t},p_{o,t},z,x_{h}) \right] dz   \\
 & \qquad -  \sum_{o'=1} ^ {O} \sum_{h'=1}^{H}\int_{\underline{z}} ^{\overline{z}} \beta ^{t+1} \Big[c\rho(z', x_h', o', A_{t+1}, p_{t+1}, \mathcal{G}^r_{t+1}) u_{t+1}^{r}(z',x'_{h},o') \non \\
 & \qquad \qquad \qquad \qquad \qquad  + k v_{t+1}(z', x_h', o', A_{t+1}, p_{t+1}, \mathcal{G}^m_{t+1})\Big] dz' \bigg), \non
\end{align}
subject to the initial conditions $(A_0, p_{0}, \mathcal{G}^{p}_0)$, the laws of motion for unemployed and employed workers described in Section C.2 of this appendix (with corresponding state space $\Omega^j$ for the decision rules), and the choice variables $\rho(.)$ and $d(.)$ being continuous variables in $[0,1]$, as the planner can decide on the proportion of workers in labor market $(z,x_{h})$ to separate from their jobs or to change occupations.

Note that implicitly the social planner is constrained in the search technology across occupations: it faces the same restrictions as an individual worker (in occupation $o \in O$), on the proportion of time that can be devoted to obtain a $z$-productivity from occupation $\tilde{o}\neq o$. Namely, $s_{\tilde{o}}(.)\in[0,1]$, $\sum_{\tilde{o} \in \vec{O}^{-}_{o}} s_{\tilde{o}}(.)=1$ and $\sum_{\tilde{o} \in \vec{O}^{-}_{o}} \alpha(s_{\tilde{o}}(.),o) \leq 1$, where $\vec{O}^{-}_{o}$ denotes the set of remaining occupations relative to $o$. The latter notation highlights that, as in the decentralised problem, once the occupational mobility decision has been taken the new $z$-productivity cannot be obtain from the departing occupation.

Rewriting the planners' problem in recursive form as the fixed point of the mapping $T^{SP}$ and letting next period's values be denoted by a prime yields
\begin{align}
&T^{SP}W^{SP}(\Omega^p) =\max_{\substack{\big\{d(\Omega^{s'}),\rho(\Omega^{r'}), \\\mathcal{S}(\Omega^{r'}), v(\Omega^{m'})\big\}}} \sum_{o=1}^{O} \sum_{h=1}^{H} \int_{\underline{z}} ^{\overline{z}}(u^{p}(z,x_{h},o)b+e^{p}(z,x_{h},o)y(A,p_{o},z,x_{h}))dz \label{e: SP_VF_contraction_mapping}\\
& \quad +\beta \mathbb{E}_{\Omega^{s}{'}|\Omega^{p}}\left[-\left( c \sum_{o=1} ^ {O} \sum_{h=1}^{H} \int_{\underline{z}} ^{\overline{z}} \rho(\Omega^{r'})u^{r}{'}( z', x_{h}' ,o)dz' +k \sum_{o=1} ^ {O} \sum_{h=1}^{H} \int_{\underline{z}} ^{\overline{z}} v(\Omega^{m'})dz' \right) \right.  +W^{SP}(\Omega^{p}{'})\Bigg] \non,
\end{align}%
subject to the same restrictions described above. Our aim is to show that this mapping is a contraction that maps functions $W^{SP}(.)$ from the space of functions linear with respect to the distribution $\mathcal{G}^p$ into itself, such that
\begin{align}
W^{SP}(\Omega^p)=\sum_{o=1}^{O} \sum_{h=1}^{H} \int_{\underline{z}} ^{\overline{z}}\!\!\!\!\left( u^{p}(z,x_h,o) W^{u,SP}(\omega)\!+e^{p}(z,x_h,o)M^{SP}(\omega)\right)dz,\label{e:linear_SP_VF_guess}
\end{align}
for some functions $W^{u,SP}(\omega)$ and $M^{SP}(\omega)$, where $\omega=(z,x_h,o, A, p)$, and $u^{p}(.)$ and $e^{p}(.)$ are implied by $\mathcal{G}^p$.

In Section C.2 of this appendix, we decomposed the next period's measure of unemployed workers at the production stage, $u^{p'}$, into three additive terms (see equation \eqref{uflow_p_t+1}). The first term refers to those unemployed workers that were unsuccessful in matching, either due to not finding a posted vacancy in their submarket or because the planner chose not to post vacancies in their submarket. The second term refers to those workers who separated from employment and hence were restricted from the search and matching stage within the period. The third term refers to those who changed occupations and came from other markets into submarket $(z,x_h,o)$ and were restricted from the search and matching stage within the period. Likewise, we decomposed the next period's measure of employed workers at the production stage into two additive terms (see equation \eqref{eflow_p_t+1}). The first one refers to the survivors in employment from the previous separation stage, while the second terms refers to new hires. Given a continuation value function $W^{SP}(\Omega^{p'})$ that is linear in $\mathcal{G}^p$, as in \eqref{e:linear_SP_VF_guess}, we now show that we can also decompose the expression in \eqref{e: SP_VF_contraction_mapping} into different additive components.

Consider first those unemployed workers who are eligible to participate in the search and matching stage. Note that the matching technology implies $v(\Omega^{m})=\theta(\Omega^{m})(1-\rho(\Omega^{r}))u^{m}(z,x_{h},o)$. Therefore we can consider $\theta(\Omega^{m'})$ as the planner's choice in equation \eqref{e: SP_VF_contraction_mapping} instead of $v(\Omega^{m'})$. Next we isolate the terms of $W^{SP}(\Omega^{p'})$ that involve workers who went through the search and matching stage: the first term of the unemployed worker flow equation \eqref{uflow_p_t+1} and the second term of the employed worker flow equation \eqref{eflow_p_t+1}. We then combine these terms with the cost of vacancy posting in \eqref{e: SP_VF_contraction_mapping}. Noting that the dependence of $u^m(\Omega^{m'})$ captures a potential dependence of the planner's occupational mobility decisions made in the previous reallocation stage, we can express the terms in the mapping \eqref{e: SP_VF_contraction_mapping} that involve $\theta(\Omega^{m'})$ as
\begin{align}
\sum_{o=1}^{O} \sum_{h=1}^{H} \int_{\underline{z}} ^{\overline{z}} u^m{'}(\Omega^{m'}) \Big\{-k \theta(.) + \lambda(\theta(.))M^{SP}(\omega') + (1-\lambda(\theta(.))W^{u,SP}(\omega') \Big\}dz'.  \label{e:SP_max_searchstage}
\end{align}
Note that by backwards induction, the planner's optimal decisions for $\theta(\Omega^{m'})$ maximize \eqref{e:SP_max_searchstage}. Given that the terms within the curly brackets only depend on $\omega'$ when the continuation value is linear as in \eqref{e:linear_SP_VF_guess}, maximizing with respect to $\theta$ implies an optimal $\theta(\omega')$. Using this result we let $W^{m,SP}(\omega')$ denote the sum of the maximized terms inside the curly brackets in \eqref{e:SP_max_searchstage} .

Next consider the search-direction choice across occupations for those workers who the planner has decided to move across occupations. The following expression summarizes the terms of \eqref{e: SP_VF_contraction_mapping} that involve this choice,
\begin{align}
\sum_{o'} \sum_{h'=1}^{H} \int_{\underline{z}} ^ {\overline{z}} \bigg[ u_{t+1}^{r}(\Omega^{r'}) \rho(\Omega^{r'})\bigg\{\sum_{\hat{o} \neq o'}\alpha(s_{\hat{o}}(.),o') \int_{\underline{z}} ^ {\overline{z}} \Big[ W^{u,SP}(\hat{z}, x_1,\hat{o}, A', p') \Big]  dF(\hat{z}) \non \\
+ (1- \sum_{\hat{o} \neq o'}\alpha(s_{\hat{o}}(.),o'))[b+\beta \mathbb{E}_{\omega''} R^{SP}(\omega'')] \bigg\}\bigg] dz' , \label{e:SP_max_reallstage1}
\end{align}
where we have applied the properties of the $z$-productivity process to change the order of integration relative to the third term of flow equation \eqref{uflow_p_t+1} and have integrated over all of next period's production stage states. Since we allow previous decisions to depend on the entire state space, $u_{t+1}^{r}(.)$ and $\rho(.)$ are written as a function of $\Omega^{r'}$ as they potentially depend on the distribution $\mathcal{G}^r$. Note, however, that the term inside the curly brackets in \eqref{e:SP_max_reallstage1} can be maximized separately for each $(z',x{_h}{'},o')$ and can therefore be summarized by $R^{gross,SP}(\omega')=R^{SP}(\omega')-c$, while the search intensity decision $\mathcal{S}(.)$ has solution vector $\{s(\omega', \tilde{o})\}$.

Noting that $u^s(.)=u^r(.)$ and $u^m(.)=(1-\rho(.))u^r(.)$ (see Section C.2 of this appendix), the above results imply we can express the term on the second line of equation \eqref{e: SP_VF_contraction_mapping} as
\begin{align}
 \beta \mathbb{E}_{\Omega^{s}{'}|\Omega^{p}}\Bigg[\Bigg(  \sum_{o=1} ^ {O} \sum_{h=1}^{H} \int_{\underline{z}} ^{\overline{z}}
& \bigg[u^{s}{'}( z', x_{h}' ,o)\Big\{\rho(.)R^{SP}(\omega')+(1-\rho(.))W^{m,SP}(\omega')\Big\} \\
& \quad + e^{s}{'}( z', x_{h}' ,o)\Big\{(1-d(.))M^{SP}(\omega')+d(.)W^{u,SP}(\omega')\Big\} \bigg]dz  \Bigg) \Bigg], \non
\end{align}
where it follows that the maximizing decisions $\rho(.)$ and $d(.)$ are also functions of $\omega'$, as these are the only other dependencies within the curly brackets.

Finally, given the properties of the shock processes, note that $u^s_{t+1}(.)$ and $e^s_{t+1}(.)$ are linear functions of $u^p(.)$ and $e^p(.)$, as demonstrated by equations \eqref{uflow_s_t+1} and \eqref{eflow_s_t+1} in Section C.2 of this appendix. This implies that we can express $T^{SP}W^{SP}(\Omega^p)$ as
\begin{equation}
T^{SP}W^{SP}(\Omega^p)=\sum_{o=1} ^ {O}\sum_{h=1}^{H} \int_{\underline{z}} ^{\overline{z}} \left[TW_{max}^{U,SP}(z,x_{h},o, A, p) u(z,x_{h},o) +TM^{SP}_{max}(z,x_{h},o, A, p)e(z,x_{h},o) \right] dz, \label{e: SP_VF_contraction_mapping2}
\end{equation}%
where $TW_{max}^{U,SP}$ is given by
\begin{align}
TW_{max}^{U,SP}(\omega) =& \max_{\rho(\omega'), \theta(\omega')}\bigg\{b+\beta \mathbb{E}_{\omega'|\omega}\bigg[\rho(\omega')\Big( \int_{\underline{z}} ^{\overline{z}} \max_{\mathcal{S(\omega')}} \Big[\sum_{\tilde{o}\neq o} \alpha(s_{\tilde{o}}(\omega', o))W^{U, SP}_{max}(\tilde{z},x_{1}, \tilde{o},A', p')dF(\tilde{z}) \non \\
& \ \ \qquad + (1-\sum_{\tilde{o}\neq o} \alpha(s_{\tilde{o}}(\omega', o)))[b+\beta \mathbb{E}_{\omega''|\omega'}R^{U, SP}_{max}(\omega'')\Big]-c\Big)  \label{e:TWmaxSP_contractionmapping} \\
& \ \ \qquad  +(1-\rho(\omega'))\Big[ \lambda (\theta(\omega'))( M^{SP}_{max}(\omega')-W^{U, SP}_{max}(\omega')) -\theta(\omega')k+W^{U,SP}_{max}(\omega')\Big]\bigg]\bigg\}, \non
\end{align}
and $TM_{max}^{SP}$ is given by
\begin{align}
TM_{max}^{SP}(p,x, z) & =\max_{d(\omega')}\bigg\{y(z,x_{h},A,p_o) +\beta \mathbb{E}_{\omega'|\omega}\bigg[\left( d(\omega')W^{U,SP}_{max}(\omega')+(1-d(\omega'))M^{SP}_{max}(\omega')\right) %
\bigg]\bigg\}. \label{e:TMmaxSP_contractionmapping}
\end{align}%
Hence we have established that the mapping $T^{SP}$ described in equation \eqref{e: SP_VF_contraction_mapping} maps functions $W^{SP}(.)$ from the space of functions linear with respect to the distribution $\mathcal{G}^p$ (of the form \eqref{e:linear_SP_VF_guess}) into itself.

It is now straightforward to show that if $TW_{max}^{U,SP}$, $TR_{max}^{SP}$ and $TM_{max}^{SP}$ are contraction mappings, then \eqref{e: SP_VF_contraction_mapping2} (and thereby \eqref{e: SP_VF_contraction_mapping}) is also a contraction mapping. Given the regularity properties assumed on the shock processes and following the proof of the decentralised case in Section C.4.1.1 of this appendix, we can show that a fixed point exists for \eqref{e:TWmaxSP_contractionmapping} and \eqref{e:TMmaxSP_contractionmapping}. Using equation \eqref{e: SP_VF_contraction_mapping2}, we then can construct the fixed point of the expression in $\eqref{e: SP_VF_contraction_mapping}$. It then follows from \eqref{e:TWmaxSP_contractionmapping} and \eqref{e:TMmaxSP_contractionmapping} that if the Hosios' condition holds, allocations of the fixed point of $T$ are allocations of the fixed point of $T^{SP}$, and hence the equilibrium allocations in the decentralised setting are also the efficient allocations.

\subsubsection{BRE gives the \emph{unique} equilibrium allocation} We now show that in any equilibrium, decisions and value functions only depend on $\omega=(z,x_h,o, A, p)$. We proceed using a contradiction argument. Suppose there is an alternative equilibrium in which values and decisions do not depend only on $\omega$, but also on an additional factor like the entire distribution of workers over employment status and $z$-productivities $\mathcal{G}$, or its entire history of observables, $H_{t}$. Consider the associated value functions in this alternative equilibrium, where the relevant state vector of the alternative equilibrium is given by $(\omega,H_{t})$. Our aim is to show that such an equilibrium cannot exist.

First suppose that in the alternative equilibrium all value functions are the same as in the BRE, but decisions differ at the same $\omega$. This violates the property that, in our setting, all maximizers in the BRE value functions are unique, leading to a contradiction. Now suppose that in the alternative equilibrium at least one value function differs from the corresponding BRE value function at the same $\omega$. It is straightforward to show that the expected values of unemployment must differ in both equilibria. Let $W^U(\omega,H_t)$ denote the value function for unemployed workers in the alternative equilibrium, and let $W^U(\omega)$ denote the corresponding value function in the BRE for the same $\omega$. Since in the proof of efficiency of a BRE we did not rely on the uniqueness of the BRE in the broader set of all equilibria, we can use the proved results of Section C.3.1.2 of this appendix here.

In particular, recall that the social planner's problem is entirely linear in the distribution of workers across states and hence $W^U(\omega)$ is the best the unemployed worker with $\omega$ can do (without transfers), including in the market equilibrium. Likewise, $M(\omega)$ is the highest value of the joint value of a match, including in the market equilibrium. Since value functions are bounded from above and from below and are continuous in their state variables, there exists a supremum of the difference between $W^U(\omega)$ and the candidate market equilibrium's $W^U(\omega, H_t)$, $\sup (W^U(\omega)-W^U(\omega,H_t))=\epsilon_u>0$. Similarly, there also exists a supremum for the difference between $M(\omega)$ and $M(\omega,H_t)$, $\sup (M(\omega)-M(\omega, H_t))=\epsilon_m>0$. In what follows, we will show that a difference in the value functions for unemployed workers (or the value functions for the joint value of the match) arbitrarily close to $\epsilon_u>0$ (or $\epsilon_m>0$) cannot occur. Otherwise, this will require that the difference in \emph{tomorrow's} values to be larger than $\epsilon_u$ (or $\epsilon_m$). In turn, this implies that an alternative equilibrium cannot exist.

From the above definition of supremum, it follows that
\[\max\{M(\omega), W^U(\omega)\}-\max\{M(\omega, H_t), W^U(\omega, H_t)\}<\max\{\epsilon_u, \epsilon_m\}.\]
Since
\[M(\omega)=y(z,x_{h},A,p_o)+\beta \Exp [\max\{M(\omega'),W^U(\omega')\}],\] and likewise for $M(\omega, H_t)$ it follows that at any $(\omega, H_t)$,
\begin{equation}\label{e:app_towards_contradiction}
M(\omega)-M(\omega, H_t))<\beta \max\{\epsilon_u, \epsilon_m\}.
 \end{equation}
Consider first the case in which $\epsilon_m\geq \epsilon_u$. For $(\omega, H_t)$ achieving a difference $M(\omega)-M(\omega, H_t)> \beta \epsilon_m$ is not possible since this will lead to a contradiction in equation \eqref{e:app_towards_contradiction} when $\epsilon_m>0$.

Next consider the case in which $\epsilon_m<\epsilon_u$. Simplifying notation by dropping $\omega$ and using the prime instead of $(\omega, H_t) $, we first establish an intermediate step. At \emph{any} $(\omega, H_t)$ it holds that
\begin{align}\label{e:app_towards_contriduction2}
\lambda(\theta)(1-\eta)M + (1-\lambda(\theta)(1-\eta))W<\lambda(\theta')(1-\eta)M' + (1-\lambda(\theta')(1-\eta))W'+\epsilon_u.
\end{align}
There are two cases to be analysed to show the above relationship.

\noindent {\bf{Case 1:}} Suppose that $(M'-W')\geq M-W$, then $\lambda(\theta')\geq \lambda(\theta)$. Define $K=(1-\eta)(\lambda(\theta')-\lambda(\theta))(M'-W')\geq 0$. Combining the latter with $\lambda(\theta)(1-\eta)(M-M') + (1-\lambda(\theta)(1-\eta))(W-W') \leq \epsilon_u$, it must be true that
\begin{align}\label{e:app_towards_contriduction3}
\lambda(\theta)(1-\eta)(M-M') + (1-\lambda(\theta)(1-\eta))(W-W')-K\leq \epsilon_u,
\end{align}
from which \eqref{e:app_towards_contriduction2} follows.

\noindent{\bf{Case 2:}} Suppose that $(M-W)>(M'-W')$, then $\lambda(\theta)>\lambda(\theta')$. From the derivative of $\frac{d}{d(M-W)}(\lambda(\theta(M-W))(1-\eta)(M-W))=\lambda(\theta(M-W))$, we can establish that, if $(M-W)>(M'-W')$,
\[\lambda(\theta)((M-W)-(M'-W'))>\lambda(\theta)(1-\eta)(M-W)-\lambda(\theta')(1-\eta)(M'-W')>\lambda(\theta')((M-W)-(M'-W')).\]
Since in any equilibrium (not only in the BRE) $\theta'$ depends only on $(M'-W')$ and constant parameters, we can use this relationship to establish that
\[W+\lambda(\theta)(M-W)-(W'+\lambda(\theta)(M'-W'))<\epsilon_u,\]
from which \eqref{e:app_towards_contriduction2} follows.

The final step is to consider an $(\omega, H_t)$ such that $\max\{\epsilon_m, \beta^{-1}\epsilon_u\}<W^U(\omega)-W^U(\omega,H_t)<\epsilon_u$, where such a $(\omega, H_t)$ exists by the definition of supremum. With this in hand it is straightforward to check that the difference in \emph{tomorrow}'s value (under the expectation sign), between $W^U(\omega)$ and $W^U(\omega, H_t)$ will not exceed $\epsilon_u$, since term-by-term, the difference is bounded by $\epsilon_u$. This also implies that today's difference, $W^U(\omega)-W^U(\omega,H_t)$, cannot be more than $\beta \epsilon_u>0$, which contradicts our premise. This establishes that a difference in the value functions for unemployed workers (or the value functions for the joint value of the match) arbitrarily close to $\epsilon_u>0$ (or $\epsilon_m>0$) cannot occur. Hence the BRE is the unique equilibrium.

This completes the proof of Proposition 2.

\subsection{Proof of Existence of a Reallocation and Separation cutoff}

\subsubsection{Reservation property of occupational mobility decisions, $z^{r}$} Here we show that $M(\omega)$ and $W^U(\omega)$ as derived in the proof of Proposition 2 are increasing in $z$. If $M(\omega)$ and $W^{U}(\omega)$ are continuous and bounded functions increasing in $z$, $T$ maps them into increasing (bounded and continuous) functions. For employed workers $(T(\Gamma(\omega,0))$, this follows since both $\max \{M(\omega^{'}),W^{U}(\omega^{'})\}$ and $y(.)$ are increasing in $z$, while the stochastic dominance of the $z$-productivity transition law implies higher expected $z$'s tomorrow. For unemployed workers $(T(\Gamma(\omega,1))$, note that the value of changing occupations $R(\omega)$ does not depend on the current $z$ of the worker, while equation \eqref{e:incr_z} implies that $D(M(\omega)-W^U(\omega))+W^U(\omega)$ is increasing in $z$. Again, given stochastic dominance of the tomorrow's $z$ when today's $z$ is higher, $T(\Gamma(\omega,1))$ is also increasing in $z$. The reservation property follows immediately, since $R(\omega)$ is constant in $z$ and $D(M(\omega)-W^U(\omega))+W^U(\omega)$ is increasing in $z$.

\subsubsection{Reservation property of job separation decisions, $z^{s}$} We now show that $M(\omega)-W^U(\omega)$ is increasing in $z$ when $\delta+\lambda(\theta(\omega))<1$ for the case of no human capital accumulation and occupational-wide shocks. In the calibration we show that this property holds also for the case of human capital accumulation and occupational-wide shocks.

Consider the same operator $T$ defined in the proof of Proposition 2, but now the relevant state space is given by $(A,z)$. Note that the value functions describing the worker's and the firm's problem, do not change, except for the fact that we are using a smaller state space. It is straightforward to verify that the derived properties of $T$ in Lemma A.1 also apply in this case. We now want to show that this operator maps the subspace of functions $\Gamma$ into itself with $M(A,z)$ increasing weakly faster in $z$ than $W^{U}(A,z)$. To show this, take $M(A,z)$ and $W^U(A,z)$ such that $M(A,z)-W^{U}(A,z)$ is weakly increasing in $z$ and let $z^s$ denote a reservation productivity such that for $z<z^s$ a firm-worker match decide to terminate the match. Using $\lambda (\theta)(M-W^{U})-\theta k=\lambda (\theta)(M-W^{U})-\lambda ^{\prime}(\theta)(M-W^{U})\theta=\lambda (\theta)(1-\eta)(M-W^{U})$, we construct the following difference
\begin{align}
T\Gamma&(A,z,0)-T\Gamma(A,z,1) = \label{e:appdiff}\\
 & y(A,z)-b+ \beta \mathbb{E}_{A,z} \Big[(1-\delta )\max \{M(A^{\prime }, z^{\prime })-W^{U}(A^{\prime }, z^{\prime },0\} - \nonumber \\
&  \max \left\{ \int W^{U}(A^{\prime },\tilde{z})dF(\tilde{z})-c-W^{U}(A^{\prime }, z^{\prime }),\lambda (\theta)(1-\eta )\big(M(A^{\prime }, z^{\prime })-W^{U}(A^{\prime }, z^{\prime })\big)\right\} \Big]. \nonumber
\end{align}
\noindent The first part of the proof shows the conditions under which $T\Gamma(A,z,0)-T\Gamma(A,z,1)$ is weakly increasing in $z$. Because the elements of the our relevant domain are restricted to have $W^U(A,z)$ increasing in $z$, and $M(A,z)-W^U(A,z)$ increasing in $z$, we can start to study the value of the term under the expectation sign, by cutting a number of different cases to consider depending on where $z'$ is relative to the implied reservation cutoffs.

\noindent \emph{-- Case 1.} Consider the range of tomorrow's $z'\in [\underline{z}(A'),z^r(A'))$, where $z^r(A')<z^s(A')$. In this case, the term under the expectation sign in the above equation reduces to $\mathbf{-}\int W^{U}(A^{\prime },\tilde{z})dF(\tilde{z})+c+W^{U}(A^{\prime }, z^{\prime })$, which is increasing in $z'$.

\noindent \emph{ -- Case 2.} Now suppose tomorrow's $z'\in [z^r(A'),z^s(A'))$. In this case, the term under the expectation sign becomes zero (as $M(A^{\prime },z^{\prime })-W^{U}(A^{\prime},z^{\prime })=0$), and is therefore constant in $z'$.

\noindent \emph{-- Case 3.} Next suppose that $z'\in [z^s(A'),z^r(A'))$. In this case, the entire term under the
expectation sign reduces to
\[(1-\delta)(M(A^{\prime }, z^{\prime})-W^{U}(A^{\prime},z^{\prime }))\mathbf{-}\int W^{U}(A^{\prime },\tilde{z})dF(\tilde{z})+c+W^{U}(A^{\prime }, z^{\prime }) ,\] and, once again, is weakly increasing in $z'$, because by supposition  $M(A',z')-W^{U}(A',z')$ is weakly increasing in $z'$, and so is $W^U(A', z')$ by Lemma A.1.

\noindent \emph{-- Case 4.} Finally consider the range of $z'\geq \max\{z^r(A'), z^s(A')\}$, such that in this range employed workers do not quit nor reallocate. In this case the term under the expectation sign equals %
\begin{equation}\label{e:appcase4}
(1-\delta )[M(A^{\prime }, z^{\prime })-W^{U}(A^{\prime }, z^{\prime})]-\lambda (\theta (A', z'))(1-\eta )[M(A^{\prime }, z^{\prime})-W^{U}(A^{\prime }, z^{\prime })].
\end{equation}
It is easy to show using the free entry condition that $\frac{d(\lambda (\theta ^{\ast }(A', z'))(1-\eta )[M(A^{\prime }, z^{\prime
})-W^{U}(A^{\prime }, z^{\prime })])}{d(M-W) }=\lambda(\theta(A',z'))$, and hence that the derivative of \eqref{e:appcase4} with respect to $z'$ is positive whenever $1-\delta-\lambda(\theta)\geq 0$.

Given $F(z'|z)<F(z'|\tilde{z})\ $for\ all $z,z'$ when $z>\tilde{z}$, the independence of $z$ of $A$, and that the term under the expectation sign are increasing in $z'$, given any $A'$, it follows that the integral in \eqref{e:appdiff} is increasing in today's $z$. Together with $y(A,z)$ increasing in $z$, it must be that $T\Gamma(A,z,1)-T\Gamma(A,z,0)$ is also increasing in $z$.

To establish that the fixed point also has increasing differences in $z$ between the first and second coordinate, we have to show that the space of this functions is closed in the space of bounded and continuous functions. In particular, consider the set of functions $\mathbb{F}\overset{def}{=}\{f\in \mathcal{C}|f:X\times Y\rightarrow \mathbb{R}^{2},|f(x,y,1)-f(x,y,2)|\text{ increasing in }y\},$ where $f(.,.,1),f(.,.,2)$ denote the first and second coordinate, respectively, and $\mathcal{C}$ the metric space of bounded and continuous functions endowed with the sup-norm.

The next step in the proof is to show that fixed point of $T\Gamma(A,z,0)-T\Gamma(A,z,1)$ is also weakly increasing in $z$. To show we first establish the following result.

\paragraph{Lemma B.1:}

\textit{$\mathbb{F}$ is a closed set in $\mathcal{C}$}
\begin{proof}
Consider an $f' \notin \mathbb{F}$ that is the limit of a sequence $\{f_n\}, f_n \in \mathbb{F}, \forall n \in \mathbb{N}$. Then there exists an $y_1< y$ such that $f'(x,y_1, 1)-f'(x,y_1, 2)>f'(x,y,1)-f'(x,y,2),$ while $ f_n(x,y_1,1)-f_n(x,y_1,2) \leq  f_n(x,y,1)-f_n(x,y,2) , $ for every $n$. Define a sequence $\{s_n\}$ with $s_n= f_n(x,y_1,1)-f_n(x,y_1,2) - (f_n(x,y,1)-f_n(x,y,2)) $. Then $s_n\geq 0,\ \forall n \in \mathbb{N}$. A standard result in real analysis guarantees that for any limit $s$ of this sequence, $s_n \to s$, it holds that $s\geq 0$. Hence $ f'(x,y_1, 1)-f'(x,y_1, 2) \leq  f'(x,y,1)-f'(x,y,2) $, contradicting the premise.
\end{proof}

Thus, the fixed point exhibits this property as well and the optimal quit policy is a reservation-$z$ policy given $1-\delta -\lambda (\theta)>0$. Since $y(A,z)$ is strictly increasing in $z$, the fixed point difference $M-W^U$ must also be strictly increasing in $z$. Furthermore, since $\lambda(\theta)$ is concave and positively valued, $\lambda^{\prime}(\theta)(M-W^U)=k$ implies that job finding rate is also (weakly) increasing in $z$.

\subsection{Proofs of the ``Model Implications and Comparative Statics''}

We now turn to the proofs of Lemmas 1, 2 and 3 presented in Section C.1. Recall that these results were derived using a simplified version of the model without occupational human capital accumulation, setting $x_{h}=1$ for all $h$, and without occupational-wide productivity shocks, setting $p_{o}=1$ for all $o$. These restrictions imply that the relevant state space is $(z,A)$. We further assumed that the $z$-productivity is redrawn randomly with probability $0<(1-\gamma)<1$ each period from cdf $F(z)$ and $A$ is held constant. In this stationary environment, the expected value of unemployment for a worker currently with productivity $z$ in occupation $o$ is given by
\begin{eqnarray}\label{a:wu}
W^U(A,z)&=&\gamma \Big(b + \beta \max \Big\{R(A), W^U(A,z)+ \max \{\lambda(\theta(A,z))(1-\eta)(M(A,z)-W^U(A,z)), 0\}\Big\}\Big)  \notag \\
 &+ &(1-\gamma)\Exp_{z}[W^U(A,z)].
\end{eqnarray}
where the expected value of occupational mobility is given by $R(A) = - c + \Exp_{z}[W^U(A,z)]$.
The values of employment at wage $w(A,z)$ for a worker currently with productivity $z$ and a firm employing this worker are given by
 \begin{equation}\label{a:we}
 W^E(A,z)=\gamma \big[ w(A,z) + \beta[ (1-\delta) W^E(A,z) + \delta W^U(A,z)] \big] +(1-\gamma)\Exp_{z} [W^E(A,z)],
\end{equation}
\begin{equation}\label{a:j}
J(A,z)=\gamma \big[ y(A,z)-w(A,z) + \beta[ (1-\delta) J(A,z))] \big] +(1-\gamma)\Exp_{z} [J(A,z)].
\end{equation}
The joint value of a match is then
\begin{align}\label{a:m}
M(A,z)=\gamma \big[ y(A,z) + \beta[ (1-\delta) M(A,z) + \delta W^U(A,z)] \big] +(1-\gamma)\Exp_{z} [M(A,z)].
\end{align}

\paragraph{Proof of Lemma 1} We state the detailed version of this lemma as Lemma H.1. To simplify the analysis and without loss of generality, in what follows we let $\delta=0$. Since we do not vary aggregate productivity to proof this lemma, we let $A=1$ and abusing notation we refer to $z$ as total output. This is done without loss of generality as now output only depends (and is strictly increasing and continuous) on the worker-occupation match-specific productivity. To abbreviate notation defined $W^s\equiv W^U(z^s)$.
\paragraph{Lemma H.1:} \textit{The differences $W^s-R$ and $z^r-z^s$, respond to changes in $c$, $b$ and $\gamma$ as follows}
\begin{enumerate}
\item \textit{(i) $W^s-R$ is strictly increasing in $c$ and (ii) $z^r-z^s$ is decreasing in $c$, strictly if $z^r>z^s$.}
\item \textit{(i) $W^s-R$ is strictly increasing in $b$ and (ii) $z^r-z^s$ is strictly decreasing in $b$.}
\item \textit{ (i) $W^s-R$ is strictly decreasing in $\gamma$ and (ii) $z^r-z^s$ is strictly increasing in $\gamma$.}
\end{enumerate}

We first consider the link between $W^s-R$ and $z^r-z^s$ as the parameter of interest $\kappa$ (i.e. $c$, $b$ or $\gamma$) changes. The reservation productivities for job separation and occupational mobility implicitly satisfy
\begin{align}
M(\kappa,z^s(\kappa))-W^s(\kappa)& =0  \label{e:l7_reservation_zs} \\
W^U(\kappa,z^r(\kappa))\!-\!R(\kappa)=  \lambda(\theta(\kappa,z^r(\kappa)))(1\!-\!\eta)(M(\kappa,z^r(\kappa))-W^s(\kappa))+(W^s(\kappa)\!-\!R(\kappa))&=0 \label{e:l7_reservation_zr1},
\end{align}
where \eqref{e:l7_reservation_zr1} only applies when $R(\kappa)\geq W^s(\kappa)$ and hence only when $z^r(\kappa) \geq z^s(\kappa)$. In the case of $W^s(\kappa)>R(\kappa)$, the assumed stochastic process for $z$ implies that $z^r(\kappa)=\underline{z}$. The latter follows as $W^U(z)=W^s(\kappa)$ for all $z<z^s$. Also note that we can use $W^s(\kappa)$ instead of $W(\kappa, z^r(\kappa))$ in \eqref{e:l7_reservation_zr1} as the assumed stochastic process for $z$ also implies that $W^U(z)=W^s(\kappa)$ for all $z<z^r$ when $R(\kappa)>W^s(\kappa)$.

Note that by defining
\begin{align*}
Z(\kappa) \equiv \Exp_z[\max\{M(\kappa,z),\!W^s(\kappa)\}-\max\{W^U(\kappa,z)\!+\!\lambda(\theta(\kappa,z))(1\!-\!\eta)(M(\kappa,\!z)\!-\!W^U(\kappa,\!z)),R(\kappa)\}],
\end{align*}
we can rewrite equations \eqref{e:l7_reservation_zs} and \eqref{e:l7_reservation_zr1} as
\begin{align*}
M(\kappa,z^s(\kappa))-W^s(\kappa)=z^s-b+\beta(1-\gamma)Z(\kappa)+\beta \gamma (W^s(\kappa)-R(\kappa))=0.
\end{align*}
Further note that
\begin{align*}
M(\kappa,z^r(\kappa))-W^s(\kappa) &=z^r-b+\beta(1-\gamma)Z(\kappa)+\beta \gamma (1-\lambda(\theta(\kappa, z^r))(1-\eta))(M(\kappa,z^r(\kappa))-W^s(\kappa)),
\end{align*}
which is strictly positive and changes with $\kappa$. Using \eqref{e:l7_reservation_zr1} leads to
\begin{align*}
(1-\beta \gamma)(M(\kappa,z^r(\kappa))-W^s(\kappa))=z^r-b+\beta(1-\gamma)Z(\kappa)+\beta \gamma(W^s(\kappa)-R(\kappa))
\end{align*}
Leaving implicit the dependency on $\kappa$, we obtain that
\[
\frac{d(M-W^s)}{d(W^s-R)}= \frac{d(M-W^s)}{d(\lambda(\theta)(1-\eta)(M-W^s))}\cdot\frac{d(\lambda)(1-\eta)(M-W^s)}{d(W^s-R)},
\]
where the latter term equals $-1$, and the former term equals $\lambda^{-1}(\theta)$. This follows as Nash Bargaining and free-entry imply $q(\theta)\eta(M-W^U)=c$ and hence $\frac{d\theta}{dM-W^U}=\frac{\theta}{(1-\eta)(M-W^U)}$.
Then $\lambda'(\theta)\frac{d\theta}{dM-W^U}(1-\eta)(M-W^U)+\lambda(\theta)(1-\eta)$ reduces to  $\lambda(\theta)$.
It then follows that when $z^r(\kappa) \geq z^s(\kappa)$, the derivative of $z^r$ with respect to $\kappa$ is given by
\begin{align}   \label{e:l7_derivative_z^r}
  \frac{dz^r(\kappa)}{d\kappa}&=-\frac{d}{d\kappa}\bigg[-b+\beta(1-\gamma)Z(\kappa)\bigg]-\left(\beta\gamma+ \frac{1-\beta\gamma}{\lambda(\theta(\kappa,z^r(\kappa)))}\right)\frac{d(W^s(\kappa)-R(\kappa))}{d\kappa}.
\end{align}
We can similarly obtain that when $z^r(\kappa) \geq z^s(\kappa)$, the derivative of $z^s$ with respect to $\kappa$ is given by
\begin{align} \label{e:l7_derivative_z^s}
  \frac{d z^s(\kappa)}{d\kappa}&=-\frac{d}{d\kappa}\bigg[-b+\beta(1-\gamma)Z(\kappa)\bigg] -   \beta \gamma \frac{d(W^s(\kappa)-R(\kappa))}{d\kappa}. \end{align}
Equations \eqref{e:l7_derivative_z^r} and \eqref{e:l7_derivative_z^s} then imply that when $z^r(\kappa) \geq z^s(\kappa)$ (i.e. $z^r$ and $z^s(\kappa)$ are interior), the derivative of $z^r(\kappa)-z^s(\kappa)$ with respect to $\kappa$ has the opposite sign to the derivative of $W^s(\kappa)-R(\kappa)$ with respect to $\kappa$.

As mentioned above, in the case of $W^s(\kappa)>R(\kappa)$ our simplified model implies $z^r(\kappa)=\underline{z}$. Hence changes in $\kappa$ can only affect $z^s$, although will affect $W^s(\kappa)$ and $R(\kappa)$. Using \eqref{a:wu} and \eqref{a:m} we obtain that for $W^s(\kappa)>R(\kappa)$ the reservation separation cutoff is given by
\begin{align*}
z^s(\kappa)=b-\frac{1-\gamma}{\gamma} \Exp_{z} [M(\kappa,z)-W^{U}(\kappa,z) ].
\end{align*}
Taking derivative with respect to $\kappa$ then lead to
\begin{align} \label{e:zs_rest}
\frac{d z^s(\kappa)}{d\kappa} &= \frac{d b}{d\kappa} - \frac{d ((1-\gamma)/\gamma)}{d\kappa} \Exp_{z} [M(\kappa,z)-W^{U}(\kappa,z) ] + \frac{1-\gamma}{\gamma} \frac{d \Exp_{z} [M(\kappa,z)-W^{U}(\kappa,z) ]}{d\kappa}.
\end{align}

To complete the lemma we now obtain the derivatives of $W^s(\kappa)-R(\kappa)$ and $\Exp_{z} [M(\kappa,z)-W^{U}(\kappa,z) ]$ with respect to $\kappa$.

\paragraph{Comparative Statics with respect to $c$:}

Consider the difference $W^s-R$ and values of $c$ such that $R\geq W^s$. In this case we have that
\begin{align*}
W^s=(1-\gamma)(R+c)+\gamma(b+\beta R), \\
W^s-R=-\gamma(1-\beta)R+(1-\gamma)c+\gamma b.
\end{align*}
Suppose towards a contradiction that $d(W^s-R)/dc<0$. The above equations imply that $\frac{dR}{dc}>\frac{(1-\gamma)}{\gamma(1-\beta)}>0$. We will proceed by showing that under $d(W^s-R)/dc<0$ both the expected match surplus (after a $z$-shock) and the match surplus for active labor markets (those with productivities that entail positive match surplus) decrease with $c$, which implies that the value of unemployment decreases with $c$, which in turn implies $\frac{dR}{dc}<0$, which is our contradiction.

Consider an active labor market with $W^U(z)>R$, the surplus on this labor market is then given by
\begin{align} \label{e:l7_1}
M(z)-W^U(z)&=\gamma(z-b + \beta (1-\lambda(\theta(z))(1-\eta))(M(z)-W^U(z))) \non \\
& \qquad +(1-\gamma)(\Exp[M(z)-W^U(z)]+(z-\Exp[z])),
\end{align}
where $\Exp[M(z)-W^U(z)]$ describes the expected surplus after a $z$-shock (after the search stage). Note that
\begin{equation}\label{e:espen_rule}
\frac{d}{d(M(z)-W^U(z))}(\lambda(\theta(z))(1-\eta)(M(z)-W^U(z)))=\lambda(\theta(z)).
\end{equation}
As described before, this result follows as (dropping the $z$ argument for brevity) our assumptions of Nash Bargaining and free-entry imply $(1-\eta)(M-W^U)=\frac{(1-\eta)}{\eta} J = \frac{1-\eta}{\eta}\frac{k}{q(\theta)}$, and hence $\lambda(\theta)(1-\eta)(M-W^U)=\frac{1-\eta}{\eta} k \theta$. Further, since $\frac{d\theta}{d(M-W^U)}=\frac{\eta}{1-\eta}\frac{\lambda(\theta)}{k}$, the chain rule then yields a derivative equal to $\lambda(\theta)$.
This result implies that from \eqref{e:l7_1} we obtain
\begin{align}\label{e:l7_2a}
0<\frac{d(M(z)-W^U(z))}{d(\Exp[M(z)-W^U(z)])}=\frac{1-\gamma}{1-\gamma \beta (1-\lambda(\theta(z)))}<1.
\end{align}

We now show that under $d(W^s-R)/dc<0$, the expected match surplus decreases in $c$. Note that the expected match surplus measured after the search stage is given by
\begin{align}\label{e:l7_3}
\Exp[M(z)-W^U(z)]=&\int_{z^r} z-b + \beta(1- \lambda(\theta(z))(1-\eta))(M(z)-W^U(z))dF(z) \non \\
& \ + \int_{z^s}^{z^r} z-b + \beta (M(z)-R)dF(z) + \int^{z^s} z-b + \beta (W^s-R)dF(z),
\end{align}
where the $(1-\gamma)$ shock integrates out. Our contradiction supposition implies that the third term of this expression is decreasing in $c$. The second term, $\int_{z^s}^{z^r}[M(z)-W^U(z)]dF(z)$, can be rewritten as
\[M(z)-W^s=\gamma(z-b+\beta(M(z)-W^s+W^s-R))+ (1-\gamma)(\Exp[M(z)-W^U(z)]+z-\Exp[z]),\]
and rearranging yields
\[M(z)-W^s=\frac{\gamma}{1-\gamma \beta} (z-b+\beta (W^s-R))+ \frac{1-\gamma}{1-\gamma \beta} \Exp[M(z)-W^U(z)],\]
where $\frac{\gamma}{1-\gamma \beta} (z-b+\beta (W^s-R))$ is decreasing in $c$ under our contradiction supposition. In the case of the first term in \eqref{e:l7_3}, note that \eqref{e:l7_2a} implies $M(z)-W^U(z)$ depends on $c$ through $\Exp[M(z)-W^U(z)]$. Combining all the elements, \eqref{e:l7_2a}, \eqref{e:l7_3} and the last two equations, we find that
\begin{align}\label{e:7_exp[m-wu]}
\frac{d\Exp[M(z)-W^U(z)]}{dc}&=\int_{z^r} \frac{(1-\gamma)\beta(1-\lambda(\theta(z)))}{1-\gamma \beta(1-\lambda(\theta(z)))}dF(z)\frac{d\Exp[M(z)-W^U(z)]}{dc} \non \\
& +(F(z^r)-F(z^s))\left(\frac{\gamma\beta}{1-\gamma \beta}\frac{d(W^s-R)}{dc} +\frac{1-\gamma}{1-\gamma \beta} \frac{d\Exp[M(z)-W^U(z)]}{dc}\right) \non  \\
& +F(z^s)\beta\frac{d(W^s-R)}{dc} \non \\
\iff \frac{d\Exp[M(z)-W^U(z)]}{dc} & = C \cdot \frac{d(W^s-R)}{dc}<0,
\end{align}
where $C$ is a positive constant. Equation \eqref{e:l7_2a} then implies that $\frac{d[M(z)-W^U(z)]}{dc}<0$.

Next consider $\frac{dW^U(z)}{dc}$ and $\frac{d\Exp[W^U(z)]}{dc}$. Note that for the case in which $z\leq z^r$, $W^U(z)=W^s=(1-\gamma)\Exp[W^U(z)]+\gamma (b+\beta \Exp[W^U(z)]-\beta c)$; while for $z>z^r$, $W^U(z)=(1-\gamma)\Exp[W^U(z)]+\gamma(b+\beta(\lambda(\theta(z))(1-\eta)(M(z)-W^U(z))+\beta W^U(z)))$. It then follows that
\[
\Exp[W^U(z)]=F(z^r)(b+\beta \Exp[W^U(z)]-\beta c)+\int_{z^r}(b+\beta \lambda(\theta(z))(1-\eta)(M(z)-W^U(z))+\beta W^U(z))dF(z),
\]
which combined with
\[
W^U=\frac{1-\gamma}{1-\beta \gamma}\Exp[W^U(z)]+\frac{\gamma }{1-\beta \gamma}(b+\beta\lambda(\theta(z))(1-\eta)(M(z)-W^U(z))),
\]
leads to
\begin{align*}
 & \left(1-\beta F(z^r)-\beta \frac{1-\gamma}{1-\beta \gamma}(1-F(z^r))\right)\Exp[W^U(z)] \non \\
   & \qquad=F(z^r)(b-\beta c) + \int_{z^r}\frac{b+\beta\lambda(\theta(z))(1-\eta)(M(z)-W^U(z))}{1-\beta \gamma}dF(z).
\end{align*}
Taking the derivative with respect to $c$, we find that both the first and second terms on the RHS of the last expression decrease with $c$, the latter because we have established before that $\frac{d(M(z)-W^U(z))}{dc}<0$. It then follows that $\frac{d\Exp[W^U(z)]}{dc}<0$ and hence $\frac{dR}{dc}=\frac{d\Exp[W^U(z)]}{dc}-1<0$, which yields our desired contradiction. Equations \eqref{e:l7_derivative_z^r} and \eqref{e:l7_derivative_z^s} then imply that $\frac{d (z^r(c)-z^s(c))}{dc}<0$ when $z^r>z^s$.

Now consider the difference $W^s-R$ and values of $c$ such that $R<W^s$. In this case rest unemployment implies $W^s=\gamma(b+\beta W^s)+(1-\gamma)\Exp[W^U(z)]$. Note that here $\frac{dW^s}{dc}=0$, since workers with productivities $z \leq z^s$ will never change occupations. Doing so implies paying a cost $c>0$ and randomly drawing a new productivity, while by waiting a worker obtains (with probability $1-\gamma$) a free draw from the productivity distribution. Hence, $d(W^s-R)/dc= -dR/dc$. Since workers with $z>z^s$ prefer employment in their current occupation, the above arguments imply that when $R<W^s$ the expected value of unemployment, $W^U(z)$, is independent of the value the worker obtains from sampling a new $z$ in a different occupation. It then follows that $\frac{dR}{dc}=\frac{d\Exp[W^U(z)]}{dc}-1=-1<0$, which once again yields are desired contradiction. Further note that in this case $\frac{d \Exp_{z} [M(c,z)-W^{U}(c,z) ]}{dc}=0$ and hence \eqref{e:zs_rest} implies that $\frac{d z^s(c)}{dc}=0$ and $\frac{d (z^r(c)-z^s(c))}{dc}=0$ when $z^s>z^r$.

\paragraph{Comparative Statics with respect to $b$:}
We proceed in the same way as in the previous case. Consider the difference $W^s-R$ such that $R\geq W^s$. Expressing $W^s$ and $W^U$, for $z>z^s$, as
\begin{align} \label{e:l7_7a}
W^s &= (1-\gamma)\Exp[W^U(z)]+\gamma(b+\beta (R-W^s)) + \gamma \beta W^s \\
W^U(z) &=  (1-\gamma)\Exp[W^U(z)]+\gamma(b+\beta (\lambda(\theta)(1-\eta)(M(z)-W^U(z)))) + \gamma \beta W^U(z), \label{e:l7_7b}
\end{align}
we find that $W^s - \Exp[W^U(z)]=\int_{z^r} (W^s-W^U(z))dF(z)$, which in turn implies
\begin{align}\label{e:l7_b_ws-r}
W^s-R=\frac{1}{1-\gamma \beta F(z^r)} \left(-\beta \gamma \int_{z^r} \lambda(\theta(z))(1-\eta)(M(z)-W^U(z))dF(z)+(1-\gamma \beta)c \right).
\end{align}
That is, the difference $W^s-R$ decomposes into (i) the forgone option of searching for a job in the new occupation next period (first term in the brackets) and (ii) a sampling cost that only has to be incurred next period with probability $\gamma$, and discounted at rate $\beta$ (second term in the brackets).

Next consider the relationship between $M(z)-W^U(z)$ and $\Exp[M(z)-W^U(z)]$. From \eqref{e:espen_rule} and \eqref{e:l7_2a}, we find that
\begin{align}\label{e:l7_b_m-wu_e(m-wu)}
\frac{d(M(z)-W^U(z))}{db}=\frac{1-\gamma}{1-\gamma \beta (1-\lambda(\theta(z)))}\frac{d\Exp[M(z)-W^U(z)]}{db} - \frac{\gamma}{1-\gamma \beta (1-\lambda(\theta(z)))}.
\end{align}
Note that $\frac{d(M(z)-W^U(z))}{db}$ must have the same sign for all $z$, which is positive if and only if \[\frac{d\Exp[M(z)-W^U(z)]}{db} >\frac{\gamma}{1-\gamma}.\]

Towards a contradiction, suppose $d(W^s-R)/db<0$. Then, we have $\frac{d(W^s-R)}{db}=\frac{d(W^s-\Exp[W^U(z)])}{db}$, which equals $\frac{d}{db} \left(-\int_{z^r} \max\{W^U(z)-W^s,0\}dF(z)\right)$. By the envelope condition, the effect $\frac{dz^r}{db}$ disappears. By the previous argument and \eqref{e:l7_7a} subtracted by \eqref{e:l7_7b}, it follows that $\frac{d(W^s-R)}{db}<0$ implies $\frac{d(M(z)-W^U(z))}{db}>0$ and by (\ref{e:l7_b_m-wu_e(m-wu)}) that $\frac{d\Exp[M(z)-W^U(z)]}{db}>0$.

Using the the same arguments as in \eqref{e:7_exp[m-wu]} we find that
\begin{align}\label{e:7_exp[m-wu]2}
\frac{d\Exp[M(z)-W^U(z)]}{db}&=-1 + \int_{z^r} \frac{\beta(1-\lambda(\theta(z)))-\gamma\beta(1-\lambda(\theta(z)))}{1-\gamma \beta(1-\lambda(\theta(z)))}dF(z)\frac{d\Exp[M(z)-W^U(z)]}{db} \non \\
& -\int_{z^r} \frac{\gamma\beta(1-\lambda(\theta(z)))}{1-\gamma \beta(1-\lambda(\theta(z)))}dF(z)\frac{d\Exp[M(z)-W^U(z)]}{db} \non \\
&  +(F(z^r)-F(z^s))\left(\frac{\gamma\beta^2}{1-\gamma \beta}\frac{d(W^s-R)}{db} +\frac{\beta(1-\gamma)}{1-\gamma \beta} \frac{d\Exp[M(z)-W^U(z)]}{db}\right) \non  \\
& -(F(z^r)-F(z^s))\frac{\gamma \beta}{1-\gamma \beta} +F(z^s)\beta\frac{d(W^s-R)}{db} \\
\Longrightarrow \frac{d\Exp[M(z)-W^U(z)]}{db} & = C_2 \cdot \frac{d(W^s-R)}{db} -C_3<0,  \non
\end{align}
where $C_2$ and $C_3$ are positive constants. The fact that the last expression leads to $\frac{d\Exp[M(z)-W^U(z)]}{db}<0$ yields our desired contradiction. Equations \eqref{e:l7_derivative_z^r} and \eqref{e:l7_derivative_z^s} then imply that $\frac{d (z^r(b)-z^s(b))}{db}<0$ when $z^r>z^s$.

Next we consider the case in which $W^s>R$ and note that in this case equation \eqref{e:l7_b_ws-r} becomes
\begin{align*}
  W^s-\Exp[W^U(z)]=-\frac{\beta \gamma}{1-\beta \gamma} \int_{z^s} \lambda(\theta(z))(1-\eta)(M(z)-W^U(z))dF(z).
\end{align*}
As before, if we start from the premise that $d(W^s-R)/db<0$, this will imply, by virtue of \eqref{e:l7_b_m-wu_e(m-wu)}, that $\frac{d\Exp[M(z)-W^U(z)]}{db}>0$. Noting that in this case equation \eqref{e:l7_3} reduces to
\begin{align*}
\Exp[M(z)-W^U(z)]=&\int_{z^s} z-b + \beta \lambda(\theta(z))(1-\eta)(M(z)-W^U(z))dF(z),
\end{align*}
and \eqref{e:7_exp[m-wu]2} reduces to
\begin{align*}
\frac{d\Exp[M(z)-W^U(z)]}{db}&=-1 + \int_{z^s} \frac{\beta(1-\lambda(\theta(z)))-\gamma\beta(1-\lambda(\theta(z)))}{1-\gamma \beta(1-\lambda(\theta(z)))}dF(z)\frac{d\Exp[M(z)-W^U(z)]}{db} \non \\
& \qquad -\int_{z^s} \frac{\gamma\beta(1-\lambda(\theta(z)))}{1-\gamma \beta(1-\lambda(\theta(z)))}dF(z)\frac{d\Exp[M(z)-W^U(z)]}{db},
\end{align*}
we obtain that $\frac{d\Exp[M(z)-W^U(z)]}{db}<0$, once again yielding our desired contradiction. Equation \eqref{e:zs_rest} then imply that $\frac{d (z^r(b)-z^s(b))}{db}<0$ also when $z^s>z^r$.

\paragraph{Comparative Statics with respect to $\gamma$:} Here we also proceed in the same way as before by assuming (towards a contradiction) that $d(W^s-R)/d\gamma>0$. We start with the case in which $R\geq W^s$. Using equation \eqref{e:l7_b_ws-r} we find that
\begin{align*}
  \frac{d(W^s-R)}{d\gamma}=&\frac{\beta F(z^r)}{1-\beta \gamma}(W^s-R)-\frac{1}{1-\beta \gamma F(z^r)}\left(\int_{z^r} \lambda(\theta(z))(1-\eta)(M(z)-W^U(z))dF(z)+\beta c\right) \non \\
  & \qquad -\int_{z^r}\beta \gamma \lambda(\theta(z))\frac{d(M(z)-W^U(z))}{d\gamma}dF(z).
\end{align*}
It then follows that
\begin{align}\label{e:l7_c_premise_derivative}
  &-\int_{z^r}\beta \gamma \lambda(\theta(z))\frac{d(M(z)-W^U(z))}{d\gamma}dF(z) \geq \frac{\beta F(z^r)}{1-\beta \gamma}(R-W^s)\non \\
  &  +\frac{1}{1-\beta \gamma F(z^r)}\left(\int_{z^r} \lambda(\theta(z))(1-\eta)(M(z)-W^U(z))dF(z)+\beta c\right)>0.
\end{align}

Using the above expressions we turn to investigate the implications of assuming $d(W^s-R)/d\gamma>0$ for $\frac{d\Exp[M(z)-W^U(z)]}{d\gamma}$. We can rewrite \eqref{e:l7_3}, bringing next period's continuation values to the LHS, as
\begin{align*}
(1-\beta)\Exp[M(z)-W^U(z)]=&\int_{z^r} z-b - \beta \lambda(\theta(z))(1-\eta)(M(z)-W^U(z))dF(z) \non \\
&  + \int_{z^s}^{z^r} z-b + \beta (W^s-R)dF(z) \non \\
&  + \int^{z^s} z-b + \beta (W^s-R)-\beta (M(z)-W^s)dF(z).
\end{align*}
Taking derivatives with respect to $\gamma$, we find
\begin{align}\label{e:l7_em-wu_c_derivative}
(1-\beta)\frac{d\Exp[M(z)-W^U(z)]}{d\gamma}=&- \int_{z^r}\beta\lambda(\theta(z))(1-\eta)\frac{d(M(z)-W^U(z))}{d\gamma}dF(z) \non \\
&  + \int_{z^s}^{z^r} \beta \frac{d(W^s-R)}{d\gamma}dF(z) \non \\
&  +  \int^{z^s} \beta \Big(\frac{d(W^s-R)}{d\gamma}- \frac{d(M(z)-W^s)}{d\gamma}  \Big)dF(z),
\end{align}
where the first term is positive by virtue of \eqref{e:l7_c_premise_derivative} and the second term is positive by assumption. For the third term it holds that
{\small
\begin{align*}
  \frac{d(M(z)-W^s)}{d\gamma}=&(1-\gamma)\frac{d\Exp[M(z)-W^U(z)]}{d\gamma}+\gamma \beta \frac{d(W^s-R)}{d\gamma}  +(z-b+\beta (W^s-R)-\Exp[M(z)-W^U(z)]).
\end{align*}}
Substituting out this expression in the third line of the RHS of \eqref{e:l7_em-wu_c_derivative} and re-arranging implies that $\frac{d\Exp[M(z)-W^U(z)]}{d\gamma}>0$.

From $M(z)-W^U(z)=(1-\gamma)\Exp[M(z)-W^U(z)]+\gamma(z-b +\beta(1-\lambda(\theta(z))(1-\eta))(M(z)-W^U(z))$, it follows that for $z>z^r$
\begin{align}
  \beta \gamma \lambda(\theta) \frac{dM(z)-W^U(z)}{d\gamma}=& \frac{\beta \gamma \lambda(\theta(z))}{1-\beta \gamma (1-\lambda(\theta(z)))}\bigg(\Big(z-b +\beta(1-\lambda(\theta(z))(1-\eta))(M(z)-W^U(z))\non \\
  & \qquad \qquad -\Exp[M(z)-W^U(z)]\Big)+(1-\gamma)\frac{d\Exp[M(z)-W^U(z)]}{d\gamma} \bigg), \label{e:l7_c_M(y)-w^u(y)_derivative}
\end{align}
where we have used \eqref{e:espen_rule} and \eqref{e:l7_2a}. Integrating this term over all $z>z^r$, we have
{\small
\begin{align}
  \int_{z^r}  \beta \gamma \lambda(\theta(z))\frac{d(M(z)-W^U(z))}{d\gamma}dF(z)&\geq
  \frac{\beta \gamma \lambda(\theta(z^r))}{1-\beta \gamma (1- \lambda(\theta(z^r)))} \bigg(\int_{z^r} (1-\gamma)\frac{d\Exp[M(z)-W^U(z)]}{d\gamma}dF(z) \non \\
  &   + \frac{1}{\gamma}\int_{z^r} (M(z)-W^U(z)-\Exp[M(z)-W^U(z)])dF(z)\bigg) > 0, \label{e:l7_c_int derivative m(y)-w^u}
\end{align}}
where the last inequality follows from the fact that $M(z)-W^U(z)-\Exp[M(z)-W^U(z)]$ and $\frac{\beta \lambda(\theta(z))}{1-\beta\gamma+\beta \gamma \lambda(\theta(z))}$ are increasing in $z$. Then $\int_{z^r} (M(z)-W^U(z)-\Exp[M(z)-W^U(z)])dF(z)>0$ and hence the LHS of \eqref{e:l7_c_int derivative m(y)-w^u} is positive as stated, contradicting our premise in \eqref{e:l7_c_premise_derivative}. Equations \eqref{e:l7_derivative_z^r} and \eqref{e:l7_derivative_z^s} then imply that $\frac{d (z^r(\gamma)-z^s(\gamma))}{d \gamma}>0$ when $z^r>z^s$.

Next we turn to investigate the implications of assuming $d(W^s-R)/d\gamma>0$ on $\frac{d\Exp[M(z)-W^U(y)]}{d\gamma}$ for the case in which $W^s>R$ to obtain a contradiction. Here a relationship between $\frac{d(M(z)-W^U(z))}{d\gamma}$ and $\frac{d\Exp[M(z)-W^U(z)]}{d\gamma}$ can be derived directly:
\begin{align}\label{e:l7_em-wu_c_derivative_2}
(1-\beta)\frac{d\Exp[M(z)-W^U(z)]}{d\gamma}=&-\beta \int_{z^r}\lambda(\theta(z))\frac{d(M(z)-W^U(z))}{d\gamma}dF(z).
\end{align}
Using the above and \eqref{e:l7_c_M(y)-w^u(y)_derivative} we obtain that
{\small
\begin{align*}
  -(1-\beta)\frac{d\Exp[M(z)-W^U(z)]}{d\gamma}=& \int_{z^s}\bigg(\frac{\beta \gamma \lambda(\theta(z))}{1-\beta \gamma (1-\lambda(\theta(z)))}\bigg(\Big(z-b +\beta(1-\lambda(\theta(z))(1-\eta))(M(z)-W^U(z))\non \\
  & \qquad \qquad -\Exp[M(z)-W^U(z)]\Big)+(1-\gamma)\frac{d\Exp[M(z)-W^U(z)]}{d\gamma}\bigg)\bigg)dF(z),
  \end{align*}}
which in turn can be expressed as
\begin{align}\label{e:l7_c_M(y)-w^u(y)_derivative_3}
  \frac{d\Exp[M(z)-W^U(z)]}{d\gamma}&\Bigg((1-\beta) +\frac{F(z^s)\beta \gamma \lambda(\theta(z))(1-\gamma)}{1-\beta\gamma+\beta \gamma \lambda(\theta(z))}\Bigg)= \non \\ & \qquad -\int_{z^s}\bigg(\frac{\beta \gamma \lambda(\theta(z))}{1-\beta \gamma (1-\lambda(\theta(z)))}\Big(z-b +\beta(1-\lambda(\theta(z))(1-\eta))(M(z)-W^U(z))  \non \\
  & \qquad -\Exp[M(z)-W^U(z)]\Big)\Big)dF(z) < 0.
\end{align}
From \eqref{e:l7_c_M(y)-w^u(y)_derivative_3}, it then follows that $\frac{d\Exp[M(z)-W^U(z)]}{d\gamma}<0$. Using equation \eqref{e:l7_em-wu_c_derivative_2}, we obtain that $W^s-R$ is decreasing in $\gamma$, where
\begin{align*}
\frac{d(W^  s-R)}{d\gamma}=\frac{d(W^s-\Exp[W^U(z)])}{d\gamma}=-\frac{1}{1-\beta}\int_{z^s}\gamma \lambda(\theta(z)) \frac{d(M(z)-W^U(z))}{d\gamma}<0,
\end{align*}
which leads to our required contradiction. Equation \eqref{e:zs_rest} then imply that $\frac{d (z^r(\gamma)-z^s(\gamma))}{d\gamma}>0$ also when $z^s>z^r$. This complete the proof for Lemma H.1.

\paragraph{Proof of Lemma 2}
Using the same setting as in Lemma H.1 we now introduce human capital $x$, assuming it enters in a multiplicative way in the production function. Keeping the same notation as in Lemma H.1, let $A=1$ such that total output is given by $zx$. Without loss of generality for the results derived in this lemma, normalize $x=1$. If we have an incremental improvement in $x$ that is occupational specific, the value of sampling will stay constant, at $R=\Exp[W^U(1,z)]-c$. However, the value of $W^s(x)$ will increases with $x$. Since $W^s(x)=(1-\gamma)\Exp[W^U(x,z)]+\gamma(b+\beta \max\{R, W^s(x)\})$ it follow that $W^s(x)$ is increasing in $x$ through $\Exp[W^U(x,z)]$.

To investigate the dependence between $W^s(x)$ and $\Exp[W^U(x,z)]$ when $x$ changes we first suppose that $R\geq W^s(x)$. The value of unemployment when $z\geq z^r(x)$ is given by
\begin{align*}
    W^U(x,z)&=(1-\gamma)\Exp[W^U(x,z)]+\gamma(b+\beta\lambda(\theta(x,z)))(1-\eta)(M(x,z)-W^U(x,z))+\beta W^U(x,z)),
\end{align*}
while for $z^r(x)>z$ it is given by
\begin{align*}
     W^s(x)&=(1-\gamma)\Exp[W^U(x,z)]+\gamma(b+\beta R).
\end{align*}
When comparing the expected value of separating with the expected value of moving to another occupations (reseting $x=1$), the difference is given by
\begin{align*}
W^s(x)-\Exp[W^U(1,z)]& = (1-\gamma) \Exp[W^U(x,z)]+C_{4},
\end{align*}
where $C_{4}$ denotes those terms that are constant in $x$.

As a result, $\frac{d(W^s(x)-\Exp[W^U(1,z)])}{dx}=(1-\gamma) \frac{d\Exp[W^U(x,z)]}{dx}$, where $\Exp[W^U(x,z)]$ can be expressed as
{\small
\begin{align}
\Exp[W^U(x,z)]=F(z^r(x))(b+\beta R)+\int_{z^r(x)} \Big[b+\beta \lambda(\theta(x,z))(1-\eta)(M(x,z)-W^U(x,z)) + \beta W^U(x,z)\Big]dF(z). \non
\end{align}}
Using the expression $W^U(x,z)$, it then follows that
\begin{align*}
 & \left(1- \frac{\beta(1-\gamma)}{1-\beta \gamma}(1-F(z^r(x)))\right)\Exp[W^U(x,z)] \non \\
   & \qquad=F(z^r(x))(b+\beta R) + \frac{1-F(z^r(x))}{1-\beta \gamma} b + \int_{z^r(x)}\frac{\beta\lambda(\theta(x,z))(1-\eta)(M(x,z)-W^U(x,z))}{1-\beta \gamma}dF(z).
\end{align*}
Taking the derivative with respect to $x$ and using the envelope condition, which implies that the term premultiplying $dz^r(x)/dx$ equal zero, yields
\begin{align}\label{e:l7_x_dew^u/dx}
&\frac{d\Exp[W^U(x,z)]}{dx}= \frac{\beta\frac{d}{dx}\big(\int_{z^r(x)} [\lambda(\theta(x,z)) (1-\eta)(M(x,z)-W^U(x,z))]dF(z)\big)}{1-\beta(1-(1-\gamma)F(z^r(x)))}.
\end{align}
Since the denominator is positive, the sign of $d\Exp[W^U(x,z)]/dx$ is given by the sign of its numerator. By virtue of the envelope condition and equation \eqref{e:espen_rule} in the proof of Lemma 1, the latter is given \[\int_{z^r(x)} \beta \lambda(\theta(x,z)) \frac{d(M(x,z)-W^U(x,z))}{dx}dF(z),\] where
\begin{align}
\frac{d(M(x,z)-W^U(x,z))}{dx}&=(1-\gamma)\frac{d\Exp[M(x,z)-W^U(x,z)]}{dx}+(1-\gamma)(z-\Exp[z])+\gamma z \non \\&
\qquad \qquad  \qquad + \beta \gamma (1-\lambda(\theta(x,z)))\frac{d(M(x,z)-W^U(x,z))}{dx}. \label{e:l7_dm-w^u/dx_y greater yr}
\end{align}
Hence the numerator of \eqref{e:l7_x_dew^u/dx} is given by
{\small
\begin{align}\label{e:derv_num}
  \int_{z^r(x)}\left(\frac{\beta \lambda(\theta(x,z))(1-\gamma)}{1-\beta \gamma (1-\lambda(\theta(x,z)))}\left(\frac{d\Exp[M(x,z)-W^U(x,z)]}{dx}+z-\Exp[z]\right)+\frac{\beta \lambda(\theta(x,z)) \gamma }{1-\beta \gamma (1-\lambda(\theta(x,z)))}z \ \right) \ dF(z).
\end{align}}
It is then clear that the sign of $d\Exp[W^U(x,z)]/dx$ is the same as the sign of $d\Exp[M(x,z)-W^U(x,z)]/dx$. To investigate the latter note that
{\small
\begin{align}\label{e:l7_x_expected surplus full}
\Exp[M(x,z)-W^U(x,z)]&=\int_{\underline{z}}^{\bar{z}} (xz-b) dF(z) + \int_{z^r(x)} \beta \lambda(\theta(x,z))(1-\eta)(M(x,z)-W^U(x,z))dF(z)  \non \\
&\ \  +\int_{z^s(x)}^{z^r(x)}\beta (M(x,z)-W^s(z))dF(z) +  \int^{z^r(x)}\beta(W^s(x)-R)dF(z).
\end{align}}
We will now take the derivative of this expression with respect to $x$ to investigate its sign. For this purpose it is useful to note that
{\small
 \begin{align*}
\int_{z^s(x)}^{z^r(x)} \frac{\beta d(M(x,y)-W^s(x))}{dx} dF(z)&= \beta \int_{z^s(x)}^{z^r(x)} \Big [(1-\gamma)\frac{d\Exp[M(x,y)-W^U(x,y)]}{dx}+(1-\gamma)(y-\Exp[y])+\gamma y \non \\
&  \qquad \qquad + \beta \gamma \frac{d(M(x,y)-W^s(x))}{dx} +\beta \gamma \frac{d(W^s(x)-R)}{dx} \Big ]dF(z),
\end{align*}}
and that
{\small
\begin{align}
\int^{z^r(x)}\frac{\beta d(W^s(x)-R))}{dx} dF(z)&=\frac{\beta(1-\gamma)F(z^r(x))}{1-\beta(1-(1-\gamma)F(z^r(x)))} \times  \label{e:l7_dm-w^u/dx_y smaller ys} \\
& \ \ \int_{z^r}\bigg [ \frac{\beta \lambda(\theta(x,z))(1-\gamma)}{1-\beta \gamma (1-\lambda(\theta(x,z)))}\left(\frac{d\Exp[M(x,z)-W^U(x,z)]}{dx}+z-\Exp[z]\right) \non \\
&+\frac{\beta \lambda(\theta(x,z)) \gamma }{1-\beta \gamma (1-\lambda(\theta(x,z)))}z \bigg]dF(z). \non
\end{align}}
Using \eqref{e:derv_num} and the above equations we obtain that  $\frac{d\Exp[M(x,z)-W^U(x,z)]}{dx}=$
{\small
\begin{align}
&\Exp(z)+ \int_{z^r(x)}\left(\frac{\beta \lambda(\theta(x,z))(1-\gamma)}{1-\beta \gamma (1-\lambda(\theta(x,z)))}\left(\frac{d\Exp[M(x,z)-W^U(x,z)]}{dx}+z-\Exp[z]\right)+\frac{\beta \lambda(\theta(x,z)) \gamma }{1-\beta \gamma (1-\lambda(\theta(x,z)))}z \ \right)dF(z) \non \\
&+ \frac{\beta}{1-\beta\gamma}  \int_{z^s(x)}^{z^r(x)} \Big[(1-\gamma)\left(\frac{d\Exp[M(x,z)-W^U(x,z)]}{dx}+z-\Exp[z]\right)  +\gamma z \Big]dF(z)\non \\
&+\left(\frac{\beta}{1-\beta \gamma}\frac{ (1-\gamma)F(z^r(x))}{1-\beta(1- (1-\gamma)F(z^r(x)))} \times \right. \label{e:l7_dm-w^u/dx_y between ys and yr} \\
& \ \ \left. \int_{z^r(x)}\left(\frac{\beta \lambda(\theta(x,z))(1-\gamma)}{1-\beta \gamma (1-\lambda(\theta(x,z)))}\left(\frac{d\Exp[M(x,z)-W^U(x,z)]}{dx}+z-\Exp[z]\right)+\frac{\beta \lambda(\theta(x,z)) \gamma }{1-\beta \gamma (1-\lambda(\theta(x,z)))}z \ \right) \ dF(z)\right). \non
\end{align}}
The next step is to investigate whether the sum of all terms premultiplying $\frac{d\Exp[M(x,z)-W^U(x,z)]}{dx}$ in \eqref{e:l7_dm-w^u/dx_y between ys and yr} is less than 1. If this is the case, grouping all these terms and solving for $d\Exp[M(x,z)-W^U(x,z)]/dx$ will imply that $d\Exp[M(x,z)-W^U(x,z)]/dx>0$ as the reminder terms in the RHS of \eqref{e:l7_dm-w^u/dx_y between ys and yr} are positive because the integrating terms $z-\Exp[z]$ will also yield a positive term.

We proceed by noting that the terms premultiplying $\frac{d\Exp[M(x,z)-W^U(x,z)]}{dx}$ in the last three lines of \eqref{e:l7_dm-w^u/dx_y between ys and yr} are larger than in \eqref{e:l7_dm-w^u/dx_y smaller ys}. By replacing the premultiplication term in \eqref{e:l7_dm-w^u/dx_y smaller ys} with the corresponding term in \eqref{e:l7_dm-w^u/dx_y between ys and yr} we will show that the entire term premultiplying $\frac{d\Exp[M(x,z)-W^U(x,z)]}{dx}$ in \eqref{e:l7_dm-w^u/dx_y between ys and yr} is less than one.
Some algebra establishes that to show the latter we need to verify that
\begin{align}\label{e:l7_desideratum}
  \beta \left(\frac{d(M(x,y)-W^U(x,y))}{dx}+\frac{d(W^s-R)}{dx}\right)F(z^r(x))<1-\beta(1-\gamma)(1-F(z^r(x))).
\end{align}
By collecting the terms premultiplying $\frac{d\Exp [M(x,z) - W^U(x,z)] }{dx}+ z-\Exp[z]$, and substituting these into the LHS of \eqref{e:l7_desideratum}, we obtain that
\begin{align}
&F(z^r)\left(\frac{\beta(1-\gamma)(1-\beta+\beta(1-\gamma)F(z^r(x)))}{(1-\gamma\beta)(1-\beta+\beta(1-\gamma)F(z^r(x)))}
+\frac{\beta(1-\gamma)(\beta(1-\gamma)(1-F(z^r(x))))}{(1-\gamma\beta)(1-\beta+\beta(1-\gamma)F(z^r(x)))} \right) \non \\
& \quad = \frac{\beta(1-\gamma)(1-\beta+\beta(1-\gamma))}{(1-\gamma\beta)(1-\beta+\beta(1-\gamma)F(z^r(x)))}F(z^r(x)). \label{e:l7_x_toprove_LHS}
\end{align}
The RHS of \eqref{e:l7_desideratum} can be rewritten as $1-\beta+\beta \gamma+\beta(1-\gamma)F(z^r(x))$. Noting that
\begin{align}
  \beta \gamma&> \frac{\beta \gamma (\beta(1-\gamma))\Big(1-\beta+\beta(1-\gamma)\Big)F(z^r(x))}{(1-\gamma \beta)(1-\beta+\beta (1-\gamma) F(z^r(x)))}\label{e:l7_x_addup1}\\
  \beta(1-\gamma)F(z^r(x))&>\frac{\beta(1-\gamma)(1-\gamma \beta)\big(1-\beta+\beta(1-\gamma)\big)F(z^r(x))}{(1-\gamma \beta)(1-\beta+\beta F(z^r(x)))}, \label{e:l7_x_addup2}
\end{align}
and adding them up we find that the RHS is precisely the term in \eqref{e:l7_x_toprove_LHS}. Therefore $\beta \gamma + \beta(1-\gamma)F(z^r(x))$ is larger than \eqref{e:l7_x_toprove_LHS}, from which the desired result follows as the remaining term, $1-\beta$, is larger than zero and the desired inequality is slack. This yields $d\Exp [M(x,z) - W^U(x,z)]/dx>0$. It then follows from \eqref{e:l7_dm-w^u/dx_y greater yr} that $d(M(x,z)-W^U(x,z))/dx>0$ and therefore, by \eqref{e:l7_x_dew^u/dx}, that $d\Exp[W^U(x,z)]/dx>0$ and $d(W^s(x)-R)/dx>0$.

We now need to consider the case in which $W^s(x)>R$. Here we once again obtain that
\[
\frac{d(W^s(x)-R)}{dx}=(1-\gamma)\frac{d\Exp[W^U(x,z)]}{dx},
\]
where
\begin{align}
(1-\beta)\frac{d\Exp[W^U(x,z)]}{dx}=\int_{z^s(x)} \beta \lambda(\theta(x,z))\frac{d(M(x,z)-W^U(x,z))}{dx}dF(z) \label{e:l7_x_case2_1}
\end{align}
and
\begin{align}
  \frac{d(M(x,z)-W^U(x,z))}{dx}&=(1-\gamma)\frac{d\Exp[M(x,z)-W^U(x,z)]}{dx}+(1-\gamma)(\Exp[z]-z) \non \\
  &\qquad \qquad +\gamma(z +\beta (1-\lambda(\theta(x,z))))\left(\frac{d(M(x,z)-W^U(x,z))}{dx}\right),\label{e:l7_x_case2_2}
  \end{align}
while the expected surplus evolves according to
\begin{align}\label{e:l7_x_case2_3}
  \frac{d\Exp[M(x,z)-W^U(x,z)]}{dx}&= \int_{z^s(x)}z +\beta (1-\lambda(\theta(x,z)))\frac{d(M(x,z)-W^U(x,z))}{dx}.
\end{align}
Substituting \eqref{e:l7_x_case2_2} into \eqref{e:l7_x_case2_3}, it follows that $\frac{d\Exp[M(x,z)-W^U(x,z)]}{dx}>0$, from which in turn it follows that \eqref{e:l7_x_case2_1} is also positive.

Finally, we investigate the implications of a change in $x$ on $z^s(x)$ and $z^r(x)$. When $z^r(x)>z^s(x)$, these reservation cutoff functions are given by
\begin{align*}
  M(x,z^s(x))-W^s(x)&=0 \\
  \lambda(\theta(x,z^r(x)))(1-\eta)(M(x,z^r(x))-W^U(x,z^r(x)))+(W^s(x)-R)&=0.
\end{align*}
Further we can obtain that
\begin{align*}
  M(x,z^s(x))-W^s(x)&=x z^s(x)-b+\beta (1-\gamma)\Exp[\max\{M(x,z)-W^U(x,z), W^s(x)-R\}]\non \\
  & \qquad \qquad +\beta \gamma (W^s(x)-R) \\
  M(x,z^r(x))-W^s(x)&=x z^r(x)-b+\beta (1-\gamma)\Exp[\max\{M(x,z)-W^U(x,z), W^s-R\}]\non \\
  & \qquad \qquad +\beta \gamma (1-\lambda(\theta(x,z^r(x))(1-\eta)(M(x,z^r(x))-W^s(x)).
\end{align*}
Taking derivatives with respect to $x$ we find that
{\small
\begin{align*}
&  z^s(x)+\beta(1-\gamma) \frac{d}{dx}\Big(\Exp[\max\{M(x,z)-W^U(x,z), W^s(x)-R\}]\Big)+ \beta \gamma \frac{d(W^s(x)-R)}{dx}+ x\frac{d z^s(x)}{dx}=0 \\
& \frac{\lambda(\theta(x,z))}{1-\beta \gamma (1-\lambda(\theta(x,z)))}\bigg(z^r(x)+\beta(1-\gamma) \frac{d}{dx}\Big(\Exp[\max\{M(x,z)-W^U(x,z), W^s(x)-R\}]\Big)+x\frac{dz^r(x)}{dx}\bigg) \non \\
& \qquad + \frac{d(W^s(x)-R)}{dx}=0
\end{align*}}
Since $\frac{d(W^s(x)-R)}{dx}>0$, this implies that
\begin{align}
z^s(x)+x\frac{d z^s(x)}{dx}\geq z^r(x)+x\frac{d z^r(x)}{dx}+ \frac{1-\beta \gamma }{\lambda(\theta(x,z))}\frac{d(W^s(x)-R)}{dx},
\end{align}
which implies that, evaluated at $x=1$,
\[\frac{d z^s(x)}{dx}-\frac{d z^r(x)}{dx}>z^r(x)-z^s(x).\]
The above result then yields that for $z^r>z^s$, more occupational human capital brings closer together the two cutoffs. For $z^r<z^s$, it holds in this simplified setting that $z^r$ jumps to the corner, $z^r=\underline{z}$, while $z^s$ decreases with $x$. This completes the proof of Lemma 2.

\paragraph{Proof of Lemma 3}

To proof this lemma we use the equations \eqref{a:wu} - \eqref{a:m}, where we have assumed no human capital accumulation. Further, to simplify we let $\gamma=1$ such that the $z$-productivity does not change. We also focus on the case in which $z^r>z^s$ such that $z^r, z^s\in(\underline{z},\bar{z})$ and without loss of generality let $\delta=0$. In this stationary environment, described by $A$ and $z$, note that at labor markets whose $z$-productivities equal $z^{r}$ it holds that
\begin{equation}
\int_{\underline{z}}^{\bar{z}}W^{U}(A,z)dF(z)-c=W^{U}(A,z^{r})+\lambda(\theta (A,z^{r}))(W^{E}(A,z^{r})-W^{U}(A,z^{r})).  \label{delay resqual}
\end{equation}
Further, the expected value of unemployment for workers with $z<z^{r}$ is given by $W^{U}(A,z)=W^{U}(A,z^{r})$. This follows since over this range of $z$'s,
 \[
 \int_{\underline{z}}^{\bar{z}}W^{U}(A,z)dF(z)-c\geq W^{U}(A,z)+\lambda(\theta (A,z))(W^{E}(A,z)-W^{U}(A,z)))
 \]
and unemployed workers prefer change occupations the period after arrival. On the other hand, the value of unemployment for workers with $z\geq z^{r}$ is given by
\begin{equation*}
W^{U}(A,z)=\frac{b+\beta \lambda (\theta (A,z))(W^{E}(A,z)-W^{U}(A,z))}{1-\beta }.
\end{equation*}%
Equation (\ref{delay resqual}) can then be expressed as
\begin{align*}
\beta & \int_{\underline{z}}^{\bar{z}}\Big(\max \{\lambda (\theta(A,z))(W^{E}(A,z)-W^{U}(A,z)),\lambda (\theta(A,z^{r}))(W^{E}(A,z^{r})-W^{U}(A,z^{r}))\}\Big)dF(z) \\
& \quad \hspace{1cm}=\lambda (\theta(A,z^{r}))(W^{E}(A,z^{r})-W^{U}(A,z^{r}))+c(1-\beta ).
\end{align*}%
Using $\eta \lambda(\theta(A,z))(W^{E}(A,z)-W^{U}(A,z))=(1-\eta )\lambda(\theta(A,z))J(A,z)=(1-\eta )\theta(A,z)k$, we have that $R(A)=W^U(A, z^r(A))$ can be expressed as
\begin{equation}
\frac{(1-\eta )k}{\eta }\Bigg(\beta \int_{\underline{z}}^{\bar{z}}\max\{\theta (A,z),\theta (A,z^{r})\}dF(z)\Bigg)-c(1-\beta )=\frac{(1-\eta )k}{\eta }\theta (A,z^{r}),  \label{res eq thetas discounted case}
\end{equation}%
where the LHS describes the net benefit of moving to a different occupation and the RHS the benefit of staying in the same occupation. With this derivation we now analyse under what conditions $dz^{r}/dA>0$ and compare it to the competitive case.

To obtain the $dz^{r}/dA$ from \eqref{res eq thetas discounted case} we first use the free-entry condition and the Cobb-Douglas specification for the matching function to obtain an implicit function that solves for $\theta$,
\begin{equation*}
\theta (A,z)^{\eta -1}\frac{\eta (y(A,z)-b)-\beta (1-\eta )\theta (A,z)k}{1-\beta}-k\equiv E(\theta ;A,z)=0,
\end{equation*}%
where differentiation then implies that $\theta $ is increasing in both $A$ and $z$,
\begin{equation*}
\frac{d\theta (A,z)}{dj}=\frac{\theta (A,z)}{w(A,z)-b}\frac{dy_{j}(A,z)}{dj}, \ \text{ for } j=A,z,
\end{equation*}%
and it is straightforward to show that in this stationary environment the wage equation is given by
\begin{align*}
w(A,z)=(1-\eta )y(A,z)+\eta b+\beta (1-\eta )\theta (A,z)k.
\end{align*}

Next, to make precise the comparison with an economy in which occupations are segmented in many competitive labor markets, consider the same environment as above, with the exception that workers can match instantly with firms. As before, we assume free entry (without vacancy costs), and constant returns to scale production. This implies that every worker will earn his marginal product $y(A,z)$. Importantly, we keep the reallocation frictions the same: workers who change occupations have to forgo production for a period, and arrive at a random labor market in a different occupation at the end of the period. In the simple case of permanent productivity $(A,z)$, the value of being in a labor market with $z$, conditional on $y(A,z)>b$, is $W^{c}(A,z)=y(A,z)/(1-\beta)$, where to simplify we have not considered job destruction shocks.

Block recursiveness, given the free entry condition, is preserved and decisions are only functions of $(A,z)$. Unemployed workers optimally choose to change occupations, and the optimal policy is a reservation quality, $z^r_c$, characterised by the following equation
\begin{equation*}
\beta \int \max \{y(A,z), y(A,z^r_c)\} dF(z) + (b-c)(1-\beta) =y(A,z^r_c).
\end{equation*}
The LHS describes the net benefit of switching occupations, while the RHS the value of of staying employed earning $y$ in the (reservation) labor market.

Rearranging the above equations, the reservation $z$-productivities for the competitive and frictional case satisfy, respectively,
\begin{align*}
&b+ \beta \int_{\underline{z}}^{\bar{z}} \frac{\max \{y(A,z),y(A,z^r_c)\}}{1-\beta}dF(z)-\frac{y(A,z^r_c)}{1-\beta}-c_{c}=0 \\
&\frac{(1-\eta)k}{\eta} \left(\beta \int_{\underline{z}}^{\bar{z}} \frac{\max \{\theta(A,z), \theta(A,z^r)\}}{1-\beta}dF(z)-\frac{\theta(A,z^r)}{1-\beta}\right)-c_{s}=0.
\end{align*}
Using these equations, the response of the reservation $z$-productivity, for the competitive, and the frictional case is then given by
\begin{align*}
\frac{d z^r_c}{d A}& = \frac{\beta F(z^r_c)\frac{y_A(A,z^r_c)}{y_z(A,z^r_c)}+\beta \int_{z^r_c}^{\bar{z}} \frac{y_A(A,z)}{y_z(A,z^r_c)}dF(z)-\frac{y_A(A,z^r_c)}{y_z(A,z^r_c)}} {1-\beta F(z^r_c)} \\
\frac{d z^r}{d A}& = \frac{\beta F(z^r)\frac{y_A(A,z^r)}{y_z(A,z^r)}+\beta \int_{z^r}^{\bar{z}} \frac{\theta(A,z)(w(A,z^r)-b)}{\theta(A,z^r) (w(A,z)-b)} \frac{y_A(A,z)}{y_z(A,z^r)}dF(z)-\frac{y_A(A,z^r)}{y_z(A,z^r)}} {1-\beta F(z^r)}
\end{align*}
These are the expression shown in Lemma 3, where we have used the fact that $\frac{\theta(A,z)}{(w(A,z)-b)}\frac{(1-\eta)k}{\eta}=\frac{\lambda(\theta(A,z))}{1-\beta+\beta \lambda(\theta(A,z))}$ by virtue of
\[
\eta \frac{w(A,z)-b}{1-\beta+\beta \lambda (\theta (A,z))}=\frac{(1-\eta )k}{q(\theta (p,z))},
\]
which follows from the combination of the free entry condition and the Hosios. This completes the proof of Lemma 3.

\paragraph{Implications of Lemma 3}

We now show two implications of Lemma 3. First we show that search frictions adds a procyclical force to occupational mobility decisions. Choosing $c_{c}, c_{s}$ appropriately such that $z^r_{c}=z^r$, the above expressions imply that $\frac{d z^r}{d A}>\frac{d z^r_c}{d A}$ if $\frac{\theta(A,z)}{w(A,z)-b}>\frac{\theta(A,z^r)}{w(A,z^r)-b}, \ \ \forall \ z>z^r$. Hence we now need to show that $\frac{\theta(A,z)}{w(A,z)-b}$ is increasing in $z$.
\begin{equation*}  \label{p2:eq2}
\frac{d\left(\frac{\theta(A,z)}{w(A,z)-b}\right)}{dz}=\frac{\theta y_z(A, z)}{(w(A,z)-b)^2}-\theta\left(\frac{(1-\eta)+(1-\eta)\beta \frac{\theta}{w(A,z)-b}k}{(w(A,z)-b)^2}\right) y_z(A,z),
\end{equation*}
which has the same sign as $\eta-(1-\eta)\beta k \frac{\theta}{w(A,z)-b}$ and the same sign as
\begin{align*}
\eta(1-\eta)&(y(A,z)-b)+\eta(1-\eta)\beta \theta k -(1-\eta)\beta \theta k
\notag \\
& \hspace{1in}= (1-\eta)(\eta(y(A,z)-b)-(1-\eta)\beta \theta k).
\end{align*}
But $\eta(y(A,z)-b)-(1-\eta)\beta \theta k=y(A,z)-w(A,z)>0$ and we have established that search frictions within labor markets make occupational mobility decisions more procyclical relative to the competitive benchmark, given the same $F(z)$ and the same initial reservation productivity $z^r=z^r_c$.

Second, we show that impact of the production function on the procyclicality of occupational mobility decisions. Here we want to show that with search frictions, if the production function is modular or supermodular (i.e. $y_{Az}\geq 0$), there exists a $c \geq 0$ under which occupational mobility decisions are procyclical. With competitive markets, if the production function is modular, occupational mobility decisions are countercyclical, for any $\beta<1$ and $c \geq 0$.

Note that modularity implies that $y_A(A,z)=y_A(A,\tilde{z}), \ \forall z>\tilde{z}$; while supermodularity implies $y_A(A,z) \geq y_A(A,\tilde{z}), \ \forall z>\tilde{z}$. Hence modularity implies
\begin{equation*}
\frac{d z^r_c}{d A}=\frac{1}{1-\beta F(z^r_c)}\frac{y_A(A,z^r_c)}{y_z(A,z^r_c)} \left(\beta F(z^r_c)+\beta \int_{z^r_c}^{\bar{z}} \frac{y_A(A,z)}{y_A(A,z^r_c)}dF(z)-1\right)<0, \ \forall \ \beta<1.
\end{equation*}
In the case with frictions,
\begin{equation*}
\frac{d z^r}{d A}=\frac{1}{1-\beta F(z^r)}\frac{y_A(A,z^r)}{y_z(A,z^r)}\left(\beta F(z^r)+\beta \int_{z^r}^{\bar{z}} \frac{\theta(A,z)(w(A,z^r)-b)}{\theta(A,z^r) (w(A,z)-b)} \frac{y_A(A,z)}{y_A(A,z^r)}dF(z)-1\right).
\end{equation*}
If we can show that the integral becomes large enough, for $c$ large enough, to dominate the other terms, we have established the claim. First note that $\frac{y_A(A,z)}{y_A(A,z^r)}$ is weakly larger than 1, for $z>z^r$ by the (super)modularity of the production function. Next consider the term $\frac{\theta(A,z)(w(A,z^r)-b)}{\theta(A,z^r) (w(A,z)-b)} $. Note that
\begin{equation*}
\lim_{z \downarrow y^{-1}(b;A)} \frac{\theta(A,z)}{w(A,z)-b}=\frac{\lambda(\theta(A,z))}{1-\beta+\beta \lambda(\theta(A,z))}=0,
\end{equation*}
because $\theta(A,z)\downarrow 0$, as $y(A,z^r) \downarrow b$. Hence, fixing a $z$ such that $y(A,z)>b$, $\frac{\theta(A,z)(w(A,z^r)-b)}{\theta(A,z^r) (w(A,z)-b)}\to \infty, \text{ as } y(a,z^r) \downarrow b $. Since this holds for any $z$ over which is integrated, the integral term becomes unboundedly large, making $d z^r/dA$ strictly positive if reservation $z^r$ is low enough. Since the integral rises continuously but slower in $z^r$ than the also continuous term $\frac{\theta(A,z^r)}{1-\beta}$, it can be readily be established that $z^r$ depends continuously on $c$, and strictly negatively so as long as $y(A,z^r)>b$ and $F(z)$ has full support. Moreover, for some $\bar{c}$ large enough, $y(A,\underline{z}^r)=b$, where $\underline{z}^r$ is a lower bound for $z^r$. Hence, as $c \uparrow \underline{z}^r$, $dz^r/dA>0$.

\paragraph{Job Separations} Here we show the derivation of the slope of $z^s$ for the case $z^r(A)>z^s(A)$ for all $A$ described in Section C.1.2. Note that $R(A)=\frac{b+\beta \theta(A,z^r(A)) k (1-\eta)/\eta}{1-\beta}$. The derivative of this function with respect to $A$ equals
\begin{align}  \label{z^s lemma 1}
\frac{\beta k(1-\eta)}{(1-\beta)\eta}\frac{\theta}{w(A, z^r(A))-b}\left(y_A(A, z^r(A))+y_z(A, z^r(A))\frac{dz^r(A)}{dA}\right).
\end{align}
Since $w(A, z^r(A))-b=(W^E(A, z^r(A))-W^U(A,z^r(A)))(1-\beta(1-\delta)+\beta \lambda (\theta(A, z^r(A))))$ and $\frac{\theta \beta k (1-\eta)}{(1-\beta)\eta}=\beta \lambda (\theta(A,z^r(A)))(W^E(A, z^r(A))-W^U(A, z^r(A))$, we find that \eqref{z^s lemma 1} reduces to
\begin{align}  \label{z^s lemma 2}
\frac{\beta \lambda(\theta(A, z^r(A)))}{1-\beta(1-\delta)+\beta \lambda(\theta(A, z^r(A))}\left(y_A(A, z^r(A))+y_z(A, z^r(A))\frac{dz^r(A)}{dA}%
\right).
\end{align}
From the cutoff condition for separation, we find $(1-\beta)R(A)=y(A,z^{s}(A))$. Taking the derivative with respect to $A$ implies the left side equals \eqref{z^s lemma 2} and the right side equals $y_{A}(A,z^{s}(A))+y_{z}(A,z^{s}(A))\frac{dz^{s}(A)}{dA}$. Rearranging yields the equation in Section C.1.2.

\subsection{Proofs of Competitive Search Model}

\paragraph{Proof of Lemma 4}

Fix any occupation $o$ and consider a firm that promised $W\geq W^{U}(A,z)$ to the worker with productivity $z$, delivers this value in such a way that his profit  $J(A,z,W)$ is maximized, i.e. solving \eqref{filledjob}. Now consider an alternative offer $\hat{W}\neq W$, which is also acceptable to the unemployed worker, and likewise maximizes the profit given $\hat{W}$ for the firm, $J(A,z, \hat{W})$. Then an alternative policy that delivers $W$ by using the optimal policy for $\hat{W}$, but transfers additionally $W-\hat{W}$ to the worker in the first period must be weakly less optimal, which using the risk neutrality of the worker, results in
\begin{equation*}
J(A,z,W)\geq J(A,z,\hat{W})-(W-\hat{W})
\end{equation*}%
Likewise, an analogue reasoning implies $J(A,z,\hat{W})\geq J(A,z,W)-(\hat{W}-W)$, which together with the previous equation implies
\begin{equation*}
J(A,z,W)\geq J(A,z,\hat{W})-(W-\hat{W})\geq J(A,z,W)-(\hat{W}-W)-(W-\hat{W}),
\end{equation*}%
and hence it must be that $J(A,z,W)= J(A,z,\hat{W})-(W-\hat{W})$, for \emph{all} $M(A,z)\geq W,\hat{W}\geq W^{U}$. Differentiability of $J$ with slope -1 follows immediately. Moreover, $M(A,z,W)=W+J(A,z,\hat{W})+\hat{W}-W=M(A,z,\hat{W})\equiv M(A,z)$. Finally, if $W^{\prime }(A^{\prime},z^{\prime}) < W^{U}(A^{\prime},z^{\prime})$ is offered tomorrow while $M(A^{\prime},z^{\prime})>W^{U}(A^{\prime},z^{\prime})$, it is a profitable deviation to offer $W^{U}(A^{\prime},z^{\prime})$, since $M(A^{\prime},z^{\prime})-W^{U}(A^{\prime},z^{\prime})=J(A^{\prime},z^{\prime},W^{U}(A^{\prime},z^{\prime}))>0$ is feasible. This completes the proof of Lemma 4.

\paragraph{Proof of Lemma 5}

Fix any occupation $o$ and consider productivity $z$, such that $M(A,z)-W^U(A,z)>0$. Since we confine ourselves to this productivity, with known continuation values $J(A,z,W)$ and $W^{U}(A,z)$ in the production stage, we drop the dependence on $A,z$ for ease of notation. Free entry implies $k=q(\theta)J(W)\Rightarrow \frac{dW}{d\theta }<0$. Notice that it follows that the maximand of workers in \eqref{e:application_max}, subject to \eqref{unfilled job fe} is continuous in $W$, and provided $M>W^{U}$, has a zero at $W=M$ and at $W=W^{U}$, and a strictly positive value for intermediate $W$: hence the problem has an interior maximum on $[W^{U},M]$. What remains to be shown is that the first order conditions are sufficient for the maximum, and the set of maximizers is singular.

Solving the worker's problem of posting an optimal value subject to tightness implied by the free entry condition yields the following first order conditions (with multiplier $\mu$):
\begin{align*}
\lambda^{\prime}(\theta)[W-W^U]-\mu q^{\prime}(\theta)J(W)&=0 \\
\lambda(\theta)-\mu q(\theta) J^{\prime}(W) &=0 \\
k-q(\theta)J(W)&=0
\end{align*}
Using the constant returns to scale property of the matching function, one has $q(\theta)=\lambda (\theta)/\theta$. This implies, combining the three equations above, to solve out $\mu$ and $J(W)$,
\begin{equation*}  \label{e:Gfunction lemma}
0=\lambda ^{\prime}(\theta)[W(\theta)-W^U]+\frac{\theta q^{\prime}(\theta)}{%
q(\theta)}k\equiv G(\theta),
\end{equation*}
where we have written $W$ as a function of $\theta$, as implied by the free entry condition. Then, one can derive $G^{\prime}(\theta)$ as
\begin{equation*}
G^{\prime}(\theta)=\lambda ^{\prime\prime}(\theta)[W(\theta)-W^U]+\lambda
^{\prime}(\theta)W^{\prime}(\theta)+ \frac{d\varepsilon_{q,\theta}(\theta)}{%
d\theta},
\end{equation*}
where $\varepsilon_{q,\theta}(\theta)$ denotes the elasticity of the vacancy filling rate with respect to $\theta$ and
\begin{equation*}
\frac{d\varepsilon_{q,\theta}(\theta)}{d\theta}=\frac{q^{\prime}(\theta)k}{%
q(\theta)}+\frac{\theta[q^{\prime\prime}(\theta)q(\theta)-q^{\prime}(%
\theta)^2] k}{q(\theta)^2}.
\end{equation*}
Since the first two terms in the RHS are strictly negative, $G^{\prime}$ is strictly negative when $\varepsilon_{q,\theta}(\theta)\leq 0$. The latter then guarantees there is a unique $\tilde{W}_f$ and corresponding $\theta$ that maximizes the worker's problem. This completes the proof of Lemma 5.

\end{document}